\documentclass[12pt]{caltech_thesis}
\pdfoutput=1
\usepackage[hyphens]{url}
\usepackage{graphicx}
\usepackage{todonotes}

\usepackage[utf8]{inputenc}
\usepackage[T1]{fontenc}
\usepackage{amsthm,amsfonts}
\usepackage{newtxtext,newtxmath}

\usepackage[numbers,sort&compress]{natbib}
\renewcommand{\bibsection}{}
\usepackage{bibunits}
\defaultbibliographystyle{h-physrev}
\usepackage{amsmath,verbatim,mathrsfs,mathtools,bbm,bm,url,cancel,subcaption,indentfirst,newfloat}
\usepackage[shortlabels]{enumitem}
\usepackage[font=small,labelfont=bf,format=plain,justification=raggedright,singlelinecheck=false]{caption}
\usepackage{doi}

\usepackage{graphicx}
\usepackage{epstopdf}
\newcommand{\caphead}[1]{{\bf #1}}

\usepackage{tikz}
\usepackage{pgfplots}
\pgfplotsset{compat=1.3}
\pgfplotsset{width=0.5*\textwidth}

\newtheorem{theorem}{Theorem}
\newtheorem{lemma}{Lemma}
\newtheorem{corollary}{Corollary}
\newtheorem{definition}{Definition}
\newtheorem{example}{Example}
\newtheorem{property}{Property}

\newtheorem{proposition}{Proposition}

\makeatletter
\newcommand\footnoteref[1]{\protected@xdef\@thefnmark{\ref{#1}}\@footnotemark}
\makeatother

\usepackage{soul}

\usepackage{color}

\newcommand{\Min}{ {\text{min} } }   
\newcommand{\Max}{ {\text{max} } }   
\newcommand{\inter}{ {\text{int} } }   
\newcommand{\hc}{ {\text{h.c.} } }
\newcommand{\cc}{ {\text{c.c.} } }
\newcommand{\tot}{ {\text{tot}} }
\def\const{ {\text{const.}} }   
\newcommand{\Tr}{{\text{Tr}}}   
\newcommand{\id}{\mathbbm{1}}
\newcommand{\kB}{k_\mathrm{B}}  

\newcommand{\worst}{\mathrm{worst}}
\newcommand{\cost}{\mathrm{cost}}

\newcommand{\Hil}{\mathcal{H}}  
\newcommand{\Basis}{\mathcal{B}}  
\newcommand{\0}{ {(0)} }
\newcommand{\1}{ {(1)} }
\newcommand{\2}{ {(2)} }
\newcommand{\3}{ {(3)} }
\newcommand{\LParen}{ \bm{(} }
\newcommand{\RParen}{ \bm{)} }
\newcommand*{\Set}[1]{\left\{  #1  \right\}}
\gdef\rm#1{\text{#1}}
\gdef\bf#1{#1}

\renewcommand\th{ {\text{th}} }

\newcommand*{\bra}[1]{\langle #1\rvert}
\newcommand*{\ket}[1]{\lvert #1 \rangle}
\newcommand*{\braket}[2]{\langle #1 \lvert #2 \rangle}
\newcommand*{\ketbra}[2]{\lvert #1 \rangle\!\langle #2 \rvert}
\newcommand*{\expval}[1]{\left\langle  #1  \right\rangle}

\DeclareFloatingEnvironment[name={Supplementary Figure}]{suppfigure}
\DeclareFloatingEnvironment[name={Figure}]{mainfigure}
\DeclareUnicodeCharacter{00A0}{ }

\begin{document}

\title{Quantum steampunk: \\ Quantum information, thermodynamics,
their intersection, and applications thereof across physics}
\author{Nicole Yunger Halpern}

\degreeaward{Doctor of Philosophy in Physics}                 
\university{California Institute of Technology}    
\address{Pasadena, California}                     
\unilogo{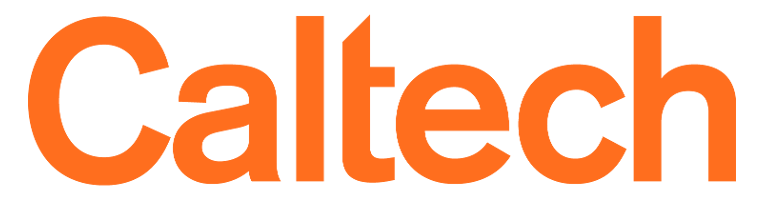}                                 
\copyyear{2018}  
\defenddate{May 25, 2018}          

\orcid{0000-0001-8670-6212}

\rightsstatement{All rights reserved except where otherwise noted}

\maketitle[logo]

%
%
\begin{acknowledgements}

I am grateful to need to acknowledge many contributors.
I thank my parents for the unconditional support and love,
and for the sacrifices, that enabled me to arrive here.
Thank you for communicating values that include 
diligence, discipline, love of education, and security in one's identity.
For a role model who embodies these virtues,
I thank my brother.

I thank my advisor, John Preskill, for your time, for mentorship,
for the communication of scientific values and scientific playfulness, 
and for investing in me.
I have deeply appreciated the time and opportunity that you've provided
to learn and create.
Advice about ``thinking big''; taking risks; prioritizing; 
embracing breadth and exhibiting nimbleness in research;
and asking, ``Are you having fun?''
will remain etched in me.
Thank you for bringing me to Caltech.

Thank you to my southern-California family for welcoming me into your homes
and for sharing holidays and lunches with me.
You've warmed the past five years.

The past five years have seen the passing of both my grandmothers:
Dr. Rosa Halpern during year one
and Mrs. Miriam Yunger during year four.
Rosa Halpern worked as a pediatrician until in her 80s.
Miriam Yunger yearned to attend college but lacked the opportunity.
She educated herself, to the point of erudition on Russian and American history,
and amassed a library.
I'm grateful for these role models who shared 
their industriousness, curiosity, and love.

I'm grateful to my research collaborators 
for sharing time, consideration, and expertise:
Ning Bao,
Daniel Braun,
Lincoln Carr,
Mahn-Soo Choi,
Elizabeth Crosson,
Oscar Dahlsten,
Justin Dressel,
Philippe Faist,
Andrew Garner,
Jos\'e Ra\'ul Gonzalez Alonso,
Sarang Gopalakrishnan,
Logan Hillberry,
Andrew Keller,
Chris Jarzynski,
Jonathan Oppenheim,
Patrick Rall,
Gil Refael,
Joe Renes,
Brian Swingle,
Vlatko Vedral,
Mordecai Waegell,
Sara Walker,
Christopher White, and
Andreas Winter.
I'm grateful to informal advisors for sharing experiences and guidance:
Michael Beverland,
Sean Carroll,
Ian Durham, 
Alexey Gorshkov, 
Daniel Harlow, 
Jim Slinkman,
Dave Kaiser, 
Shaun Maguire, 
Spiros Michalakis, 
Jenia Mozgunov, 
Renato Renner, 
Barry Sanders, 
many of my research collaborators,
and many other colleagues and peers.
Learning and laughing with my quantum-information/-thermodynamics colleagues 
has been a pleasure and a privilege:
\'Alvaro Mart\'in Alhambra,
L\'idia del R\'io,
John Goold,
David Jennings,
Matteo Lostaglio,
Nelly Ng,
Mischa Woods,
aforementioned collaborators,
and many others.

I'm grateful to Caltech's Institute for Quantum Information and Matter (IQIM)
for conversations, collaborations, financial support, 
an academic and personal home, and more.
I thank especially
Fernando Brand\~{a}o, 
Xie Chen, 
Manuel Endres,
David Gosset,
Stacey Jeffery,
Alexei Kitaev,
Alex Kubica,
Roger Mong,
Oskar Painter, 
Fernando Pastawski,
Kristan Temme,
and the aforementioned IQIM members.
Thanks to my administrators for logistical assistance,
for further logistical assistance,
for hallway conversations that counterbalanced the rigors of academic life,
for your belief in me,
and for more logistical assistance:
Marcia Brown, 
Loly Ekmekjian, 
Ann Harvey, 
Bonnie Leung, 
Jackie O'Sullivan, and
Lisa Stewart.

For more such conversations,
and for weekend lunches in the sun on Beckman Lawn,
I'm grateful to too many friends to name.
Thank you for your camaraderie, candidness, and sincerity.
Also too many to name are the mentors and teachers
I encountered before arriving at Caltech.
I recall your guidance and encouragement more often than you realize.

Time ranks amongst the most valuable resources a theorist can hope for.
I deeply appreciate the financial support 
that has offered freedom to focus on research.
Thanks to
Caltech's Graduate Office; 
the IQIM; 
the Walter Burke Institute;
the Kavli Institute for Theoretical Physics (KITP); 
and Caltech's Division of Physics, Mathematics, and Astronomy for 
a Virginia Gilloon Fellowship,
an IQIM Fellowship,
a Walter Burke Graduate Fellowship,
a KITP Graduate Fellowship, and
a Barbara Groce Fellowship.
Thanks to John Preskill and Gil Refael for help with securing funding.
Thanks to many others (especially the Foundational Questions Institute's Large Grant for "Time and the Structure of Quantum Theory", Jon Barrett, and Oscar Dahlsten) for financial support for research visits.
NSF grants PHY-0803371, PHY-1125565, and PHY-1125915
have supported this research.
The IQIM is an NSF Physics Frontiers Center with support from 
the Gordon and Betty Moore Foundation (GBMF-2644).

\end{acknowledgements}

%
%
\begin{abstract}
Combining quantum information theory with thermodynamics unites 21st-century technology with 19th-century principles. The union elucidates the spread of information, the flow of time, and the leveraging of energy. This thesis contributes to the theory of quantum thermodynamics, particularly to quantum-information-theoretic thermodynamics. The thesis also contains applications of the theory, wielded as a toolkit, across physics. Fields touched on include atomic, molecular, and optical physics; nonequilibrium statistical mechanics; condensed matter; high-energy physics; and chemistry. I propose the name \emph{quantum steampunk} for this program. The term derives from the steampunk genre of literature, art, and cinema that juxtaposes futuristic technologies with 19th-century settings.

\end{abstract}

\extrachapter{Published Content}



The following publications form the basis for this thesis.
The multi-author papers resulted from collaborations
to which all parties contributed equally.

\begin{publishedcontent}[iknowwhattodo]
\nocite{NYH_15_Introducing,NYH_16_Beyond,NYH_18_Beyond,Bao_17_Quantum,Dahlsten_17_Entropic,NYH_15_What's,NYH_17_Toward,NYH_16_Number,NYH_16_Micro,NYH_17_Jarzynski,NYH_17_Quasi,NYH_17_MBL,NYH_17_Quantum,Swingle_18_Resilience,Dressel_18_Strengthening}
\putbib[OwnPubs]
\end{publishedcontent}

\tableofcontents

%
%
\mainmatter

%
%
\chapter{Introduction}
\label{section:Intro}
\begin{bibunit}

The steampunk movement has invaded
literature, film, and art over the past three decades.\footnote{
\emph{Parts this introduction were adapted from~\cite{NYH_13_Steampunk,NYH_16_Cal,NYH_16_Bringing}.
}}
Futuristic technologies mingle, in steampunk works,
with Victorian and wild-west settings.
Top hats, nascent factories, and grimy cities
counterbalance time machines, airships, and automata.
The genre 
arguably originated in 1895, with the H.G. Wells novel \emph{The Time Machine}.
Recent steampunk books include
the best-selling \emph{The Invention of Hugo Cabret};
films include the major motion picture \emph{Wild Wild West};
and artwork ranges from painting to jewelry to sculpture.

Steampunk captures the romanticism of fusing the old with the cutting-edge.
Technologies proliferated during the Victorian era:
locomotives, Charles Babbage's analytical engine, factories, and more.
Innovation facilitated exploration.
Add time machines, and the spirit of adventure sweeps you away.
Little wonder that fans flock to steampunk conventions, 
decked out in overcoats, cravats, and goggles.

What steampunk fans dream,
quantum-information thermodynamicists live.

Thermodynamics budded during the late 1800s,
when steam engines drove the Industrial Revolution.
Sadi Carnot, Ludwig Boltzmann, and other thinkers
wondered how efficiently engines could operate.
Their practical questions led to fundamental insights---about
why time flows;
how much one can know about a physical system;
and how simple macroscopic properties, like temperature,
can capture complex behaviors, like collisions by steam particles.
An idealization of steam---the classical ideal gas---exemplifies 
the conventional thermodynamic system.
Such systems contain many particles, behave classically,
and are often assumed to remain in equilibrium.

But thermodynamic concepts---such as heat, work, and equilibrium---characterize 
small scales, quantum systems, and out-of-equilibrium processes.
Today's experimentalists probe these settings,
stretching single DNA strands with optical tweezers~\cite{MossaMFHR09},
cooling superconducting qubits to build quantum computers~\cite{Gambetta_17_Building,Neill_17_Blueprint},
and extracting work from single-electron boxes~\cite{Saira12}.
These settings demand reconciliation with 19th-century thermodynamics.
We need a toolkit for fusing the old with the new.

Quantum information (QI) theory provides such a toolkit.
Quantum phenomena serve as resources for processing information
in ways impossible with classical systems.
Quantum computers can solve 
certain computationally difficult problems quickly;
quantum teleportation transmits information as telephones cannot;
quantum cryptography secures messages;
and quantum metrology centers on high-precision measurements.
These applications rely on entanglement
(strong correlations between quantum systems), 
disturbances by measurements, quantum uncertainty, 
and discreteness.

Technological promise has driven fundamental insights,
as in thermodynamics.
QI theory has blossomed into a mathematical toolkit that includes
entropies, uncertainty relations, and resource theories.
These tools are reshaping fundamental science,
in applications across physics, computer science, and chemistry. 

QI is being used to update thermodynamics,
in the field of \emph{quantum thermodynamics} (QT)~\cite{Goold_16_Role,Vinjanampathy_16_Quantum}.
QT features entropies suited to small scales; quantum engines; 
the roles of coherence in thermalization and transport;
and the transduction of information into work,
\`{a} la Maxwell's demon~\cite{MaruyamaNV09}.

This thesis (i) contributes to the theory of QI thermodynamics
and (ii) applies the theory, as a toolkit, across physics.
Spheres touched on include 
atomic, molecular, and optical (AMO) physics; 
nonequilibrium statistical mechanics; 
condensed matter; 
chemistry; and high-energy physics.
I propose the name \emph{quantum steampunk} for this program.
The thesis contains samples of 
the research performed during my PhD. 
See~\cite{NYH_15_Introducing,NYH_16_Beyond,NYH_18_Beyond,Bao_17_Quantum,Dahlsten_17_Entropic,NYH_15_What's,NYH_17_Toward,NYH_16_Number,NYH_16_Micro,NYH_17_Jarzynski,NYH_17_Quasi,NYH_17_MBL,NYH_17_Quantum,Swingle_18_Resilience} for a complete catalog.

Three vertebrae form this research statement's backbone.
I overview the contributions here;
see the chapters for more context, including related literature.
First, the \emph{out-of-time-ordered correlator} 
signals the scrambling of information  
in quantum many-body systems that thermalize internally.
Second, athermal systems serve as resources in thermodynamic tasks,
such as work extraction and information storage.
Examples include \emph{many-body-localized} systems,
for which collaborators and I designed a quantum many-body engine cycle.
Third, consider a small quantum system thermalizing with a bath.
The systems could exchange quantities,
analogous to heat and particles,
that fail to commute with each other.
The small system would approach a \emph{non-Abelian thermal state}.

Related PhD research is mentioned where relevant.
One paper has little relevance to thermodynamics,
so I will mention it here: 
\emph{Quantum voting} illustrates the power of nonclassical resources,
in the spirit of quantum game theory, through elections~\cite{Bao_17_Quantum}.
%

%
%
%
\paragraph{Information scrambling and quantum thermalization:}
Chaotic evolution scrambles information stored in quantum many-body systems,
such as spin chains and black holes.
QI spreads throughout many degrees of freedom via entanglement.
The \emph{out-of-time-ordered correlator} (OTOC) 
registers this spread---loosely speaking,
the equilibration of QI~\cite{Kitaev_15_Simple}.

Chaos and information scrambling smack of
time's arrow and the second law of thermodynamics.
So do \emph{fluctuation relations} in nonequilibrium statistical mechanics.
The best-known fluctuation relations include Jarzynski's equality,
$\langle e^{ - \beta W } \rangle = e^{ - \beta \Delta F }$~\cite{Jarzynski97}.
$W$ represents the work required to perform a protocol,
such as pushing an electron onto a charged island in a circuit~\cite{Saira_12_Test}.
$\langle . \rangle$ denotes an average over nonequilibrium pushing trials;
$\beta$ denotes the inverse temperature at which 
the electron begins;
and $\Delta F$ denotes a difference between equilibrium free energies.
Chemists and biologists use $\Delta F$;
but measuring $\Delta F$ proves difficult.
Jarzynski's equality suggests a measurement scheme:
One measures the work $W$ in each of many finite-time trials
(many pushings of the electron onto the charged island).
One averages $e^{ - \beta W }$ over trials, 
substitutes into the equation's left-hand side,
and solves for $\Delta F$.
Like $\Delta F$, the OTOC is useful but proves difficult to measure.

I developed a fluctuation relation, analogous to Jarzynski's equality,
for the OTOC~\cite{NYH_17_Jarzynski} (Ch.~\ref{ch:Jarz_like}). 
The relation has three significances.
First, the equality unites two disparate, yet similar-in-spirit concepts: 
the OTOC of AMO, condensed matter, and high energy
with fluctuation relations of nonequilibrium statistical mechanics.
Second, the equality suggests a scheme for inferring the OTOC experimentally.
The scheme hinges on \emph{weak measurements},
which fail to disturb the measured system much.
Third, the equality unveils a quantity more fundamental than the OTOC:
a quasiprobability.

\emph{Quasiprobability distributions} represent quantum states
as phase-space densities represent classical statistical-mechanical states.
But quasiprobabilities assume nonclassical values 
(e.g., negative and nonreal values)
that signal nonclassical physics 
(e.g., the capacity for superclassical computation~\cite{Spekkens_08_Negativity}).
Many classes of quasiprobabilities exist.
Examples include the well-known Wigner function
and its obscure little sibling, the \emph{Kirkwood-Dirac (KD) quasiprobability}.

An extension of the KD quasiprobability, I found,
underlies the OTOC~\cite{NYH_17_Jarzynski}.
Collaborators and I characterized this quasiprobability in~\cite{NYH_17_Quasi}
(Ch.~\ref{ch:OTOC_Quasi}).
We generalized KD theory, 
proved mathematical properties of the OTOC quasiprobability,
enhanced the weak-measurement scheme,
and calculated the quasiprobability numerically and analytically in examples.
The quasiprobability, we found, strengthens the parallel between 
OTOCs and chaos: Plots of the quasiprobability bifurcate,
as in classical-chaos pitchfork diagrams. 
QI scrambling, the plots reveal, breaks a symmetry in the quasiprobability.

The Jarzynski-like equality for the OTOC (Ch.~\ref{ch:Jarz_like})
broadens my earlier work on fluctuation relations.
Collaborators and I merged fluctuation relations with two QI toolkits:
\emph{resource theories} (QI-theoretic models, discussed below,
including for thermodynamics) 
and \emph{one-shot information theory}
(a generalization of Shannon theory to small scales)~\cite{NYH_15_Introducing,NYH_15_What's,Dahlsten_17_Entropic}.
We united mathematical tools 
from distinct disciplines, nonequilibrium statistical mechanics and QI.
The union describes small-scale thermodynamics,
such as DNA strands and ion traps.

I applied our results with Christopher Jarzynski~\cite{NYH_16_Number}.
We bounded, in terms of an entropy, 
the number of trials required
to estimate $\Delta F$ with desired precision.
Our work harnesses QI for experiments.

Experimental imperfections can devastate
OTOC-measurement schemes (e.g.,~\cite{Swingle_16_Measuring,Yao_16_Interferometric,Zhu_16_Measurement,NYH_17_Jarzynski,NYH_17_Quasi}).
Many schemes require experimentalists to effectively reverse time,
to negate a Hamiltonian $H$.
An attempted negation could map
$H$  to $-H + \varepsilon$
for some small perturbation $\varepsilon$.
Also, environments can decohere quantum systems.
Brian Swingle and I proposed a scheme
for mitigating such errors~\cite{Swingle_18_Resilience}.
The measured OTOC signal is renormalized by
data from easier-to-implement trials.
The scheme improves the weak-measurement scheme
and other OTOC-measurement schemes~\cite{Swingle_16_Measuring,Yao_16_Interferometric,Zhu_16_Measurement},
for many classes of Hamiltonians.

The weak-measurement scheme was improved alternatively in~\cite{Dressel_18_Strengthening}.
Collaborators and I focused on observables $O_j$ 
that square to the identity operator: $ ( O_j )^2  =  \id$.
Examples include qubit Pauli operators.
Consider time-evolving such an observable in the Heisenberg picture,
forming $O_j (t_j)$.
Define a correlator
$C  =  \expval{ O_1 (t_1) O_2(t_2)  \ldots  O_m(t_m) }$
from $m$ observables. 
$C$ can be inferred from
a sequence of measurements
interspersed with time evolutions.
Each measurement requires an ancilla qubit
coupled to the system locally.
The measurements can be of arbitrary strengths, we showed,
``strengthening'' the weak-measurement protocol.

\paragraph{Athermal states as resources in thermodynamic tasks:
work extraction and information processing} 

\emph{Many-body localization} (MBL) defines a phase of 
quantum many-body systems.
The phase can be realized with
ultracold atoms, trapped ions, and nitrogen-vacancy centers.
MBL behaves athermally:
Consider measuring the positions of MBL particles.
The particles stay fixed for a long time afterward.
For contrast, imagine measuring the positions of equilibrating gas particles.
The particles thereafter random-walk throughout their container.

Athermal systems serve as resources in thermodynamic tasks:
Consider a hot bath in a cool environment.
The hot bath is athermal relative to the atmosphere.
You can connect the hot bath to the cold,
let heat flow, and extract work.
As work has thermodynamic value, so does athermality.

MBL's athermality facilitates thermodynamic tasks,
I argued with collaborators~\cite{NYH_17_MBL}
(Ch.~\ref{ch:MBL_Mobile}).
We illustrated by formulating an engine cycle
for a quantum many-body system.
The engine is tuned between deep MBL and a ``thermal'' regime.
``Thermal'' Hamiltonians exhibit level repulsion:
Any given energy gap has a tiny probability of being small.
Energy levels tend to lie far apart.
MBL energy spectra lack level repulsion.

The athermality of MBL energy spectra
curbs worst-case trials,
in which the engine would output net negative work $W_{\text{tot}} < 0$;
constrains fluctuations in $W_{\text{tot}}$; 
and offers flexibility in choosing the engine's size,
from mesoscale to macroscopic.
We calculated the engine's power and efficiency;
numerically simulated a spin-chain engine; 
estimated diabatic corrections to results,
using adiabatic perturbation theory; 
and modeled interactions with a bosonic bath.

This project opens MBL---a newly characterized phase 
realized recently in experiments---to applications.
Possible applications include engines, energy-storing ratchets, and dielectrics.
These opportunities should point to new physics.
For example, formulating an engine cycle led
us to define and calculate heat and work quantities
that, to our knowledge, had never been defined for MBL.
Just as quantum thermodynamics provided a new lens onto MBL,
MBL fed back on QT.
Quantum states $\rho \neq  e^{ - \beta H } / Z$
are conventionally regarded as athermal resources.
Also gap statistics, we showed, offer athermal tools.

The benefits of athermality may extend to biomolecules.
Matthew Fisher recently proposed that 
\emph{Posner} biomolecules store QI protected from thermalization
for long times~\cite{Fisher15}.
Elizabeth Crosson and I assessed how efficiently 
these molecules could process QI~\cite{NYH_17_Quantum}.
We abstracted out the logical operations
from Fisher's physics, defining
the model of \emph{Posner quantum computation}.
Operations in the model, we showed, 
can be used to teleport QI imperfectly.
We also identified quantum error-detecting codes 
that could flag whether the molecules' QI has degraded.
Additionally, we identified molecular states
that can serve as universal resources in
measurement-based quantum computation~\cite{Raussendorf_03_Measurement}.
Finally, we established a framework for quantifying
Fisher's conjecture that
entanglement can influence molecular-binding rates.
This work opens the door to the QI-theoretic analysis and applications
of Posner molecules.

\paragraph{Non-Abelian thermal state:}
Consider a small quantum system $S$
equilibrating with a bath $B$.
$S$ exchanges quantities, such as heat, with $B$.
Each quantity is conserved globally; so it may be called a \emph{charge}.
If exchanging just heat and particles, $S$ equilibrates to
a grand canonical ensemble $e^{ - \beta (H - \mu N) } / Z$.
$S$ can exchange also electric charge, angular momentum, etc.:
$m$ observables $Q_1, \ldots Q_m$.
Renes and I incorporated thermodynamic exchanges
of commuting quantities
into resource theories~\cite{NYH_16_Beyond,NYH_18_Beyond}.

What if the $Q_j$'s fail to commute? 
Can $S$ thermalize?
What form would the thermal state $\gamma$ have?
These questions concern truly quantum thermodynamics~\cite{NYH_18_Beyond}.
Collaborators and I used QI to characterize $\gamma$,
which we dubbed the \emph{non-Abelian thermal state} 
(NATS)~\cite{NYH_16_Micro} (Ch.~\ref{ch:Noncommq}).
Parallel analyses took place in~\cite{Lostaglio_17_Thermodynamic,Guryanova_16_Thermodynamics}.

We derived the form of $\gamma$ in three ways.
First, we invoked \emph{typical subspaces},
a QI tool used to quantify data compression.
Second, thermal states are the fixed points of ergodic dynamics.
We modeled ergodic dynamics with a random unitary.
Randomly evolved states have been characterized
with another QI tool, \emph{canonical typicality}~\cite{Goldstein_06_Canonical,Gemmer_04_18,Popescu_06_Entanglement,Linden_09_Quantum}. 
We applied canonical typicality to our system's time-evolved state.
The state, we concluded, lies close to the expected
$e^{ - \sum_{j = 1}^m  \mu_j  Q_j } / Z$.

Third, thermal states are \emph{completely passive}:
Work cannot be extracted even from infinitely many copies
of a thermal state~\cite{Pusz_78_Passive}.
We proved the complete passivity of
$e^{ - \sum_{j = 1}^m  \mu_j  Q_j } / Z$, 
using a thermodynamic resource theory.

\emph{Resource theories} are QI models
for agents who transform quantum states,
using a restricted set of operations.
The first law of thermodynamics and the ambient temperature $T$
restrict thermodynamic operations.
Restrictions prevent agents from preparing certain states,
e.g., pure nonequilibrium states.
Scarce states have value,
as work can be extracted from nonequilibrium systems.
Resource theories help us to quantify states' usefulness,
to identify allowed and forbidden transformations between states,
and to quantify the efficiencies with which tasks (e.g., work extraction) 
can be performed outside the large-system limit
(e.g.,~\cite{LiebY99,Janzing00,Brandao_13_Resource,FundLimits2}.
The efficiencies are quantified with 
quantum entropies for small scales~\cite{Tomamichel_16_Quantum}.
Most of my PhD contributions were mentioned above~\cite{NYH_16_Beyond,NYH_18_Beyond,NYH_15_Introducing,NYH_16_Micro}.
Such theoretical results require testing.
I outlined experimental challenges and opportunities in~\cite{NYH_17_Toward}.

\putbib[OwnPubs,Intro_bib]  
\end{bibunit}

%
%
\chapter{Jarzynski-like equality for the out-of-time-ordered correlator}
\label{ch:Jarz_like}
\begin{bibunit}  

\noindent \emph{This chapter was published as~\cite{NYH_17_Jarzynski}.}



\begingroup
\newcommand{\W}{ \mathcal{W} }  
\newcommand{\C}{ F }  
\newcommand{\RegC}{ \C_\reg }  
\newcommand{\Dim}{ d }  
\newcommand{\TW}{ \tilde{W} } 
\newcommand{\Protocol}{ \mathcal{P} }  
\newcommand{\Protocoll}{ \tilde{ \mathcal{P} } }  
\newcommand{\RegProtocollOne}{ \tilde{ \mathcal{P} }_{\reg, 1} }
\newcommand{\RegProtocollTwo}{ \tilde{ \mathcal{P} }_{\reg, 2} }
\newcommand{\reg}{ {\text{reg}} }
\newcommand{\GW}{ G_\W }  
\newcommand{\GV}{ G_V }  
\newcommand{\gw}{ g_w }  
\newcommand{\gv}{ g_v }  
\newcommand{\gwP}{ g_{w'} }  
\newcommand{\gvP}{ g_{v'} }  
\newcommand{\DegenW}{ \alpha }  
\newcommand{\DegenV}{ \lambda }  
\newcommand{\Coupling}{ c }  
\newcommand{\Charac}{ \mathcal{G} }  
\newcommand{\NondegW}{ \tilde{\W} } 
\newcommand{\NondegV}{ \tilde{V} } 
\newcommand{\U}{ \mathcal{U} } 
\newcommand{\Prob}{ \mathscr{P} } 
\newcommand{\TProb}{ \tilde{ \Prob } } 
\newcommand{\weak}{ {\text{weak}} } 
\newcommand{\PWeak}{ \mathscr{P}_\weak } 
\newcommand{\WeakInt}{ \mathcal{I} }

The out-of-time-ordered correlator (OTOC) $\C(t)$ 
diagnoses the scrambling of quantum information~\cite{Shenker_Stanford_14_BHs_and_butterfly,Shenker_Stanford_14_Multiple_shocks,Shenker_Stanford_15_Stringy,Roberts_15_Localized_shocks,Roberts_Stanford_15_Diagnosing,Maldacena_15_Bound}:
Entanglement can grow rapidly in a many-body quantum system,
dispersing information throughout many degrees of freedom.
$\C(t)$ quantifies the hopelessness of 
attempting to recover the information 
via local operations.

Originally applied to superconductors~\cite{LarkinO_69},
$\C(t)$ has undergone a revival recently.
$\C(t)$ characterizes quantum chaos, holography, black holes, and condensed matter.
The conjecture that black holes scramble quantum information
at the greatest possible rate
has been framed in terms of $\C(t)$~\cite{Maldacena_15_Bound,Sekino_Susskind_08_Fast_scramblers}.
The slowest scramblers include disordered systems~\cite{Huang_16_MBL_OTOC,Swingle_16_MBL_OTOC,Fan_16_MBL_OTOC,He_16_MBL_OTOC,Chen_16_MBL_OTOC}.
In the context of quantum channels,
$\C(t)$ is related to the tripartite information~\cite{HosurYoshida_16_Chaos}.
Experiments have been proposed~\cite{Swingle_16_Measuring,Yao_16_Interferometric,Zhu_16_Measurement}
and performed~\cite{Li_16_Measuring,Garttner_16_Measuring}
to measure $\C(t)$ with cold atoms and ions, 
with cavity quantum electrodynamics,
and with nuclear-magnetic-resonance quantum simulators.

$\C(t)$ quantifies sensitivity to initial conditions, a signature of chaos.
Consider a quantum system $S$ governed by a Hamiltonian $H$.
Suppose that $S$ is initialized to a pure state $\ket{ \psi }$
and perturbed with a local unitary operator $V$.
$S$ then evolves forward in time under the unitary
$U = e^{ - i H t}$ for a duration $t$,
is perturbed with a local unitary operator $\W$,
and evolves backward under $U^\dag$.
The state $\ket{ \psi' }  :=  U^\dag  \W  U  V  \ket{\psi}
= \W (t) V \ket{\psi}$ 
results.
Suppose, instead, that $S$ is perturbed with $V$
not at the sequence's beginning, but at the end:
$\ket{\psi}$ evolves forward under $U$,
is perturbed with $\W$,
evolves backward under $U^\dag$,
and is perturbed with $V$.
The state $\ket{ \psi'' }  :=  V U^\dag \W U \ket{ \psi } 
= V \W(t) \ket{\psi}$ results.
The overlap between the two possible final states equals the correlator:
$\C(t)  :=  \expval{ \W^\dag(t) \, V^\dag \, \W(t) \, V }  
=  \langle \psi'' | \psi' \rangle$.
The decay of $\C(t)$ reflects the growth
of $[\W(t), \, V]$~\cite{Maldacena_16_Comments,Polchinski_16_Spectrum}.

Forward and reverse time evolutions,
as well as information theory and diverse applications,  
characterize not only the OTOC, but also fluctuation relations.
Fluctuation relations have been derived 
in quantum and classical
nonequilibrium statistical mechanics~\cite{Jarzynski97,Crooks99,Tasaki00,Kurchan00}.
Consider a Hamiltonian $H(t)$
tuned from $H_i$ to $H_f$ at a finite speed.
For example, electrons may be driven within a circuit~\cite{Saira_12_Test}. 
Let $\Delta F  :=  F (H_f)  -  F( H_i )$
denote the difference between the equilibrium free energies
at the inverse temperature $\beta$:\footnote{
$F ( H_\ell )$ denotes the free energy in statistical mechanics,
while $F(t)$ denotes the OTOC in high energy and condensed matter.}
$F( H_\ell )  =  - \frac{1}{\beta} \ln Z_{\beta, \ell}$,
wherein the partition function is 
$Z_{\beta, \ell}  :=  \Tr ( e^{ - \beta H_\ell } )$
and $\ell = i, f$.
The free-energy difference
has applications in chemistry, biology, and pharmacology~\cite{Chipot_07_Free}.
One could measure $\Delta F$, in principle, by measuring
the work required to tune $H(t)$ from $H_i$ to $H_f$
while the system remains in equilibrium.
But such quasistatic tuning would require an infinitely long time.

$\Delta F$ has been inferred in a finite amount of time 
from Jarzynski's fluctuation relation,
$\expval{ e^{ - \beta W} }  =  e^{ - \beta \Delta F}$.
The left-hand side can be inferred from data about experiments
in which $H(t)$ is tuned from $H_i$ to $H_f$
arbitrarily quickly.
The work required to tune $H(t)$ during some particular trial
(e.g., to drive the electrons) is denoted by $W$.
$W$ varies from trial to trial 
because the tuning can eject the system arbitrarily far from equilibrium.
The expectation value $\langle \, . \, \rangle$ 
is with respect to the probability distribution $P(W)$
associated with any particular trial's
requiring an amount $W$ of work.
Nonequilibrium experiments have been combined
with fluctuation relations to estimate $\Delta F$~\cite{CollinRJSTB05,Douarche_05_Experimental,Blickle_06_Thermo,Harris_07_Experimental,MossaMFHR09,ManosasMFHR09,Saira_12_Test,Batalhao_14_Experimental,An_15_Experimental}:
\begin{align}
   \label{eq:DeltaF}
   \Delta F  =  - \frac{ 1 }{ \beta }  \:   
   \log  \expval{ e^{ - \beta W} }  \, .
\end{align}

Jarzynski's Equality, with the exponential's convexity, 
implies $\expval{ W }   \geq  \Delta F$.
The average work $\expval{W}$ required to tune $H(t)$
according to any fixed schedule
equals at least the work $\Delta F$ required to tune $H(t)$ quasistatically.
This inequality has been regarded as a manifestation of
the Second Law of Thermodynamics.
The Second Law governs information loss~\cite{Maruyama_09_Colloquium},
similarly to the OTOC's evolution.

I derive a Jarzynski-like equality, analogous to Eq.~\eqref{eq:DeltaF}, for $\C(t)$
(Theorem~\ref{theorem:OTOC_FT}).
The equality unites two powerful tools 
that have diverse applications
in quantum information, high-energy physics,
statistical mechanics, and condensed matter.
The union sheds new light on
both fluctuation relations and the OTOC,
similar to the light shed 
when fluctuation relations were introduced 
into ``one-shot'' statistical mechanics~\cite{Aberg_13_Truly,YungerHalpern_15_Introducing,Salek_15_Fluctuations,YungerHalpern_15_What,Dahlsten_15_Equality,Alhambra_16_Fluctuating_Work}.
The union also relates the OTOC,
known to signal quantum behavior in high energy and condensed matter,
to a quasiprobability,
known to signal quantum behavior in optics.
The Jarzynski-like equality suggests
a platform-nonspecific protocol for measuring $\C(t)$ indirectly.
The protocol can be implemented with weak measurements
or with interference.
The time evolution need not be reversed in any interference trial.
First, I present the set-up and definitions.
I then introduce and prove the Jarzynski-like equality for $\C(t)$.

\section{Set-up}

Let $S$ denote a quantum system
associated with a Hilbert space $\mathcal{H}$
of dimensionality $\Dim$.
The simple example of a spin chain~\cite{Yao_16_Interferometric,Zhu_16_Measurement,Li_16_Measuring,Garttner_16_Measuring} informs this paper:
Quantities will be summed over, as spin operators have discrete spectra.
Integrals replace the sums
if operators have continuous spectra.

Let $\W  =  \sum_{ w_\ell, \DegenW_{w_\ell} } w_\ell 
\ketbra{ w_\ell, \DegenW_{w_\ell} }{ w_\ell, \DegenW_{w_\ell} }$ and 
$V  =  \sum_{ v_\ell, \DegenV_{v_\ell} }  v_\ell  
\ketbra{ v_\ell, \DegenV_{v_\ell} }{ v_\ell, \DegenV_{v_\ell}}$ 
denote local unitary operators.
The eigenvalues are denoted by $w_\ell$ and $v_\ell$;
the degeneracy parameters, by $\DegenW_{w_\ell}$ and $\DegenV_{v_\ell}$.
$\W$ and $V$ may commute.
They need not be Hermitian.
Examples include single-qubit Pauli operators
localized at opposite ends of a spin chain.

We will consider measurements
of eigenvalue-and-degeneracy-parameter tuples
$(w_\ell, \DegenW_{w_\ell} )$ and $(v_\ell, \DegenV_{v_\ell} )$.
Such tuples can be measured as follows. 
A Hermitian operator
$\GW  =  \sum_{ w_\ell, \DegenW_{w_\ell} } 
   g ( w_\ell ) 
   \ketbra{ w_\ell, \DegenW_{w_\ell} }{ w_\ell, \DegenW_{w_\ell} }$ 
generates the unitary $\W$.
The generator's eigenvalues are labeled by the unitary's eigenvalues:
$w = e^{i g ( w_\ell ) }$.
Additionally, there exists a Hermitian operator
that shares its eigenbasis with $\W$
but whose spectrum is nondegenerate:
$\tilde{G}_{ \W }  =  \sum_{w_\ell, \DegenW_{w_\ell}}  
\tilde{g} ( \DegenW_{w_\ell} )
\ketbra{ w_\ell, \DegenW_{w_\ell} }{ w_\ell, \DegenW_{w_\ell} }$,
wherein $\tilde{g} ( \DegenW_{w_\ell} )$ denotes a real one-to-one function.
I refer to a collective measurement of $\GW$ and $\tilde{G}_{ \W }$
as a $\NondegW$ measurement.
Analogous statements concern $V$.
If $\Dim$ is large, measuring $\NondegW$ and $\NondegV$ 
may be challenging but is possible in principle.
Such measurements may be reasonable if $S$ is small.
Schemes for avoiding measurements of 
the $\DegenW_{w_\ell}$'s and $\DegenV_{v_\ell}$'s
are under investigation~\cite{BrianDisc}.

Let $H$ denote a time-independent Hamiltonian.
The unitary $U = e^{ - i H t}$ evolves $S$ forward in time for an interval $t$.
Heisenberg-picture operators are defined as
$\W(t) :=  U^\dag \W U$ and
$\W^\dag(t)  =  [ \W(t) ]^\dag  =  U^\dag \W^\dag U$.

The OTOC is conventionally evaluated on
a Gibbs state $e^{ - H / T } / Z$,
wherein $T$ denotes a temperature:
$\C(t)  =  \Tr \left( \frac{ e^{ - H / T } }{ Z } \W^\dag (t) V^\dag \W(t) V \right)$.
Theorem~\ref{theorem:OTOC_FT} generalizes beyond 
$e^{ - H / T } / Z$ to arbitrary density operators
$\rho = \sum_j  p_j  \ketbra{j}{j}  \in  \mathcal{D} ( \mathcal{H} )$.
[$\mathcal{D} ( \mathcal{H} )$ denotes the set of density operators
defined on $\mathcal{H}$.]

\section{Definitions}

Jarzynski's Equality concerns thermodynamic work, $W$.
$W$ is a random variable
calculated from measurement outcomes.
The out-of-time-ordering in $\C(t)$ requires two such random variables.
I label these variables $W$ and $W'$.

Two stepping stones connect $\W$ and $V$ to $W$ and $W'$.
First, I define a complex probability amplitude
$A_\rho ( w_2  , \DegenW_{w_2}  ;  v_1  ,  \DegenV_{v_1} ;
w_1  , \DegenW_{w_1}  ; j)$
associated with a quantum protocol.
I combine amplitudes $A_\rho$
into a $\tilde{A}_\rho$ inferable
from weak measurements and from interference.
$\tilde{A}_\rho$ resembles a quasiprobability,
a quantum generalization of a probability.
In terms of the $w_\ell$'s and $v_\ell$'s in $\tilde{A}_\rho$,
I define the measurable random variables $W$ and $W'$.

Jarzynski's Equality involves a probability distribution $P(W)$
over possible values of the work.
I define a complex analog $P(W, W')$.
These definitions are designed to parallel
expressions in~\cite{TLH_07_Work}.
Talkner, Lutz and H\"{a}nggi cast Jarzynski's Equality
in terms of a time-ordered correlation function.
Modifying their derivation will lead to the OTOC Jarzynski-like equality.

\subsection{Quantum probability amplitude $A_\rho$}

The probability amplitude $A_\rho$
is defined in terms of the following protocol, $\Protocol$:
\begin{enumerate}
   \item Prepare $\rho$.
   \item Measure the eigenbasis of $\rho$, $\{ \ketbra{j}{j} \}$. 
   \item \label{item:FirstU}
            Evolve $S$ forward in time under $U$.
   \item Measure $\NondegW$. 
   \item Evolve $S$ backward in time under $U^\dag$.
   \item Measure $\NondegV$. 
   \item Evolve $S$ forward under $U$.
   \item Measure $\NondegW$. 
\end{enumerate}
An illustration appears in Fig.~\ref{fig:Protocoll_Trial1}.
Consider implementing $\Protocol$ in one trial.
The complex probability amplitude
associated with the measurements' yielding $j$,
then $( w_1, \DegenW_{w_1} )$, then $(v_1, \DegenV_{v_1} )$, 
then $( w_2, \DegenW_{w_2} )$ is
\begin{align}
   \label{eq:ADef}
   & A_\rho( w_2  , \DegenW_{w_2}  ;  v_1  ,  \DegenV_{v_1} ;
w_1  , \DegenW_{w_1}  ; j)
   :=  \langle w_2, \DegenW_{w_2} | U | v_1, \DegenV_{v_1} \rangle
   \nonumber \\ & \qquad \times
   \langle v_1, \DegenV_{v_1}  |  U^\dag  |  w_1, \DegenW_{w_1}  \rangle
   \langle w_1,  \DegenW_{w_1}  |  U | j \rangle
   \sqrt{ p_j } \, .
\end{align}
The square modulus $ | A_\rho ( . ) |^2$ equals
the joint probability that these measurements yield these outcomes.

Suppose that $[\rho, \, H] = 0$.
For example, suppose that 
$S$ occupies the thermal state $\rho = e^{ - H / T } / Z$.
(I set Boltzmann's constant to one: $\kB = 1$.)
Protocol $\Protocol$ and Eq.~\eqref{eq:ADef} simplify:
The first $U$ can be eliminated,
because $[\rho, \, U] = 0$.
Why $[\rho, \, U] = 0$ obviates the unitary
will become apparent when 
we combine $A_\rho$'s into $\tilde{A}_\rho$.

The protocol $\Protocol$ defines $A_\rho$;
$\Protocol$ is not a prescription measuring $A_\rho$.
Consider implementing $\Protocol$ many times
and gathering statistics about the measurements' outcomes.
From the statistics, one can infer
the probability $| A_\rho |^2$,
not the probability amplitude $A_\rho$.
$\Protocol$ merely is the process 
whose probability amplitude equals $A_\rho$.
One must calculate combinations of $A_\rho$'s
to calculate the correlator.
These combinations, labeled $\tilde{A}_\rho$,
can be inferred from weak measurements and interference.

%
%
\begin{figure}[h]
\centering
\begin{subfigure}{0.4\textwidth}
\centering
\includegraphics[width=.9\textwidth]{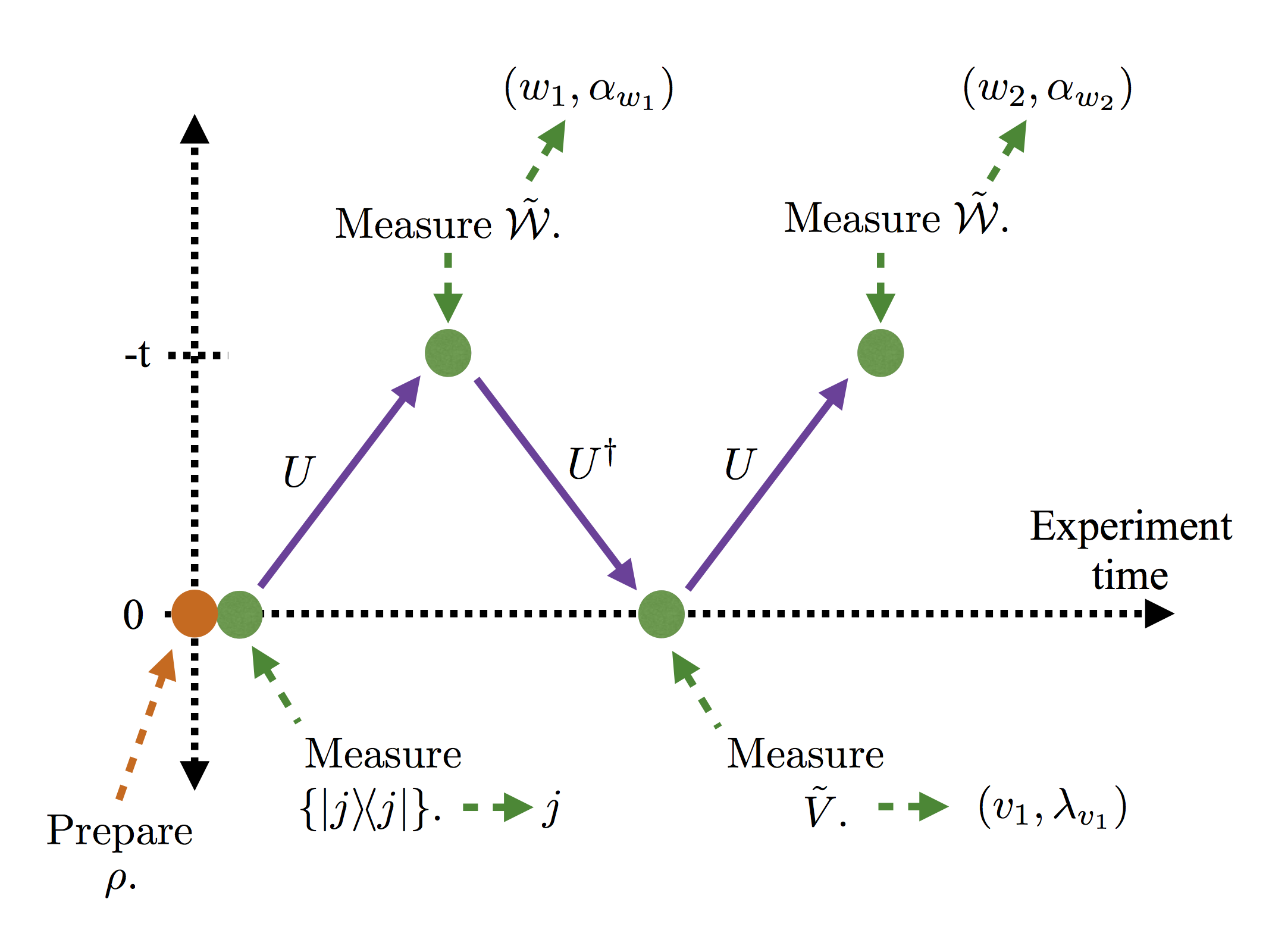}
\caption{}
\label{fig:Protocoll_Trial1}
\end{subfigure}
\begin{subfigure}{.4\textwidth}
\centering
\includegraphics[width=.9\textwidth]{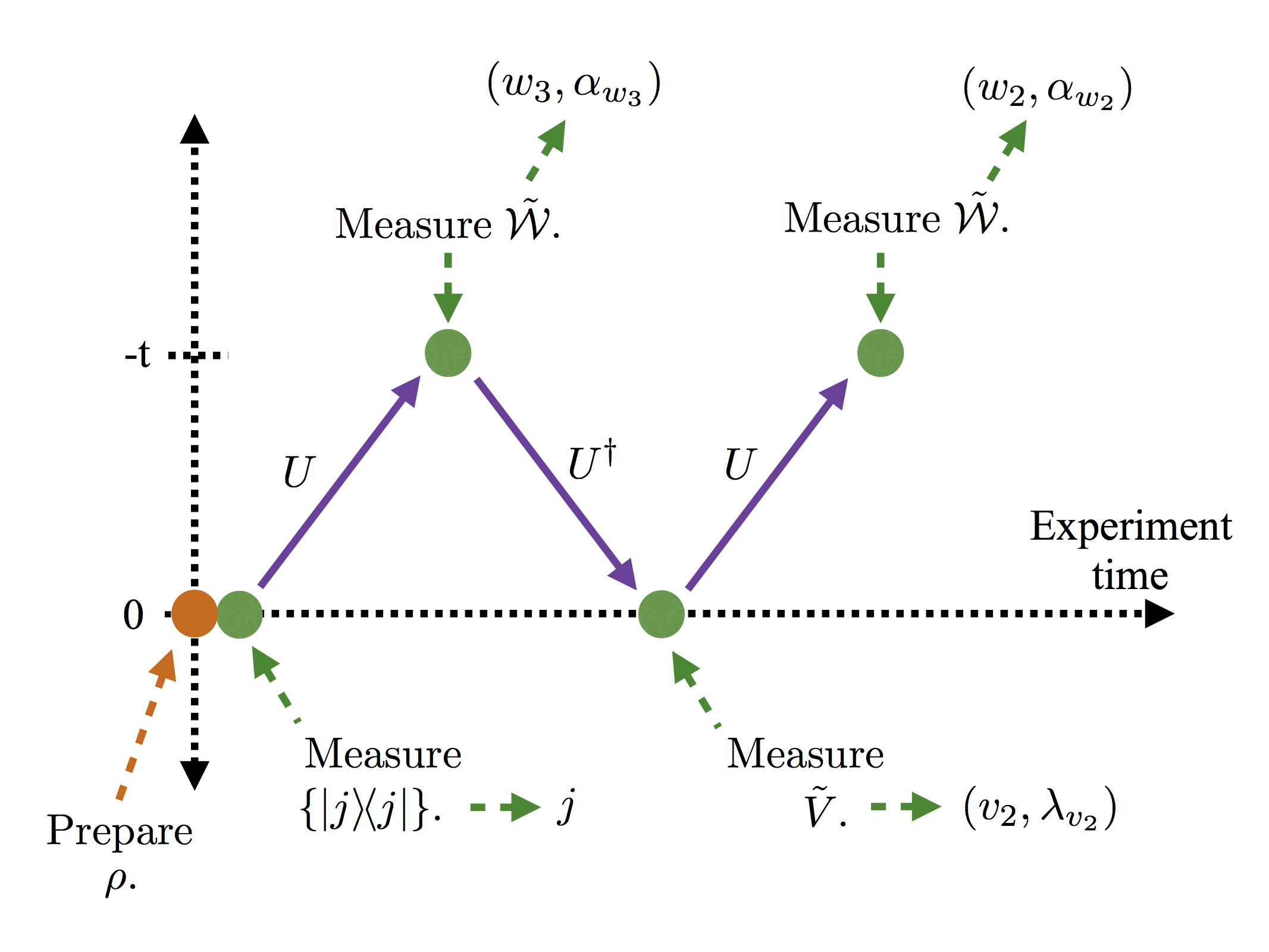}
\caption{}
\label{fig:Protocoll_Trial2}
\end{subfigure}
\caption{\caphead{Quantum processes described by 
the complex amplitudes in
the Jarzynski-like equality for the out-of-time-ordered correlator (OTOC):}
Theorem~\ref{theorem:OTOC_FT} shows that the OTOC
depends on a complex distribution $P(W, W')$.
This $P(W, W')$ parallels the probability distribution
over possible values of thermodynamic work 
in Jarzynski's Equality.
$P(W, W')$ results from summing products 
$A_\rho^*( . ) A_\rho( . )$.
Each $A_\rho( . )$ denotes a probability amplitude [Eq.~\eqref{eq:ADef}],
so each product resembles a probability.
But the amplitudes' arguments differ,
due to the OTOC's out-of-time ordering:
The amplitudes correspond to different quantum processes.
Figure~\ref{fig:Protocoll_Trial1} illustrates 
the process associated with the $A_\rho( . )$;
and Fig.~\ref{fig:Protocoll_Trial2}, the process associated with the $A_\rho^*( . )$.
Time runs from left to right.
Each process begins with the preparation of 
the state $\rho = \sum_j p_j \ketbra{j}{j}$ 
and a measurement of the state's eigenbasis.
Three evolutions ($U$, $U^\dag$, $U$) then alternate with 
three measurements of observables ($\NondegW$, $\NondegV$, $\NondegW$).
If the initial state commutes with the Hamiltonian $H$
(e.g., if $\rho = e^{ - H / T } / Z$),
the first $U$ can be omitted.
Figures~\ref{fig:Protocoll_Trial1} and~\ref{fig:Protocoll_Trial2}
are used to define $P(W, W')$,
rather than illustrating protocols for measuring $P(W, W')$.
$P(W, W')$ can be inferred from weak measurements
and from interferometry.}
\label{fig:Protocoll}
\end{figure}
\subsection{Combined quantum amplitude $\tilde{A}_\rho$}
\label{section:ADirac}

Combining quantum amplitudes $A_\rho$ 
yields a quantity $\tilde{A}_\rho$
that is nearly a probability
but that differs due to the OTOC's out-of-time ordering.
I first define $\tilde{A}_\rho$, 
which resembles the Kirkwood-Dirac quasiprobability~\cite{Kirkwood_33_Quantum,Dirac_45_On,Dressel_15_Weak,BrianDisc}.
We gain insight into $\tilde{A}_\rho$ 
by supposing that $[\rho, \, \W] = 0$,
e.g., that $\rho$ is the infinite-temperature Gibbs state $\id / \Dim$.
$\tilde{A}_\rho$ can reduce to a probability in this case,
and protocols for measuring $\tilde{A}_\rho$ simplify.
I introduce weak-measurement and interference schemes
for inferring $\tilde{A}_\rho$ experimentally.

\subsubsection{Definition of the combined quantum amplitude $\tilde{A}_\rho$}
\label{section:TildeA}

Consider measuring the probability amplitudes $A_\rho$
associated with all the possible measurement outcomes.
Consider fixing an outcome septuple 
$( w_2  , \DegenW_{w_2}  ;  v_1  ,  \DegenV_{v_1} ;
w_1  , \DegenW_{w_1}  ; j)$.
The amplitude $A_\rho( w_2  , \DegenW_{w_2}  ;  v_1  ,  \DegenV_{v_1} ;
w_1  , \DegenW_{w_1}  ; j)$
describes one realization, illustrated in Fig.~\ref{fig:Protocoll_Trial1},
of the protocol $\Protocol$.
Call this realization $a$.

Consider the $\Protocol$ realization, labeled $b$,
illustrated in Fig.~\ref{fig:Protocoll_Trial2}.
The initial and final measurements
yield the same outcomes as in $a$
[outcomes $j$ and $(w_2, \DegenW_{w_2} )$].
Let $(w_3, \DegenW_{w_3} )$ and $( v_2, \DegenV_{v_2} )$
denote the outcomes of
the second and third measurements in $b$.
Realization $b$ corresponds to the probability amplitude
$A_\rho( w_2  , \DegenW_{w_2}  ;  v_2  ,  \DegenV_{v_2} ;
w_3  , \DegenW_{w_3}  ; j)$.

Let us complex-conjugate the $b$ amplitude
and multiply by the $a$ amplitude.
We marginalize over $j$ and over $(w_1,  \DegenW_{w_1} )$,
forgetting about the corresponding measurement outcomes:
\begin{align}
   \label{eq:TildeADef}
   \tilde{A}_\rho ( w, v, \DegenW_w, \DegenV_v )   & :=  
   \sum_{ j,  (w_1,  \DegenW_{w_1} ) }
   A^*_\rho( w_2  , \DegenW_{w_2}  ;  v_2  ,  \DegenV_{v_2} ;
w_3  , \DegenW_{w_3}  ; j)
   \nonumber \\ &  \; \times
   A_\rho( w_2  , \DegenW_{w_2}  ;  v_1  ,  \DegenV_{v_1} ;
w_1  , \DegenW_{w_1}  ; j ) \, .
\end{align}
The shorthand $w$ encapsulates the list $(w_2, w_3)$.
The shorthands $v$, $\DegenW_w$ and $\DegenV_v$
are defined analogously. 

Let us substitute in from Eq.~\eqref{eq:ADef}
and invoke $\langle A | B \rangle^*  =  \langle B | A \rangle$.
The sum over $(w_1,  \DegenW_{w_1} )$ evaluates
to a resolution of unity.
The sum over $j$ evaluates to $\rho$:
\begin{align}
   \label{eq:TildeAExp}
   & \tilde{A}_\rho ( w, v, \DegenW_w, \DegenV_v )
   =    \langle  w_3,  \DegenW_{w_3}  |  U  |  v_2,  \DegenV_{v_2}  \rangle
   \langle  v_2,  \DegenV_{v_2}  |  U^\dag  |  w_2,  \DegenW_{w_2}  \rangle
   \nonumber \\ & \qquad \times
   \langle  w_2,  \DegenW_{w_2}  |  U  |  v_1,  \DegenV_{v_1}  \rangle
   \langle  v_1,  \DegenV_{v_1}  |  \rho  U^\dag  |  w_3,  \DegenW_{w_3}  \rangle   \, .
\end{align}

This $\tilde{A}_\rho$ resembles the Kirkwood-Dirac quasiprobability~\cite{Dressel_15_Weak,BrianDisc}.
Quasiprobabilities surface in quantum optics and quantum foundations~\cite{Carmichael_02_Statistical,Ferrie_11_Quasi}.
Quasiprobabilities generalize probabilities to quantum settings.
Whereas probabilities remain between 0 and 1,
quasiprobabilities can assume negative and nonreal values.
Nonclassical values signal quantum phenomena such as entanglement.
The best-known quasiprobabilities include
the Wigner function, the Glauber-Sudarshan $P$ representation,
and the Husimi $Q$ representation.
Kirkwood and Dirac defined another quasiprobability
in 1933 and in 1945~\cite{Kirkwood_33_Quantum,Dirac_45_On}.
Interest in the Kirkwood-Dirac quasiprobability has revived recently.
The distribution can assume nonreal values,
obeys Bayesian updating, and has been measured experimentally~\cite{Lundeen_11_Direct,Lundeen_12_Procedure,Bamber_14_Observing,Mirhosseini_14_Compressive}.

The Kirkwood-Dirac distribution for a state 
$\sigma \in \mathcal{D}( \mathcal{H} )$ has the form
$\langle f | a \rangle   \langle a |  \sigma  | f \rangle$,
wherein $\Set{ \ketbra{f}{f} }$ and $\Set{ \ketbra{a}{a} }$
denote bases for $\mathcal{H}$~\cite{Dressel_15_Weak}.
Equation~\eqref{eq:TildeAExp} has the same form
except contains more outer products.
Marginalizing $\tilde{A}_\rho$ over every variable
except one $w_\ell$ [or one $v_\ell$, 
one $( w_\ell,  \, \DegenW_{w_\ell} )$,
or one $( v_\ell,  \,  \DegenV_{v_\ell} )$]
yields a probability,
as does marginalizing the Kirkwood-Dirac distribution
over every variable except one.
The precise nature of the relationship between 
$\tilde{A}_\rho$ and the Kirkwood-Dirac quasiprobability
is under investigation~\cite{BrianDisc}.
For now, I harness the similarity 
to formulate a weak-measurement scheme for $\tilde{A}_\rho$
in Sec.~\ref{section:WeakMain}.

$\tilde{A}_\rho$ is nearly a probability:
$\tilde{A}_\rho$ results from multiplying
a complex-conjugated probability amplitude $A^*_\rho$
by a probability amplitude $A_\rho$.
So does the quantum mechanical probability density
$p(x) = \psi^*(x) \psi(x)$.
Hence the quasiprobability resembles a probability.
Yet the argument of the $\psi^*$ equals
the argument of the $\psi$.
The argument of the $A^*_\rho$ does not equal
the argument of the $A_\rho$.
This discrepancy stems from the OTOC's out-of-time ordering.
$\tilde{A}_\rho$ can be regarded
as like a probability,
differing due to the out-of-time ordering.
$\tilde{A}_\rho$ reduces to a probability 
under conditions discussed in Sec.~\ref{section:SimpleTildeA}.
The reduction reinforces the parallel between
Theorem~\ref{theorem:OTOC_FT} and
the fluctuation-relation work~\cite{TLH_07_Work},
which involves a probability distribution that resembles $\tilde{A}_\rho$.

\subsubsection{Simple case, 
reduction of $\tilde{A}_\rho$ to a probability} 
\label{section:SimpleTildeA}

Suppose that $\rho$ shares the $\NondegW(t)$ eigenbasis:
$\rho  =  \rho_{ \W(t) }
:=  \sum_{w_\ell, \DegenW_{w_\ell} }  p_{ w_\ell, \DegenW_{w_\ell} }
U^\dag   \ketbra{ w_\ell, \DegenW_{w_\ell} }{ w_\ell, \DegenW_{w_\ell} }   U$.
For example, $\rho$ may be 
the infinite-temperature Gibbs state $\id / \Dim$.
Equation~\eqref{eq:TildeAExp} becomes
\begin{align}
   \label{eq:TildeASimple}
   & \tilde{A}_{ \rho_{ \W(t) } } ( w, v, \DegenW_w,  \DegenV_v )
   =  \langle w_3, \DegenW_{w_3}  |  U  |  v_2, \DegenV_{v_2}  \rangle
   \nonumber \\  & \qquad \times
   \langle v_2, \DegenV_{v_2}  |  U^\dag  |  w_2,  \DegenW_{w_2}  \rangle
   \langle  w_2,  \DegenW_{w_2}  |  U  |  v_1,  \DegenV_{v_1}  \rangle
   \nonumber \\  & \qquad \times
   \langle  v_1,  \DegenV_{v_1}  |  U^\dag  |  w_3, \DegenW_{w_3}  \rangle
   \, p_{w_3, \DegenW_{w_3} } \, .
\end{align}
The weak-measurement protocol simplifies,
as discussed in Sec.~\ref{section:WeakMain}.

Equation~\eqref{eq:TildeASimple}
reduces to a probability if 
$(w_3, \DegenW_{w_3})  =  ( w_2,  \DegenW_{w_2})$ or if
$( v_2,  \DegenV_{v_2} )  =  ( v_1,  \DegenV_{v_1} )$.
For example, suppose that
$(w_3, \DegenW_{w_3})  =  ( w_2,  \DegenW_{w_2})$:
\begin{align}
   \label{eq:TildeASimple2}
   & \tilde{A}_{ \rho_{ \W(t) } } \LParen  
   ( w_2, w_2 ),  v ,  ( \DegenW_{w_2} ,  \DegenW_{w_2} ),  \DegenV_v  \RParen
   = | \langle v_2,  \DegenV_{v_2}  | U^\dag | w_2,  \DegenW_{w_2}  \rangle |^2
   \nonumber \\ & \qquad \qquad \qquad  \times
   | \langle  v_1,  \DegenV_{v_1}  |  U^\dag  |  w_2,  \DegenW_{w_2}  \rangle  |^2 \,
   p_{ w_2, \DegenW_{w_2} } \\
   \label{eq:ReduceToProb}
    & \qquad = 
    p ( v_2, \DegenV_{v_2} | w_2, \DegenW_{w_2} ) \,
    p ( v_1, \DegenV_{v_1} | w_2, \DegenW_{w_2} ) \,
    p_{ w_2, \DegenW_{w_2} } \, .
\end{align}
The $p_{ w_2, \DegenW_{w_2} }$ denotes the probability that
preparing $\rho$ and measuring $\NondegW$
will yield $(w_2, \DegenW_{w_2})$.
Each $p( v_\ell, \DegenV_{v_\ell} | w_2, \DegenW_{w_2} )$ denotes
the conditional probability that
preparing $\ket{w_2, \DegenW_{w_2} }$, backward-evolving under $U^\dag$, 
and measuring $\NondegV$ 
will yield $(v_\ell, \DegenV_{v_\ell})$.
Hence the combination $\tilde{A}_\rho$ of probability amplitudes
is nearly a probability:
$\tilde{A}_\rho$ reduces to a probability under simplifying conditions.

Equation~\eqref{eq:ReduceToProb} strengthens the analogy 
between Theorem~\ref{theorem:OTOC_FT} and
the fluctuation relation in~\cite{TLH_07_Work}.
Equation~(10) in~\cite{TLH_07_Work} contains 
a conditional probability $p( m, t_f | n )$
multiplied by a probability $p_n$. 
These probabilities parallel the $p ( v_1, \DegenV_{v_1} | w_1, \DegenW_{w_1} )$
and $p_{ w_1, \DegenW_{w_1} }$ in Eq.~\eqref{eq:ReduceToProb}.
Equation~\eqref{eq:ReduceToProb} contains another conditional probability,
$p ( v_2, \DegenV_{v_2} | w_1, \DegenW_{w_1} )$,
due to the OTOC's out-of-time ordering.

\subsubsection{Weak-measurement scheme for 
the combined quantum amplitude $\tilde{A}_\rho$}
\label{section:WeakMain}

$\tilde{A}_\rho$ is related to the Kirkwood-Dirac quasiprobability,
which has been inferred from weak measurements~\cite{Dressel_14_Understanding,Kofman_12_Nonperturbative,Lundeen_11_Direct,Lundeen_12_Procedure,Bamber_14_Observing,Mirhosseini_14_Compressive}.
I sketch a weak-measurement scheme for inferring $\tilde{A}_\rho$.
Details appear in Appendix~\ref{section:Jarz_like_App_A}.

Let $\Protocol_\weak$ denote the following protocol:
\begin{enumerate}
   \item Prepare $\rho$.
   \item \label{item:FirstVWeak}
            Couple the system's $\NondegV$ weakly to an ancilla $\mathcal{A}_{ a }$. 
            Measure $\mathcal{A}_{ a }$ strongly.
   \item \label{item:FirstUWeak}
            Evolve $S$ forward under $U$.
   \item Couple the system's $\NondegW$ weakly to an ancilla $\mathcal{A}_{ b }$. 
            Measure $\mathcal{A}_{ b }$ strongly.
   \item Evolve $S$ backward under $U^\dag$.
   \item Couple the system's $\NondegV$ weakly to an ancilla $\mathcal{A}_c$.
            Measure $\mathcal{A}_c$ strongly.
   \item Evolve $S$ forward under $U$.
   \item Measure $\NondegW$ strongly (e.g., projectively). 
\end{enumerate}
Consider performing $\Protocol_\weak$ many times.
From the measurement statistics, one can infer the form of 
$\tilde{A}_\rho ( w, v, \DegenW_w, \DegenV_v )$.

$\Protocol_\weak$ offers an experimental challenge:
Concatenating weak measurements raises
the number of trials required to infer a quasiprobability.
The challenge might be realizable 
with modifications to existing set-ups (e.g.,~\cite{White_16_Preserving,Dressel_14_Implementing}).
Additionally, $\Protocol_\weak$ simplifies
in the case discussed in Sec.~\ref{section:SimpleTildeA}---if
$\rho$ shares the $\NondegW(t)$ eigenbasis,
e.g., if $\rho = \id / \Dim$.
The number of weak measurements reduces from three to two.
Appendix~\ref{section:Jarz_like_App_A} contains details.

\subsubsection{Interference-based measurement of $\tilde{A}_\rho$}

$\tilde{A}_\rho$ can be inferred
not only from weak measurement, but also from interference.
In certain cases---if $\rho$ shares neither 
the $\NondegW(t)$ nor the $\NondegV$  eigenbasis---also
quantum state tomography is needed.
From interference, one infers the inner products 
$\langle a | \U | b \rangle$ in $\tilde{A}_\rho$.
Eigenstates of $\NondegW$ and $\NondegV$
are labeled by $a$ and $b$;
and $\U = U, U^\dag$.
The matrix element 
$\langle v_1, \DegenV_{v_1}  |  \rho  U^\dag  
|  w_3,  \DegenW_{w_3}  \rangle$
is inferred from quantum state tomography in certain cases.

The interference scheme proceeds as follows.
An ancilla $\mathcal{A}$ is prepared in
a superposition $\frac{1}{ \sqrt{2} } \; ( \ket{0} + \ket{1} )$.
The system $S$ is prepared in a fiducial state $\ket{f}$.
The ancilla controls a conditional unitary on $S$:
If $\mathcal{A}$ is in state $\ket{0}$, 
$S$ is rotated to $\U \ket{b}$.
If $\mathcal{A}$ is in $\ket{1}$, $S$ is rotated to $\ket{a}$.
The ancilla's state is rotated about
the $x$-axis [if the imaginary part $\Im ( \langle a | \U | b \rangle )$
is being inferred]
or about the $y$-axis [if the real part
$\Re ( \langle a | \U | b \rangle )$ is being inferred].
The ancilla's $\sigma_z$ and the system's $\Set{ \ket{a} }$ are measured.
The outcome probabilities imply the value of $\langle a | \U | b \rangle$.
Details appear in Appendix~\ref{section:Interfere}.

The time parameter $t$ need not be negated
in any implementation of the protocol.
The absence of time reversal has been regarded as beneficial
in OTOC-measurement schemes~\cite{Yao_16_Interferometric,Zhu_16_Measurement},
as time reversal can be difficult to implement.

Interference and weak measurement have been performed with cold atoms~\cite{Smith_04_Continuous},
which have been proposed as platforms for realizing scrambling and quantum chaos~\cite{Swingle_16_Measuring,Yao_16_Interferometric,Danshita_16_Creating}.
Yet cold atoms are not necessary for measuring $\tilde{A}_\rho$.
The measurement schemes in this paper are platform-nonspecific.

\subsection{Measurable random variables $W$ and $W'$}

The combined quantum amplitude $\tilde{A}_\rho$
is defined in terms of 
two realizations of the protocol $\Protocol$.
The realizations yield measurement outcomes
$w_2$, $w_3$, $v_1$, and $v_2$.
Consider complex-conjugating two outcomes:
$w_3  \mapsto  w^*_3$, and $v_2  \mapsto  v^*_2$.
The four values are combined into
\begin{align}
   \label{eq:WDefs}
   W := w_3^*  v_2^*
   \quad \text{and} \quad
   W'  :=  w_2 v_1 \, .
\end{align}

Suppose, for example, that $\W$ and $V$ denote
single-qubit Paulis. $( W, W')$ can equal
$( 1, 1 ) , ( 1, - 1 ) , ( -1, 1 )$, or $( -1, -1 )$.
$W$ and $W'$ function analogously to 
the thermodynamic work in Jarzynski's Equality:
$W$, $W'$, and work are random variables
calculable from measurement outcomes.

%
%
\subsection{Complex distribution function $P(W, W')$}
\label{section:PW}

Jarzynski's Equality depends on a probability distribution $P(W)$.
I define an analog $P(W, W')$
in terms of the combined quantum amplitude $\tilde{A}_\rho$.

Consider fixing $W$ and $W'$.
For example, let $(W, W' )  =  ( 1 , -1 )$.
Consider the set of all possible outcome octuples
$(w_2,  \DegenW_{w_2} ; w_3, \DegenW_{w_3} ;
    v_1 ,  \DegenV_{v_1} ;  v_2, \DegenV_{v_2}  )$
that satisfy the constraints $W = w_3^*  v_2^*$ and $W' = w_2 v_1$.
Each octuple corresponds to 
a set of combined quantum amplitudes
$\tilde{A}_\rho ( w, v, \DegenW_w, \DegenV_v )$.
These $\tilde{A}_\rho$'s are summed,
subject to the constraints:
\begin{align}
   \label{eq:PDef}
   P (W, W')  & :=  \sum_{ w, v, \DegenW_w, \DegenV_v }
   \tilde{A}_\rho  ( w, v, \DegenW_w, \DegenV_v )  
   \nonumber \\ & \qquad \times
   \delta_{ W ( w_3^*  v_2^* ) } \,
   \delta_{ W' ( w_2 v_1 ) } \, .
\end{align}
The Kronecker delta is denoted by $\delta_{ab}$.

The form of Eq.~\eqref{eq:PDef} is analogous to 
the form of the $P(W)$ in~\cite{TLH_07_Work} [Eq.~(10)],
as $\tilde{A}_\rho$ is nearly a probability.
Equation~\eqref{eq:PDef}, however, encodes interference
of quantum probability amplitudes.

$P(W, W')$ resembles a joint probability distribution.
Summing any function $f(W, W')$ with weights $P(W, W')$ 
yields the average-like quantity
\begin{align}
   \label{eq:ExpValDef}
   \expval{ f(W, W') }  :=
   \sum_{W, W'}  f(W, W')  \,  P(W, W') \, .
\end{align}

\section{Result} 

The above definitions feature in the Jarzynski-like equality for the OTOC.

\begin{theorem}
\label{theorem:OTOC_FT}
The out-of-time-ordered correlator
obeys the Jarzynski-like equality
\begin{align}
   \label{eq:Result}
   \C(t) =  \frac{ \partial^2 }{ \partial \beta  \,  \partial \beta' }  \:
      \expval{ e^{ - ( \beta W + \beta' W' ) }  }  
      \Big\lvert_{ \beta,  \beta' = 0 }   \, ,
\end{align}
wherein $\beta, \beta'  \in  \mathbb{R}$.
\end{theorem}
\noindent 

\begin{proof}

The derivation of Eq.~\eqref{eq:Result} 
is inspired by~\cite{TLH_07_Work}.
Talkner \emph{et al.} cast Jarzynski's Equality in terms of 
a time-ordered correlator of two exponentiated Hamiltonians.
Those authors invoke the characteristic function
\begin{align}
   \label{eq:CharDef1}
   \Charac(s)  :=  \int dW \; e^{i s W} \, P(W) \, ,
\end{align}
the Fourier transform of the probability distribution $P(W)$.
The integration variable $s$ is regarded as
an imaginary inverse temperature: $i s = - \beta$.
We analogously invoke the (discrete) Fourier transform of $P(W, W')$:
\begin{align}
   \label{eq:CharDef2}
   \Charac(s, s')  :=  \sum_W e^{i s W}   \sum_{W'} e^{i s' W'}
   P(W, W') \, ,
\end{align}
wherein $is = - \beta$ and $is'  =  - \beta'$.

$P(W, W')$ is substituted in from 
Eqs.~\eqref{eq:PDef} and~\eqref{eq:TildeAExp}.
The delta functions are summed over:
\begin{align}
   & \Charac (s, s')  =  
   \sum_{ w, v, \DegenW_w, \DegenV_v  }   
   e^{i s w_3^* v_2^*} \: e^{i s' w_2 v_1 } \:
   \langle  w_3,  \DegenW_{w_3}  |  U  |  v_2, \DegenV_{v_2}  \rangle
   \nonumber \\ & \qquad \times
   \langle  v_2,  \DegenV_{v_2}  |  U^\dag  |  w_2 ,  \DegenW_{w_2}  \rangle
   \langle     w_2 ,  \DegenW_{w_2}   |  U  |  v_1,  \DegenV_{v_1}  \rangle
   \nonumber \\ & \qquad \times
   \langle  v_1,  \DegenV_{v_1}  |  U^\dag  \rho(t)  |  w_3,  \DegenW_{w_3}  \rangle  \, .
\end{align}
The $\rho U^\dag$ in Eq.~\eqref{eq:TildeAExp} 
has been replaced with $U^\dag \rho(t)$,
wherein $\rho(t)  :=  U \rho U^\dag$.

The sum over $( w_3 ,  \DegenW_{w_3 }  )$ is recast as a trace.
Under the trace's protection, $\rho(t)$ is shifted
to the argument's left-hand side.
The other sums and the exponentials
are distributed across the product:
\begin{align}
   \Charac  & (s, s')  =
   \Tr \Bigg(  \rho(t) 
   \Bigg[ \sum_{ w_3, \DegenW_{w_3} }   
             \ketbra{ w_3, \DegenW_{w_3} }{ w_3, \DegenW_{w_3} }
             \nonumber \\ & \qquad \times
             U  \sum_{v_2, \DegenV_{v_2} }  
                      e^{is { {w_3}^* } v_2^*}    
                      \ketbra{ v_2, \DegenV_{v_2} }{ v_2, \DegenV_{v_2} }  U^\dag \Bigg]
   \nonumber \\ &  \times
   \Bigg[ \sum_{ w_2, \DegenW_{w_2} }  
             \ketbra{ w_2, \DegenW_{w_2} }{ w_2, \DegenW_{w_2} } 
             \nonumber \\ & \qquad \times
             U  \sum_{ v_1, \DegenV_{v_1} } 
                      e^{ i s' w_2 v_1 }  
                      \ketbra{  v_1, \DegenV_{v_1}  }{  v_1, \DegenV_{v_1}  }  \,  U^\dag
   \Bigg] \Bigg)  \, .
\end{align}

The $v_\ell$ and $\DegenV_{v_\ell}$ sums are eigendecompositions
of exponentials of unitaries:
\begin{align}
   \Charac  & (s, s')  =
   \Tr \Bigg(   \rho(t) 
   \Bigg[  \sum_{ w_3, \DegenW_{w_3} }   
             \ketbra{ w_3, \DegenW_{w_3} }{ w_3, \DegenW_{w_3} }     
             U  \,  e^{is w_3^* V^\dag}  \, U^\dag   \Bigg]
   \nonumber \\ & \qquad   \times
   \Bigg[  \sum_{ w_2,  \DegenW_{w_2} }  
              \ketbra{ w_2,  \DegenW_{w_2} }{ w_2,  \DegenW_{w_2} }   
              U  \,e^{ i s' w_2 V}  \, U^\dag  \Bigg]
   \Bigg)  \, .
\end{align}
The unitaries time-evolve the $V$'s:
\begin{align}
   \Charac  & (s, s')  =
   \Tr \Bigg(   \rho(t)
   \Bigg[  \sum_{  w_3, \DegenW_{w_3}  }   
            \ketbra{ w_3, \DegenW_{w_3} }{ w_3, \DegenW_{w_3} }   
            e^{is w_2^* V^\dag( -t )}  \Bigg]
   \nonumber \\ &   \qquad \times
   \Bigg[  \sum_{ w_2,  \DegenW_{w_2} }  
              \ketbra{ w_2,  \DegenW_{w_2} }{ w_2,  \DegenW_{w_2} }   
              e^{ i s' w_2 V( -t )}  \Bigg]  
   \Bigg)  \, .
\end{align}

We differentiate with respect to $is' = - \beta'$
and with respect to $is = - \beta$.
Then, we take the limit as $\beta, \beta' \to 0$:
\begin{align}
   \label{eq:GEq}
   & \frac{ \partial^2 }{ \partial \beta \, \partial \beta' }
   \Charac  \left( i \beta,  i \beta'  \right)  
   \Big\lvert_{\beta, \beta' = 0}
   \\  & 
   =   \Tr \Bigg(   \rho(t)
   \Bigg[  \sum_{  w_3, \DegenW_{w_3} }    w_3^* 
              \ketbra{ w_3, \DegenW_{w_3} }{ w_3, \DegenW_{w_3} }  
              V^\dag( -t )  \Bigg]
   \\ \nonumber & \qquad \qquad \times
   \Bigg[  \sum_{ w_2,  \DegenW_{w_2} }  w_2  
              \ketbra{ w_2,  \DegenW_{w_2} }{ w_2,  \DegenW_{w_2} }
              V( -t )  \Bigg]  \Bigg)  \\
   &  =   \label{eq:GEq2}
    \Tr \LParen  \rho(t)  \,
    \W^\dag  \,  V^\dag(-t)  \,  \W  \,  V( -t )  \RParen  \, .
\end{align}

Recall that $\rho(t)  :=  U \rho U^\dag$.
Time dependence is transferred 
from $\rho(t)$, $V(-t) = U V^\dag U^\dag$, 
and $V^\dag(t) = U V U^\dag$ to $\W^\dag$ and $\W$,
under the trace's cyclicality:
\begin{align}
   & \frac{ \partial^2 }{ \partial \beta \, \partial \beta' }
   \Charac  ( i \beta,  i \beta')  \Big\lvert_{\beta, \beta' = 0}
   \label{eq:GEq3}
   =   \Tr \left(  \rho  \,
   \W^\dag(t)  \,  V^\dag  \,  \W(t)  \,  V  \right)  \\
   & \qquad \qquad \qquad \qquad
   =  \expval{ \W^\dag(t)  \,  V^\dag  \,  \W(t)  \,  V } 
   =  \C(t) \, .
\end{align}
By Eqs.~\eqref{eq:ExpValDef} and~\eqref{eq:CharDef2},
the left-hand side equals 
\begin{align}
   \frac{ \partial^2 }{ \partial \beta \, \partial \beta' }
   \expval{ e^{ - ( \beta W + \beta' W') } }  
   \Big\lvert_{\beta, \beta' = 0} \, .
\end{align}
\end{proof} 

Theorem~\ref{theorem:OTOC_FT} resembles 
Jarzynski's fluctuation relation in several ways.
Jarzynski's Equality encodes a scheme for measuring 
the difficult-to-calculate $\Delta F$ from realizable nonequilibrium trials.
Theorem~\ref{theorem:OTOC_FT} encodes a scheme for measuring
the difficult-to-calculate $\C(t)$ from realizable nonequilibrium trials.
$\Delta F$ depends on just 
a temperature and two Hamiltonians.
Similarly, the conventional $\C(t)$ 
(defined with respect to $\rho = e^{ - H / T } / Z$) 
depends on just a temperature, a Hamiltonian, 
and two unitaries.
Jarzynski relates $\Delta F$ to 
the characteristic function of a probability distribution.
Theorem~\ref{theorem:OTOC_FT} relates $\C(t)$ to
(a moment of) the characteristic function
of a (complex) distribution.

The complex distribution, $P(W, W')$,
is a combination of probability amplitudes $\tilde{A}_\rho$
related to quasiprobabilities.
The distribution in Jarzynski's Equality
is a combination of probabilities.
The quasiprobability-vs.-probability contrast  
fittingly arises from the OTOC's out-of-time ordering.
$\C(t)$ signals quantum behavior (noncommutation), 
as quasiprobabilities signal quantum behaviors (e.g., entanglement).
Time-ordered correlators similar to $\C(t)$ 
track only classical behaviors
and are moments of (summed) classical probabilities~\cite{BrianDisc}.
OTOCs that encode more time reversals than $\C(t)$
are moments of combined quasiprobability-like distributions
lengthier than $\tilde{A}_\rho$~\cite{BrianDisc}.

\section{Conclusions}

The Jarzynski-like equality for the out-of-time correlator
combines an important tool from nonequilibrium statistical mechanics
with an important tool from 
quantum information, high-energy theory, and condensed matter.
The union opens all these fields to new modes of analysis.

For example, Theorem~\ref{theorem:OTOC_FT} relates the OTOC
to a combined quantum amplitude $\tilde{A}_\rho$.
This $\tilde{A}_\rho$ is closely related to a quasiprobability.
The OTOC and quasiprobabilities have signaled nonclassical behaviors
in distinct settings---in high-energy theory and condensed matter 
and in quantum optics, respectively.
The relationship between OTOCs and quasiprobabilities merits study:
What is the relationship's precise nature?
How does $\tilde{A}_\rho$ behave over time scales
during which $\C(t)$ exhibits known behaviors
(e.g., until the dissipation time
or from the dissipation time to the scrambling time~\cite{Swingle_16_Measuring})?
Under what conditions does $\tilde{A}_\rho$ behave nonclassically
(assume negative or nonreal values)?
How does a chaotic system's $\tilde{A}_\rho$ look?
These questions are under investigation~\cite{BrianDisc}.

As another example, fluctuation relations have been used
to estimate the free-energy difference $\Delta F$
from experimental data.
Experimental measurements of $\C(t)$ are possible for certain platforms,
in certain regimes~\cite{Swingle_16_Measuring,Yao_16_Interferometric,Zhu_16_Measurement,Li_16_Measuring,Garttner_16_Measuring}.
Theorem~\ref{theorem:OTOC_FT} expands the set of platforms and regimes.
Measuring quantum amplitudes, as via weak measurements~\cite{Lundeen_11_Direct,Lundeen_12_Procedure,Bamber_14_Observing,Mirhosseini_14_Compressive},
now offers access to $\C(t)$.
Inferring small systems' $\tilde{A}_\rho$'s with existing platforms~\cite{White_16_Preserving}
might offer a challenge for the near future.

Finally, Theorem~\ref{theorem:OTOC_FT} can provide 
a new route to bounding $\C(t)$.
A Lyapunov exponent $\lambda_{\text{L}}$ governs the chaotic decay of $\C(t)$.
The exponent has been bounded, 
including with Lieb-Robinson bounds and complex analysis~\cite{Maldacena_15_Bound,Lashkari_13_Towards,Kitaev_15_Simple}.
The right-hand side of Eq.~\eqref{eq:Result} can provide
an independent bounding method that offers new insights.

\endgroup

%
%

\putbib[Jarz_like_bib] 
\end{bibunit}

%
%
\chapter{The quasiprobability behind the out-of-time-ordered correlator}
\label{ch:OTOC_Quasi}
\begin{bibunit}  

\noindent \emph{This chapter was published as~\cite{NYH_17_Quasi}.}

\begingroup


\newcommand{\W}{ \mathcal{W} }  
\newcommand{\Dim}{ d }  
\newcommand{\DegenW}{ \alpha }  
\newcommand{\DegenV}{ \lambda }  
\newcommand{\Sites}{N}  
\newcommand{\Sys}{S}  
\newcommand{\td}{t_{\text{d}}}  
\newcommand{\Lyap}{\lambda_{\text{L}}}  
\newcommand{\KD}{ \tilde{p}_{\text{KD}} }  
\newcommand*{\OurKD}[1]{\tilde{A}_{#1}}  
\newcommand*{\SumKD}[1]{\tilde{ \mathscr{A} }_{#1}}  
\newcommand*{\ProjW}[1]{\Pi^{ \W }_{#1}}  
\newcommand*{\ProjWt}[1]{\Pi^{ \W(t) }_{#1}}  
\newcommand*{\ProjV}[1]{\Pi^{ V }_{#1}}  

\newcommand{\NondegW}{ \tilde{\W} }
\newcommand{\NondegV}{ \tilde{V} }
\newcommand{\reg}{ {\text{reg}} } 
\newcommand{\TOC}{F_\toc} 
\newcommand{\GW}{ G_\W }
\newcommand{\GV}{ G_V }
\newcommand{\gw}{ g_w }
\newcommand{\gv}{ g_v }
\newcommand{\gwP}{ g_{w'} }
\newcommand{\gvP}{ g_{v'} }
\newcommand{\Coupling}{c}
\newcommand{\Charac}{ \mathcal{G} }  
\newcommand{\U}{ \mathcal{U} }  
\newcommand{\weak}{ {\text{weak}} }
\newcommand{\target}{ {\text{target}} }
\newcommand{\WeakInt}{ \mathcal{I} }
\newcommand{\ParenW}{{(\W)}}
\newcommand{\ParenV}{{(V)}}
\newcommand{\Protocol}{\mathcal{P}}  
\newcommand{\Protocoll}{\mathcal{P}_{\W}}  
\newcommand{\ProtocolA}{\mathscr{P}_\Amp}  
\newcommand{\toc}{{\text{TOC}}}
\newcommand*{\TOCKD}[1]{\tilde{A}^\toc_{#1}}  
\newcommand{\A}{\mathcal{A}} 
\newcommand{\B}{ \mathcal{B} } 
\newcommand{\C}{ \mathcal{C} } 
\newcommand{\K}{ \mathcal{K} } 
\newcommand{\Oper}{\mathcal{O}} 
\newcommand{\Ops}{\mathscr{K}} 
\newcommand{\Opsb}{\bar{\mathscr{K}}} 
\newcommand{\Gest}{ \Gamma_{\text{est}} } 
\newcommand{\TU}{ \tilde{U} } 
\newcommand{\TRho}{ \tilde{\rho} } 
\newcommand{\Amp}{A} 

\newcommand*{\Unit}[1]{ \bm{ \hat{ #1 }} }  
\newcommand{\ParenA}{{(a)}}
\newcommand{\ParenB}{{(b)}}
\newcommand{\ParenK}{{(\Ops)}}
\newcommand{\ParenKB}{{( \Opsb )}}

%
%
Two topics have been flourishing independently:
the out-of-time-ordered correlator (OTOC)
and the Kirkwood-Dirac (KD) quasiprobability distribution.
The OTOC signals chaos,
and the dispersal of information through entanglement,
in quantum many-body systems~\cite{Shenker_Stanford_14_BHs_and_butterfly,Shenker_Stanford_14_Multiple_shocks,Shenker_Stanford_15_Stringy,Roberts_15_Localized_shocks,Roberts_Stanford_15_Diagnosing,Maldacena_15_Bound}.
Quasiprobabilities represent quantum states
as phase-space distributions represent statistical-mechanical states~\cite{Carmichael_02_Statistical}.
Classical phase-space distributions are restricted to positive values;
quasiprobabilities are not.
The best-known quasiprobability is the Wigner function.
The Wigner function can become negative;
the KD quasiprobability, negative and nonreal~\cite{Kirkwood_33_Quantum,Dirac_45_On,Lundeen_11_Direct,Lundeen_12_Procedure,Bamber_14_Observing,Mirhosseini_14_Compressive,Dressel_15_Weak}.
Nonclassical values flag contextuality,
a resource underlying quantum-computation speedups~\cite{Spekkens_08_Negativity,Ferrie_11_Quasi,Kofman_12_Nonperturbative,Dressel_14_Understanding,Howard_14_Contextuality,Dressel_15_Weak,Delfosse_15_Wigner}.
Hence the KD quasiprobability, like the OTOC,
reflects nonclassicality.

Yet disparate communities use these tools:
The OTOC $F(t)$ features in quantum information theory,
high-energy physics, and condensed matter. Contexts include black holes within AdS/CFT duality~\cite{Shenker_Stanford_14_BHs_and_butterfly,Maldacena_98_AdSCFT,Witten_98_AdSCFT,Gubser_98_AdSCFT}, weakly interacting field theories~\cite{Stanford_15_WeakCouplingChaos,Patel_16_ChaosCritFS,Chowdhury_17_ONChaos,Patel_17_DisorderMetalChaos}, spin models~\cite{Shenker_Stanford_14_BHs_and_butterfly,HosurYoshida_16_Chaos}, and the Sachdev-Ye-Kitaev model~\cite{Sachdev_93_Gapless,Kitaev_15_Simple}. The KD distribution features in quantum optics.
Experimentalists have inferred the quasiprobability
from weak measurements of photons~\cite{Bollen_10_Direct,Lundeen_11_Direct,Lundeen_12_Procedure,Bamber_14_Observing,Mirhosseini_14_Compressive,Suzuki_16_Observation,Piacentini_16_Measuring,Thekkadath_16_Direct}
and superconducting qubits~\cite{White_16_Preserving,Groen_13_Partial}.

The two tools were united in~\cite{YungerHalpern_17_Jarzynski}.
The OTOC was shown to equal a moment of
a summed quasiprobability, $\OurKD{\rho}$:
\begin{align}
   \label{eq:JarzLike}
   F (t)  =  \frac{ \partial^2 }{ \partial \beta  \,  \partial \beta' }
   \expval{ e^{ - ( \beta W + \beta' W' ) } }
   \Bigg\rvert_{ \beta, \beta' = 0 }   \, .
\end{align}
$W$ and $W'$ denote measurable random variables
analogous to thermodynamic work; and $\beta, \beta'  \in  \mathbb{R}$.
The average $\expval{ . }$ is with respect to
a sum of quasiprobability values $\OurKD{\rho} ( . )$.
Equation~\eqref{eq:JarzLike} resembles Jarzynski's Equality,
a fluctuation relation in nonequilibrium statistical mechanics~\cite{Jarzynski_97_Nonequilibrium}.
Jarzynski cast a useful, difficult-to-measure free-energy difference $\Delta F$
in terms of the characteristic function of a probability.
Equation~\eqref{eq:JarzLike} casts the useful, difficult-to-measure OTOC
in terms of the characteristic function of a summed quasiprobability.\footnote{
For a thorough comparison of Eq.~\eqref{eq:JarzLike}
with Jarzynski's equality,
see the two paragraphs that follow the proof in~\cite{YungerHalpern_17_Jarzynski}.
}
The OTOC has recently been linked to thermodynamics also in~\cite{Campisi_16_Thermodynamics,Tsuji_16_Out}.

Equation~\eqref{eq:JarzLike} motivated 
definitions of quantities
that deserve study in their own right.
The most prominent quantity is
the quasiprobability $\OurKD{\rho}$.
$\OurKD{\rho}$ is more fundamental than $F(t)$:
$\OurKD{\rho}$ is a distribution that consists of many values.
$F(t)$ equals a combination of those values---a
derived quantity, a coarse-grained quantity.
$\OurKD{\rho}$ contains more information than $F(t)$.
This paper spotlights $\OurKD{\rho}$
and related quasiprobabilities ``behind the OTOC.''

$\OurKD{\rho}$, we argue, is an extension of the KD quasiprobability.
Weak-measurement tools used to infer KD quasiprobabilities
can be applied to infer $\OurKD{\rho}$ from experiments~\cite{YungerHalpern_17_Jarzynski}.
Upon measuring $\OurKD{\rho}$, one can recover the OTOC.
Alternative OTOC-measurement proposals rely on
Lochshmidt echoes~\cite{Swingle_16_Measuring}, interferometry~\cite{Swingle_16_Measuring,Yao_16_Interferometric,YungerHalpern_17_Jarzynski,Bohrdt_16_Scrambling},
clocks~\cite{Zhu_16_Measurement},
particle-number measurements of ultracold atoms~\cite{Danshita_16_Creating,Tsuji_17_Exact,Bohrdt_16_Scrambling},
and two-point measurements~\cite{Campisi_16_Thermodynamics}.
Initial experiments have begun the push
toward characterizing many-body scrambling:
OTOCs of an infinite-temperature four-site NMR system
have been measured~\cite{Li_16_Measuring}.
OTOCs of symmetric observables have been measured
with infinite-temperature trapped ions~\cite{Garttner_16_Measuring} and in nuclear spin chains~\cite{Wei_16_NuclearSpinOTOC}.
Weak measurements offer a distinct toolkit,
opening new platforms and regimes to OTOC measurements.
The weak-measurement scheme in~\cite{YungerHalpern_17_Jarzynski}
is expected to provide a near-term challenge for
superconducting qubits~\cite{White_16_Preserving,Hacohen_16_Quantum,Rundle_16_Quantum,Takita_16_Demonstration,Kelly_15_State,Heeres_16_Implementing,Riste_15_Detecting},
trapped ions~\cite{Gardiner_97_Quantum,Choudhary_13_Implementation,Lutterbach_97_Method,Debnath_16_Nature,Monz_16_Realization,Linke_16_Experimental,Linke_17_Experimental},
ultracold atoms~\cite{Browaeys_16_Experimental}, cavity quantum electrodynamics (QED)~\cite{Guerlin_07_QND,Murch_13_SingleTrajectories}, and perhaps NMR~\cite{Xiao_06_NMR,Dawei_14_Experimental}.

We investigate the quasiprobability $\OurKD{\rho}$
that ``lies behind'' the OTOC.
The study consists of three branches:
We discuss experimental measurements,
calculate (a coarse-grained) $\OurKD{\rho}$,
and explore mathematical properties.
Not only does quasiprobability theory shed new light on the OTOC.
The OTOC also inspires questions about quasiprobabilities
and motivates weak-measurement experimental challenges.

The paper is organized as follows.
In a technical introduction, we review the KD quasiprobability, the OTOC,
the OTOC quasiprobability $\OurKD{\rho}$,
and schemes for measuring $\OurKD{\rho}$.
We also introduce our set-up and notation.
All the text that follows the technical introduction is new
(never published before, to our knowledge).

Next, we discuss experimental measurements.
We introduce a coarse-graining $\SumKD{\rho}$ of $\OurKD{\rho}$.
The coarse-graining involves a ``projection trick'' that decreases,
exponentially in system size, the number of trials required
to infer $F(t)$ from weak measurements.
We evaluate pros and cons of
the quasiprobability-measurement schemes in~\cite{YungerHalpern_17_Jarzynski}.
We also compare our schemes
with alternative $F(t)$-measurement schemes~\cite{Swingle_16_Measuring,Yao_16_Interferometric,Zhu_16_Measurement}.
We then present a circuit for weakly measuring
a qubit system's $\SumKD{\rho}$.
Finally, we show how to infer the coarse-grained $\SumKD{\rho}$
from alternative OTOC-measurement schemes
(e.g.,~\cite{Swingle_16_Measuring}).

Sections~\ref{section:Numerics} and~\ref{section:Brownian} feature
calculations of $\SumKD{\rho}$.
First, we numerically simulate a transverse-field Ising model.
$\SumKD{\rho}$ changes significantly, we find,
over time scales relevant to the OTOC.
The quasiprobability's behavior distinguishes
nonintegrable from integrable Hamiltonians.
The quasiprobability's negativity and nonreality remains robust
with respect to substantial quantum interference.
We then calculate an average, over Brownian circuits, of $\SumKD{\rho}$.
Brownian circuits model chaotic dynamics:
The system is assumed to evolve, at each time step,
under random two-qubit couplings~\cite{Brown_13_Scrambling,Hayden_07_Black,Sekino_08_Fast,Lashkari_13_Towards}.

A final ``theory'' section concerns mathematical properties
and physical interpretations of $\OurKD{\rho}$.
$\OurKD{\rho}$ shares some, though not all,
of its properties with the KD distribution.
The OTOC motivates a generalization of
a Bayes-type theorem obeyed by the KD distribution~\cite{Aharonov_88_How,Johansen_04_Nonclassical,Hall_01_Exact,Hall_04_Prior,Dressel_15_Weak}.
The generalization exponentially shrinks the memory required
to compute weak values, in certain cases.
The OTOC also motivates a generalization of
decompositions of quantum states $\rho$.
This decomposition property may help experimentalists assess
how accurately they prepared the desired initial state
when measuring $F(t)$.
A time-ordered correlator $F_\toc(t)$ analogous to $F(t)$, we show next,
depends on a quasiprobability that can reduce to a probability.
The OTOC quasiprobability lies farther from classical probabilities
than the TOC quasiprobability,
as the OTOC registers quantum-information scrambling
that $F_\toc(t)$ does not.
Finally, we recall that the OTOC encodes three time reversals.
OTOCs that encode more are 
moments of sums of ``longer'' quasiprobabilities.
We conclude with theoretical and experimental opportunities.

We invite readers to familiarize themselves with the technical review,
then to dip into the sections that interest them most.
The technical review is intended to introduce
condensed-matter, high-energy, and quantum-information readers
to the KD quasiprobability
and to introduce quasiprobability and weak-measurement readers
to the OTOC.
Armed with the technical review,
experimentalists may wish to focus on Sec.~\ref{section:Measuring}
and perhaps Sec.~\ref{section:Numerics}.
Adherents of abstract theory may prefer Sec.~\ref{section:Theory}.
The computationally minded may prefer
Sections~\ref{section:Numerics} and~\ref{section:Brownian}.
The paper's modules (aside from the technical review) are independently accessible.


%
%
%
\section{Technical introduction}
\label{section:Tech_intro}

This review consists of three parts.
In Sec.~\ref{section:Intro_to_KD}, we overview the KD quasiprobability.
Section~\ref{section:SetUp} introduces our set-up and notation.
In Sec.~\ref{section:OTOC_review}, we review the OTOC and
its quasiprobability $\OurKD{\rho}$.
We overview also the weak-measurement and interference schemes
for measuring $\OurKD{\rho}$ and $F(t)$.

The quasiprobability section (\ref{section:Intro_to_KD}) provides background for
quantum-information, high-energy, and condensed-matter readers.
The OTOC section (\ref{section:OTOC_review}) targets
quasiprobability and weak-measurement readers.
We encourage all readers to study
the set-up (\ref{section:SetUp}), as well as
$\OurKD{\rho}$ and the schemes for measuring $\OurKD{\rho}$ (\ref{section:Review_OTOC_quasiprob}).

\subsection{The KD quasiprobability in quantum optics}
\label{section:Intro_to_KD}

The Kirkwood-Dirac quasiprobability is defined as follows.
Let $\Sys$ denote a quantum system
associated with a Hilbert space $\Hil$.
Let $\Set{ \ket{ a } }$ and $\Set{ \ket{ f } }$
denote orthonormal bases for $\Hil$.
Let $\mathcal{B} ( \Hil )$ denote the set of bounded operators
defined on $\Hil$, and let $\Oper  \in  \mathcal{B} ( \Hil )$.
The KD quasiprobability
\begin{align}
   \label{eq:KD_rho}   
   \OurKD{\Oper}^\1 ( a, f )  :=
   \langle f | a \rangle  \langle a |  \Oper  | f \rangle  \, ,
\end{align}
regarded as a function of $a$ and $f$,
contains all the information in $\Oper$,
if $\langle a | f \rangle \neq 0$ for all $a, f$.
Density operators $\Oper = \rho$ are often focused on
in the literature and in this paper.
This section concerns the context, structure,
and applications of $\OurKD{\Oper}^\1 ( a, f )$.

We set the stage with phase-space representations of quantum mechanics,
alternative quasiprobabilities, and historical background.
Equation~\eqref{eq:KD_rho}
facilitates retrodiction, or inference about the past,
reviewed in Sec.~\ref{section:KD_Retro}.
How to decompose an operator $\Oper$
in terms of KD-quasiprobability values
appears in Sec.~\ref{section:KD_Coeffs}.
The quasiprobability has mathematical properties
reviewed in Sec.~\ref{section:KDProps}.

Much of this section parallels Sec.~\ref{section:Theory},
our theoretical investigation of the OTOC quasiprobability.
More background appears in~\cite{Dressel_15_Weak}.

\subsubsection{Phase-space representations, alternative quasiprobabilities, and history}

Phase-space distributions form a mathematical toolkit
applied in Liouville mechanics~\cite{Landau_80_Statistical}.
Let $\Sys$ denote a system of $6 \Sites$ degrees of freedom (DOFs).
An example system consists of $\Sites$ particles,
lacking internal DOFs,
in a three-dimensional space.
We index the particles with $i$ and let $\alpha = x, y, z$.
The $\alpha^\th$ component $q_i^\alpha$
of particle $i$'s position
is conjugate to
the $\alpha^\th$ component $p_i^\alpha$
of the particle's momentum.
The variables $q_i^\alpha$ and $p_i^\alpha$ label
the axes of \emph{phase space}.

Suppose that the system contains many DOFs: $\Sites \gg 1$.
Tracking all the DOFs is difficult.
Which phase-space point $\Sys$ occupies,
at any instant, may be unknown.
The probability that, at time $t$, $\Sys$ occupies
an infinitesimal volume element
localized at $(q_1^x, \ldots, p_N^z)$ is
$\rho( \{ q_i^\alpha \} , \{ p_i^\alpha \}; t ) \, d^{3N} q  \:  d^{3N} p$.
The \emph{phase-space distribution}
$\rho( \{ q_i^\alpha \} , \{ p_i^\alpha \}; t )$
is a probability density.

$q_i^\alpha$ and $p_i^\alpha$
seem absent from quantum mechanics (QM), \emph{prima facie}.
Most introductions to QM cast quantum states in terms of
operators, Dirac kets $\ket{ \psi }$, and wave functions $\psi (x)$.
Classical variables are relegated to measurement outcomes
and to the classical limit.
Wigner, Moyal, and others represented QM
in terms of phase space~\cite{Carmichael_02_Statistical}.
These representations are used most in quantum optics.

In such a representation, a \emph{quasiprobability} density replaces
the statistical-mechanical probability density $\rho$.\footnote{
We will focus on discrete quantum systems,
motivated by a spin-chain example.
Discrete systems are governed by quasiprobabilities,
which resemble probabilities.
Continuous systems are governed by
quasiprobability densities,
which resemble probability densities.
Our quasiprobabilities can be replaced with quasiprobability densities,
and our sums can be replaced with integrals,
in, e.g., quantum field theory.}
Yet quasiprobabilities violate axioms of probability~\cite{Ferrie_11_Quasi}.
Probabilities are nonnegative, for example.
Quasiprobabilities can assume negative values,
associated with nonclassical physics such as contextuality~\cite{Spekkens_08_Negativity,Ferrie_11_Quasi,Kofman_12_Nonperturbative,Dressel_14_Understanding,Dressel_15_Weak,Delfosse_15_Wigner},
and nonreal values.
Relaxing different axioms leads to different quasiprobabilities.
Different quasiprobabilities correspond also to
different orderings of noncommutative operators~\cite{Dirac_45_On}.
The best-known quasiprobabilities include
the Wigner function, the Glauber-Sudarshan $P$ representation,
and the Husimi $Q$ function~\cite{Carmichael_02_Statistical}.

The KD quasiprobability resembles a little brother of theirs,
whom hardly anyone has heard of~\cite{Banerji_07_Exploring}.
Kirkwood and Dirac defined the quasiprobability independently
in 1933~\cite{Kirkwood_33_Quantum} and 1945~\cite{Dirac_45_On}.
Their finds remained under the radar for decades.
Rihaczek rediscovered the distribution in 1968,
in classical-signal processing~\cite{Rihaczek_68_Signal,Cohen_89_Time}.
(The KD quasiprobability is sometimes called
``the Kirkwood-Rihaczek distribution.'')
The quantum community's attention has revived recently.
Reasons include experimental measurements, mathematical properties,
and applications to retrodiction and state decompositions.

\subsubsection{Bayes-type theorem and retrodiction with
the KD quasiprobability}
\label{section:KD_Retro}

Prediction is inference about the future.
\emph{Retrodiction} is inference about the past. 
One uses the KD quasiprobability to infer about a time $t'$,
using information about an event that occurred before $t'$
and information about an event that occurred after $t'$.
This forward-and-backward propagation
evokes the OTOC's out-of-time ordering.

We borrow notation from,
and condense the explanation in,~\cite{Dressel_15_Weak}.
Let $\Sys$ denote a discrete quantum system.
Consider preparing $\Sys$ in
a state $\ket{ i }$ at time $t = 0$.
Suppose that $\Sys$ evolves under a time-independent Hamiltonian
that generates the family $U_t$ of unitaries.
Let $F$ denote an observable
measured at time $t'' > 0$.
Let $F = \sum_f  f  \ketbra{f }{ f }$ be the eigendecomposition,
and let $f$ denote the outcome.

Let $\A  =  \sum_a a  \ketbra{a}{a}$ be
the eigendecomposition of an observable
that fails to commute with $F$.
Let $t'$ denote a time in $(0, t'')$.
Which value can we most reasonably attribute
to the system's time-$t'$ $\A$,
knowing that $\Sys$ was prepared in $\ket{i}$
and that the final measurement yielded $f$?

Propagating the initial state forward to time $t'$ yields
$\ket{ i' }  :=  U_{t'}  \ket{i}$.
Propagating the final state backward yields
$\ket{f'}  :=  U^\dag_{t'' - t'} \ket{f}$.
Our best guess
about $\A$ is the \emph{weak value}~\cite{Ritchie_91_Realization,Hall_01_Exact,Johansen_04_Nonclassical,Hall_04_Prior,Pryde_05_Measurement,Dressel_11_Experimentals,Groen_13_Partial}
\begin{align}
   \label{eq:WeakVal}
   \A_\weak (i, f )  :=  \Re  \left(
   \frac{ \langle f' | \A | i' \rangle }{  \langle f' | i' \rangle }
   \right)  \, .
\end{align}
The real part of a complex number $z$
is denoted by $\Re (z)$.
The guess's accuracy is quantified with
a distance metric (Sec.~\ref{section:TA_retro})
and with comparisons to weak-measurement data.

Aharonov \emph{et al.} discovered weak values
in 1988~\cite{Aharonov_88_How}.
Weak values be \emph{anomalous}, or \emph{strange}:
$\A_\weak$ can exceed the greatest eigenvalue $a_\Max$ of $\A$
and can dip below the least eigenvalue $a_\Min$.
Anomalous weak values concur with
negative quasiprobabilities and nonclassical physics~\cite{Kofman_12_Nonperturbative,Dressel_14_Understanding,Pusey_14_Anomalous,Dressel_15_Weak,Waegell_16_Confined}.
Debate has surrounded weak values' role in quantum mechanics~\cite{Ferrie_14_How,Vaidman_14_Comment,Cohen_14_Comment,Aharonov_14,Sokolovski_14_Comment,Brodutch_15_Comment,Ferrie_15_Ferrie}.

The weak value $\A_\weak$, we will show,
depends on the KD quasiprobability.
We replace the $\A$ in Eq.~\eqref{eq:WeakVal}
with its eigendecomposition.
Factoring out the eigenvalues yields
\begin{align}
  \label{eq:WeakVal2}
   \A_\weak( i , f )  =   \sum_a  a  \,  \Re  \left(
   \frac{ \langle f' | a \rangle \langle a | i' \rangle }{
            \langle f' | i' \rangle }  \right) \, .
\end{align}
The weight $\Re ( . )$ is a \emph{conditional quasiprobability}.
It resembles a conditional probability---the
likelihood that, if $\ket{i}$ was prepared
and the measurement yielded $f$,
$a$ is the value most reasonably attributable to $\A$.
Multiplying and dividing the argument
by $\langle i' | f' \rangle$ yields
\begin{align}
   \label{eq:CondQuasi}
   \tilde{p} ( a | i, f )  :=   \frac{
   \Re  \left(  \langle f' | a \rangle \langle a | i' \rangle  \langle i' | f' \rangle
                      \right) }{
   |  \langle f' | i' \rangle  |^2 }     \, .
\end{align}
Substituting  into Eq.~\eqref{eq:WeakVal2} yields
\begin{align}
   \label{eq:WeakVal3}
   \A_\weak( i , f )  =   \sum_a  a  \,
   \tilde{p} ( a | i, f )  \, .
\end{align}

Equation~\eqref{eq:WeakVal3} illustrates why
negative quasiprobabilities concur with anomalous weak values.
Suppose that $\tilde{p} ( a | i, f )  \geq  0  \;  \:  \forall a$.
The triangle inequality, followed by the Cauchy-Schwarz inequality, implies
\begin{align}
   | \A_\weak( i , f ) |  
   & \leq  \left\lvert  \sum_a  a  \,  \tilde{p} ( a | i, f )  \right\rvert \\
   & \leq  \sum_a  |a |  \cdot  |  \tilde{p} ( a | i, f )  |  \\
   \label{eq:Anom_help1}
   & \leq  |  a_\Max |  \sum_a    |  \tilde{p} ( a | i, f )  |  \\
   \label{eq:Anom_help2}
   & =  |  a_\Max |  \sum_a    \tilde{p} ( a | i, f ) \\
   & =  | a_\Max |  \, .
\end{align}
The penultimate equality follows from $\tilde{p}( a | i, f )  \geq  0$.
Suppose, now, that the quasiprobability contains 
a negative value $\tilde{p} ( a_- | i, f ) < 0$.
The distribution remains normalized.
Hence the rest of the $\tilde{p}$ values sum to $>1$.
The RHS of~\eqref{eq:Anom_help1} 
exceeds $|  a_\Max |$.
%

The numerator of Eq.~\eqref{eq:CondQuasi} is
the \emph{Terletsky-Margenau-Hill (TMH) quasiprobability}~\cite{Terletsky_37_Limiting,Margenau_61_Correlation,Johansen_04_Nonclassical,Johansen_04_Nonclassicality}.
The TMH distribution is the real part of a complex number.
That complex generalization,
\begin{align}
   \label{eq:KD}
   \langle f' | a \rangle    \langle a | i' \rangle   \langle i' | f' \rangle \, ,
\end{align}
is the KD quasiprobability~\eqref{eq:KD_rho}.

We can generalize the retrodiction argument to arbitrary states $\rho$~\cite{Wiseman_02_Weak}.
Let $\mathcal{D} ( \Hil )$ denote the set of density operators
(unit-trace linear positive-semidefinite operators)
defined on $\mathcal{H}$.
Let $\rho  =  \sum_i  p_i  \ketbra{ i }{ i }
\in  \mathcal{D} ( \Hil )$
be a density operator's eigendecomposition.
Let $\rho'  :=  U_{t'} \rho U_{t'}^\dag$.
The weak value Eq.~\eqref{eq:WeakVal} becomes
\begin{align}
   \label{eq:WeakVal_rho}
   \A_\weak ( \rho, f )  :=  \Re \left(  \frac{
   \langle f' | \A \rho' | f' \rangle }{
   \langle f' | \rho' | f' \rangle }  \right) \, .
\end{align}
Let us eigendecompose $\A$ and factor out $\sum_a a$.
The eigenvalues are weighted by the conditional quasiprobability
\begin{align}
   \tilde{p} ( a | \rho, f )  =  \frac{  \Re \left(
   \langle f' | a \rangle \langle a | \rho' | f' \rangle  \right) }{
   \langle f' | \rho' | f' \rangle } \, .
\end{align}
The numerator is the TMH quasiprobability for $\rho$.
The complex generalization
\begin{align}
   \label{eq:KD_rho_2}
   \OurKD{\rho}^\1 ( a, f )  =
   \langle f' | a \rangle  \langle a | \rho' | f' \rangle
\end{align}
is the KD quasiprobability~\eqref{eq:KD_rho} for $\rho$.\footnote{
The $A$ in the quasiprobability $\OurKD{\rho}$
should not be confused with
the observable $\A$.}
We rederive~\eqref{eq:KD_rho_2}, via
an operator decomposition, next.

\subsubsection{Decomposing operators in terms of
KD-quasiprobability coefficients}
\label{section:KD_Coeffs}

The KD distribution can be interpreted
not only in terms of retrodiction,
but also in terms of operation decompositions~\cite{Lundeen_11_Direct,Lundeen_12_Procedure}.
Quantum-information scientists decompose
qubit states in terms of Pauli operators.
Let $\bm{ \sigma } =
\sigma^x \hat{ \mathbf{ x } }  +  \sigma^y  \hat{ \mathbf{ y } }
+  \sigma^z  \hat{ \mathbf{ z } }$
denote a vector of the one-qubit Paulis.
Let $\hat{ \mathbf{n} }  \in  \mathbb{R}^3$ denote a unit vector.
Let $\rho$ denote any state of a \emph{qubit},
a two-level quantum system.
$\rho$ can be expressed as
$\rho  =  \frac{1}{2}
\left( \id  +  \hat{ \mathbf{n} }  \cdot  \bm{ \sigma }  \right) \, .$
The identity operator is denoted by $\id$.
The $\Unit{n}$ components $n_\ell$ constitute decomposition coefficients.
The KD quasiprobability consists of coefficients
in a more general decomposition.

Let $\Sys$ denote a discrete quantum system
associated with a Hilbert space $\mathcal{H}$.
Let $\Set{ \ket{ f } }$ and
$\Set{  \ket{a} }$ denote orthonormal bases for $\mathcal{H}$.
Let $\Oper \in \mathcal{B} ( \mathcal{H} )$ denote a bounded operator
defined on $\mathcal{H}$.
Consider operating on each side of $\Oper$
with a resolution of unity:
\begin{align}
   \Oper  & =  \id  \Oper  \id
   =  \left(  \sum_a  \ketbra{ a }{ a }  \right)  \Oper
       \left(  \sum_f  \ketbra{ f }{ f }  \right) \\
   & \label{eq:DecomposeHelp1}
   =  \sum_{a, f }  \ketbra{ a }{ f }  \;  \langle a | \Oper | f \rangle \, .
\end{align}
Suppose that every element of $\Set{ \ket{a} }$ has a nonzero overlap
with every element of $\Set{ \ket{f} }$:
\begin{align}
   \label{eq:OverlapCond}
   \langle f | a \rangle  \neq  0  \qquad \forall a, f \, .
\end{align}
Each term in Eq.~\eqref{eq:DecomposeHelp1}
can be multiplied and divided by the inner product:
\begin{align}
   \label{eq:StateDecomp}
   \Oper  =  \sum_{a , f }
   \frac{ \ketbra{ a }{ f } }{  \langle f | a \rangle }  \;
   \langle f | a \rangle   \langle a | \Oper | f \rangle \, .
\end{align}

Under condition~\eqref{eq:OverlapCond},
$\Set{ \frac{ \ketbra{ a }{ f } }{  \langle f | a \rangle } }$
forms an orthonormal basis for $\mathcal{B} ( \mathcal{ H } ) \, .$
[The orthonormality is with respect to
the Hilbert-Schmidt inner product.
Let $\Oper_1 ,  \Oper_2  \in  \mathcal{B} ( \Hil )$.
The operators have the Hilbert-Schmidt inner product
$( \Oper_1 ,  \,  \Oper_2  )  =  \Tr ( \Oper_1^\dag \Oper_2 )$.]
The KD quasiprobability
$\langle f | a \rangle   \langle a | \Oper | f \rangle$
consists of the decomposition coefficients.

Condition~\eqref{eq:OverlapCond} is usually assumed to hold~\cite{Lundeen_11_Direct,Lundeen_12_Procedure,Thekkadath_16_Direct}.
In~\cite{Lundeen_11_Direct,Lundeen_12_Procedure}, for example,
$\Set{ \ketbra{ a }{ a } }$ and $\Set{ \ketbra{ f }{ f } }$ manifest as
the position and momentum eigenbases
$\Set{ \ket{ x } }$ and $\Set{ \ket{ p } }$.
Let $\ket{ \psi }$ denote a pure state.
Let $\psi(x)$ and $\tilde{\psi} (p)$ represent $\ket{ \psi }$
relative to the position and momentum eigenbases.
The KD quasiprobability for
$\rho = \ketbra{ \psi }{ \psi }$ has the form
\begin{align}
   \label{eq:KD_ex}
   \OurKD{ \ketbra{\psi}{\psi} }^\1 ( p, x )
   & = \langle x | p \rangle  \langle p | \psi \rangle \langle \psi | x \rangle \\
   & =  \frac{ e^{ - i x p / \hbar } }{ \sqrt{ 2 \pi \hbar } }  \;
   \tilde{\psi} ( p )  \,  \psi^* ( x )  \, .
\end{align}
The OTOC motivates a relaxation of condition~\eqref{eq:OverlapCond}
(Sec.~\ref{section:TA_Coeffs}).
[Though assumed 
in the operator decomposition~\eqref{eq:StateDecomp},
and assumed often in the literature,
condition~\eqref{eq:OverlapCond} need not hold
in arbitrary KD-quasiprobability arguments.]

\subsubsection{Properties of the KD quasiprobability}
\label{section:KDProps}

The KD quasiprobability shares some, but not all,
of its properties with other quasiprobabilities.
The notation below is defined as it has been
throughout Sec.~\ref{section:Intro_to_KD}.

\begin{property}   \label{prop:Complex}
The KD quasiprobability $\OurKD{\Oper}^\1 ( a, f )$ maps
$\mathcal{B} ( \mathcal{H} )   \times  \Set{ a }  \times  \Set{ f }$
to $\mathbb{C} \, .$
The domain is a composition of the set
$\mathcal{B} ( \mathcal{H} )$ of bounded operators
and two sets of real numbers.
The range is the set $\mathbb{C}$ of complex numbers,
not necessarily the set $\mathbb{R}$ of real numbers.
\end{property}

The Wigner function assumes only real values.
Only by dipping below zero can the Wigner function
deviate from classical probabilistic behavior.
The KD distribution's negativity has the following physical significance:
Imagine projectively measuring two (commuting) observables,
$\A$ and $\B$, simultaneously.
The measurement has some probability $p(a ; b)$
of yielding the values $a$ and $b$.
Now, suppose that $\A$ does not commute with $\B$.
No joint probability distribution $p(a ; b)$ exists.
Infinitely precise values cannot be ascribed
to noncommuting observables simultaneously.
Negative quasiprobability values are not observed directly:
Observable phenomena are modeled by
averages over quasiprobability values.
Negative values are visible only on scales
smaller than the physical coarse-graining scale.
But negativity causes observable effects,
visible in sequential measurements.
Example effects include anomalous weak values~\cite{Aharonov_88_How,Kofman_12_Nonperturbative,Dressel_14_Understanding,Pusey_14_Anomalous,Dressel_15_Weak,Waegell_16_Confined}
and violations of Leggett-Garg inequalities~\cite{Leggett_85_Quantum,Emary_14_LGI}.

Unlike the Wigner function, the KD distribution can assume nonreal values.
Consider measuring two noncommuting observables sequentially.
How much does the first measurement
affect the second measurement's outcome?
This disturbance is encoded in
the KD distribution's imaginary component~\cite{Hofmann_12_Complex,Dressel_12_Significance,Hofmann_14_Derivation,Hofmann_14_Sequential}.

\begin{property}
\label{prop:SumToProb}
Summing $\OurKD{\rho}^\1 ( a, f )$ over $a$
yields a probability distribution.
So does summing $\OurKD{\rho}^\1 ( a, f )$ over $f$.
\end{property} \noindent
Consider substituting $\Oper = \rho$ into Eq.~\eqref{eq:KD_rho}.
Summing over $a$ yields $\langle f | \rho | f \rangle$.
This inner product equals a probability, by Born's Rule.

\begin{property}
\label{prop:Discrete}
The KD quasiprobability is defined as in Eq.~\eqref{eq:KD_rho}
regardless of whether $\Set{ a }$ and $\Set{ f }$ are discrete.
\end{property} \noindent
The KD distribution and the Wigner function
were defined originally for continuous systems.
Discretizing the Wigner function
is less straightforward~\cite{Ferrie_11_Quasi,Delfosse_15_Wigner}.

\begin{property}
\label{prop:Bayes}
The KD quasiprobability obeys an analog of Bayes' Theorem,
Eq.~\eqref{eq:CondQuasi}.
\end{property}

Bayes' Theorem governs the conditional probability $p(f | i)$
that an event $f$ will occur, given that an event $i$ has occurred.
$p(f | i)$ is expressed in terms of
the conditional probability $p(i | f )$
and the absolute probabilities $p(i)$ and $p(f)$:
\begin{align}
   \label{eq:BayesThm}
   p( f | i )  =  \frac{ p ( i | f )  \:  p ( f ) }{ p ( i ) }  \, .
\end{align}

Equation~\eqref{eq:BayesThm} can be expressed in terms of
jointly conditional distributions.
Let $p ( a | i , f )$ denote the probability that an event $a$ will occur,
given that an event $i$ occurred
and that $f$ occurred subsequently.
$p( a, f | i )$ is defined similarly.
What is the joint probability $p( i, f, a )$ that $i$, $f$, and $a$ will occur?
We can construct two expressions:
\begin{align}
   \label{eq:ToBayes}
   p( i , f , a )  =  p ( a | i, f ) \,  p( i , f )
   =  p ( a , f | i )  \,  p ( i )  \, .
\end{align}
The joint probability $p ( i , f )$ equals $p ( f | i )  \,  p ( i )$.
This $p( i )$ cancels with the $p ( i )$ on
the right-hand side of Eq.~\eqref{eq:ToBayes}.
Solving for $p( a | i, f )$ yields
Bayes' Theorem for jointly conditional probabilities,
\begin{align}
   \label{eq:BayesThm2}
   p( a | i, f )  =  \frac{ p ( a, f | i ) }{ p ( f | i ) } \, .
\end{align}

Equation~\eqref{eq:CondQuasi} echoes Eq.~\eqref{eq:BayesThm2}.
The KD quasiprobability's Bayesian behavior~\cite{Hofmann_14_Derivation,Bamber_14_Observing}
has been applied to quantum state tomography~\cite{Lundeen_11_Direct,Lundeen_12_Procedure,Hofmann_14_Sequential,Salvail_13_Full,Malik_14_Direct,Howland_14_Compressive,Mirhosseini_14_Compressive}
and to quantum foundations~\cite{Hofmann_12_Complex}.

Having reviewed the KD quasiprobability,
we approach the extended KD quasiprobability behind the OTOC.
We begin by concretizing our set-up,
then reviewing the OTOC.

\subsection{Set-up}
\label{section:SetUp}

This section concerns the set-up and notation
used throughout the rest of this paper.
Our framework is motivated by the OTOC,
which describes quantum many-body systems.
Examples include black holes~ \cite{Shenker_Stanford_14_BHs_and_butterfly,Kitaev_15_Simple},
the Sachdev-Ye-Kitaev model~\cite{Sachdev_93_Gapless,Kitaev_15_Simple},
other holographic systems~\cite{Maldacena_98_AdSCFT,Witten_98_AdSCFT,Gubser_98_AdSCFT}
and spin chains.
We consider a system $\Sys$ associated with
a Hilbert space $\Hil$ of dimensionality $\Dim$.
The system evolves under a Hamiltonian $H$
that might be nonintegrable or integrable.
$H$ generates the time-evolution operator $U  :=  e^{ - i H t } \, .$

We will have to sum or integrate over spectra.
For concreteness, we sum, supposing that $\Hil$ is discrete.
A spin-chain example, discussed next, motivates our choice.
Our sums can be replaced with integrals
unless, e.g., we evoke spin chains explicitly.

We will often illustrate with a one-dimensional (1D) chain
of spin-$\frac{1}{2}$ degrees of freedom.
Figure~\ref{fig:Spin_chain} illustrates the chain,
simulated numerically in Sec.~\ref{section:Numerics}.
Let $\Sites$ denote the number of spins.
This system's $\Hil$ has dimensionality $\Dim  =  2^\Sites$.

%
%
\begin{figure}[h]
\centering
\includegraphics[width=.35\textwidth]{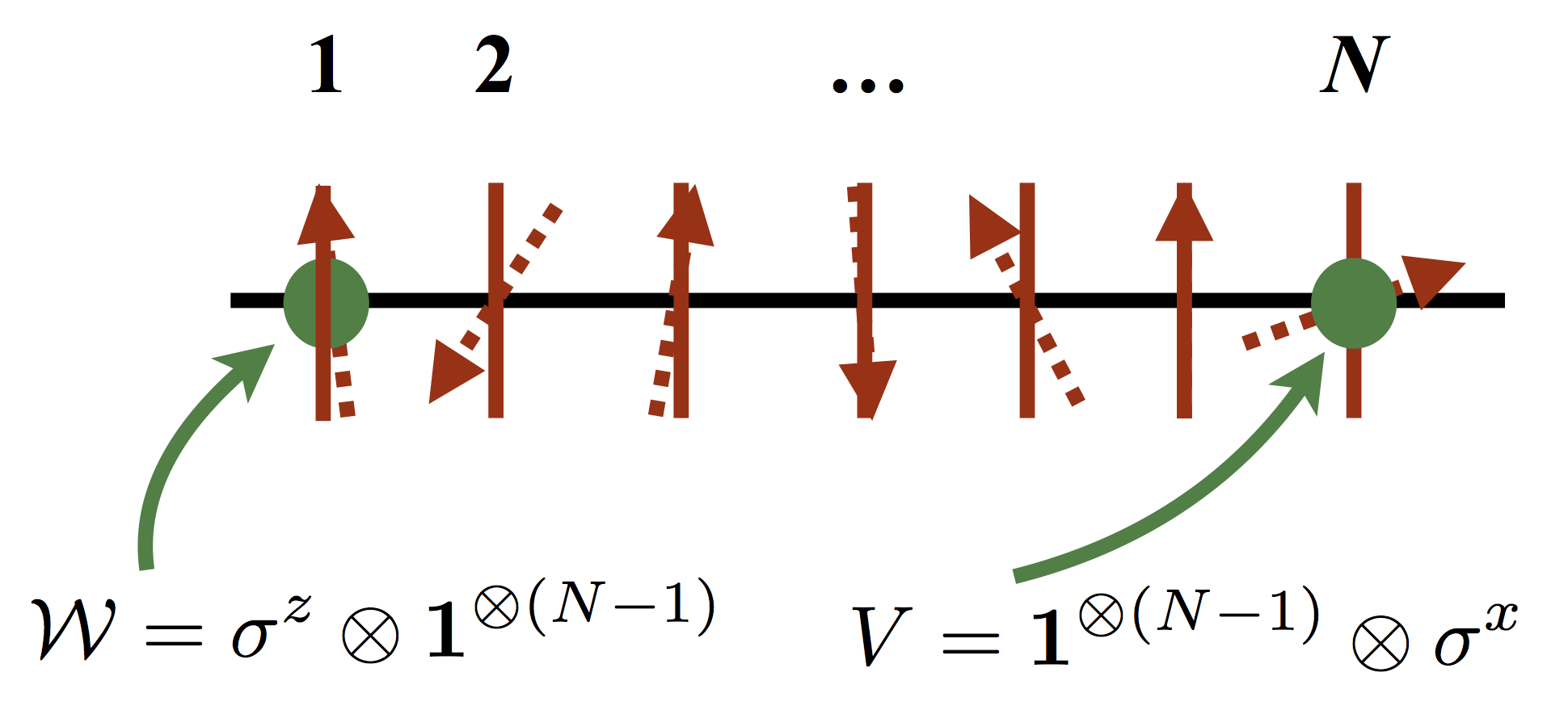}
\caption{\caphead{Spin-chain example:}
A spin chain exemplifies the quantum many-body systems
characterized by the out-of-time-ordered correlator (OTOC).
We illustrate with a one-dimensional chain
of $\Sites$ spin-$\frac{1}{2}$ degrees of freedom. 
The vertical red bars mark the sites.
The dotted red arrows illustrate how spins can point in arbitrary directions.
The OTOC is defined in terms of
local unitary or Hermitian operators $\W$ and $V$.
Example operators include single-qubit Paulis $\sigma^x$ and $\sigma^z$
that act nontrivially on opposite sides of the chain.}
\label{fig:Spin_chain}
\end{figure}

We will often suppose that $\Sys$ occupies, or is initialized to, a state
\begin{align}
   \label{eq:Rho}
   \rho  =  \sum_j  p_j  \ketbra{j}{j}   \in  \mathcal{D} (\Hil)  \, .
\end{align}
The set of density operators defined on $\Hil$
is denoted by $\mathcal{D} ( \Hil )$, as in Sec.~\ref{section:Intro_to_KD}.
Orthonormal eigenstates are indexed by $j$;
eigenvalues are denoted by $p_j$.
Much literature focuses on
temperature-$T$ thermal states $e^{ - H / T } / Z$.
(The partition function $Z$ normalizes the state.)
We leave the form of $\rho$ general,
as in~\cite{YungerHalpern_17_Jarzynski}.

The OTOC is defined in terms of local operators $\W$ and $V$.
In the literature, $\W$ and $V$ are assumed to be unitary and/or Hermitian.
Unitarity suffices for deriving
the results in~\cite{YungerHalpern_17_Jarzynski},
as does Hermiticity.
Unitarity and Hermiticity are assumed there,
and here, for convenience.\footnote{
Measurements of $\W$ and $V$ are discussed
in~\cite{YungerHalpern_17_Jarzynski} and here.
Hermitian operators $\GW$ and $\GV$ generate $\W$ and $V$.
If $\W$ and $V$ are not Hermitian,
$\GW$ and $\GV$ are measured instead of $\W$ and $V$.}
In our spin-chain example,
the operators manifest as one-qubit Paulis
that act nontrivially on opposite sides of the chain, e.g.,
$\W  =  \sigma^z \otimes \id^{ \otimes ( \Sites - 1 ) }$,
and $V  =  \id^{ \otimes ( \Sites - 1 ) } \otimes \sigma^x$.
In the Heisenberg Picture, $\W$ evolves as
$\W(t)  :=  U^\dag  \W  U \, .$

The operators eigendecompose as
\begin{align}
   \W  =  \sum_{w_\ell,  \DegenW_{w_\ell} }
   w_\ell  \ketbra{ w_\ell,  \DegenW_{w_\ell} }{ w_\ell,  \DegenW_{w_\ell} }
\end{align}
and
\begin{align}
   V  =  \sum_{ v_\ell,  \DegenV_{v_\ell} }
   v_\ell  \ketbra{ v_\ell,  \DegenV_{v_\ell} }{ v_\ell,  \DegenV_{v_\ell} } \, .
\end{align}
The eigenvalues are denoted by $w_\ell$ and $v_\ell$.
The degeneracy parameters are denoted by
$\DegenW_{ w_\ell }$ and $\DegenV_{v_\ell}$.
Recall that $\W$ and $V$ are local.
In our example, $\W$ acts nontrivially on just
one of $\Sites \gg 1$ qubits.
Hence $\W$ and $V$ are
exponentially degenerate in $\Sites$.
The degeneracy parameters can be measured:
Some nondegenerate Hermitian operator $\NondegW$
has eigenvalues in a one-to-one correspondence with
the $\DegenW_{ w_\ell }$'s.
A measurement of $\W$ and $\NondegW$
outputs a tuple $(w_\ell, \DegenW_{ w_\ell } )$.
We refer to such a measurement as ``a $\NondegW$ measurement,''
for conciseness.
Analogous statements concern $V$ and a Hermitian operator $\NondegV$.
Section~\ref{section:ProjTrick} introduces a trick
that frees us from bothering with degeneracies.

\subsection{The out-of-time-ordered correlator}
\label{section:OTOC_review}

Given two unitary operators $\mathcal{W}$ and $V$, the
out-of-time-ordered correlator is defined as
\begin{align}
   \label{eq:OTOC_Def}
   F(t) := \langle \mathcal{W}^\dagger(t) V^\dagger \mathcal{W}(t) V \rangle
   \equiv \text{Tr} \LParen  \rho \mathcal{W}^\dagger(t) V^\dagger
   \mathcal{W}(t) V  \RParen  \, .
\end{align}
This object reflects the degree of noncommutativity of $V$ and the Heisenberg operator $\mathcal{W}(t)$. More precisely, the OTOC appears in the expectation value of the squared magnitude of the commutator $[\mathcal{W}(t),V]$,
\begin{align}
C(t) := \langle [\mathcal{W}(t),V]^\dagger [\mathcal{W}(t),V] \rangle
= 2 - 2 \Re \LParen  F(t)  \RParen  \, .
\end{align}
Even if $\mathcal{W}$ and $V$ commute, the Heisenberg operator $\mathcal{W}(t)$ generically does not commute with $V$ at sufficiently late times.

An analogous definition involves Hermitian $\mathcal{W}$ and $V$.
The commutator's square magnitude becomes
\begin{align}
C(t) = - \langle [\mathcal{W}(t),V]^2\rangle.
\end{align}
This squared commutator involves TOC (time-ordered-correlator) and OTOC terms. The TOC terms take the forms
$\langle V \mathcal{W}(t) \mathcal{W}(t) V \rangle$ and
$\langle \mathcal{W}(t) V V \mathcal{W}(t) \rangle$.
[Technically, $\langle V \mathcal{W}(t) \mathcal{W}(t) V \rangle$
is time-ordered. $\langle \mathcal{W}(t) V V \mathcal{W}(t) \rangle$
behaves similarly.]

The basic physical process reflected by the OTOC is the spread of Heisenberg operators with time. Imagine starting with a simple $\mathcal{W}$, e.g., an operator acting nontrivially on just one spin in a many-spin system. Time-evolving yields $\mathcal{W}(t)$. The operator has grown if $\mathcal{W}(t)$ acts nontrivially on more spins than $\W$ does.
The operator $V$ functions as a probe for testing whether the action of $\mathcal{W}(t)$ has spread to the spin on which $V$ acts nontrivially.

Suppose $\mathcal{W}$ and $V$ are unitary and commute.
At early times, $\mathcal{W}(t)$ and $V$ approximately commute.
Hence $F(t) \approx 1$, and $C(t) \approx 0$.
Depending on the dynamics, at later times, $\mathcal{W}(t)$ may significantly fail to commute with $V$.
In a chaotic quantum system, $\W(t)$ and $V$ generically do not commute at late times, for most choices of $\W$ and $V$.

The analogous statement for Hermitian $\mathcal{W}$ and $V$ is that $F(t)$ approximately equals the TOC terms at early times. At late times, depending on the dynamics, the commutator can grow large. The time required for the TOC terms to approach their equilibrium values is called the \emph{dissipation time} $t_{\text{d}}$. This time parallels the time required for a system to reach local thermal equilibrium. The time scale on which the commutator grows to be order-one is called the \emph{scrambling time} $t_*$. The scrambling time parallels the time over which a drop of ink spreads across a container of water.

Why consider the commutator's square modulus?
The simpler object $\langle [\mathcal{W}(t),V]\rangle$ 
often vanishes at late times, 
due to cancellations between states
in the expectation value. Physically, the vanishing of $\langle [\mathcal{W}(t),V]\rangle$ signifies that perturbing the system with $V$ does not significantly change the expectation value of $\mathcal{W}(t)$. This physics is expected for a chaotic system, which effectively loses its memory of its initial conditions. In contrast, $C(t)$ is the expectation value of a positive operator (the magnitude-squared commutator). The cancellations that zero out $\langle [\mathcal{W}(t),V]\rangle$ cannot zero out $\expval{ | [ \W(t) , V ] |^2 }$. 

Mathematically, the diagonal elements of
the matrix that represents $[\mathcal{W}(t),V]$
relative to the energy eigenbasis
can be small.
$\expval{ [\mathcal{W}(t),V] }$, evaluated on a thermal state, 
would be small.
Yet the matrix's off-diagonal elements can boost
the operator's Frobenius norm,
$\sqrt{  \Tr \left( | [\mathcal{W}(t),V] |^2 \right)  }$,
which reflects the size of $C(t)$.

We can gain intuition about the manifestation of chaos in $F(t)$
from a simple quantum system
that has a chaotic semiclassical limit.
Let $\mathcal{W} = q$ and $\mathcal{V} = p$
for some position $q$ and momentum $p$:
\begin{align}
C(t) = - \langle [q(t),p]^2 \rangle \sim \hbar^2 e^{2 \Lyap t}  \, .
\end{align}
This $\Lyap$ is a classical Lyapunov exponent. The final expression follows from the Correspondence Principle:
Commutators are replaced with $i \hbar$ times the corresponding Poisson bracket. The Poisson bracket of $q(t)$ with $p$ equals the derivative of the final position with respect to the initial position. This derivative reflects the butterfly effect in classical chaos, i.e., sensitivity to initial conditions. The growth of $C(t)$, and the deviation of $F(t)$ from the TOC terms, provide a quantum generalization of the butterfly effect.

Within this simple quantum system, the analog of the dissipation time may be regarded as $t_{\text{d}} \sim \Lyap^{-1}$. The analog of the scrambling time is $t_* \sim \Lyap^{-1} \ln \frac{\Omega}{\hbar}$. The $\Omega$ denotes some measure of the accessible phase-space volume. Suppose that the phase space is large in units of $\hbar$. The scrambling time is much longer than the dissipation time: $t_*  \gg  t_{\text{d}}$.
Such a parametric separation between the time scales
characterizes the systems that interest us most.

In more general chaotic systems, the value of $t_*$ depends on whether the interactions are geometrically local and on $\mathcal{W}$ and $V$. Consider, as an example, a spin chain
 governed by a local Hamiltonian.
Suppose that $\mathcal{W}$ and $V$ are local operators
that act nontrivially on spins separated by a distance $\ell$. The scrambling time is generically proportional to $\ell$. For this class of local models,
$\ell/t_*$ defines a velocity $v_{\text{B}}$ called the \emph{butterfly velocity}. Roughly, the butterfly velocity reflects how quickly initially local Heisenberg operators grow in space.

Consider a system in which $\td$ is separated parametrically from $t_*$.
The rate of change of $F(t)$
[rather, a regulated variation on $F(t)$]
was shown to obey a nontrivial bound.
Parameterize the OTOC as
$F(t) \sim \text{TOC} - \epsilon  \, e^{\Lyap t}$.
The parameter $\epsilon \ll 1$ encodes the separation of scales.
The exponent $\Lyap$ obeys $\Lyap \leq 2 \pi \kB T$ in thermal equilibrium at temperature $T$~\cite{Maldacena_15_Bound}. $\kB$ denotes Boltzmann's constant. Black holes in the AdS/CFT duality saturate this bound, exhibiting maximal chaos~\cite{Shenker_Stanford_14_BHs_and_butterfly,Kitaev_15_Simple}.

More generally, $\Lyap$ and $v_{\text{B}}$ control the operators' growth and the spread of chaos. The OTOC has thus attracted attention for a variety of reasons, including (but not limited to) the possibilities of nontrivial bounds on quantum dynamics, a new probe of quantum chaos, and a signature of black holes in AdS/CFT.

\subsection{Introducing the quasiprobability
behind the OTOC}
\label{section:Review_OTOC_quasiprob}

$F(t)$ was shown, in~\cite{YungerHalpern_17_Jarzynski},
to equal a moment of a summed quasiprobability.
We review this result, established in four steps:
A quantum probability amplitude $A_\rho$
is reviewed in Sec.~\ref{section:Review_A} .
Amplitudes are combined to form the quasiprobability $\OurKD{\rho}$
in Sec.~\ref{section:Review_OTOC_quasiprob_sub}.
Summing $\OurKD{\rho}( . )$ values,
with constraints, yields a complex distribution
$P (W, W')$ in Sec.~\ref{section:Intro_PWWPrime}.
Differentiating $P(W, W')$ yields the OTOC.
$\OurKD{\rho}$ can be inferred experimentally
from a weak-measurement scheme
and from interference.
We review these schemes in Sec.~\ref{section:Intro_weak_meas}.

A third quasiprobability is introduced in Sec.~\ref{section:ProjTrick},
the \emph{coarse-grained quasiprobability} $\SumKD{\rho}$.
$\SumKD{\rho}$ follows from summing values of $\OurKD{\rho}$.
$\SumKD{\rho}$ has a more concise description than $\OurKD{\rho}$.
Also, measuring $\SumKD{\rho}$ requires fewer resources
(e.g., trials)
than measuring $\OurKD{\rho}$.
Hence Sections~\ref{section:Measuring}-\ref{section:Brownian}
will spotlight $\SumKD{\rho}$.
$\OurKD{\rho}$ returns to prominence in 
the proofs of Sec.~\ref{section:Theory} and in
opportunities detailed in Sec.~\ref{section:Outlook}.
Different distributions suit different investigations.
Hence the presentation of three distributions in this thorough study:
$\OurKD{\rho}$, $\SumKD{\rho}$, and $P(W, W')$.

\subsubsection{Quantum probability amplitude $A_\rho$}
\label{section:Review_A}

The OTOC quasiprobability $\OurKD{\rho}$
is defined in terms of probability amplitudes $A_\rho$.
The $A_\rho$'s are defined in terms of the following process, $\ProtocolA$:
\begin{enumerate}[(1)]
   \item   Prepare $\rho$.
   \item   Measure the $\rho$ eigenbasis, $\Set{ \ketbra{j}{j} }$.
   \item   Evolve $\Sys$ forward in time under $U$.
   \item   Measure $\NondegW$.
   \item   Evolve $\Sys$ backward under $U^\dag$.
   \item   Measure $\NondegV$.
   \item   Evolve $\Sys$ forward under $U$.
   \item   Measure $\NondegW$.
\end{enumerate}
Suppose that the measurements yield the outcomes
$j$,  $(w_1, \DegenW_{w_1})$,  $(v_1, \DegenV_{v_1})$,
and  $(w_2, \DegenW_{w_2})$.
Figure~\ref{fig:Protocoll_Trial1} illustrates this process.
The process corresponds to the probability amplitude\footnote{
We order the arguments of $A_\rho$ differently
than in~\cite{YungerHalpern_17_Jarzynski}.
Our ordering here parallels
our later ordering of the quasiprobability's argument.
Weak-measurement experiments motivate
the quasiprobability arguments' ordering.
This motivation is detailed in Footnote~\ref{footnote:Order}.}
\begin{align}
   \label{eq:Amp}
   & A_\rho( j  ;  w_1,  \DegenW_{w_1}  ;  v_1, \DegenV_{v_1}  ;
                   w_2, \DegenW_{w_2}  )
   :=  \langle w_2, \DegenW_{w_2} | U | v_1, \DegenV_{v_1} \rangle
   \nonumber \\ & \qquad \times
   \langle  v_1, \DegenV_{v_1}  |  U^\dag  |  w_1,  \DegenW_{w_1}   \rangle
   \langle  w_1,  \DegenW_{w_1}   |  U  |  j  \rangle
   \sqrt{ p_j } \, .
\end{align}

We do not advocate for performing $\ProtocolA$ in any experiment.
$\ProtocolA$ is used to define $A_\rho$
and to interpret $A_\rho$ physically.
Instances of $A_\rho$ are combined into $\OurKD{\rho}$.
A weak-measurement protocol can be used
to measure $\OurKD{\rho}$ experimentally.
An interference protocol can be used to measure
$A_\rho$ (and so $\OurKD{\rho}$) experimentally.

\subsubsection{The fine-grained OTOC quasiprobability $\OurKD{\rho}$}
\label{section:Review_OTOC_quasiprob_sub}

The quasiprobability's definition is constructed as follows.
Consider a realization of $\ProtocolA$ that yields the outcomes
$j$, $( w_3, \DegenW_{w_3} )$, $(  v_2, \DegenV_{v_2} )$, and
$( w_2 , \DegenW_{w_2} )$.
 Figure~\ref{fig:Protocoll_Trial2} illustrates this realization.
The initial and final measurements yield
the same outcomes as in the~\eqref{eq:Amp} realization.
We multiply the complex conjugate of
the second realization's amplitude
by the first realization's probability amplitude.
Then, we sum over $j$ and $(w_1, \DegenW_{w_1})$:\footnote{
Familiarity with tensors might incline one to sum
over the $(w_2,  \DegenW_{w_2})$
shared by the trajectories.
But we are not invoking tensors.
More importantly, summing over $(w_2,  \DegenW_{w_2})$
introduces a $\delta_{ v_1 v_2 }  \delta_{ \DegenV_{v_1} \DegenV_{v_2} }$
that eliminates one $( v_\ell,  \DegenV_{v_\ell} )$ degree of freedom.
The resulting quasiprobability would not ``lie behind'' the OTOC.
One could, rather than summing over $( w_1,  \DegenW_{w_1} )$,
sum over $(w_3,  \DegenW_{w_3})$.
Either way, one sums over one trajectory's first $\NondegW$ outcome.
We sum over $(w_1,  \DegenW_{w_1})$ to maintain consistency with~\cite{YungerHalpern_17_Jarzynski}.}$^,$\footnote{   \label{footnote:Order}
In~\cite{YungerHalpern_17_Jarzynski},
the left-hand side's arguments are ordered differently
and are condensed into the shorthand
$(w, v,  \DegenW_w, \DegenV_v)$.
Experiments motivate our reordering:
Consider inferring $\OurKD{\rho} ( a , b , c , d )$ from
experimental measurements.
In each trial, one (loosely speaking) weakly measures
$a$, then $b$, then $c$;
and then measures $d$ strongly.
As the measurements are ordered, so are the arguments.}
\begin{align}
   \label{eq:TADef}
   & \OurKD{\rho} ( v_1,  \DegenV_{v_1} ;  w_2,  \DegenW_{w_2} ;
   v_2,  \DegenV_{v_2}  ;  w_3,  \DegenW_{w_3}  )
   \nonumber  \\ &
   :=  \sum_{j , (w_1, \DegenW_{w_1} ) }
   A_{\rho}^* ( j  ;  w_3,  \DegenW_{w_3}  ;  v_2, \DegenV_{v_2}  ;
                   w_2, \DegenW_{w_2} )
  \nonumber  \\ & \qquad \qquad \qquad \times
  A_{\rho} ( j  ;  w_1,  \DegenW_{w_1}  ;  v_1, \DegenV_{v_1}  ;
                   w_2, \DegenW_{w_2} ) \, .
\end{align}

Equation~\eqref{eq:TADef} resembles a probability
but differs due to the noncommutation of $\W(t)$ and $V$.
We illustrate this relationship in two ways.

Consider a 1D quantum system, e.g., a particle on a line.
We represent the system's state with
a wave function $\psi(x)$.
The probability density at point $x$ equals $\psi^* (x)  \,  \psi(x)$.
The $A^*_\rho  \,  A_\rho$ in Eq.~\eqref{eq:TADef} echoes $\psi^* \psi$.
But the argument of the $\psi^*$ equals
the argument of the $\psi$.
The argument of the $A^*_\rho$ differs from
the argument of the $A_\rho$,
because $\W(t)$ and $V$ fail to commute.

Substituting into Eq.~\eqref{eq:TADef} from Eq.~\eqref{eq:Amp} yields
\begin{align}
   \label{eq:TAForm}
   & \OurKD{\rho} ( v_1,  \DegenV_{v_1} ;  w_2,  \DegenW_{w_2} ;
   v_2,  \DegenV_{v_2}  ;  w_3,  \DegenW_{w_3} )
   \nonumber \\ &
   =  \langle  w_3 , \DegenW_{w_3}  |  U  |
                   v_2, \DegenV_{v_2}  \rangle
   \langle  v_2, \DegenV_{v_2}  |  U^\dag  |
               w_2,  \DegenW_{w_2}  \rangle
   \nonumber \\ & \qquad \times
   \langle  w_2,  \DegenW_{w_2}  |  U  |  v_1,  \DegenV_{v_1}  \rangle
   \langle  v_1,  \DegenV_{v_1}  |  \rho  U^\dag  |  w_3 , \DegenW_{w_3}  \rangle \, .
\end{align}
A simple example illustrates how
$\OurKD{\rho}$ nearly equals a probability.
Suppose that an eigenbasis of $\rho$ coincides with
$\Set{ \ketbra{ v_\ell,  \DegenV_{v_\ell} }{ v_\ell,  \DegenV_{v_\ell} } }$
or  with  $\Set{ U^\dag  \ketbra{ w_\ell , \DegenW_{w_\ell} }{
                        w_\ell , \DegenW_{w_\ell} } U }$.
Suppose, for example, that
\begin{align}
   \label{eq:WRho}
   \rho =  \rho_{V}  :=
   \sum_{ v_\ell , \DegenV_{v_\ell} }
   p_{ v_\ell , \DegenV_{v_\ell} }
   \ketbra{ v_\ell , \DegenV_{v_\ell} }{ v_\ell , \DegenV_{v_\ell} }  \, .
\end{align}
One such $\rho$ is the infinite-temperature Gibbs state
$\id / \Dim$.
Another example is easier to prepare:
Suppose that $\Sys$ consists of $\Sites$ spins
and that $V = \sigma^x_\Sites$.
One $\rho_V$ equals a product of $\Sites$ $\sigma^x$ eigenstates.
Let $( v_2 ,  \DegenV_{v_2} )  =  ( v_1,  \DegenV_{v_1} )$.
[An analogous argument follows from
$( w_3 , \DegenW_{w_3} ) = ( w_2, \DegenW_{w_2} )$.]
Equation~\eqref{eq:TAForm} reduces to
\begin{align}
   \label{eq:Reduce_to_p}
   & | \langle  w_2,  \DegenW_{w_2}  |  U  |
        v_1,  \DegenV_{v_1}  \rangle |^2  \,
   |  \langle  w_3,  \DegenW_{w_3}  |  U  |
       v_1,  \DegenV_{v_1}  \rangle |^2  \,
   p_{ v_1 , \DegenV_{v_1} } \, .
\end{align}
Each square modulus equals a conditional probability.
$p_{ v_1,  \DegenV_{v_1} }$ equals the probability that,
if $\rho$ is measured with respect to
$\Set{ \ketbra{ v_\ell , \DegenV_{v_\ell} }{
                        v_\ell , \DegenV_{v_\ell} } }$,
outcome $( v_1,  \DegenV_{v_1} )$ obtains.

In this simple case,
certain quasiprobability values 
equal probability values---the 
quasiprobability values that satisfy 
$( v_2,   \DegenV_{v_2} )  =  ( v_1,  \DegenV_{v_1} )$
or $( w_3 ,  \DegenW_{w_3} )  =  ( w_2,  \DegenW_{w_2} )$.
When both conditions are violated, typically, 
the quasiprobability value does not
equal a probability value. 
Hence not all the OTOC quasiprobability's values
reduce to probability values.
Just as a quasiprobability lies behind the OTOC,
quasiprobabilities lie behind
time-ordered correlators (TOCs).
Every value of a TOC quasiprobability
reduces to a probability value
in the same simple case (when $\rho$ equals, e.g., a $V$ eigenstate)
(Sec.~\ref{section:OTOC_TOC}).

\subsubsection{Complex distribution $P(W, W')$}
\label{section:Intro_PWWPrime}

$\OurKD{\rho}$ is summed, in~\cite{YungerHalpern_17_Jarzynski},
to form a complex distribution $P(W, W')$.
Let $W :=  w_3^*  v_2^*$ and $W'  :=  w_2  v_1$ denote
random variables calculable from measurement outcomes.
If $\W$ and $V$ are Paulis,
$(W, W')$ can equal $(1, 1), (1, -1), (-1, 1),$ or $(-1, -1)$.

$W$ and $W'$ serve, in the Jarzynski-like equality~\eqref{eq:JarzLike},
analogously to thermodynamic work $W_\th$ in Jarzynski's equality.
$W_\th$ is a random variable,
inferable from experiments,
that fluctuates from trial to trial.
So are $W$ and $W'$.
One infers a value of $W_\th$
by performing measurements and processing the outcomes.
The \emph{two-point measurement scheme} (TPMS)
illustrates such protocols most famously.
The TPMS has been used to derive 
quantum fluctuation relations~\cite{Tasaki00}.
One prepares the system in a thermal state,
measures the Hamiltonian, $H_i$, projectively;
disconnects the system from the bath;
tunes the Hamiltonian to $H_f$;
and measures $H_f$ projectively.
Let $E_i$ and $E_f$ denote the measurement outcomes.
The work invested in the Hamiltonian tuning
is defined as $W_\th  :=  E_f  -  E_i$.
Similarly, to infer $W$ and $W'$,
one can measure $\W$ and $V$
as in Sec.~\ref{section:Intro_weak_meas},
then multiply the outcomes.

Consider fixing the value of $(W, W')$.
For example, let  $(W, W')  =  (1, -1)$.
Consider the octuples
$( v_1,  \DegenV_{v_1} ;  w_2,  \DegenW_{w_2}  ;
    v_2,  \DegenV_{v_2}   ;  w_3,  \DegenW_{w_3} )$
that satisfy the constraints $W = w_3^*  v_2^*$
and $W'  =  w_2  v_1$.
Each octuple corresponds to
a quasiprobability value $\OurKD{\rho} ( . )$.
Summing these quasiprobability values yields
\begin{align}
   \label{eq:PWWPrime}
   & P(W, W')  :=  \sum_{ \substack{
   ( v_1,  \DegenV_{v_1} ) , ( w_2,  \DegenW_{w_2} ) ,
   ( v_2,  \DegenV_{v_2} ) , ( w_3,  \DegenW_{w_3} ) } }
   \\  \nonumber  &
   \OurKD{\rho} ( v_1,  \DegenV_{v_1} ;  w_2,  \DegenW_{w_2} ;
   v_2,  \DegenV_{v_2}  ;  w_3,  \DegenW_{w_3} )  \:
   \delta_{W ( w_3^*  v_2^* ) }    \delta_{ W' (w_2  v_1) } \, .
\end{align}
The Kronecker delta is represented by $\delta_{ab}$.
$P(W, W')$ functions analogously to
the probability distribution, in the fluctuation-relation paper~\cite{Jarzynski_97_Nonequilibrium},
over values of thermodynamic work.

The OTOC equals a moment of $P(W, W')$ [Eq.~\eqref{eq:JarzLike}],
which equals a constrained sum over $\OurKD{\rho}$~\cite{YungerHalpern_17_Jarzynski}.
Hence our labeling of $\OurKD{\rho}$ as a
``quasiprobability behind the OTOC.''
Equation~\eqref{eq:PWWPrime} expresses
the useful, difficult-to-measure $F(t)$
in terms of a characteristic function of a (summed) quasiprobability,
as Jarzynski~\cite{Jarzynski_97_Nonequilibrium} expresses
a useful, difficult-to-measure
free-energy difference $\Delta F$
in terms of a characteristic function of a probability.
Quasiprobabilities reflect nonclassicality (contextuality)
as probabilities do not;
so, too, does $F(t)$ reflect nonclassicality (noncommutation)
as $\Delta F$ does not.

The definition of $P$ involves arbitrariness:
The measurable random variables, and $P$,
may be defined differently.
Alternative definitions, introduced in Sec.~\ref{section:HigherOTOCs},
extend more robustly to
OTOCs that encode more time reversals.
All possible definitions share two properties:
(i) The arguments $W$, etc. denote random variables
inferable from measurement outcomes.
(ii) $P$ results from summing $\OurKD{\rho}( . )$ values
subject to constraints $\delta_{ab}$.

$P(W, W')$ resembles a work distribution
constructed by Solinas and Gasparinetti (S\&G)~\cite{Jordan_chat,Solinas_chat}.
They study fluctuation-relation contexts,
rather than the OTOC.
S\&G propose a definition for
the work performed on a quantum system~\cite{Solinas_15_Full,Solinas_16_Probing}.
The system is coupled weakly to detectors
at a protocol's start and end.
The couplings are represented by constraints
like $\delta_{W ( w_3^* v_2^*) }$ and $\delta_{W' ( w_2 v_1 ) }$.
Suppose that the detectors measure the system's Hamiltonian.
Subtracting the measurements' outcomes
yields the work performed during the protocol.
The distribution over possible work values
is a quasiprobability.
Their quasiprobability is a Husimi $Q$-function,
whereas the OTOC quasiprobability is a KD distribution~\cite{Solinas_16_Probing}.
Related frameworks appear in~\cite{Alonso_16_Thermodynamics,Miller_16_Time,Elouard_17_Role}.
The relationship between those thermodynamics frameworks
and our thermodynamically motivated OTOC framework
merits exploration.

\subsubsection{Weak-measurement and interference schemes
for inferring $\OurKD{\rho}$}
\label{section:Intro_weak_meas}

$\OurKD{\rho}$ can be inferred from weak measurements
and from interference, as shown in~\cite{YungerHalpern_17_Jarzynski}.
Section~\ref{section:MeasSumFromOTOC} shows
how to infer a coarse-graining of $\OurKD{\rho}$
from other OTOC-measurement schemes (e.g.,~\cite{Swingle_16_Measuring}).
We focus mostly on the weak-measurement scheme here.
The scheme is simplified
in Sec.~\ref{section:Measuring}.
First, we briefly review the interference scheme.

The interference scheme in~\cite{YungerHalpern_17_Jarzynski}
differs from other interference schemes for measuring $F(t)$~\cite{Swingle_16_Measuring,Yao_16_Interferometric,Bohrdt_16_Scrambling}:
From the~\cite{YungerHalpern_17_Jarzynski} interference scheme,
one can infer not only $F(t)$, but also $\OurKD{\rho}$.
Time need not be inverted ($H$ need not be negated) in any trial.
The scheme is detailed in Appendix~B of~\cite{YungerHalpern_17_Jarzynski}.
The system is coupled to an ancilla prepared in a superposition
$\frac{1}{ \sqrt{2} } \, ( \ket{0}  +  \ket{1} )$.
A unitary, conditioned on the ancilla, rotates the system's state.
The ancilla and system are measured projectively.
From many trials' measurement data,
one infers $\langle a | \U | b \rangle$,
wherein $\U = U$ or $U^\dag$ and
$a, b =  ( w_\ell, \DegenW_{w_\ell} ), ( v_m , \DegenV_{v_m} )$.
These inner products are multiplied together
to form $\OurKD{\rho}$ [Eq.~\eqref{eq:TAForm}].
If $\rho$ shares neither the $\NondegV$
nor the $\NondegW(t)$ eigenbasis,
quantum-state tomography is needed to infer
$\langle v_1 ,  \DegenV_{v_1}  |  \rho  U^\dag  |
   w_3,  \DegenW_{w_3}  \rangle$.

The weak-measurement scheme
is introduced in Sec. II B 3 of~\cite{YungerHalpern_17_Jarzynski}.
A simple case, in which $\rho = \id / \Dim$,
is detailed in Appendix~A of~\cite{YungerHalpern_17_Jarzynski}.
Recent weak measurements~\cite{Bollen_10_Direct,Lundeen_11_Direct,Lundeen_12_Procedure,Bamber_14_Observing,Mirhosseini_14_Compressive,White_16_Preserving,Piacentini_16_Measuring,Suzuki_16_Observation,Thekkadath_16_Direct},
some used to infer KD distributions,
inspired our weak $\OurKD{\rho}$-measurement proposal.
We review weak measurements,
a Kraus-operator model for measurements,
and the $\OurKD{\rho}$-measurement scheme.

\emph{Review of weak measurements:}
Measurements can alter quantum systems' states.
A weak measurement barely disturbs the measured system's state.
In exchange, the measurement
provides little information about the system.
Yet one can infer much by
performing many trials and processing the outcome statistics.

Extreme disturbances result from strong measurements~\cite{NielsenC10}.
The measured system's state collapses onto a subspace.
For example, let $\rho$ denote the initial state.
Let $\A  =  \sum_a  a  \ketbra{a}{a}$ denote
the measured observable's eigendecomposition.
A strong measurement has a probability $\langle a | \rho | a \rangle$
of projecting $\rho$ onto $\ket{a}$.

One can implement a measurement with an ancilla.
Let $X  =  \sum_x  x \ketbra{x}{x}$ denote an ancilla observable.
One correlates $\A$ with $X$ via an interaction unitary.
Von Neumann modeled such unitaries with
$V_\inter  :=  e^{ - i  \tilde{g} \,  \A \otimes X }$~\cite{vonNeumann_32_Mathematische,Dressel_15_Weak}.
The parameter $\tilde{g}$ signifies the interaction strength.\footnote{
$\A$ and $X$ are dimensionless: To form them,
we multiply dimensionful observables by
natural scales of the subsystems.
These scales are incorporated into $\tilde{g}$.}
An ancilla observable---say,
$Y  =  \sum_y y \ketbra{y}{y}$---is measured strongly.

The greater the $\tilde{g}$, the stronger 
the correlation between $\A$ and $Y$.
$\A$ is measured strongly if 
it is correlated with $Y$ maximally,
if a one-to-one mapping interrelates 
the $y$'s and the $a$'s.
Suppose that the $Y$ measurement yields $y$.
We say that an $\A$ measurement has yielded
some outcome $a_y$.

Suppose that $\tilde{g}$ is small. 
$\A$ is correlated imperfectly with $Y$.
The $Y$-measurement outcome, $y$,
provides incomplete information about $\A$.
The value most reasonably attributable to $\A$ remains $a_y$.
But a subsequent measurement of $\A$
would not necessarily yield $a_y$.
In exchange for forfeiting information about $\A$,
we barely disturb the system's initial state.
We can learn more about $\A$ by measuring $\A$ weakly
in each of many trials, then processing measurement statistics.

\emph{Kraus-operator model for measurement:}
Kraus operators~\cite{NielsenC10} model
the system-of-interest evolution
induced by a weak measurement.
Let us choose the following form for $\A$.
Let $V  =  \sum_{ v_\ell,  \DegenV_{v_\ell} }  v_\ell
\ketbra{ v_\ell,  \DegenV_{v_\ell} }{ v_\ell,  \DegenV_{v_\ell} }
=  \sum_{ v_\ell }  v_\ell  \,  \ProjV{v_\ell}$
denote an observable of the system.
$\ProjV{v_\ell}$ projects onto the $v_\ell$ eigenspace.
Let $\A  =  \ketbra{ v_\ell,  \DegenV_{v_\ell} }{ v_\ell,  \DegenV_{v_\ell} }$.
Let $\rho$ denote the system's initial state,
and let $\ket{ D }$ denote the detector's initial state.

Suppose that the $Y$ measurement yields $y$.
The system's state evolves under the Kraus operator
\begin{align}
    \label{eq:Kraus_form00}
    M_y  & =  \langle y | V_\inter | D \rangle  \\
    & =  \langle y | 
    \exp \left( - i \tilde{g} 
                    \ketbra{ v_\ell,  \DegenV_{v_\ell} }{ v_\ell,  \DegenV_{v_\ell} }
                    \otimes  X  \right)
   | D \rangle      \\
    & = \label{eq:Kraus_form0}
    \langle y | D \rangle  \,  \id
    \nonumber \\ & \quad
    +  \langle y |
        \left( e^{ - i \tilde{g} X }  -  \id  \right) 
        | D \rangle  \,
    \ketbra{ v_\ell,  \DegenV_{v_\ell} }{ v_\ell,  \DegenV_{v_\ell} }  \,
\end{align}
as $\rho  \mapsto
\frac{ M_y  \rho  M_y^\dag }{ 
         \Tr \left(  M_y \rho  M_y^\dag  \right) }  \, .$
The third equation follows from Taylor-expanding the exponential,
then replacing the projector's square with the projector.\footnote{
\label{footnote:MeasPauli}
Suppose that each detector observable 
(each of $X$ and $Y$) has
at least as many eigenvalues as $V$.
For example, let $Y$ represent a pointer's position
and $X$ represent the momentum.
Each $X$ eigenstate can be coupled to
one $V$ eigenstate.
$\A$ will equal $V$, and
$V_\inter$ will have the form $e^{ - i \tilde{g} V \otimes X }$.
Such a coupling makes efficient use of the detector:
Every possible final pointer position $y$ correlates with
some $(v_\ell ,  \DegenV_{v_\ell} )$.
Different $\ketbra{ v_\ell,  \DegenV_{v_\ell} }{ v_\ell,  \DegenV_{v_\ell} }$'s 
need not couple to different detectors.
Since a weak measurement of $V$ provides information about
one $( v_\ell,  \DegenV_{v_\ell} )$
as well as a weak measurement of
$\ketbra{ v_\ell,  \DegenV_{v_\ell} }{ v_\ell,  \DegenV_{v_\ell} }$ does,
we will sometimes call a weak measurement of
$\ketbra{ v_\ell,  \DegenV_{v_\ell} }{ v_\ell,  \DegenV_{v_\ell} }$
``a weak measurement of $V$,'' for conciseness.

The efficient detector use
trades off against mathematical simplicity,
if $\A$ is not a projector:
Eq.~\eqref{eq:Kraus_form00} fails to simplify to
Eq.~\eqref{eq:Kraus_form0}.
Rather, $V_\inter$ should be approximated to 
some order in $\tilde{g}$.
The approximation is (i) first-order
if a KD quasiprobability is being inferred
and (ii) third-order
if the OTOC quasiprobability is being inferred.

If $\A$ is a projector,
Eq.~\eqref{eq:Kraus_form00} simplifies to
Eq.~\eqref{eq:Kraus_form0} even if
$\A$ is degenerate,
e.g., $\A  =  \ProjV{v_\ell}$.
Such an $\A$ assignment will prove natural
in Sec.~\ref{section:Measuring}:
Weak measurements of eigenstates
$\ketbra{ v_\ell,  \DegenV_{v_\ell} }{ v_\ell,  \DegenV_{v_\ell} }$
are replaced with less-resource-consuming
weak measurements of $\ProjV{v_\ell}$'s.

Experimentalists might prefer measuring Pauli operators $\sigma^\alpha$
(for $\alpha = x, y, z$)
to measuring projectors $\Pi$ explicitly.
Measuring Paulis suffices, as
the eigenvalues of $\sigma^\alpha$ map,
bijectively and injectively, onto
the eigenvalues of $\Pi$
(Sec.~\ref{section:Measuring}).
Paulis square to the identity, rather than to themselves:
$\left( \sigma^\alpha \right)^2  =  \id$.
Hence Eq.~\eqref{eq:Kraus_form0} becomes
\begin{align}
   \langle y | \cos \left(  \tilde{g} X \right)  |  D \rangle  \,  \id
   - i \langle y | 
     \sin \left( \tilde{g} X \right)  
     | D \rangle  \,  \sigma^\alpha  \,   .
\end{align}
}
We reparameterize the coefficients as
$\langle y | D \rangle  \equiv  p(y)  \,  e^{ i \phi }$,
wherein $p(y)  :=  |  \langle y | D \rangle |$, and
$\langle y |  \left( e^{ - i \tilde{g} X }  -  \id  \right) | D \rangle
\equiv   g(y)  \,   e^{ i \phi }$.
An unimportant global phase is denoted by $e^{ i \phi }$.
We remove this global phase from the Kraus operator,
redefining $M_y$ as
\begin{align}
   \label{eq:Kraus_form}
    M_y  & =    \sqrt{ p( y ) }  \:  \id
    +  g( y )  \,  \ketbra{ v_\ell,  \DegenV_{v_\ell} }{
                                 v_\ell,  \DegenV_{v_\ell} }  \, .
\end{align}

The coefficients have the following significances.
Suppose that the ancilla did not couple to the system.
The $Y$ measurement would have
a baseline probability $p( y )$ of outputting $y$.
The dimensionless parameter $g( y )  \in  \mathbb{C}$
is derived from $\tilde{g}$.
We can roughly interpret $M_y$ statistically:
In any given trial, the coupling has a probability $p( y )$
of failing to disturb the system
(of evolving $\rho$ under $\id$)
and a probability $| g( y ) |^2$ of projecting $\rho$ onto
$\ketbra{ v_\ell,  \DegenV_{v_\ell} }{
                                 v_\ell,  \DegenV_{v_\ell} }$.

\emph{Weak-measurement scheme for inferring
the OTOC quasiprobability $\OurKD{\rho}$:}
Weak measurements have been used to measure
KD quasiprobabilities~\cite{Bollen_10_Direct,Lundeen_11_Direct,Lundeen_12_Procedure,Bamber_14_Observing,Mirhosseini_14_Compressive,White_16_Preserving,Suzuki_16_Observation,Thekkadath_16_Direct}.
These experiments' techniques can be applied
to infer $\OurKD{\rho}$ and, from $\OurKD{\rho}$, the OTOC.
Our scheme involves
three sequential weak measurements per trial
(if $\rho$ is arbitrary) or two
[if $\rho$ shares the $\NondegV$ or the $\NondegW(t)$ eigenbasis,
e.g., if $\rho = \id / \Dim$].
The weak measurements alternate with time evolutions
and precede a strong measurement.

We review the general and simple-case protocols.
A projection trick, introduced in Sec.~\ref{section:ProjTrick},
reduces exponentially the number of trials required
to infer about $\OurKD{\rho}$ and $F(t)$.
The weak-measurement and interference protocols
are analyzed in Sec.~\ref{section:Advantages}.
A circuit for implementing the weak-measurement scheme
appears in Sec.~\ref{section:Circuit}.

Suppose that $\rho$ does not share the $\NondegV$
or the $\NondegW(t)$ eigenbasis.
One implements the following protocol, $\Protocol$:
\begin{enumerate}[(1)]
   \item  Prepare $\rho$.
   \item  Measure $\NondegV$ weakly.
             (Couple the system's $\NondegV$ weakly to
             some observable $X$ of a clean ancilla.
             Measure $X$ strongly.)
   \item  Evolve the system forward in time under $U$.
   \item  Measure $\NondegW$ weakly.
             (Couple the system's $\NondegW$ weakly to
             some observable $Y$ of a clean ancilla.
             Measure $Y$ strongly.)
   \item  Evolve the system backward under $U^\dag$.
   \item  Measure $\NondegV$ weakly.
             (Couple the system's $\NondegV$ weakly to
             some observable $Z$ of a clean ancilla.
             Measure $Z$ strongly.)
   \item  Evolve the system forward under $U$.
   \item  Measure $\NondegW$ strongly.
\end{enumerate}
$X$, $Y$, and $Z$ do not necessarily denote Pauli operators.
Each trial yields three ancilla eigenvalues ($x$, $y$, and $z$)
and one $\NondegW$ eigenvalue ($w_3, \DegenW_{w_3}$).
One implements $\Protocol$ many times.
From the measurement statistics, one infers the probability
$\mathscr{P}_\weak ( x ; y ; z ; w_3, \DegenW_{w_3} )$
that any given trial will yield the outcome quadruple
$( x ; y ; z ; w_3, \DegenW_{w_3} )$.

From this probability, one infers the quasiprobability
$\OurKD{\rho} ( v_1,  \DegenV_{v_1} ;  w_2,  \DegenW_{w_2} ;
  v_2,  \DegenV_{v_2}  ;  w_3,  \DegenW_{w_3}  )$.
The probability has the form
\begin{align}
   \label{eq:P_weak}
   & \mathscr{P}_\weak ( x ; y ; z ; w_3, \DegenW_{w_3} )
   =  \bra{ w_3,  \DegenW_{w_3} }  U
   M_z  U^\dag  M_y  U  M_x
   \nonumber \\ & \qquad \qquad \qquad \quad \times
   \rho  M_x^\dag  U^\dag  M_y^\dag  U  M_z^\dag  U^\dag
   \ket{ w_3,  \DegenW_{w_3} } \, .
\end{align}
We integrate over $x$, $y$, and $z$,
to take advantage of all measurement statistics.
We substitute in for the Kraus operators from Eq.~\eqref{eq:Kraus_form},
then multiply out.
The result appears in Eq.~(A7) of~\cite{YungerHalpern_17_Jarzynski}.
Two terms combine into $\propto \Im \LParen \OurKD{\rho} ( . ) \RParen$.
The other terms form independently measurable ``background'' terms.
To infer $\Re \LParen \OurKD{\rho} ( . ) \RParen$,
one performs $\Protocol$ many more times,
using different couplings
(equivalently, measuring different detector observables).
Details appear in Appendix~A of~\cite{YungerHalpern_17_Jarzynski}.

To infer the OTOC, one multiplies
each quasiprobability value
$\OurKD{\rho} ( v_1,  \DegenV_{v_1} ;  w_2,  \DegenW_{w_2}  ;
    v_2,  \DegenV_{v_2}   ;  w_3,  \DegenW_{w_3} )$
by the eigenvalue product
$v_1 w_2 v_2^* w_3^*$.
Then, one sums over the eigenvalues
and the degeneracy parameters:
\begin{align}
   \label{eq:RecoverF1}
   F(t)   & =  \sum_{ ( v_1,  \DegenV_{v_1} ) ,
   ( w_2,  \DegenW_{w_2} ) , 
   ( v_2,  \DegenV_{v_2} ) ,  ( w_3,  \DegenW_{w_3} ) }
   %
   v_1 w_2 v_2^* w_3^*
   \\  \nonumber &  \quad  \times
   \OurKD{\rho} ( v_1,  \DegenV_{v_1}  ;  w_2,  \DegenW_{w_2}  ;
   v_2,  \DegenV_{v_2}  ;  w_3,  \DegenW_{w_3} )   \, .
\end{align}
Equation~\eqref{eq:RecoverF1} follows from Eq.~\eqref{eq:JarzLike}.
Hence inferring the OTOC from the weak-measurement scheme---inspired by Jarzynski's equality---requires a few steps more than
inferring a free-energy difference $\Delta F$ from
Jarzynski's equality~\cite{Jarzynski_97_Nonequilibrium}.
Yet such quasiprobability reconstructions are performed
routinely in quantum optics.

$\W$ and $V$ are local.
Their degeneracies therefore scale with the system size.
If $\Sys$ consists of $\Sites$ spin-$\frac{1}{2}$ degrees of freedom,
$| \DegenW_{w_\ell} |,  | \DegenV_{v_\ell} |  \sim  2^\Sites$.
Exponentially many $\OurKD{\rho}( . )$ values must be inferred.
Exponentially many trials must be performed.
We sidestep this exponentiality in Sec.~\ref{section:ProjTrick}:
One measures eigenprojectors of the degenerate $\W$ and $V$,
rather than of the nondegenerate $\NondegW$ and $\NondegV$.
The one-dimensional
$\ketbra{ v_\ell,  \DegenV_{v_\ell} }{ v_\ell,  \DegenV_{v_\ell} }$
of Eq.~\eqref{eq:Kraus_form0} is replaced with $\ProjV{v_\ell}$.
From the weak measurements, one infers the coarse-grained quasiprobability
$\sum_{\text{degeneracies}} \OurKD{\rho}( . )
=:  \SumKD{\rho} ( . )$.
Summing $\SumKD{\rho}( . )$ values yields the OTOC:
\begin{align}
          \label{eq:RecoverF2}  
          F(t)  & =  \sum_{ v_1, w_2, v_2, w_3 }
          v_1 w_2 v_2^* w_3^*  \;
          \SumKD{\rho} ( v_1, w_2, v_2, w_3 ) \, .
\end{align}
Equation~\eqref{eq:RecoverF2} follows from
performing the sums over the degeneracy parameters
$\DegenW$ and $\DegenV$ in Eq.~\eqref{eq:RecoverF1}.

Suppose that $\rho$ shares the $\NondegV$ or the $\NondegW(t)$ eigenbasis.
The number of weak measurements reduces to two.
For example, suppose that $\rho$ is
the infinite-temperature Gibbs state $\id / \Dim$.
The protocol $\Protocol$ becomes
\begin{enumerate}[(1)]
   \item  Prepare a $\NondegW$ eigenstate
            $\ket{ w_3,  \DegenW_{w_3} }$.
   \item  Evolve the system backward under $U^\dag$.
   \item  Measure $\NondegV$ weakly.
   \item  Evolve the system forward under $U$.
   \item  Measure $\NondegW$ weakly.
   \item  Evolve the system backward under $U^\dag$.
   \item  Measure $\NondegV$ strongly.
\end{enumerate}

In many recent experiments, only one weak measurement
is performed per trial~\cite{Bollen_10_Direct,Lundeen_11_Direct,Bamber_14_Observing}.
A probability $\mathscr{P}_\weak$ must be approximated
to first order in the coupling constant $g(x)$.
Measuring $\OurKD{\rho}$ requires
two or three weak measurements per trial.
We must approximate $\mathscr{P}_\weak$ to second or third order.
The more weak measurements performed sequentially,
the more demanding the experiment.
Yet sequential weak measurements have been performed recently~\cite{Piacentini_16_Measuring,Suzuki_16_Observation,Thekkadath_16_Direct}.
The experimentalists aimed to reconstruct density matrices
and to measure non-Hermitian operators.
The OTOC measurement provides
new applications for their techniques.

\section{Experimentally measuring $\OurKD{\rho}$
and the coarse-grained $\SumKD{\rho}$}
\label{section:Measuring}

Multiple reasons motivate measurements of
the OTOC quasiprobability $\OurKD{\rho}$.
$\OurKD{\rho}$ is more fundamental than the OTOC $F(t)$,
$F(t)$ results from combining values of $\OurKD{\rho}$.
$\OurKD{\rho}$ exhibits behaviors
not immediately visible in $F(t)$,
as shown in Sections~\ref{section:Numerics} and~\ref{section:Brownian}.
$\OurKD{\rho}$ therefore holds interest in its own right.
Additionally, $\OurKD{\rho}$ suggests
new schemes for measuring the OTOC.
One measures the possible values of $\OurKD{\rho} ( . )$, then
combines the values to form $F(t)$.
Two measurement schemes are detailed in~\cite{YungerHalpern_17_Jarzynski}
and reviewed in Sec.~\ref{section:Intro_weak_meas}.
One scheme relies on weak measurements;
one, on interference.
We simplify, evaluate, and augment these schemes.

First, we introduce a ``projection trick'':
Summing over degeneracies turns one-dimensional projectors
(e.g., $\ketbra{ w_\ell,  \DegenW_{w_\ell} }{ w_\ell,  \DegenW_{w_\ell} }$)
into projectors onto degenerate eigenspaces (e.g., $\ProjW{w_\ell}$).
The \emph{coarse-grained OTOC quasiprobability} $\SumKD{\rho}$ results.
This trick decreases exponentially the number of trials required
to infer the OTOC from weak measurements.\footnote{
The summation preserves interesting properties of the quasiprobability---nonclassical negativity and nonreality, as well as intrinsic time scales.
We confirm this preservation via numerical simulation
in Sec.~\ref{section:Numerics}.}
Section~\ref{section:Advantages} concerns pros and cons of
the weak-measurement and interference schemes
for measuring $\OurKD{\rho}$ and $F(t)$.
We also compare those schemes with
alternative schemes for measuring $F(t)$.
Section~\ref{section:Circuit} illustrates a circuit for implementing
the weak-measurement scheme.
Section~\ref{section:MeasSumFromOTOC}
shows how to infer $\SumKD{\rho}$ not only from
the measurement schemes in Sec.~\ref{section:Intro_weak_meas},
but also with alternative OTOC-measurement proposals
(e.g.,~\cite{Swingle_16_Measuring})
(if the eigenvalues of $\W$ and $V$ are $\pm 1$).

\subsection{The coarse-grained OTOC quasiprobability $\SumKD{\rho}$
and a projection trick}
\label{section:ProjTrick}

$\W$ and $V$ are local.
They manifest, in our spin-chain example, as one-qubit Paulis
that nontrivially transform opposite ends of the chain.
The operators' degeneracies grows exponentially with
the system size $\Sites$:
$| \DegenW_{w_\ell} | ,  \:  | \DegenV_{v_m} |  \sim  2^\Sites$.
Hence the number of $\OurKD{\rho} ( . )$ values grows exponentially.
One must measure exponentially many numbers
to calculate $F(t)$ precisely via $\OurKD{\rho}$.
We circumvent this inconvenience
by summing over the degeneracies in $\OurKD{\rho} ( . ) $,
forming the coarse-grained quasiprobability $\SumKD{\rho} ( . )$.
$\SumKD{\rho} ( . )$ can be measured in numerical simulations,
experimentally via weak measurements,
and (if the eigenvalues of $\W$ and $V$ are $\pm 1$)
experimentally with other $F(t)$-measurement set-ups
(e.g.,~\cite{Swingle_16_Measuring}).

   The \emph{coarse-grained OTOC quasiprobability}
   results from marginalizing $\OurKD{\rho} ( . )$ over its degeneracies:
   \begin{align}
      \label{eq:Sum_KD_def}
      & \SumKD{\rho} ( v_1, w_2, v_2, w_3 )  :=
      \sum_{ \DegenV_{v_1},  \DegenW_{w_2} ,
                  \DegenV_{v_2} ,  \DegenW_{w_3} }
      \nonumber \\ & \quad
      \OurKD{\rho} ( v_1,  \DegenV_{v_1} ;  w_2,  \DegenW_{w_2} ;
                              v_2,  \DegenV_{v_2}  ;  w_3,  \DegenW_{w_3} )   \, .
   \end{align}

Equation~\eqref{eq:Sum_KD_def} reduces to a more practical form.
Consider substituting into Eq.~\eqref{eq:Sum_KD_def}
for $\OurKD{\rho} ( . )$ from Eq.~\eqref{eq:TAForm}.
The right-hand side of Eq.~\eqref{eq:TAForm} equals a trace.
Due to the trace's cyclicality,
the three rightmost factors can be shifted leftward:
\begin{align}
   & \SumKD{\rho} ( v_1, w_2, v_2, w_3 )
   =  \sum_{ \substack{  \DegenV_{v_1},  \DegenW_{w_2} ,  \\
                    \DegenV_{v_2} ,  \DegenW_{w_3} } }
   \Tr \Big( \rho U^\dag
   \ketbra{ w_3,  \DegenW_{w_3} }{ w_3,  \DegenW_{w_3} }  U
   \nonumber \\ & \times
   \ketbra{ v_2,  \DegenV_{v_2} }{  v_2,  \DegenV_{v_2} }
   U^\dag  \ketbra{ w_2,  \DegenW_{w_2} }{  w_2,  \DegenW_{w_2} } U
   \ketbra{ v_1,  \DegenV_{v_1} }{ v_1,  \DegenV_{v_1} }
   \Big) \, .
\end{align}
The sums are distributed throughout the trace:
\begin{align}
   \label{eq:Corase_help1}
   & \SumKD{\rho} ( v_1, w_2, v_2, w_3 )
   =  \Tr \Bigg( \rho
   \Bigg[  U^\dag  \sum_{ \DegenW_{w_3} }
      \ketbra{ w_3,  \DegenW_{w_3} }{ w_3,  \DegenW_{w_3} }  U  \Bigg]
   \nonumber \\ & \times
   \Bigg[  \sum_{ \DegenV_{v_2} }
      \ketbra{ v_2,  \DegenV_{v_2} }{  v_2,  \DegenV_{v_2} }  \Bigg]
   \Bigg[  U^\dag  \sum_{ \DegenW_{w_2} }
      \ketbra{ w_2,  \DegenW_{w_2} }{  w_2,  \DegenW_{w_2} }  U  \Bigg]
   \nonumber \\ & \times
   \Bigg[  \sum_{ \DegenV_{v_1} }
      \ketbra{ v_1,  \DegenV_{v_1} }{ v_1,  \DegenV_{v_1} }  \Bigg]
   \Bigg)  \, .
\end{align}
Define
 \begin{align}
      \label{eq:ProjW}
      \ProjW{w_\ell}  :=  \sum_{ \DegenW_{w_\ell} }
      \ketbra{ w_\ell,  \DegenW_{w_\ell} }{ w_\ell,  \DegenW_{w_\ell} }
   \end{align}
   as the projector onto the $w_\ell$ eigenspace of $\W$,
   \begin{align}
      \label{eq:ProjWt}
      \ProjWt{ w_\ell }  :=  U^\dag  \ProjW{ w_\ell }  U
   \end{align}
   as the projector onto the $w_\ell$ eigenspace of $\W(t)$, and
   \begin{align}
      \label{eq:ProjV}
      \ProjV{v_\ell}  :=  \sum_{ \DegenV_{v_\ell} }
      \ketbra{ v_\ell,  \DegenV_{v_\ell} }{ v_\ell,  \DegenV_{v_\ell} }
   \end{align}
   as the projector onto the $v_\ell$ eigenspace of $V$.
We substitute into Eq.~\eqref{eq:Corase_help1},
then invoke the trace's cyclicality:
\begin{align}
      \label{eq:SumKD_simple2} \boxed{
      \SumKD{\rho} ( v_1, w_2, v_2, w_3 )
      =  \Tr  \Big(  \ProjWt{ w_3 }  \ProjV{v_2}
                            \ProjWt{w_2}  \ProjV{v_1}  \rho  \Big)  }  \, .
\end{align}

Asymmetry distinguishes Eq.~\eqref{eq:SumKD_simple2}
from Born's Rule and from expectation values.
Imagine preparing $\rho$,
measuring $V$ strongly, evolving $\Sys$ forward under $U$,
measuring $\W$ strongly, evolving $\Sys$ backward under $U^\dag$,
measuring $V$ strongly, evolving $\Sys$ forward under $U$,
and measuring $\W$.
The probability of obtaining the outcomes
$v_1, w_2, v_2$, and $w_3$, in that order, is
\begin{align}
   \label{eq:BornCompare}
   \Tr \Big(  & \ProjWt{ w_3 }  \ProjV{ v_2 }   \ProjWt{ w_2 }  \ProjV{ v_1 }
              \rho  \ProjV{ v_1 }  \ProjWt{ w_2 }   \ProjV{ v_2 }  \ProjWt{ w_3 }  \Big) \, .
\end{align}
The operator $\ProjWt{ w_3 }  \ProjV{ v_2 }  \ProjWt{ w_2 }  \ProjV{ v_1 }$
conjugates $\rho$ symmetrically.
This operator multiplies $\rho$ asymmetrically in Eq.~\eqref{eq:SumKD_simple2}.
Hence $\SumKD{\rho}$ does not obviously equal a probability.

Nor does $\SumKD{\rho}$ equal an expectation value.
Expectation values have the form
$\Tr ( \rho \A )$, wherein $\A$ denotes a Hermitian operator.
The operator leftward of the $\rho$ in Eq.~\eqref{eq:SumKD_simple2}
is not Hermitian.
Hence $\SumKD{\rho}$ lacks two symmetries of
familiar quantum objects:
the symmetric conjugation in Born's Rule
and the invariance, under Hermitian conjugation,
of the observable $\A$ in an expectation value.

The right-hand side of Eq.~\eqref{eq:SumKD_simple2} can be measured
numerically and experimentally.
We present numerical measurements in Sec.~\ref{section:Numerics}.
The weak-measurement scheme follows
from Appendix~A of~\cite{YungerHalpern_17_Jarzynski},
reviewed in Sec.~\ref{section:Intro_weak_meas}:
Section~\ref{section:Intro_weak_meas} features projectors onto
one-dimensional eigenspaces, e.g.,
$\ketbra{ v_1, \DegenV_{v_1} }{ v_1, \DegenV_{v_1} }$.
Those projectors are replaced with
$\Pi$'s onto higher-dimensional eigenspaces.
Section~\ref{section:MeasSumFromOTOC} details
how $\SumKD{\rho}$ can be inferred
from alternative OTOC-measurement schemes.

\subsection{Analysis of the quasiprobability-measurement schemes and
comparison with other OTOC-measurement schemes}
\label{section:Advantages}

\begin{figure}[h]
\centering
\includegraphics[width=.9\textwidth]{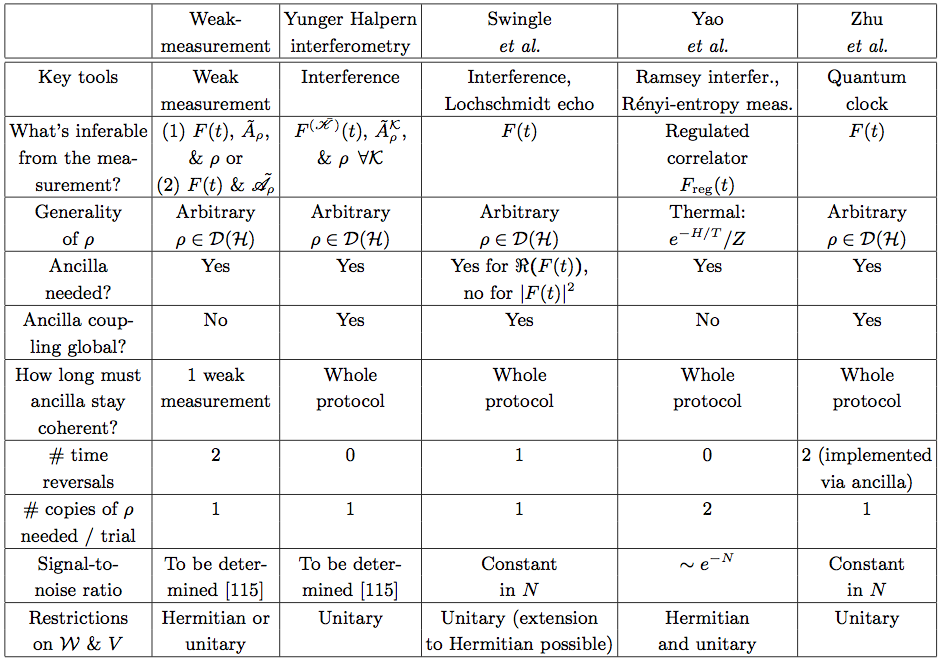}
\caption{\caphead{Comparison of our measurement schemes with alternatives:}
This paper focuses on the weak-measurement and interference schemes
for measuring the OTOC quasiprobability $\OurKD{\rho}$
or the coarse-grained quasiprobability $\SumKD{\rho}$.
From $\OurKD{\rho}$ or $\SumKD{\rho}$,
one can infer the OTOC $F(t)$.
These schemes appear in~\cite{YungerHalpern_17_Jarzynski},
are reviewed in Sec.~\ref{section:Intro_weak_meas},
and are assessed in Sec.~\ref{section:Advantages}.
We compare our schemes with
the OTOC-measurement schemes in~\cite{Swingle_16_Measuring,Yao_16_Interferometric,Zhu_16_Measurement}.
More OTOC-measurement schemes appear in~\cite{Danshita_16_Creating,Li_16_Measuring,Garttner_16_Measuring,Tsuji_17_Exact,Campisi_16_Thermodynamics,Bohrdt_16_Scrambling}.
Each row corresponds to a desirable quantity
or to a resource potentially challenging to realize experimentally.
The regulated correlator $F_\reg(t)$ [Eq.~\eqref{eq:RegOTOC_def}]
is expected to behave similarly to $F(t)$~\cite{Maldacena_15_Bound,Yao_16_Interferometric}.
$\mathcal{D} ( \Hil )$ denotes the set of density operators defined on
the Hilbert space $\Hil$.
$\rho$ denotes the initially prepared state.
Target states $\rho_\target$ are never prepared perfectly;
$\rho$ may differ from $\rho_\target$.
Experimentalists can reconstruct $\rho$ by trivially processing
data taken to infer $\OurKD{\rho}$~\cite{YungerHalpern_17_Jarzynski}
(Sec.~\ref{section:TA_retro}).
$F^\ParenKB(t)$ denotes the $\Opsb$-fold OTOC,
which encodes $\Ops  =  2 \Opsb - 1$ time reversals.
The conventional OTOC corresponds to $\Ops = 3$.
The quasiprobability behind $F^\ParenKB(t)$ is
$\OurKD{\rho}^\ParenK$ (Sec.~\ref{section:HigherOTOCs}).
$\Sites$ denotes the system size, e.g., the number of qubits.
The Swingle \emph{et al.} and Zhu \emph{et al.} schemes
have constant signal-to-noise ratios (SNRs)
in the absence of environmental decoherence.
The Yao \emph{et al.} scheme's SNR varies inverse-exponentially with
the system's entanglement entropy, $S_{\text{vN}}$.
The system occupies a thermal state $e^{ - H / T } / Z$,
so $S_{\text{vN}}  \sim  \log ( 2^\Sites )  =  \Sites$.}
\label{table:Compare}
\end{figure}
%

Section~\ref{section:Intro_weak_meas} reviews
two schemes for inferring $\OurKD{\rho}$:
a weak-measurement scheme and an interference scheme.
From $\OurKD{\rho}$ measurements,
one can infer the OTOC $F(t)$.
We evaluate our schemes' pros and cons.
Alternative schemes for measuring $F(t)$ have been proposed~\cite{Swingle_16_Measuring,Danshita_16_Creating,Yao_16_Interferometric,Zhu_16_Measurement,Tsuji_17_Exact,Campisi_16_Thermodynamics,Bohrdt_16_Scrambling},
and two schemes have been realized~\cite{Li_16_Measuring,Garttner_16_Measuring}.
We compare our schemes with alternatives,
as summarized in Table~\ref{table:Compare}.
For specificity, we focus on~\cite{Swingle_16_Measuring,Yao_16_Interferometric,Zhu_16_Measurement}.

The weak-measurement scheme augments
the set of techniques and platforms
with which $F(t)$ can be measured.
Alternative schemes rely on interferometry~\cite{Swingle_16_Measuring,Yao_16_Interferometric,Bohrdt_16_Scrambling}, controlled unitaries~\cite{Swingle_16_Measuring,Zhu_16_Measurement},
ultracold-atoms tools~\cite{Danshita_16_Creating,Tsuji_17_Exact,Bohrdt_16_Scrambling},
and strong two-point measurements~\cite{Campisi_16_Thermodynamics}.
Weak measurements, we have shown, belong in the OTOC-measurement toolkit.
Such weak measurements are expected
to be realizable, in the immediate future,
with superconducting qubits~\cite{White_16_Preserving,Hacohen_16_Quantum,Rundle_16_Quantum,Takita_16_Demonstration,Kelly_15_State,Heeres_16_Implementing,Riste_15_Detecting},
trapped ions~\cite{Gardiner_97_Quantum,Choudhary_13_Implementation,Lutterbach_97_Method,Debnath_16_Nature,Monz_16_Realization,Linke_16_Experimental,Linke_17_Experimental},
cavity QED~\cite{Guerlin_07_QND,Murch_13_SingleTrajectories}, 
ultracold atoms~\cite{Browaeys_16_Experimental},
and perhaps NMR~\cite{Xiao_06_NMR,Dawei_14_Experimental}.
Circuits for weakly measuring qubit systems
have been designed~\cite{Groen_13_Partial,Hacohen_16_Quantum}.
Initial proof-of-principle experiments might not require
direct access to the qubits:
The five superconducting qubits available from IBM,
via the cloud, might suffice~\cite{IBM_QC}.
Random two-qubit unitaries could simulate chaotic Hamiltonian evolution.

In many weak-measurement experiments,
just one weak measurement is performed per trial~\cite{Lundeen_11_Direct,Lundeen_12_Procedure,Bamber_14_Observing,Mirhosseini_14_Compressive}.
Yet two weak measurements
have recently been performed sequentially~\cite{Piacentini_16_Measuring,Suzuki_16_Observation,Thekkadath_16_Direct}.
Experimentalists aimed to ``directly measure general quantum states''~\cite{Lundeen_12_Procedure}
and to infer about non-Hermitian observable-like operators.
The OTOC motivates a new application
of recently realized sequential weak measurements.

Our schemes furnish not only the OTOC $F(t)$,
but also more information:
\begin{enumerate}[(1)]

   \item From the weak-measurement scheme in~\cite{YungerHalpern_17_Jarzynski},
   we can infer the following:
   \begin{enumerate}[(A)]
      \item  The OTOC quasiprobability $\OurKD{\rho}$.
      The quasiprobability is more fundamental than $F(t)$,
as combining $\OurKD{\rho} ( . )$ values yields $F(t)$
[Eq.~\eqref{eq:RecoverF1}].

      \item  The OTOC $F(t)$.
      
      \item The form $\rho$ of the state prepared.
      Suppose that we wish to evaluate $F(t)$ on
a target state $\rho_\target$.
$\rho_\target$ might be difficult to prepare, e.g., might be thermal.
The prepared state $\rho$ approximates $\rho_\target$.
Consider performing
the weak-measurement protocol $\Protocol$
with $\rho$. One infers $\OurKD{\rho}$.
Summing $\OurKD{\rho}( . )$ values yields the form of $\rho$.
We can assess the preparation's accuracy
without performing tomography independently.
Whether this assessment meets experimentalists' requirements for precision
remains to be seen.
Details appear in Sec.~\ref{section:TA_Coeffs}.
   \end{enumerate}

   \item The weak-measurement protocol $\Protocol$
            is simplified later in this section.
            Upon implementing the simplified protocol,
            we can infer the following information:
   \begin{enumerate}[(A)]
      \item The coarse-grained OTOC quasiprobability $\SumKD{\rho}$.
      Though less fundamental than the fine-grained $\OurKD{\rho}$,
      $\SumKD{\rho}$ implies the OTOC's form
      [Eq.~\eqref{eq:RecoverF2}].
      \item  The OTOC $F(t)$.
   \end{enumerate}

   \item Upon implementing the interferometry scheme in~\cite{YungerHalpern_17_Jarzynski},
   we can infer the following information:
   \begin{enumerate}[(A)]
      \item  The OTOC quasiprobability $\OurKD{\rho}$.
      \item  The OTOC $F(t)$.
      \item  The form of the state $\rho$ prepared.
      \item  All the $\Opsb$-fold OTOCs $F^\ParenKB(t)$,
      which generalize the OTOC $F(t)$.
                $F(t)$ encodes three time reversals.
                $F^\ParenKB(t)$ encodes $\Ops = 2 \Opsb - 1 = 3, 5, \ldots$
                time reversals.
                Details appear in Sec.~\ref{section:HigherOTOCs}.
      \item The quasiprobability $\OurKD{\rho}^\ParenK$
               behind $F^\ParenKB(t)$, for all $\Ops$
               (Sec.~\ref{section:HigherOTOCs}).
   \end{enumerate}

\end{enumerate}

We have delineated the information inferable from
the weak-measurement and interference schemes
for measuring $\OurKD{\rho}$ and $F(t)$.
Let us turn to other pros and cons.

The weak-measurement scheme's ancillas
need not couple to the whole system.
One measures a system weakly by coupling an ancilla to the system,
then measuring the ancilla strongly.
Our weak-measurement protocol requires
one ancilla per weak measurement.
Let us focus, for concreteness, on
an $\SumKD{\rho}$ measurement for a general $\rho$.
The protocol involves three weak measurements and so three ancillas.
Suppose that $\W$ and $V$ manifest as one-qubit Paulis
localized at opposite ends of a spin chain.
Each ancilla need interact with only one site (Fig.~\ref{fig:Circuit}).
In contrast, the ancilla in~\cite{Zhu_16_Measurement}
couples to the entire system.
So does the ancilla in our interference scheme
for measuring $\OurKD{\rho}$.
Global couplings can be engineered in some platforms,
though other platforms pose challenges.
Like our weak-measurement scheme,~\cite{Swingle_16_Measuring} and~\cite{Yao_16_Interferometric}
require only local ancilla couplings.

In the weak-measurement protocol,
each ancilla's state must remain coherent
during only one weak measurement---during
the action of one (composite) gate in a circuit.
The first ancilla may be erased, then reused in the third weak measurement.
In contrast, each ancilla in~\cite{Swingle_16_Measuring,Yao_16_Interferometric,Zhu_16_Measurement}
remains in use throughout the protocol.
The Swingle \emph{et al.} scheme
for measuring $\Re \LParen F(t) \RParen$, too,
requires an ancilla that remains coherent throughout the protocol~\cite{Swingle_16_Measuring}.
The longer an ancilla's ``active-duty'' time,
the more likely the ancilla's state is to decohere.
Like the weak-measurement sheme,
the Swingle \emph{et al.} scheme for measuring $| F (t) |^2$
requires no ancilla~\cite{Swingle_16_Measuring}.

Also in the interference scheme for measuring $\OurKD{\rho}$~\cite{YungerHalpern_17_Jarzynski},
an ancilla remains active throughout the protocol.
That protocol, however, is short: Time need not be reversed in any trial.
Each trial features exactly one $U$ or $U^\dag$, not both.
Time can be difficult to reverse in some platforms, for two reasons.
Suppose that a Hamiltonian $H$ generates a forward evolution.
A perturbation $\varepsilon$ might lead
$- (H + \varepsilon)$ to generate the reverse evolution.
Perturbations can mar
long-time measurements of $F(t)$~\cite{Zhu_16_Measurement}.
Second, systems interact with environments.
Decoherence might not be completely reversible~\cite{Swingle_16_Measuring}.
Hence the lack of a need for time reversal,
as in our interference scheme and in~\cite{Yao_16_Interferometric,Zhu_16_Measurement},
has been regarded as an advantage.

Unlike our interference scheme,
the weak-measurement scheme requires that time be reversed.
Perturbations $\varepsilon$ threaten the weak-measurement scheme
as they threaten the Swingle \emph{et al.} scheme~\cite{Swingle_16_Measuring}.
$\varepsilon$'s might threaten the weak-measurement scheme more,
because time is inverted twice in our scheme.
Time is inverted only once in~\cite{Swingle_16_Measuring}.
However, our error might be expected to have roughly the size
of the Swingle \emph{et al.} scheme's error~\cite{Swingle_Resilience}.
Furthermore, tools for mitigating
the Swingle \emph{et al.} scheme's inversion error
are being investigated~\cite{Swingle_Resilience}.
Resilience of the Swingle \emph{et al.} scheme to decoherence
has been analyzed~\cite{Swingle_16_Measuring}.
These tools may be applied to the weak-measurement scheme~\cite{Swingle_Resilience}.
Like resilience, our schemes' signal-to-noise ratios
require further study.

As noted earlier, as the system size $\Sites$ grows,
the number of trials required to infer $\OurKD{\rho}$ grows exponentially.
So does the number of ancillas required to infer $\OurKD{\rho}$:
Measuring a degeneracy parameter $\DegenW_{w_\ell}$ or $\DegenV_{v_m}$
requires a measurement of each spin.
Yet the number of trials, and the number of ancillas,
required to measure the coarse-grained $\SumKD{\rho}$
remains constant as $\Sites$ grows.
One can infer $\SumKD{\rho}$ from weak measurements
and, alternatively, from other $F(t)$-measurement schemes
(Sec.~\ref{section:MeasSumFromOTOC}).
$\SumKD{\rho}$ is less fundamental than $\OurKD{\rho}$,
as $\SumKD{\rho}$ results from coarse-graining $\OurKD{\rho}$.
$\SumKD{\rho}$, however, exhibits
nonclassicality and OTOC time scales (Sec.~\ref{section:Numerics}).
Measuring $\SumKD{\rho}$ can balance
the desire for fundamental knowledge with practicalities.

The weak-measurement scheme for inferring $\SumKD{\rho}$
can be rendered more convenient.
Section~\ref{section:ProjTrick} describes measurements
of projectors $\Pi$.
Experimentalists might prefer measuring
Pauli operators $\sigma^\alpha$.
Measuring Paulis suffices
for inferring a multiqubit system's $\SumKD{\rho}$:
The relevant $\Pi$ projects onto
an eigenspace of a $\sigma^\alpha$.
Measuring the $\sigma^\alpha$ yields $\pm 1$.
These possible outcomes map bijectively onto
the possible $\Pi$-measurement outcomes.
See Footnote~\ref{footnote:MeasPauli} for mathematics.

Our weak-measurement and interference schemes
offer the advantage of involving general operators.
$\W$ and $V$ must be Hermitian or unitary,
not necessarily one or the other.
Suppose that $\W$ and $V$ are unitary.
Hermitian operators $\GW$ and $\GV$ generate $\W$ and $V$,
as discussed in Sec.~\ref{section:SetUp}.
$\GW$ and $\GV$ may be measured in place of $\W$ and $V$.
This flexibility expands upon the measurement opportunities of, e.g.,~\cite{Swingle_16_Measuring,Yao_16_Interferometric,Zhu_16_Measurement}, 
which require unitary operators.

Our weak-measurement and interference schemes offer leeway
in choosing not only $\W$ and $V$, but also $\rho$.
The state can assume any form  $\rho \in \mathcal{D} ( \Hil )$.
In contrast, infinite-temperature Gibbs states $\rho = \id / \Dim$
were used in~\cite{Li_16_Measuring,Garttner_16_Measuring}.
Thermality of $\rho$ is assumed in~\cite{Yao_16_Interferometric}.
Commutation of $\rho$ with $V$ is assumed in~\cite{Campisi_16_Thermodynamics}.
If $\rho$ shares a $V$ eigenbasis or the $\W(t)$ eigenbasis,
e.g., if $\rho = \id / \Dim$,
our weak-measurement protocol simplifies
from requiring three sequential weak measurements
to requiring two.

\subsection{Circuit for inferring $\SumKD{\rho}$
from weak measurements}
\label{section:Circuit}

Consider a 1D chain $\Sys$ of $\Sites$ qubits.
A circuit implements the weak-measurement scheme
reviewed in Sec.~\ref{section:Intro_weak_meas}.
We exhibit a circuit for measuring $\SumKD{\rho}$.
One subcircuit implements each weak measurement.
These subcircuits result from augmenting
Fig.~1 of~\cite{Dressel_14_Implementing}.

Dressel \emph{et al.} use the \emph{partial-projection formalism},
which we review first.
We introduce notation, then review
the weak-measurement subcircuit of~\cite{Dressel_14_Implementing}.
Copies of the subcircuit are embedded into
our $\SumKD{\rho}$-measurement circuit.

\subsubsection{Partial-projection operators}
\label{section:PartialProj_main}

Partial-projection operators update a state after a measurement
that may provide incomplete information.
Suppose that $\Sys$ begins in a state $\ket{ \psi }$.
Consider performing a measurement that could output $+$ or $-$.
Let $\Pi_+$ and $\Pi_-$ denote
the projectors onto the $+$ and $-$ eigenspaces.
Parameters $p, q  \in  [0, 1]$ quantify the correlation
between the outcome and the premeasurement state.
If $\ket{ \psi }$ is a $+$ eigenstate, the measurement has
a probability $p$ of outputting $+$.
If $\ket{ \psi }$ is a $-$ eigenstate, the measurement has
a probability $q$ of outputting $-$.

Suppose that outcome $+$ obtains.
We update $\ket{ \psi }$ using the \emph{partial-projection operator}
$D_+  :=  \sqrt{p}  \;  \Pi_+
+  \sqrt{1 - q} \;  \Pi_-$:
$\ket{\psi}  \mapsto
   \frac{ D_{+} \ket{ \psi } }{ || D_+ \ket{ \psi } ||^2 } \, .$
If the measurement yields $-$,
we update $\ket{ \psi }$ with
$D_-  :=  \sqrt{1 - p}  \;  \Pi_+
+  \sqrt{q}  \;  \Pi_-$.

The measurement is strong if
$(p, q) = (0, 1)$ or $(1, 0)$.
$D_+$ and $D_-$ reduce to projectors.
The measurement collapses $\ket{ \psi }$ onto an eigenspace.
The measurement is weak if $p$ and $q$ lie close to $\frac{1}{2}$:
$D_\pm$ lies close to the normalized identity, $\frac{\id}{ \Dim }$.
Such an operator barely changes the state.
The measurement provides hardly any information.

We modeled measurements with Kraus operators $M_x$
in Sec.~\ref{section:Intro_weak_meas}.
The polar decomposition of $M_x$~\cite{Preskill_15_Ch3}
is a partial-projection operator.
Consider measuring a qubit's $\sigma^z$.
Recall that $X$ denotes a detector observable.
Suppose that, if an $X$ measurement yields $x$,
a subsequent measurement of the spin's $\sigma^z$
most likely yields $+$.
The Kraus operator $M_x  =  \sqrt{ p(x) }  \:  \id  +  g(x)  \,  \Pi_+$
updates the system's state.
$M_x$ is related to $D_+$ by
   $D_+  =  U_x  \sqrt{ M_x^\dag  M_x }$
for some unitary $U_x$.
The form of $U_x$ depends on the system-detector coupling
and on the detector-measurement outcome.

The imbalance $| p - q |$ can be tuned experimentally.
Our scheme has no need for a nonzero imbalance.
We assume that $p$ equals $q$.

\subsubsection{Notation}
\label{section:Circuit_notation}

Let $\bm{\sigma}  :=  \sigma^x  \,  \hat{ \mathbf{ x } }
+  \sigma^y  \,  \hat{ \mathbf{ y } }
+  \sigma^z \,  \hat{ \mathbf{ z } }$
denote a vector of one-qubit Pauli operators.
The $\sigma^z$ basis serves as
the computational basis in~\cite{Dressel_14_Implementing}.
We will exchange the $\sigma^z$ basis with
the $\W$ eigenbasis, or with the $V$ eigenbasis,
in each weak-measurement subcircuit.

In our spin-chain example, $\W$ and $V$ denote one-qubit Pauli operators
localized on opposite ends of the chain $\Sys$:
$\W  =  \sigma^{ \W }  \otimes  \id^{ \otimes (\Sites - 1) }$,
and $V  =  \id^{ \otimes (\Sites - 1) }  \otimes  \sigma^{ V }$.
Unit vectors $\hat{\W}, \hat{V}  \in  \mathbb{R}^3$
are chosen such that
$\sigma^{ n }  :=  \bm{\sigma}  \cdot  \hat{ \bm{n} }$,
for $n  =  \W,  V$.

The one-qubit Paulis eigendecompose as
$\sigma^{ \W }  =  \ketbra{ + \W }{ + \W }
-  \ketbra{ - \W }{ - \W }$ and
$\sigma^{ V }  =  \ketbra{ +V }{ +V }  -  \ketbra{ -V }{ -V }$.
The whole-system operators eigendecompose as
$\W  =  \ProjW{+}  -  \ProjW{-}$  and
$V  =  \ProjV{+}  -  \ProjV{-}$.
A rotation operator $R_{ n }$ maps
the $\sigma^z$ eigenstates to the $\sigma^{ n }$ eigenstates:
$R_{ n }  \ket{ +z }  =  \ket{ + n }$, and
$R_{ n }  \ket{ -z }  =  \ket{ - n }$.

We model weak $\W$ measurements
with the partial-projection operators
\begin{align}
   \label{eq:PartialProjW}
   & D_+^{ \W }  :=  \sqrt{p_{\W}}  \;  \ProjW{+}
   +  \sqrt{1 - p_{\W} }  \;  \ProjW{-}
   \;  \:  \text{and}  \\  &
   D_-^{ \W }  :=  \sqrt{ 1 - p_{\W} }  \;  \ProjW{+}
   +  \sqrt{p_{\W}}  \;  \ProjW{-}  \, .
\end{align}
The $V$ partial-projection operators
are defined analogously:
\begin{align}
   \label{eq:PartialProjV}
   & D_+^{ V }  :=  \sqrt{p_V}  \;  \ProjV{+}
   +  \sqrt{1 - p_V }  \;  \ProjV{-}
   \;  \:  \text{and}  \\  &
   D_-^{ V }  :=  \sqrt{ 1 - p_V }  \;  \ProjV{+}
   +  \sqrt{p_V}  \;  \ProjV{-}  \, .
\end{align}

\subsubsection{Weak-measurement subcircuit}
\label{section:Weak_subcircuit}

%
%
\begin{figure}[h]
\centering
\begin{subfigure}{0.5\textwidth}
\centering
\includegraphics[width=.95\textwidth]{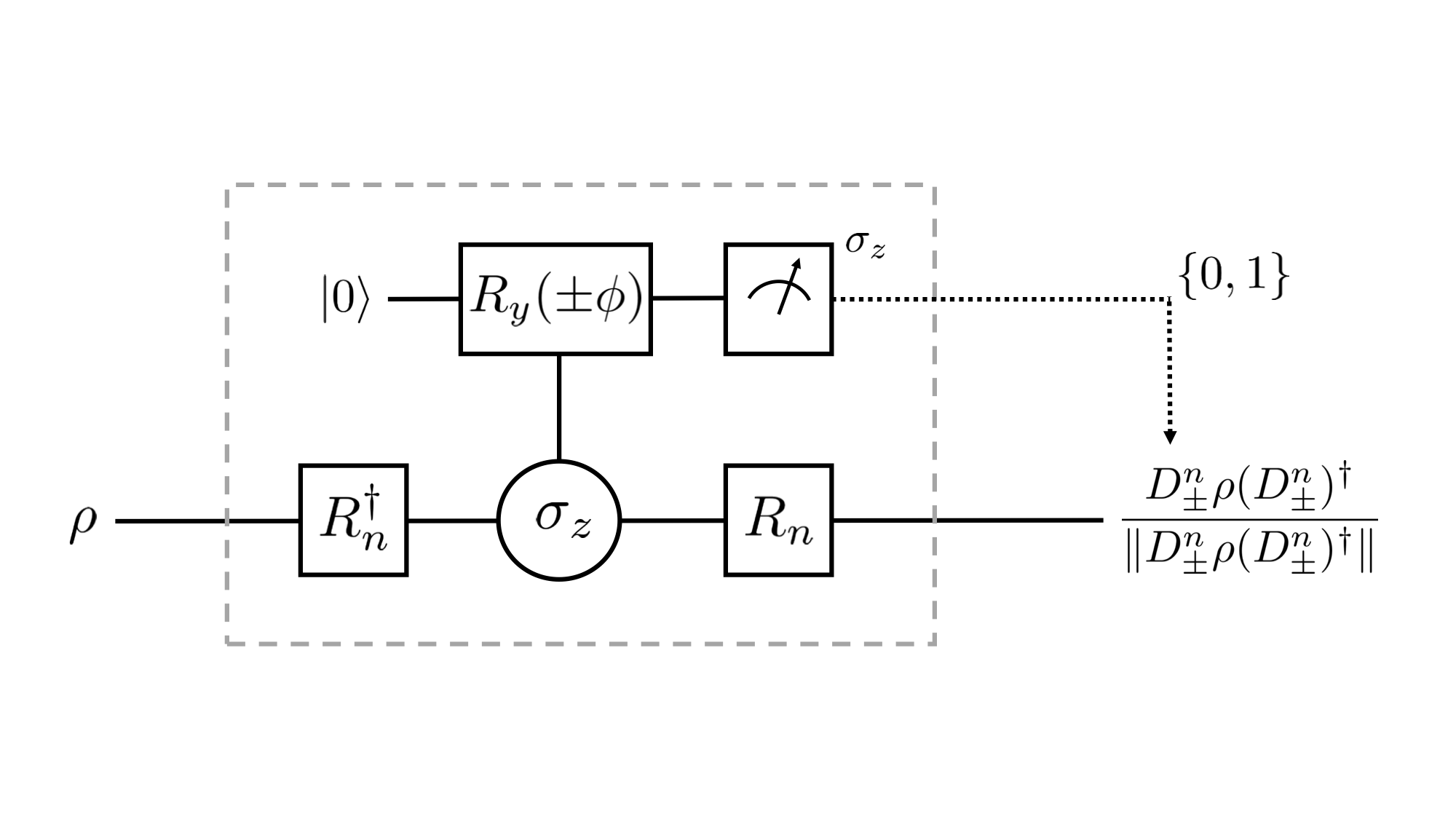}
\caption{}
\label{fig:D_subcircuit}
\end{subfigure}
\begin{subfigure}{.5\textwidth}
\centering
\includegraphics[width=.95\textwidth]{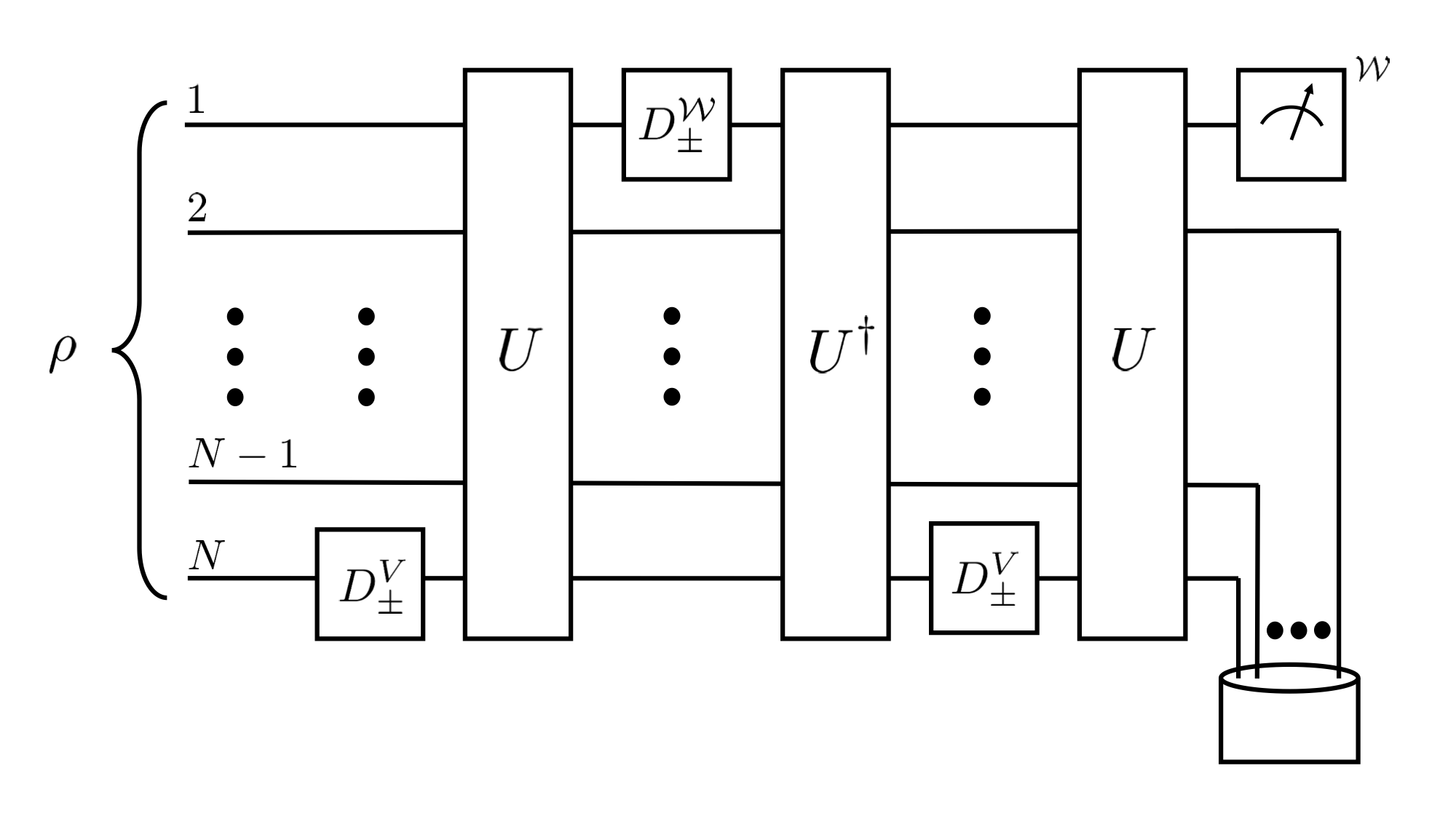}
\caption{}
\label{fig:Full_circuit}
\end{subfigure}
\caption{\caphead{Quantum circuit for inferring
the coarse-grained OTOC quasiprobability $\SumKD{\rho}$
from weak measurements:}
We consider a system of $\Sites$ qubits
prepared in a state $\rho$.
The local operators $\W = \sigma^{ \W }  \otimes
\id^{ \otimes (\Sites - 1) }$ and
$V  = \id^{ \otimes (\Sites - 1) }  \otimes
\sigma^{ V }$
manifest as one-qubit Paulis.
Weak measurements can be used to infer
the coarse-grained quasiprobability $\SumKD{\rho}$.
Combining values of $\SumKD{\rho}$ yields the OTOC $F(t)$.
Figure~\ref{fig:D_subcircuit} depicts a subcircuit
used to implement a weak measurement
of $n = \W$ or $V$.
An ancilla is prepared in a fiducial state $\ket{0}$.
A unitary $R_{ n }^\dag$ rotates
the qubit's $\sigma^{ n }$ eigenbasis
into its $\sigma^z$ eigenbasis.
$R_y ( \pm \phi )$ rotates the ancilla's state counterclockwise
about the $y$-axis through a small angle $\pm \phi$,
controlled by the system's $\sigma^z$.
The angle's smallness guarantees the measurement's weakness.
$R_{ n }$ rotates the system's $\sigma^z$ eigenbasis
back into the $\sigma^{ n }$ eigenbasis.
The ancilla's $\sigma^z$ is measured strongly.
The outcome, $+$ or $-$, dictates
which partial-projection operator $D_{\pm}^n$
updates the state.
Figure~\ref{fig:Full_circuit} shows the circuit
used to measure $\SumKD{\rho}$.
Three weak measurements, interspersed with
three time evolutions ($U$, $U^\dag$, and $U$),
precede a strong measurement.
Suppose that the initial state, $\rho$, commutes with $\W$ or $V$,
e.g., $\rho = \id / \Dim$.
Figure~\ref{fig:Full_circuit} requires only two weak measurements.
}
\label{fig:Circuit}
\end{figure}

Figure~\ref{fig:D_subcircuit} depicts a subcircuit
for measuring $n = \W$ or $V$ weakly.
To simplify notation, we relabel $p_{n}$ as $p$.
Most of the subcircuit appears in Fig.~1 of~\cite{Dressel_14_Implementing}.
We set the imbalance parameter $\epsilon$ to 0.
We sandwich Fig.~1 of~\cite{Dressel_14_Implementing}
between two one-qubit unitaries.
The sandwiching interchanges the computational basis
with the $n$ eigenbasis.

The subcircuit implements the following algorithm:
\begin{enumerate}[(1)]

   \item   Rotate the $n$ eigenbasis into the $\sigma^z$ eigenbasis,
   using $R_{ n }^\dag$.

   \item   Prepare an ancilla in a fiducial state $\ket{0}  \equiv   \ket{ +z }$.

   \item   Entangle $\Sys$ with the ancilla via a $Z$-controlled-$Y$:
   If $\Sys$ is in state $\ket{0}$, rotate the ancilla's state
   counterclockwise (CCW) through a small angle
   $\phi  \ll  \frac{\pi}{2}$ about the $y$-axis.
   Let $R_y ( \phi )$ denote the one-qubit unitary
   that implements this rotation.
   If $\Sys$ is in state $\ket{1}$, rotate the ancilla's state CCW
   through an angle $- \phi$, with $R_y ( - \phi )$.

   \item   Measure the ancilla's $\sigma^z$.
   If the measurement yields outcome $+$,
   $D_+$ updates the system's state;
   and if $-$, then $D_-$.

   \item   Rotate the $\sigma^z$ eigenbasis
   into the $n$ eigenbasis, using $R_{ n }$.

\end{enumerate} \noindent
The measurement is weak because $\phi$ is small.
Rotating through a small angle precisely
can pose challenges~\cite{White_16_Preserving}.

\subsubsection{Full circuit for weak-measurement scheme}
\label{section:Full_circuit}

Figure~\ref{fig:Full_circuit} shows the circuit
for measuring $\SumKD{\rho}$.
The full circuit contains three weak-measurement subcircuits.
Each ancilla serves in only one subcircuit.
No ancilla need remain coherent throughout the protocol,
as discussed in Sec.~\ref{section:Advantages}.
The ancilla used in the first $V$ measurement
can be recycled for the final $V$ measurement.

The circuit simplifies in a special case.
Suppose that $\rho$ shares an eigenbasis with $V$ or with $\W(t)$,
e.g., $\rho = \id / \Dim$.
Only two weak measurements are needed,
as discussed in Sec.~\ref{section:Intro_weak_meas}.

We can augment the circuit to measure $\OurKD{\rho}$,
rather than $\SumKD{\rho}$:
During each weak measurement,
every qubit will be measured.
The qubits can be measured individually:
The $\Sites$-qubit measurement can be
a product of local measurements.
Consider, for concreteness, the first weak measurement.
Measuring just qubit $\Sites$ would yield
an eigenvalue $v_1$ of $V$.
We would infer whether qubit $\Sites$ pointed upward or downward
along the $\hat{V}$ axis.
Measuring all the qubits would yield
a degeneracy parameter $\DegenV_{v_1}$.
We could define $\DegenV_{v_\ell}$ as encoding
the $\hat{V}$-components of
the other $\Sites - 1$ qubits' angular momenta.

%
%
%
\subsection{How to infer $\SumKD{\rho}$
from other OTOC-measurement schemes}
\label{section:MeasSumFromOTOC}

$F(t)$ can be inferred, we have seen, from
the quasiprobability $\OurKD{\rho}$
and from the coarse-grained $\SumKD{\rho}$.
$\SumKD{\rho}$ can be inferred from $F(t)$-measurement schemes,
we show, if the eigenvalues of $\W$ and $V$ equal $\pm 1$.
We assume, throughout this section, that they do.
The eigenvalues equal $\pm 1$ if $\W$ and $V$ are Pauli operators.

The projectors~\eqref{eq:ProjW} and~\eqref{eq:ProjV}
can be expressed as
\begin{align}
   \label{eq:ProjW2}
   \ProjW{ w_\ell }  =  \frac{1}{2} ( \id  +  w_\ell \W )
   \quad \text{and} \quad
   \ProjV{ v_\ell }  =  \frac{1}{2}  ( \id  +  v_\ell V ) \, .
\end{align}
Consider substituting from Eqs.~\eqref{eq:ProjW2}
into Eq.~\eqref{eq:SumKD_simple2}.
Multiplying out yields sixteen terms.
If $\expval{ . }  :=  \Tr (  .  \, . )$,
\begin{align}
   \label{eq:ProjTrick}
   & \SumKD{\rho} ( v_1, w_2, v_2, w_3 )
   =  \frac{1}{16}  \Big[  1  +  ( w_2 + w_3 )  \expval{ \W (t) }
   \nonumber \\ &
   +  ( v_1  +  v_2 )  \expval{ V }  +  w_2 w_3  \expval{ \W^2 (t) }
   +  v_1 v_2  \expval{ V^2 }
   \nonumber \\ &
   +  ( w_2 v_1  +  w_3 v_1  +  w_3 v_2 )  \expval{ \W (t) V }
   +  w_2  v_2  \expval{ V \W(t) }
   \nonumber \\ &
   +  w_2 w_3 v_1  \expval{ \W^2 (t) V }
   +  w_3  v_1  v_2  \expval{ \W(t)  V^2  }
   \nonumber \\ &
   +  w_2  w_3  v_2  \expval{ \W(t)  V  \W(t) }
   +  w_2  v_1  v_2  \expval{ V \W(t)  V }
   \nonumber \\ &
   +  w_2  w_3  v_1  v_2  \,  F(t)  \Big]  \, .
\end{align}
If $\W(t)$ and $V$ are unitary, they square to $\id$.
Equation~\eqref{eq:ProjTrick} simplifies to
\begin{align}
   \label{eq:ProjTrick2}
   & \SumKD{\rho} ( v_1, w_2, v_2, w_3 )
   =  \frac{1}{16}  \Big\{  ( 1  +  w_2 w_3  +  v_1 v_2 )
   \nonumber  \\ &
   +  [ w_2  +  w_3 ( 1  +  v_1 v_2 ) ]  \expval{ \W(t) }
   +  [  v_1 ( 1 + w_2 w_3 )  +  v_2 ]  \expval{ V }
   \nonumber \\ &
   +  ( w_2  v_1  +  w_3  v_1  +  w_3  v_2 )  \expval{ \W(t) V }
   +  w_2 v_2  \expval{ V \W(t) }
    \nonumber \\ &
   +  w_2  w_3 v_2  \expval{ \W (t) V \W(t) }
   +  w_2  v_1 v_2  \expval{ V \W(t) V }
   \nonumber \\ &+  w_2 w_3 v_1 v_2  \, F (t)  \Big\}  \, .
\end{align}

The first term is constant.
The next two terms are single-observable expectation values.
The next two terms are two-point correlation functions.
$\expval{ V \W(t)  V }$ and $\expval{ \W(t)  V  \W(t) }$
are time-ordered correlation functions.
$F(t)$ is the OTOC.
$F(t)$ is the most difficult to measure.
If one can measure it, one likely has the tools to infer $\SumKD{\rho}$.
One can measure every term, for example,
using the set-up in~\cite{Swingle_16_Measuring}.

\section{Numerical simulations}
\label{section:Numerics}

We now study the OTOC quasiprobability's physical content in two simple models. In this section, we study a geometrically local 1D model, an Ising chain with transverse and longitudinal fields. In Sec.~\ref{section:Brownian}, we study a geometrically nonlocal model known as the \emph{Brownian-circuit model.} This model effectively has a time-dependent Hamiltonian.

We compare the physics of $\SumKD{\rho}$ with that of the OTOC.
The time scales inherent in $\SumKD{\rho}$,
as compared to the OTOC's time scales, particularly interest us.
We study also nonclassical behaviors---negative and nonreal values---of
$\SumKD{\rho}$.
Finally, we find a parallel with classical chaos:
The onset of scrambling breaks a symmetry.
This breaking manifests in bifurcations of $\SumKD{\rho}$,
reminiscent of pitchfork diagrams.

The Ising chain is defined on a Hilbert space of $\Sites$ spin-$\frac{1}{2}$ degrees of freedom.
The total Hilbert space has dimensionality $\Dim = 2^\Sites$.
The single-site Pauli matrices are labeled $\{\sigma^x_i,\sigma^y_i,\sigma^z_i\}$, for $i=1,...,\Sites$. The Hamiltonian is
\begin{align}\label{eq:IsingH}
  H = - J \sum_{i=1}^{\Sites-1} \sigma_i^z \sigma_{i+1}^z - h \sum_{i=1}^{\Sites} \sigma_i^z - g \sum_{i=1}^{\Sites} \sigma^x_i  \, .
\end{align}
The chain has open boundary conditions. Energies are measured in units of $J$.
Times are measured in units of $1/J$. The interaction strength is thus set to one, $J=1$, henceforth. We numerically study this model for $\Sites=10$ by exactly diagonalizing $H$. This system size suffices for probing
the quasiprobability's time scales.
However, $\Sites = 10$ does not necessarily illustrate the thermodynamic limit.

When $h = 0$, this model is integrable and can be solved with noninteracting-fermion variables. When $h\neq 0$, the model appears to be reasonably chaotic. These statements' meanings are clarified in the data below.
As expected, the quasiprobability's qualitative behavior
is sensitive primarily to
whether $H$ is integrable,
as well as to the initial state's form.
We study two sets of parameters,
\begin{align}
  \text{Integrable:} \;  & h=0,  \:  g=1.05
  \quad \text{and} \nonumber \\
  \text{Nonintegrable:}  \; & h=.5,   \:  g=1.05  \, .
\end{align}
We study several classes of initial states $\rho$,
including thermal states, random pure states, and product states.

For $\W$ and $V$, we choose single-Pauli operators
that act nontrivially on just the chain's ends.
We illustrate with $\mathcal{W} = \sigma_1^{x}$ or $\mathcal{W}
= \sigma_1^z$
and $V= \sigma_\Sites^x$ or $\sigma_\Sites^z$.
These operators are unitary and Hermitian.
They square to the identity, enabling us to use Eq.~\eqref{eq:ProjTrick2}.
We calculate the coarse-grained quasiprobability directly:
\begin{align}
   \label{eq:SumKD_Num}
   \SumKD{\rho}( v_1 , w_2 , v_2 , w_3 )
   = \text{Tr}\left( \rho \Pi^{\mathcal{W}(t)}_{w_3} \Pi^V_{v_2} \Pi^{\mathcal{W}(t)}_{w_2}    \Pi^{V}_{v_1} \right)  \, .
\end{align}
For a Pauli operator $\Oper$,
$\Pi^\Oper_a = \frac{1}{2} \: ( 1 + a \Oper )$ projects onto
the $a \in \{1,-1\}$ eigenspace.
We also compare the quasiprobability with the OTOC,
Eq.~\eqref{eq:RecoverF2}.

$F(t)$ deviates from one at roughly
the time needed for information to propagate
from one end of the chain to the other.
This onset time, which up to a constant shift is also approximately the scrambling time, lies approximately between $t=4$ and $t=6$,
according to our the data.
The system's length and the butterfly velocity $v_{\text{B}}$
set the scrambling time (Sec.~\ref{section:OTOC_review}).
Every term in the Hamiltonian~\eqref{eq:IsingH} is order-one.
Hence $v_{\text{B}}$ is expected to be order-one, too.
In light of our spin chain's length,
the data below are all consistent with a $v_{\text{B}}$ of approximately two.

\subsection{Thermal states}

We consider first thermal states $\rho \propto e^{-H/T}$.
Data for the infinite-temperature ($T = \infty$) state,
with $\mathcal{W} = \sigma_1^z$, $V = \sigma_\Sites^z$,
and nonintegrable parameters, appear in
Figures \ref{fig:TInf_zz_F}, \ref{fig:TInf_zz_AR}, and \ref{fig:TInf_zz_AI}.
The legend is labeled such that $abcd$ corresponds to
$w_3 = (-1)^a$, $v_2 = (-1)^b$, $w_2 = (-1)^c$, and $v_1 = (-1)^d$. This labelling corresponds to the order in which the operators appear in Eq.~\eqref{eq:SumKD_Num}.

Three behaviors merit comment.
Generically, the coarse-grained quasiprobability is a complex number:
$\SumKD{\rho}( . )  \in  \mathbb{C}$.
However, $\SumKD{( \id / \Dim) }$ is real.
The imaginary component $\Im \left(  \SumKD{( \id / \Dim) }  \right)$
might appear nonzero in Fig.~\ref{fig:TInf_zz_AI}.
Yet $\Im \left(  \SumKD{( \id / \Dim) }  \right)  \leq  10^{ -16 }$.
This value equals zero, to within machine precision.
The second feature to notice is that
the time required for $\SumKD{ ( \id / \Dim) }$ to deviate from its initial value
equals approximately the time required for the OTOC to deviate from its initial value. Third, although $\SumKD{ ( \id / \Dim) }$ is real, it is negative and hence nonclassical for some values of its arguments.

What about lower temperatures?
Data for the $T=1$ thermal state are shown in Figures \ref{fig:T1_zz_F}, \ref{fig:T1_zz_AR}, and \ref{fig:T1_zz_AI}.
The coarse-grained quasiprobability is no longer real.
Here, too, the time required for $\SumKD{\rho}$ to deviate significantly from its initial value is comparable with the time scale of changes in $F(t)$.
This comparability characterizes
the real and imaginary parts of $\SumKD{\rho}$.
Both parts oscillate at long times. In the small systems considered here, such oscillations can arise from finite-size effects,
including the energy spectrum's discreteness.
With nonintegrable parameters, this model has an energy gap $\Delta_{\Sites=10} = 2.92$ above the ground state.
The temperature $T=1$ is smaller than the gap.
Hence lowering $T$ from $\infty$ to 1
brings the thermal state close to the ground state.

What about long-time behavior?
At infinite temperature, $\SumKD{ (\id / \Dim ) }$ approaches a limiting form
after the scrambling time
but before any recurrence time.
Furthermore, $\SumKD{ (\id / \Dim ) }$ can approach one of
only a few possible limiting values,
depending on the function's arguments.
This behavior follows from the terms in Eq.~\eqref{eq:ProjTrick2}.
At infinite temperature, $\langle \mathcal{W} \rangle = \langle V \rangle = 0$.
Also the 3-point functions vanish, due to the trace's cyclicity.
We expect the nontrivial 2- and 4-point functions
to be small at late times.
(Such smallness is visible in the 4-point function in Fig.~\ref{fig:TInf_zz_F}.) Hence Eq.~\eqref{eq:ProjTrick2} reduces as
\begin{align}\label{eq:infiniteTlatetime}
  \SumKD{\rho}( v_1 , w_2 , v_2 , w_3 )
  \underbrace{\longrightarrow}_{t\rightarrow \infty} \frac{1 + w_2 w_3 + v_1 v_2}{16}  \, .
\end{align}

According to Eq.~\eqref{eq:infiniteTlatetime},
the late-time values of $\SumKD{ (\id / \Dim ) }$
should cluster around $3/16$, $1/16$, and $-1/16$.
This expectation is roughly consistent with Fig.~\ref{fig:TInf_zz_AR},
modulo the upper lines' bifurcation.

A bifurcation of $\SumKD{\rho}$ signals
the breaking of a symmetry at the onset of scrambling. Similarly, pitchfork plots signal the breaking of a symmetry
in classical chaos~\cite{Strogatz_00_Non}.
The symmetry's mathematical form follows from Eq.~\eqref{eq:ProjTrick2}.
At early times, $\W(t)$ commutes with $V$, and $F(t) \approx 1$.
Suppose, for simplicity, that $\rho = \id / \Dim$.
The expectation values $\expval{ \W(t) }$ and $\expval{V}$ vanish,
because every Pauli has a zero trace.
Equation~\eqref{eq:ProjTrick2} becomes
\begin{align}
   \label{eq:ProjTrick_early}
   & \SumKD{\rho} ( v_1 , w_2 , v_2 , w_3 )
   =  \frac{1}{16}  \Big[ (1 + w_2 w_3  +  v_1 v_2  +  w_2 w_3 v_1 v_2 )
   \nonumber \\ & \qquad \qquad \qquad
   +  ( w_2  +  w_3 )  ( v_1  +  v_2 )  \expval{ \W(t)  V }  \Big]  \, .
\end{align}

Suppose that $w_2 = - w_3$ and/or $v_1 = - v_2$,
as in the lower lines in Fig.~\ref{fig:TInf_zz_AR}.
$\SumKD{\rho} (.)$ reduces to the constant
\begin{align}
   \label{eq:Early_square}
   & \frac{1}{16}  ( 1 + w_2 w_3  +  v_1 v_2  +  w_2 w_3 v_1 v_2 )
   \\ \nonumber &
   =  \frac{1}{32}  \Big[  ( 1 + w_2 w_3  +  v_1 v_2  )^2
   -  ( w_2 w_3 )^2  -  ( v_1 v_2 )^2  +  1  \Big] \, .
\end{align}
The right-hand side depends on the eigenvalues $w_\ell$ and $v_m$
only through squares.
$\SumKD{\rho} ( . )$ remains invariant
under the interchange of $w_2$ with $w_3$,
under the interchange of $v_1$ with $v_2$,
under the simultaneous negations of $w_2$ and $w_3$,
and under the simultaneous negations of $v_1$ and $v_2$.
These symmetries have operational significances:
$\OurKD{\rho}$ remains constant under permutations and negations
of measurement outcomes in
the weak-measurement scheme (Sec.~\ref{section:Intro_weak_meas}).
Symmetries break as the system starts scrambling:
$F(t)$ shrinks, shrinking the final term in Eq.~\eqref{eq:Early_square}.
$\SumKD{\rho}$ starts depending not only on
squares of $w_\ell$-and-$v_m$ functions,
but also on the eigenvalues individually.

Whereas the shrinking of $F(t)$ bifurcates
the lower lines in Fig.~\ref{fig:TInf_zz_AR},
the shrinking does not bifurcate the upper lines.
The reason is that each upper line corresponds to $w_2 w_3  =  v_1 v_2  =  1$.
[At early times, $| F(t) |$ is small enough that
any $F(t)$-dependent correction would fall within the lines' widths.]
Hence the final term in Eq.~\eqref{eq:ProjTrick_early} is proportional to
$\pm \expval{ \mathcal{W}(t) V }$.
This prediction is consistent with the observed splitting.
The $\expval{ \mathcal{W}(t) V }$ term does not split the lower lines:
Each lower line satisfies $w_2 = - w_3$ and/or $v_1 = - v_2$.
Hence the $\expval{ \mathcal{W}(t) V }$ term vanishes.
We leave as an open question
whether these pitchforks can be understood in terms of equilibria,
like classical-chaos pitchforks~\cite{Strogatz_00_Non}.

In contrast with the $T=\infty$ data, the $T=1$ data
oscillate markedly at late times
(after the quasiprobability's initial sharp change).
We expect these oscillations to decay to zero at late times,
if the system is chaotic, in the thermodynamic limit.
Unlike at infinite temperature, $\mathcal{W}$ and $V$
can have nonzero expectation values.
But, if all nontrivial connected correlation functions have decayed,
Eq.~\eqref{eq:ProjTrick2} still implies a simple dependence on
the $w_\ell$ and $v_m$ parameters at late times.

Finally, Figures \ref{fig:TInf_zz_int_F} and \ref{fig:TInf_zz_int_AR} show
the coarse-grained quasiprobability at infinite temperature,
$\SumKD{ (\id / \Dim ) }$, with integrable parameters.
The imaginary part remains zero, so we do not show it.
The difference from the behavior in
Figures \ref{fig:TInf_zz_F} and \ref{fig:TInf_zz_AR}
(which shows $T = \infty$, nonintegrable-$H$ data) is obvious.
Most dramatic is the large revival that occurs
at what would, in the nonintegrable model, be a late time.
Although this is not shown, the quasiprobability depends significantly
on the choice of operator.
This dependence is expected, since different Pauli operators have different degrees of complexity in terms of the noninteracting-fermion variables.


\begin{figure}
\begin{center}
\includegraphics[width=.49\textwidth]{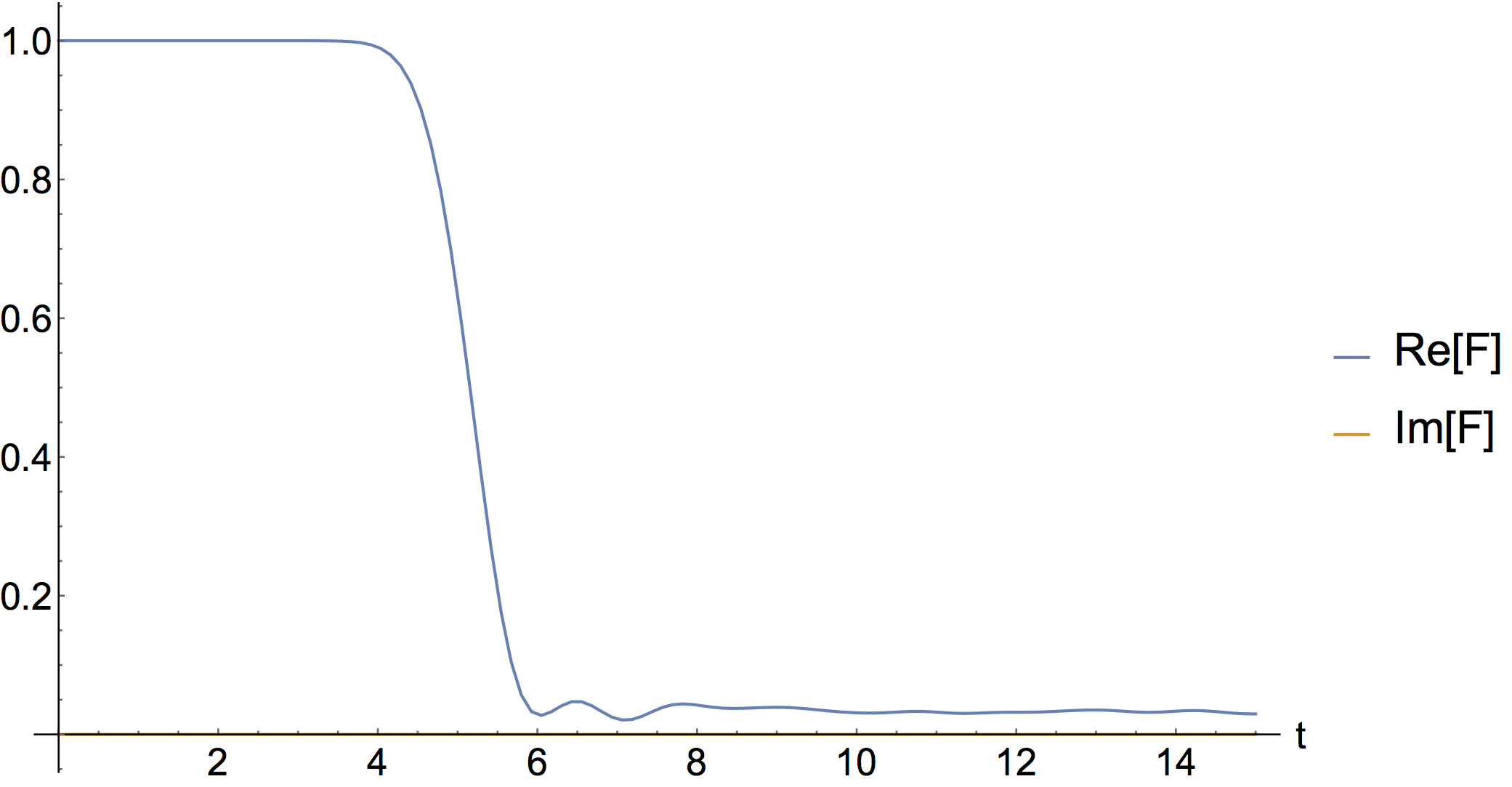}
\end{center}
\caption{Real and imaginary parts of $F(t)$ as a function of time. $T=\infty$ thermal state. Nonintegrable parameters, $\Sites=10$, $\mathcal{W}=\sigma_1^z$, $V=\sigma_\Sites^z$. }
\label{fig:TInf_zz_F}
\end{figure}

\begin{figure}
\begin{center}
\includegraphics[width=.49\textwidth]{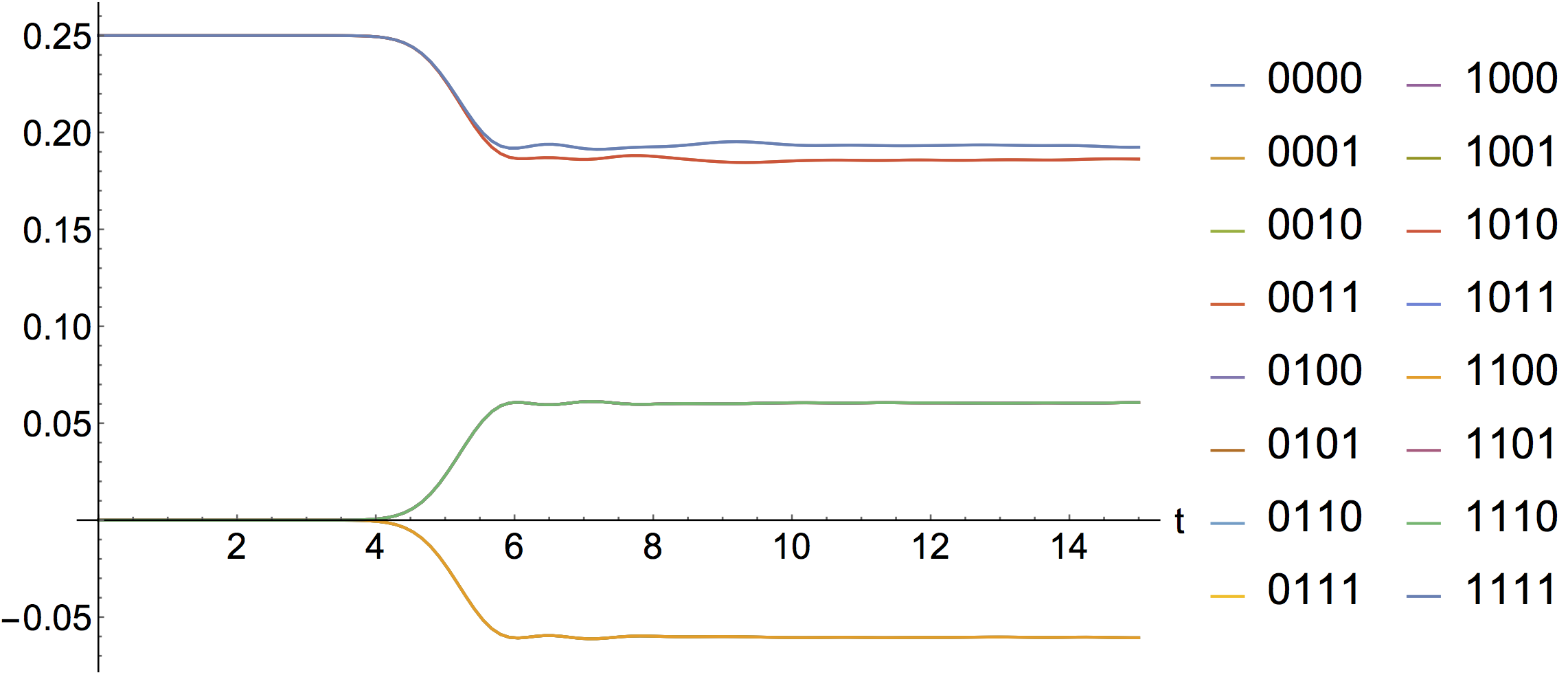}
\end{center}
\caption{Real part of $\SumKD{\rho}$ as a function of time. $T=\infty$ thermal state. Nonintegrable parameters, $\Sites=10$, $\mathcal{W}=\sigma_1^z$, $V=\sigma_\Sites^z$. }
\label{fig:TInf_zz_AR}
\end{figure}

\begin{figure}
\begin{center}
\includegraphics[width=.49\textwidth]{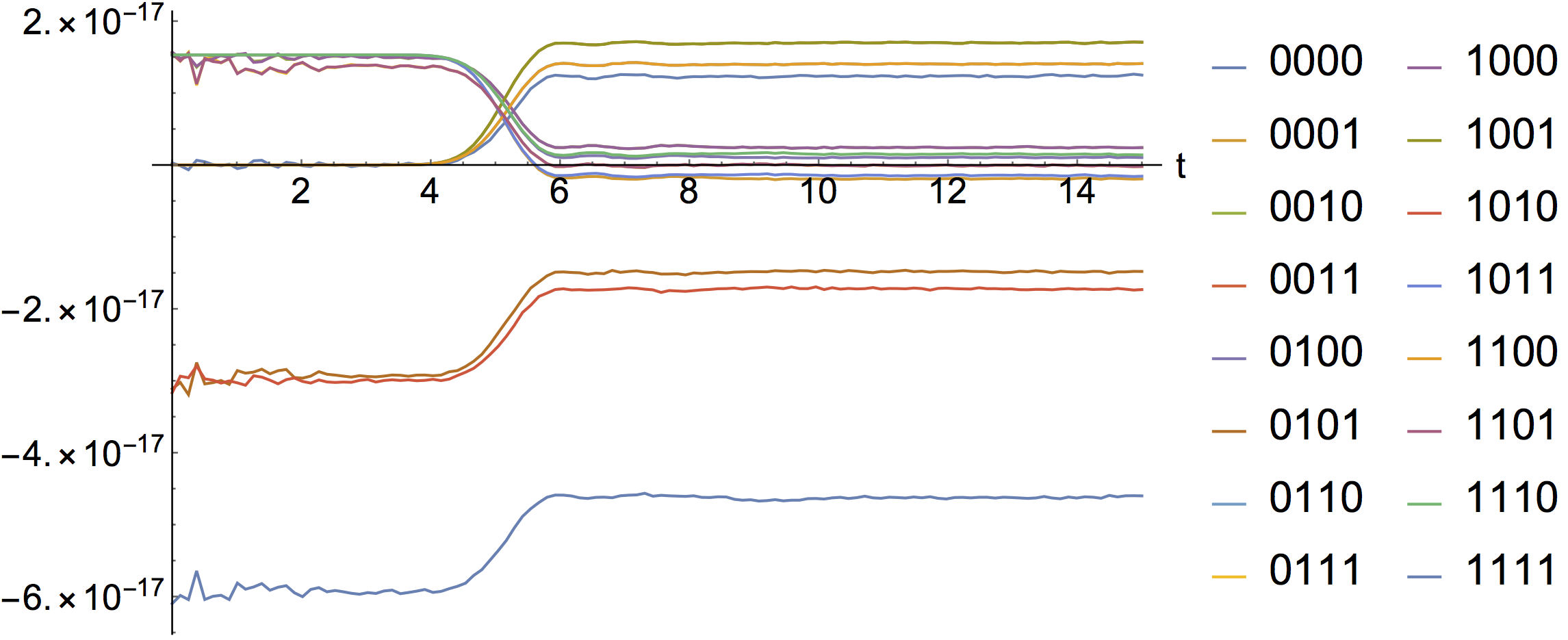}
\end{center}
\caption{Imaginary part of $\SumKD{\rho}$ as a function of time. $T=\infty$ thermal state. Nonintegrable parameters, $\Sites=10$, $\mathcal{W}=\sigma_1^z$, $V=\sigma_\Sites^z$.
To within machine precision, $\Im \left(  \OurKD{\rho}  \right)$ vanishes
for all values of the arguments.}
\label{fig:TInf_zz_AI}
\end{figure}


\begin{figure}
\begin{center}
\includegraphics[width=.49\textwidth]{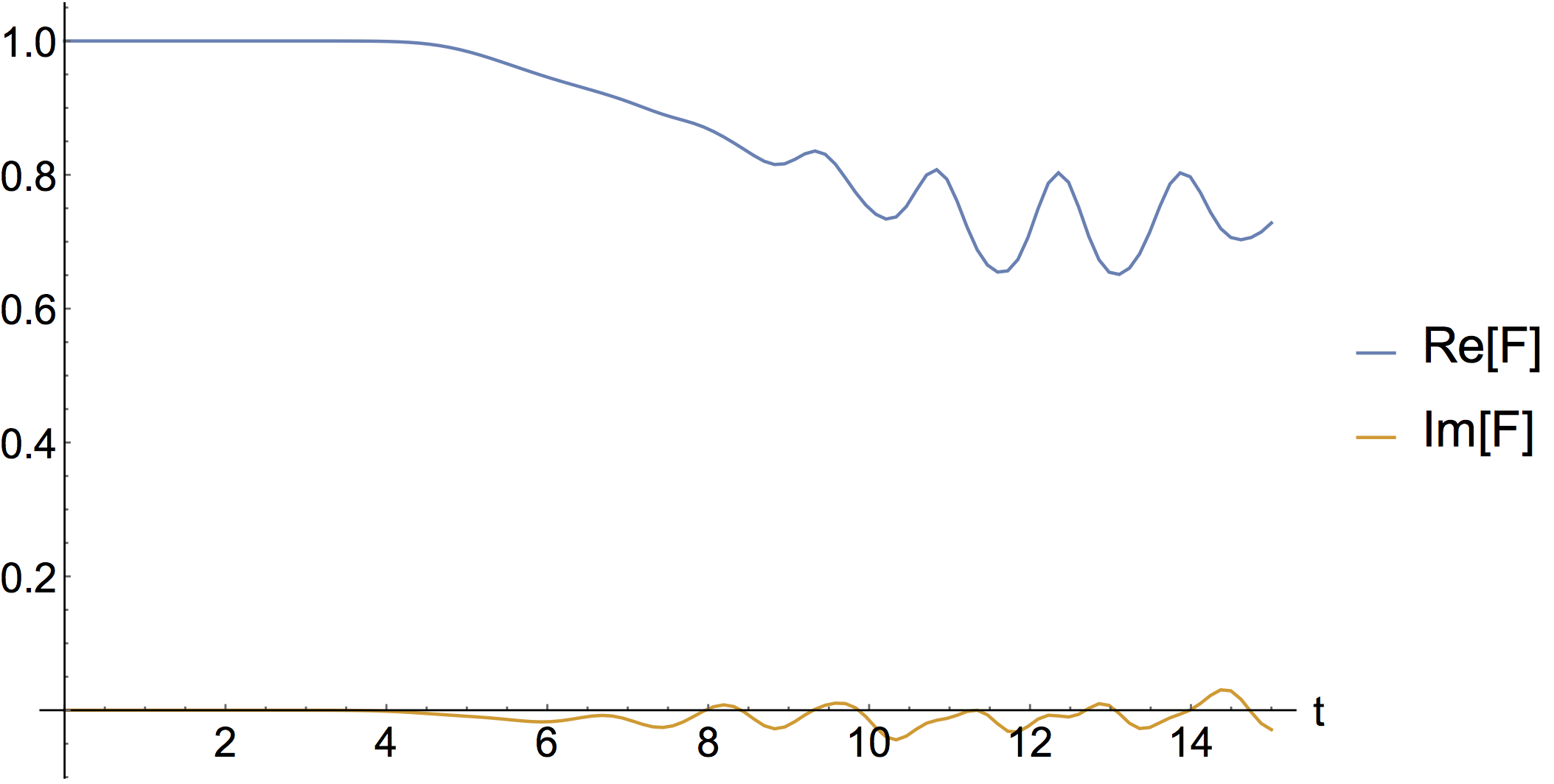}
\end{center}
\caption{Real and imaginary parts of $F(t)$ as a function of time. $T=1$ thermal state. Nonintegrable parameters, $\Sites=10$, $\mathcal{W}=\sigma_1^z$, $V=\sigma_\Sites^z$. }
\label{fig:T1_zz_F}
\end{figure}

\begin{figure}
\begin{center}
\includegraphics[width=.49\textwidth]{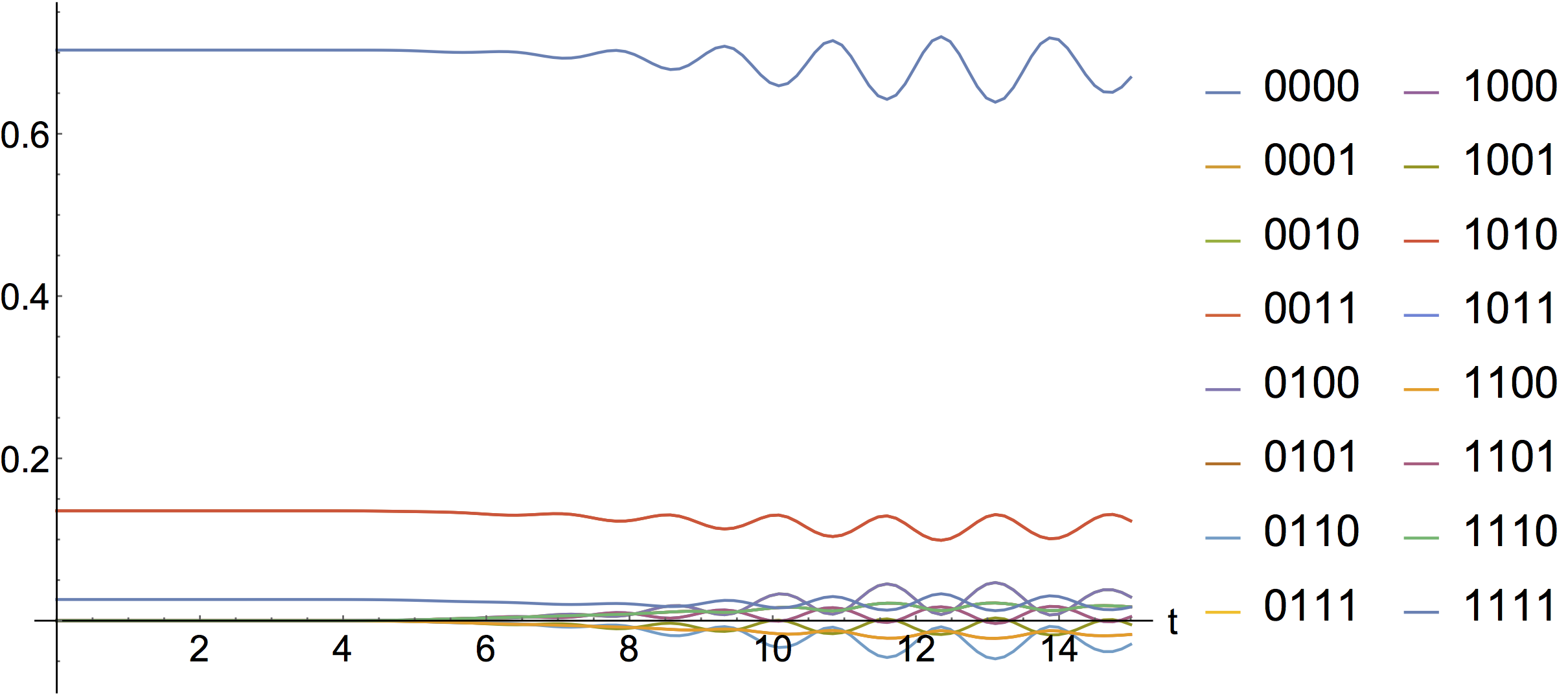}
\end{center}
\caption{Real part of $\SumKD{\rho}$ as a function of time. $T=1$ thermal state. Nonintegrable parameters, $\Sites=10$, $\mathcal{W}=\sigma_1^z$, $V=\sigma_\Sites^z$. }
\label{fig:T1_zz_AR}
\end{figure}

\begin{figure}
\begin{center}
\includegraphics[width=.49\textwidth]{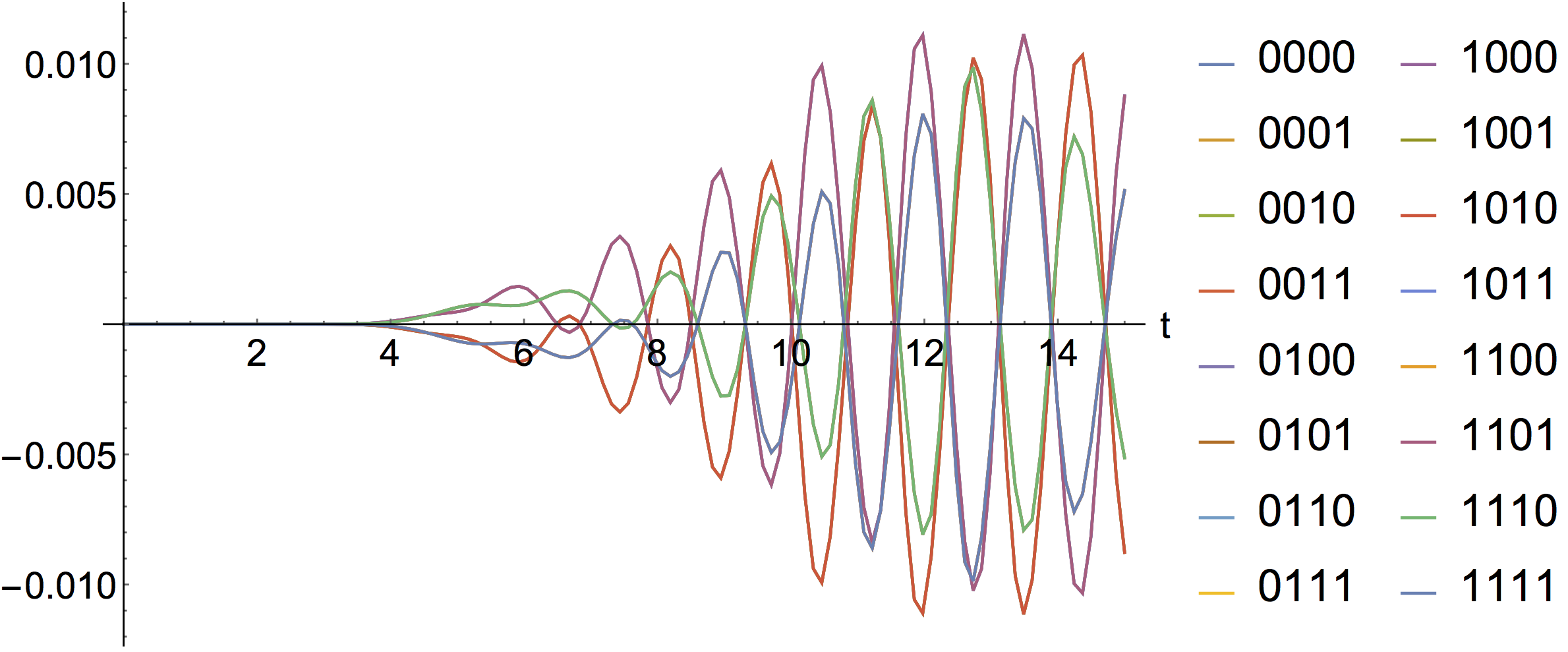}
\end{center}
\caption{Imaginary part of $\SumKD{\rho}$ as a function of time. $T=1$ thermal state. Nonintegrable parameters, $\Sites=10$, $\mathcal{W}=\sigma_1^z$, $V=\sigma_\Sites^z$. }
\label{fig:T1_zz_AI}
\end{figure}


\begin{figure}
\begin{center}
\includegraphics[width=.49\textwidth]{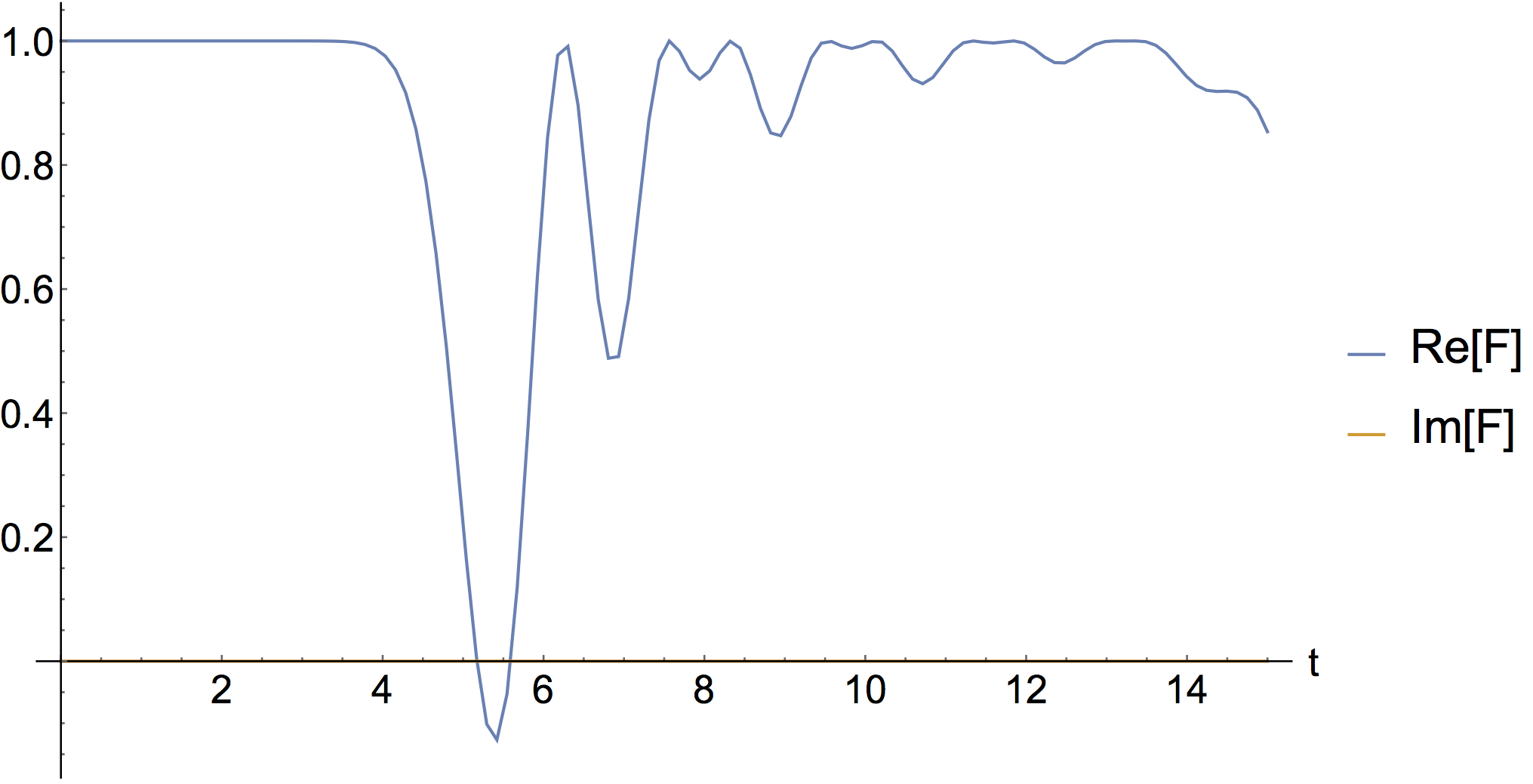}
\end{center}
\caption{Real and imaginary parts of $F(t)$ as a function of time. $T=\infty$ thermal state. Integrable parameters, $\Sites=10$, $\mathcal{W}=\sigma_1^z$, $V=\sigma_\Sites^z$. }
\label{fig:TInf_zz_int_F}
\end{figure}

\begin{figure}
\begin{center}
\includegraphics[width=.49\textwidth]{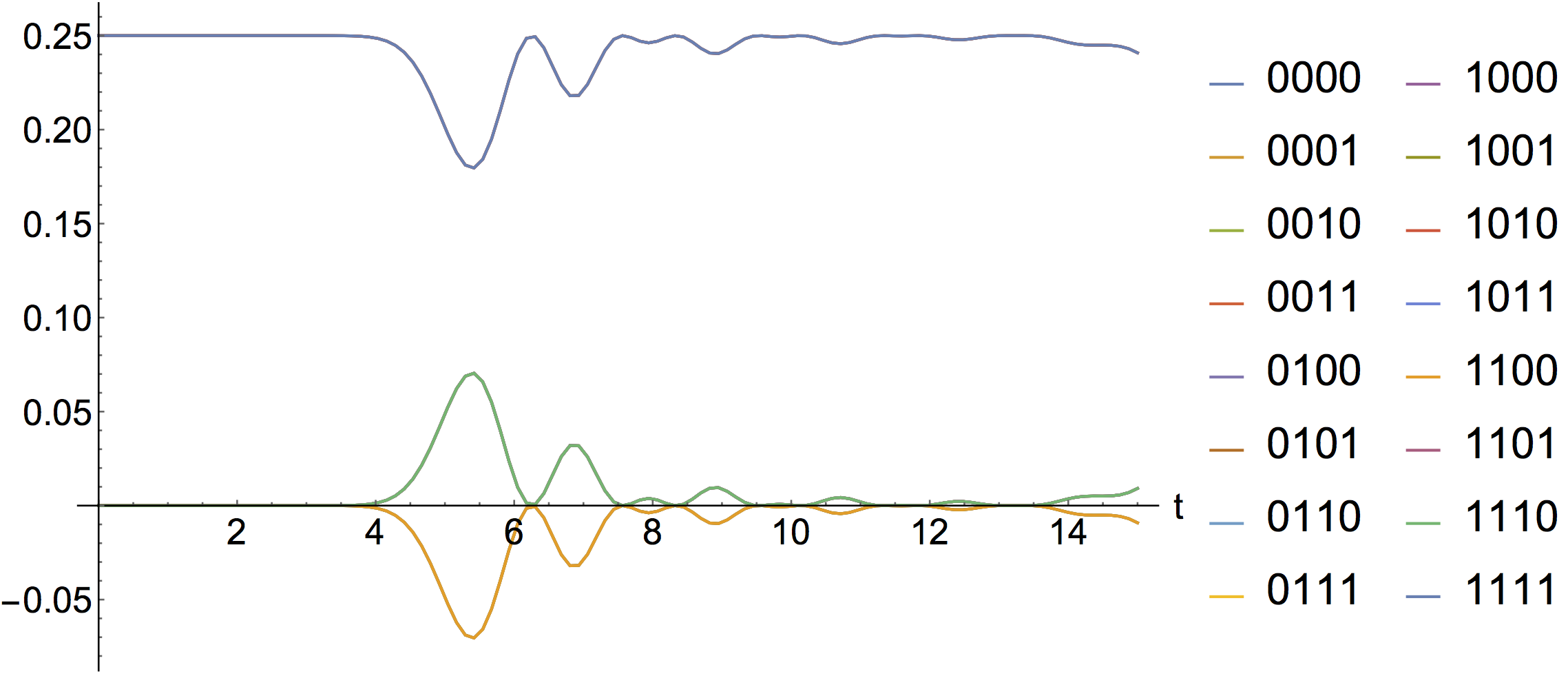}
\end{center}
\caption{Real part of $\SumKD{\rho}$ as a function of time. $T=\infty$ thermal state. Integrable parameters, $\Sites=10$, $\mathcal{W}=\sigma_1^z$, $V=\sigma_\Sites^z$. }
\label{fig:TInf_zz_int_AR}
\end{figure}


\subsection{Random states}

We now consider random pure states
$\rho \propto \ketbra{ \psi }{ \psi }$ and nonintegrable parameters.
Figures \ref{fig:rand_zz_F}, \ref{fig:rand_zz_AR}, and \ref{fig:rand_zz_AI}
show $F(t)$ and $\SumKD{\rho}$
for the operator choice $\mathcal{W} = \sigma_1^z$ and $V = \sigma_\Sites^z$
in a randomly chosen pure state.
The pure state is drawn according to the Haar measure.
Each figure shows a single shot 
(contains data from just one pure state).
Broadly speaking, the features are similar to those exhibited by
the infinite-temperature $\rho = \id / \Dim$, with additional fluctuations.

The upper branch of lines in Fig.~\ref{fig:rand_zz_AR}
exhibits dynamics before the OTOC does.
However, lines' average positions move significantly
(the lower lines bifurcate, and the upper lines shift downward)
only after the OTOC begins to evolve.
The early motion must be associated with
the early dynamics of the 2- and 3-point functions in Eq.~\eqref{eq:ProjTrick2}.
The late-time values are roughly consistent with
those for $\rho = \id / \Dim$ but fluctuate more pronouncedly.

The agreement between random pure states
and the $T = \infty$ thermal state is expected,
due to closed-system thermalization~\cite{D'Alessio_16_From,Gogolin_16_Equilibration}.
Consider assigning a temperature to a pure state
by matching its energy density with
the energy density of the thermal state $e^{ - H / T } / Z$,
cast as a function of temperature.
With high probability, any given random pure state
corresponds to an infinite temperature.
The reason is the thermodynamic entropy's
monotonic increase with temperature.
Since the thermodynamic entropy gives the density of states,
more states correspond to higher temperatures.
Most states correspond to infinite temperature.

For the random states and system sizes $\Sites$ considered,
if $H$ is nonintegrable,
the agreement with thermal results is not complete.
However, the physics appears qualitatively similar.

\begin{figure}
\begin{center}
\includegraphics[width=.49\textwidth]{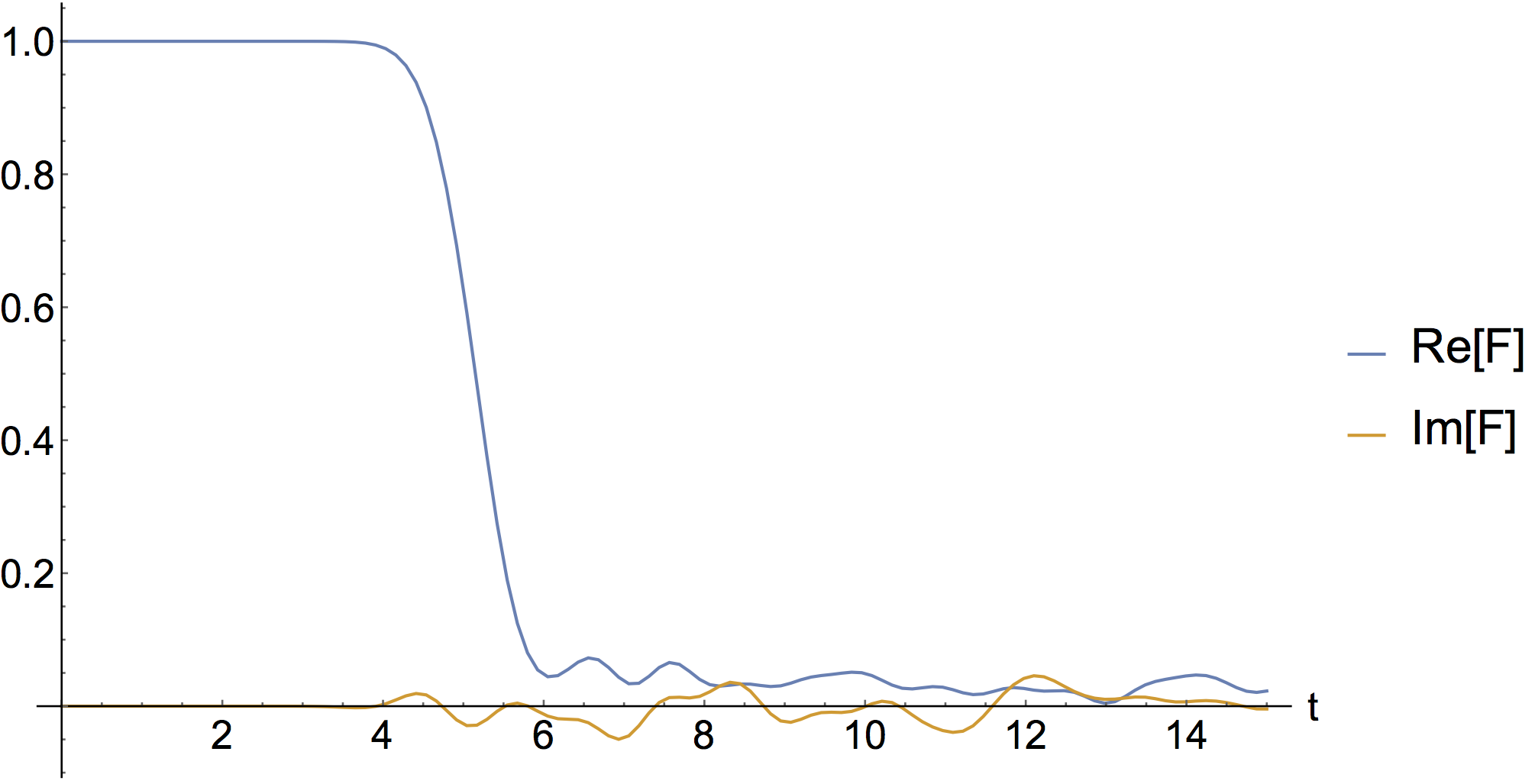}
\end{center}
\caption{Real and imaginary parts of $F(t)$ as a function of time. Random pure state. Nonintegrable parameters, $\Sites=10$, $\mathcal{W}=\sigma_1^z$, $V=\sigma_\Sites^z$. }
\label{fig:rand_zz_F}
\end{figure}

\begin{figure}
\begin{center}
\includegraphics[width=.49\textwidth]{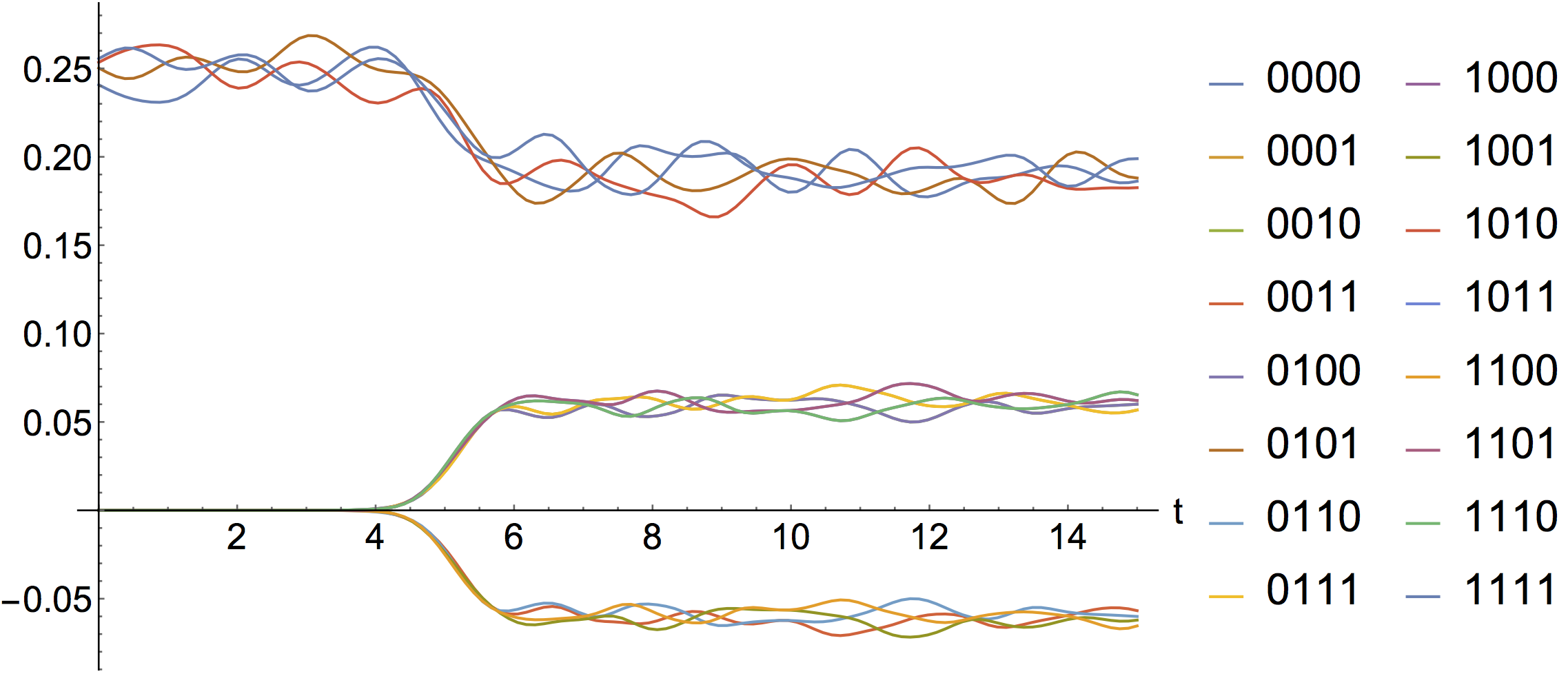}
\end{center}
\caption{Real part of $\SumKD{\rho}$ as a function of time. Random pure state. Nonintegrable parameters, $\Sites=10$, $\mathcal{W}=\sigma_1^z$, $V=\sigma_\Sites^z$. }
\label{fig:rand_zz_AR}
\end{figure}

\begin{figure}
\begin{center}
\includegraphics[width=.49\textwidth]{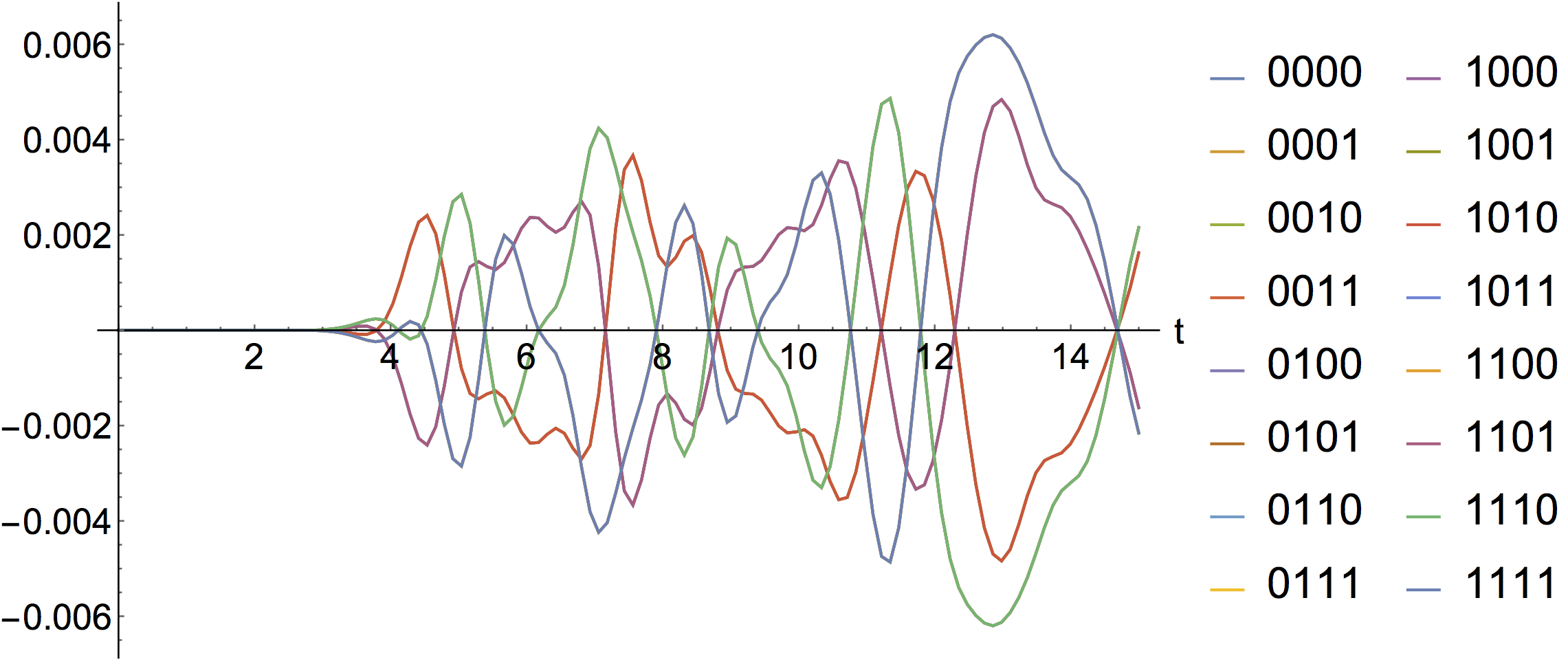}
\end{center}
\caption{Imaginary part of $\SumKD{\rho}$ as a function of time. Random pure state. Nonintegrable parameters, $\Sites=10$, $\mathcal{W}=\sigma_1^z$, $V=\sigma_\Sites^z$. }
\label{fig:rand_zz_AI}
\end{figure}

\subsection{Product states}

Finally, we consider the product $\ket{ +x }^{ \otimes \Sites}$
of $\Sites$ copies of the $+1$ $\sigma^x$ eigenstate
(Figures~\ref{fig:xup_zz_F}--\ref{fig:xup_zz_AI}).
We continue to use $\mathcal{W} = \sigma_1^z$ and $V = \sigma_\Sites^z$.
For the Hamiltonian parameters chosen,
this state lies far from the ground state.
The state therefore should correspond to
a large effective temperature.
Figures \ref{fig:xup_zz_F}, \ref{fig:xup_zz_AR}, and \ref{fig:xup_zz_AI} show
$F (t)$ and $\SumKD{\rho}$ for nonintegrable parameters.

The real part of $F(t)$ decays significantly from its initial value of one.
The imaginary part of $F(t)$ is nonzero but remains small.
These features resemble the infinite-temperature features.
However, the late-time $F(t)$ values are substantially larger than
in the $T = \infty$ case and oscillate significantly.

Correspondingly, the real and imaginary components of $\SumKD{\rho}$
oscillate significantly.
$\Re \left( \SumKD{\rho} \right)$ exhibits dynamics before scrambling begins,
as when $\rho$ is a random pure state.
The real and imaginary parts of $\SumKD{\rho}$ differ more from
their $T = \infty$ counterparts than
$F(t)$ differs from its counterpart.
Some of this differing is apparently washed out
by the averaging needed to construct $F(t)$ 
[Eq.~\eqref{eq:RecoverF2}].

We expected pure product states to behave
roughly like random pure states.
The data support this expectation very roughly, at best.
Whether finite-size effects cause this deviation,
we leave as a question for further study.

\begin{figure}
\begin{center}
\includegraphics[width=.49\textwidth]{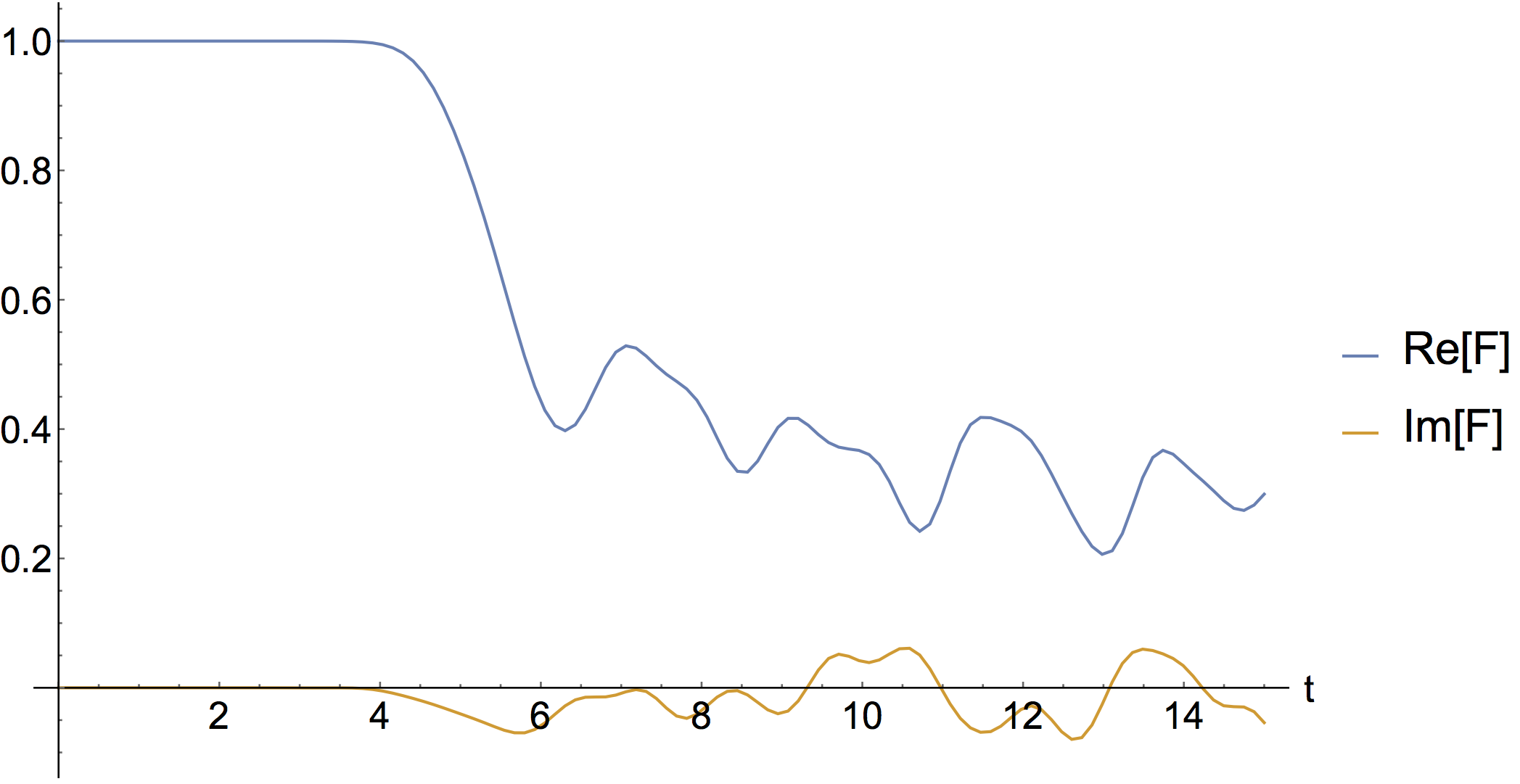}
\end{center}
\caption{Real and imaginary parts of $F(t)$ as a function of time.
Product $\ket{ +x }^{ \otimes \Sites}$ of $\Sites$ copies of
the $+1$ $\sigma^x$ eigenstate.
Nonintegrable parameters, $\Sites=10$, $\mathcal{W}=\sigma_1^z$, $V=\sigma_\Sites^z$. }
\label{fig:xup_zz_F}
\end{figure}

\begin{figure}
\begin{center}
\includegraphics[width=.49\textwidth]{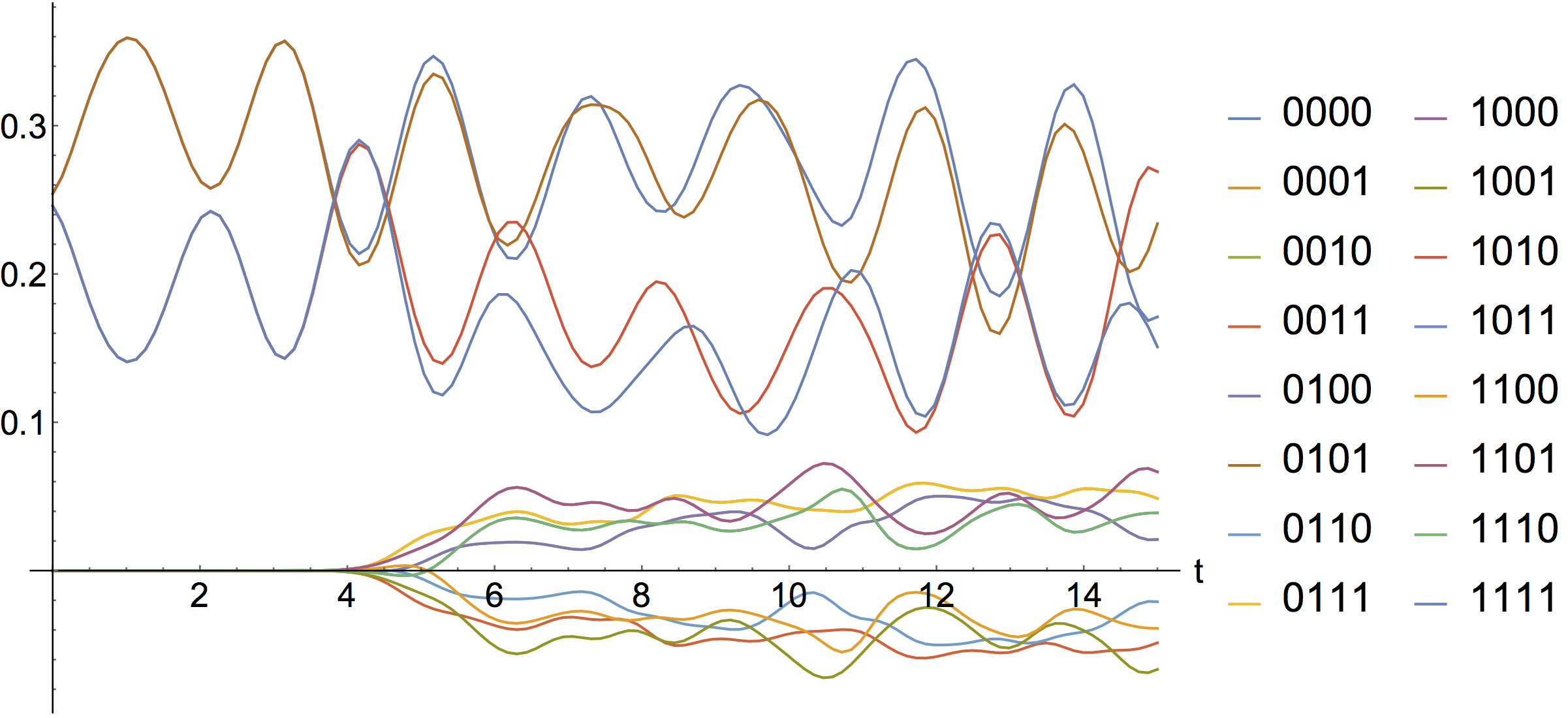}
\end{center}
\caption{Real part of $\SumKD{\rho}$ as a function of time.
Product $\ket{ +x }^{ \otimes \Sites}$ of $\Sites$ copies of
the $+1$ $\sigma^x$ eigenstate.
Nonintegrable parameters, $\Sites=10$, $\mathcal{W}=\sigma_1^z$, $V=\sigma_\Sites^z$. }
\label{fig:xup_zz_AR}
\end{figure}

\begin{figure}
\begin{center}
\includegraphics[width=.49\textwidth]{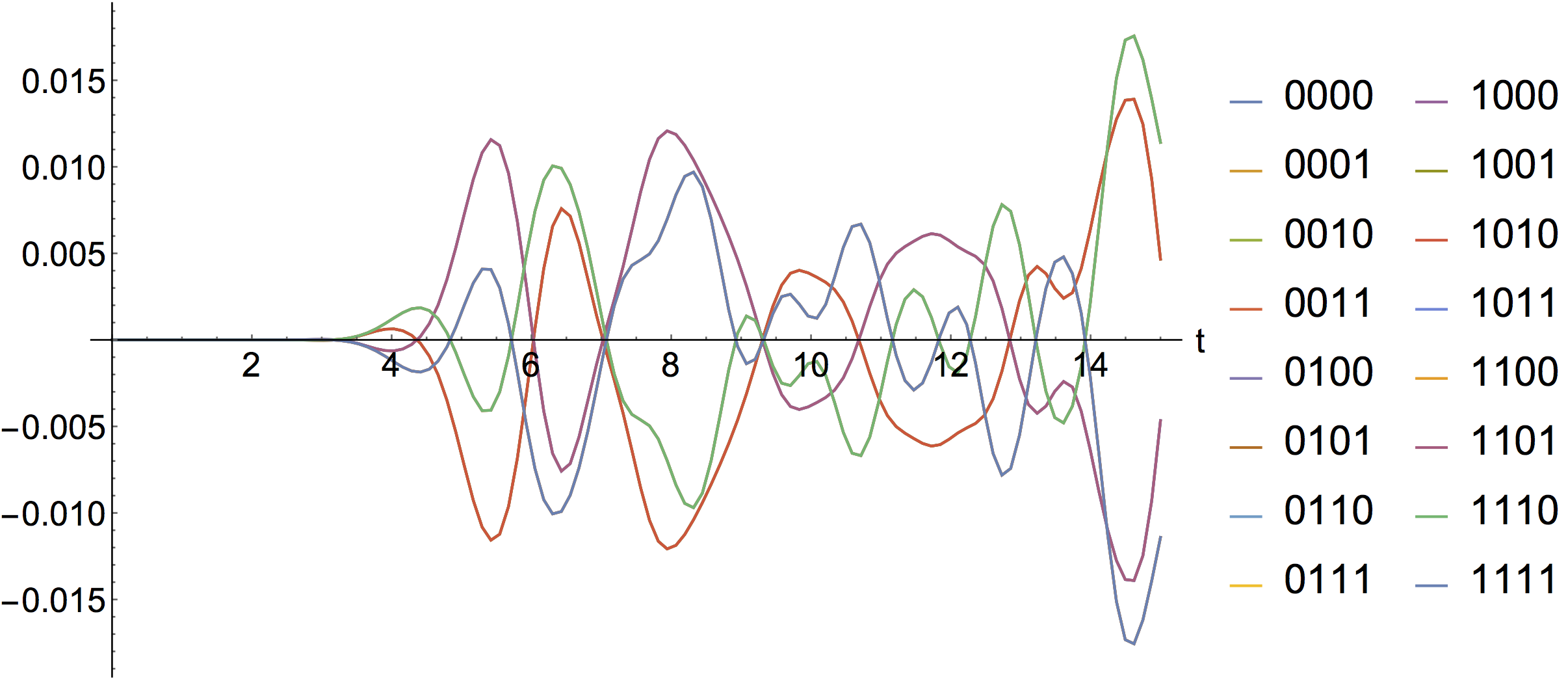}
\end{center}
\caption{Imaginary part of $\SumKD{\rho}$ as a function of time.
Product $\ket{ +x }^{ \otimes \Sites}$ of $\Sites$ copies of
the $+1$ $\sigma^x$ eigenstate.
Nonintegrable parameters, $\Sites=10$, $\mathcal{W}=\sigma_1^z$, $V=\sigma_\Sites^z$. }
\label{fig:xup_zz_AI}
\end{figure}

\subsection{Summary}

The main messages from this study are the following.
\begin{enumerate}[(1)]

\item The coarse-grained quasiprobability $\SumKD{\rho}$ is generically complex.
Exceptions include the $T = \infty$ thermal state $\id / \Dim$
and states $\rho$ that share
an eigenbasis with $V$ or with $\W(t)$
[e.g., as in Eq.~\eqref{eq:WRho}].
Recall that the KD distribution's nonreality
signals nonclassical physics (Sec.~\ref{section:Intro_to_KD}).

\item The derived quantity $P(W,W')$ is generically complex,
our results imply.\footnote{
The relevant plots are not shown,
so that this section maintains
a coherent focus on $\SumKD{\rho}$.
This result merits inclusion, however,
as $P(W, W')$ plays important roles in
(i)~\cite{YungerHalpern_17_Jarzynski}
and (ii) connections between the OTOC
and quantum thermodynamics (Sec.~\ref{section:Outlook}).
}
Nonclassicality thus survives even
the partial marginalization that defines $P$ [Eq.~\eqref{eq:PWWPrime}].
In general, marginalization can cause interference to dampen nonclassicality.
(We observe such dampening in Property~\ref{prop:MargOurKD}
of Sec.~\ref{section:TA_Props}
and in Property~\ref{prop:MargP} of Appendix~\ref{section:P_Properties}.)

\item Random pure states' quasiprobabilities resemble
the $T = \infty$ thermal state's quasiprobability
but fluctuate more.

\item Certain product states' quasiprobabilities
display anomalously large fluctuations.
We expected these states to resemble random states more.

\item The $\SumKD{\rho}$'s generated by integrable Hamiltonians
differ markedly from
the $\SumKD{\rho}$'s generated by nonintegrable Hamiltonians.
Both types of $\SumKD{\rho}$'s achieve nonclassical values, however.
We did not clearly observe a third class of behavior.

\item The time scale after which $\SumKD{\rho}$ changes significantly
is similar to the OTOC time scale.
$\SumKD{\rho}$ can display nontrivial early-time dynamics
not visible in $F(t)$.
This dynamics can arise, for example, because of
the 2-point function contained in the expansion of $\SumKD{\rho}$
[see Eq.~\eqref{eq:ProjTrick2}].

\item $\SumKD{\rho}$ reveals that scrambling breaks a symmetry.
Operationally, the symmetry consists of invariances of $\SumKD{\rho}$
under permutations and negations of measurement outcomes
in the weak-measurement scheme (Sec.~\ref{section:Intro_weak_meas}).
The symmetry breaking manifests in bifurcations of $\SumKD{\rho}$.
These bifurcations evoke classical-chaos pitchfork diagrams,
which also arise when a symmetry breaks.
One equilibrium point splits into three
in the classical case~\cite{Strogatz_00_Non}.
Perhaps the quasiprobability's pitchforks can be recast
in terms of equilibria.

\end{enumerate}

\section{Calculation of $\SumKD{\rho}$ averaged over Brownian circuits}
\label{section:Brownian}

We study a geometrically nonlocal model---the \emph{Brownian-circuit model}---governed by a time-dependent Hamiltonian~\cite{Lashkari_13_Towards}.
We access physics qualitatively different from
the physics displayed in the numerics of Sec.~\ref{section:Numerics}.
We also derive results for large systems
and compare with the finite-size numerics.
Since the two models' locality properties differ,
we do not expect agreement at early times.
The late-time scrambled states, however,
may be expected to share similarities.
We summarize our main findings at the end of the section.

We consider a system of $\Sites$ qubits
governed by the random time-dependent Hamiltonian
\begin{align}
H(t) \propto \sum_{i < j} \sum_{\alpha_i ,\alpha_j}
J^{\alpha_i,\alpha_j}_{i,j}(t)  \,
\sigma_i^{\alpha_i} \sigma_j^{\alpha_j}  \, .
\end{align}
The couplings $J$ are time-dependent random variables.
We denote the site-$i$ identity operator and Pauli operators by
$\sigma_i^\alpha$, for $\alpha=0,1,2,3$.
According to the model's precise formulation,
the time-evolution operator $U(t)$ is a random variable that obeys
\begin{align}\label{eq:browncirc}
   U(t+dt) & - U(t)  = - \frac{\Sites}{2} U(t) dt
   - i  \,  dB(t)  \, .
\end{align}
The final term's $dB(t)$ has the form
\begin{align}
   \label{eq:dB}
   dB(t) = \sqrt{\frac{1}{8(\Sites-1)}} \sum_{i < j} \sum_{\alpha_i, \alpha_j}
   \sigma_i^{\alpha_i} \sigma_j^{\alpha_j} dB^{\alpha_i,\alpha_j}_{i,j}(t)  \, .
\end{align}
We will sometimes call Eq.~\eqref{eq:dB} ``$dB$.''
$dB$ is a Gaussian random variable
with zero mean and with variance
\begin{align} \label{eq:dBdB}
   \mathbf{E}_B
   \left\{ dB^{\alpha,\beta}_{i,j}  \,
   dB^{\alpha',\beta'}_{i',j'} \right\} = \delta_{\alpha,\alpha'}\delta_{\beta,\beta'}
   \delta_{i,i'} \delta_{j,j'}  \,  dt.
\end{align}
The expectation value $\mathbf{E}_B$ is an average
over realizations of the noise $B$.
We demand that $dt  \,  dt =0$ and $dB  \,  dt = 0$,
in accordance with the standard Ito calculus.
$dB(t)$ is independent of $U(t)$,
i.e., of all previous $dB$'s.

We wish to compute the average,
over the ensemble defined by Eq.~\eqref{eq:browncirc},
of the coarse-grained quasiprobability:
\begin{align}
\mathfrak{A}( v_1 , w_2 , v_2 , w_3 )
= \mathbf{E}_B\left\{ \SumKD{\rho} ( v_1 , w_2 , v_2 , w_3 ) \right\}  \, .
\end{align}

\subsection{Infinite-temperature thermal state $\id / 2^\Sites$}

We focus here on the infinite-temperature thermal state,
$\rho = \id/2^{\Sites}$, for two reasons.
First, a system with a time-dependent Hamiltonian generically heats to infinite temperature with respect to any Hamiltonian in the ensemble.
Second, the $T = \infty$ state is convenient for calculations.
A discussion of other states follows.

The ensemble remains invariant under single-site rotations,
and all qubits are equivalent.
Therefore, all possible choices of
single-site Pauli operators for $\mathcal{W}$ and $V$ are equivalent.
Hence we choose
$\mathcal{W} = \sigma_1^z$ and $V = \sigma_2^z$
without loss of generality.

Let us return to Eq.~\eqref{eq:ProjTrick}.
Equation~\eqref{eq:ProjTrick} results from
substituting in for the projectors in $\SumKD{\rho}$.
The sum contains $16$ terms.
To each term, each projector contributes
the identity $\id$ or a nontrivial Pauli ($\mathcal{W}$ or $V$).
The terms are
\begin{enumerate}[(1)]
  \item $\id \id \id \id $: $ \text{Tr}\left\{\frac{ \id }{2^\Sites} \right\} =  1 $,

  \item $\mathcal{W}\id \id \id $, $\id V\id \id $,
  $\id \id \mathcal{W} \id $, $\id \id \id V$: $ 0 $,

  \item $\mathcal{W}V\id \id $, $\mathcal{W} \id \id V$,
  $ \id V\mathcal{W} \id $,  $ \id  \id \mathcal{W}V$: \\
  $ \text{Tr}\left\{\frac{\sigma_1^z(t) \sigma_2^z}{2^\Sites} \right\} =: G(t) $,

  \item $\mathcal{W} \id \mathcal{W} \id $, $ \id V \id V$:
  $ \text{Tr}\left\{\frac{ \id }{2^\Sites} \right\} =  1 $,

  \item $\mathcal{W}V\mathcal{W} \id $,
  $\mathcal{W}V \id V$, $\mathcal{W} \id \mathcal{W}V$,
  $ \id V\mathcal{W}V$: $ 0 $,  \quad and

  \item $\mathcal{W}V\mathcal{W}V$:
  $ \text{Tr}\left\{\frac{\sigma_1^z(t)
  \sigma_2^z \sigma_1^z(t) \sigma_2^z}{2^\Sites} \right\} = F(t) $.

\end{enumerate}
These computations rely on $\rho =  \id /2^\Sites$.
Each term that contains an odd number of Pauli operators vanishes,
due to the trace's cyclicality and to the Paulis' tracelessness.
We have introduced a 2-point function $G(t)$.
An overall factor of $1/16$ comes from the projectors' normalization.

Combining all the ingredients,
we can express $\SumKD{\rho}$ in terms of $G$ and $F$. The result is
\begin{align}
& 16  \,  \SumKD{\rho} ( v_1 , w_2 , v_2 , w_3 )
= (1 + w_2 w_3 + v_1 v_2)  \\  \nonumber
& \qquad + (w_2+w_3)(v_1+v_2)  \,  G
+ w_2 w_3 v_1 v_2  \, F.
\end{align}
This result depends on $\rho = \id / 2^\Sites$,
not on the form of the dynamics.
But to compute $\mathfrak{A}$, we must compute
\begin{align}
\mathfrak{G} = \mathbf{E}_B\left\{G \right\}
\end{align}
and
\begin{align}
\mathfrak{F} = \mathbf{E}_B\left\{F \right\}.
\end{align}

The computation of $\mathfrak{F}$ appears in the literature~\cite{Shenker_Stanford_15_Stringy}.
$\mathfrak{F}$ initially equals unity.
It decays to zero around 
$t_* = \frac{1}{3} \log \Sites$, the scrambling time.
The precise functional form of $\mathfrak{F}$ is not crucial.
The basic physics is captured in a phenomenological form
inspired by AdS/CFT computations \cite{Shenker_Stanford_15_Stringy},
\begin{align}
\mathfrak{F} \sim \left(\frac{1+c_1}{1+c_1 e^{3 t}}\right)^{c_2},
\end{align}
wherein $c_1 \sim 1/\Sites$ and $c_2 \sim 1$.

To convey a sense of the physics, we review the simpler calculation of $\mathfrak{G}$. The two-point function evolves according to
\begin{align}
G(t+dt) &= \frac{1}{2^\Sites}  \:  \text{Tr}\bigg\{
\left[U(t) - \frac{\Sites}{2 } U(t) dt - i  \,  dB  \,  U(t) \right]
\sigma_1^z
\nonumber \\ & \times
\left[U(t)^\dagger - \frac{\Sites}{2} U(t)^\dagger dt
+ i  \,  U(t)^\dagger dB^\dagger \right]  \sigma_2^z \bigg\} \, .
\end{align}
Using the usual rules of Ito stochastic calculus,
particularly Eq.~\eqref{eq:dBdB} and $dt  \, dt = dB  \, dt =0$,
we obtain
\begin{align}
   & \mathfrak{G}(t+dt)  - \mathfrak{G}(t)
   = - \Sites  \,  dt  \,  \mathfrak{G}(t)
   + dt  \,   \frac{1}{8(\Sites-1)}
   \nonumber \\ & \times
   \sum_{i < j}  \sum_{\alpha_i,\alpha_j} \frac{1}{2^\Sites}
   \mathbf{E}_B\left\{ \text{Tr}\left\{
   \sigma_1^z(t) \sigma_i^{\alpha_i} \sigma_j^{\alpha_j} \sigma_2^z     \sigma_i^{\alpha_i} \sigma_j^{\alpha_j}   \right\} \right\}.
\end{align}
We have applied the trace's cyclicality in the second term.

The second term's value depends on whether $i$ and/or $j$ equals $2$.
If $i$ and/or $j$ equals $2$,
the second term vanishes because
$\sum_{\alpha=0}^3 \sigma^\alpha \sigma^z \sigma^\alpha = 0$.
If neither $i$ nor $j$ is $2$,
$\sigma_i^{\alpha_i} \sigma_j^{\alpha_j}$ commutes with $\sigma_2^z$.
The second term becomes proportional to $G$.
In $(\Sites-1)(\Sites-2)/2$ terms, $i, j \neq 2$.
An additional factor of $4^2 = 16$
comes from the two sums over Pauli matrices. Hence
\begin{align}
\mathfrak{G}(t+dt) - \mathfrak{G}(t) = - 2 dt  \, \mathfrak{G}  \, ,
\end{align}
or
\begin{align}
\frac{d \mathfrak{G}}{dt} = - 2 \mathfrak{G} .
\end{align}

This differential equation implies that $\mathfrak{G}$
exponentially decays from its initial value.
The initial value is zero: $\mathfrak{G}(0) = G(0) = 0$.
Hence $\mathfrak{G}(t)$ is identically zero.

Although it does not arise when we consider $\mathfrak{A}$,
the ensemble-average autocorrelation function
$\mathbf{E}_B\left\{\langle \sigma_1^z(t) \sigma_1^z\rangle \right\}$
obeys a differential equation similar to
the equation obeyed by $\mathfrak{G}$.
In particular, the equation decays exponentially with
an order-one rate.

By the expectation value's linearity and the vanishing of $\mathfrak{G}$,
\begin{align} \label{avgA}
\mathfrak{A} = \frac{(1+w_2 w_3 + v_1 v_2)
+ w_2 w_3 v_1 v_2  \,  \mathfrak{F} }{16}.
\end{align}
This simple equation states that
the ensemble-averaged quasiprobability depends only on
the ensemble-averaged OTOC $F(t)$,
at infinite temperature.
The time scale of $\mathfrak{F}$'s decay is $t_* = \frac{1}{3} \log \Sites$.
Hence this is the time scale of changes in $\mathfrak{A}$.

Equation~\eqref{avgA} shows (as intuition suggests)
that $\mathfrak{A}$ depends only on
the combinations $w_2 w_3$ and $v_1 v_2$.
At $t=0$, $\mathfrak{F}(0)=1$. Hence $\mathfrak{A}$ is
\begin{align}
\mathfrak{A}_{t=0} = \frac{1 + w_2 w_3 + v_1 v_2 +  w_2 w_3 v_1 v_2}{16}.
\end{align}
The cases are
\begin{enumerate}[(1)]
\item $w_2 w_3 =1, v_1 v_2 =1$: $\mathfrak{A} = 1/4$,
\item $w_2 w_3=1, v_1 v_2=-1$: $\mathfrak{A} = 0$,
\item $w_2 w_3=-1, v_1 v_2=1$: $\mathfrak{A} = 0$,
\quad \text{and}
\item $w_2 w_3=-1, v_1 v_2=-1$: $\mathfrak{A} = 0$.
\end{enumerate}
These values are consistent with Fig.~\ref{fig:TInf_zz_AR} at $t=0$.
These values' degeneracies are consistent with
the symmetries discussed in Sec.~\ref{section:Numerics}
and in Sec.~\ref{section:TA_Props} (Property~\ref{property:Syms}).

At long times, $\mathfrak{F}(\infty) = 0$, so $\mathfrak{A}$ is
\begin{align}
\mathfrak{A}_{t=\infty} = \frac{1 + w_2 w_3 + v_1 v_2}{16}.
\end{align}
The cases are
\begin{enumerate}[(1)]
\item $w_2 w_3=1, v_1 v_2=1$: $\mathfrak{A} = 3/16$,
\item $w_2 w_3=1, v_1 v_2=-1$: $\mathfrak{A} = 1/16$,
\item $w_2 w_3=-1, v_1 v_2=1$: $\mathfrak{A} = 1/16$,
\quad \text{and}
\item $w_2 w_3=-1, v_1 v_2=-1$: $\mathfrak{A} = -1/16$.
\end{enumerate}
Modulo the splitting of the upper two lines,
this result is broadly consistent with
the long-time behavior in Fig.~\ref{fig:TInf_zz_AR}.
As the models in Sec.~\ref{section:Numerics} and this section differ,
the long-time behaviors need not agree perfectly.
However, the models appear to achieve
qualitatively similar scrambled states at late times.

\subsection{General state}

Consider a general state $\rho$, such that
$\SumKD{\rho}$ assumes the general form in Eq.~\eqref{eq:ProjTrick}.
We still assume that $\mathcal{W}=\sigma_1^z$ and $V = \sigma_2^z$.
However, the results will, in general, now depend on these choices
via the initial condition $\rho$.
We still expect that, at late times,
the results will not depend on the precise choices.
Below, we use the notation $\langle . \rangle \equiv \text{Tr}( \rho \, . )$.

We must consider 16 terms again.
The general case involves fewer simplifications. The terms are
\begin{enumerate}[(1)]
  \item $\id \id \id \id $: $ 1 $,

  \item $\mathcal{W} \id \id \id$, $ \id V \id  \id $,
  $ \id  \id \mathcal{W} \id $, $ \id  \id  \id V$:
  $\langle \sigma_1^z(t) \rangle $ , $\langle \sigma_2^z \rangle$,

  \item $\mathcal{W}V \id  \id $, $\mathcal{W} \id  \id V$,
  $ \id V\mathcal{W} \id $, $ \id  \id \mathcal{W}V$: \\
  $ \langle \sigma_1^z(t)  \,  \sigma_2^z \rangle $,
  $\langle \sigma_2^z  \,   \sigma_1^z(t) \rangle$,

  \item $\mathcal{W} \id \mathcal{W} \id $, $ \id V \id V$: $ 1 $,

  \item  \label{item:ThreeIs}
  $\mathcal{W}V\mathcal{W} \id $, $\mathcal{W}V \id V$,
  $\mathcal{W} \id \mathcal{W}V$, $ \id V\mathcal{W}V$:
  $\langle \sigma_1^z(t)  \,   \sigma_2^z  \,   \sigma_1^z(t)  \rangle$, $
  \langle \sigma_1^z(t)  \,   \rangle$,
  $\langle \sigma_2^z \rangle$,
  $\langle \sigma_2^z  \,   \sigma_1^z(t)  \,   \sigma_2^z \rangle$,
  \quad \text{and}

  \item $\mathcal{W}V\mathcal{W}V$:
  $ \langle \sigma_1^z(t)  \,   \sigma_2^z  \,   \sigma_1^z(t)  \,   \sigma_2^z \rangle = F(t) $.

\end{enumerate}

Consider first the terms of the form
$\mathfrak{q}_i(t) := \mathbf{E}_B\{\langle \sigma_i^z(t) \rangle\}$.
The time derivative is
\begin{align}
&\frac{d \mathfrak{q}_i}{dt} = - \Sites \mathfrak{q}_i  \\ \nonumber
& + \frac{1}{8(\Sites-1)} \sum_{j<k} \sum_{\alpha_j,\alpha_k} \mathbf{E}_B \{ \langle \sigma_j^{\alpha_j} \sigma_k^{\alpha_k} U(t) \sigma_i^z U(t)^\dagger \sigma_j^{\alpha_j} \sigma_k^{\alpha_k} \rangle \}.
\end{align}
To simplify the second term, we use a trick. Since
\begin{align}
\label{eq:BrwnTrick}
\sigma_j^{\alpha_j} \sigma_k^{\alpha_k} \sigma_m^{\alpha_m} \sigma_n^{\alpha_n} \sigma_j^{\alpha_j} \sigma_k^{\alpha_k} = \pm \sigma_m^{\alpha_m} \sigma_n^{\alpha_n},
\end{align}
we may pass the factors of $\sigma_j^{\alpha_j} \sigma_k^{\alpha_k}$
through $U(t)$, at the cost of changing some Brownian weights.
We must consider a different set of $dB$'s,
related to the originals by minus signs.
This alternative set of Brownian weights
has the original set's ensemble probability.
Hence the ensemble average gives the same result. Therefore,
\begin{align}
& \mathbf{E}_B\{\langle \sigma_j^{\alpha_j} \sigma_k^{\alpha_k}
U(t) \, \sigma_i^z U(t)^\dagger \sigma_j^{\alpha_j} \sigma_k^{\alpha_k} \rangle \} \nonumber \\
& = \mathbf{E}_B\{\langle U(t) \sigma_j^{\alpha_j} \sigma_k^{\alpha_k}  \sigma_i^z \sigma_j^{\alpha_j} \sigma_k^{\alpha_k}  U(t)^\dagger \rangle \}.
\end{align}

If $i = j$ and/or $i = k$,
the sum over $\alpha_{j}$
and/or the sum over $\alpha_k$ vanishes.
If $i$ equals neither $j$ nor $k$, the Pauli operators commute.
The term reduces to $\mathfrak{q}_i$.
$i$ equals neither $j$ nor $k$ in
$(\Sites-1)(\Sites-2)/2$ terms.
A factor of $16$ comes from the sums over $\alpha_j$ and $\alpha_k$. Hence
\begin{align}
\frac{d \mathfrak{q}_i}{dt} = - \Sites \mathfrak{q}_i + (\Sites-2) \mathfrak{q}_i = - 2 \mathfrak{q}_i.
\end{align}

Consider the terms of the form $\mathfrak{q}_{ij}(t) := \langle \sigma_i^z(t) \sigma_j^z \rangle$.
Note that $\langle \sigma_j^z \sigma_i^z(t) \rangle = \mathfrak{q}_{ij}^*$.
We may reuse the trick introduced above.
[This trick fails only when more than two copies of $U$ appear, as in $F(t)$].
To be precise,
\begin{align}
& \mathbf{E}_B\{\langle \sigma_m^{\alpha_m} \sigma_n^{\alpha_n} U(t) \sigma_i^z U(t)^\dagger \sigma_m^{\alpha_m} \sigma_n^{\alpha_n} \sigma_j^z \rangle \} \nonumber \\
& = \mathbf{E}_B\{\langle U(t) \sigma_m^{\alpha_m} \sigma_n^{\alpha_n}  \sigma_i^z \sigma_m^{\alpha_m} \sigma_n^{\alpha_n} U(t)^\dagger \sigma_j^z \rangle \}.
\end{align}
As before, the sums over $\alpha$
kill the relevant term in
the time derivative of $\mathfrak{q}_{ij}$, unless $i \neq m,n$. Hence
\begin{align}
\frac{d \mathfrak{q}_{ij}}{dt} = - 2 \mathfrak{q}_{ij} ,
\end{align}
as at infinite temperature.

Item~\ref{item:ThreeIs}, in the list above, concerns
products of three $\W$'s and $V$'s.
We must consider four expectation values of Pauli products.
As seen above, two of these terms reduce to $\mathfrak{q}_i$ terms.
By the trick used earlier,
\begin{align}
& \mathbf{E}_B \{ \langle \sigma_2^z U(t) \sigma_1^z U(t)^\dagger \sigma_2^z \} \nonumber \\
& = \mathbf{E}_B \{ \langle U(t) \sigma_2^z  \sigma_1^z \sigma_2^z U(t)^\dagger  \} = \mathfrak{q}_1(t).
\end{align}
The other term we must consider is
$\mathbf{E}_B \{ \langle \sigma_i^z(t) \sigma_j^z \sigma_i^z(t) \rangle \}
=:  \mathfrak{f}_{ij}$.
Our trick will not work, because
there are multiple copies of $U(t)$
that are not all simultaneously switched
as operators are moved around.
At early times, when $\sigma_i^z(t)$ and $\sigma_j^z$ approximately commute, this term approximately equals $\langle \sigma_j^z \rangle = \mathfrak{q}_j(0)$.
At later times, including around the scrambling time,
this term decays to zero.

The general expression for $\mathfrak{A}$ becomes
\begin{eqnarray}
  16  \,  \mathfrak{A} &=& 1 + w_3 w_2 + v_1 v_2 \nonumber \\
    &+& (w_3 +w_2)  \, \mathfrak{q}_1(t)
    + (v_1 + v_2)  \, \mathfrak{q}_2(0) \nonumber \\
    &+& (w_3 v_2 + w_3 v_1 + w_2 v_1)  \,  \mathfrak{q}_{12}(t)
    + v_2 w_2  \,  \mathfrak{q}_{12}(t)^* \nonumber \\
    &+& w_3 v_2 w_2  \,  \mathfrak{f}_{12}(t)
    + (w_3 v_1 v_2 + w_2 v_1 v_2)  \,  \mathfrak{q}_1(t)
    \nonumber \\ \label{eq:Brown_help}
    &+& w_3 w_2 v_1  \,  \mathfrak{q}_2(0)
    + w_3 w_2 v_1 v_2  \,  \mathfrak{F}(t).
\end{eqnarray}
All these $\mathfrak{q}$ functions obey known differential equations.
The functions decay after a time of order one.
We do not have explicit expressions for the $\mathfrak{f}$ functions that appear.
They are expected to vary after a time $\sim  \log \Sites$.

\subsubsection{Special case: $\sigma_2^z$ eigenstate}

In a concrete example, we suppose that $\rho$ is
a $+1$ eigenstate of $\sigma_2^z$.
Expressions simplify:
\begin{align}
\mathfrak{q}_2(0) = 1,
\end{align}
\begin{align}
\mathfrak{q}_{12}(t) = \mathfrak{q}_1(t) = \mathfrak{q}_{12}(t)^*,
\end{align}
and
\begin{align}
\mathfrak{f}_{12} = \mathfrak{F}.
\end{align}
Hermiticity of the Pauli operators implies that $\mathfrak{f}_{12}$ is real.
Hence the ensemble-averaged OTOC $\mathfrak{F}$ is real for this choice of $\rho$.
The ensemble-averaged $\OurKD{\rho}$ has the form
\begin{align}
\label{eq:Brown_ex0}
\mathfrak{A} = \frac{k_1 + k_2 \mathfrak{q}_1 + k_3 \mathfrak{F}  }{16} ,
\end{align}
wherein
\begin{align}
k_1 = (1+v_1)(1 + v_2  + w_3 w_2),
\end{align}
\begin{align}
k_2 =  (1+v_1)(w_3 +w_2)(1+v_2),
\end{align}
and
\begin{align}
\label{eq:Brown_ex3}
k_3 =  (1+v_1) w_3 v_2 w_2.
\end{align}
Equations~\eqref{eq:Brown_ex0}--\eqref{eq:Brown_ex3} imply that
$\mathfrak{A} = 0$ unless $v_1=1$.

The time scale after which $\mathfrak{q}_1$ decays
is order-one.
The time required for $\mathfrak{F}$ to decay is of order $\log \Sites$ (although not necessarily exactly the same as for the $T = \infty$ state).
Therefore, the late-time value of $\mathfrak{A}$ is well approximated by
\begin{align}
\mathfrak{A}_{t \gg 1} = \frac{k_1 + k_3  \,  \mathfrak{F}}{16}.
\end{align}

\subsection{Summary}

This study has the following main messages.
\begin{enumerate}[(1)]
\item In this model, the ensemble-averaged quasiprobability varies
on two time scales.
The first time scale is an order-one relaxation time.
At later times, the OTOC controls the physics entirely.
$F(t)$ varies after a time of order $\log \Sites$.
\item While the late-time physics of $\SumKD{\rho}$ is controlled entirely by
the ensemble-averaged $F(t)$,
the negative values of $\SumKD{\rho}$ show a nonclassicality
that might not be obvious from $F(t)$ alone.
Furthermore, we computed only the first moment of $\SumKD{\rho}$.
The higher moments are likely not determined by $F(t)$ alone.
\item For $T = \infty$, the late-time physics
is qualitatively similar to the late-time physics of
the geometrically local spin chain in Sec.~\ref{section:Numerics}.
\item Nonclassicality, as signaled by negative values of $\SumKD{\rho}$,
is extremely robust.
It survives the long-time limit and the ensemble average.
One might have expected thermalization and interference
to stamp out nonclassicality.
On the other hand, we expect the circuit average
to suppress the imaginary part of $\SumKD{\rho}$ rapidly.
We have no controlled examples in which
$\Im \left( \SumKD{\rho} \right)$ remains nonzero at long times.
Finding further evidence for or against this conjecture
remains an open problem.
\end{enumerate}

\section{Theoretical study of $\OurKD{\rho}$}
\label{section:Theory}

We have discussed experimental measurements,
numerical simulations, and analytical calculations
of the OTOC quasiprobability $\OurKD{\rho}$.
We now complement these discussions with
mathematical properties and physical interpretations.
First, we define an \emph{extended Kirkwood-Dirac distribution}
exemplified by $\OurKD{\rho}$.
We still denote by $\mathcal{B} ( \Hil )$ the set of
bounded operators defined on $\Hil$.

\begin{definition}[$\Ops$-extended Kirkwood-Dirac quasiprobability]
\label{definition:Extend_KD}
Let $\Set{ \ket{a} },  \ldots,  \Set{ \ket{k} }$
and $\Set{ \ket{f} }$ denote orthonormal bases
for the Hilbert space $\mathcal{H}$.
Let $\Oper \in \mathcal{B}( \mathcal{H} )$ denote
a bounded operator defined on $\Hil$.
A \emph{$\Ops$-extended Kirkwood-Dirac quasiprobability}
for $\Oper$ is defined as\footnote{
\label{footnote:ImplicitTime}
Time evolutions may be incorporated into the bases.
For example, Eq.~\eqref{eq:KD_rho_2} features
the 1-extended KD quasiprobability
$\langle f' | a \rangle  \langle a | \rho' | f' \rangle$.
The $\rho'  :=  U_{t'} \rho U_{t'}^\dag$ results from
time-evolving a state $\rho$.
The $\ket{ f' }  :=  U_{ t'' - t' }^\dag  \ket{f}$ results from
time-evolving an eigenket $\ket{f}$ of $F = \sum_f  f  \ketbra{f}{f}$.
We label~\eqref{eq:KD_rho_2} as
$\OurKD{\rho}^\1 ( \rho, a , f )$,
rather than as $\OurKD{\rho}^\1 ( \rho', a , f' )$.
Why? One would measure~\eqref{eq:KD_rho_2}
by preparing $\rho$, evolving the system,
measuring $\A$ weakly, inferring outcome $a$,
evolving the system, measuring $F$, and obtaining outcome $f$.
No outcome $f'$ is obtained.
Our notation is that in~\cite{Dressel_15_Weak}
and is consistent with the notation in~\cite{YungerHalpern_17_Jarzynski}.}
\begin{align}
   \label{eq:Extend_KD}
   \OurKD{\Oper}^\ParenK ( a,  \ldots,  k, f )  :=
   \langle f | k \rangle \langle k | \ldots | a \rangle \langle a | \Oper | f \rangle \, .
\end{align}
\end{definition}

This quasiprobability can be measured 
via an extension of the protocol in Sec.~\ref{section:Intro_weak_meas}.
Suppose that $\Oper$ denotes a density matrix.
In each trial, one prepares $\Oper$,
weakly measures the bases sequentially 
(weakly measures $\Set{ \ket{a} }$, and so on,
until weakly measuring $\Set{ \ket{k} }$),
then measures $\ketbra{f}{f}$ strongly.

We will focus mostly on density operators
$\Oper = \rho \in \mathcal{D} ( \Hil )$.
One infers $\OurKD{\rho}^\ParenK$ by performing
$2 \Ops - 1$ weak measurements,
and one strong measurement, per trial.
The order in which the bases are measured is
the order in which the labels $a , \ldots, k, f$ appear in
the argument of $\OurKD{\Oper}^\ParenK ( . )$.
The conventional KD quasiprobability is 1-extended.
The OTOC quasiprobability $\OurKD{\rho}$ is 3-extended.

Our investigation parallels the exposition, in Sec.~\ref{section:Intro_to_KD},
of the KD distribution.
First, we present basic mathematical properties.
$\OurKD{\rho}$, we show next,
obeys an analog of Bayes' Theorem.
Our analog generalizes the known analog~\eqref{eq:CondQuasi}.
Our theorem reduces exponentially (in system size)
the memory needed to compute weak values, in certain cases.
Third, we connect $\OurKD{\rho}$ with
the operator-decomposition argument in Sec.~\ref{section:KD_Coeffs}.
$\OurKD{\rho}$ consists of coefficients
in a decomposition of an operator $\rho'$
that results from asymmetrically decohering $\rho$.
Summing $\OurKD{\rho} ( . )$ values yields
a KD representation for $\rho$.
This sum can be used, in experimental measurements
of $\OurKD{\rho}$ and the OTOC, to evaluate
how accurately the desired initial state was prepared.
Fourth, we explore the relationship between
out-of-time ordering and quasiprobabilities.
Time-ordered correlators are moments of quasiprobabilities
that clearly reduce to classical probabilities.
Finally, we generalize beyond the OTOC,
which encodes $\Ops = 3$ time reversals.
Let $\Opsb  :=  \frac{1}{2} ( \Ops + 1 )$.
A \emph{$\Opsb$-fold OTOC} $F^\ParenKB(t)$
encodes $\Ops$ time reversals~\cite{Roberts_16_Chaos,Hael_17_Classification}.
The quasiprobability behind $F^\ParenKB(t)$, we find,
is $\Ops$-extended.

Recent quasiprobability advances involve out-of-time ordering,
including in correlation functions~\cite{Manko_00_Lyapunov,Bednorz_13_Nonsymmetrized,Oehri_16_Time,Hofer_17_Quasi,Lee_17_On}.
Merging these works with the OTOC framework
offers an opportunity for further research (Sec.~\ref{section:Outlook}).

\subsection{Mathematical properties of $\OurKD{\rho}$}
\label{section:TA_Props}

$\OurKD{\rho}$ shares some of its properties
with the KD quasiprobability (Sec.~\ref{section:KDProps}).
Properties of $\OurKD{\rho}$ imply properties of $P(W, W')$,
presented in Appendix~\ref{section:P_Properties}.

\begin{property}  \label{prop:TA_Complex}
The OTOC quasiprobability is a map
$\OurKD{\rho}  \:  :  \:
\mathcal{D} ( \mathcal{H} )  \times
\Set{ v_1 }    \times   \Set{ \DegenV_{v_1} }    \times
\Set{ w_2 }    \times  \Set{ \DegenW_{w_2} }    \times
\Set{ v_2 }    \times   \Set{ \DegenV_{v_2} }      \times
\Set{ w_3 }    \times   \Set{ \DegenW_{w_3} }    \times
\to \mathbb{C} \, .$
The domain is a composition of
the set $\mathcal{D} ( \Hil )$ of density operators defined on $\Hil$
and eight sets of complex numbers.
The range is not necessarily real:
$\mathbb{C}  \supset \mathbb{R}$.
\end{property}

$\OurKD{\rho}$ depends on $H$ and $t$ implicitly through $U$.
The KD quasiprobability in~\cite{Dressel_15_Weak}
depends implicitly on time similarly
(see Footnote~\ref{footnote:ImplicitTime}).
Outside of OTOC contexts, $\mathcal{D} ( \Hil )$
may be replaced with $\mathcal{B} ( \Hil )$.
$\Ops$-extended KD distributions represent bounded operators,
not only quantum states.
$\mathbb{C}$, not necessarily $\mathbb{R}$,
is the range also of
the $\Ops$-fold generalization $\OurKD{\rho}^\ParenK$.
We expound upon the range's complexity
after discussing the number of arguments of $\OurKD{\rho}$.

\emph{Five effective arguments of $\OurKD{\rho}$:}
On the left-hand side of Eq.~\eqref{eq:TADef},
semicolons separate four tuples.
Each tuple results from a measurement, e.g., of $\NondegW$.
We coarse-grained over the degeneracies
in Sections~\ref{section:ProjTrick}--\ref{section:Brownian}.
Hence each tuple often functions as one degree of freedom.
We treat $\OurKD{\rho}$ as a function of
four arguments (and of $\rho$).
The KD quasiprobability has just two arguments (apart from $\Oper$).
The need for four arises from
the noncommutation of $\W(t)$ and $V$.

\emph{Complexity of $\OurKD{\rho}$:}
The ability of $\OurKD{\rho}$ to assume nonreal values
mirrors Property~\ref{prop:Complex} of the KD distribution.
The Wigner function, in contrast, is real.
The OTOC quasiprobability's real component, $\Re ( \OurKD{\rho} )$,
parallels the Terletsky-Margenau-Hill distribution.
We expect nonclassical values of $\OurKD{\rho}$
to reflect nonclassical physics,
as nonclassical values of the KD quasiprobability do
(Sec.~\ref{section:Intro_to_KD}).

Equations~\eqref{eq:TADef} and~\eqref{eq:TAForm}
reflect the ability of $\OurKD{\rho}$ to assume nonreal values.
Equation~\eqref{eq:TADef} would equal
a real product of probabilities
if the backward-process amplitude $A_\rho^*$
and the forward-process amplitude $A_\rho$
had equal arguments.
But the arguments typically do not equal each other.
Equation~\eqref{eq:TAForm} reveals conditions under which
$\OurKD{\rho} ( . )  \in  \mathbb{R}$ and $\not\in \mathbb{R}$.
We illustrate the $\in$ case with two examples
and the $\not\in$ case with one example.

\begin{example}[Real $\OurKD{\rho}$ \#1:
$t = 0$, shared eigenbasis, arbitrary $\rho$]
\label{ex:Real1}
Consider $t  =  0$, at which $U  =  \id$.
The operators $\W(t) = \W$ and $V$ share an eigenbasis,
under the assumption that $[ \W ,  \,  V ]  =  0$:
$\Set{ \ket{ w_\ell,  \DegenW_{w_\ell} } }
=  \Set{ \ket{ v_\ell,  \DegenV_{v_\ell} } }$.
With respect to that basis,
\begin{align}
   \label{eq:Ex1}
   & \OurKD{\rho} ( v_1,  \DegenV_{v_1}  ;  w_2,  \DegenW_{w_2} ;
   v_2,  \DegenV_{v_2}  ;  w_3,  \DegenW_{w_3}  )
   \nonumber \\ &
   =  \left(  \delta_{w_3 v_2}  \delta_{ \DegenW_{w_3} \DegenV_{v_2} }  \right)
   \left(  \delta_{v_2 w_2}  \delta_{ \DegenV_{v_2}  \DegenW_{w_2} }  \right)
   \left(  \delta_{w_2  v_1}  \delta_{ \DegenW_{w_2} \DegenV_{v_1} }   \right)
   \nonumber \\ &  \quad  \times
   \sum_j  p_j
   | \langle w_3,  \DegenW_{w_3}  |  j  \rangle |^2
    \nonumber \\ &
   \in  \mathbb{R}  \, .
\end{align}
We have substituted into Eq.~\eqref{eq:TAForm}.
We substituted in for $\rho$ from Eq.~\eqref{eq:Rho}.
\end{example}

Example~\ref{ex:Real1} is consistent with
the numerical simulations in Sec.~\ref{section:Numerics}.
According to Eq.~\eqref{eq:Ex1}, at $t = 0$,
$\sum_{\text{degeneracies}} \OurKD{\rho}  =:  \SumKD{\rho}  \in \mathbb{R}$.
In Figures~\ref{fig:T1_zz_AI},~\ref{fig:rand_zz_AI}, and~\ref{fig:xup_zz_AI},
the imaginary parts $\Im ( \SumKD{\rho} )$
clearly vanish at $t = 0$.
In Fig.~\ref{fig:TInf_zz_AI}, $\Im ( \SumKD{\rho} )$ vanishes
to within machine precision.\footnote{
The $\Im ( \SumKD{\rho} )$ in Fig.~\ref{fig:TInf_zz_AI}
equals zero identically, if $w_2 = w_3$ and/or if $v_1  =  v_2$.
For general arguments,
\begin{align}
   \label{eq:Imag_zero}
   \Im  \LParen  \SumKD{\rho} ( v_1, w_2, v_2, w_3 ) \RParen
   & = \frac{1}{ 2 i }  \:  \Big[  \OurKD{\rho} ( v_1, w_2, v_2, w_3 )
   \nonumber \\ & \qquad
   -  \OurKD{\rho}^* ( v_1, w_2, v_2, w_3 )  \Big]  \, .
\end{align}
The final term equals
\begin{align}
   & \left[  \Tr \left(  \ProjWt{ w_3 }  \ProjV{ v_2 }  \ProjWt{ w_2 }
                     \ProjV{ v_1 }  \right)  \right]^*
   =  \Tr \left(  \ProjV{ v_1 }  \ProjWt{ w_2 }  \ProjV{ v_2 }
                      \ProjWt{ w_3 }  \right)
   \\ & \quad  \label{eq:Imag_zero_help}
   =  \Tr \left(  \ProjWt{ w_2 }  \ProjV{ v_2 } \ProjWt{ w_3 }
                     \ProjV{ v_1 }  \right)
   =  \SumKD{\rho}  ( v_1 , w_3 , v_2 , w_2 )  \, .
\end{align}
The first equality follows from projectors' Hermiticity;
and the second, from the trace's cyclicality.
Substituting into Eq.~\eqref{eq:Imag_zero} shows that
$\SumKD{\rho} ( . )$ is real if $w_2 = w_3$.
$\SumKD{\rho} ( . )$ is real if $v_1 = v_2$,
by an analogous argument.}

Consider a $\rho$ that lacks coherences relative to
the shared eigenbasis, e.g., $\rho  =  \id / \Dim$.
Example~\ref{ex:Real1} implies that
$\Im \left( \OurKD{( \id / \Dim)} \right)$ at $t = 0$.
But $\Im \left( \OurKD{( \id / \Dim)} \right)$ remains zero for all $t$
in the numerical simulations.
Why, if time evolution deforms the $\W(t)$ eigenbasis
from the $V$ eigenbasis?
The reason appears to be a cancellation, as in Example~\ref{ex:Real2}.

Example~\ref{ex:Real2} requires more notation.
Let us focus on a chain of $\Sites$ spin-$\frac{1}{2}$
degrees of freedom.
Let $\sigma^\alpha$ denote the $\alpha = x, y, z$ Pauli operator.
Let $\ket{ \sigma^\alpha,  \pm }$ denote
the $\sigma^\alpha$ eigenstates, such that
$\sigma^\alpha  \ket{ \sigma^\alpha,  \pm }
=  \pm  \ket{ \sigma^\alpha,  \pm }$.
$\Sites$-fold tensor products are denoted by
$\ket{ \bm{ \sigma^\alpha,  \pm } }  :=
\ket{ \sigma^\alpha,  \pm }^{\otimes \Sites }$.
We denote by $\sigma_j^\alpha$
the $\alpha^\th$ Pauli operator
that acts nontrivially on site $j$.

\begin{example}[Real $\OurKD{\rho}$ \#2:
$t = 0$, nonshared eigenbases, $\rho = \id / \Dim$]
\label{ex:Real2}
Consider the spin chain at $t = 0$, such that $U  =  \id$.
Let $\W = \sigma_1^z$ and $V  =  \sigma_\Sites^y$.
Two $\W$ eigenstates are $\ket{ \bm{ \sigma^z, \pm } }$.
Two $V$ eigenstates are
$\ket{ \bm{ \sigma^y,  +  } }
=  \left[  \frac{1}{ \sqrt{2} }  \:
\left( \ket{ \sigma^z, + }
+  i  \ket{ \sigma^z, - }  \right)  \right]^{ \otimes \Sites }$
and  $\ket{ \bm{ \sigma^y,  -  } }
=  \left[  \frac{1}{ \sqrt{2} }  \:
\left( \ket{ \sigma^z, + }
-  i  \ket{ \sigma^z, - }  \right)  \right]^{ \otimes \Sites }$.
The overlaps between the $\W$ eigenstates
and the $V$ eigenstates are
\begin{align}
   \label{eq:InnerPs}
   & \langle \bm{ \sigma^z, + }  |  \bm{ \sigma^y, + }  \rangle
   =  \left(  \frac{ 1 }{ \sqrt{2} }  \right)^\Sites \, ,
   \nonumber \\ &
   \langle \bm{ \sigma^z, + }  |  \bm{ \sigma^y, - }  \rangle
   =  \left(  \frac{ 1 }{ \sqrt{2} }  \right)^\Sites  \, ,
   \nonumber \\ &
   \langle \bm{ \sigma^z, - }  |  \bm{ \sigma^y, + }  \rangle
   = \left(  \frac{ i }{ \sqrt{2} }  \right)^\Sites \, ,
   \; \text{and} \nonumber \\ &
   \langle \bm{ \sigma^z, - }  |  \bm{ \sigma^y, - }  \rangle
   = \left(  \frac{ - i }{ \sqrt{2} }  \right)^\Sites \, .
\end{align}

Suppose that $\rho = \id / \Dim$.
$\OurKD{ ( \id / \Dim) } ( . )$ would have a chance of being nonreal
only if some $\ket{ v_\ell ,  \DegenV_{v_\ell} }$ equaled
$\ket{ \bm{ \sigma^z,  - } }$.
That $\ket{ \bm{ \sigma^z,  - } }$ would introduce
an $i$ into Eq.~\eqref{eq:TAForm}.
But $\bra{ \bm{ \sigma^z,  - } }$ would introduce another $i$.
The product would be real.
Hence $\OurKD{ ( \id / \Dim) } ( . )  \in  \mathbb{R}$.
\end{example}
\noindent $\OurKD{\rho}$ is nonreal in the following example.

\begin{example}[Nonreal $\OurKD{\rho}$:
$t = 0$, nonshared eigenbases, $\rho$ nondiagonal relative to both]
\label{ex:Nonreal1}
Let $t$, $\W$, $V$, $\Set{ \ket{ w_\ell,  \DegenW_{w_\ell} } }$,
and $\Set{ \ket{ v_m,  \DegenV_{v_m} } }$
be as in Example~\ref{ex:Real2}.

Suppose that $\rho$ has coherences relative to
the $\W$ and $V$ eigenbases.
For instance, let
$\rho  =  \ketbra{ \bm{ \sigma^x, + } }{ \bm{ \sigma^x, + } }$.
Since $\ket{ \sigma^x, + }  =  \frac{1}{ \sqrt{2} }  \:
( \ket{ \sigma^z, + }  +  \ket{ \sigma^z, - } )$,
\begin{align}
   \rho  & =  \frac{1}{ 2^\Sites }  \:
   (  \ketbra{ \sigma^z, + }{ \sigma^z, + }
   +  \ketbra{ \sigma^z, + }{ \sigma^z, - }
   \nonumber \\ & \qquad
   +  \ketbra{ \sigma^z, - }{ \sigma^z, + }
   +  \ketbra{ \sigma^z, - }{ \sigma^z, - }  )^{ \otimes \Sites }  \, .
\end{align}

Let $\ket{ w_3, \DegenW_{w_3} } =  \ket{ \bm{ \sigma^z, - } }$,
such that its overlaps with $V$ eigenstates can contain $i$'s.
The final factor in Eq.~\eqref{eq:TAForm} becomes
\begin{align}
   \langle  v_1,  \DegenV_{v_1}  |  \rho  |  w_3,  \DegenW_{w_3}  \rangle
   & =  \frac{1}{ 2^\Sites }  \Big[
   \langle  v_1,  \DegenV_{v_1}  |
   \left(  \ket{ \sigma^z,  +  }^{ \otimes \Sites }  \right)
   \nonumber \\ & \qquad
   +   \langle  v_1,  \DegenV_{v_1}  |
   \left(  \ket{ \sigma^z,  -  }^{ \otimes \Sites }  \right)
   \Big]  \, .
\end{align}
The first inner product evaluates to
$\left(  \frac{ 1 }{ \sqrt{2} }  \right)^\Sites$,
by Eqs.~\eqref{eq:InnerPs}.
The second inner product evaluates to
$\left( \pm  \frac{ i }{ \sqrt{2} }  \right)^\Sites$.
Hence
\begin{align}
   \langle  v_1,  \DegenV_{v_1}  |  \rho  |  w_3,  \DegenW_{w_3}  \rangle
   =  \frac{1}{ 2^{2 \Sites} }
   \left[  1  +  \left(  \pm i \right)^\Sites  \right] \, .
\end{align}
This expression is nonreal if $\Sites$ is odd.
\end{example}

Example~\ref{ex:Nonreal1}, with the discussion after Example~\ref{ex:Real1},
shows how interference can eliminate nonreality from a quasiprobability.
In Example~\ref{ex:Nonreal1},
$\Im \left( \OurKD{\rho} \right)$ does not necessarily vanish.
Hence the coarse-grained $\Im \left( \SumKD{\rho} \right)$
does not obviously vanish.
But $\Im \left( \SumKD{\rho} \right) = 0$
according to the discussion after Example~\ref{ex:Real1}.
Summing Example~\ref{ex:Nonreal1}'s nonzero
$\Im \left( \OurKD{\rho}  \right)$ values
must quench the quasiprobability's nonreality.
This quenching illustrates how interference
can wash out quasiprobabilities' nonclassicality.
Yet interference does not always wash out nonclassicality.
Section~\ref{section:Numerics} depicts
$\SumKD{\rho}$'s that have nonzero imaginary components
(Figures~\ref{fig:T1_zz_AI},~\ref{fig:rand_zz_AI}, and~\ref{fig:xup_zz_AI}).

Example~\ref{ex:Nonreal1} resonates with a finding in~\cite{Solinas_15_Full,Solinas_16_Probing}.
Solinas and Gasparinetti's quasiprobability assumes nonclassical values
when the initial state
has coherences relative to the energy eigenbasis.

\begin{property}
\label{prop:MargOurKD}
Marginalizing $\OurKD{\rho} ( . )$ over all its arguments
except any one
yields a probability distribution.
\end{property}

   Consider, as an example, summing  Eq.~\eqref{eq:TAForm} over
   every tuple except $( w_3, \DegenW_{w_3} )$.
   The outer products become resolutions of unity, e.g.,
   $\sum_{ ( w_2, \DegenW_{w_2} ) }
      \ketbra{ w_2, \DegenW_{w_2} }{ w_2, \DegenW_{w_2} }
      = \id$.
   A unitary cancels with its Hermitian conjugate:
   $U^\dag U = \id$.
   The marginalization yields
   $\langle  w_3, \DegenW_{w_3}  |  U  \rho  U^\dag  |
                      w_3, \DegenW_{w_3}  \rangle$.
   This expression equals the probability that
   preparing $\rho$, time-evolving,
   and measuring the $\NondegW$ eigenbasis
   yields the outcome $( w_3, \DegenW_{w_3} )$.

This marginalization property,
with the structural and operational resemblances
between $\OurKD{\rho}$
and the KD quasiprobability,
accounts for our calling $\OurKD{\rho}$ an extended quasiprobability.
The general $\Ops$-extended $\OurKD{\rho}^\ParenK$
obeys Property~\ref{prop:MargOurKD}.

\begin{property}[Symmetries of $\OurKD{ ( \id / \Dim) }$]
\label{property:Syms}
Let $\rho$ be the infinite-temperature Gibbs state $\id / \Dim$.
The OTOC quasiprobability $\OurKD{ ( \id / \Dim) }$
has the following symmetries.
\begin{enumerate}[(A)]

   \item  \label{item:Sym1}
   $\OurKD{ ( \id / \Dim) } ( . )$ remains invariant under
   the simultaneous interchanges of
   $( w_2,  \DegenW_{w_2} )$ with $( w_3,  \DegenW_{w_3} )$
   and $( v_1,  \DegenV_{v_1} )$ with $( v_2,  \DegenV_{v_2} )$:
   $\OurKD{ ( \id / \Dim) } ( v_1,  \DegenV_{v_1} ;  w_2,  \DegenW_{w_2} ;
   v_2,  \DegenV_{v_2}  ;  w_3,  \DegenW_{w_3}  )
     =  \OurKD{ ( \id / \Dim) } ( v_2,  \DegenV_{v_2}  ;  w_3,  \DegenW_{w_3} ;
     v_1,  \DegenV_{v_1} ;  w_2,  \DegenW_{w_2} )$.

   \item  \label{item:Sym2}
   Let $t = 0$, such that
   $\Set{ \ket{ w_\ell,  \DegenW_{w_\ell} } }
   =  \Set{ \ket{ v_\ell,  \DegenV_{v_\ell} } }$
   (under the assumption that $[\W,  V]  =  0$).
   $\OurKD{ ( \id / \Dim) } ( . )$ remains invariant under
   every cyclic permutation of its arguments.

\end{enumerate}
\end{property}

Equation~\eqref{eq:TAForm} can be recast as a trace.
Property~\ref{property:Syms} follows from the trace's cyclicality.
Subproperty~\ref{item:Sym2} relies on the triviality of
the $t = 0$ time-evolution operator: $U = \id$.
The symmetries lead to degeneracies visible in numerical plots
(Sec.~\ref{section:Numerics}).

Analogous symmetries characterize a \emph{regulated} quasiprobability.
Maldacena \emph{et al.} regulated $F(t)$
to facilitate a proof~\cite{Maldacena_15_Bound}:\footnote{
The name ``regulated'' derives from quantum field theory.
$F(t)$ contains operators $\W^\dag(t)$ and $\W(t)$
defined at the same space-time point
(and operators $V^\dag$ and $V$ defined at the same space-time point).
Products of such operators encode divergences.
One can regulate divergences
by shifting one operator to another space-time point.
The inserted $\rho^{1/4}  =  \frac{1}{Z^{1/4} }  \;  e^{ - H / 4 T }$
shifts operators along an imaginary-time axis.}
\begin{align}
   \label{eq:RegOTOC_def}
   F_\reg (t)  :=  \Tr \left(  \rho^{1/4}  \W(t)  \rho^{1/4}  V
   \rho^{1/4}  \W(t)  \rho^{1/4}  V  \right)  \, .
\end{align}
$F_\reg(t)$ is expected to behave roughly like $F(t)$~\cite{Maldacena_15_Bound,Yao_16_Interferometric}.
Just as $F(t)$ equals a moment of
a sum over $\OurKD{\rho}$,
$F_\reg(t)$ equals a moment of a sum over
\begin{align}
   \label{eq:RegKD}
   &\OurKD{\rho}^\reg  ( v_1,  \DegenV_{v_1} ; w_2,  \DegenW_{w_2} ;
   v_2,  \DegenV_{v_2}  ;  w_3,  \DegenW_{w_3}  )
   \\ \nonumber &
   :=  \langle  w_3,  \DegenW_{w_3}  |  U  \rho^{1/4}  |
        v_2,  \DegenV_{v_2}  \rangle
        \langle  v_2,  \DegenV_{v_2}  |  \rho^{1/4}  U^\dag  |
        w_2,  \DegenW_{w_2}  \rangle
        \\ \nonumber  & \qquad \times
        \langle  w_2,  \DegenW_{w_2}  |  U  \rho^{1/4}  |
        v_1,  \DegenV_{v_1}  \rangle
        \langle  v_1,  \DegenV_{v_1}  |  \rho^{1/4}  U^\dag  |
        w_3,  \DegenW_{w_3}  \rangle \\
   &   \label{eq:RegKD2}
   \equiv \langle  w_3,  \DegenW_{w_3}  |  \tilde{U}  |
        v_2,  \DegenV_{v_2}  \rangle
        \langle  v_2,  \DegenV_{v_2}  |  \tilde{U}^\dag  |
        w_2,  \DegenW_{w_2}  \rangle
        \\ \nonumber  & \qquad \times
        \langle  w_2,  \DegenW_{w_2}  |  \tilde{U}  |
        v_1,  \DegenV_{v_1}  \rangle
        \langle  v_1,  \DegenV_{v_1}  |  \tilde{U}^\dag  |
        w_3,  \DegenW_{w_3}  \rangle  \, .
\end{align}
The proof is analogous to the proof of Theorem~1 in~\cite{YungerHalpern_17_Jarzynski}.
Equation~\eqref{eq:RegKD2} depends on
$\tilde{U}  :=  \frac{1}{Z}  \:  e^{ - i H \tau }$,
which propagates in the complex-time variable
$\tau :=  t  -  \frac{i}{4 T}$.
The Hermitian conjugate $\tilde{U}^\dag  =  \frac{1}{Z}  \:  e^{ i H \tau^* }$
propagates along $\tau^* = t + \frac{i }{ 4 T }$.

$\OurKD{ \left( e^{ - H / T } / Z \right) }^\reg$ has
the symmetries of $\OurKD{ ( \id / \Dim) }$
(Property~\ref{property:Syms}) for arbitrary $T$.
One might expect $\OurKD{\rho}^\reg$ to behave
similarly to $\OurKD{\rho}$,
as $F_\reg(t)$ behaves similarly to $F(t)$.
Numerical simulations largely support this expectation.
We compared $\SumKD{\rho} ( . )$ with
$\SumKD{ \rho }^\reg ( . )
:=  \sum_{\text{degeneracies}} \OurKD{\rho}^{ \reg } ( . ) \, .$
The distributions vary significantly over similar time scales
and have similar shapes.
$\SumKD{\rho}^\reg$ tends to have a smaller imaginary component
and, as expected, more degeneracies.

The properties of $\OurKD{\rho}$ imply properties of $P(W, W')$.
We discuss these properties in Appendix~\ref{section:P_Properties}.

%
%
\subsection{Bayes-type theorem and retrodiction with $\OurKD{\rho}$}
\label{section:TA_retro}

We reviewed, in Sec.~\ref{section:KD_Retro},
the KD quasiprobability's role in retrodiction.
The KD quasiprobability $\OurKD{\rho}^\1$ generalizes the nontrivial part
$\Re ( \langle f' | a \rangle  \langle a | \rho' | f' \rangle )$
of a conditional quasiprobability $\tilde{p} ( a | \rho, f )$
used to retrodict about an observable $\A$.
Does $\OurKD{\rho}$ play a role similar to $\OurKD{\rho}^\1$?

It does. To show so, we generalize Sec.~\ref{section:KD_Retro}
to composite observables.
Let $\A, \B, \ldots, \K$ denote $\Ops$ observables.
$\K  \ldots  \B  \A$ might not be Hermitian but can be symmetrized.
For example, $\Gamma  :=  \K \ldots  \A  +  \A \ldots \K$
is an observable.\footnote{
So is $\tilde{\Gamma}  :=  i ( \K  \ldots  \A  -  \A  \ldots  \K )$.
An operator can be symmetrized in multiple ways.
Theorem~\ref{theorem:RetroK} governs $\Gamma$.
Appendix~\ref{section:RetroK2} contains
an analogous result about $\tilde{\Gamma}$.
Theorem~\ref{theorem:RetroK} extends trivially to
Hermitian (already symmetrized) instances of $\K \ldots \A$.
Corollary~\ref{corollary:RetroOurKD} illustrates this extension.}
Which value is most reasonably attributable to
$\Gamma$ retrodictively?
A weak value $\Gamma_\weak$ given by Eq.~\eqref{eq:WeakVal}.
We derive an alternative expression for $\Gamma_\weak$.
In our expression, $\Gamma$ eigenvalues
are weighted by $\Ops$-extended KD quasiprobabilities.
Our expression reduces exponentially, in the system's size,
the memory required to calculate weak values,
under certain conditions.
We present general theorems about $\OurKD{\rho}^\ParenK$,
then specialize to the OTOC $\OurKD{\rho}$.

%
%
\begin{theorem}[Retrodiction about composite observables]
\label{theorem:RetroK}
Consider a system $\Sys$ associated with a Hilbert space $\Hil$.
For concreteness, we assume that $\Hil$ is discrete.
Let $\A  =  \sum_a  a  \ketbra{a}{a} \, ,  \ldots ,
\K  =  \sum_k  k  \ketbra{k}{k}$ denote
$\Ops$ observables defined on $\mathcal{H}$.
Let $U_t$ denote the family of unitaries
that propagates the state of $\Sys$
along time $t$.

Suppose that $\Sys$ begins in the state
$\rho$ at time $t = 0$,
then evolves under $U_{t''}$ until $t = t''$.
Let $F = \sum_f  f  \ketbra{f}{f}$ denote an observable
measured at $t = t''$.
Let $f$ denote the outcome.
Let $t'  \in  (0,  t'')$ denote an intermediate time.
Define $\rho'  :=  U_{t'}  \rho  U_{t'}^\dag$   and
$\ket{f'}  :=  U^\dag_{t'' - t'}  \ket{f}$ as time-evolved states.

The value most reasonably attributable retrodictively to the time-$t'$
$\Gamma  :=  \K  \ldots  \A  +  \A \ldots \K$
is the weak value
\begin{align}
   \label{eq:GammaW}
   \Gamma_\weak ( \rho , f )
   & = \sum_{a, \ldots, k }  (a \ldots k)
   \Big[ \tilde{p}_\rightarrow (a, \ldots, k | \rho, f )
           \nonumber \\ & \qquad \qquad  +
           \tilde{p}_\leftarrow ( k, \ldots, a | \rho, f )
   \Big] \, .
\end{align}
The weights are joint conditional quasiprobabilities.
They obey analogs of Bayes' Theorem:
\begin{align}
   \tilde{p}_\rightarrow ( a, \ldots, k | \rho , f )
   & =  \label{eq:QuasiBayesLeft1}
   \frac{ \tilde{p}_\rightarrow ( a, \ldots, k , f | \rho ) }{
             p ( f | \rho ) }   \\
   & \equiv  \label{eq:QuasiBayesLeft2}
   \frac{ \Re ( \langle f' | k \rangle \langle k |  \ldots
                      | a \rangle \langle a | \rho' | f' \rangle ) }{
             \langle f'  | \rho' | f' \rangle }\, ,
\end{align}
and
\begin{align}
    \tilde{p}_\leftarrow ( k, \ldots, a | \rho , f )
    & =  \label{eq:QuasiBayesRt1}
            \frac{ \tilde{p}_\leftarrow ( k, \ldots, a, f | \rho ) }{
                      p ( f | \rho ) } \\
    & \equiv  \label{eq:QuasiBayesRt2}
    \frac{ \Re ( \langle f' | a \rangle \langle a |  \ldots
                      | k \rangle \langle k | \rho' | f' \rangle ) }{
             \langle f'  | \rho' | f' \rangle }    \, .
\end{align}
Complex generalizations of the weights' numerators,
\begin{align}
   \label{eq:Extend_KD2}
   \OurKD{ \rho ,  \rightarrow }^\ParenK ( a, \ldots, k, f )
   :=  \langle f' | k \rangle  \langle k |  \ldots  | a \rangle
         \langle a | \rho' | f' \rangle
\end{align}
and
\begin{align}
   \label{eq:Extend_KD1}
   \OurKD{ \rho ,  \leftarrow }^\ParenK  ( k, \ldots, a, f )
   :=  \langle f' | a \rangle  \langle a |  \ldots  | k \rangle
        \langle k |  \rho' | f'  \rangle   \, ,
\end{align}
are $\Ops$-extended KD distributions.
\end{theorem} \noindent
A rightward-pointing arrow $\rightarrow$
labels quantities in which the outer products,
$\ketbra{k}{k}$,  \ldots,  $\ketbra{a}{a}$, are ordered analogously to
the first term $\K \ldots \A$ in $\Gamma$.
A leftward-pointing arrow $\leftarrow$
labels quantities in which
reading the outer products $\ketbra{a}{a}$, \ldots, $\ketbra{k}{k}$
backward---from right to left---parallels
reading $\K \ldots \A$ forward.

\begin{proof}
The initial steps come from~\cite[Sec. II A]{Dressel_15_Weak},
which recapitulates~\cite{Johansen_04_Nonclassical,Hall_01_Exact,Hall_04_Prior}.
For every measurement outcome $f$, we assume,
some number $\gamma_f$ is
the guess most reasonably attributable to $\Gamma$.
We combine these best guesses into the effective observable
$\Gest  :=  \sum_f  \gamma_f  \ketbra{f'}{f'}$.
We must optimize our choice of $\Set{ \gamma_f }$.
We should quantify the distance between
(1) the operator $\Gest$ we construct and
(2) the operator $\Gamma$ we wish to infer about.
We use the weighted trace distance
\begin{align}
   \label{eq:OprDist}
   \mathscr{D}_{\rho'} ( \Gamma,  \Gest )  =
   \Tr  \left(  \rho'  [  \Gamma  -  \Gest ]^2  \right) \, .
\end{align}
$\rho'$ serves as a ``positive prior bias''~\cite{Dressel_15_Weak}.

Let us substitute in for the form of $\Gest$.
Expanding the square, then invoking the trace's linearity, yields
\begin{align}
   \label{eq:DHelp1}
   & \mathscr{D}_{ \rho' } ( \Gamma,  \Gest )  =
  \Tr ( \rho' \Gamma^2 )
  +  \sum_f  \Big[
      \gamma_f^2    \langle  f'  |  \rho'  |  f'  \rangle
      \nonumber \\ & \qquad \quad
       -  \gamma_f  ( \langle f' | \rho' \Gamma | f' \rangle
           +  \langle f' | \Gamma \rho' | f' \rangle )
   \Big] \, .
\end{align}
The parenthesized factor equals
$2 \Re ( \langle f' | \Gamma \rho' | f' \rangle )$.
Adding and subtracting
\begin{align}
   \sum_f  \langle f' | \rho' | f' \rangle
   [  \Re ( \langle f' | \Gamma \rho' | f' \rangle )  ]^2
\end{align}
to and from Eq.~\eqref{eq:DHelp1},
we complete the square:
\begin{align}
   \label{eq:DResult}
   & \mathscr{D}_{ \rho' } ( \Gamma,  \Gest )  =
   \Tr ( \rho' \Gamma^2 )
   -  \sum_f  \langle f' | \rho' | f' \rangle
   [  \Re ( \langle f' | \Gamma \rho' | f' \rangle )  ]^2
   \nonumber \\ &  \qquad \qquad
   +  \sum_f  \langle f' | \rho' | f' \rangle
   \Bigg(  \gamma_f  -
   \frac{ \Re ( \langle f' | \Gamma \rho' | f' \rangle ) }{
   \langle f' | \rho' | f' \rangle }    \Bigg)^2  .
\end{align}

Our choice of $\Set{ \gamma_f }$ should minimize the distance~\eqref{eq:DResult}.
We should set the square to zero:
\begin{align}
   \label{eq:Choose}
   \gamma_f  =
   \frac{ \Re ( \langle f' | \Gamma \rho' | f' \rangle ) }{
   \langle f' | \rho' | f' \rangle }  \, .
\end{align}

Now, we deviate from~\cite{Johansen_04_Nonclassical,Hall_01_Exact,Hall_04_Prior,Dressel_15_Weak}.
We substitute the definition of $\Gamma$ into Eq.~\eqref{eq:Choose}.
Invoking the linearity of $\Re$ yields
\begin{align}
   \label{eq:Choose2}
   \gamma_f  =  \frac{
   \Re ( \langle f' | \K  \ldots  \A \rho' | f' \rangle ) }{
   \langle f' | \rho' | f' \rangle }
   + \frac{  \Re ( \langle f' | \A  \ldots  \K \rho' | f' \rangle ) }{
   \langle f' | \rho' | f' \rangle }  \, .
\end{align}
We eigendecompose $\A, \ldots, \K$.
The eigenvalues, being real,
can be factored out of the $\Re$'s.
Defining the eigenvalues' coefficients
as in Eqs.~\eqref{eq:QuasiBayesLeft2} and~\eqref{eq:QuasiBayesRt2},
we reduce Eq.~\eqref{eq:Choose2} to the form in Eq.~\eqref{eq:GammaW}.
\end{proof}

Theorem~\ref{theorem:RetroK} reduces exponentially,
in system size, the space required
to calculate $\Gamma_\weak$, in certain cases.\footnote{
``Space'' means ``memory,'' or ``number of bits,'' here.}
For concreteness, we focus on a multiqubit system
and on $l$-local operators $\A, \ldots, \K$.
An operator $\mathcal{O}$ is \emph{$l$-local} if
$\mathcal{O} =  \sum_j  \mathcal{O}_j$,
wherein each $\mathcal{O}_j$ operates nontrivially on,
at most, $l$ qubits.
Practicality motivates this focus:
The lesser the $l$, the more easily
$l$-local operators can be measured.

We use asymptotic notation from computer science:
Let $f \equiv f( \Sites )$ and $g \equiv g (\Sites)$ denote
any functions of the system size.
If $g  =  O( f )$, $g$ grows no more quickly than
(is upper-bounded by)
a constant multiple of $f$
in the asymptotic limit, as $\Sites \to \infty$.
If $g  =  \Omega ( f )$, $g$ grows at least as quickly as
(is lower-bounded by) a constant multiple of $f$
in the asymptotic limit.
If $g  =  \Theta ( f )$, $g$ is upper- and lower-bounded by $f$:
$g  =  O ( f )$,  and  $g = \Omega ( f )$.
If $g  =  o ( f )$, $g$ shrinks strictly more quickly than $f$
in the asymptotic limit.

\begin{theorem}[Weak-value space saver]
\label{theorem:Space}

Let $\Sys$ denote a system of $\Sites$ qubits.
Let $\Hil$ denote the Hilbert space associated with $\Sys$.
Let $\ket{ f' }  \in  \Hil$ denote a pure state
and $\rho' \in \mathcal{D} ( \Hil )$ denote a density operator.
Let $\Basis$ denote any fixed orthonormal basis for $\Hil$
in which each basis element equals a tensor product
of $\Sites$ factors, each of which operates nontrivially on
exactly one site.
$\Basis$ may, for example, consist of tensor products of
$\sigma^z$ eigenstates.

Let $\Ops$ denote any polynomial function of $\Sites$:
$\Ops  \equiv  \Ops( \Sites )  =  {\text{poly}}( \Sites )$.
Let $\A , \ldots, \K$ denote
$\Ops$ traceless $l$-local observables defined on $\Hil$,
for any constant $l $.
Each observable may, for example, be a tensor product of
$\leq l$ nontrivial Pauli operators and $\geq \Sites  -  l$ identity operators.
The composite observable $\Gamma :=  \A \ldots \K  +  \K \ldots \A$
is not necessarily $l$-local.
Let $\A = \sum_a a \ketbra{a}{a} \, ,  \ldots, \K = \sum_k k \ketbra{k}{k}$
denote eigenvalue decompositions of the local observables.
Let $\mathcal{O}_{ \Basis }$ denote the matrix that represents
an operator $\mathcal{O}$ relative to $\Basis$.

Consider being given the matrices $\A_\Basis, \ldots, \K_\Basis$,
$\rho'_\Basis$, and $\ket{ f' }_\Basis$.
From this information, the weak value $\Gamma_\weak$
can be computed in two ways:
\begin{enumerate}[(1)]

   \item \label{item:Conven}
            \textbf{Conventional method}
   \begin{enumerate}[(A)]
      \item Multiply and sum given matrices to form
      $\Gamma_{ \Basis }  =  \K_\Basis \ldots \A_\Basis
                                            +  \A_\Basis \ldots \K_\Basis$.
      \item Compute $\langle f' | \rho' | f' \rangle
               =  \langle f' |_\Basis  \: \rho'_\Basis  \: | f' \rangle_\Basis$.
      \item Substitute into $\Gamma_\weak  =  \Re \left(
      \frac{ \langle f' |_\Basis  \:  \Gamma_\Basis  \:   \rho'_\Basis  \:
               | f' \rangle_\Basis }{ \langle f' | \rho' | f' \rangle }
               \right) \, .$
   \end{enumerate}

   \item \label{item:KFac}
            \textbf{$\Ops$-factored method}
   \begin{enumerate}[(A)]
      \item Compute $\langle f' | \rho' | f' \rangle$.
      \item \label{eq:TermStep}
               For each nonzero term in Eq.~\eqref{eq:GammaW},
               \begin{enumerate}[(i)]
                  \item  calculate $\tilde{p}_\rightarrow ( . )$ and $\tilde{p}_\leftarrow ( . )$
               from Eqs.~\eqref{eq:QuasiBayesLeft2} and~\eqref{eq:QuasiBayesRt2}.
                  \item  substitute into Eq.~\eqref{eq:GammaW}.
               \end{enumerate}
   \end{enumerate}

\end{enumerate}

Let $\Sigma_{(n)}$ denote the space required to compute $\Gamma_\weak$,
aside from the space required to store $\Gamma_\weak$,
with constant precision,
using method $(n) = \ref{item:Conven}, \ref{item:KFac}$,
in the asymptotic limit.
Method~\ref{item:Conven} requires a number of bits at least
exponential in the number $\Ops$ of local observables:
\begin{align}
   \label{eq:Space1}
   \Sigma_{ \ref{item:Conven} }  =  \Omega \left( 2^\Ops \right) \, .
\end{align}
Method~\ref{item:KFac} requires a number of bits linear in $\Ops$:
\begin{align}
   \label{eq:Space2}
   \Sigma_{ \ref{item:KFac} }  =  O ( \Ops )  \, .
\end{align}
Method~\ref{item:KFac} requires exponentially---in $\Ops$ and so in $\Sites$---less memory than Method~\ref{item:Conven}.
\end{theorem}
\begin{proof}
Using Method~\ref{item:Conven}, one computes $\Gamma_\Basis$.
$\Gamma_\Basis$ is a $2^\Sites \times 2^\Sites$ complex matrix.
The matrix has $\Omega ( 2^\Ops )$ nonzero elements:
$\A , \ldots, \K$ are traceless,
so each of $\A_\Basis, \ldots, \K_\Basis$ contains
at least two nonzero elements.
Each operator at least doubles
the number of nonzero elements in $\Gamma_\Basis$.
Specifying each complex number with constant precision
requires $\Theta(1)$ bits.
Hence Method~\ref{item:Conven} requires $\Omega \left( 2^\Ops \right)$ bits.

Let us turn to Method~\ref{item:KFac}.
We can store $\langle f' | \rho' | f' \rangle$
in a constant number of bits.

Step~\ref{eq:TermStep} can be implemented with
a counter variable $C_\Oper$ for each local operator $\Oper$,
a running-total variable $G$, and a ``current term'' variable $T$.
$C_\Oper$ is used to iterate through
the nonzero eigenvalues of $\Oper$
(arranged in some fiducial order).
$\Oper$ has $O ( 2^l )$ nonzero eigenvalues.
Hence $C_\Oper$  requires $O ( l )$ bits.
Hence the set of $\Ops$ counters $C_\Oper$
requires $O ( l \Ops )  =  O ( \Ops )$ bits.

The following algorithm implements Step~\ref{eq:TermStep}:
\begin{enumerate}[(i)]

   \item   \label{item:CNormal}
   If $C_\K <$ its maximum possible value, proceed as follows:
   \begin{enumerate}[(a)]

      \item  For each $\Oper = \A, \ldots, \K$,
      compute the $(2^{ C_\Oper} )^\th$ nonzero eigenvalue
      (according to the fiducial ordering).

      \item  Multiply the eigenvalues to form $a \ldots k$.
      Store the product in $T$.

      \item  For each $\Oper  =  \A, \ldots, \K$,
      calculate the $(2^{ C_\Oper} )^\th$ eigenvector column
      (according to some fiducial ordering).

      \item  Substitute the eigenvector columns into
      Eqs.~\eqref{eq:QuasiBayesLeft2} and~\eqref{eq:QuasiBayesRt2},
      to compute $\tilde{p}_\rightarrow ( . )$ and $\tilde{p}_\leftarrow ( . )$.

      \item  Form $(a \ldots k) \Big[ \tilde{p}_\rightarrow (a, \ldots, k | \rho, f )
           +  \tilde{p}_\leftarrow ( k, \ldots, a | \rho, f )$.
           Update $T$ to this value.

      \item  Add $T$ to $G$.

      \item  Erase $T$.

      \item  Increment $C_\K$.
   \end{enumerate}

   \item If $C_\K$ equals its maximum possible value,
   increment the counter of the preceding variable, $\mathcal{J}$, in the list;
   reset $C_\K$ to one;
   and, if $\mathcal{J}$ has not attained its maximum possible value,
   return to Step~\ref{item:CNormal}.
   Proceed in this manner---incrementing counters;
   then resetting counters, incrementing preceding counters,
   and returning to Step~\ref{item:CNormal}---until $C_\A$ reaches its maximum possible value. Then, halt.
\end{enumerate}

The space needed to store $G$ is
the space needed to store $\Gamma_\weak$.
This space does not contribute to $\Sigma_{ \ref{item:KFac} }$.

How much space is needed to store $T$?
We must calculate $\Gamma_\weak$ with constant precision.
$\Gamma_\weak$ equals a sum of $2^{ l \Ops }$ terms.
Let $\varepsilon_j$ denote the error in term $j$.
The sum $\sum_{j = 1}^{ 2^{ l \Ops } }  \varepsilon_j$
must be $O(1)$.
This requirement is satisfied if
$2^{ l \Ops } \,  \left( \max_j | \varepsilon_j | \right)  =  o (1)$, which implies
$\max_j | \varepsilon_j |  =  o \left(  2^{ - l \Ops }  \right)$.
We can specify each term, with a small-enough roundoff error,
using $O ( l \Ops )  =  O ( \Ops )$ bits.

Altogether, the variables require $O( \Ops )$ bits.
As the set of variables does, so does the $\Oper$-factored method.
\end{proof}

Performing Method~\ref{item:KFac} requires slightly more time
than performing Method~\ref{item:Conven}.
Yet Theorem~\ref{theorem:Space} can benefit computations
about quantum many-body systems.
Consider measuring a weak value of a quantum many-body system.
One might wish to predict the experiment's outcome
and to compare the outcome with the prediction.
Alternatively, consider simulating quantum many-body systems
independently of laboratory experiments,
as in Sec.~\ref{section:Numerics}.
One must compute weak values numerically,
using large matrices.
The memory required to store these matrices
can limit computations.
Theorem~\ref{theorem:Space} can free up space.

Two more aspects of retrodiction deserve exposition:
related studies and the physical significance of $\K \ldots \A$.

\emph{Related studies:}
Sequential weak measurements have been proposed~\cite{Lundeen_12_Procedure}
and realized recently~\cite{Piacentini_16_Measuring,Suzuki_16_Observation,Thekkadath_16_Direct}.
Lundeen and Bamber proposed a ``direct measurement''
of a density operator~\cite{Lundeen_12_Procedure}.
Let $\rho$ denote a density operator
defined on a dimension-$\Dim$ Hilbert space $\Hil$.
Let $\Basis_a  :=  \Set{ \ket{ a_\ell } }$ and
$\Basis_b  :=  \Set{ \ket{ b_\ell } }$ denote orthonormal
\emph{mutually unbiased bases} (MUBs) for $\Hil$.
The interbasis inner products have constant magnitudes:
$| \langle a_\ell | b_m \rangle |  =  \frac{1}{ \sqrt{ \Dim } }
\;  \forall \ell, m$.
Consider measuring $\Basis_a$ weakly,
then $\Basis_b$ weakly, then $\Basis_a$ strongly,
in each of many trials.
One can infer (1) a KD quasiprobability for $\rho$
and (2) a matrix that represents $\rho$ relative to $\Basis_a$~\cite{Lundeen_12_Procedure}.

KD quasiprobabilities are inferred from experimental measurements in~\cite{Piacentini_16_Measuring,Thekkadath_16_Direct}.
Two weak measurements are performed sequentially also in~\cite{Suzuki_16_Observation}.
Single photons are used in~\cite{Piacentini_16_Measuring,Suzuki_16_Observation}.
A beam of light is used in~\cite{Thekkadath_16_Direct}.
These experiments indicate the relevance
of Theorem~\ref{theorem:RetroK}
to current experimental capabilities.
Additionally, composite observables $\A \B  +  \B \A$
accompany KD quasiprobabilities in e.g.,~\cite{Halliwell_16_Leggett}.

%
\emph{Physical significance of $\K \ldots \A$:}
Rearranging Eq.~\eqref{eq:GammaW} offers insight
into the result:
\begin{align}
   \label{eq:GammaW2}
   \Gamma_\weak ( \rho , f )    & =
   \sum_{ k, \ldots, a }  ( k \ldots a )
   \tilde{p}_\rightarrow  ( k, \ldots, a | \rho , f )
   \nonumber \\ & \qquad   +
   \sum_{a, \ldots, k }  (a \ldots k)
   \tilde{p}_\leftarrow ( a, \ldots, k | \rho , f )   \, .
\end{align}
Each sum parallels the sum in Eq.~\eqref{eq:WeakVal3}.
Equation~\eqref{eq:GammaW2} suggests that we are retrodicting
about $\K \ldots \A$ independently of $\A \ldots \K$.
But neither $\K \ldots \A$ nor $\A \ldots \K$ is Hermitian.
Neither operator seems measurable.
Ascribing a value to neither
appears to have physical significance, \emph{prima facie}.

Yet non-Hermitian products $\B \A$
have been measured weakly~\cite{Piacentini_16_Measuring,Suzuki_16_Observation,Thekkadath_16_Direct}.
Weak measurements associate a value with
the supposedly unphysical $\K \ldots \A$,
just as weak measurements enable us to infer
supposedly unphysical probability amplitudes $\Amp_\rho$.
The parallel between $\K \ldots \A$ and $\Amp_\rho$
can be expanded.
$\K \ldots \A$ and $\A \ldots \K$, being non-Hermitian,
appear to lack physical significance independently.
Summing the operators forms an observable.
Similarly, probability amplitudes $\Amp_\rho$ and $\Amp_\rho^*$
appear to lack physical significance independently.
Multiplying the amplitudes forms a probability.
But $\Amp_\rho$ and $\K \ldots \A$
can be inferred individually from weak measurements.

We have generalized Sec.~\ref{section:KD_Retro}.
Specializing to $k = 3$,
and choosing forms for $\A , \ldots \K$,
yields an application of $\OurKD{\rho}$ to retrodiction.

%
%
\begin{corollary}[Retrodictive application of $\OurKD{\rho}$]
   \label{corollary:RetroOurKD}
   Let $\Sys$, $\Hil$, $\rho$, $\W(t)$, and $V$ be defined
   as in Sec.~\ref{section:SetUp}.
   Suppose that $\Sys$ is in state $\rho$ at time $t = 0$.
   Suppose that the observable
   $F = \W
   =  \sum_{ w_3, \DegenW_{w_3} }  w_3
       \ketbra{ w_3, \DegenW_{w_3} }{ w_3, \DegenW_{w_3} }$
   of $\Sys$ is measured at time $t'' = t$.
   Let $( w_3, \DegenW_{w_3} )$ denote the outcome.
   Let $\A  =  V  =  \sum_{ v_1,  \DegenV_{v_1} }  v_1
                       \ketbra{ v_1,  \DegenV_{v_1} }{ v_1,  \DegenV_{v_1} }$,
   $\B  =  \W(t)  =  \sum_{ w_2,  \DegenW_{w_2} }  w_2  \,
                            U^\dag \ketbra{ w_2,  \DegenW_{w_2} }{
                                                      w_2,  \DegenW_{w_2} }  U$,
   and $\C  =  V  =  \sum_{ v_2,  \DegenV_{v_2} }  v_2
                              \ketbra{ v_2,  \DegenV_{v_2} }{ v_2,  \DegenV_{v_2} }  \, .$
   Let the composite observable $\Gamma =  \A \B \C  =  V  \W(t)  V$.
   The value most reasonably attributable to
   $\Gamma$ retrodictively is the weak value
   \begin{align}
      \label{eq:OTOC_retro1}
      & \Gamma_\weak ( \rho ; w_3, \DegenW_{w_3} )
      =  \sum_{ ( v_1 ,  \DegenV_{v_1} ) ,  ( v_2 ,  \DegenV_{v_2} ),
                      ( w_2 ,  \DegenW_{w_2} ) }
      v_1  w_2  v_2
      \nonumber \\ & \times
      \tilde{p}_{\leftrightarrow}
      ( v_2,  \DegenV_{v_2} ; w_2,  \DegenW_{w_2}  ;
        v_1,  \DegenV_{v_1}   |
      \rho  ;  w_3 , \DegenW_{w_3} )  \, .
   \end{align}
   The weights are joint conditional quasiprobabilities
   that obey an analog of Bayes' Theorem:
   \begin{align}
      \label{eq:OTOC_retro2}
      & \tilde{p}_{\leftrightarrow}
      ( v_1,  \DegenV_{v_1} ; w_2,  \DegenW_{w_2}  ;
        v_2,  \DegenV_{v_2}  |
      \rho  ;  w_3, \DegenW_{w_3}  )
      \nonumber \\ &
      =  \frac{ \tilde{p}_\leftrightarrow
      ( v_1,  \DegenV_{v_1} ; w_2,  \DegenW_{w_2}  ;
        v_2,  \DegenV_{v_2}  ;  w_3, \DegenW_{w_3}   |  \rho )   }{
      p ( w_3, \DegenW_{w_3}  | \rho ) }  \\
      & \equiv  \Re (
      \langle w_3, \DegenW_{w_3} | U | v_2,  \DegenV_{v_2} \rangle
      \langle v_2,  \DegenV_{v_2} | U^\dag | w_2,  \DegenW_{w_2} \rangle
      \nonumber \\ & \qquad \times
      \langle w_2,  \DegenW_{w_2} | U | v_1,  \DegenV_{v_1} \rangle
      \langle v_1,  \DegenV_{v_1} | \rho U^\dag |
        w_3, \DegenW_{w_3} \rangle )
      \nonumber \\ & \qquad  \;  /
      \langle w_3, \DegenW_{w_3} | \rho | w_3, \DegenW_{w_3} \rangle \, .
   \end{align}
   A complex generalization of the weight's numerator
   is the OTOC quasiprobability:
   \begin{align}
      \label{eq:OTOC_retro3}
      & \OurKD{\rho, \leftrightarrow }^\3 (
      v_1,  \DegenV_{v_1}  ;  w_2,  \DegenW_{w_2}  ;
      v_2,  \DegenV_{v_2}  ;  w_3 , \DegenW_{w_3}  )
      \nonumber \\ &
      =  \OurKD{\rho}
      ( v_1,  \DegenV_{v_1}  ;  w_2,  \DegenW_{w_2}  ;
      v_2,  \DegenV_{v_2}  ;  w_3 , \DegenW_{w_3} )  \, .
   \end{align}
\end{corollary}

The OTOC quasiprobability, we have shown,
assists with Bayesian-type inference,
similarly to the KD distribution.
The inferred-about operator is $V \W(t) V$,
rather than the $\W(t) V \W(t) V$ in the OTOC.
The missing $\W(t)$ plays the role of $F$.
This structure parallels the weak-measurement scheme in
the main text of~\cite{YungerHalpern_17_Jarzynski}:
$V$, $\W(t)$, and $V$ are measured weakly.
$\W(t)$ is, like $F$, then measured strongly.

\subsection{$\OurKD{\rho} ( . )$ values as coefficients
in an operator decomposition}
\label{section:TA_Coeffs}

Let $\Basis$ denote any orthonormal operator basis for $\mathcal{H}$.
Every state $\rho  \in  \mathcal{D} ( \mathcal{H} )$ can be decomposed
in terms of $\Basis$, as in Sec.~\ref{section:KD_Coeffs}.
The coefficients form a KD distribution.
Does $\OurKD{\rho}$ consist of
the coefficients in a state decomposition?

Summing $\OurKD{\rho} ( . )$ values yields
a coefficient in a decomposition of an operator $\rho'$.\footnote{
This $\rho'$ should not be confused with
the $\rho'$ in Theorem~\ref{theorem:RetroK}.}
$\rho'$ results from asymmetrically ``decohering'' $\rho$.
This decoherence relates to time-reversal asymmetry.
We expect $\rho'$ to tend to converge to $\rho$
after the scrambling time $t_*$.
By measuring $\OurKD{\rho}$ after $t_*$,
one may infer how accurately one prepared
the target initial state.

\begin{theorem}   \label{theorem:TA_Decomp}
Let
\begin{align}
   \label{eq:RhoPrime}
   \rho'  & :=  \rho  -  \sum_{ \substack{
   ( v_2,  \DegenV_{v_2} ) ,  ( w_3, \DegenW_{w_3} )
   \: : \:   \\
   \langle w_3, \DegenW_{w_3} | U | v_2,  \DegenV_{v_2} \rangle
   \neq 0 } }
   \ketbra{ v_2,  \DegenV_{v_2} }{ w_3, \DegenW_{w_3} }  U
   \nonumber \\ & \qquad \qquad \qquad \qquad \quad \times
   \langle v_2,  \DegenV_{v_2} |  \rho  U^\dag  |
                w_3, \DegenW_{w_3}  \rangle
\end{align}
denote the result of removing, from $\rho$,
the terms that connect the ``input state''
$U^\dag  \ket{ w_3, \DegenW_{w_3} }$
to the ``output state'' $\ket{ v_2,  \DegenV_{v_2} }$.
We define the set
\begin{align}
   \Basis  :=  \Set{  \frac{
   \ketbra{ v_2,  \DegenV_{v_2} }{ w_3, \DegenW_{w_3} }  U }{
   \langle  w_3, \DegenW_{w_3}  |  U  |  v_2,  \DegenV_{v_2}  \rangle }
   }_{ \langle w_3, \DegenW_{w_3} | U | v_2,  \DegenV_{v_2} \rangle
         \neq 0 }
\end{align}
of trace-one operators.
$\rho'$ decomposes in terms of $\Basis$ as
\begin{align}
   \sum_{ \substack{
   ( v_2,  \DegenV_{v_2} ) ,  ( w_3, \DegenW_{w_3} )
   \: : \:   \\
   \langle w_3, \DegenW_{w_3} | U | v_2,  \DegenV_{v_2} \rangle
   \neq 0 } }
   C^{ ( w_3, \DegenW_{w_3} ) }_{ ( v_2,  \DegenV_{v_2} ) }  \;
   \frac{
   \ketbra{ v_2,  \DegenV_{v_2} }{ w_3, \DegenW_{w_3} }  U }{
   \langle  w_3, \DegenW_{w_3}  |  U  |  v_2,  \DegenV_{v_2}  \rangle }  \, .
\end{align}
The coefficients follow from summing values
of the OTOC quasiprobability:
\begin{align}
   \label{eq:TA_decomp_coeff}
   & C^{ ( w_3, \DegenW_{w_3} ) }_{ ( v_2,  \DegenV_{v_2} ) }
   :=
   \sum_{ \substack{ ( w_2,  \DegenW_{w_2} ), \\  ( v_1,  \DegenV_{v_1} ) } }
   \OurKD{\rho}
   ( v_1,  \DegenV_{v_1}  ;  w_2,  \DegenW_{w_2} ;
     v_2,  \DegenV_{v_2}  ;  w_3,  \DegenW_{w_3} )  \, .
\end{align}
\end{theorem}

\begin{proof}
We deform the argument in Sec.~\ref{section:KD_Coeffs}.
Let the $\Set{ \ket{a} }$ in Sec.~\ref{section:KD_Coeffs} be
$\Set{ \ket{ v_2,  \DegenV_{v_2} } }$.
Let the $\Set{ \ket{f} }$ be
$\Set{ U^\dag \ket{ w_3,  \DegenW_{w_3} } }$.
We sandwich $\rho$ between resolutions of unity:
$\rho = \left( \sum_a  \ketbra{a}{a}  \right)  \rho
\left(  \sum_f  \ketbra{f}{f}  \right)$.
Rearranging yields
\begin{align}
   \label{eq:TA_Decomp_1}
   \rho  & =  \sum_{ ( v_2,  \DegenV_{v_2} ) ,  ( w_3,  \DegenW_{w_3} ) }
   \ketbra{ v_2,  \DegenV_{v_2} }{ w_3,  \DegenW_{w_3} } U
   \nonumber \\ & \qquad \qquad \qquad \qquad \times
   \langle  v_2,  \DegenV_{v_2}  |  \rho U^\dag  |
               w_3,  \DegenW_{w_3}  \rangle \, .
\end{align}

We wish to normalize the outer product,
by dividing by its trace.
We assumed, in Sec.~\ref{section:KD_Coeffs}, that
no interbasis inner product vanishes.
But inner products could vanish here.
Recall Example~\ref{ex:Real1}: When $t = 0$,
$\W(t)$ and $V$ share an eigenbasis.
That eigenbasis can have orthogonal states
$\ket{\psi}$ and $\ket{\phi}$.
Hence $\langle  w_3,  \DegenW_{w_3}  |  U  |
                          v_2,  \DegenV_{v_2}  \rangle$
can equal  $\langle \psi | \phi \rangle  =  0$.
No such term in Eq.~\eqref{eq:TA_Decomp_1}
can be normalized.

We eliminate these terms from the sum with
the condition $\langle  w_3,  \DegenW_{w_3}  |  U  |
                          v_2,  \DegenV_{v_2}  \rangle  \neq 0$.
The left-hand side of Eq.~\eqref{eq:TA_Decomp_1}
is replaced with the $\rho'$ in Eq.~\eqref{eq:RhoPrime}.
We divide and multiply by
the trace of each $\Basis$ element:
\begin{align}
   \rho' & =
   \sum_{ \substack{
   ( v_2,  \DegenV_{v_2} )  ,   ( w_3, \DegenW_{w_3} )  \: : \:   \\
   \langle w_3, \DegenW_{w_3} | U | v_2,  \DegenV_{v_2} \rangle
   \neq 0 } }
   \frac{
   \ketbra{ v_2,  \DegenV_{v_2} }{ w_3, \DegenW_{w_3} }  U }{
   \langle  w_3, \DegenW_{w_3}  |  U  |  v_2,  \DegenV_{v_2}  \rangle }
   \nonumber \\ &  \quad \times
   \langle  w_3, \DegenW_{w_3}  |  U  |  v_2,  \DegenV_{v_2}  \rangle
   \langle  v_2,  \DegenV_{v_2}  |  \rho  U^\dag  |
                w_3, \DegenW_{w_3}  \rangle \, .
\end{align}
The coefficients are KD-quasiprobability values.

Consider inserting, just leftward of the $\rho$,
the resolution of unity
\begin{align}
   \id  & =
   \left(  U^\dag   \sum_{ w_2, \DegenW_{w_2} }
            \ketbra{ w_2, \DegenW_{w_2} }{ w_2, \DegenW_{w_2} }  U  \right)
   \nonumber \\ & \qquad \times
   \left(  \sum_{ v_1,  \DegenV_{v_1} }
            \ketbra{ v_1,  \DegenV_{v_1} }{ v_1,  \DegenV_{v_1} }  \right) \, .
\end{align}
In the resulting $\rho'$ decomposition,
the $\sum_{ w_2, \DegenW_{w_2} }
\sum_{ v_1,  \DegenV_{v_1} }$
is pulled leftward, to just after the
$\frac{
   \ketbra{ v_2,  \DegenV_{v_2} }{ w_3, \DegenW_{w_3} }  U }{
   \langle  w_3, \DegenW_{w_3}  |  U  |  v_2,  \DegenV_{v_2}  \rangle }$.
This double sum becomes a sum of $\OurKD{\rho}$'s.
The $\rho'$ weights have the form in Eq.~\eqref{eq:TA_decomp_coeff}.
\end{proof}

Theorem~\ref{theorem:TA_Decomp} would hold
if $\rho$ were replaced with
any bounded operator $\Oper \in \mathcal{B} ( \Hil )$.
Four more points merit discussion.
We expect that, after the scrambling time $t_*$,
there tend to exist parameterizations
$\Set{ \DegenW_{w_\ell} }$ and $\Set{ \DegenV_{v_m} }$
such that $\Basis$ forms a basis.
Such a tendency could facilitate error estimates:
Suppose that $\OurKD{\rho}$ is measured after $t_*$.
One can infer the form of the state $\rho$ prepared
at the trial's start.
The target initial state may be difficult to prepare, e.g., thermal.
The preparation procedure's accuracy can be assessed at a trivial cost.
Third, the physical interpretation of $\rho'$ merits investigation.
The asymmetric decoherence relates to time-reversal asymmetry.
Fourth, the sum in Eq.~\eqref{eq:TA_decomp_coeff} relates to
a sum over trajectories,
a marginalization over intermediate-measurement outcomes.

\emph{Relationship between scrambling and
completeness of $\Basis$:}
The $\Set{ \frac{ \ketbra{ a }{ f } }{ \langle f | a \rangle } }$
in Sec.~\ref{section:KD_Coeffs} forms a basis for $\mathcal{D} ( \Hil )$.
But suppose that $\rho'  \neq  \rho$.
$\Basis$ fails to form a basis.

What does this failure imply about $\W(t)$ and $V$?
The failure is equivalent to the existence of a vanishing
$\xi  :=  | \langle w_3, \DegenW_{w_3} | U |
               v_2,  \DegenV_{v_2} \rangle |$.
Some $\xi$ vanishes if
some degenerate eigensubspace $\mathcal{H}_0$ of $\W(t)$
is a degenerate eigensubspace of $V$:
Every eigenspace of every Hermitian operator
has an orthogonal basis.
$\mathcal{H}_0$ therefore has an orthogonal basis.
One basis element can be labeled $U^\dag \ket{ w_3, \DegenW_{w_3} }$;
and the other, $\ket{ v_2,  \DegenV_{v_2} }$.

The sharing of an eigensubspace is equivalent to
the commutation of some component of $\W(t)$ with
some component of $V$.
The operators more likely commute
before the scrambling time $t_*$
than after.
Scrambling is therefore expected to magnify the similarity between
the OTOC quasiprobability $\OurKD{\rho}$
and the conventional KD distribution.

Let us illustrate with an extreme case.
Suppose that all the $\xi$'s lie as far from zero as possible:
\begin{align}
   \label{eq:MUB}
   \xi  =  \frac{1}{ \sqrt{ \Dim } }  \quad  \forall \xi \, .
\end{align}
Equation~\eqref{eq:MUB} implies that
$\mathcal{W}(t)$ and $V$ eigenbases are
\emph{mutually unbiased biases} (MUBs)~\cite{Durt_10_On}.
MUBs are eigenbases of operators
that maximize the lower bound
in an uncertainty relation~\cite{Coles_15_Entropic}.
If you prepare any eigenstate of one operator
(e.g., $U^\dag  \ket{ w_\ell,  \DegenW_{w_\ell} }$)
and measure the other operator (e.g., $V$),
all the possible outcomes have equal likelihoods.
You have no information with which to predict the outcome;
your ignorance is maximal.
$\W(t)$ and $V$ are maximally incompatible,
in the quantum-information (QI) sense of entropic uncertainty relations.
Consistency between this QI sense
of ``mutually incompatible'' and the OTOC sense
might be expected:
$\W(t)$ and $V$ eigenbases might be expected
to form MUBs after the scrambling time $t_*$.
We elaborate on this possibility in Sec.~\ref{section:TheoryOpps}.

KD quasiprobabilities are typically evaluated
on MUBs, such as position and momentum eigenbases~\cite{Lundeen_11_Direct,Lundeen_12_Procedure,Thekkadath_16_Direct}.
One therefore might expect $\OurKD{\rho}$
to relate more closely the KD quasiprobability
after $t_*$ than before.
The OTOC motivates a generalization of KD studies
beyond MUBs.

\emph{Application: Evaluating a state preparation's accuracy:}
Experimentalists wish to measure the OTOC $F(t)$
at each of many times $t$.
One may therefore wish to measure $\OurKD{\rho}$ after $t_*$.
Upon doing so, one may be able to infer not only $F(t)$,
but also the accuracy with which one prepared the target initial state.

Suppose that, after $t_*$, some $\Basis$
that forms a basis for $\Hil$.
Consider summing late-time $\OurKD{\rho}( . )$ values
over $( w_2,  \DegenW_{w_2} )$
and $( v_1, \DegenV_{v_1} )$.
The sum equals a KD quasiprobability for $\rho$.
The quasiprobability encodes all the information in $\rho$~\cite{Lundeen_11_Direct,Lundeen_12_Procedure}.
One can reconstruct the state that one prepared~\cite{Piacentini_16_Measuring,Suzuki_16_Observation,Thekkadath_16_Direct}.

The prepared state $\rho$ might differ from
the desired, or target, state $\rho_\target$.
Thermal states $e^{ - H / T } / Z$
are difficult to prepare, for example.
How accurately was $\rho_\target$ prepared?
One may answer by comparing $\rho_\target$
with the KD quasiprobability $\OurKD{\rho}$ for $\rho$.

Reconstructing the KD quasiprobability requires a trivial sum
over already-performed measurements
[Eq.~\eqref{eq:TA_decomp_coeff}].
One could reconstruct $\rho$ independently
via conventional quantum-state tomography~\cite{Paris_04_Q_State_Estimation}.
The $\rho$ reconstruction inferred from $\OurKD{\rho}$
may have lower precision,
due to the multiplicity of weak measurements
and to the sum.
But independent tomography would likely require extra measurements,
exponentially many in the system size.
Inferring $\OurKD{\rho}$ requires
exponentially many measurements, granted.\footnote{
One could measure, instead of $\OurKD{\rho}$,
the coarse-grained quasiprobability
$\SumKD{\rho}  =:  \sum_{\text{degeneracies}} \OurKD{\rho}$
(Sec.~\ref{section:ProjTrick}).
From $\SumKD{\rho}$, one could infer the OTOC.
Measuring $\SumKD{\rho}$ would require
exponentially fewer measurements.
But from $\SumKD{\rho}$, one could not infer the KD distribution.
One could infer a coarse-grained KD distribution, akin to
a block-diagonal matrix representation for $\rho$.}
But, from these measurements,
one can infer $\OurKD{\rho}$, the OTOC, and $\rho$.
Upon reconstructing the KD distribution for $\rho$,
one can recover a matrix representation for $\rho$
via an integral transform~\cite{Lundeen_12_Procedure}.

\emph{The asymmetrically decohered $\rho'$:}
What does the decomposed operator $\rho'$ signify?
$\rho'$ has the following properties:
The term subtracted off in Eq.~\eqref{eq:RhoPrime}
has trace zero.
Hence $\rho'$ has trace one, like a density operator.
But the subtracted-off term is not Hermitian.
Hence $\rho'$ is not Hermitian,
unlike a density operator.
Nor is $\rho'$ anti-Hermitian, necessarily unitarity,
or necessarily anti-unitary.

$\rho'$ plays none of the familiar roles---of
state, observable, or time-evolution operator---in quantum theory.
The physical significance of $\rho'$ is not clear.
Similar quantities appear in weak-measurement theory:
First, non-Hermitian products $\B \A$ of observables
have been measured weakly
(see Sec.~\ref{section:TA_retro} and~\cite{Piacentini_16_Measuring,Suzuki_16_Observation,Thekkadath_16_Direct}).
Second, nonsymmetrized correlation functions
characterize quantum detectors
of photon absorptions and emissions~\cite{Bednorz_13_Nonsymmetrized}.
Weak measurements imbue these examples with physical significance.
We might therefore expect $\rho'$ to have physical significance.
Additionally, since $\rho'$ is non-Hermitian,
non-Hermitian quantum mechanics
might offer insights~\cite{Moiseyev_11_Non}.

The subtraction in Eq.~\eqref{eq:RhoPrime}
constitutes a removal of coherences.
But the subtraction is not equivalent to a decohering channel~\cite{NielsenC10},
which outputs a density operator.
Hence our description of the decoherence as asymmetric.

The asymmetry relates to the breaking time-reversal invariance.
Let $U^\dag  \ket{  w_3,  \DegenW_{w_3}  }
=:  \ket{  \tilde{w}_3  }$
be fixed throughout the following argument
(be represented, relative to any given basis, by a fixed list of numbers).
Suppose that $\rho = e^{ - H / T } / Z$.
The removal of $\langle v_2,  \DegenV_{v_2}  |  \rho  |
\tilde{w}_3  \rangle$     terms from $\rho$
is equivalent to the removal of
$\langle v_2,  \DegenV_{v_2}  |  H  |  \tilde{w}_3  \rangle$
terms from $H$:
$\rho  \mapsto  \rho'  \;  \Leftrightarrow  \;  H  \mapsto  H'$.
Imagine, temporarily, that $H'$ could represent a Hamiltonian
without being Hermitian.
$H'$ would generate a time evolution under which
$\ket{ \tilde{w}_3 }$ could not evolve into
$\ket{ v_2,  \DegenV_{v_2} }$.
But $\ket{ v_2,  \DegenV_{v_2} }$ could evolve into
$\ket{ \tilde{w}_3 }$.
The forward process would be allowed;
the reverse would be forbidden.
Hence $\rho \mapsto \rho'$ relates to
a breaking of time-reversal symmetry.

\emph{Interpretation of the sum in Eq.~\eqref{eq:TA_decomp_coeff}:}
Summing $\OurKD{\rho} ( . )$ values, in Eq.~\eqref{eq:TA_decomp_coeff},
yields a decomposition coefficient $C$ of $\rho'$.
Imagine introducing that sum into Eq.~\eqref{eq:OTOC_retro3}.
The OTOC quasiprobability $\OurKD{\rho} ( . )$
would become a KD quasiprobability.
Consider applying this summed Eq.~\eqref{eq:OTOC_retro3}
in Eq.~\eqref{eq:OTOC_retro1}.
We would change from retrodicting about $V \W(t) V$
to retrodicting about the leftmost $V$.

\subsection{Relationship between out-of-time ordering
and quasiprobabilities}
\label{section:OTOC_TOC}

The OTOC has been shown to equal
a moment of the complex distribution $P(W, W')$~\cite{YungerHalpern_17_Jarzynski}.
This equality echoes Jarzynski's~\cite{Jarzynski_97_Nonequilibrium}.
Jarzynski's equality governs
out-of-equilibrium statistical mechanics.
Examples include a quantum oscillator
whose potential is dragged quickly~\cite{An_15_Experimental}.
With such nonequilibrium systems, one can associate
a difficult-to-measure, but useful,
free-energy difference $\Delta F$.
Jarzynski cast $\Delta F$ in terms of
the characteristic function $\expval{ e^{ - \beta W } }$
of a probability distribution $P(W)$.\footnote{
Let $P(W)$ denote a probability distribution
over a random variable $W$.
The characteristic function $\Charac (s)$
equals the Fourier transform:
$\Charac (s)  :=  \int dW  \;  e^{ i s W }$.
Defining $s$ as an imaginary-time variable,
$is \equiv - \beta$, yields $\expval{ e^{ - \beta W } }$.
Jarzynski's equality reads,
$\expval{ e^{ - \beta W } }  =  e^{ - \beta \Delta F }$.}
Similarly, the difficult-to-measure, but useful, OTOC $F(t)$
has been cast in terms of
the characteristic function $\expval{ e^{ - (\beta W + \beta' W') } }$
of the summed quasiprobability $P(W, W')$~\cite{YungerHalpern_17_Jarzynski}.

Jarzynski's classical probability must be replaced with a quasiprobability
because $[\W(t),  V]  =  0$.
This replacement appeals to intuition:
Noncommutation and quasiprobabilities reflect nonclassicality
as commuting operators and probabilities do not.
The OTOC registers quantum-information scrambling
unregistered by \emph{time-ordered correlators} (TOCs).
One might expect TOCs to equal
moments of coarse-grained quasiprobabilities
closer to probabilities than $\OurKD{\rho}$ is.

We prove this expectation.
First, we review the TOC $\TOC (t)$.
Then, we introduce the TOC analog $\Amp_\rho^\toc$ of
the probability amplitude $\Amp_\rho$ [Eq.~\eqref{eq:Amp}].
$\Amp_\rho$ encodes no time reversals, as expected.
Multiplying a forward amplitude $\Amp_\rho^\toc$
by a backward amplitude $\left( \Amp_\rho^\toc \right)^*$
yields the TOC quasiprobability $\TOCKD{\rho}$.
Inferring $\TOCKD{\rho}$ requires
only two weak measurements per trial.
$\TOCKD{\rho}$ reduces to a probability
if $\rho = \rho_{V}$ [Eq.~\eqref{eq:WRho}].
In contrast, under no known condition on $\rho$
do all $\OurKD{\rho}( . )$ values
reduce to probability values.
Summing $\TOCKD{\rho}$ under constraints
yields a complex distribution $P_\toc (W, W')$.
The TOC $\TOC (t)$ equals a moment of $P_\toc (W, W')$.

%
%
%
\subsubsection{Time-ordered correlator $\TOC(t)$}
\label{section:TOC_def}

The OTOC equals a term in
the expectation value $\expval{ . }$ of
the squared magnitude $| . |^2$ of a commutator $[ . \, , \: . ]$~\cite{Kitaev_15_Simple,Maldacena_15_Bound},
\begin{align}
   \label{eq:CommSquared}
   C(t)  & :=  \expval{ [ \W(t) ,  V ]^\dag  [ \W(t),  V ] } \\
   \nonumber \\ &
   =  - \expval{ \W^\dag (t)  V^\dag  V   \W (t)  }
   -  \expval{ V^\dag   \W^\dag (t)  \W(t)  V }
   \nonumber \\ &  \qquad
   +  2 \Re \LParen F(t) \RParen \, .
\end{align}
The second term is a time-ordered correlator (TOC),
\begin{align}
   \label{eq:TOC_def}
   \TOC (t)  :=   \expval{ V^\dag   \W^\dag (t)  \W(t)  V } \, .
\end{align}
The first term, $\expval{ \W^\dag (t)  V^\dag  V   \W (t)  }$,
exhibits similar physics.
Each term evaluates to one if $\W$ and $V$ are unitary.
If $\W$ and $V$ are nonunitary Hermitian operators,
the TOC reaches its equilibrium value by the dissipation time $t_d < t_*$
(Sec.~\ref{section:OTOC_review}).
The TOC fails to reflect scrambling,
which generates the OTOC's Lyapunov-type behavior
at $t \in (t_d,  \,  t_*)$.

\subsubsection{TOC probability amplitude $\Amp_\rho^\toc$}
\label{section:TOC_Amp}

We define
\begin{align}
   \label{eq:TOC_Amp}
   & \Amp_\rho^\toc  ( j ;  v_1,  \DegenV_{v_1} ;  w_1, \DegenW_{w_1} )
   \nonumber \\ &
   :=  \langle  w_1,  \DegenW_{w_1}  | U  |  v_1   \DegenV_{v_1}  \rangle
   \langle  v_1   \DegenV_{v_1}  |  j  \rangle \,
   \sqrt{ p_j }
\end{align}
as the \emph{TOC probability amplitude}.
$\Amp_\rho^\toc$ governs a quantum process $\ProtocolA^\toc$.
Figure~\ref{fig:TOC_amp1}, analogous to Fig.~\ref{fig:Protocoll_Trial1},
depicts $\ProtocolA^\toc$,
analogous to the $\ProtocolA$ in Sec.~\ref{section:Review_A}:
\begin{enumerate}[(1)]

   \item Prepare $\rho$.

   \item Measure the $\rho$ eigenbasis, $\Set{ \ketbra{j}{j} }$.

   \item Measure $\NondegV$.

   \item Evolve the system forward in time under $U$.

   \item Measure $\NondegW$.

\end{enumerate}
Equation~\eqref{eq:TOC_Amp} represents
the probability amplitude associated with
the measurements' yielding the outcomes
$j,  ( v_1,  \DegenV_{v_1} )$, and $( w_1,  \DegenW_{w_1} )$,
in that order.
All the measurements are strong.
$\ProtocolA^\toc$ is not a protocol for measuring $\Amp_\rho^\toc$.
Rather, $\ProtocolA^\toc$ facilitates
the physical interpretation of $\Amp_\rho^\toc$.

%
%
\begin{figure}[h]
\centering
\begin{subfigure}{0.4\textwidth}
\centering
\includegraphics[width=.9\textwidth]{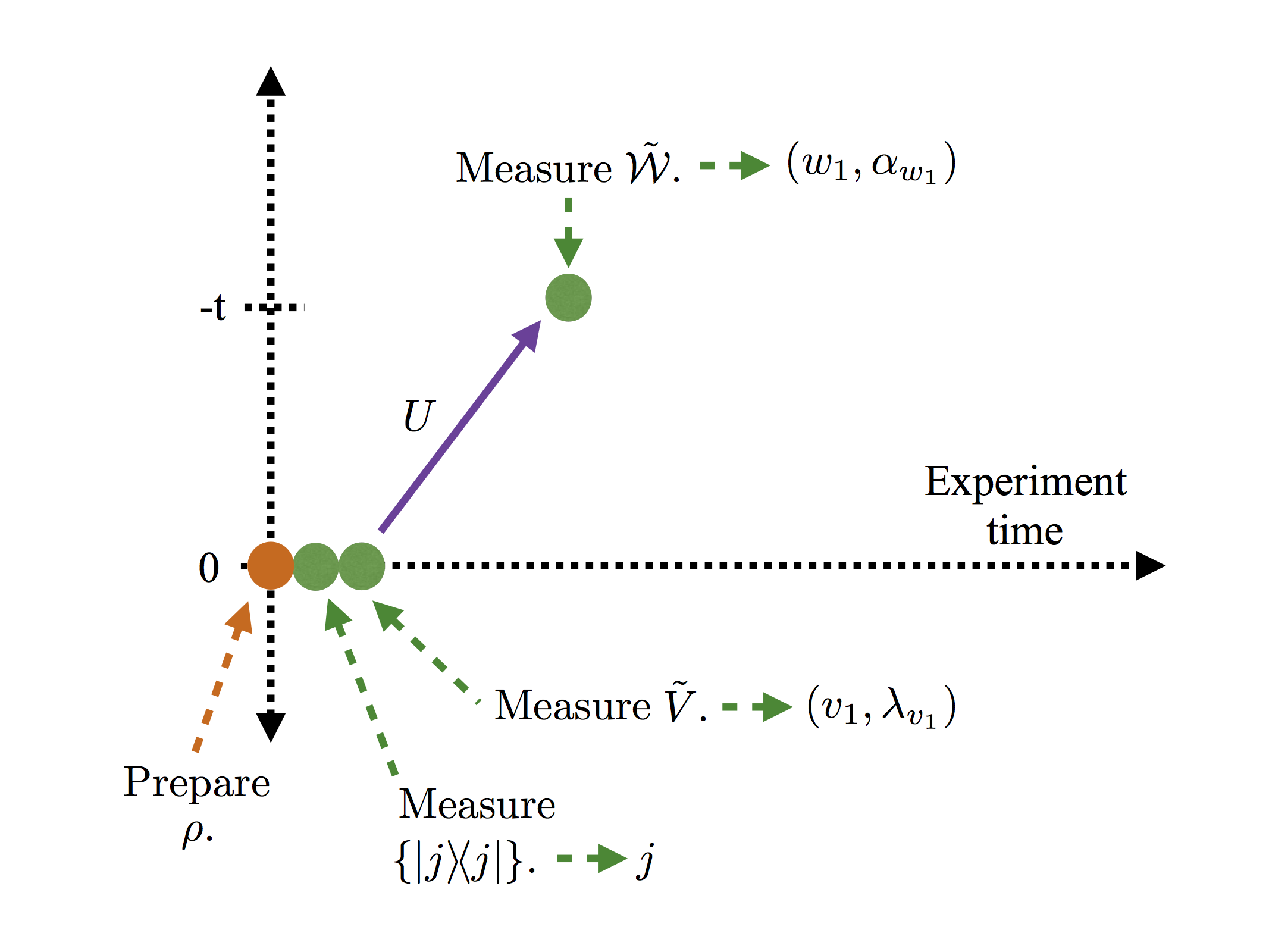}
\caption{}
\label{fig:TOC_amp1}
\end{subfigure}
\begin{subfigure}{.4\textwidth}
\centering
\includegraphics[width=.9\textwidth]{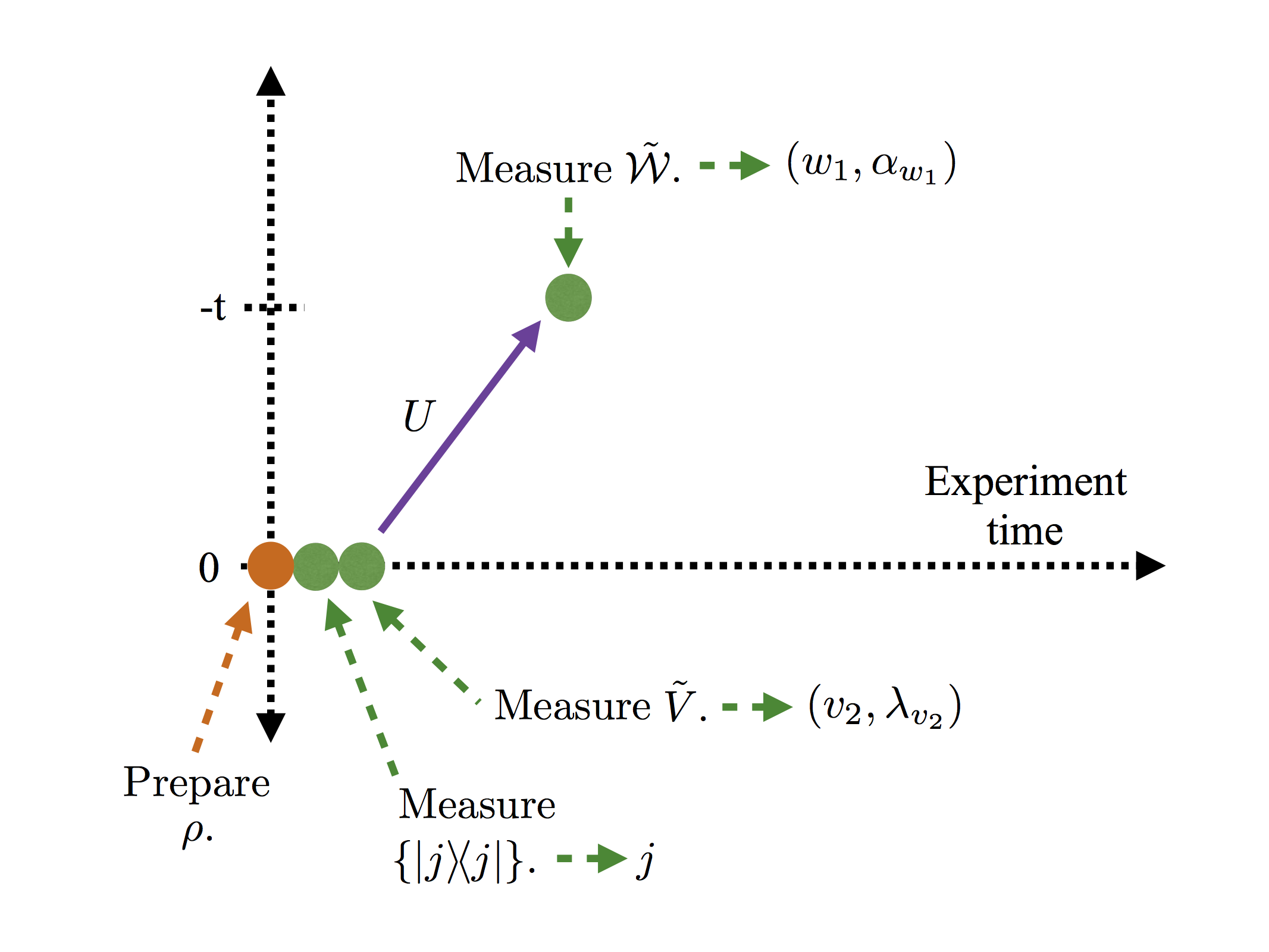}
\caption{}
\label{fig:TOC_amp2}
\end{subfigure}
\caption{\caphead{Quantum processes described by
the probability amplitudes $\Amp_\rho^\toc$
in the time-ordered correlator (TOC) $\TOC (t)$:}
$\TOC (t)$, like $F(t)$, equals a moment of
a summed quasiprobability (Theorem~\ref{theorem:TOC_Jarz}).
The quasiprobability, $\OurKD{\rho}^\toc$, equals
a sum of multiplied probability amplitudes $\Amp_\rho^\toc$
[Eq.~\eqref{eq:TOCKD_def}].
Each product contains two factors:
$\Amp_\rho^\toc ( j  ;  v_1,  \DegenV_{v_1}  ;  w_1,  \DegenW_{w_1} )$
denotes the probability amplitude
associated with the ``forward'' process in Fig.~\ref{fig:TOC_amp1}.
The system, $\Sys$, is prepared in a state $\rho$.
The $\rho$ eigenbasis $\Set{ \ketbra{j}{j} }$ is measured,
yielding outcome $j$.
$\NondegV$ is measured,
yielding outcome $( v_1,  \DegenV_{v_1} )$.
$\Sys$ is evolved forward in time under the unitary $U$.
$\NondegW$ is measured,
yielding outcome $( w_1,  \DegenW_{w_1} )$.
Along the abscissa runs the time
measured by a laboratory clock.
Along the ordinate runs the $t$ in $U := e^{ - i H t }$.
The second factor in each $\OurKD{\rho}^\toc$ product is
$\Amp_\rho^\toc ( j  ;  v_2,  \DegenV_{v_2}  ;  w_1,  \DegenW_{w_1} )^*$.
This factor relates to the process in Fig.~\ref{fig:TOC_amp2}.
The operations are those in Fig.~\ref{fig:TOC_amp1}.
The processes' initial measurements yield the same outcome.
So do the final measurements.
The middle outcomes might differ.
Complex-conjugating $\Amp_\rho^\toc$ yields
the probability amplitude associated with the \emph{reverse} process.
Figures~\ref{fig:TOC_amp1} and~\ref{fig:TOC_amp2}
depict no time reversals.
Each analogous OTOC figure
(Fig.~\ref{fig:Protocoll_Trial1} and Fig.~\ref{fig:Protocoll_Trial2}) depicts two.}
\label{fig:TOC_amps}
\end{figure}

$\ProtocolA^\toc$ results from eliminating, from $\ProtocolA$,
the initial $U$, $\NondegW$ measurement, and $U^\dag$.
$\Amp_\rho$ encodes two time reversals.
$\Amp_\rho^\toc$ encodes none,
as one might expect.

%
%
%
\subsubsection{TOC quasiprobability $\TOCKD{\rho}$}
\label{section:TOC_KD}

Consider a $\ProtocolA^\toc$ implementation that yields the outcomes
$j$, $( v_2,  \DegenV_{v_2} )$, and $( w_1,  \DegenW_{w_1} )$.
Such an implementation appears in Fig.~\ref{fig:TOC_amp2}.
The first and last outcomes [$j$ and $( w_1,  \DegenW_{w_1} )$]
equal those in Fig.~\ref{fig:TOC_amp1}, as in the OTOC case.
The middle outcome can differ.
This process corresponds to the probability amplitude
\begin{align}
   \label{eq:Amp_TOC_rev}
   & \Amp_\rho^\toc ( j ; v_2,  \DegenV_{v_2}  ;  w_1,  \DegenW_{w_1} )
   \nonumber \\ & \quad
   =  \langle w_1 , \DegenW_{w_1}  |  U  |  v_2,  \DegenV_{v_2}
   \rangle \langle   v_2,  \DegenV_{v_2}  |  j  \rangle  \:
   \sqrt{p_j} \, .
\end{align}
Complex conjugation reverses the inner products,
yielding the reverse process's amplitude.

We multiply this reverse amplitude by the forward amplitude~\eqref{eq:TOC_Amp}.
Summing over $j$ yields the \emph{TOC quasiprobability}:
\begin{align}
   \label{eq:TOCKD_def}
   & \TOCKD{\rho}
   ( v_1 ,  \DegenV_{v_1}  ;  w_1,  \DegenW_{w_1} ;  v_2 ,  \DegenV_{v_2}  )
   \nonumber \\ &
   :=  \sum_j
   \Amp_\rho^\toc ( j ; v_2,  \DegenV_{v_2}  ;  w_1,  \DegenW_{w_1} )^*
    \Amp_\rho^\toc  ( j ; v_1,  \DegenV_{v_1}  ;  w_1,  \DegenW_{w_1}  ) \\
   \label{eq:TOCKD_form}
   & =  \langle  v_2 ,  \DegenV_{v_2}  |  U^\dag  |  w_1,  \DegenW_{w_1}  \rangle
   \langle  w_1,  \DegenW_{w_1}  |  U  |  v_1 ,  \DegenV_{v_1}  
   \rangle 
   \nonumber \\ & \qquad \times
   \langle v_1 ,  \DegenV_{v_1} |  \rho  |   v_2 ,  \DegenV_{v_2}  \rangle  \, .
\end{align}

Like $\OurKD{\rho}$, $\TOCKD{\rho}$ is
an extended Kirkwood-Dirac quasiprobability.
$\TOCKD{\rho}$ is 2-extended, whereas $\OurKD{\rho}$ is 3-extended.
$\TOCKD{\rho}$ can be inferred from
a weak-measurement protocol $\Protocol^\toc$:
\begin{enumerate}[(1)]
   \item Prepare $\rho$.

   \item Measure $\NondegV$ weakly.

   \item Evolve the system forward under $U$.

   \item Measure $\NondegW$ weakly.

   \item Evolve the system backward under $U^\dag$.

   \item Measure $\NondegV$ strongly.

\end{enumerate}
$\Protocol^\toc$ requires just two weak measurements.
The weak-measurement protocol $\Protocol$
for inferring $\OurKD{\rho}$ requires three.
$\Protocol^\toc$ requires one time reversal;
$\Protocol$ requires two.

In a simple case, every $\Amp_\rho^\toc ( . )$ value
reduces to a probability value.
Suppose that $\rho$ shares the $\NondegV$ eigenbasis,
as in Eq.~\eqref{eq:WRho}.
The $( v_2 ,  \DegenV_{v_2} )$ in Eq.~\eqref{eq:TOCKD_form}
comes to equal $( v_1,  \DegenV_{v_1} )$;
Figures~\ref{fig:TOC_amp1} and~\ref{fig:TOC_amp2}
become identical.
Equation~\eqref{eq:TOCKD_form} reduces to
\begin{align}
   & \Amp_{ \rho_{V} }^\toc
   ( v_1,  \DegenV_{v_1}  ;  w_1,  \DegenW_{w_1} ;  v_2,  \DegenV_{v_2} ) \\
   & =  |  \langle  w_1,  \DegenW_{w_1}  |  U  |
             v_1,  \DegenV_{v_1}  \rangle  |^2  \,
   p_{ v_1,  \DegenV_{v_1} }  \,
   \delta_{ v_1  v_2 }  \,  \delta_{ \DegenV_{v_1}  \DegenV_{v_2} }  \\
   &  =  p (  w_1,  \DegenW_{w_1}  |  v_1,  \DegenV_{v_1} )  \,
   p_{ v_1,  \DegenV_{v_1} }  \,
  \delta_{ v_1  v_2 }  \,  \delta_{ \DegenV_{v_1}  \DegenV_{v_2} }  \\
   & =  p ( v_1,  \DegenV_{v_1} ;  w_1,  \DegenW_{w_1} )  \,
   \delta_{ v_1  v_2 }  \,  \delta_{ \DegenV_{v_1}  \DegenV_{v_2} }  \, .
\end{align}
The $p( a | b)$ denotes the conditional probability that,
if $b$ has occurred, $a$ will occur.
$p( a ; b )$ denotes the joint probability that $a$ and $b$ will occur.

All values $\OurKD{\rho_{ V } }^\toc ( . )$ of the TOC quasiprobability
have reduced to probability values.
Not all values of $\OurKD{ \rho_V }$ reduce:
The values associated with
$( v_2 ,  \DegenV_{v_2 } )  =  ( v_1 ,  \DegenV_{v_1} )$ or
$( w_3 ,  \DegenW_{w_3} )  =  ( w_2 ,  \DegenW_{w_2} )$
reduce to products of probabilities.
[See the analysis around Eq.~\eqref{eq:Reduce_to_p}.]
The OTOC quasiprobability encodes nonclassicality---violations
of the axioms of probability---more resilient than the TOC quasiprobability's.

\subsubsection{Complex TOC distribution $P_\toc (W_\toc, W'_\toc)$}
\label{section:P_TOC}

Let $W_\toc$ and $W'_\toc$ denote random variables
analogous to thermodynamic work.
We fix the constraints $W_\toc  =  w_1  v_2$ and
$W'_\toc  =  w_1  v_1$.
($w_1$ and $v_2$ need not be complex-conjugated because
they are real, as $\W$ and $V$ are Hermitian.)
Multiple outcome sextuples
$( v_2,  \DegenV_{v_2} ; w_1,  \DegenW_{w_1} ; v_1,  \DegenV_{v_1} )$
satisfy these constraints.
Each sextuple corresponds to a quasiprobability
$\TOCKD{\rho} ( . )$.
We sum the quasiprobabilities that satisfy the constraints:
\begin{align}
   \label{eq:P_TOC}
   & P_\toc ( W_\toc, W'_\toc )  :=
   \sum_{ ( v_1,  \DegenV_{v_1} ) ,  ( w_1,  \DegenW_{w_1} ) ,
                ( v_2,  \DegenV_{v_2} ) }
   \nonumber \\ &  \times
   \OurKD{\rho}^\toc ( v_1,  \DegenV_{v_1}  ;  w_1,  \DegenW_{w_1} ;
                                   v_2,  \DegenV_{v_2} )  \,
   \delta_{ W ( w_1^* v_2^* ) }  \,  \delta_{W' ( w_1 v_1 ) }  \, .
\end{align}

$P_\toc$ forms a complex distribution.
Let $f$ denote any function of $W_\toc$ and $W'_\toc$.
The $P_\toc$ average of $f$ is
\begin{align}
   \label{eq:TOC_avg}
   & \expval{ f ( W_\toc ,  W'_\toc ) }
   \\ \nonumber &
   :=  \sum_{ W_\toc ,  W'_\toc }  f ( W_\toc ,  W'_\toc )  P_\toc ( W_\toc, W'_\toc )  \, .
\end{align}

\subsubsection{TOC as a moment of the complex distribution}
\label{eq:TOC_Jarz}

The TOC obeys an equality
analogous to Eq.~(11) in~\cite{YungerHalpern_17_Jarzynski}.
%
%
\begin{theorem}[Jarzynski-like theorem for the TOC]
   \label{theorem:TOC_Jarz}
   The time-ordered correlator~\eqref{eq:TOC_def}
   equals a moment of the complex distribution~\eqref{eq:P_TOC}:
   \begin{align}
      \label{eq:TOC_Jarz}
      \TOC (t)  =  \frac{ \partial^2 }{ \partial \beta \, \partial \beta' }
      \expval{ e^{ - ( \beta W_\toc  +  \beta' W'_\toc ) } }
      \Bigg\rvert_{ \beta, \beta' = 0 }   ,
   \end{align}
   wherein $\beta, \beta' \in \mathbb{R}$.
\end{theorem}
%
%
\begin{proof}
The proof is analogous to the proof of Theorem~1 in~\cite{YungerHalpern_17_Jarzynski}.
\end{proof}
\noindent 
Equation~\eqref{eq:TOC_Jarz} can be recast as
$\TOC (t)  =  \expval{ W_\toc  W'_\toc }  \, ,$
along the lines of Eq.~\eqref{eq:RecoverF2}.

\subsection{Higher-order OTOCs as moments of
longer (summed) quasiprobabilities}
\label{section:HigherOTOCs}

Differentiating a characteristic function again and again
yields higher- and higher-point correlation functions.
So does differentiating $P(W, W')$ again and again.
But each resulting correlator encodes
just $\Ops = 3$ time reversals.
Let $\Opsb  =  \frac{1}{2} ( \Ops + 1 )  =  2, 3, \ldots$,
for $\Ops = 3, 5, \ldots$
A \emph{$\Opsb$-fold OTOC} has been defined~\cite{Roberts_16_Chaos,Hael_17_Classification}:
\begin{align}
   \label{eq:k_OTOC}
   F^\ParenKB (t)  :=
   \langle \underbrace{ \W(t)  V  \ldots  \W(t) V }_{2 \Opsb } \rangle
   \equiv  \Tr \LParen  \rho
   \underbrace{ \W(t)  V  \ldots  \W(t) V }_{2 \Opsb }  \RParen \, .
\end{align}
Each such correlation function contains $\Opsb$
Heisenberg-picture operators $\W(t)$
interleaved with $\Opsb$ time-0 operators $V$.
$F^\ParenKB (t)$ encodes $2 \Opsb - 1  =  \Ops$ time reversals,
illustrated in Fig.~\ref{fig:k_OTOC}.
We focus on Hermitian $\W$ and $V$,
as in~\cite{Maldacena_15_Bound,HosurYoshida_16_Chaos},
for simplicity.

The conventional OTOC corresponds to $\Ops = 3$ and $\Opsb = 2$:
$F(t)  =  F^\2 (t)$.
If $\Ops < 3$, $F^\ParenKB (t)$ is not OTO.

%
%
\begin{figure}[h]
\centering
\includegraphics[width=.45\textwidth]{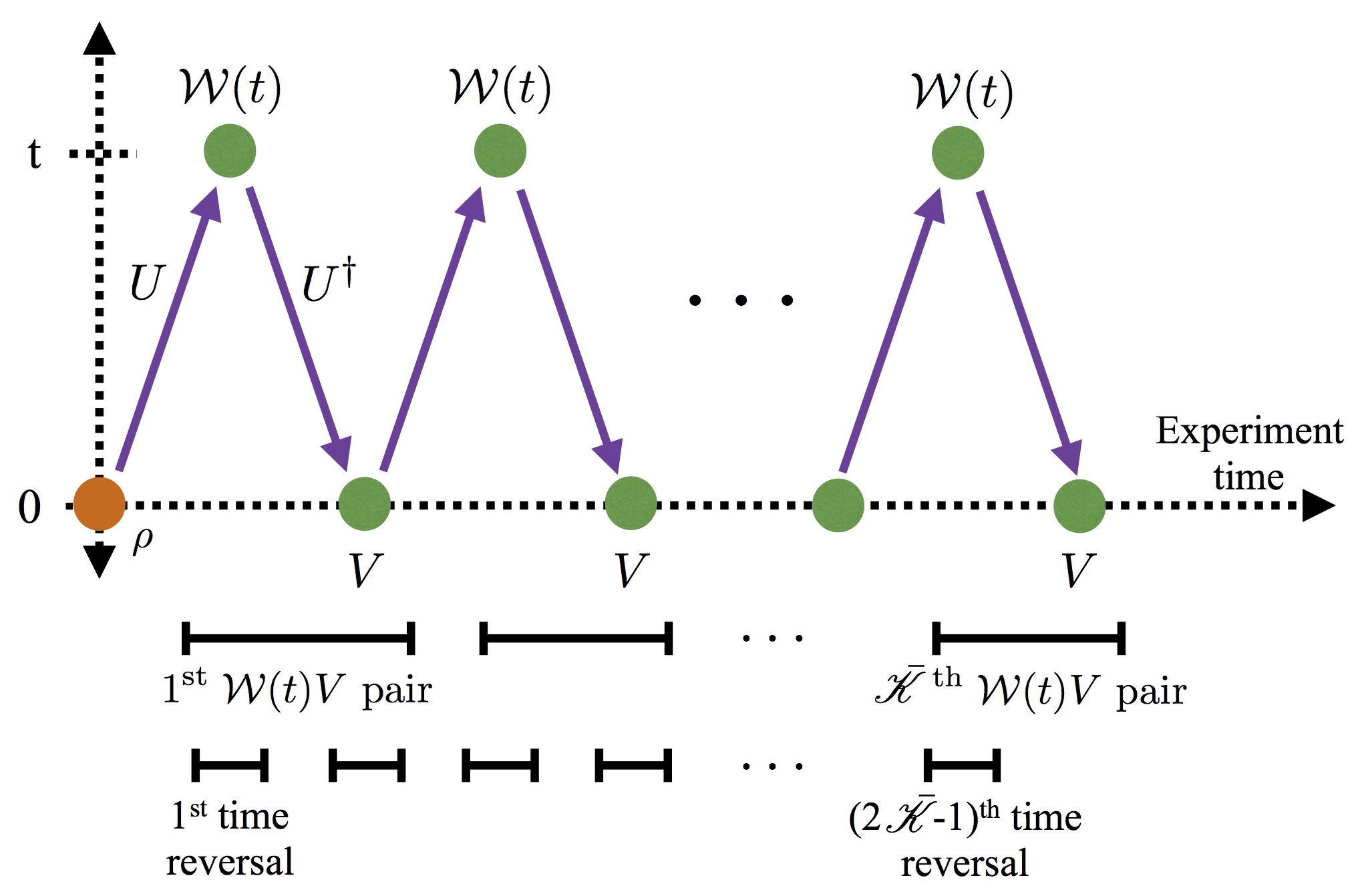}
\caption{\caphead{$\Opsb$-fold out-of-time-ordered correlator (OTOC):}
The conventional OTOC [Eq.~\eqref{eq:OTOC_Def}],
encodes just three time reversals.
The \emph{$\Opsb$-fold OTOC} $F^\ParenKB (t)$ encodes
$2 \Opsb - 1  =  \Ops   =  3, 5, \ldots$ time reversals.
The time measured by a laboratory clock
runs along the abscissa.
The ordinate represents the time parameter $t$,
which may be inverted in experiments.
The orange, leftmost dot represents
the state preparation $\rho$.
Each green dot represents a $\W(t)$ or a $V$.
Each purple line represents a unitary time evolution.
The diagram, scanned from left to right,
represents $F^\ParenKB (t)$, scanned from left to right.
}
\label{fig:k_OTOC}
\end{figure}

The greater the $\Ops$, the longer the distribution $P^\ParenK$
of which $F^\ParenKB (t)$ equals a moment.
We define $P^\ParenK$ in three steps:
We recall the $\Ops$-extended quasiprobability $\OurKD{\rho}^\ParenK$
[Eq.~\eqref{eq:Extend_KD}].
We introduce measurable random variables
$W_\ell$ and $W'_{\ell'}$.
These variables participate in constraints
on sums of $\OurKD{\rho}^\ParenK ( . )$ values.

Let us evaluate Eq.~\eqref{eq:Extend_KD}
on particular arguments:
\begin{align}
   \label{eq:k_OTO_quasi}
   & \OurKD{\rho}^\ParenK (
   v_1,  \DegenV_{v_1}  ;  w_2 ,  \DegenW_{w_2} ;
   \ldots ; v_{\Opsb} ,  \DegenV_{ v_{\Opsb} } ;
   w_{\Opsb + 1},  \DegenW_{ w_{\Opsb + 1} } )
   \nonumber \\ &
   =  \langle  w_{\Opsb + 1},  \DegenW_{w_{\Opsb + 1}}  |  U  |
                    v_{\Opsb}, \DegenV_{v_{\Opsb}}  \rangle
   \langle  v_{\Opsb}, \DegenV_{v_{\Opsb}}  |  U^\dag  |
               w_{\Opsb},  \DegenW_{w_{\Opsb}}  \rangle
   \nonumber \\ &  \times \ldots  \times
   \langle w_2,  \DegenW_{w_2}  |  U  |
               v_1,  \DegenV_{v_1}  \rangle
   \langle  v_1,  \DegenV_{v_1}  |  \rho  U^\dag  |
               w_{\Opsb + 1},  \DegenW_{w_{\Opsb+1}}  \rangle  \, .
\end{align}
One can infer $\OurKD{ \rho }^\ParenK$
from the interferometry scheme in~\cite{YungerHalpern_17_Jarzynski}
and from weak measurements.
Upon implementing one batch of the interferometry trials,
one can infer $\OurKD{\rho}^\ParenK$ for all $\Ops$-values:
One has measured all the inner products $\langle a | \U | b \rangle$.
Multiplying together arbitrarily many inner products yields
an arbitrarily high-$\Ops$ quasiprobability.
Having inferred some $\OurKD{\rho}^\ParenK$,
one need not perform new experiments
to infer $\OurKD{\rho}^{( \Ops + 2 )}$.
To infer $\OurKD{\rho}^\ParenK$ from weak measurements,
one first prepares $\rho$.
One performs $\Ops  =  2 \Opsb - 1$ weak measurements
interspersed with unitaries.
(One measures $\NondegV$ weakly,
evolves with $U$, measures $\NondegW$ weakly,
evolves with $U^\dag$, etc.)
Finally, one measures $\NondegW$ strongly.
The strong measurement corresponds to
the anomalous index $\Opsb + 1$ in
$( w_{\Opsb + 1},    \DegenW_{w_{\Opsb + 1}} )$.

We define $2 \Opsb$ random variables
\begin{align}
   \label{eq:W_ell}
   & W_\ell  \in \{ w_\ell \}   \qquad  \forall \ell =  2, 3, \ldots, \Opsb +1
   \qquad \text{and} \\
   & W'_{\ell'}  \in  \{ v_{\ell'} \}  \qquad \forall  \ell'  =  1, 2, \ldots, \Opsb \, .
\end{align}
Consider fixing the values of the $W_\ell$'s and the $W'_{\ell'}$'s.
Certain quasiprobability values $\OurKD{\rho}^\ParenK ( . )$
satisfy the constraints $W_\ell = w_\ell$ and $W'_{\ell'}  =  v_{\ell'}$
for all $\ell$ and $\ell'$.
Summing these quasiprobability values yields
\begin{align}
   \label{eq:P_k}
   & P^\ParenK ( W_2,  W_3,  \ldots,  W_{\Opsb +1},
   W'_1,  W'_2,  \ldots,  W'_{\Opsb} )
   \\ \nonumber &
   :=  \sum_{ W_2,  W_3,  \ldots,  W_{\Opsb +1} }
   \sum_{ W'_1,  W'_2,  \ldots,  W'_{\Opsb} } \\ &
   \OurKD{\rho}^\ParenK (
   v_1,  \DegenV_{v_1}  ;  w_2 ,  \DegenW_{w_2} ;
   \ldots ; v_{\Opsb} ,  \DegenV_{v_{\Opsb}} ;  w_{\Opsb+1} ,  \DegenW_{w_{\Opsb+1}  } )
   \nonumber \\ &  \nonumber  \times
   \left(  \delta_{ W_2 w_2 }  \times  \ldots
            \times  \delta_{ W_{\Opsb+1} w_{\Opsb+1} }  \right)
   \left(  \delta_{ W'_1  v_1 }  \times  \ldots
            \times  \delta_{ W'_{\Opsb}  v_{\Opsb} }  \right)    .
\end{align}


\begin{theorem}[The $\Opsb$-fold OTOC as a moment]
\label{theorem:k_OTOC}
The $\Opsb$-fold OTOC equals a $2\Opsb^\th$ moment of
the complex distribution~\eqref{eq:P_k}:
\begin{align}
   \label{eq:k_OTOC_thm}
   & F^\ParenKB (t)  =
   \frac{ \partial^{2\Opsb} }{
            \partial \beta_2  \ldots  \partial \beta_{\Opsb + 1 }  \,
            \partial \beta'_1  \ldots  \partial \beta'_{\Opsb} }
   \nonumber \\ &
   \expval{ \exp \left(  -  \left[
                 \sum_{ \ell = 2}^{\Opsb + 1 }  \beta_\ell  W_\ell
                 +   \sum_{ \ell' = 1 }^{\Opsb}  \beta'_{\ell'}  W'_{\ell'}
                 \right]  \right)  }
   \Bigg\lvert_{ \beta_\ell, \beta'_{\ell'}   =  0  \;  \forall \ell, \ell' }  \, ,
\end{align}
wherein $\beta_\ell, \beta'_\ell  \in  \mathbb{R}$.
\end{theorem}

\begin{proof}
The proof proceeds in analogy with
the proof of Theorem~1 in~\cite{YungerHalpern_17_Jarzynski}.
   %
   %
\end{proof}

The greater the $\Ops$, the ``longer''
the quasiprobability $\OurKD{\rho}^\ParenK$.
The more weak measurements are required to infer
$\OurKD{\rho}^\ParenK$.
Differentiating $\OurKD{\rho}^\ParenK$ more
does not raise the number of time reversals encoded in the correlator.

Equation~\eqref{eq:k_OTOC_thm} can be recast as
$F^\ParenKB (t)  =  \expval{ 
\left(  \prod_{ \ell = 2}^{\Opsb + 1 }  W_\ell  \right)
\left(  \prod_{ \ell' = 1 }^{\Opsb}  W'_{\ell'}   \right)
}  \, ,$
along the lines of Eq.~\eqref{eq:RecoverF2}.

\section{Outlook}
\label{section:Outlook}

We have characterized the quasiprobability $\OurKD{\rho}$
that ``lies behind'' the OTOC $F(t)$.
$\OurKD{\rho}$, we have argued, is an extension of
the Kirkwood-Dirac distribution used in quantum optics.
We have analyzed and simplified measurement protocols for $\OurKD{\rho}$,
calculated $\OurKD{\rho}$ numerically and on average over Brownian circuits,
and investigated mathematical properties.
This work redounds upon quantum chaos,
quasiprobability theory, and weak-measurement physics.
As the OTOC equals a combination of $\OurKD{\rho}( . )$ values,
$\OurKD{\rho}$ provides more-fundamental information about scrambling.
The OTOC motivates generalizations of,
and fundamental questions about, KD theory.
The OTOC also suggests a new application
of sequential weak measurements.

At this intersection of fields lie many opportunities.
We classify the opportunities by the tools
that inspired them: experiments, calculations, and abstract theory.

\subsection{Experimental opportunities}

We expect the weak-measurement scheme for $\OurKD{\rho}$ and $F(t)$
to be realizable in the immediate future.
Candidate platforms include
superconducting qubits, trapped ions, ultracold atoms, 
and perhaps NMR.
Experimentalists have developed key tools required to implement the protocol~\cite{Bollen_10_Direct,Lundeen_11_Direct,Lundeen_12_Procedure,Bamber_14_Observing,Mirhosseini_14_Compressive,White_16_Preserving,Hacohen_16_Quantum,Browaeys_16_Experimental,Piacentini_16_Measuring,Suzuki_16_Observation,Thekkadath_16_Direct}.


Achievable control and dissipation must be compared with
the conditions needed to infer the OTOC.
Errors might be mitigated with
tools under investigation~\cite{Swingle_Resilience}.

\subsection{Opportunities motivated by calculations}

Numerical simulations and analytical calculations
point to three opportunities.

Physical models' OTOC quasiprobabilities may be evaluated.
The Sachdev-Ye-Kitaev model, for example, scrambles quickly~\cite{Sachdev_93_Gapless,Kitaev_15_Simple}.
The quasiprobability's functional form
may suggest new insights into chaos.
Our Brownian-circuit calculation (Sec.~\ref{section:Brownian}),
while a first step, involves averages over unitaries.
Summing quasiprobabilities can cause interference
to dampen nonclassical behaviors~\cite{Dressel_15_Weak}.
Additionally, while unitary averages model chaotic evolution,
explicit Hamiltonian evolution might provide different insights.
Explicit Hamiltonian evolution would also preclude the need
to calculate higher moments of the quasiprobability.

In some numerical plots, the real part $\Re ( \SumKD{\rho} )$ bifurcates.
These bifurcations resemble classical-chaos pitchforks~\cite{Strogatz_00_Non}.
Classical-chaos plots bifurcate when
a differential equation's equilibrium point
branches into three.
The OTOC quasiprobability $\OurKD{\rho}$ might be recast
in terms of equilibria.
Such a recasting would strengthen the parallel between
classical chaos and the OTOC.

Finally, the Brownian-circuit calculation has untied threads.
We calculated only the first moment of $\SumKD{\rho}$.
Higher moments may encode physics
less visible in $F(t)$.
Also, evaluating certain components of $\SumKD{\rho}$
requires new calculational tools.
These tools merit development,
then application to $\SumKD{\rho}$.
An example opportunity is discussed after Eq.~\eqref{eq:Brown_help}.

\subsection{Fundamental-theory opportunities}
\label{section:TheoryOpps}

Seven opportunities concern
the mathematical properties
and physical interpretations of $\OurKD{\rho}$.

The KD quasiprobability prompts the question,
``Is the OTOC definition of `maximal noncommutation' consistent with
the mutually-unbiased-bases definition?''
Recall Sec.~\ref{section:TA_Coeffs}:
We decomposed an operator $\rho'$
in terms of a set $\Basis  =  \Set{
\frac{ \ketbra{ a }{ f } }{  \langle f | a \rangle } }_{
\langle f | a \rangle  \neq 0 }$
of operators.
In the KD-quasiprobability literature, the bases
$\Basis_a =  \Set{ \ket{a} }$ and $\Basis_f  =  \Set{ \ket{f} }$
tend to be mutually unbiased (MU):
$| \langle f | a \rangle |  =  \frac{1}{ \sqrt{ \Dim } }  \;  \forall a, f$.
Let $\A$ and $\B$ denote operators that have MU eigenbases.
Substituting $\A$ and $\B$ into an uncertainty relation
maximizes the lower bound on an uncertainty~\cite{Coles_15_Entropic}.
In this quantum-information (QI) sense,
$\A$ and $\B$ noncommute maximally.

In Sec.~\ref{section:TA_Coeffs},
$\Basis_a  =  \Set{ \ket{v_2,  \DegenV_{v_2} } }$, and
$\Basis_f  =  \Set{  U^\dag \ket{ w_3, \DegenW_{w_3} } }$.
These $\Basis$'s are eigenbases of $V$ and $\W(t)$.
When do we expect these eigenbases to be MU,
as in the KD-quasiprobability literature?
After the scrambling time $t_*$---after $F(t)$ decays to zero---when
$\W(t)$ and $V$ noncommute maximally in the OTOC sense.

The OTOC provides one definition of ``maximal noncommutation.''
MUBs provide a QI definition.
To what extent do these definitions overlap?
Initial results show that, in some cases,
the distribution over possible values of
$| \langle v_2,  \DegenV_{v_2} | U | w_3, \DegenW_{w_3} \rangle |$
peaks at $\frac{1}{ \sqrt{ \Dim } }$.
But the distribution approaches this form before $t_*$.
Also, the distribution's width seems constant in $\Dim$.
Further study is required.
The overlap between OTOC and two QI definitions of scrambling
have been explored already:
(1) When the OTOC is small, a tripartite information is negative~\cite{HosurYoshida_16_Chaos}.
(2) An OTOC-like function is proportional to
a \emph{frame potential} that quantifies pseudorandomness~\cite{Roberts_16_Chaos}.
The relationship between the OTOC
and a third QI sense of incompatibility---MUBs and entropic uncertainty relations---merits investigation.

Second, $\OurKD{\rho}$ effectively has four arguments, apart from $\rho$
(Sec.~\ref{section:TA_Props}).
The KD quasiprobability has two.
This doubling of indices parallels the Choi-Jamiolkowski (CJ) representation
of quantum channels~\cite{Preskill_15_Ch3}.
Hosur \emph{et al.} have, using the CJ representation,
linked $F(t)$ to the tripartite information~\cite{HosurYoshida_16_Chaos}.
The extended KD distribution might be linked
to information-theoretic quantities similarly.

Third, our $P(W, W')$ and weak-measurement protocol
resemble analogs in~\cite{Solinas_15_Full,Solinas_16_Probing}.
\{See~\cite{Alonso_16_Thermodynamics,Miller_16_Time,Elouard_17_Role}
for frameworks similar to Solinas and Gasparinetti's (S\&G's).\}
Yet~\cite{Solinas_15_Full,Solinas_16_Probing}
concern quantum thermodynamics, not the OTOC.
The similarity between the quasiprobabilities in~\cite{Solinas_15_Full,Solinas_16_Probing}
and those in~\cite{YungerHalpern_17_Jarzynski},
their weak-measurement protocol and ours,
and the thermodynamic agendas in~\cite{Solinas_15_Full,Solinas_16_Probing}
and~\cite{YungerHalpern_17_Jarzynski}
suggest a connection between the projects~\cite{Jordan_chat,Solinas_chat}.
The connection merits investigation and might yield new insights.
For instance, S\&G calculate
the heat dissipated by an open quantum system
that absorbs work~\cite[Sec. IV]{Solinas_15_Full}.
OTOC theory focuses on closed systems.
Yet experimental systems are open.
Dissipation endangers measurements of $F(t)$.
Solinas and Gasparinetti's toolkit might facilitate predictions about,
and expose interesting physics in, open-system OTOCs.

Fourth, $W$ and $W'$ suggest understudies for work
in quantum thermodynamics.
Thermodynamics sprouted during the 1800s,
alongside steam engines and factories.
How much work a system could output---how
much ``orderly'' energy one could reliably draw---held practical importance.
Today's experimentalists draw energy from power plants.
Quantifying work may be less critical
than it was 150 years ago.
What can replace work in the today's growing incarnation of thermodynamics,
quantum thermodynamics?
Coherence relative to the energy eigenbasis is being quantified~\cite{Lostaglio_15_Description,Narasimhachar_15_Low}.
The OTOC suggests alternatives:
$W$ and $W'$ are random variables, analogous to work,
natural to quantum-information scrambling.
The potential roles of $W$ and $W'$ within quantum thermodynamics
merit exploration.

Fifth, relationships amongst three ideas were identified recently:
\begin{enumerate}[(1)]
   \item
We have linked quasiprobabilities with the OTOC,
following~\cite{YungerHalpern_17_Jarzynski}.
   \item
Aleiner \emph{et al.}~\cite{Aleiner_16_Microscopic}
and Haehl \emph{et al.}~\cite{Haehl_16_Schwinger_I,Haehl_16_Schwinger_II}
have linked the OTOC with Schwinger-Keldysh path integrals.
   \item
Hofer has linked Schwinger-Keldysh path integrals with quasiprobabilities~\cite{Hofer_17_Quasi}.
\end{enumerate}
The three ideas---quasiprobabilities, the OTOC, and Schwinger-Keldysh path integrals---form the nodes of the triangle in Fig.~\ref{fig:OTOC_path_quasi}.
The triangle's legs were discovered recently;
their joinings can be probed further.
For example, Hofer focuses on single-timefold path integrals.
OTOC path integrals contain multiple timefolds~\cite{Aleiner_16_Microscopic,Haehl_16_Schwinger_I,Haehl_16_Schwinger_II}.
Just as Hofer's quasiprobabilities involve fewer timefolds
than the OTOC quasiprobability $\OurKD{\rho}$,
the TOC quasiprobability $\TOCKD{\rho}$~\eqref{eq:TOCKD_def}
can be inferred from fewer weak measurements than $\OurKD{\rho}$ can.
One might expect Hofer's quasiprobabilities to relate to
$\TOCKD{\rho}$.
Kindred works, linking quasiprobabilities with out-of-time ordering, include~\cite{Manko_00_Lyapunov,Bednorz_13_Nonsymmetrized,Oehri_16_Time,Hofer_17_Quasi,Lee_17_On}.

%
%
\begin{figure}[h]
\centering
\includegraphics[width=.45\textwidth]{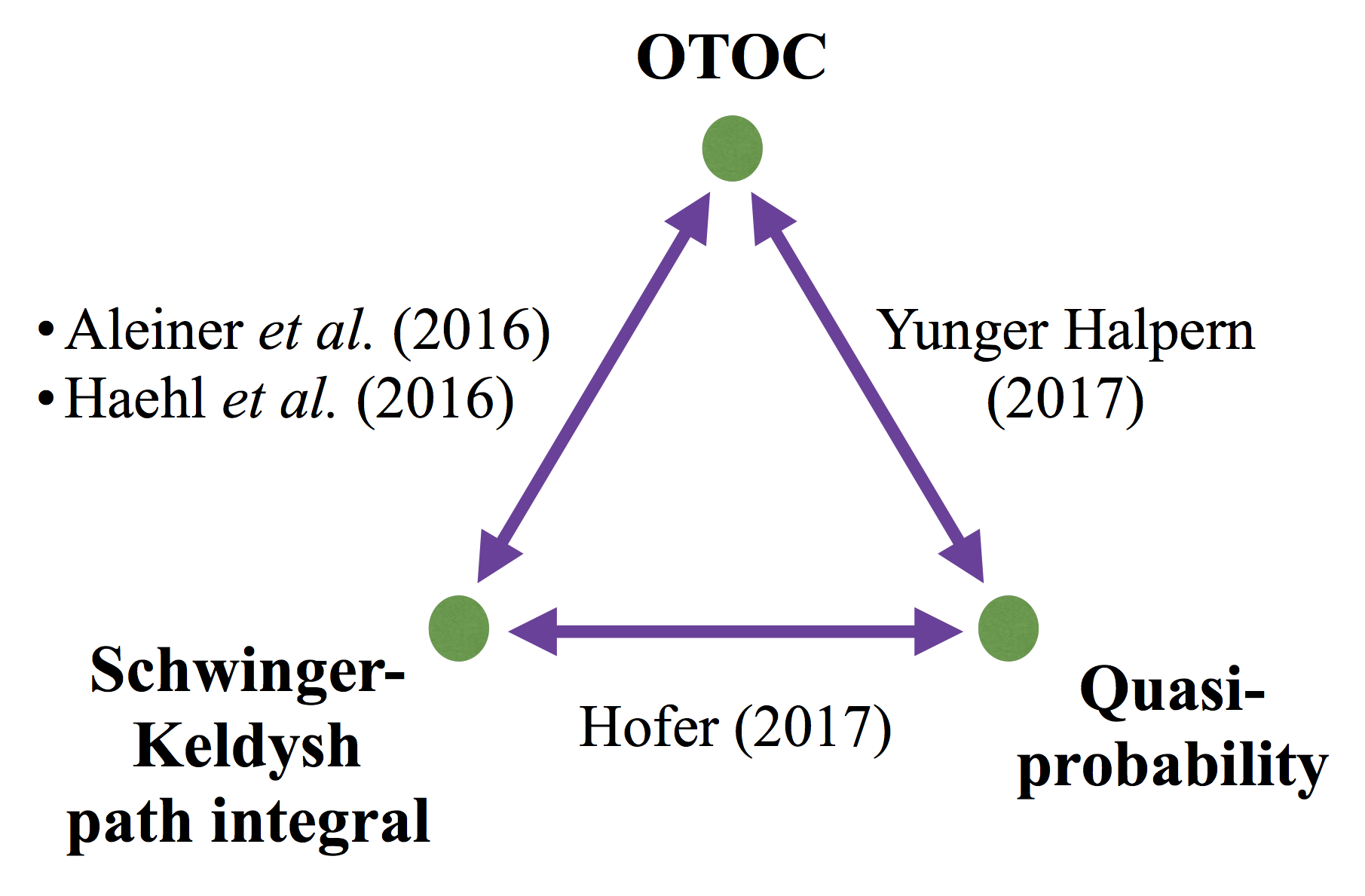}
\caption{\caphead{Three interrelated ideas:}
Relationships amongst the out-of-time-ordered correlator, quasiprobabilities,
and Schwinger-Keldysh path integrals
were articulated recently.}
\label{fig:OTOC_path_quasi}
\end{figure}

Sixth, the OTOC equals a moment of
the complex distribution $P( W, W')$~\cite{YungerHalpern_17_Jarzynski}.
The OTOC has been bounded with general-relativity
and Lieb-Robinson tools~\cite{Maldacena_15_Bound,Lashkari_13_Towards}.
A more information-theoretic bound might follow
from the Jarzynski-like equality in~\cite{YungerHalpern_17_Jarzynski}.

Finally, the KD distribution consists of the coefficients
in a decomposition of a quantum state
$\rho  \in  \mathcal{D} ( \mathcal{H} )$~\cite{Lundeen_11_Direct,Lundeen_12_Procedure}
(Sec.~\ref{section:KD_Coeffs}).
$\rho$ is decomposed in terms of a set
$\Basis  :=  \Set{  \frac{ \ketbra{ a }{ f } }{ \langle f | a \rangle }  }$
of operators.
$\Basis$ forms a basis for $\Hil$
only if $\langle f | a \rangle  \neq  0  \;  \forall a, f$.
The inner product has been nonzero in experiments,
because $\Set{ \ket{ a } }$ and $\Set{ \ket{f} }$
are chosen to be mutually unbiased bases (MUBs):
They are eigenbases of ``maximally noncommuting'' observables.
The OTOC, evaluated before the scrambling time $t = t_*$,
motivates a generalization beyond MUBs.
What if, $F(t)$ prompts us to ask,
$\langle f | a \rangle  =  0$ for some $a, f$
(Sec.~\ref{section:TA_Coeffs})?
The decomposition comes to be of
an ``asymmetrically decohered'' $\rho'$.
This decoherence's physical significance
merits investigation.
The asymmetry appears related to time irreversibility.
Tools from non-Hermitian quantum mechanics
might offer insight~\cite{Moiseyev_11_Non}.

\endgroup

%
%

\putbib[OTOC_Quasi_bib] 
\end{bibunit}

%
%
\chapter{MBL-Mobile: Many-body-localized engine}
\label{ch:MBL_Mobile}
\begin{bibunit}

\noindent \emph{This chapter appeared, in an earlier form, in~\cite{NYH_17_MBL}.}



\begingroup

\newcommand{\JAvg}{ \langle J \rangle } 
\newcommand{\dTyp}{ \delta_{\text{typ}} } 
\newcommand{\VeryDeloc}{ {\text{v. \; deloc}} } 
\newcommand{\Deloc}{ {\text{deloc}} } 
\newcommand{\VeryLoc}{<} 
\newcommand{\Loc}{>} 
\newcommand{\HDim}{\mathcal{N}} 
\newcommand{\Sites}{N} 
\newcommand{\SitesTot}{ \Sites_\tot }
\newcommand{\NCross}{ n_{\text{cross}} }
\newcommand{\Cross}{{\text{Cross}}}
\newcommand{\Poisson}{ {\text{Poisson}} }
\newcommand{\etaLarge}{ \eta_{ {\text{large}} } }
\newcommand{\etaSmall}{ \eta_{ {\text{small}} } }
\newcommand{\high}{ {\text{high}} } 
\newcommand{\low}{ {\text{low}} } 
\newcommand{\Power}{ \mathscr{P} } 
\newcommand{\In}{ {\text{in}} } 
\newcommand{\zd}{ {\text{ZD}} } 
\newcommand{\diab}{{\text{diab}}}
\newcommand{\adiab}{{\text{adiab}}}
\newcommand{\loc}{ {\text{loc}} }
\newcommand{\MBL}{{\text{MBL}}}
\newcommand{\ETH}{{\text{GOE}}}
\newcommand{\Anderson}{ {\text{And}} }
\newcommand{\class}{ {\text{class}} }
\newcommand{\shallow}{ {\text{shallow}} }
\newcommand{\disorder}{ {\text{disorder}} }
\newcommand{\particle}{ {\text{particle}} }
\newcommand{\Otto}{{\text{Otto}}}
\newcommand{\QHO}{{\text{QHO}}}
\newcommand{\ideal}{ {\text{ideal}} }
\newcommand{\therm}{{\text{th}}}
\newcommand{\coupling}{g}
\newcommand{\tune}{{\text{tune}}}
\newcommand{\imperfect}{ {\text{imp}} }
\newcommand{\opt}{ {\text{opt}} }
\newcommand{\Carnot}{\text{Carnot}} 
\newcommand{\Wb}{W_{\text{b}}}  
\newcommand{\HTemp}{{\text{H}}} 
\newcommand{\CTemp}{ {\text{C}} }  
\newcommand{\THot}{T_\HTemp}  
\newcommand{\TCold}{T_\CTemp}  
\newcommand{\betaH}{\beta_\HTemp}  
\newcommand{\betaC}{\beta_\CTemp}  
\newcommand{\gap}{ \Delta }
\newcommand{\HScale}{\mathcal{E}}
\newcommand{\TGap}{ \tilde{\varepsilon} }  
\newcommand{\LZ}{{\text{LZ}}} 
\newcommand{\HalfLZ}{\text{frac-LZ}}  
\newcommand{\DB}{{\text{DB}}}  
\newcommand{\APT}{{\text{APT}}}  
\newcommand{\PDown}{ P_\downarrow }  
\newcommand{\PUp}{ P_\uparrow }  
\newcommand{\PDownn}{ \PDownnn^{\text{true}} }
\newcommand{\PDownnn}{ \mathcal{P}_\downarrow }
\newcommand{\PUppp}{ \mathcal{P}_\uparrow }  
\newcommand{\EjE}{ E_j }  
\newcommand{\EjM}{ E'_j }  
\newcommand{\disp}{ {\text{displ}} }  
\newcommand{\qubit}{{\text{qubit}}}  
\newcommand{\meso}{{\text{meso}}}  
\newcommand{\Sim}{{\text{sim}}}  
\newcommand{\ZH}{Z}    
\newcommand{\ZC}{Z'}
\newcommand{\dAvg}{\expval{\delta}}
\newcommand{\dAvgSub}{\dAvg}
\newcommand{\macro}{{\text{macro}}}
\newcommand{\deltaMBL}{\delta_-}
\newcommand{\Sys}{S}
\newcommand{\J}{ \mathcal{J} }
\newcommand{\JFar}{\J_{L \gg \xi}}
\newcommand{\JClose}{\J_{L \leq \xi}}
\newcommand{\Cp}{C_{\text{P}}} 
\newcommand{\Cv}{C_{\text{v}}}  
\newcommand{\A}{{\text{A}}}
\newcommand{\B}{{\text{B}}} 
\newcommand{\DOS}{\mu}  
\newcommand{\Err}{\epsilon}  
\newcommand{\Li}{{\text{Li}}}  
\newcommand{\bath}{{\text{bath}}}  
\newcommand{\LBath}{L_\bath}  
\newcommand{\cycle}{{\text{cycle}}}  
\newcommand{\HighOrd}{{\text{high-ord.}}}  
\newcommand{\cold}{{\text{cold}}}  

\newcommand{\Ell}{ { (\ell) } }
\newcommand{\ParenJ}{ { (j) } } 
\newcommand{\ParenE}{ { (E) } } 
\newcommand{\LL}{ { (L) } } 


%
%
Many-body localization (MBL) has emerged as a unique phase in which
an isolated interacting quantum system does not thermalize internally.
MBL systems are integrable and have local integrals of motion~\cite{Huse_14_phenomenology},
which retain information about initial conditions
for long times, or even indefinitely \cite{KjallIsing}. 
This and other aspects of MBL were recently observed experimentally \cite{Schreiber_15_Observation,Kondov_15_Disorder,Ovadia_15_Evidence,Choi_16_Exploring,Luschen_17_Signatures,Kucsko_16_Critical,Smith_16_Many,Bordia_17_Probing}.
In contrast, in thermalizing isolated quantum systems, information and energy can easily diffuse. Such systems obey the eigenstate thermalization hypothesis (ETH)~\cite{Deutsch_91_Quantum,Srednicki_94_Chaos,Rigol_07_Relaxation}.

A tantalizing question is whether the unique properties of MBL phases could be utilized. So far, MBL was proposed to be used for robust quantum memories~\cite{Nandkishore_15_MBL}. We believe, however, that the potential of MBL is much greater.  MBL systems behave athermally, and athermality (lack of thermal equilibrium) facilitates thermodynamic tasks. 
When a cold bath is put in contact with a hot environment, for instance, 
work can be extracted from the heat flow. 
More generally, athermal systems serve as thermodynamic resources~\cite{Janzing_00_Thermodynamic,Dahlsten_11_Inadequacy,Aberg_13_Truly,Brandao_13_Resource,Horodecki_13_Fundamental,Egloff_15_Measure,Goold_15_review,Gour_15_Resource,YungerHalpern_16_Beyond,YungerHalpern14,Deffner_16_Quantum,Wilming_17_Third}.
Could MBL's athermality have thermodynamic applications?

We present a thermodynamic application of MBL:
We formulate, analyze, and numerically simulate an Otto engine cycle
for a quantum many-body system that has an MBL phase.
The engine contacts a hot bath and a narrow-band cold bath, 
as sketched in Fig.~\ref{fig:Artist_conceptn}.
This application unites the growing fields of 
quantum thermal machines~\cite{Geusic_maser_67,del_Campo_14_Super,Brunner_15_Ent_fridge,Binder_15_Quantacell,Woods_15_Maximum,Gelbwaser_15_Strongly,Song_16_polariton_engine,Tercas_16_Casimir,PerarnauLlobet_16_Work,Kosloff_17_QHO,Lekscha_16_Quantum,Jaramillo_16_Quantum}
and MBL~\cite{BAA,Oganesyan_07_Level_stats,Pal_Huse_10_MBL,Huse_14_phenomenology,Nandkishore_15_MBL,serbynmoore}. 
Our proposal could conceivably be explored in cold-atom~\cite{Schreiber_15_Observation,Kondov_15_Disorder,Choi_16_Exploring,Luschen_17_Signatures,Bordia_17_Probing}; 
nitrogen-vacancy-center~\cite{Kucsko_16_Critical};
trapped-ion~\cite{Smith_16_Many}; and possibly
doped-semiconductor~\cite{Kramer_93_Localization} experiments.

%
%
\begin{figure}[tb]
\centering
\includegraphics[width=.45\textwidth, clip=true]{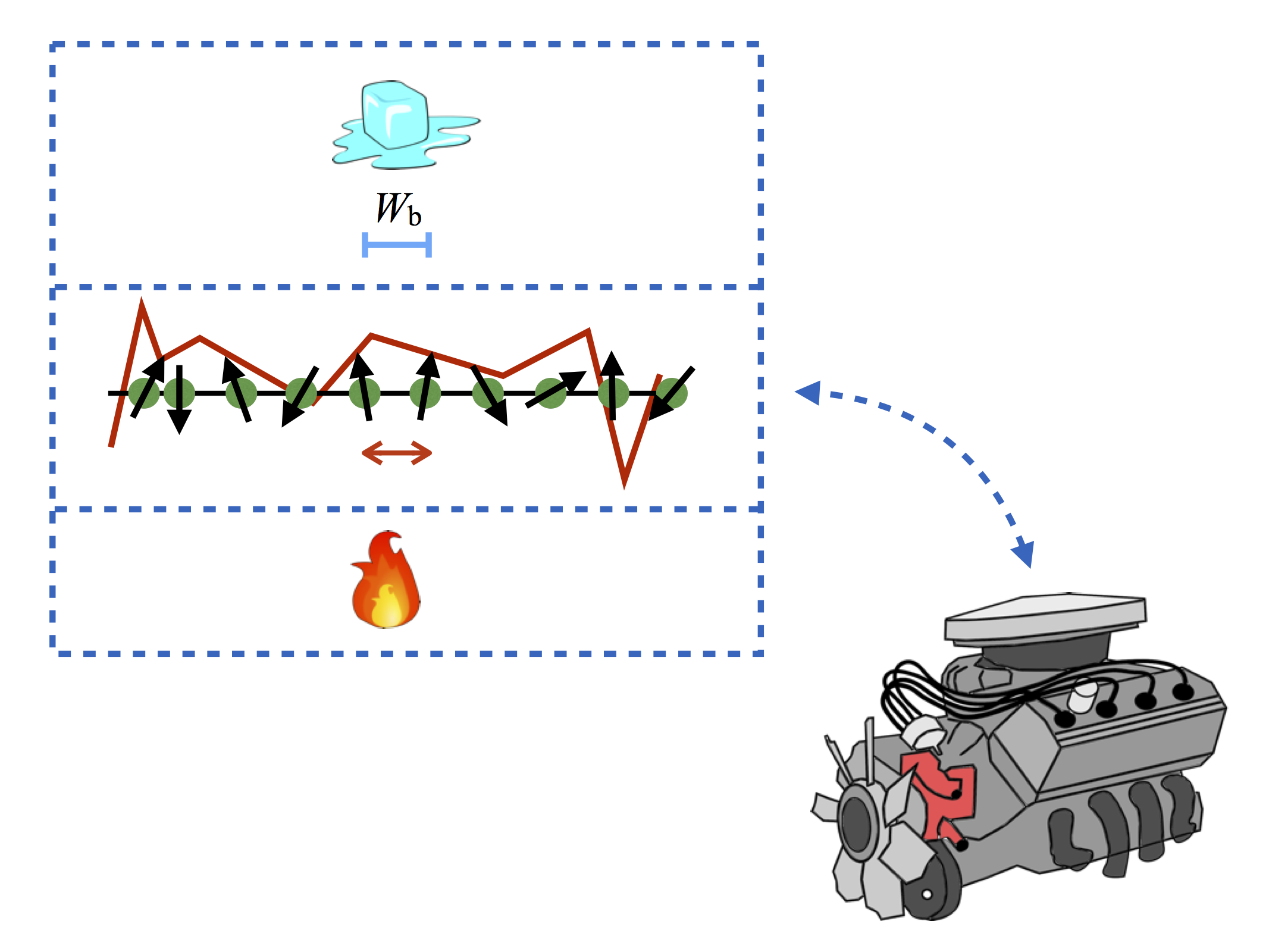}
\caption{\caphead{Schematic of many-body-localized (MBL) engine:}
We formulate an Otto engine cycle for a many-body quantum system
that exhibits an MBL phase.
The system is exemplified by the spin chain
illustrated by the green dots and black arrows.
A random disorder potential (the jagged red line) localizes the particles.
Particles interact and hop between sites
(as suggested by the horizontal red arrows).
Consider strengthening the interactions and the hopping frequency.
The system transitions from strong localization
to a thermal phase (which obeys the eigenstate thermalization hypothesis),
or at least to weak localization.
The engine thermalizes with a hot bath (represented by the flames)
and with a cold bath (represented by the ice cube).
The cold bath has a small bandwidth $\Wb$,
to take advantage of small energy gaps' greater prevalence
in the highly localized regime.}
\label{fig:Artist_conceptn}
\end{figure}

Our engine relies on the spectral-correlation properties 
that distinguish MBL from thermal systems~\cite{Sivan_87_Energy,serbynmoore}. 
Take an interacting finite spin chain as an example.
Consider the statistics of gaps 
between consecutive energy eigenvalues 
far from the energy band's edges.
A gap distribution $P (\delta)$ encodes
the probability that any given gap
has size $\delta$.
The MBL gap distribution enables small (and large) gaps to appear
much more often than in ETH spectra~\cite{D'Alessio_16_From}. 
This difference enables MBL to enhance 
our quantum many-body Otto cycle.

Let us introduce the MBL and ETH distributions in greater detail.
Let $\dAvg_E$ denote the average gap at the energy $E$.
MBL gaps approximately obey Poisson statistics~\cite{Oganesyan_07_Level_stats,D'Alessio_16_From}:
\begin{align}
   \label{eq:P_MBL_Main}
   P_\MBL^\ParenE(\delta)  
   \approx  \frac{1}{\dAvg_E}  e^{-\delta / \dAvg_E }  \, .
\end{align}
Any given gap has a decent chance of being small: 
As $\delta  \to  0$,
$P_\MBL^\ParenE (\delta)  \to  \frac{1}{\dAvg_E}  >  0$.
Neighboring energies have finite probabilities of lying close together: 
MBL systems' energies do not repel each other, 
unlike thermal systems' energies.
Thermalizing systems governed by real Hamiltonians 
obey the level statistics of random matrices drawn from 
the Gaussian orthogonal ensemble (GOE)~\cite{Oganesyan_07_Level_stats}: 
\begin{align}
   \label{eq:P_ETH_Main}
   P_\ETH^\ParenE (\delta)  
   \approx   \frac{\pi}{2}   \frac{\delta}{\dAvg_E^2}  \:  
   e^{-\frac{\pi}{4}  \delta^2  /  \dAvg_E^2}  \, .
\end{align}
Unlike in MBL spectra, small gaps rarely appear:
As $\delta\to0$,  $P_\ETH^\ParenE (\delta)  \to  0$.

%

MBL's athermal gap statistics should be construed as 
a thermodynamic resource as athermal quantum states are~\cite{Janzing_00_Thermodynamic,Dahlsten_11_Inadequacy,Aberg_13_Truly,
Brandao_13_Resource,Horodecki_13_Fundamental,Egloff_15_Measure,
Goold_15_review,Gour_15_Resource,YungerHalpern_16_Beyond,YungerHalpern14,
Deffner_16_Quantum,Wilming_17_Third}.  
In particular, MBL's athermal gap statistics improve our engine's reliability:
The amount $W$ of work extracted by our engine 
fluctuates relatively little from successful trial to successful trial.
Athermal statistics also lower the probability of worst-case trials,
in which the engine outputs net negative work, $W_\tot < 0$.
Furthermore, MBL's localization 
enables the engine to scale robustly: 
Mesoscale ``subengines'' can run in parallel 
without disturbing each other much,
due to the localization inherent in MBL. 
Even in the thermodynamic limit, 
an MBL system behaves like an ensemble of finite,  
mesoscale quantum systems, due to its  
\emph{local level correlations}~\cite{Sivan_87_Energy,imryma,Syzranov_17_OTOCs}. 
Any local operator can probe only
a discrete set of sharp energy levels, 
which emerge from its direct environment.



This paper is organized as follows.
Section~\ref{section:Thermo_backgrnd_main} contains background about
the Otto cycle and quantum work and heat.
We present the MBL Otto engine in three steps in Sec.~\ref{section:Intro_cycle}. 
In Sec.~\ref{section:Qubit_main}, we introduce the basic idea 
using a single qubit (two-level quantum system). 
In Sec.~\ref{section:Meso_main}, we scale the engine up to 
a mesoscopic chain tuned between MBL and ETH.
In Sec.~\ref{section:Thermo_limit_main}, we show that 
the mesoscopic segments could be combined into 
a macroscopic MBL system, 
while operating in parallel.
Our analytic calculations are tested in Sec.~\ref{section:Numerics_main},
with numerical simulations of disordered spin chains.
In Sec.~\ref{section:Order_main}, we provide order-of-magnitude 
estimates for a localized semiconductor engine's power and power density.
We compare the localized engine with more traditional alternatives in Sec. \ref{section:Compare_main}.
Background information, intuitive examples, and extensive calculations
appear in~\cite{NYH_17_MBL}.

\section{Thermodynamic background}
\label{section:Thermo_backgrnd_main}

The classical Otto engine (see e.g.,~\cite{MIT_Otto}) 
consists of a gas that expands, cools, contracts, and heats.
During the two isentropic (constant-entropy) strokes,  
the gas's volume is tuned 
between values $V_1$ and $V_2 < V_1$. 
The \emph{compression ratio} is defined as $r := \frac{ V_1 }{ V_2 }$ . 
The heating and cooling are isochoric (constant-volume). 
 The engine outputs a net amount $W_\tot$
of work per cycle, absorbing heat $Q_\In > 0$
during the heating isochore.

A general engine's thermodynamic efficiency is
\begin{align}
   \label{eq:Eff_main}
   \eta  :=  \frac{ W_\tot }{ Q_\In }  \, .
\end{align}
The Otto engine operates at the efficiency
\begin{align}
   \label{eq:Otto_eff_main}
   \eta_\Otto  =  1  -  \frac{1}{  r^{ \gamma  -  1 }  }
   <  \eta_\Carnot  \, .
\end{align}
$\gamma:=\frac{\Cp}{\Cv}$ denotes
a ratio of the gas's 
constant-pressure and constant-volume specific heats.
The Carnot efficiency $\eta_\Carnot$ upper-bounds
the efficiency of every thermodynamic engine that involves just two heat baths.

A quantum Otto cycle~\cite{Kosloff_17_QHO}
for harmonic oscillators has been formulated~\cite{Scully_02_Quantum,Abah_12_Single,Deng_13_Boosting,del_Campo_14_Super,Zheng_14_Work,Karimi_16_Otto,V_Anders_15_review,Kosloff_17_QHO}.
The quantum harmonic oscillator's (QHO's) gap plays the role
of the classical Otto engine's volume.
Let $\omega_1$ and $\omega_2 > \omega_1$ denote 
the values between which
the angular frequency is tuned.
The ideal QHO Otto cycle operates at the efficiency
\begin{align}
   \label{eq:Eff_QHO}
   \eta_\QHO  =  1  -  \frac{ \omega_1 }{ \omega_2 }  \, .
\end{align}
This oscillator model resembles the qubit toy model
that informs our MBL Otto cycle (Sec.~\ref{section:Qubit_main}).

The heat and work exchanged by slowly tuned systems are defined as
\begin{align}
   & \label{eq:Work_def_main}
   W:=\int_0^\tau dt\;\Tr   \left(  \rho  \:  \frac{dH}{dt}  \right)
   \, , \quad \text{and} \\ 
   &  \label{eq:Heat_def_main}
   Q:=\int_0^\tau dt\;\Tr\left(  \frac{d\rho}{dt}  \:  H  \right)  
\end{align}
in quantum thermodynamics~\cite{V_Anders_15_review}.
This $Q$ definition is narrower than the definition
prevalent in the MBL literature~\cite{Lin_16_Ginzburg,Corboz_16_Variational,Gopalakrishnan_16_Regimes,D'Alessio_16_From}: 
Here, all energy exchanged during unitary evolution 
counts as work.

\section{The MBL Otto cycle}
\label{section:Intro_cycle}

%
%
\begin{figure}[tb]
\centering
\includegraphics[width=.45\textwidth, clip=true]{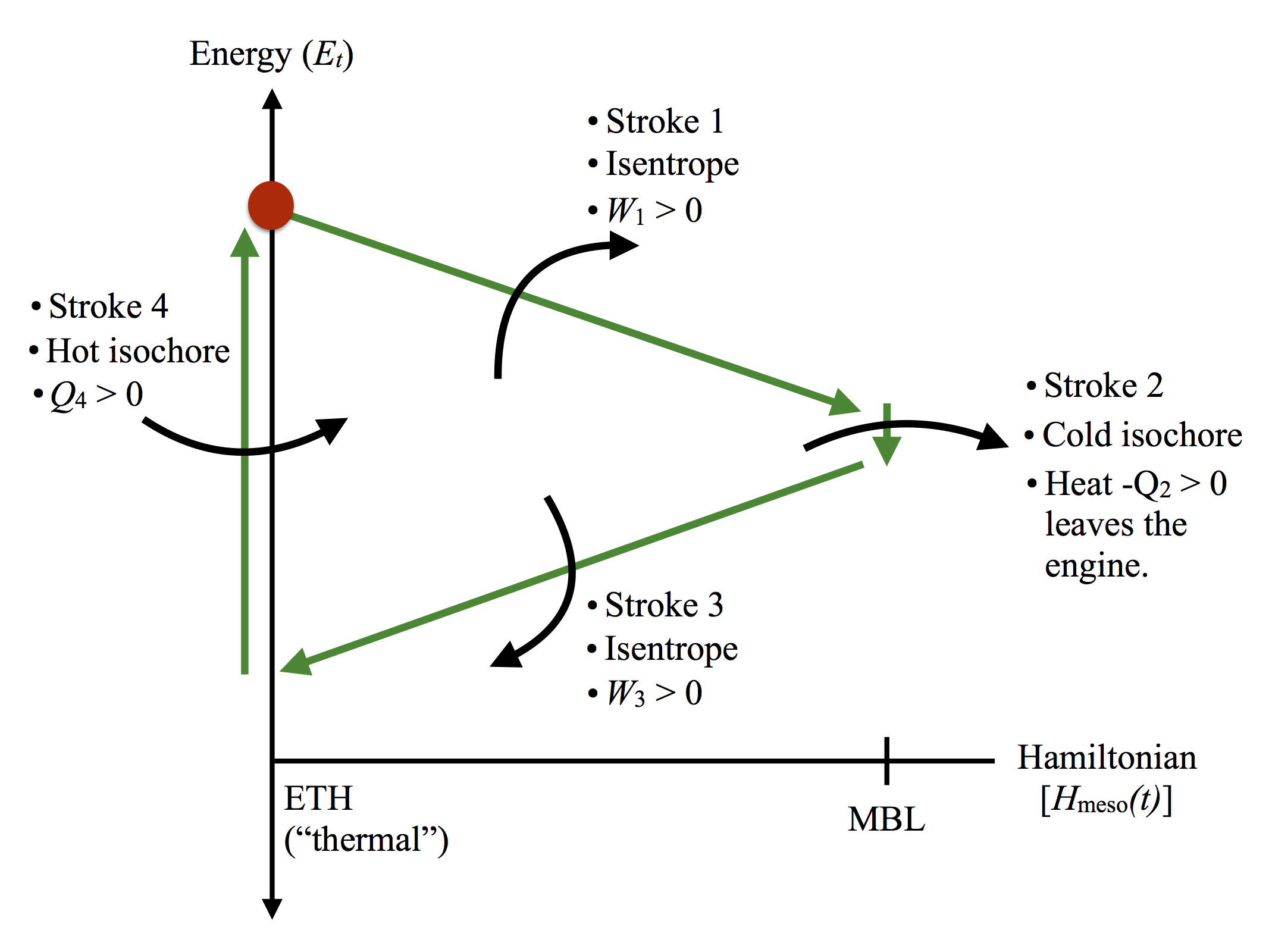}
\caption{\caphead{Otto engine cycle for
a mesoscale many-body-localized (MBL) system:}
Two energies in the many-body spectrum 
capture the cycle's basic physics.
The engine can be regarded as beginning each trial
in an energy eigenstate drawn from a Gibbs distribution.
Let the red dot denote the engine's starting state 
in some trial of interest.
The cycle consists of four strokes:
During stroke 1, the Hamiltonian $H_\meso(t)$ 
is tuned from ``thermal'' (obeying the eigenstate thermalization hypothesis, or ETH) 
to MBL.
During stroke 2, the engine thermalizes with a cold bath.
$H_\meso(t)$ returns from MBL to thermal during stroke 3.
Stroke 4 resets the engine, which thermalizes with a hot bath.
The tunings (strokes 1 and 3) map onto
the thermodynamic Otto cycle's isentropes.
The thermalizations (strokes 2 and 4) map onto isochores.
The engine outputs work $W_1$ and $W_3$
during the tunings
and absorbs heat $Q_2$ and $Q_4$ during thermalizations.
The engine benefits from the discrepancy between
MBL and thermal gap statistics:
Energies have a greater probability of lying close together
in the MBL phase than in the thermal phase.
This discrepancy leads the engine to 
``slide down'' the lines that represent tunings.
During downward slides, the engine loses energy outputted as work.}
\label{fig:Compare_thermo_Otto_fig}
\end{figure}

During the MBL Otto cycle, a quantum many-body system
is cycled between two disorder strengths and so between
two level-repulsion strengths and
two localization lengths.
The system begins in the less localized regime,
in thermal equilibrium with a hot bath
at a temperature $\THot  \equiv  1 /  \betaH$.
(We set Boltzmann's constant to one: $\kB = 1$.) 
Next, disorder is effectively increased,
suppressing level suppression. 
The system then thermalizes with a finite-size cold bath
that has a narrow bandwidth
at a temperature $\TCold  \equiv  1 / \betaC  <  \betaH$.
Finally, the disorder is decreased.
The system then returns to its initial state
by thermalizing with the hot bath.

Below, we introduce the MBL Otto cycle in three steps:
(1) A qubit toy model illustrates the basic physics. 
(2) A mesoscale engine (Fig.~\ref{fig:Compare_thermo_Otto_fig}) 
is tuned between MBL and ETH phases.
(3) Mesoscale subengines operate in parallel
in a macroscopic MBL engine.
Table~\ref{table:Notation} summarizes parameters 
of the mesoscale and macroscopic MBL engines.

\subsection{Qubit toy model}
\label{section:Qubit_main}

At the MBL Otto engine's basis lies a qubit Otto engine
whose energy eigenbasis transforms during the cycle~\cite{Kosloff_02_Discrete,Kieu_04_Second,Kosloff_10_Optimal,Cakmak_17_Irreversible}.
Consider a 2-level system evolving under 
the time-varying Hamiltonian 
\begin{align}
   H_\qubit(t)  :=   (1-\alpha_t)  h\sigma^x
   +   \alpha_t h'\sigma^z  \, .
\end{align}
$\sigma^x$and $\sigma^z$ denote the Pauli $x$- and $z$-operators.
$\alpha_t$ denotes a parameter tuned between 0 and 1.

The engine begins in thermal equilibrium 
at the temperature $\THot$.
During stroke 1, the engine is thermally isolated,
and $\alpha_t$ is tuned from 0 to 1. 
During stroke 2, the engine thermalizes 
to the temperature $\TCold$. 
During stroke 3, the engine is thermally isolated, 
and $\alpha_t$ returns from 1 to 0.
During stroke 4, the engine resets 
by thermalizing with a hot bath.

Let us make two simplifying assumptions
(see~\cite[App.~C]{NYH_17_MBL} for a generalization):
First, let $\THot=\infty$ and $\TCold=0$.
Second, assume that the engine is tuned slowly enough
to satisfy the quantum adiabatic theorem.
We also choose\footnote{
The gaps' labels are suggestive: A qubit, having only one gap,
obeys neither $\ETH$ nor $\MBL$ gap statistics.
But, when large, the qubit gap apes a typical $\ETH$ gap;
and, when small, the qubit gap apes a useful $\MBL$ gap.
This mimicry illustrates how the mesoscopic engine benefits from
the greater prevalence of small gaps in MBL spectra
than in $\ETH$ spectra.}
\[
h=  \frac{ \delta_\ETH }{2},\,\,
h'=  \frac{ \delta_\MBL }{2}\]
and $\delta_\ETH\gg\delta_\MBL$.

Let us analyze the cycle's energetics.
The system begins with $\expval{ H_\qubit (t) }  =  0$. 
Stroke 1 preserves the $T = \infty$ state $\id / 2$.
Stroke 2 drops the energy to $-\frac{\delta_\MBL}{2}$.
The energy drops to $-\frac{\delta_\ETH}{2}$ during stroke 3.
During stroke 4, the engine resets to zero average energy,
absorbing heat
$\expval{Q_4}=  \frac{ \delta_\ETH }{ 2 }$, on average.

The energy exchanged during the tunings (strokes 1 and 3)
constitutes work [Eq.~\eqref{eq:Work_def_main}], 
while the energy exchanged during the thermalizations
(strokes 2 and 4) is heat
[Eq.~\eqref{eq:Heat_def_main}].
The engine outputs the \emph{per-cycle power,} 
or average work outputted per cycle,
$\expval{W_\tot}=\frac{1}{2}(\delta_\ETH-\delta_\MBL)$.
The efficiency is 
$\eta_\qubit   =   \frac{\expval{W_\tot}}{\expval{Q_4}}
=   1-\frac{\delta_\MBL}{\delta_\ETH}$. 
This result is equivalent to
the efficiency $\eta_\Otto$ of
a thermodynamic Otto engine [Eq.~\eqref{eq:Otto_eff_main}].
The gap ratio $\frac{\delta_\MBL}{\delta_\ETH}$
plays the role of $r^{\gamma-1}$.
$\eta_\qubit$ also equals the efficiency $\eta_\QHO$
[Eq.~\eqref{eq:Eff_QHO}],
if the frequency ratio $\omega / \Omega$ is chosen to equal
the gap ratio $\delta_\MBL / \delta_\ETH$.
As shown in Sections~\ref{section:Meso_main}-\ref{section:Thermo_limit_main}, however,
the qubit engine can scale to 
a large composite engine
of densely packed qubit subengines operating in parallel.
The dense packing is possible if
the qubits are encoded in
the MBL system's localized degrees of freedom 
($\ell$-bits, roughly speaking~\cite{Huse_14_phenomenology}).

\subsection{Level-statistics engine for a mesoscale system}
\label{section:Meso_main}

%
%
\begin{figure}[tb]
\centering
\includegraphics[width=.99\textwidth, clip=true]{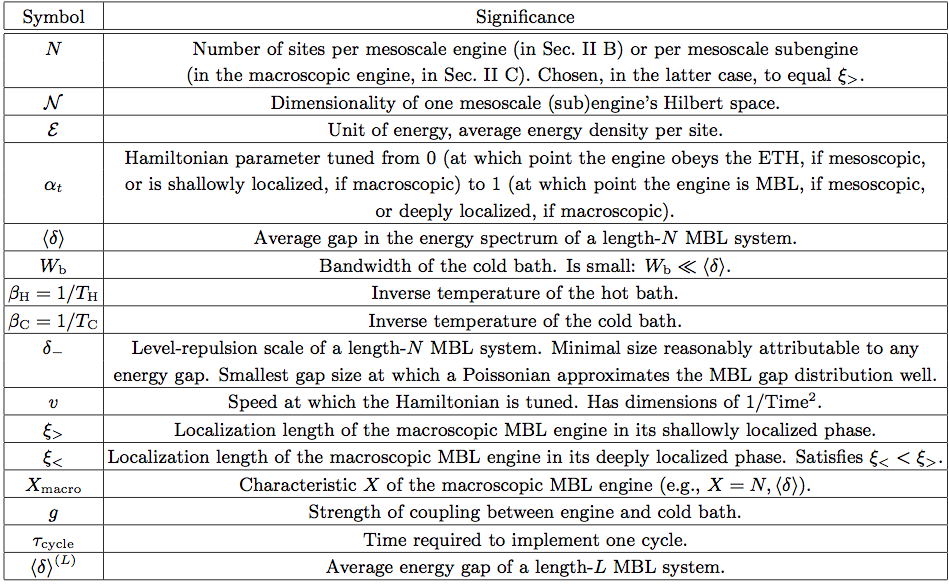}
\caption{\caphead{Parameters of
the mesoscopic and macroscopic MBL engines:}
Introduced in Sections~\ref{section:Meso_main} and~\ref{section:Thermo_limit_main}.
Boltzmann's constant is set to one: $\kB = 1$.}
\label{table:Notation}
\end{figure}

The next step is an interacting finite-size system 
tuned between MBL and ETH phases. 
Envision a mesoscale engine as a one-dimensional (1D) system 
of $\Sites  \approx  10$ sites.
This engine will ultimately model one region in 
a thermodynamically large MBL engine. 
We will analyze 
the mesoscopic engine's per-trial power $\expval{ W_\tot }$, 
the efficiency $\eta_\MBL$,
and work costs $\expval{ W_\diab }$ of undesirable diabatic transitions.

\subsubsection{Set-up for the mesoscale MBL engine}
\label{section:Meso_setup}

The mesoscopic engine evolves under the Hamiltonian
\begin{align}
   \label{eq:H_meso_main}
   H_\meso(t)  :=  \frac{ \HScale }{ Q ( \alpha_t ) }  \left[
   (1-\alpha_t )H_\ETH   +   \alpha_t  \, H_\MBL  \right]  \, .
\end{align}
The unit of energy, or average energy density per site,
is denoted by $\HScale$.
The tuning parameter $\alpha_t  \in  [0, 1]$.
When $\alpha_t  =  0$, the system evolves under
a random Hamiltonian $H_\ETH$
whose gaps $\delta$ are distributed according to 
$P^\ParenE_\ETH ( \delta )$
[Eq.~\eqref{eq:P_ETH_Main}].
When $\alpha_t = 1$, $H_\meso(t) = H_\MBL$, 
a Hamiltonian whose gaps are distributed according to 
$P^\ParenE_\MBL ( \delta )$
[Eq.~\eqref{eq:P_MBL_Main}].
We simulate $H_\ETH$ and $H_\MBL$ using
a disordered Heisenberg model in Sec.~\ref{section:Numerics_main}.
There, $H_\ETH$ and $H_\MBL$ differ only in
their ratios of hopping frequency to disorder strength.


The mesoscale engine's cycle
is analogous to the qubit cycle,
including initialization at $\alpha_t = 0$, 
tuning of $\alpha_t$ to one, 
thermalization with a temperature-$\TCold$ bath,
tuning of $\alpha_t$ to zero,
and thermalization~\cite{Huse_15_Localized,DeLuca_15_Dynamic,Levi_16_Robustness,Fischer_16_Dynamics} 
with a temperature-$\THot$ bath.
To highlight the role of level statistics in the cycle,
we hold the average energy gap, $\dAvg$, constant.\footnote{
\label{footnote:dAvg}
$\dAvg$ is defined as follows.
Let $\DOS(E)  =   \approx  
\frac{ \HDim }{  \sqrt{ 2 \pi \Sites }  \,  \HScale  }  \,
e^{ - E^2 / 2 \Sites  \HScale^2 }$ 
denote the density of states at energy $E$.
Inverting $\DOS (E)$ yields
the \emph{local average gap}:
$\dAvg_E  :=  \frac{1}{  \DOS (E) }$.
Inverting \emph{the average of} $\DOS (E)$ yields
the \emph{average gap}:
\begin{align}
   \label{eq:dAvg_def}
   \dAvg  :=  \frac{ 1 }{ \expval{  \DOS (E) }_{\text{energies}} } 
   =  \frac{ \HDim }{   \int_{ - \infty }^\infty  dE  \;  \DOS^2 (E)  }
   =   \frac{ 2 \sqrt{ \pi \Sites }  }{ \HDim }  \,  \HScale    \, .
\end{align}
}
We do so using renormalization factor $Q ( \alpha_t )$.\footnote{
Imagine removing $Q(\alpha_t)$ from Eq.~\eqref{eq:H_meso_main}.
One could increase $\alpha_t$---could 
tune the Hamiltonian from ETH to MBL~\cite{serbynmoore}---by
strengthening a disorder potential.
This strengthening would expand the energy band.
Tuning from MBL to ETH would compress the band.
Expanding and compressing would generate an accordion-like motion.
By interspersing the accordion motion with thermalizations,
one could extract work.
Such an engine would benefit little from properties of MBL,
whose thermodynamic benefits we wish to highlight.
Hence we ``zero out'' the accordion-like motion, 
by fixing $\dAvg$ through $Q( \alpha_t )$.}
Section~\ref{section:Numerics_main} details how we define $Q (\alpha_t)$
in numerical simulations.

The key distinction between 
GOE level statistics~\eqref{eq:P_ETH_Main} 
and Poisson (MBL) statistics~\eqref{eq:P_MBL_Main}
is that small gaps (and large gaps)
appear more often in Poisson spectra.
A toy model illuminates these level statistics' physical origin:
An MBL system can be modeled as 
a set of noninteracting quasilocal qubits~\cite{Huse_14_phenomenology}.
Let $g_j$ denote the $j^\th$ qubit's gap.
Two qubits, $j$ and $j'$, may have nearly equal gaps: $g_j  \approx  g_{j'}$. 
The difference $| g_j  -  g_{j'} |$ equals a gap
in the many-body energy spectrum.
Tuning the Hamiltonian from MBL to ETH
couples the qubits together,
producing matrix elements between the nearly degenerate states.
These matrix elements force energies apart.

To take advantage of the phases' distinct level statistics,
we use a cold bath that has a small bandwidth $\Wb$.
According to Sec.~\ref{section:Qubit_main},
net positive work is extracted from the qubit engine because 
$\delta_\MBL  <  \delta_\ETH$.
The mesoscale analog of $\delta_\ETH$ is $\sim \dAvg$,
the typical gap ascended during hot thermalization. 
During cold thermalization,
the system must not emit energy on the scale of
the energy gained during cold thermalization. 
Limiting $\Wb$ ensures that cold thermalization
relaxes the engine only across gaps 
$\delta\leq \Wb  \ll  \dAvg$.
Such anomalously small gaps appear more often 
in MBL energy spectra than in ETH spectra
~\cite{Khaetskii_02_Electron,Gopalakrishnan_14_Mean,Parameswaran_17_Spin}.


This level-statistics argument holds only 
within superselection sectors.
Suppose, for example, that $H_\meso(t)$ conserves particle number.
The level statistics arguments apply only if 
the particle number remains constant 
throughout the cycle~\cite[App.~F]{NYH_17_MBL}.
Our numerical simulations (Sec.~\ref{section:Numerics_main}) 
take place at half-filling,
in a subspace of dimensionality $\HDim$ 
of the order of magnitude of
the whole space's dimensionality:
$\HDim  \sim  \frac{ 2^\Sites }{ \sqrt{ \Sites } }$.


%
%
%

We are now ready to begin analyzing the mesoscopic engine Otto cycle. 
The engine begins in the thermal state 
$\rho(0)=e^{-\betaH H_\ETH}/\ZH$, wherein 
$\ZH:=\Tr\left(e^{-\betaH H_\ETH}\right)$.
The engine can be regarded as starting each trial
in some energy eigenstate $j$
drawn according to the Gibbs distribution
(Fig.~\ref{fig:Compare_thermo_Otto_fig}).
During stroke 1, $H_\meso(t)$ is tuned from $H_\ETH$ to $H_\MBL$.
We approximate the tuning as quantum-adiabatic.
(Diabatic corrections are modeled in Sec.~\ref{section:Diab_main}.)
Stroke 2, cold thermalization, depends on 
the gap $\delta'_j$ between the $j^\th$ and $(j-1)^\th$ MBL levels.
This gap typically exceeds $\Wb$.
If it does, cold thermalization preserves the engine's energy,
and the cycle outputs $W_\tot=0$. 
With probability $\sim  \frac{\Wb}{\dAvg}$,
the gap is small enough to thermalize: $\delta'_j  <  \Wb$. 
In this case, cold thermalization drops the engine to level $j-1$.
Stroke 3 brings the engine to level $j-1$ of $H_\ETH$. 
The gap $\delta_j$ between the $(j-1)^\th$ and $j^\th$ $H_\ETH$ levels 
is $\dAvg  \gg  \Wb$, with the high probability
$\sim  1 -  ( \Wb / \dAvg )^2$.
Hence the engine likely outputs $W_\tot>0$.
Hot thermalization (stroke 4) returns the engine to $\rho(0)$.

%
%
%
\subsubsection{Quantitative analysis of the mesoscale engine}
\label{section:Quant_main}

How well does the mesoscale Otto engine perform?
We calculate average work $\expval{ W_\tot }$
outputted per cycle
and the efficiency $\eta_\MBL$.
Details appear in Suppl. Mat.~\ref{section:PowerApp}.

We focus on the parameter regime in which 
the cold bath is very cold, 
the cold-bath bandwidth $\Wb$ is very small, and
the hot bath is very hot:
$\TCold   \ll   \Wb  \ll  \dAvg$, and
$\sqrt{ \Sites }  \:  \betaH \HScale  \ll  1$.
The mesoscale engine resembles a qubit engine
whose state and gaps are averaged over.
The gaps, $\delta_j$ and $\delta'_j$,
obey the distributions $P_\ETH^\ParenE(\delta_j)$ and 
$P_\MBL^\ParenE(\delta'_j)$
[Eqs.~\eqref{eq:P_ETH_Main} and~\eqref{eq:P_MBL_Main}].
Correlations between the $H_\ETH$ and $H_\MBL$ spectra 
can be neglected.

We make three simplifying assumptions, generalizing later:
(i) The engine is assumed to be tuned quantum-adiabatically.
Diabatic corrections are calculated in Sec.~\ref{section:Diab_main}.
(ii) The hot bath is at $\THot = \infty$.
We neglect finite-temperature corrections, which scale as 
$\Sites ( \betaH \HScale )^2  \left( \frac{ \Wb }{ \dAvg }  \right)^2  \dAvg$.
(iii) The gap distributions vary negligibly with energy:
$P_\ETH^\ParenE(\delta_j)  \approx  P_\ETH ( \delta_j )$, and 
$P_\MBL^\ParenE(\delta'_j)  \approx  P_\MBL ( \delta'_j )$,
while $\dAvg_E  \approx  \dAvg$.


%
%
%
\textbf{Average work $\expval{ W_\tot }$ per cycle:} 
The crucial question is whether 
the cold bath manages to relax the engine 
across the MBL-side gap $\delta'  \equiv  \delta'_j$. 
This gap obeys the Poisson distribution
$P_\MBL ( \delta' )$. 
If $\delta' < \Wb$, the engine has a probability
$1  /  (1  +  e^{-  \delta  \betaC}  )$
of thermalizing.
Hence the overall probability of relaxation by the cold bath is 
\begin{align}
   \label{eq:PCold}
   p_\cold  \approx  \int\limits_{0}^{\Wb}  d\delta' \;
   \frac{1}{\dAvg }   \frac{  e^{-\delta'  /  \dAvg}  }{ 1  +  e^{- \betaC \delta' }  }  \, .
\end{align}
In a simple, illustrative approximation,
we Taylor-expand to order $\frac{ \Wb }{ \dAvg }$
and $e^{ - \betaC \delta' }$:
$p_\cold  \approx  
\frac{ \Wb }{ \dAvg }  -  \frac{ 1 }{ \betaC \dAvg }$.
A more sophisticated analysis tweaks the multiplicative constants
(Suppl. Mat.~\ref{section:PowerApp}):
$p_\cold   \approx  
\frac{\Wb}{ \dAvg}  
-  \frac{ 2 \ln 2 }{ \betaC \dAvg }$.

Upon thermalizing with the cold bath,
the engine gains heat $\expval{Q}_4  \approx  \dAvg$, on average,
during stroke 4.
Hence the cycle outputs work
\begin{equation}
   \expval{ W_\tot } 
   \approx  p_\cold  \dAvg  +  \expval{ Q_2 }  
   \label{eq:WTotApprox2_Main}
   \approx   \Wb  \left(  1  -  \frac{ 2 \ln 2 }{ \betaC }  \right)  \, ,
\end{equation} 
on average.
$\expval{ Q_2 }$ denotes 
the average heat absorbed by the engine during cold thermalization:
\begin{equation}
   \expval{ Q_2 }   \approx   - \int\limits_{0}^{\Wb}   d\delta' \;
    \frac{ \delta' }{\dAvg}   
   \frac{  e^{-\delta' / \dAvg}  }{ 1  +  e^{-  \betaC  \delta' }}
   \approx  - \frac{ ( \Wb )^2}{2 \dAvg}  \, .
\end{equation}
In $\expval{ W_\tot }$, $\expval{ Q_2 }$ cancels with 
terms, in $\expval{ Q_4 }$, that come from high-order processes.
We have excluded the processes 
from Eq.~\eqref{eq:PCold} for simplicity.
See App.~\eqref{section:Q4} for details.
   
This per-cycle power scales with the system size $\Sites$ as\footnote{
The \emph{effective bandwidth} is defined as follows.
The many-body system has a Gaussian density of states:
$\DOS(E)  \approx  
\frac{ \HDim }{  \sqrt{ 2 \pi \Sites }  \,  \HScale  }  \,
e^{ - E^2 / 2 \Sites  \HScale^2 }$.
The states within a standard deviation $\HScale \sqrt{ \Sites }$
of the mean obey 
Eqs.~\eqref{eq:P_MBL_Main} and~\eqref{eq:P_ETH_Main}.
These states form the effective band,
whose width scales as $\HScale  \sqrt{ \Sites }$.}
$\Wb  \ll  \dAvg  
\sim  \frac{ \text{effective bandwidth} }{ \text{\# energy eigenstates} }
\sim  \frac{ \HScale \sqrt{ \Sites } }{\HDim} $. 
\textbf{Efficiency $\eta_\MBL$:}
The efficiency is
\begin{align}
   \label{eq:Eff_MBL}  
    \eta_\MBL 
    = \frac{ \expval{ W_\tot } }{ \expval{ Q_4 } }
    =  \frac{ \expval{ Q_4 }  +  \expval{ Q_2 } }{ \expval{ Q_4 } }
    \approx  1  -  \frac{ \Wb }{ 2 \dAvg } \, .
\end{align}
The imperfection is small, $\frac{\Wb}{ 2 \dAvg}  \ll  1$,
because the cold bath has a small bandwidth. 
This result mirrors the qubit-engine efficiency $\eta_\qubit$.\footnote{
$\eta_\MBL$ is comparable also to $\eta_\QHO$ [Eq.~\eqref{eq:Eff_QHO}].
Imagine operating an ensemble 
of independent QHO engines.
Let the $j^\th$ QHO frequency be tuned between 
$\Omega_j$ and $\omega_j$, distributed according to 
$P_\ETH ( \Omega_j )$ and
$P_\MBL ( \omega_j )$.
The average MBL-like gap $\omega_j$,
conditioned on $\omega_j  \in  [ 0 , \Wb ]$, is
$\expval{ \omega_j }     \sim \frac{ 1 }{ \Wb / \dAvg }
   \int_0^{ \Wb }  d \omega_j  \,  \omega_j  \,
   P_\MBL ( \omega_j )
   \approx  \frac{1}{ \Wb }  \int_0^{ \Wb }  d \omega_j  \;   \omega_j  
   =  \frac{ \Wb }{ 2 }  \, .$
Averaging the efficiency over the QHO ensemble yields
$\expval{ \eta_\QHO }
   :=  1  -  \frac{ \expval{ \omega } }{ \expval{ \Omega} }
   \approx  1 - \frac{ \Wb }{ 2 \dAvg }  
   \approx  \eta_\MBL  \, .$
The mesoscale MBL engine operates at the ideal average efficiency
of an ensemble of QHO engines.
But MBL enables qubit-like engines to pack together densely
in a large composite engine.}
But our engine is a many-body system
of $\Sites$ interacting sites.
MBL will allow us to employ
segments of the system as 
independent qubit-like subengines despite interactions.
In the absence of MBL, each subengine's effective $\dAvg = 0$.
With $\dAvg$ vanishes the ability to extract $\expval{ W_\tot } > 0$ 
using a local cold bath.


%
%
%
\subsubsection{Diabatic corrections to the per-cycle power}
\label{section:Diab_main}

\begin{figure}[tb]
\centering
\includegraphics[width=.45\textwidth, clip=true]{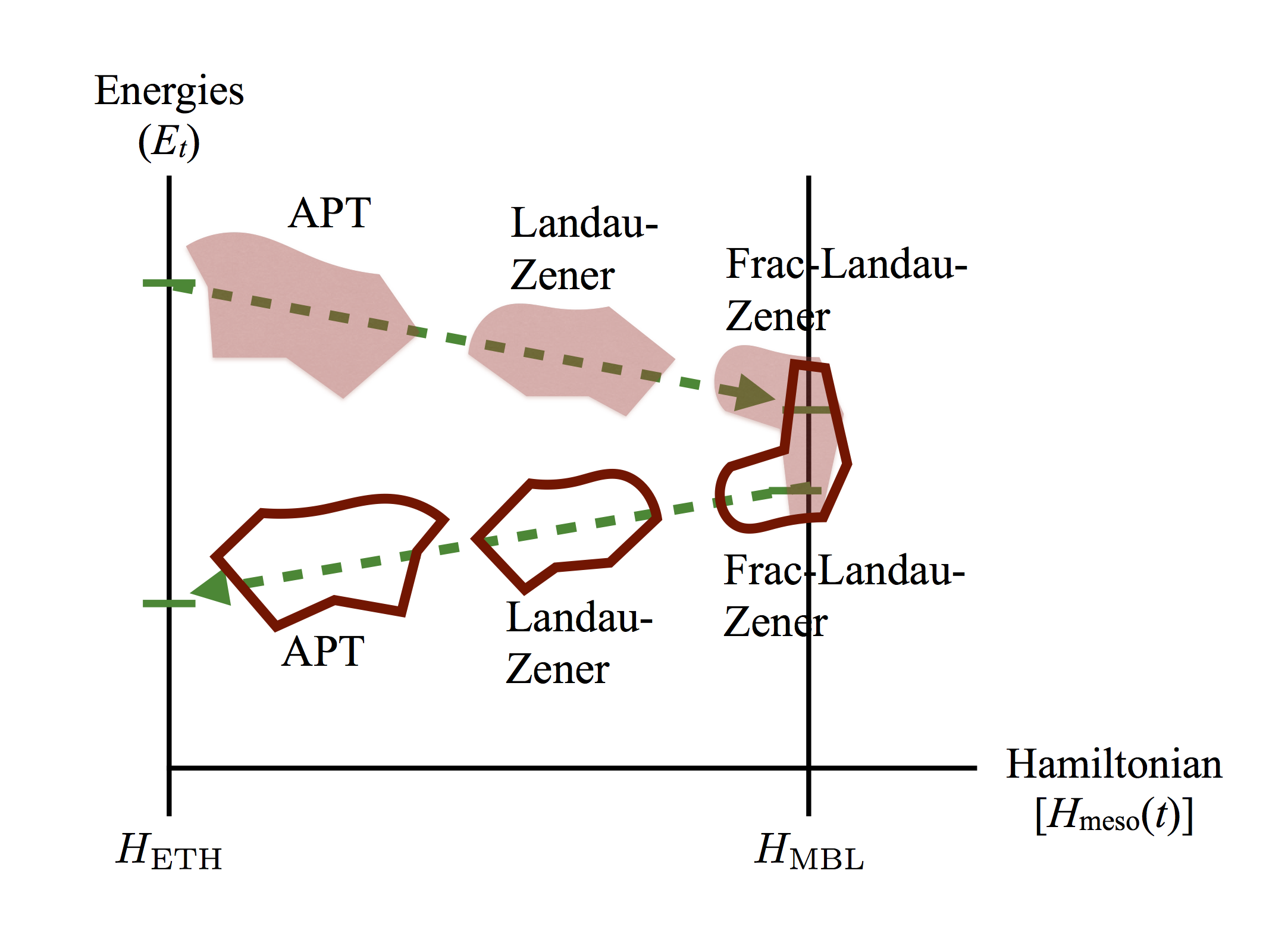}
\caption{\caphead{Three (times two) classes of diabatic transitions:}
Hops to arbitrary energy levels, modeled with
general adiabatic perturbation theory (APT),
plague the ETH regime.
Landau-Zener transitions and fractional-Landau-Zener transitions
plague the many-body-localized regime.}
\label{fig:Diab_Types}
\end{figure}

We have modeled the Hamiltonian tuning
as quantum-adiabatic.
Realistic tuning speeds 
$v  :=  \HScale \left\lvert  \frac{ d \alpha_t }{ dt }  \right\rvert$ 
are finite, inducing diabatic hops:
Suppose that the engine starts some trial
in the $j^\th$ energy eigenstate, with energy $\EjE$.
Suppose that $H_\meso(t)$ is measured at the end of stroke 1,
e.g., by the cold bath.
The measurement's outcome may be
the energy $E'_\ell$ of some MBL level
other than the $j^\th$. 
The engine will be said to have undergone 
a diabatic transition.
Transitions of three types can occur during stroke 1
and during stroke 3 (Fig.~\ref{fig:Diab_Types}).

If the engine jumps diabatically, its energy changes.
Heat is not entering, as the engine is not interacting with any bath.
The energy comes from the battery
used to tune the Hamiltonian, e.g., to strengthen a magnetic field.
Hence the energy change consists of work.
In addition to depleting the battery,
diabatic transitions can derail trials 
that would otherwise have outputted $W_\tot > 0$.

We estimate, to lowest order in small parameters, 
the average per-cycle work costs $\expval{ W_\diab }$ of diabatic jumps.
Supplementary Materials~\ref{section:App_Diab} contain
detailed derivations.
Numerics in Sec.~\ref{section:Numerics_main} support the analytics:
\begin{enumerate}[leftmargin=*]

   \item \emph{Thermal-regime transitions modeled by 
   general adiabatic perturbation theory 
   (APT transitions):}
Tuning $H_\meso(t)$ within the ETH phase
ramps a perturbation.
A matrix $\mathcal{M}$ represents the perturbation
relative to the original energy eigenbasis.
Off-diagonal elements of $\mathcal{M}$
may couple the engine's state
to arbitrary eigenstates of the original Hamiltonian.
We model such couplings with general adiabatic perturbation theory (APT)~\cite{DeGrandi_10_APT},
calling the induced transitions \emph{APT transitions}
(Suppl. Mat.~\ref{section:App_Diab_APT}).

APT transitions mimic thermalization with
an infinite-temperature bath:
The probability of transitioning across a size-$\delta$ gap 
does not depend on
whether the gap lies above or below the engine's initial state.
More levels exist above the initial state than below,
if the initial state is selected
according to a Gibbs distribution at $\THot < \infty$.
Hence APT transitions tend to hop the engine upward,
costing an amount
\begin{align}
   \label{eq:APT_main}
   \expval{ W_{ \APT } }  & \sim
   \frac{ 1 }{ \sqrt{\Sites} }  \:
   \frac{ v^2  \betaH }{ \HScale  \dAvg }  \:
   \log \left( \frac{ \dAvg^2 }{ v }  \right)  \,
   e^{ - \Sites ( \betaH \HScale )^2 / 4 }  
\end{align}
of work per trial, on average.

Suppose that the engine starts at $\THot = \infty$.
APT transitions have no work to do during stroke 1, on average,
by the argument above.
As expected, the right-hand side of Eq.~\eqref{eq:APT_main} vanishes.

The logarithm in Eq.~\eqref{eq:APT_main} is 
a regulated divergence.
Let $P_\APT ( n | m )$ denote the probability
of the engine's hopping from level $m$ to level $n$.
The probability diverges as the difference $| E_n  -  E_m |$ 
between the levels' energies shrinks:
$P_\APT ( n | m )  \to  \infty$  as $| E_n  -  E_m |  \to  0$.
The consequent divergence in $\expval{ W_\APT }$ is logarithmic.
We cut off the $\expval{ W_\APT }$ integral at
the greatest energy difference that contributes significantly to the integral,
$| E_n  -  E_m |  \sim  \dAvg$.
The logarithm diverges in the adiabatic limit, as $v \to 0$.
Yet the $v^2$ in Eq.~\eqref{eq:APT_main} vanishes more quickly,
sending $\expval{ W_\APT }$ to zero, as expected.

The exponential in Eq.~\eqref{eq:APT_main} results from
averaging over the thermal initial state,
$e^{ - \betaH H_\ETH } / Z$.
Since the hot bath is hot, $\sqrt{ \Sites }  \:  \betaH \HScale  \ll  1$,
the exponential $\sim 1$. 
The $\frac{1}{ \sqrt{ \Sites } }$ and the logarithm
scale subdominantly in the system size.

Let us recast the dominant factors in terms of 
small dimensionless parameters:
$\expval{ W_{\APT} }  
\sim \left( \frac{ \sqrt{ v } }{ \dAvg }  \right)^4
( \sqrt{ \Sites }  \,  \betaH  \HScale )  
\left(  \frac{ \dAvg }{ \HScale }  \right)^2
\dAvg$.
The average work cost is suppressed 
fourfold in $\frac{ \sqrt{v} }{ \dAvg }  \ll  1$,
is suppressed linearly in $\sqrt{ \Sites }  \:  \betaH \HScale  \ll  1$,
and is twofold large in $\frac{ \dAvg }{ \HScale }  \gg  1$.

   \item \emph{Landau-Zener transitions:}
   Landau-Zener-type transitions overshadow APT transitions 
in the MBL phase.
Consider tuning the Hamiltonian parameter $\alpha_t$ within the MBL regime
but at some distance from the deep-localization value $1$.
Energies drift close together and separate.
When the energies are close together,
the engine can undergo a Landau-Zener transition~\cite{Landau_Zener_Shevchenko_10}
(Suppl. Mat.~\ref{section:WLZ}).
Landau-Zener transitions hop the engine from one energy level to a nearby level.
(General APT transitions hop the engine to arbitrary levels.)

Landau-Zener transitions cost zero average work, due to symmetries:
$\expval{ W_\LZ } = 0$.
The $j^\th$ level as likely wiggles upward, toward the $(j + 1)^\th$ level,
as it wiggles downward, toward the $(j - 1)^\th$ level.
The engine as likely consumes work $W > 0$, during a Landau-Zener transition,
as it outputs work $W > 0$.
The consumption cancels the output, on average.

   \item \emph{Fractional-Landau-Zener transitions:} 
   At the beginning of stroke 3, 
   nonequilbrium effects could excite the system 
   back across the small gap to energy level $j$. 
   The transition would cost work
   and would prevent the trial from outputting $W_\tot > 0$.
   We dub this excitation a fractional-Landau-Zener (frac-LZ) transition.
   It could be suppressed by 
   a sufficiently slow drive \cite{DeGrandi_10_APT}.
   The effects, and the resultant bound on $v$,
   are simple to derive
   (see Suppl. Mat.~\ref{section:HalfLZ} for details).
   
   Let the gap start stroke 3 at size $\delta$
   and grow to a size $\Delta > \delta$.
   The probability of a frac-LZ transition between 
   a small gap and a large gap $\delta  <  \Delta$ is~\cite{DeGrandi_10_APT}
\begin{align}
   \label{eq:HalfLZ_mainP}
   p_{ \HalfLZ }(\delta)  
   \approx   \frac{ v^2 ( \deltaMBL )^2 }{ 16}    
    \left(   \frac{1}{ \delta^6 }  +  \frac{1}{\Delta^6}  \right)
    \approx    \frac{  v^2 ( \deltaMBL )^2  }{ 16 \delta^6 }  \, .
\end{align}
$\deltaMBL$ denotes the MBL level-repulsion scale, 
the characteristic matrix element 
introduced, by a perturbation, between 
eigenstates of an unperturbed Hamiltonian. 
This mode of failure must be factored into 
the success probability $p_\cold$ of stroke-2 cooling. 
To suppress the probability of
a frac-LZ transition, the gap must satisfy 
$\delta  >  \delta_\Min
:=  (v  \deltaMBL  /  4)^{1/3}$. 
Neglecting $\TCold  >  0$ and $\Wb / \dAvg$ corrections, 
we modify Eq.~\eqref{eq:PCold}:
\begin{align}
   p_\cold 
   \approx    \int\limits_{\delta_\Min}^{\Wb} d\delta  \; 
   P_\MBL (\delta)  \left[  1  -  p_{\HalfLZ}(\delta)  \right]
   \approx \frac{\Wb - \delta_\Min }{\dAvg}  \, .
\end{align} 
To avoid frac-LZ costs, 
we must have $\Wb$ must  $\gg \delta_m$, and 
\begin{align}
   v  \ll  \frac{4  ( \Wb )^3}{\deltaMBL}
\end{align}
Since $\Wb  /  \deltaMBL  \gg 1$,
$v  <  ( \Wb )^2$.  

\end{enumerate}

\subsection{MBL engine in the thermodynamic limit}
\label{section:Thermo_limit_main}

The mesoscale engine has two drawbacks.
Consider increasing the system size $\Sites$.
The average gap declines exponentially:
$\dAvg  \sim  \frac{  \HScale  \sqrt{ \Sites }  }{  2^\Sites  }$.
Hence the average work extracted per trial,
$\expval{W_\tot}  \sim  \Wb  \ll  \dAvg$,
declines exponentially.
Additionally, the tuning speed $v$ must shrink exponentially:
$H_\meso(t)$ is ideally tuned quantum-adiabatically.
The time per tuning stroke must far exceed 
$\dAvg^{ - 1 }$.
The mesoscale engine scales poorly,
but properties of MBL offer a solution.

We introduce a thermodynamically large, or macroscopic, MBL Otto engine.
The engine consists of mesoscale subengines
that operate mostly independently.
This independence hinges on \emph{local level correlations} of the MBL phase,
detailed in Sec.~\ref{section:Local_level_corr_main}:
Energy eigenstates localized near each other spatially
tend to correspond to far-apart energies and vice versa.
Local level correlations inform the engine 
introduced in Sec.~\ref{section:Describe_macro_main}.
The engine cycle lasts for a time $\tau_\cycle$ that obeys three constraints,
introduced in Sec.~\ref{section:Times_main}.
We focus on exponential scaling behaviors.

%
%
%
\subsubsection{Local level correlations}
\label{section:Local_level_corr_main}

Consider subsystems, separated by a distance $L$,
of an MBL system.
The subsystems evolve roughly independently 
until times exponential in $L$,
due to the localization~\cite{Nandkishore_15_MBL}.
We apply this independence to parallelize mesoscale engines
in different regions of a large MBL system.
This application requires us to shift focus
from whole-system energy-level statistics
to \emph{local level correlations}~\cite{Sivan_87_Energy,imryma,Syzranov_17_OTOCs}.
We review local level correlations here.


An MBL system has a complete set of quasilocal integrals of motion~\cite{Nandkishore_15_MBL}.\footnote{
``Local'' refers to spatial locality here.
``Quasilocal'' means that each integral of motion
 can be related to a local operator 
 via a finite-depth unitary transformation 
 that consists only of local unitaries, up to exponentially small corrections.}
Thus, each integral of motion can be associated with a lattice site. 
This association is unique, other than for 
a small fraction of the integrals of motion. 

Let $\ket{ \psi_1 }$ and $\ket{ \psi_2 }$ denote
many-body energy eigenstates associated with 
the eigenvalues $E_1$ and $E_2$.
$\ket{ \psi_1 }$ and $\ket{ \psi_2 }$ are eigenstates of 
every integral of motion~\cite{Nandkishore_15_MBL}.
Let $O$ denote a generic strictly local operator.
$O$ is represented, relative to the energy eigenbasis, 
by matrix elements
$O_{21}  :=  \bra{ \psi_2 }  O  \ket{ \psi_1 }$.
Local level correlations interrelate
(1) the matrix-element size $| O_{21} |$ and
(2) the difference $| E_1 - E_2 |$ between the states' energies. 

%
%

Suppose that $\ket{ \psi_1 }$ and $\ket{ \psi_2 }$
correspond to the same configurations of the integrals of motion,
of energy, and of particle density
everywhere except in a size-$L$ region.
Such eigenstates are said to be ``close together,'' or
``a distance $L$ apart.''
Let $\xi$ denote the system's localization length.
If the eigenfunctions lie close together ($L \ll \xi$), 
the matrix-element size scales as 
\begin{align}
   \label{eq:O_Far}
   | O_{21} |  \sim  2^{ -L }  \, .
\end{align}
All lengths appear in units of the lattice spacing, set to one. 
If the states are far apart ($L \gg \xi$), 
\begin{align}
   \label{eq:O_Close}
   | O_{21} | \sim  
   e^{ - L / \xi }  \,  2^{ - L }  \, .
\end{align}

%

Having related the matrix-element size $| O_{21} |$
to the spatial separation $L$,
we relate $L$ to the energy difference $| E_1  -  E_2 |$.
Spatially close-together wave functions 
($L  \leq  \xi$) hybridize.
Hybridization prevents $E_1$ and $E_2$
from having an appreciable probability of lying
within $\HScale e^{-L/ \xi } \,  2^{ - L }$ of one another
(see~\cite{Anderson_58_Absence,Sivan_87_Energy,BAA,Nandkishore_15_MBL}
and Suppl. Mat.~\ref{section:ThermoLimitApp}).
Hence small energy differences 
correlate with rearrangements of particles
across large distances,
which correlate with
small matrix elements:\footnote{
These features are consistent with globally Poisson level statistics:
Suppose that $E_1$ and $E_2$ denote 
large nearest-neighbor energies.
$\ket{ \psi_1 }$ and $\ket{ \psi_2 }$ typically represent configurations 
that differ at extensively many sites.
Hence $| O_{21} |  \sim  e^{-L/ \xi } \,  2^{ - L }$.
This matrix element is exponentially smaller, in $L$, than
the average gap $2^{-L }$ implied by Poisson statistics.}
\begin{align}
   \label{eq:Small_EDiff}
   & | E_1  -  E_2 |  \ll  \HScale  e^{-L/ \xi } \,  2^{ - L }
   \quad  \leftrightarrow  \quad
   L  \gg  \xi
   \quad  \leftrightarrow  \nonumber \\ &
   | O_{21} |  \sim  e^{-L/ \xi } \,  2^{ - L }  \, .
\end{align}
Conversely, large energy differences 
correlate with rearrangements of particles
across small distances,
which correlate with
large matrix elements:
\begin{align}
   \label{eq:Large_EDiff}
   | E_1  -  E_2 |  \gg  \HScale  2^{ - L }
   \quad  \leftrightarrow  \quad
   L  \ll  \xi
   \quad  \leftrightarrow  \quad
   | O_{21} |  \sim  2^{ - L }  \, .
\end{align}


%
%
%
\subsubsection{Application of local level correlations
in the macroscopic MBL engine}
\label{section:Describe_macro_main}

We apply local level correlations in constructing
a scalable generalization of the mesoscale Otto engine.
We denote properties of the macroscopic, composite engine
with the subscript ``macro.''
(For example, as $\Sites$ denoted the number of sites in
a mesoscale engine, 
$\Sites_\macro$ denotes 
the number of sites in the macroscopic engine.)
Strokes 1 and 3 require modification:
The Hamiltonian $H_\macro(t)$ is tuned within the MBL phase, 
between a point analogous to $H_\ETH$
and a point analogous to $H_\MBL$.

The $H_\ETH$-like Hamiltonian has 
a localization length  $\xi_\Loc$;
and $H_\MBL$-like Hamiltonian, 
$\xi_\VeryLoc  \ll  \xi_\Loc$.
We illustrate with $\xi_\Loc  =  1$ and $\xi_\VeryLoc  =  12$
in Suppl. Mat.~\ref{section:Optimize}.
Particles mostly remain in regions of, at most, length $\xi_\Loc$.
Such regions function as ``subengines,''
instances of the mesoscale engine.
What happens in a subengine stays in a subengine.

This subdivision boosts the engine's power.
A length-$\Sites$ mesoscale engine operates at
the per-cycle power
$\expval{ W_\tot }  \sim  \Wb  \ll  \dAvg
\sim  \frac{ \HScale \sqrt{\Sites} }{ 2^\Sites }$
(Sec.~\ref{section:Quant_main}).
Suppose that the whole system
consisted of one length-$\Sites_\macro$ engine.
The power would scale as
$\sim  \frac{ \HScale \sqrt{\Sites_\macro} }{ 2^{\Sites_\macro} }$.
This quantity $\to 0$ in the thermodynamic limit,
as $\Sites_\macro  \to  \infty$.
But our engine consists of length-$\xi_\Loc$ subengines.
Local level correlations give each subengine
an effective average gap
\begin{align}
   \label{eq:DAvg_Subeng}
   \dAvg  
   \sim  \frac{  \HScale  \sqrt{ \xi_\Loc }  }{  
                      2^{ \xi_\Loc }  }  
   \sim  \frac{  \HScale  }{   2^{ \xi_\Loc }  }
\end{align}
The composite-engine power $\expval{ W_\tot }_\macro$
is suppressed not in $\Sites_\macro$, 
but in the subengine length $\xi_\Loc$:
\begin{align}
   \label{eq:Composite_W_Scale}
   \expval{ W_\tot }_\macro
   \sim   \Sites_\macro    \,
   \frac{  \sqrt{ \xi_\Loc }  }{     2^{ \xi_\Loc }  }  \,  
   \HScale \, .
\end{align}

\subsubsection{Time scales of the macroscopic MBL engine}
\label{section:Times_main}

Three requirements constrain the time for which a cycle is implemented:
(1) Subengines must operate mostly independently.
Information propagates between subengines,
albeit slowly due to localization.
$H_\macro(t)$ must be tuned too quickly
for much information to cross-pollinate subengines 
(Suppl. Mat.~\ref{section:Optimize}).
(2) Tuning at a finite speed $v > 0$
induces diabatic transitions between energy levels (Sec.~\ref{section:Diab_main}).
$v$ must be small enough to suppress the average work cost,
$\expval{ W_\diab }$, of undesirable diabatic transitions:
$\expval{ W_\diab }  \ll  \expval{ W_\tot }$
(Suppl. Mat.~\ref{section:Optimize}).
(3) The cold bath has a small bandwidth, $\Wb  \ll  \dAvg$;
couples to the engine with a small strength $g$;
and interacts locally.
Stroke 2 must last long enough 
to thermalize each subengine nonetheless.
We detail these requirements and
bound the cycle time, $\tau_\cycle$.
$\tau_\cycle$ may be optimized via, 
e.g., shortcuts to adiabaticity~\cite{Chen_10_Fast,Kosloff_10_Optimal,Torrontegui_13_Shortcuts,Deng_13_Boosting,del_Campo_14_Super,Abah_16_Performance,Cakmak_17_Irreversible}.

\paragraph{Lower bound on the tuning speed $v$ 
from the subengines' (near) independence:}
The price paid for scalability is the impossibility of adiabaticity. 
Suppose that $H_\macro(t)$ were tuned infinitely slowly.
Information would have time to propagate 
from one subengine to every other.
The slow spread of information through MBL~\cite{Khemani_15_NPhys_Nonlocal}
lower-bounds the tuning speed.
We introduce notation, then sketch the derivation,
detailed in Suppl. Mat.~\ref{section:Optimize}.

Let $\J_L$ denote the level-repulsion scale---the 
least width reasonably attributable to any gap---of 
a length-$L$ MBL system.
(The $\deltaMBL$ introduced earlier equals $\J_\Sites = \J_{\xi_\Loc}$.)
The time-$t$ localization length is denoted by $\xi(t)$.
A length-$L$ MBL system's average gap
is denoted by $\dAvg^\LL$. 
(The average subengine gap $\dAvg$, introduced earlier, 
equals $\dAvg^{ ( \xi_\Loc ) }$.)

The engine must not lose too much work to
undesirable adiabatic transitions.
During tuning, energy levels approach each other.
Typically, if such a ``close encounter'' results in an adiabatic transition, 
many particles shift across the engine.
Subengines effectively interact, consuming a total amount
$\sim \Sites_\macro \expval{ W_\adiab^\cost }$ of work, on average.
Undesirable adiabatic transitions must cost less than
the average work~\eqref{eq:Composite_W_Scale}
outputted by  ideal (independent) subengines:
\begin{align}
   \label{eq:Lower_v_condn_main}
   \expval{ W_\adiab^\cost }  \ll  \expval{ W_\tot }  \, .
\end{align}

We approximate the left-hand side with
\begin{align}
   \label{eq:W_adiab_cost_main}
   & \expval{ W_\adiab^\cost }  \approx
   \left(  \frac{ \text{Work cost} }{ \text{1 undesirable adiab. transition} }  \right)
   \\ \nonumber & \qquad \times
   \left( \frac{ \text{Prob. of undesirable adiab. transition} }{ 
                    \text{1 close encounter} } \right)
   \\ \nonumber &  \times
   \left(  \frac{ \text{\# close encounters} }{ \text{1 tuning stroke} }  \right)
   \\ \nonumber & \qquad \times
   \left(  \frac{ \text{Avg. \# strokes during which can lose work} }{ 
   \text{1 cycle} }  \right)  \, .
\end{align}
The first factor $\sim \dAvg$.
The second factor follows from the Landau-Zener probability 
$P_\LZ = e^{ - 2 \pi \J^2 / v }  \sim  1 - \frac{ \J^2 }{ v }$
that any given close encounter induces a diabatic transition.
The Hamiltonian-matrix element
that couples the approaching states 
has the size $\J  \sim  \J_{1.5 \xi_\Loc}$.
The $1.5 \xi_\Loc$ encodes nearest-neighbor subengines' isolation:
Information should not propagate from the left-hand side of one subengine
rightward, across a distance $1.5 \xi_\Loc$, 
to the neighbor's center.
We estimate the third factor in Eq.~\eqref{eq:W_adiab_cost_main} as
$\frac{ \dAvg }{ \dAvg^{ (1.5) } }$.
This 1.5 has the same origin as the 1.5 in the $\J_{1.5 \xi_\Loc}$.
The final factor in Eq.~\eqref{eq:W_adiab_cost_main} 
$\sim \frac{ \Wb }{ \dAvg }$,
the fraction of the cycles that would, in the absence of undesirable transitions,
output $W_\tot > 0$.

Upon substituting into Eq.~\eqref{eq:W_adiab_cost_main},
we substitute into Ineq.~\eqref{eq:Lower_v_condn_main}.
The right-hand side $\sim \Wb$ [Eq.~\eqref{eq:WTotApprox2_Main}].
Solving for $v$ yields 
\begin{align}
   \label{eq:LowerVBound_Main}
   v   & \gg 
   (  \J_{1.5 \xi_\Loc } )^2   \:  
   \frac{ \dAvgSub }{ \dAvg^{ ( 1.5 \xi_\Loc ) } }  \\
   & \label{eq:LowerVBound_Main2}
   \sim  \HScale^2  \:  e^{ - 3 \xi_\Loc  /  \xi(t) }  
   \;  2^{ - 2.5 \xi_\Loc }
\end{align}

\paragraph{Upper bound on $v$ from 
the work cost $\expval{ W_\diab }$ of
undesirable diabatic transitions:}
Tuning at a finite speed $v > 0$ induces diabatic transitions,
(Sec.~\ref{section:Diab_main}).
Diabatic hops cost a subengine 
an amount $\expval{ W_\diab }$ of work per cycle, on average.
For clarity, we relabel as $\expval{ W_\tot^\adiab }$ 
the average work outputted by one ideal subengine,
tuned adiabatically, per cycle.
The requirement $\expval{ W_\diab }  \ll  \expval{ W_\tot^\adiab }$
upper-bounds $v$ 
(Suppl. Mat.~\ref{section:Optimize}).

When the engine is shallowly localized,
APT transitions dominate $\expval{ W_\diab }$ 
[Eq.~\eqref{eq:APT_main}].
They pose little risk if
the speed is small, compared to the typical gap:
\begin{align}
   v   \ll  \dAvg^2  
   & \sim  \frac{ \HScale^2 }{ \HDim^2 }  
   \label{eq:v_small_param_main2}
   \sim  \HScale^2  2^{ - 2 \xi_\Loc }  \, .
\end{align}
The third expression follows from
(i) the text below Eq.~\eqref{eq:WTotApprox2_Main} and
(ii) the subdominance of $\sqrt{ \Sites }$ in
our scaling analysis.
The final expression approximates $\dAvg^2$ because
the tuning rearranges particles across each subengine,
across a distance $L  \sim  \xi$.
Such rearrangements induce the energy changes
in~\eqref{eq:Large_EDiff}.

When the engine is very localized, fractional-Landau-Zener transitions 
dominate $\expval{ W_\diab }$. 
Equation~\eqref{eq:HalfLZ_main} approximates,
under $\Err  \approx  \frac{1}{3}$, to
$\expval{ W_\HalfLZ }  
\sim  \frac{ v^2  \left(  \deltaMBL  \right)^2  }{  \left(  \Wb  \right)^5  }
+  \frac{1}{3}  \,  \Wb$.
This work cost must be far less than
the work $\expval{ W_\tot }$ extracted adiabatically:
$\expval{ W_\HalfLZ }  \ll  \expval{ W_\tot }$.
Solving for $v$ yields
\begin{align}
   \label{eq:Lower_v_bd_main_HalfLZ}
   v  & \ll  \frac{ ( \Wb )^3 }{ \deltaMBL }
   \sim \frac{ 1 }{ 10^3 }   \;
   e^{ \xi_\Loc / \xi_\VeryLoc }  \:  2^{ - 2 \xi_\Loc }
   \HScale^2  \, .
\end{align}
The final expression follows if $\Wb  \sim  \frac{ \dAvg }{ 10 }$.
Both upper bounds,~\eqref{eq:v_small_param_main2}
and~\eqref{eq:Lower_v_bd_main_HalfLZ},
lie above the lower bound~\eqref{eq:LowerVBound_Main2},
in an illustrative example in which
$\xi_\Loc  =  12$,  $\xi_\VeryLoc  =  1$, and
$\xi (t)  \sim  \xi_\Loc$.

\paragraph{Lower bound on the cycle time $\tau_\cycle$ from cold thermalization:}
Thermalization with the cold bath (stroke 2) 
bounds $\tau_\cycle$ more stringently
than the Hamiltonian tunings do.
The reasons are (1) the slowness with which MBL thermalizes and
(2) the restriction $\Wb  \ll  \dAvg$ on the cold-bath bandwidth.
We elaborate after introducing our cold-thermalization model
(see~\cite[App.~I]{NYH_17_MBL} for details).

We envision the cold bath as a bosonic system
that couples to the engine locally, as via the Hamiltonian
\begin{align}
   \label{eq:H_Inter_Main}
   H_\inter  & =  \coupling  
   \int_{ - \Wb /  \xi_\Loc }^{ \Wb / \xi_\Loc } d \omega   
   \sum_{j = 1}^{ \Sites_\macro }
   \left(  c_j^\dag c_{ j + 1 }   +  \hc  \right)
   \left(  b_\omega  +  b_\omega^\dag  \right)
   \nonumber \\ & \qquad \times
   \delta \LParen  \langle 0 | c_j  H_\macro ( \tau ) c_{j + 1}^\dag | 0 \rangle
                            -  \omega  \RParen  \, .
\end{align}
The coupling strength is denoted by $\coupling$.
$c_j$ and $c_j^\dag$ denote
the annihilation and creation of a fermion at site $j$.
$H_\macro (t)$ denotes the Hamiltonian that would govern the engine
at time $t$ in the bath's absence.
Cold thermalization lasts from $t = \tau$ to $t = \tau'$ (Fig.~\ref{fig:2Level_v3}).
$b_\omega$ and $b_\omega^\dag$ represent
the annihilation and creation of a frequency-$\omega$ boson in the bath.
The Dirac delta function is denoted by $\delta ( . )$.

The bath couples locally, e.g., 
to pairs of nearest-neighbor spins.
This locality prevents subengines from 
interacting with each other much through the bath.
The bath can, e.g., flip spin $j$ upward while flipping spin $j + 1$ downward.
These flips likely change a subengine's energy by an amount $E$.
The bath can effectively absorb only energy quanta of size
$\leq  \Wb$ from any subengine.
The cap is set by the bath's speed of sound~\cite{KimPRL13},
which follows from microscopic parameters in 
the bath's Hamiltonian~\cite{Lieb_72_Finite}.
The rest of the energy emitted during the spin flips, $| E - \Wb |$, 
is distributed across the subengine
as the intrinsic subengine Hamiltonian flips more spins.

Let $\tau_\therm$ denote the time required for stroke 2.
We estimate $\tau_\therm$ from Fermi's Golden Rule,
\begin{align}
   \label{eq:FGR_Main}
   \Gamma_{fi}  =  \frac{2 \pi }{ \hbar }  
   | \langle f | V | i \rangle |^2  \,  \DOS_\bath  \, .
\end{align}
Cold thermalization transitions the engine
from an energy level $\ket{ i }$ to a level $\ket{ f }$.
The bath has a density of states
$\DOS_\bath  \sim  1 / \Wb$.

We estimate the matrix-element size
$| \langle f | V | i \rangle |$ as follows.
Cold thermalization transfers energy $E_{if}  \sim  \Wb$
from the subengine to the bath.
$\Wb$ is very small.
Hence the energy change 
rearranges particles across a large distance 
$L  \gg  \xi  =  \xi_\VeryLoc$,
due to local level correlations~\eqref{eq:Small_EDiff}.
$V$ nontrivially transforms just a few subengine sites.
Such a local operator rearranges particles 
across a large distance $L$
at a rate that scales as~\eqref{eq:Small_EDiff},
$\HScale e^{ - L / \xi }  \;  2^{ - L }
\sim  \deltaMBL$.
Whereas $\HScale$ sets the scale of the level repulsion $\deltaMBL$,
$\coupling$ sets the scale of $| \langle f | V | i \rangle |$.
The correlation length $\xi = \xi_\VeryLoc$ during cold thermalization.
We approximate $L$ with the subengine length $\xi_\Loc$. Hence
$| \langle f | V | i \rangle |  \sim  \frac{ \coupling \deltaMBL }{ \HScale}$.

We substitute into Eq.~\eqref{eq:FGR_Main}.
The transition rate $\Gamma_{fi}  =  \frac{1}{ \tau_\therm }$.
Inverting yields
\begin{align}
   \label{eq:Tau_therm_main}
   \tau_\cycle  \sim  \tau_\therm  \sim
   \Wb   \left(  \frac{  \HScale }{ \coupling \deltaMBL }  \right)^2  \, .
\end{align}

To bound $\tau_\cycle$, we must bound the coupling $\coupling$.
The interaction is assumed to be Markovian:
Information leaked from the engine dissipates throughout the bath quickly.
Bath correlation functions must decay much more quickly
than the coupling transfers energy.
If $\tau_\bath$ denotes the correlation-decay time,
$\tau_\bath  <   \frac{1}{ \coupling }$.
The small-bandwidth bath's $\tau_\bath  \sim  1 / \Wb$. Hence
$\coupling  <  \Wb$.
This inequality, with Ineq.~\eqref{eq:Tau_therm_main}, implies
\begin{align}
   \label{eq:Markov_main}
   \tau_\cycle
   =  \tau_\therm  
   >  \frac{ \HScale^2 }{  \Wb  ( \deltaMBL )^2  }
   \sim  \frac{ 10 }{ \HScale }  \:
   e^{ 2 \xi_\Loc  / \xi_\VeryLoc }  \:  2^{ 3 \xi_\Loc }  \, .
\end{align}
The final expression follows if $\Wb  \sim \frac{ \dAvg }{ 10 }$.

Like Markovianity, higher-order processes bound $\tau_\therm$.
Higher-order processes occur at rates set by $g^a$,
wherein $a > 1$.
Such processes transfer energy $E > \Wb$ between
the engine and the cold bath.
These transfers must be suppressed.
The resulting bound on $\tau_\therm$ 
is less stringent than Ineq.~\eqref{eq:Markov_main}
(Suppl. Mat.~\ref{section:Tau_therm_virtual_app}).

\section{Numerical simulations}
\label{section:Numerics_main}

The engine can be implemented with a disordered Heisenberg model.
A similar model's MBL phase has been realized with cold atoms~\cite{Schreiber_15_Observation}.
We numerically simulated a 1D mesoscale chain of $\Sites = 12$
spin-$\frac{1}{2}$ degrees of freedom,
neglecting dynamical effects during strokes 1 and 3 (the Hamiltonian tunings).
The chain evolves under the Hamiltonian
\begin{align} 
      \label{eq:SpinHam}
      H_\Sim(t)  = \frac{\HScale}{ Q \LParen h ( \alpha_t ) \RParen }   \Bigg[  
      \sum_{j = 1}^{\Sites - 1}   
      \bm{\sigma}_j  \cdot  \bm{\sigma}_{j+1}
      +  h ( \alpha_t )  \sum_{j = 1}^\Sites
      h_j     \sigma_j^z  \Bigg] \, .
\end{align}
Equation~\eqref{eq:SpinHam} describes spins 
equivalent to interacting spinless fermions.
Energies are expressed in units of $\HScale$,
the average per-site energy density.
For $\gamma=x,y,z$, the $\gamma^\th$ Pauli operator 
that operates nontrivially on the $j^\th$ site is denoted by 
$\sigma_j^\gamma$.
The Heisenberg interaction
$\bm{\sigma}_j \cdot \bm{\sigma}_{j+1}$ 
encodes nearest-neighbor hopping and repulsion.

The tuning parameter $\alpha_t\in[0,1]$
determines the phase occupied by $H_\Sim(t)$.
The site-$j$ disorder potential depends on 
a random variable $h_j$ distributed uniformly across $[-1,1].$
The disorder strength $h(\alpha_t)$ varies as
$h(\alpha_t)=\alpha_t\,h_\ETH+(1-\alpha_t)h_\MBL$.
When $\alpha_t = 0$, the disorder is weak, $h = h_\ETH$,
and the engine occupies the ETH phase. 
When $\alpha_t = 1$, the disorder is strong, $h = h_\MBL  \gg  h_\ETH$,
and the engine occupies the MBL phase.


The normalization factor $Q\LParen h(\alpha_t)\RParen$
preserves the width of the density of states (DOS)
and so $\dAvg$.
$Q \LParen h(\alpha_t)\RParen$ prevents 
the work extractable via change of bandwidth 
from polluting
the work extracted with help from level statistics,
(Sec.~\ref{section:Meso_setup}).
$Q \LParen h(\alpha_t)\RParen$ is defined 
and calculated in Suppl. Mat.~\ref{section:MBLNumApp:scale}.

\begin{figure}
\centering
  \begin{subfigure}{0.45\textwidth}
    \includegraphics[width=\textwidth]{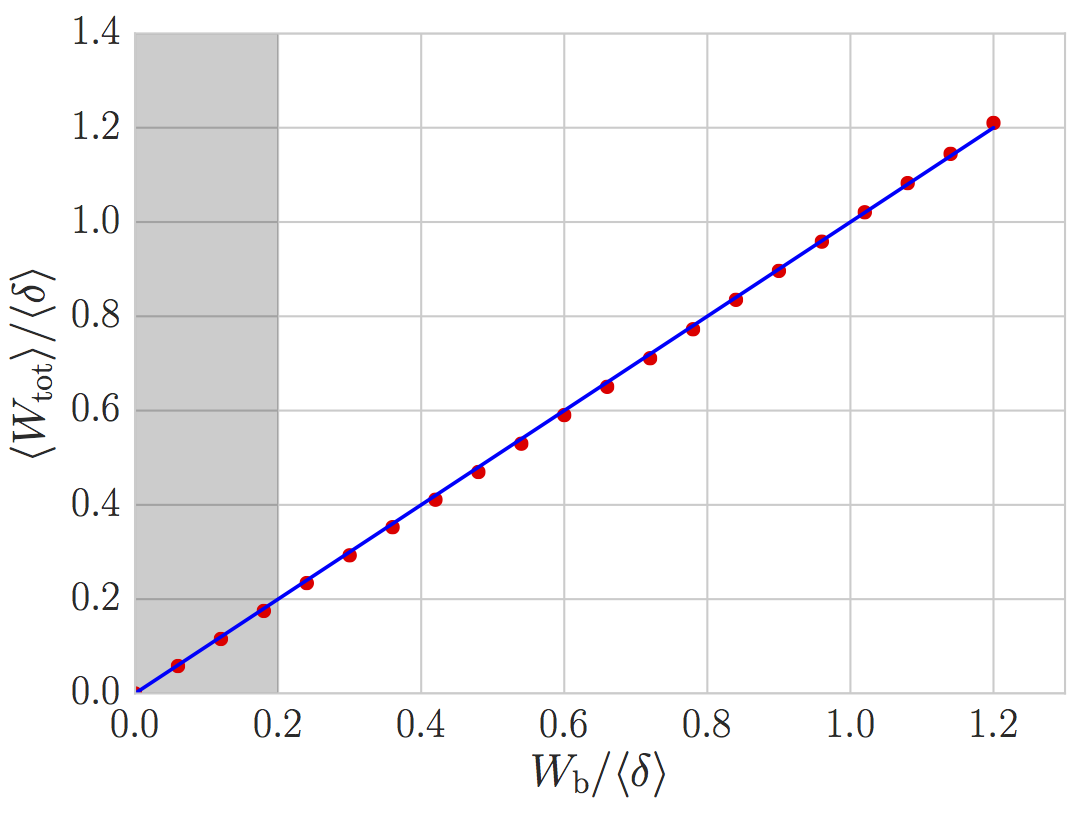}
    \caption{}
    \label{fig:Numerics_main_WTOT}
  \end{subfigure}
  %
  \begin{subfigure}{0.45\textwidth}
    \includegraphics[width=\textwidth]{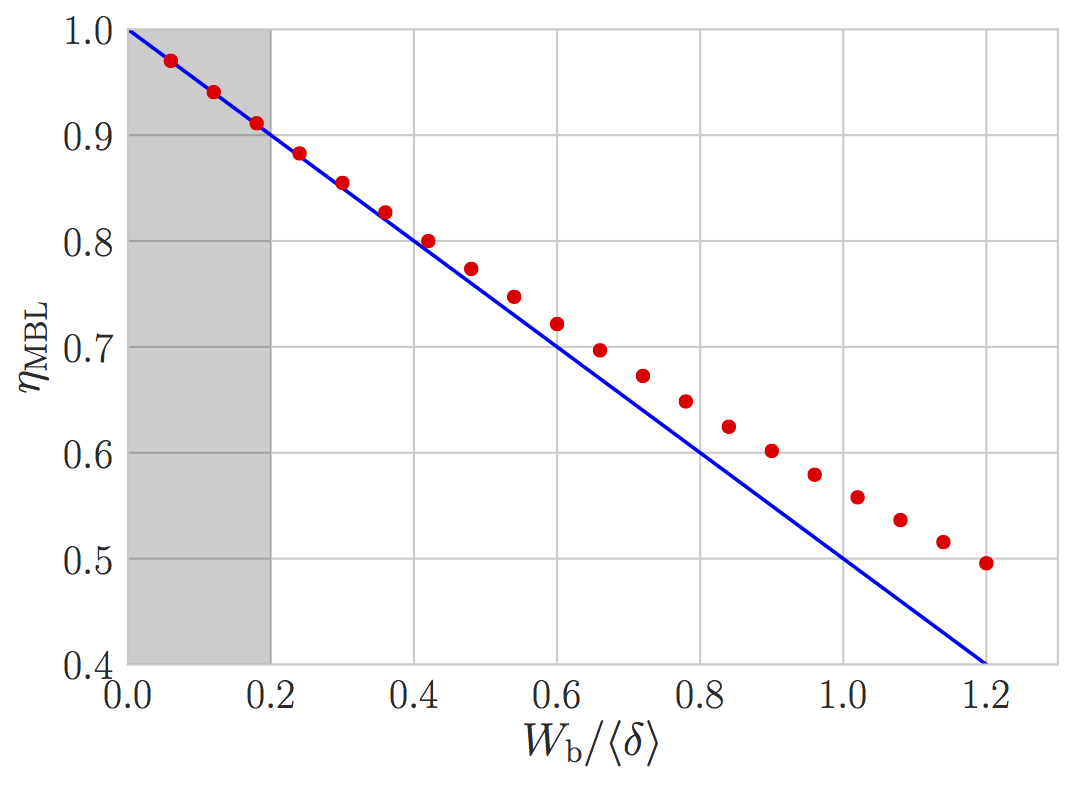}
    \caption{}
    \label{fig:Numerics_main_eta}
  \end{subfigure}
  \caption{\caphead{
    Average per-cycle power $\expval{ W_\tot }$ (top) 
    and efficiency  $\eta_\MBL$ (bottom) as functions of 
    the cold-bath bandwidth $\Wb$:}
    Each red dot represents an average over 1,000 disorder realizations of the random-field Heisenberg Hamiltonian~\eqref{eq:SpinHam}.
    The slanted blue lines represent the analytical predictions~\eqref{eq:WTotApprox2_Main} and~\eqref{eq:Eff_MBL}.
    When $\Wb \ll \dAvg$ (in the gray shaded region), 
    $\expval{ W_\tot }$ and $\eta_\MBL$ vary linearly with $\Wb$, 
    as predicted.}
  \label{fig:Numerics_main}
\end{figure} 

We simulated the spin chain using exact diagonalization,
detailed in Suppl. Mat.~\ref{section:MBLNumApp}.
The ETH-side field had a magnitude $h(0) = 2.0$, 
and the MBL-side field had a magnitude $h(1) = 20.0$. 
These $h( \alpha_t )$ values fall squarely on opposite sides 
of the MBL transition at $h \approx 7$.

\subsection{Adiabatic engine performance}
We first simulated the evolution of each state in strokes 1 and 3 as though the Hamiltonian were tuned adiabatically.
We index the energies $E_j ( \alpha_t )$
from least to greatest at each instant:
$E_j ( \alpha_t )  <  E_k ( \alpha_t )  \;  \forall j < k$.
Let $\rho_j$ denote the state's weight 
on eigenstate $j$ 
of the pre-tuning Hamiltonian $H ( \alpha_t = 0 )$. 
The engine ends the stroke with weight $\rho_j$ 
on eigenstate $j$ of the post-tuning Hamiltonian $H ( 1 )$.

The main results appear in Fig.~\ref{fig:Numerics_main}.
Figure~\ref{fig:Numerics_main_WTOT} shows 
the average work extracted per cycle, $\expval{ W_\tot }$;
and Fig.~\ref{fig:Numerics_main_eta} shows 
the efficiency, $\eta_\MBL$.

In these simulations, the baths had
the extreme temperatures $\THot = \infty$ and $\TCold = 0$.
This limiting case elucidates the $\Wb$-dependence 
of $\expval{ W_\tot }$ and of $\eta_\MBL$:
Disregarding finite-temperature corrections,
on a first pass, builds intuition.
Finite-temperature numerics appear alongside 
finite-temperature analytical calculations
in Suppl. Mat.~\ref{section:PowerApp}.

Figure~\ref{fig:Numerics_main} shows how 
the per-cycle power and the efficiency
depend on the cold-bath bandwidth $\Wb$.
As expected, $\expval{ W_\tot }  \approx  \Wb$. 
The dependence's linearity, and the unit proportionality factor,
agree with Eq.~\eqref{eq:WTotApprox2_Main}.
Also as expected, the efficiency declines as the cold-bath bandwidth rises:
$\eta_\MBL  \approx  1  -  \frac{ \Wb }{ 2 \dAvg } \, .$
The linear dependence and the proportionality factor
agree with Eq.~\eqref{eq:Eff_MBL}.

The gray columns in Fig.~\ref{fig:Numerics_main} highlight the regime
in which the analytics were performed, where $\frac{ \Wb }{ \dAvg }  \ll  1$.
If the cold-bath bandwidth is small, $\Wb \lesssim \dAvg$, 
the analytics-numerics agreement is close.
But the numerics agree with the analytics even outside this regime.
If $\Wb \gtrsim \dAvg$, the analytics slightly underestimate $\eta_\MBL$:
The simulated engine operates more efficiently than predicted.
To predict the numerics' overachievement,
one would calculate higher-order corrections 
in Suppl. Mat.~\ref{section:PowerApp}:
One would Taylor-approximate to higher powers,
modeling subleading physical processes.
Such processes include the engine's dropping across
a chain of three small gaps $\delta'_1, \delta'_2, \delta'_3 < \Wb$
during cold thermalization.

The error bars are smaller than the numerical-data points.
Each error bar represents the error in the estimate of a mean
(of $\expval{ W_\tot }$ or of
$\eta_\MBL  :=  1  -  \frac{ \expval{ W_\tot } }{ \expval{ Q_\In } }$)
over 1,000 disorder realizations.
Each error bar extends a distance
$(\text{sample standard deviation})/\sqrt{\text{\# realizations}}$ 
above and below that mean.

\subsection{Diabatic engine performance}
We then simulated the evolution of each state in strokes 1 and 3 as though the Hamiltonian were tuned at finite speed for 8 sites.
  (We do not simulate larger diabatic engines: That our upper bounds on tuning speed for a mesoscopic engine go as powers of the level spacing $\dAvg \sim 2^{-L}$ means that these simulations quickly become slow to run.)
  We simulate a stepwise tuning, taking
  \begin{equation}
    \alpha(t) = (\delta t) \lfloor v t/(\delta t) \rfloor   \, .
  \end{equation}
  This protocol is considerably more violent than the protocols we
  treat analytically: In our estimates, we leave $v$ general, but we
  always assume that it is finite. In the numerics, we tune by a series of
  sudden jumps. (We do this for reasons of numerical convenience.)  We
  work at $\betaC = \infty$ and $\betaH = 0$, to capture the
  essential physics without the added confusion of finite-temperature
  corrections. In this case, we expect the engine to work \emph{well
    enough}---to output a finite fraction of its adiabatic
  work output---for
  \begin{equation}
    v \ll \frac{(\Wb)^3}{\delta_-}
  \end{equation}
  [c.f. Eq.~\eqref{eq:Lower_v_bd_main_HalfLZ}].
  
  In Fig.~\ref{fig:Numerics_main_diabatic-work}, we show work output as
  a function of speed. Despite the simulated protocol's violence,
  $W_\tot$ is a finite fraction of its adiabatic
  value for $v \lesssim \frac{(\Wb)^3}{\delta_-}$ and even for $v >
  \frac{(\Wb)^3}{\delta_-}$: Our engine is much less sensitive to
  tuning speed than our crude diabatic-corrections bounds suggest.
  
  \begin{figure}
  \begin{centering}
    \includegraphics[width=0.45\textwidth]{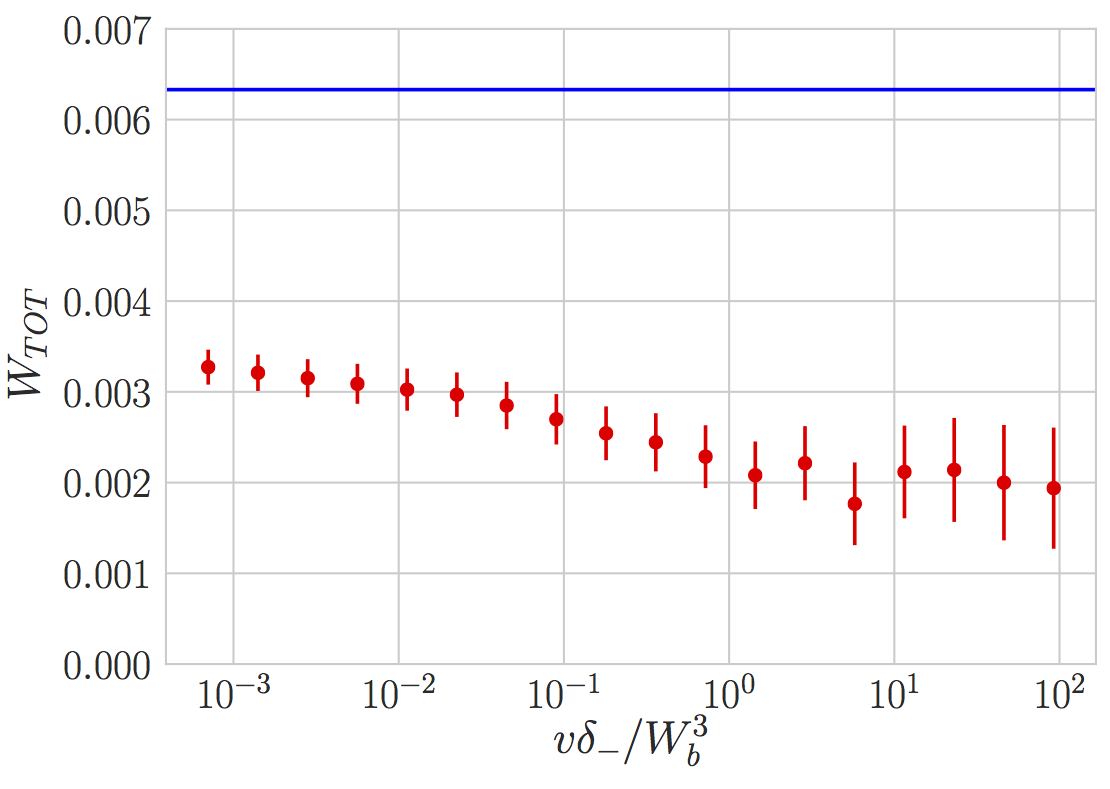}
    \caption{\caphead{Average per-cycle work as a function of tuning speed} for 995 disorder realizations of the random-field Heisenberg Hamiltonian~\eqref{eq:SpinHam} at system size $L = 8$ (red dots), compared to the analytical estimate \eqref{eq:WTotApprox2_Main} for the adiabatic work output (blue line). Each error bar represents the error in the estimate of the mean, computed as $(\text{sample standard deviation})/\sqrt(\#\ \text{realizations})$.
    }
    \label{fig:Numerics_main_diabatic-work}
  \end{centering}
  \end{figure}
  
  These numerics not only confirm the validity of our
  analytics, but also indicate the robustness of the MBL Otto engine to
  changes in the tuning protocol.

\section{Order-of-magnitude estimates}
\label{section:Order_main}

How well does the localized engine perform?
We estimate its power and power density,
then compare the values with three competitors' performances.

\textbf{Localized engine:}
Localization has been achieved in solid-state systems.\footnote{
This localization is single-particle, or Anderson~\cite{Anderson_58_Absence}, 
rather than many-body.
Section~\ref{section:Anderson_engine_main} extends
the MBL Otto engine to an Anderson-localized Otto engine.}
Consider silicon doped with phosphorus~\cite{Kramer_93_Localization}.
A distance of $\sim  10  \text{ nm}$
may separate phosphorus impurities.
Let our engine cycle's shallowly localized regime
have a localization length of $\xi_\Loc \sim 10$ sites, 
or $100  \text{ nm}$.
The work-outputting degrees of freedom will be electronic.
The localized states will correspond to energies 
$\HScale \sim 1 \text{ eV}$. 
Each subengine's half-filling Hilbert space has dimensionality 
$\HDim  =  {10 \choose 5}  \sim  10^2$.
Hence each subengine has an effective average gap
$\dAvg  \sim  \frac{ \HScale  \sqrt{\Sites} }{ \HDim }  
\sim  \frac{ 1  \text{ eV}  }{ 10^2 }  \sim  10  \text{ meV}$.
The cold-bath bandwidth must satisfy $\dAvg \gg \Wb \, .$
We set $\Wb$ to be an order of magnitude down from $\dAvg$:
$\Wb   \sim  1   \text{ meV}  \sim 10  \text{ K}$. 
The cold-bath bandwidth approximates 
the work outputted by one subengine per cycle:\footnote{
The use of semiconductors would require corrections to our results.
(Dipolar interactions would couple the impurities' spins.
Energy eigenfunctions would decay as power laws with distance.)
But we aim for just a rough estimate.}
$\expval{ W_\tot }  \sim  \Wb
\sim  1   \text{ meV}$
[Eq.~\eqref{eq:WTotApprox2_Main}].

What volume does a localized subengine fill?
Suppose that the engine is three-dimensional (3D).\footnote{
Until now, we have supposed that the engine is 1D.
Anderson localization, which has been realized in semiconductors,
exists in all dimensionalities.
Yet whether MBL exists in dimensionalities $D > 1$ 
remains an open question.
Some evidence suggests that MBL exists in $D \geq 2$~\cite{Choi_16_Exploring,Kucsko_16_Critical,Bordia_17_Probing}.
But attributing a 3D volume to the engine
facilitates comparisons with competitors.
We imagine 10-nm-long 1D strings of sites.
Strings are arrayed in a plane, separated by 10 nm.
Planes are stacked atop each other, separated by another 10 nm.}
A little room should separate the subengines.
Classical-control equipment requires more room.
Also, the subengine needs space to connect to the baths.
We therefore associate each subengine with a volume of
$V  \approx  (100 \text{ nm})^3$.

The last element needed is the cycle time, $\tau_\cycle$. 
We choose for $\deltaMBL$ to be 
a little smaller than $\Wb$---of the same order: 
$\deltaMBL  \sim  \Wb  \sim  1 \text{ meV}$.
In the extreme case allowed by Ineq.~\eqref{eq:Markov_main},
$\tau_\cycle  \sim  \frac{ \hbar \HScale^2 }{ \Wb ( \deltaMBL )^2 }
\sim  \frac{ \hbar  \HScale^2 }{ ( \Wb )^3 }
\sim  \frac{ ( 10^{ -15} \text{ eV s}  )  ( 1 \text{ eV} )^2 }{
         ( 1 \text{ meV} )^3 }
\sim  1 \text{ $\mu$s}$.

The localized engine therefore operates with a power 
$\Power  \sim  \frac{ \Wb }{ \tau_\cycle }
\sim  \frac{ 1 \text{ meV}  }{  1 \text{ $\mu$s} }
\approx 10^{-16}  \text{ W}$. 
Interestingly, this $\Power$ is one order of magnitude greater than 
a flagellar motor's~\cite{Brown_13_Bacterial} power,
according to our estimates.

We can assess the engine by calculating not only its power, 
but also its power density.
The localized engine packs a punch at 
$\frac{ \Power }{ V }  
\sim  \frac{ 10^{-16} \text{ W}  }{  ( 10^{-7} \text{ m} )^3 }
=  100 \text{ kW}/\text{m}^3$.

\textbf{Car engine:}
The quintessential Otto engine powers cars.
A typical car engine outputs
$\Power \sim 100 \text{ horsepower}  \sim  100 \text{ kW} \, .$
A car's power density is $\frac{ \Power }{ V }  
\sim  \frac{ 100 \text{ kW} }{ 100 \text{ L} }
=  1  \text{ MW} / \text{ m}^3$
(wherein L represents liters).
The car engine's $\frac{ \Power }{ V }$ exceeds 
the MBL engine's
by only an order of magnitude,
according to these rough estimates.

\textbf{Array of quantum dots:}
MBL has been modeled with
quasilocal bits~\cite{Huse_14_phenomenology,Chandran_15_Constructing}.
A string of ideally independent bits or qubits, such as quantum dots,
forms a natural competitor.
A qubit Otto engine's gap is shrunk, widened, and shrunk~\cite{Geva_92_Quantum,Geva_92_On,Feldmann_96_Heat,He_02_Quantum,Alvarado_17_Role}. 

A realization could consist of double quantum dots~\cite{Petta_05_Coherent,Petta_06_Charge}. 
The scales in~\cite{Petta_05_Coherent,Petta_06_Charge} suggest that 
a quantum-dot engine could output an amount 
$W_\tot  \sim 10  \text{ meV}$  of work per cycle. 
We approximate the cycle time $\tau_\cycle$ with
the spin relaxation time: $\tau_\cycle  \sim  1  \: \mu \text{s}$.
(The energy eigenbasis need not rotate,
unlike for the MBL engine.
Hence diabatic hops do not lower-bound
the ideal-quantum-dot $\tau_\cycle$.)
The power would be 
$\Power  \sim  \frac{ W_\tot }{ \tau_\cycle }  
\sim  \frac{ 10  \text{ meV}  }{  1  \:  \mu \text{s} }
\sim 10^{-15} \text{ W}$.
The quantum-dot engine's power exceeds the MBL engine's
by an order of magnitude.

However, the quantum dots must be separated widely.
Otherwise, they will interact, as an ETH system.
(See~\cite{Kosloff_10_Optimal} for disadvantages of interactions
in another quantum thermal machine.
Spin-spin couplings cause ``quantum friction,''
limiting the temperatures to which a refrigerator can cool.)
We compensate by attributing a volume 
$V  \sim (1 \:  \mu \text{m})^3$ to each dot.
The power density becomes 
$\frac{ \Power  }{  V } \sim 1  \text{ kW} / \text{m}^3$,
two orders of magnitude less than the localized engine's.
Localization naturally implies
near independence of the subengines.

\section{Formal comparisons with competitor engines}
\label{section:Compare_main}

The Otto cycle can be implemented with many media.
Why use MBL?
How does the ``athermality'' of MBL level correlations
advantage our engine? 
We compare our engine with five competitors:
an ideal thermodynamic gas,
a set of ideally noninteracting qubits (e.g., quantum dots),
a many-body system whose bandwidth
is compressed and expanded,
an MBL engine tuned between 
equal-disorder-strength disorder realizations,
and an Anderson-localized Otto engine.
An MBL Otto engine whose cold bath has 
an ordinary bandwidth $\Wb > \dAvg$
is discussed in Suppl. Mat.~\ref{section:CompetitorApp}.

%
%
%
\subsection{Ideal-gas Otto engine}
\label{section:Ideal_gas_main}

The conventional thermodynamic Otto engine 
consists of an ideal gas.
Its efficiency, $\eta_\Otto$, approximately equals the efficiency $\eta_\MBL$
of an ideal mesoscopic MBL engine:
$\eta_\Otto  \approx  \eta_\MBL$.
More precisely, for every MBL parameter ratio $\frac{ \Wb }{ \dAvg } \, ,$
and for every ideal-gas heat-capacity ratio $\gamma = \frac{ \Cp }{ \Cv } \, ,$
there exists a compression ratio $r  :=  \frac{ V_1 }{ V_2 }$ such that
$\eta_\Otto  =  1  -  \frac{1}{  r^{ \gamma  -  1 }  }
=  1  -  \frac{ \Wb }{ 2 \dAvg }  \approx  \eta_\MBL \, .$

However, scaling up the mesoscopic MBL engine 
to the thermodynamic limit
requires a lower bound on the tuning speed $v$ 
(Sec.~\ref{section:Times_main}).
The lower bound induces diabatic jumps that cost work $\expval{ W_\diab }$,
detracting from $\eta_\MBL$ by an amount 
$\sim \frac{ \Wb }{ \dAvg }$
(Suppl. Mat.~\ref{eq:Diab_eff}).
(For simplicity, we have assumed that 
$\TCold = 0$ and $\THot = \infty$ 
and have kept only the greatest terms.)
The ideal-gas engine suffers no such diabatic jumps.
However, the MBL engine's $\expval{ W_\diab }$ 
is suppressed in small parameters
($\frac{ \Wb }{ \dAvg },  \frac{ v }{ \sqrt{ \dAvg } } ,
\frac{ \deltaMBL }{ \dAvg }  \ll  1$).
Hence the thermodynamically large MBL engine's efficiency lies close to
the ideal-gas engine's efficiency:
$\eta_\MBL^{\text{true}}  \approx  \eta_\Otto \, .$

Moreover, the thermodynamically large MBL engine 
may be tuned more quickly than the ideal-gas engine.
The MBL engine is tuned nearly quantum-adiabatically.
The ideal-gas engine is tuned quasistatically.
The physics behind the quantum adiabatic theorem
differs from the physics behind the quasistatic condition.
Hence the engines' speeds $v$ are bounded with
different functions of the total system size $\Sites_\macro$.
The lower bound on the MBL engine's $v$
remains constant as $\Sites_\macro$ grows:
$v  \gg  \HScale^2  \:  e^{ - 3 \xi_\Loc  /  \xi(t) }  
   \;  2^{ - 2.5 \xi_\Loc }$
[Ineq.~\eqref{eq:LowerVBound_Main}].
Rather than $\Sites_\macro$, the fixed localization length $\xi_\Loc$
governs the bound on $v$.\footnote{
Cold thermalization of the MBL engine lasts longer than one tuning stroke:
$\tau_\therm  \gg  \frac{ \HScale }{ v }$
(Sec.~\ref{section:Thermo_limit_main}).
But even $\tau_\therm$ does not depend on $\Sites_\macro$.}
In contrast, we expect an ideal-gas engine's speed
to shrink: $v  \sim  \frac{1}{\Sites_\macro} \, .$
The quasistatic condition requires that the engine remain in equilibrium.
The agent changes the tuning parameter $\alpha$
by a tiny amount $\Delta \alpha$,
waits 
until the gas calms,
then changes $\alpha$ by $\Delta \alpha$.
The changes are expected to propagate as waves
with some speed $c$.
The wave reaches the engine's far edge in a time 
$\sim  \frac{ \Sites_\macro }{ c } \, .$
Hence $v  <  \HScale \frac{ c }{ \Sites_\macro } \, .$

However, the ideal-gas engine is expected to output more work per unit volume
than the MBL engine.
According to our order-of-magnitude estimates (Sec.~\ref{section:Order_main}),
the ideal-gas engine operates at a power density of
$\frac{ \Power }{ V }  \sim  1 \text{ MW} / \text{m}^3$;
and the localized engine, at
$\frac{ \Power }{ V }  \sim  100 \text{ kW} / \text{m}^3 \, .$
An order of magnitude separates the estimates.

\subsection{Quantum-dot engine}
\label{section:Q_dot_main}

Section~\ref{section:Order_main} introduced the quantum-dot engine,
an array of ideally independent bits or qubits. 
We add to the order-of-magnitude analysis 
two points about implementations' practicality.
The MBL potential's generic nature offers an advantage.
MBL requires a random disorder potential $\{ h ( \alpha_t ) h_j \}$, e.g., 
a ``dirty sample,'' a defect-riddled crystal.
This ``generic'' potential contrasts with the pristine background
required by quantum dots.
Imposing random MBL disorder is expected to be simpler.
On the other hand, a quantum-dot engine does not necessarily need
a small-bandwidth cold bath, $\Wb \ll \dAvg$.

\subsection{Bandwidth engine}
\label{section:Bandwidth_engine_main}

Imagine eliminating the scaling factor $Q \LParen h ( \alpha_t ) \RParen$
from the Hamiltonian~\eqref{eq:SpinHam}.
The energy band is compressed and expanded
as the disorder strength $h( \alpha_t )$ is ramped down and up.
The whole band, rather than a gap,
contracts and widens as in Fig.~\ref{fig:Compare_thermo_Otto_fig},
between a size $\sim \HScale \Sites_\macro \, h ( \alpha_0 )$
and a size $\sim  \HScale \Sites_\macro  \,  h ( \alpha_1 )
\gg  \HScale \Sites_\macro \, h ( \alpha_0 )$.
The engine can remain in one phase throughout the cycle.
The cycle does not benefit from 
the ``athermality'' of local level correlations.

Furthermore, this accordion-like motion requires 
no change of the energy eigenbasis's form.
Tuning may proceed quantum-adiabatically:
$v \approx 0$.
The ideal engine suffers no diabatic jumps, losing
$\expval{ W_\diab }_\macro  =  0$.
 
But this engine is impractical:
Consider any perturbation $V$ that fails to commute with 
the ideal Hamiltonian $H(t)$: $[V, H(t)]  \neq  0$.
Stray fields, for example, can taint an environment.
As another example, consider cold atoms in an optical lattice.
The disorder strength is ideally $\HScale h ( \alpha_t )$.
One can strengthen the disorder by
strengthening the lattice potential $U_{\text{lattice}}$.
Similarly, one can raise the hopping frequency (ideally $\HScale$)
by raising the pressure $p$.
Strengthening $U_{\text{lattice}}$ and $p$
while achieving the ideal disorder-to-hopping ratio
$\frac{ \HScale h(\alpha_t) }{ \HScale }  =  h ( \alpha_t )$
requires fine control.
If the ratio changes from $h ( \alpha_t )$, 
the Hamiltonian $H(t)$ acquires a perturbation $V$
that fails to commute with other terms.

This $V$ can cause diabatic jumps
that cost work $\expval{ W_\diab }_\macro$.
Jumps suppress the scaling of the average work outputted per cycle
by a factor of $\sqrt{ \Sites_\macro }$
(Suppl. Mat.~\ref{section:Bandwidth_engine_app}).
The MBL Otto engine may scale more robustly:
The net work extracted scales as $\Sites_\macro$
[Eq.~\eqref{eq:Composite_W_Scale}].
Furthermore, diabatic jumps cost work $\expval{ W_\diab }_\macro$
suppressed small parameters such as $\frac{ \sqrt{v} }{ \dAvg }$.

\subsection{Engine tuned between  
equal-disorder-strength disorder realizations}
\label{section:Equal_h_main}

The disorder strength $h(\alpha_t)$ in Eq.~\eqref{eq:SpinHam}
would remain $\gg 1$ and constant in $t$,
while the random variables $h_j$ would change.
Let $\tilde{\Sys}$ denote this constant-$h (\alpha_t)$ engine,
and let $\Sys$ denote the MBL engine.
$\tilde{\Sys}$ takes less advantage of MBL's ``athermality,''
as $\tilde{\Sys}$ is not tuned between 
level-repelling and level-repulsion-free regimes.

Yet $\tilde{\Sys}$ outputs the amount $\expval{ W_\tot }$ of work
outputted by $\Sys$ per cycle, on average.
Because $\Wb$ is small, cold thermalization drops $\tilde{\Sys}$
across only small gaps $\delta' \ll  \dAvg$.
$\tilde{\Sys}$ traverses a trapezoid, 
as in Fig.~\ref{fig:Compare_thermo_Otto_fig}, in each trial.
However, the MBL engine has two advantages:
greater reliability and fewer worst-case (negative-work-outputted) trials.

Both the left-hand gap $\delta$ 
and the right-hand gap $\delta'$ 
traversed by $\tilde{\Sys}$ are Poisson-distributed.
Poisson-distributed gaps more likely assume extreme values
than GOE-distributed gaps:
$P_\MBL^\ParenE ( \delta )  >  P_\ETH^\ParenE ( \delta )$
if $\delta \sim 0$ or $\delta \gg  \dAvg$~\cite{D'Alessio_16_From}.
The left-hand gap $\delta$ traversed by $\Sys$ is GOE-distributed.
Hence the $W_\tot$ outputted by $\tilde{\Sys}$ 
more likely assumes extreme values
than the $W_\tot$ outputted by $\Sys$.
The greater reliability of $\Sys$ may suit $\Sys$ better
to ``one-shot statistical mechanics''~\cite{Dahlsten_11_Inadequacy,delRio_11_Thermo,Aberg_13_Truly,Horodecki_13_Fundamental,Dahlsten_13_Non,Egloff_15_Measure,Brandao_15_Second,Gour_15_Resource,YungerHalpern_16_Beyond,Gour_17_Quantum,Ito_16_Optimal,vanderMeer_17_Smoothed}.
In one-shot theory, predictability of the work $W_\tot$ 
extractable in any given trial
serves as a resource.

$\Sys$ suffers fewer worst-case trials than $\tilde{\Sys}$.
We define as \emph{worst-case} a trial in which 
the engine outputs net negative work, $W_\tot < 0$.
Consider again Fig.~\ref{fig:Compare_thermo_Otto_fig}.
Consider a similar figure that depicts the trapezoid traversed
by $\tilde{\Sys}$ in some trial.
The left-hand gap, $\delta$, is distributed as the right-hand gap, $\delta'$, is,
according to $P_\MBL^\ParenE ( \delta )$.
Hence $\delta$ has a decent chance
of being smaller than $\delta'$: $\delta < \delta'$.
$\tilde{\Sys}$ would output $W_\tot < 0$
in such a trial.

We estimate worst-case trials' probabilities
in Suppl. Mat.~\ref{section:DisorderEngine}.
Each trial undergone by one constant-$h ( \alpha_t )$ subengine has a probability 
$\sim  \left( \frac{ \Wb }{ \dAvg } \right)^2$
of yielding $W_\tot < 0 \, .$
An MBL subengine has a worst-case probability 
one order of magnitude lower:
$\sim \left( \frac{ \Wb }{ \dAvg } \right)^3  \, .$
Hence the constant-$h ( \alpha_t )$ engine illustrates that 
local MBL level correlations' athermality
suppresses worst-case trials
and enhances reliability.

\subsection{Anderson-localized engine}
\label{section:Anderson_engine_main}

Anderson localization follows from
removing the interactions from MBL
(Suppl. Mat.~\ref{section:ThermoLimitApp}).
One could implement our Otto cycle with an Anderson insulator
because Anderson Hamiltonians exhibit 
Poissonian level statistics~\eqref{eq:P_MBL_Main}.
But strokes 1 and 3 would require 
the switching off and on of interactions.
Tuning the interaction, as well as the disorder-to-interaction ratio,
requires more effort than tuning just the latter.

Also, particles typically interact in many-body systems.
MBL particles interact; Anderson-localized particles do not.
Hence one might eventually expect less difficulty 
in engineering MBL engines 
than in engineering Anderson-localized engines.

%
%
\section{Outlook}
\label{section:Outlook}

The realization of thermodynamic cycles
with quantum many-body systems
was proposed very recently~\cite{PerarnauLlobet_16_Work,Lekscha_16_Quantum,Jaramillo_16_Quantum,Campisi_16_Power,Modak_17_Work,Verstraelen_17_Unitary,Ferraro_17_High,Ma_17_Quantum}.
MBL offers a natural platform, due to its ``athermality''
and to athermality's resourcefulness in thermodynamics.
We designed an Otto engine 
that benefits from the discrepancy between 
many-body-localized and ``thermal'' level statistics.
The engine illustrates how MBL
can be used for thermodynamic advantage.

Realizing the engine may provide a near-term challenge 
for existing experimental set-ups.
Possible platforms include cold atoms~\cite{Schreiber_15_Observation,Kondov_15_Disorder,Choi_16_Exploring,Luschen_17_Signatures,Bordia_17_Probing}; nitrogen-vacancy centers~\cite{Kucsko_16_Critical};
ion traps~\cite{Smith_16_Many}; and
doped semiconductors~\cite{Kramer_93_Localization},
for which we provided order-of-magnitude estimates.
Realizations will require platform-dependent corrections
due to, e.g., variable-range hopping induced by particle-phonon interactions.
As another example, semiconductors' impurities
suffer from dipolar interactions.
The interactions extend particles' wave functions
from decaying exponentially across space to decaying as power laws.

Reversing the engine may pump heat from the cold bath to the hot,
lowering the cold bath's temperature.
Low temperatures facilitate quantum computation
and low-temperature experiments.
An MBL engine cycle might facilitate state preparation
and coherence preservation
in quantum many-body experiments.

Experiments motivate explicit modeling of the battery.
We have defined as work
the energy outputted during Hamiltonian tunings.
A work-storage device, or battery, must store this energy.
We have refrained from specifying the battery's physical form,
using an \emph{implicit battery model}.
An equivalent \emph{explicit battery model} could depend on 
the experimental platform.
Quantum-thermodynamics batteries have been modeled abstractly with 
ladder-like Hamiltonians~\cite{Skrzypczyk_13_Extracting}.
An oscillator battery for our engine could manifest as a cavity mode.

MBL is expected to have thermodynamic applications
beyond this Otto engine.
A localized ratchet, which leverages information
to transform heat into work, is under investigation.
The paucity of transport in MBL may have 
technological applications beyond thermodynamics.
Dielectrics, for example, prevent particles from flowing
in certain directions.
Dielectrics break down in strong fields.
To survive, a dielectric must insulate well---as does MBL.

In addition to suggesting applications of MBL,
this work identifies an opportunity within quantum thermodynamics.
Athermal quantum states (e.g., $\rho  \neq e^{-H/T}/Z$) are usually 
regarded as resources in quantum thermodynamics~\cite{Janzing_00_Thermodynamic,Dahlsten_11_Inadequacy,Brandao_13_Resource,Horodecki_13_Fundamental,Goold_15_review,Gour_15_Resource,YungerHalpern_16_Beyond,YungerHalpern14,LostaglioJR14,Lostaglio_15_Thermodynamic,YungerHalpern_16_Microcanonical,Guryanova_16_Thermodynamics,Deffner_16_Quantum,Wilming_17_Third}.
Not only athermal states, we have argued,
but also athermal energy-level statistics,
offer thermodynamic advantages.
Generalizing the quantum-thermodynamics definition 
of ``resource''
may expand the set of goals
that thermodynamic agents can achieve.

Optimization offers another theoretical opportunity.
We have shown that the engine works,
but better protocols could be designed.
For example, we prescribe 
nearly quantum-adiabatic tunings.
Shortcuts to adiabaticity (STA) avoid both
diabatic transitions and exponentially slow tunings~\cite{Chen_10_Fast,Kosloff_10_Optimal,Torrontegui_13_Shortcuts,Deng_13_Boosting,del_Campo_14_Super,Abah_16_Performance}.
STA have been used to reduce other quantum engines' cycle times~\cite{Deng_13_Boosting,del_Campo_14_Super,Abah_16_Performance}.
STA might be applied to the many-body Otto cycle,
after being incorporated in to MBL generally.

\endgroup

\renewcommand{\bibsection}{\section*{\refname}}
\putbib[MBL_bib,MBL_bib2] 
\end{bibunit}

%
%
\chapter{Non-Abelian thermal state: The thermal state of a quantum system with noncommuting charges}
\label{ch:Noncommq}
\begin{bibunit}

\noindent \emph{This chapter was published as~\cite{NYH_16_Micro}.}

\begingroup


%
%
%
%
%
%

%
%
%

\makeatletter
\newcommand\bibalias[2]{%
  \@namedef{bibali@#1}{#2}%
}

%
%
%
%

\newtoks\biba@toks
\let\bibalias@oldcite\cite
\def\cite{%
  \@ifnextchar[{%
    \biba@cite@optarg%
  }{%
    \biba@cite{}%
  }%
}
\newcommand\biba@cite@optarg[2][]{%
  \biba@cite{[#1]}{#2}%
}
\newcommand\biba@cite[2]{%
  \biba@toks{\bibalias@oldcite#1}%
  \def\biba@comma{}%
  \def\biba@all{}%
  \@for\biba@one@:=#2\do{%
    \edef\biba@one{\expandafter\@firstofone\biba@one@\@empty}%
    \@ifundefined{bibali@\biba@one}{%
      \edef\biba@all{\biba@all\biba@comma\biba@one}%
    }{%
      \PackageInfo{bibalias}{%
        Replacing citation `\biba@one' with `\@nameuse{bibali@\biba@one}'
      }%
      \edef\biba@all{\biba@all\biba@comma\@nameuse{bibali@\biba@one}}%
    }%
    \def\biba@comma{,}%
  }%
  %
  %
  \immediate\write\@auxout{\noexpand\bgroup\noexpand\renewcommand\noexpand\citation[1]{}\noexpand\citation{#2}\noexpand\egroup}%
  %
  %
  \edef\biba@tmp{\the\biba@toks{\biba@all}}%
  \biba@tmp%
}
\makeatother

%
%

\bibalias{360221697}{thermolimit}
\bibalias{Aberg14}{aberg2014catalytic}
\bibalias{BartlettRS07}{BRS-refframe-review}
\bibalias{BrandaoHNOW14}{brandao2013second}
\bibalias{FundLimits2}{HO-limitations}
\bibalias{HO2011-qthermo}{HO-limitations}
\bibalias{Hayden-embezzling--2}{Hayden-embezzling}
\bibalias{HorodeckiHHH09}{horodecki_quantum_2009}
\bibalias{PuszW78}{pusz_passive_1978}
\bibalias{YungerHalpernR14}{halpern2014beyond}
\bibalias{aaberg2013truly}{aaberg-singleshot}
\bibalias{adler2001gss}{adler2001generalized}
\bibalias{dahlsten2011inadequacy}{workvalue}
\bibalias{faist2012quantitative}{qlandauer}
\bibalias{gour_measuring_2009}{gour_measuring_2009--2}
\bibalias{horodecki_are_2002--2}{horodecki_are_2002}
\bibalias{horodecki_locking_2005--2}{horodecki_locking_2005}
\bibalias{horodecki_reversible_2003--2}{horodecki_reversible_2003}
\bibalias{lenard_thermodynamical_1978}{lenard78}
\bibalias{lenard_thermodynamical_1978--2}{lenard78}
\bibalias{linden2010smallest}{linden2010small}
\bibalias{linden_reversibility_2005--2}{linden_reversibility_2005}
\bibalias{marcothesis}{Tomamichel-thesis}
\bibalias{pusz_passive_1978--2}{pusz_passive_1978}
\bibalias{scovil1959maser}{Scovil1959masers}
\bibalias{synak-radtke_asymptotic_2006--2}{synak-radtke_asymptotic_2006}
\bibalias{wilming2014weak--2}{wilming2014weak}
\bibalias{wyner75}{Wyner-wiretap}

\def\batt{ {\text{W}} }
\def\bath{ {\text{R}} }
\def\sys{ {\text{S}} }
\def\anc{ {\text{A}} }
\def\cat{ {\text{X}} }

\def\id{\mathbbm{1}}

\newcommand{\Wtran}[2]{W_{#1\rightarrow#2}}
\def\qtot{Q_{i_{\text{tot}}}}
\def\final{\rho_{\text{S}}'}  

\def\GTOlong{Non-Abelian Thermal Operations}
\def\GTOlongSing{Non-Abelian Thermal Operation}
\def\GTO{NATO}
\def\GGS{NATS}
\def\GGSlong{Non-Abelian Thermal State}



\def\toto{\mathop{\to}\limits^{TO}}
\def\pij{p_{i\to j}}
\def\dmin{D_{\min}}
\def\dmax{D_{\max}}
\def\<{\langle}
\def\>{\rangle}
\def\ot{\otimes}
\def\ermax{E_R^{\max}}
\def\esmax{E_S^{\max}}
\def\ecal{\mathcal{E}}
\def\hcal{\mathcal{H}}
\def\ecalr{{\ecal_R}}
\def\etaE{\eta_{E-E_S}^R}
\def\etaEprim{\eta'_{E-E_S}}
\def\ideta{\eta}
\def\rhor{\rho_R}
\def\rhos{\rho_S}
\def\rhors{\rho_{RS}}
\def\energy{E_0}

\def\rhoinr{\rho^{\text{in}}_R}
\def\rhoins{\rho^{\text{in}}_S}
\def\rhoinw{\rho^{\text{in}}_W}
\def\rhoinc{\rho^{\text{in}}_C}
\def\rhooutr{\rho^{\text{out}}_R}
\def\rhoouts{\rho^{\text{out}}_S}
\def\rhooutw{\rho^{\text{out}}_W}
\def\rhooutsc{\rho^{\text{out}}_{SC}}
\def\rhooutsw{\rho^{\text{out}}_{SW}}
\def\rhooutc{\rho^{\text{out}}_C}
\def\rhozeror{\rho^{0}_R}
\def\rhozeros{\rho^{0}_S}
\def\rhozeroc{\rho^{0}_C}
\def\sgibbs{\rho_S^\beta}
\def\rgibbs{\rho_R^\beta}
\def\cgibbs{\rho_C^\beta}
\def\ancgibbs{\rho_{\text{anc}}^\beta}
\def\rscgibbs{\rho_{\text{RSC}}^\beta}
\def\scgibbs{\rho_{\text{SC}}^\beta}

\def\twirl{\mathcal{T}}
\def\funny{{f-smooth}}
\def\funnyh{{f-smooth}}
\def\Funnyh{{F-smooth}}
\def\fh{\tilde H}
\def\Funny{{F-smooth}}
\def\frk{{\text{frk}}}
\def\ftail{{\text{ftail}}}
\def\tail{{\text{tail}}}
\def\rk{{\text{rk}}}
\def\rkd{{\text{rk}}^\delta}

\def\stipp{\bar{p}^{(\delta)}}
\def\stipq{\bar{q}^{(\delta)}}
\def\stipf{\bar{f}^{(\delta)}}
\def\flatp{\mathbf{p}^{(\delta)}}
\def\flatq{\mathbf{q}^{(\delta)}}
\def\flatf{\mathbf{f}^{(\delta)}}
\def\flattest{flattest}
\def\stippest{steeppest}

\newcommand{\be}{\begin{eqnarray} \begin{aligned}}
\newcommand{\ee}{\end{aligned} \end{eqnarray} }
\newcommand{\benn}{\begin{eqnarray*} \begin{aligned}}
\newcommand{\eenn}{\end{aligned} \end{eqnarray*} }

\newcommand{\ben}{\begin{eqnarray} \begin{aligned}}
\newcommand{\een}{\end{aligned} \end{eqnarray} }

\newcommand{\hmax}{\mathrm{H}_{\max}}
\newcommand{\hminp}{\mathrm{H}_{\min}}
\newcommand{\h}{\mathrm{H}}
\newcommand{\bc}{\begin{center}}
\newcommand{\ec}{\end{center}}
\newcommand{\half}{\frac{1}{2}}
\newcommand{\ran}{\rangle}
\newcommand{\lan}{\langle}
\newcommand{\im}{\mathbbmss{1}}
\newcommand{\idop}{\mathcal{I}}
\newcommand{\nullmat}{\mathbf{0}}
\newcommand{\fid}{\mathcal{F}}
\newcommand{\Cn}{\mathbb{C}}
\newcommand{\re}{\mathop{\mathbb{R}}\nolimits}
\newcommand{\natnum}{\mathop{\mathbb{N}}\nolimits}
\newcommand{\esp}{\mathbb{E}}
\newcommand{\cov}{\mathop{\mathrm{Cov}} \nolimits}
\newcommand{\var}{\mathop{\mathrm{Var}} \nolimits}
\newcommand{\mse}{\mathop{\mathrm{MSE}} \nolimits}
\newcommand{\rlprt}{\mathop{\mathrm{Re}} \nolimits}
\newcommand{\nth}{\mathrm{th}}
\newcommand{\imprt}{\mathop{\mathrm{Im}}\nolimits}
\newcommand{\tr}{\mathop{\mathsf{tr}}\nolimits}
\newcommand{\eval}[2]{\left. #1 \right \arrowvert _{#2}}
\newcommand{\norm}[1]{\left\| #1\right \|}
\newcommand{\smalloh}{\mathrm{o}}
\newcommand{\bigoh}{\mathrm{O}}
\newcommand{\pr}{\prime}
\newcommand{\di}{\text{d}}
\newcommand{\trans}[1]{#1^{\top}}
\newcommand{\iu}{i}
\newcommand{\vc}[1]{#1}				
\newcommand{\abs}[1]{\left|#1 \right|}				
\newcommand{\der}[1]{\frac{\partial}{\partial #1}}
\newcommand{\e}{\mathrm{e}}
\newcommand{\upa}{\uparrow}
%
\newcommand{\beq}{\begin{eqnarray} \begin{aligned}}
\newcommand{\eeq}{\end{aligned} \end{eqnarray} }
\newcommand{\bea}{\begin{array}}
\newcommand{\eea}{\end{array}}

\newcommand{\bee}{\begin{enumerate}}
\newcommand{\eee}{\end{enumerate}}
\newcommand{\bei}{\begin{itemize}}
\newcommand{\eei}{\end{itemize}}

\newcommand{\dna}{\downarrow}
\newcommand{\ra}{\rightarrow}
\newcommand{\hil}{\mathcal{H}}
\newcommand{\kil}{\mathcal{K}}
\newcommand{\allrhos}{\mathcal{M}}
\newcommand{\allrhosin}[1]{\mathcal{S}(#1)}
\newcommand{\pop}[1]{\mathcal{M}_{+}(#1)}
\newcommand{\cpmap}{\mathcal{E}}
\newcommand{\uop}{\mathcal{U}}
\newcommand{\su}{\mathfrak{su}}
\newcommand{\finproof}{$\Box$}
\newcommand{\textmath}{}
\newcommand{\wt}{\widetilde}
\newcommand{\wavyline}[3]{
\multiput(#1,#2)(4,0){#3}{
\qbezier(0,0)(1,1)(2,0)
\qbezier(2,0)(3,-1)(4,0)}}
\newcommand{\spann}{\mathop{\mathrm{span}}\nolimits}  

\newcommand{\wigglyline}[3]{
\multiput(#1,#2)(1,0){#3}{
\qbezier(0,0)(0.25,0.25)(0.5,0)
\qbezier(0.5,0)(0.75,-0.25)(1,0)}}

\newcommand{\allsigmasin}[1]{\tilde{\mathcal{S}}(#1)}
\newcommand{\instr}{\mathcal{N}}
\newcommand{\dfdas}{\stackrel{\textrm{\tiny def}}{=}}
\newcommand{\argmax}{\mathop{\mathrm{argmax}}\nolimits}
\newcommand{\bin}{\textrm{Bin}}
\newcommand{\topr}{\stackrel{P}{\to}}
\newcommand{\todst}{\stackrel{D}{\to}}
\newcommand{\lil}{\mathcal{L}}
\newcommand{\myabstract}[1]{\begin{quote}{\small  #1 }\end{quote}}
\newcommand{\myacknowledgments}{\begin{center}{\bf Acknowledgments}\end{center}\par}
\newcommand{\otherfid}{\mathsf{F}}
\newcommand{\mypacs}[1]{\begin{flushleft} {\small PACS numbers: #1}\end{flushleft}}
\newcommand{\basedon}[1]{\begin{flushleft} {\small This chapter is based on #1.}\end{flushleft}}
\newcommand{\onbased}[1]{\begin{flushleft} {\small  #1 is based on this chapter.}\end{flushleft}}
\newcommand{\ds}{\displaystyle}
\newcommand{\co}{\textrm{co}}
\newcommand{\eg}{\epsilon}
\newcommand{\zg}{\theta}
\newcommand{\ag}{\alpha}
\newcommand{\bg}{\beta}
\newcommand{\dg}{\delta}
\newcommand{\mc}{\mathcal}

\def\Real{\mathbb{R}}
\def\Complex{\mathbb{C}}
\def\Natural{\mathbb{N}}
\def\id{\mathbb{I}}

\def\01{\{0,1\}}
\newcommand{\ceil}[1]{\lceil{#1}\rceil}
\newcommand{\floor}[1]{\lfloor{#1}\rfloor}
\newcommand{\eps}{\varepsilon}
\newcommand{\outp}[2]{|#1\rangle\langle#2|}
\newcommand{\proj}[1]{|#1\rangle\langle#1|}

\newcommand{\inp}[2]{\langle{#1}|{#2}\rangle} 

\newcommand{\rank}{\operatorname{rank}}

\newcommand{\mX}{\mathcal{X}}
\newcommand{\mY}{\mathcal{Y}}
\newcommand{\mB}{\mathcal{B}}

\newenvironment{sdp}[2]{
\smallskip
\begin{center}
\begin{tabular}{ll}
#1 & #2\\
subject to
}
{
\end{tabular}
\end{center}
\smallskip
}

\newcommand{\ens}{\mathcal{E}}
\newcommand{\Y}{|\mY|}
\newcommand{\psucRAC}{P^{\text{uncert}}}
\newcommand{\psuc}{P^{\text{succ}}}
\newcommand{\pgame}{{P^{\text{game}}}}
\newcommand{\pgameMAX}{{P^{\text{game}}_{\rm max}}}
\newcommand{\cE}{\mathcal{E}}
\newcommand{\cU}{\mathcal{U}}
\newcommand{\sx}{\mathcal{S}_x}
\newcommand{\hmin}{\ensuremath{H}_{\infty}}

\newcommand{\stateSet}{\mathscr{S}}

\newcommand{\secchsh}{\ref{sec:chsh}}
\newcommand{\nn}[1]{\textcolor{magenta}{nelly: #1}}
\newcommand{\mcomment}[1]{{\sf [#1]}\marginpar[\hfill !!!]{!!!}}
\newcommand{\steph}[1]{{\textcolor{blue}{steph: #1}}}
\newcommand{\jono}[1]{\textcolor{blue}{#1}}
\newcommand{\mh}[1]{\textcolor{red}{mh: #1}}
\newcommand{\pc}[1]{\textcolor{green}{\tt Piotr: #1}}
\newcommand{\fer}[1]{\textcolor{brown}{\tt Fernando: #1}}
\newcommand{\xstr}{{\mathbf{x}_{s,a}}}
\newcommand{\str}{{\mathbf{x}}}
\newcommand{\vstr}{x}
\newcommand{\vxstr}{{x_{s,a}}}
\newcommand{\xstrsub}{s,a}
\newcommand{\probun}[1]{P^{\text{un}}(#1)}
\newcommand{\probsteer}[1]{P^{\text{st}}(#1)}

\newcommand{\prob}[1]{p(#1)}

\newcommand{\good}{\mathcal{G}}
\newcommand{\bstate}[1]{\sigma_{\xstrsub}^{#1}}
\newcommand{\xor}{XOR}
\newcommand{\avgb}{\rho}
\newcommand{\icaus}{I_{caus}}
\newcommand{\irac}{I_{rac}}
\newcommand{\unop}{Q}
\newcommand{\supl}{Appendix}
\newcommand{\stateset}{\mathcal{F}_{s}}
\newcommand{\badproj}{\Pi_{bad}}

\newcommand{\setA}{\mathcal{A}}
\newcommand{\setB}{\mathcal{B}}
\newcommand{\setS}{\mathcal{S}}
\newcommand{\setT}{\mathcal{T}}
\newcommand{\secxor}{I}

\newcommand{\dist}[2]{\mathcal{D}(#1\rangle#2)}

\def\<{\langle}
\def\>{\rangle}
\def\ot{\otimes}
\def\ermax{E_R^{\max}}
\def\esmax{E_S^{\max}}
\def\ecal{\mathcal{E}}
\def\hcal{\mathcal{H}}
\def\ecalr{{\ecal_R}}
\def\etaE{\eta_{E-E_{\text{S}}}}
\def\rhor{\rho_{\text{R}}}  
\def\rhos{\rho_{\text{S}}}  
\def\nats{\gamma_{\mathbf{v}}}
\def\gibbs{\rho_{\text{R}}^\beta}  
\def\gibbsS{\gamma_{\text{S}}}  
\def\gibbsBatt{\gamma_{ \batt }}  
\def\gibbsSBatt{\gamma_{\sys{\batt}}}  
\newcommand\gibbsParam[1]{\gamma_{#1}}
\def\final{\rho_{\text{S}}'}  
\def\initialBatt{\rho_{ \batt }}  
\def\finalBatt{{\rho'_{ \batt }}}  
\def\ergotropymin{F_\epsilon^{min}}
\def\ergotropymax{F_\epsilon^{max}}
\def\gmin{G_\epsilon^{min}}
\def\gmax{G_\epsilon^{max}}
\def\supthermomaj{C}
\def\supdistillation{D}
\def\supformation{D}
\def\supparadigm{A}
\def\supprocesses{E}

\def\rhodec{\omega}
\def\s{\,\,\,\,}
\def\dmin{D_{min}}
\def\dmax{D_{max}}
\def\dmine{D^\epsilon_{min}}
\def\dmaxe{D^{\epsilon}_{max}}
\def\wit{\psi_W}
\def\witi{0}
\def\initial{\rho_\sys}
\def\Hw{\hat{W}}
\def\Hin{H}
\def\Hout{H'}
\def\Zout{Z'}
\def\mainsection{Main Section}
\newcommand{\alfree}[1]{F_\alpha(#1,\gibbs)}
\newcommand{\qalfree}{{\hat F}}
\newcommand{\qalfreesimple}{{\tilde F}}
\def\fmin{F_{\text{min}}}
\def\fmax{F_{\text{max}}}
\def\tauout{\tau'}
\def\ep{\epsilon}
\newcommand{\sgn}{\operatorname{sgn}}
\def\qrenyi{S}
\def\qrenyisimple{\tilde S}
\def\trumpd{\text{D_{work}}}
\def\catalyst{\sigma}
\def\expbound{\exp(-\Omega(\sqrt{\log(N)}))} 

\def\gibbsin{\rho_\beta^{(0)}}
\def\gibbsout{\rho_\beta^{(1)}}

\def\genFs{generalized free energies}
\def\workf{\mathcal{W}}

\newcommand{\newreptheorem}[2]{%
\newenvironment{rep#1}[1]{%
 \def\rep@title{#2 \ref{##1} (restatement)}%
 \begin{rep@theorem}}%
 {\end{rep@theorem}}}
\makeatother

\newreptheorem{thm}{Theorem}
\newreptheorem{lem}{Lemma}

%
%
\def\topfraction{0.5}
\def\bottomfraction{0}

%
%
\let\oldparagraph\paragraph
\def\paragraph#1{%
  \smallskip%
  \par\noindent{\textbf{#1}}\quad
}

%
%
\let\tr\Tr


\def\topfraction{1}

%
%
%

Recently reignited interest in quantum thermodynamics has prompted information-theoretic approaches to fundamental questions.
have enjoyed particular interest.~\cite{gemmer2009quantum,Gogolin2015arXiv_review,Goold2015arXiv_review,Vinjanampathy2015arXiv_review}.
The role of entanglement, for example, has been clarified with canonical
typicality~\cite{goldstein2006canonical,gemmer200418,PopescuSW06,LindenPSW09}.
Equilibrium-like behaviors have been predicted~\cite{FermiPU55,KinoshitaWW06,Rigol07,PolkovnikovSSV11}
and experimentally observed in
integrable quantum gases~\cite{Langen15,LangenGS15}.

Thermodynamic resource theories offer a powerful tool for analyzing
fundamental properties of the thermodynamics of quantum systems.
Heat exchanges with a bath are modeled
with ``free states'' and ``free operations''~\cite{janzing_thermodynamic_2000,BrandaoHORS13,brandao2013second,HO-limitations}.
These resource theories have been extended to model exchanges of additional physical
quantities, such as particles and angular
momentum~\cite{FundLimits2,VaccaroB11,YungerHalpernR14,YungerHalpern14,Weilenmann2015arXiv_axiomatic}.

A central concept in thermodynamics and statistical mechanics is the thermal state.
The thermal state has several important properties.
First, typical dynamics evolve the system toward the thermal state.
The thermal state is the equilibrium state. 
Second, consider casting statistical mechanics as an inference problem. 
The thermal state is the state which maximizes the entropy 
under constraints on physical quantities~\cite{Jaynes57I,Jaynes57II}.  
Third, consider the system as interacting with a large bath.
The system-and-bath composite occupies a microcanonical state.
Physical observables of the composite,
such as the total energy and total particle number,
have sharply defined values.
The system's reduced state is the thermal state.  
Finally, in a resource theory, 
the thermal state is the only completely passive state.
No work can be extracted from any number of copies of the thermal state~\cite{PuszW78,Lenard78}.

If a small system exchanges heat and particles with a large environment,
the system's thermal state
is a grand canonical ensemble:
$e^{ - \beta( H - \mu N ) } / Z$.
The system's Hamiltonian and particle number are represented by $H$ and $N$.
$\beta$ and $\mu$ denote the environment's 
inverse temperature and chemical potential.
The partition function $Z$ normalizes the state.
The system-and-bath dynamics conserves 
the total energy and total particle number.
More generally, subsystems exchange conserved quantities, or  ``charges,''
$Q_j,  \;  \:  j = 1, 2, \ldots c$.  To these charges correspond
generalized chemical potentials $\mu_j$.
The $\mu_j$'s characterize the bath.

We address the following question.
Suppose that the charges fail to commute with each other:
$[Q_j,  Q_k]  \neq  0$.
What form does the thermal state have?
We call this state ``the \GGSlong{}'' (\GGS{}).
Jaynes applied the Principle of Maximum Entropy to this question~\cite{Jaynes57II}.
He associated fixed values $v_j$ with the charges' expectation values.
He calculated the state that,
upon satisfying these constraints,
maximizes an entropy.
This thermal state has a generalized Gibbs form:
\begin{align}
   \label{eq:GGS}
   \gamma_{ \mathbf{v} }
   :=   \frac{1}{Z}   e^{ - \sum_{j = 0}^c  \mu_j  Q_j } \ ,
\end{align}
wherein the  the $v_j$'s determine the $\mu_j$'s.

Our contribution is a mathematical, physically justified derivation of the thermal state's form
for systems whose dynamics conserve noncommuting observables.
We recover the state~\eqref{eq:GGS} 
via several approaches, demonstrating its physical importance.
We address  puzzles raised in~\cite{YungerHalpern14,Imperial15} about how
to formulate a resource theory in which thermodynamic charges fail to commute.
Closely related, independent work was performed by Guryanova \emph{et al.}~\cite{teambristol}.
We focus primarily on the nature of passive states.
Guryanova \emph{et al.}, meanwhile, focus more 
on the resource theory for multiple charges
and on tradeoffs amongst types of charge extractions.

In this paper, we derive the \GGS{}'s form from a microcanonical argument.
A simultaneous eigenspace of all the noncommuting
physical charges might not exist.
Hence we introduce the notion of an approximate microcanonical subspace.
This subspace consists of the states in which
the charges have sharply defined values.
We derive conditions under which this subspace exists.
We show that a small subsystem's reduced state
lies, on average, close to $\gamma_{ \mathbf{v} }$.
Second, we invoke canonical typicality~\cite{PopescuSW06,LindenPSW09}.
If the system-and-bath composite 
occupies a random state in the approximate microcanonical subspace, 
we argue, a small subsystem's state likely lies close to the \GGS{}. 
Typical dynamics are therefore expected to evolve 
a well-behaved system's state towards the \GGS{}.
Third, we define a resource theory for thermodynamic exchanges 
of noncommuting conserved charges.
We extend existing resource theories 
to model the exchange of noncommuting quantities. 
We show that the \GGS{} is the only possible free state 
that renders the theory nontrivial: 
 Work cannot be extracted from 
any number of copies of $\gamma_{ \mathbf{v} }$.  
We show also that the \GGS{} is the only state preserved by free operations.  
From this preservation, we derive ``second laws'' that govern state transformations.
This work provides a well-rounded, and novelly physical, perspective
on equilibrium in the presence of quantum noncommutation.
This perspective opens truly quantum avenues in thermodynamics.

%
%
%
\section{Results}
\label{section:Results}

\subsection{Overview}

We derive the \GGSlong{}'s form via three routes:
from a microcanonical argument,
from a dynamical argument
built on canonical typicality,
and from complete passivity in a resource theory.
Details appear in Appendices~\ref{section:SI_Micro}--\ref{section:SI_RT}.

\subsection{Microcanonical derivation}
In statistical mechanics, the form $e^{- \beta ( H - \mu N) } / Z$
of the grand canonical ensemble is well-known to be derivable 
as follows.
The system of interest is assumed 
to be part of a larger system.
Observables of the composite have fixed values $v_j$.
For example, the energy equals $E_0$,
and the particle number equals $N_0$.
The microcanonical ensemble
is the whole-system state spread uniformly across 
these observables' simultaneous eigenspace.
Tracing out the environmental degrees of freedom 
yields the state $e^{ - \beta (H - \mu N) } / Z$.

We derive the \GGS{}'s form similarly.
Crucially, however, we adapt the above strategy 
to allow for noncommuting observables.
Observables might not have well-defined values $v_j$ simultaneously.
Hence a microcanonical ensemble as discussed above, suitable for commuting 
observables, may not exist.
We overcome this obstacle by introducing an
approximate microcanonical ensemble $\Omega$.
We show that, for every state satisfying the conditions of
an approximate microcanonical ensemble, tracing out most of the larger system
yields, on average, a state close to the \GGS{}.
We exhibit conditions under which an approximate 
microcanonical ensemble exists.
The conditions can be satisfied when the
larger system consists of many noninteracting replicas of the system.
An important step in the proof consists of reducing the 
noncommuting case to the commuting one.
This reduction relies on a result by Ogata~\cite[Theorem 1.1]{Ogata11}.
A summary appears in Fig.~\ref{fig:Overview}.

%
%
%
%
\begin{mainfigure}
\centering
\includegraphics[width=87mm]{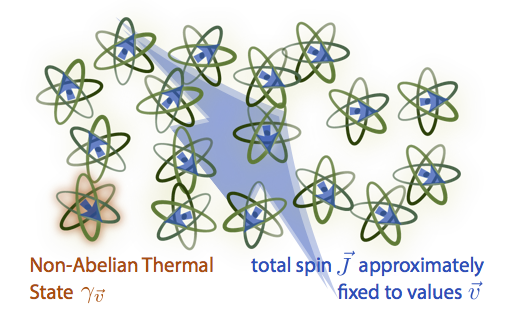}
\caption{\textbf{\GGSlong{}: }
We derive the form of the thermal state
of a system that has charges
that might not commute with each other.
Example charges include the components $J_i$
of the spin $\mathbf{J}$.
We derive the thermal state's form
by introducing an approximate microcanonical state.
An ordinary microcanonical ensemble
could lead to the thermal state's form
if the charges commuted:
Suppose, for example, that the charges were 
a Hamiltonian $H$
and a particle number $N$
that satisfied $[H, N] = 0$.
Consider many copies of the system.
The composite system could have a well-defined energy $E_{\text{tot}}$
and particle number $N_{\text{tot}}$ simultaneously.
$E_{\text{tot}}$ and $N_{\text{tot}}$ would correspond to 
some eigensubspace $\mathcal{H}_{E_{\text{tot}}, N_{\text{tot}}}$
shared by the total Hamiltonian
and the total-particle-number operator.
The (normalized) projector onto $\mathcal{H}_{E_{\text{tot}}, N_{\text{tot}}}$
would represent the composite system's microcanonical state.
Tracing out the bath would yield the system's thermal state.
But the charges $J_i$ under consideration might not commute.
The charges might share no eigensubspace.
Quantum noncommutation demands a modification
of the ordinary microcanonical argument.
We define an approximate microcanonical subspace $\mathcal{M}$.
Each state in $\mathcal{M}$ simultaneously has
almost-well-defined values of noncommuting whole-system charges:
Measuring any such whole-system charge
has a high probability of outputting a value close to an ``expected value''
analogous to $E_{\text{tot}}$ and $N_{\text{tot}}$.
We derive conditions under which
the approximate microcanonical subspace $\mathcal{M}$ exists.
The (normalized) projector onto $\mathcal{M}$
represents the whole system's state.
Tracing out most of the composite system yields
the reduced state of the system of interest.
We show that the reduced state is, on average,
close to the \GGSlong{} (\GGS{}).
This microcanonical derivation of the \GGS{}'s form
links Jaynes's information-theoretic derivation to physics.
}
\label{fig:Overview}
\end{mainfigure}

%
%
%
%
Set-up:  
Let $\mathcal{S}$ denote a system
associated with a Hilbert space $\mathcal{H}$;
with a Hamiltonian $H  \equiv  Q_0$;
and with observables (which we call ``charges'') $Q_1, Q_2, \ldots, Q_c$.
The charges do not necessarily commute with each other:
$[Q_j,  Q_k]  \neq  0$.

Consider $N$ replicas of $\mathcal{S}$, associated with 
the composite system Hilbert space $\mathcal{H}^{\otimes N}$.
We average each charge $Q_j$ over the $N$ copies:
\begin{align}
  \bar{Q}_j := \frac{1}{N}  \sum_{ \ell = 0}^{ N - 1}  
                \id^{\otimes \ell }  \otimes  Q_j  \otimes  \id^{ \otimes (N - 1 - \ell) }.
\end{align}
The basic idea is that, as $N$ grows, the averaged operators $\bar{Q}_j$
come increasingly to commute. 
Indeed, there exist operators operators $\bar{Y}_j$ 
that commute with each other
and that approximate the averages~\cite[Theorem 1.1]{Ogata11}.
An illustration appears in Fig.~\ref{fig:OgataProofSetup}.

\begin{mainfigure}
  \centering
  \includegraphics{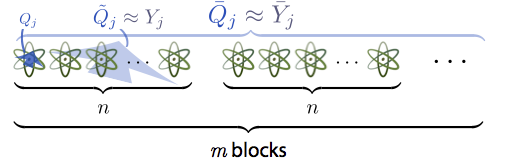}
  \caption{\textbf{Noncommuting charges:  }
We consider a thermodynamic system $\mathcal{S}$
that has conserved charges $Q_j$.
These $Q_j$'s might not commute with each other.
The system occupies a thermal state
whose form we derive.
The derivation involves an approximate microcanonical state
of a large system that contains the system of interest.
Consider a block of $n$ copies of $\mathcal{S}$.
Most copies act, jointly, similarly to a bath
for the copy of interest.
We define $\tilde Q_j$ as the average of the $Q_j$'s 
of the copies in the block. 
Applying results from Ogata~\protect\cite{Ogata11},
we find operators $\tilde{Y}_j$
that are close to the $\tilde Q_j$'s
and that commute with each other. 
Next, we consider $m$ such blocks.
This set of $m$ blocks
contains $N=mn$ copies of $\mathcal{S}$. 
Averaging the $\tilde{Q}_j$'s over the blocks, for a fixed $j$-value, 
yields a global observable $\bar Q_j$.
The $\bar Q_j$'s are approximated by $\bar Y_j$'s.
The $\bar Y_j$'s are the corresponding averages of the $\tilde{Y}_j$'s.
The approximate global charges $\bar{Y}_j$ 
commute with each other.
The commuting $\bar{Y}_j$'s
enable us to extend the concept of a microcanonical ensemble
from the well-known contexts in which all charges commute
to truly quantum systems whose charges do not necessarily commute.}
\label{fig:OgataProofSetup}
\end{mainfigure}

%
%
%
%
Derivation:
Since the $\bar{Y}_j$'s commute mutually, they can be measured
simultaneously. 
More importantly, the joint Hilbert space $\mathcal{H}^{\otimes n}$
contains a subspace
on which each $\bar{Q}_j$ has 
prescribed values close to $v_j$.
Let $\mathcal{M}$ denote the subspace.
Perhaps unsurprisingly, because the $\bar{Y}_j$'s approximate the $\bar{Q}_j$'s,  
each state in $\mathcal{M}$ 
has a nearly well-defined value of $\bar{Q}_j$ near $v_j$.
If $\bar{Q}_j$ is measured, the distribution
is sharply peaked around $v_j$. 
We can also show the opposite: every state with nearly well-defined values $v_j$ of all
$\bar{Q}_j$'s has most of its probability weight in $\mathcal{M}$.

These two properties show that $\mathcal{M}$ is
an approximate microcanonical subspace for the $\bar{Q}_j$'s
with values $v_j$. The notion of the approximate microcanonical subspace
is the first major contribution of our work. It captures the idea 
that, for large $N$, we can approximately fix the values of the
noncommuting charges $Q_j$.
An approximate microcanonical subspace $\mathcal{M}$ is any subspace
consisting of the whole-system states whose average observables $\bar{Q}_j$
have nearly well-defined values $v_j$. More precisely,
a measurement of any $\bar{Q}_j$ has a high probability of yielding 
a value near $v_j$ if and only if most of the state's probability weight lies 
in $\mathcal{M}$.

%
%
Normalizing the projector onto $\mathcal{M}$
yields an approximate microcanonical ensemble, $\Omega$.
Tracing out every copy of $\mathcal{S}$ but the $\ell^{\text{th}}$
yields the reduced state $\Omega_\ell$.
The distance between $\Omega_\ell$
and the \GGS{} $\gamma_{ \mathbf{v} }$
can be quantified by the relative entropy
\begin{align} 
   \label{eq:RelEntMain}
   D ( \Omega_\ell \| \gamma_{ \mathbf{v} } )
   :=  - S( \Omega_\ell )
        -  \Tr (  \Omega_\ell    \log \gamma_{ \mathbf{v} } ).
\end{align}
Here, $S ( \Omega_\ell )  :=  - \Tr ( \Omega_\ell  \log  \Omega_\ell )$
is the von Neumann entropy.
The relative entropy $D$ is bounded by the trace norm $\| . \|_1$,
which quantifies the distinguishability 
of $\Omega_\ell$ and $\gamma_{ \mathbf{v} }$~\cite{HiaiOT81}:
\begin{align}
  \label{eq:PinskerIneq}
  D(\Omega_\ell\Vert\gamma_{\mathbf v})
  \geq \frac12 \norm{\Omega_\ell-\gamma_{\mathbf v}}_1^2.
\end{align}

Our second main result is that, if $\Omega$ is an approximate
microcanonical ensemble, then the average, over systems $\ell$,
of the relative entropy $D$ between $\Omega_\ell$ and $\gamma_{ \mathbf{v} }$
is small:
\begin{align}
     \label{eq:AvgRelEntMain}
      \frac{1}{N} \sum_{ \ell = 0}^{ N - 1} D ( \Omega_\ell \| \gamma_{ \mathbf{v} } )
       \leq \theta + \theta'.
\end{align}
The parameter $\theta  = \mathrm{(const.)} / \sqrt{N}$
vanishes in the many-copy limit.
$\theta'$ depends on the number $c$ of charges,
on the approximate expectation values $v_j$,
on the eigenvalues of the charges $Q_j$,
and on the (small) parameters in terms of which 
$\mathcal{M}$ approximates a microcanonical subspace.

Inequality~\eqref{eq:AvgRelEntMain} capstones the derivation.
The inequality follows from bounding each term in Eq.~\eqref{eq:RelEntMain},
the definition of the relative entropy $D$.
The entropy $S( \Omega_\ell )$ is bounded with $\theta$.
This bound relies on Schumacher's Theorem,
which quantifies the size of a high-probability subspace like $\mathcal{M}$
with an entropy $S(\gamma_{ \mathbf{v} })$~\cite{Schumacher95}.
We bound the second term in the $D$ definition with $\theta'$.
This bound relies on the definition of $\mathcal{M}$:
Outcomes of measurements of the $\bar{Q}_j$'s
are predictable up to parameters
on which $\theta'$ depends.

%
%
Finally, we present conditions
under which the approximate microcanonical subspace $\mathcal{M}$ exists.
Several parameters quantify the approximation.
The parameters are shown to be interrelated
and to approach zero simultaneously as $N$ grows. In particular,
the approximate microcanonical subspace 
$\mathcal{M}$ exists if $N$ is great enough.

%
%
%
%
This microcanonical derivation offers a physical counterpoint
to Jaynes's maximum-entropy derivation of the \GGS{}'s form.
We relate the \GGS{} to the physical picture
of a small subsystem in a vast universe
that occupies an approximate microcanonical state.
This vast universe allows the Correspondence Principle
to underpin our argument.
In the many-copy limit as $N \to \infty$,
the principle implies that quantum behaviors should vanish,
as the averages of the noncommuting charges $Q_j$
come to be approximated by commuting $\bar{Y}_j$'s.
Drawing on Ogata's~\cite[Theorem 1.1]{Ogata11},
we link thermality in the presence of noncommutation 
to the physical Correspondence Principle.

\subsection{Dynamical considerations}
The microcanonical and maximum-entropy arguments
rely on kinematics and information theory.
But we wish to associate the \GGS{}
with the fixed point of dynamics.
The microcanonical argument, combined with canonical typicality,
suggests that the \GGS{} is the equilibrium state of typical dynamics.
Canonical typicality enables us to model
the universe's state with a pure state
in the approximate microcanonical subspace $\mathcal{M}$.
If a large system occupies a randomly chosen pure state,
the reduced state of a small subsystem
is close to thermal~\cite{goldstein2006canonical,gemmer200418,PopescuSW06,LindenPSW09}.

Consider, as in the previous section, $N$ copies of the system $\mathcal{S}$.
By $\Omega$, we denoted
the composite system's approximately microcanonical state.
We denoted by $\Omega_\ell$
the reduced state of the $\ell^{\text{th}}$ copy,
formed by tracing out most copies from $\Omega$.
Imagine that the whole system 
occupies a pure state $\ket{ \psi }  \in  \mathcal{M}$. 
Denote by $\rho_\ell$
the reduced state of the $\ell^{\text{th}}$ copy.
$\rho_\ell$ is close to $\Omega_\ell$, on average,
by canonical typicality~\cite{PopescuSW06}:
\begin{equation}
   \<  \|  \rho_\ell -  \Omega_\ell  \|_1 \> \leq  \frac{d}{\sqrt{D_M}}.
   \label{eq:typical}
\end{equation}
The average $\langle . \rangle$ is over pure states $\ket{ \psi } \in \mathcal{M}$.
The trace norm is denoted by $ \| . \|_1$;
$d  :=  {\text{dim}} ( \mathcal{H} )$ denotes the dimensionality
of the Hilbert space $\mathcal{H}$ of one copy of $\mathcal{S}$;
and $D_M  :=  {\text{dim}} ( \mathcal{M} )$ denotes the dimensionality
of the approximate microcanonical subspace $\mathcal{M}$.

We have bounded, using canonical typicality, 
the average trace norm between $\rho_\ell$ and $\Omega_\ell$.
We can bound the average trace norm
between $\Omega_\ell$ and the \GGS{} $\gamma_{ \mathbf{v} }$,
using our microcanonical argument.
[Equation~\eqref{supp-eq:RelEnt} bounds the average relative entropy $D$
between $\Omega_\ell$ and  $\gamma_{ \mathbf{v} }$.
Pinsker's Inequality, Ineq.~\eqref{eq:PinskerIneq},
lower bounds $D$ in terms of the trace norm.]
Combining these two trace-norm bounds via the Triangle Inequality,
we bound the average distance between $\rho_\ell$ and $\gamma_{ \mathbf{v} }$:
\begin{equation}
   \biggl\langle    \frac1N    \sum_{\ell=0}^{N-1} 
         \| \rho_\ell   -    \gamma_{\mathbf{v}}  \|_1    \biggr\rangle
   \leq    \frac{d}{ \sqrt{D} }
          +   \sqrt{ 2 (\theta + \theta') }.
\label{eq:typical-2}
\end{equation}
If the whole system occupies a random pure state $\ket{ \psi }$
in $\mathcal{M}$,
the reduced state $\rho_\ell$ of a subsystem 
is, on average, close to the \GGS{} $\gamma_{ \mathbf{v} }$.

Sufficiently ergodic dynamics
is expected to evolve the whole-system state
to a $\ket{\psi}$ that satisfies Ineq.~\eqref{eq:typical-2}:
Suppose that the whole system begins in a pure state $\ket{ \psi (t{=}0) } \in \mathcal{M}$.
Suppose that the system's Hamiltonian
commutes with the charges: $[H,Q_j]=0$ for all $j=1,\ldots,c$.
The dynamics conserves the charges.
Hence most of the amplitude of $\ket{ \psi (t) }$
remains in $\mathcal{M}$ for appreciable times.
Over sufficient times, ergodic dynamics yields 
a state $\ket{ \psi(t) }$ that can be regarded as random.
Hence the reduced state is expected be close to 
$\Omega_\ell \approx \gamma_{ \mathbf{v}}$ for most long-enough times $t$.

Exploring how the dynamics depends on the number of copies of the system offers promise for interesting future research.


\subsection{Resource theory}

A thermodynamic resource theory is an explicit characterization of 
a thermodynamic system's resources, free states, and free operations 
with a rigorous mathematical model.
The resource theory specifies what 
an experimenter considers valuable (e.g., work) and what
is considered plentiful, or free (e.g., thermal states). 
To define a resource theory, we specify allowed operations and
which states can be accessed for free.
We use this framework to quantify the resources needed to transform one state into another. 

The first resource theory was entanglement theory~\cite{HorodeckiHHH09}.
The theory's free operations are local operations and classical communication (LOCC).
The free states are the states which can be easily prepared with LOCC,
the separable states. Entangled states constitute valuable resources. 
One can quantify entanglement using this resource theory. 

We present a resource theory for thermodynamic systems 
that have noncommuting conserved charges $Q_j$.
The theory is defined by its set of free operations, which we call ``{\GTOlong}'' (\GTO).
\GTO{} generalize thermal operations~\cite{janzing_thermodynamic_2000,FundLimits2}.  
How to extend thermodynamic resource theories 
to conserved quantities other than energy was noted in~\cite{FundLimits2,YungerHalpernR14,YungerHalpern14}. 
The {\GTO} theory is related to the resource theory in \cite{Imperial15}.

We supplement these earlier approaches with two additions.
First, a battery has a work payoff function dependent on chemical potentials.
We use this payoff function to define chemical work.
Second, we consider a reference system for a non-Abelian group.
The reference system is needed to resolve the difficulty encountered in \cite{YungerHalpern14,Imperial15}: There
might be no nontrivial operations which respect all the conservation laws. 
The laws of physics require that 
any operation performed by an experimenter
commutes with all the charges. 
If the charges fail to commute with each other, there might be no nontrivial unitaries
which commute with all of them. 
In practice, one is not limited by such a stringent constraint. 
The reason is that an experimenter has access to a reference frame~\cite{aharonov-susskind,kitaev2014super,BRS-refframe-review}.  
  
A reference frame is a system $W$
prepared in a state such that, for any unitary on a system $S$ which does not
commute with the charges of $S$, some global unitary on $WS$ conserves the total
charges and approximates the unitary on $S$ to arbitrary precision.  
The reference frame relaxes the strong constraint on the unitaries.  
The reference frame can be merged with the battery, 
in which the agent stores the ability to perform work.
We refer to the composite as ``the battery.''
We denote its state by $\rho_\batt$. 
The battery has a Hamiltonian $H_\batt$ and charges $Q_{j_\batt}$, described below.

Within this resource theory, the \GGSlong{} emerges in two ways:
\begin{enumerate}
   \item 
   The \GGS{} is the unique state from which work cannot be extracted,
   even if arbitrarily many copies are available. 
   That is, the \GGS{} is completely passive.
   \item The \GGS{} is the only state of $S$ that remains invariant under the free operations during which no work is performed on $S$.
\end{enumerate}

Upon proving the latter condition, we prove second laws for
thermodynamics with noncommuting charges.
These laws provide necessary conditions for a transition to be possible.
In some cases, we show, the laws are sufficient. 
These second laws govern state transitions of 
a system $\rho_\sys$, governed by a  Hamiltonian $H_\sys$,
whose charges $Q_{j_\sys}$ can be exchanged with the surroundings.
We allow the experimenter to couple  $\rho_\sys$ to free states $\rho_\bath$.
The form of $\rho_\bath$ is determined by the Hamiltonian $H_\bath$ and the charges $Q_{j_\bath}$
attributable to the free system.
We will show that these free states have the form of the \GGS. 
As noted above, no other state could be free.
If other states were free, an arbitrarily large amount of work 
could be extracted from them.

Before presenting the second laws,  
we must define ``work.''  
In textbook examples about gases, one defines
work as $\delta W = p\,dV$, because a change in volume at a fixed pressure can be
translated into the ordinary notion of mechanical work.  
If a polymer is stretched, then
$\delta W = F\,dx$, wherein $x$ denotes the polymer's linear displacement
and $F$ denotes the restoring force.
If $B$ denotes a magnetic field and $M$ denotes a medium's magnetization,
$\delta W = B\,dM$.  
The definition of ``work'' can depend on one's
ability to transform changes in thermodynamic variables into a standard
notion of ``work,'' such as mechanical or electrical work.

Our approach is to define a notion of chemical work. 
We could do so by modelling explicitly how the change in some quantity $Q_j$ 
can be used to extract $\mu_j \, \delta Q_j$ work.
Explicit modelling would involve adding a term to the battery Hamiltonian $H_\batt$. 
Rather than considering a specific work Hamiltonian or model of chemical work, 
however, we consider a work payoff function,
\begin{align}
  \workf= \sum_{j=0}^c\mu_j Q_{j_\batt}\ .
  \label{eq:workfunc}
\end{align}
The physical situation could determine the form of this $\workf$.
For example, the $\mu_j$'s could denote the battery's chemical potentials.
In such a  case, $\workf$ would denote the battery's total Hamiltonian, 
which would depend on those potentials. 

We choose a route conceptually simpler than considering 
an explicit Hamiltonian and battery system, however.
We consider Eq.~\eqref{eq:workfunc} as a payoff function that 
defines the linear combination of charges that interests us.
We define the (chemical) work expended or distilled during a transformation
as the change
in the quantum expectation value $\langle \workf \rangle$.

The form of $\workf$ is implicitly determined by the battery's structure
and by how charges can be converted into work.
For our purposes, however, the origin of the form of $\workf$ need not be known.
$\workf$ will uniquely determine the $\mu_j$'s in the \GGS.
Alternatively, we could first imagine that the agent could access, 
for free, a particular \GGS.
This \GGS{}'s form would determine the work function's form. 
If the charges commute, the corresponding Gibbs state is known to be 
the unique state that is completely passive with respect to the  observable~\eqref{eq:workfunc}.

In App.~\ref{section:SI_RT},
we specify the resource theory for noncommuting charges in more detail.
We show how to construct allowable operations, using the reference frame and battery. 
From the allowable operations, we derive a zeroth law of thermodynamics.
 
Complete passivity and zeroth law:
This zeroth law relates to the principle of complete passivity,
discussed in~\cite{PuszW78,Lenard78}.
A state is complete passive if,
an agent cannot extract work
from arbitrarily many copies of the state.
In the resource theory for heat exchanges, 
completely passive states can be free. 
They do not render the theory trivial 
because no work can be drawn from them~\cite{brandao2013second}.

In the \GTO{} resource theory, we show, the only reasonable free states 
have the \GGS{}'s form.
The free states' chemical potentials equal 
the $\mu_j$'s in the payoff function $\workf$, 
at some common fixed temperature.
 Any other state would render the resource theory trivial: 
 From copies of any other state,
arbitrary much work could be extracted for free. 
Then, we show that the \GGS{} is preserved by \GTO{},
the operations that perform no work on the system.
  
The free states form an equivalence class.
They lead to notions of temperature and chemical potentials $\mu_j$.
This derivation of the free state's form extends
complete passivity and the zeroth law from~\cite{brandao2013second} 
to noncommuting conserved charges.
The derivation further solidifies the role of the \GGSlong{} in thermodynamics.

%
%
%
%
Second laws:
The free operations preserve the \GGS{}.
We therefore focus on contractive measures 
of states' distances from the \GGS{}.
Contractive functions decrease monotonically under the free operations. 
Monotones feature in ``second laws'' that signal whether 
\GTO{} can implement a state transformation.
For example, the $\alpha$-R\'enyi relative entropies 
between a state and the \GGS{} cannot increase.  

Monotonicity allows us to define generalized free energies as
\begin{align}
  F_\alpha   \left(\rho_\sys,      \gibbsS  \right) 
  := \kB T D_\alpha\left(\rho_\sys   \Vert\gibbsS\right) 
  - \kB T\log Z\ ,
\end{align}
wherein $\beta  \equiv  1/ (k_{\text{B}} T)$ and $\kB$ denotes Boltzmann's constant.
$\gibbsS$ denotes the \GGS\ with respect to 
the Hamiltonian $H_\sys$ and the charges $Q_{j_\sys}$ of the system $S$.
The partition function is denoted by $Z$.
Various classical and quantum definitions of the R\'enyi relative entropies $D_\alpha$ 
are known to be contractive~\cite{brandao2013second,HiaiMPB2010-f-divergences,Muller-LennertDSFT2013-Renyi,WildeWY2013-strong-converse,JaksicOPP2012-entropy}.
The free energies $F_\alpha$ decrease monotonically
if no work is performed on the system.
Hence the $F_\alpha$'s characterize 
natural second laws that govern achievable transitions.

For example, the classical R{\'e}nyi divergences $D_\alpha(\initial\|\gibbsS)$ 
are defined as 
\begin{equation}
   D_\alpha(\initial   \|   \gibbsS)
   := \frac{\sgn(\alpha)}{\alpha-1} \log 
   \left(   \sum_k    p_k^\alpha    q_k^{1-\alpha}   \right),
   \label{eq:renyidivergence}
\end{equation}
wherein $p_k$ and $q_k$ denote the probabilities 
of $\initial$ and of $\gibbsS$ in the $\workf$ basis. 
The $D_\alpha$'s lead to second laws that hold 
even in the absence of a reference frame
and even outside the context of the average work.

The $F_\alpha$'s reduce to the standard free energy
when averages are taken over large numbers.
Consider the asymptotic (``thermodynamic'') limit 
in which many copies $( \initial )^{\otimes n}$ of $\initial$ are transformed.
Suppose that the agent has some arbitrarily small probability $\varepsilon$ of failing 
to implement the desired transition.
$\varepsilon$ can be incorporated into the free energies via a technique called ``smoothing''~\cite{brandao2013second}.
The average, over copies of the state, of every smoothed 
$F^\varepsilon_\alpha$ approaches $F_1$~\cite{brandao2013second}:
\begin{align}
  \lim_{n\rightarrow\infty}    \frac{1}{n}   
  F^\varepsilon_\alpha  & \Big(  
       (\initial)^{\otimes n},   ( \gibbsS )^{\otimes n}   \Big)
  =  F_1  \\
  & =   k_B T D(\rho_\sys   \|\gibbsS) -k_BT\log ( Z )   \\
  & =    
       \< H_\sys  \>_{\initial}   -   k_B TS(\initial)   
       +   \sum_{j = 1}^c   \mu_j   \<Q_{j_\sys}\>.
  \label{eq:freeenergy}
\end{align}
We have invoked the relative entropy's definition,
\begin{align}
   D(\rho_\sys   \|   \gibbsS)  :=   \Tr     \Big(  \rho_\sys   \log ( \rho_\sys ) \Big) 
   -  \Tr     \Big(  \rho_\sys   \log (\gibbsS)  \Big).
\end{align}
 Note the similarity between the many-copy average $F_1$ 
  in Eq.~\eqref{eq:freeenergy}
and the ordinary free energy,
$F = E - T \, dS + \sum_j \mu_j \, dN_j$.
The monotonic decrease of $F_1$ 
constitutes a necessary and sufficient condition 
for a state transition to be possible in the presence of a reference system
in the asymptotic limit.

In terms of the generalized free energies, we formulate second laws:
\begin{proposition}
  \label{prop:second-laws}
  In the presence of a heat bath 
  of inverse temperature $\beta$ and chemical potentials $\mu_j$, 
  the free energies   $F_\alpha(\initial,   \gibbsS)$ 
  decrease monotonically:
  \begin{align}
     F_{\alpha}(\initial,   \gibbsS) \geq F_{\alpha}(\final,  \gibbsS')
     \; \: \forall \alpha\geq 0,
  \end{align}
  wherein $\initial$ and $\final$ denote the system's initial and final states.
  The system's Hamiltonian and charges may transform
  from $H_\sys$ and $Q_{j_\sys}$ to $H'_\sys$ and $Q_{j_\sys}'$.
  The \GGS{}s associated with the same Hamiltonians and charges
  are denoted by $\gibbsS$ and $\gibbsS'$.
If
\begin{align}
& [\mathcal{W},  \final ]  =  0
    \quad {\text{and}} \quad \nonumber\\
&   F_{\alpha}(\initial,  \gibbsS) \geq F_{\alpha}(\final, \gibbsS')
\;   \;   \forall   \alpha \geq 0,
\end{align}
some \GTO\ maps $\initial$ to $\final$.
\end{proposition}

As in~\cite{brandao2013second}, additional laws can be defined 
in terms of quantum R{\'e}nyi divergences~\cite{HiaiMPB2010-f-divergences,Muller-LennertDSFT2013-Renyi,WildeWY2013-strong-converse,JaksicOPP2012-entropy}. 
This amounts to choosing, in Proposition~\ref{prop:second-laws}, 
a definition of the R\'enyi divergence which accounts for 
the possibility that $\initial$ and $\final$ 
have coherences relative to the $\workf_\sys$ eigenbasis.
Several measures are known to be
contractive~\cite{HiaiMPB2010-f-divergences,Muller-LennertDSFT2013-Renyi,WildeWY2013-strong-converse,JaksicOPP2012-entropy}.
They, too, provide a new set of second laws.

%
%
%
%
Extractable work:
In terms of the free energies $F_\alpha$, we can bound the work 
extractable from a resource state via \GTO{}.
We consider the battery $W$ separately from the system $S$ of interest.
We assume that $W$ and $S$ occupy a product state.
(This assumption is unnecessary
if we focus on average work.)
Let $\initialBatt$ and $\finalBatt$
denote the battery's initial and final states.

For all $\alpha$,
\begin{align}
  F_\alpha(\initial   \otimes   \initialBatt,   \gibbsSBatt)
  \geq F_\alpha(\final   \otimes   \finalBatt,   \gibbsSBatt).
\end{align}
Since
$F_\alpha(   \initial   \otimes   \initialBatt,   \gibbsSBatt) 
= F_\alpha(\initial,   \gibbsS) +
F_\alpha   \left(\initialBatt,   \gibbsBatt   \right)$,
\begin{align}
     F_\alpha   \left( \finalBatt,   \gibbsBatt   \right)
     -   F_\alpha   \left( \initialBatt,   \gibbsBatt   \right)
     \leq
     F_\alpha(\initial,   \gibbsS) -  F_\alpha(\final,   \gibbsS).
     \label{eq:work-alpha-free-energyMAIN}
\end{align}
The left-hand side of Ineq.~\eqref{eq:work-alpha-free-energyMAIN} 
represents the work extractable during one implementation of $\initial   \to   \final$.
Hence the right-hand side 
bounds the  work extractable during the transition.

Consider extracting work from many copies of $\initial$
(i.e., extracting work from $\initial^{ \otimes n}$) in each of many trials.
Consider the average-over-trials extracted work, defined as
$\Tr(  \workf [ \finalBatt  -  \initialBatt ] )$.
The average-over-trials work extracted per copy of $\initial$ is
$\frac{1}{n}   \Tr(  \workf [ \finalBatt  -  \initialBatt ] )$.
This average work per copy has a high probability of lying close to
the change in the expectation value of the system's work function,
$\frac{1}{n} \Tr ( \workf  [ \finalBatt - \initialBatt ] )
\approx    \Tr(  \workf [ \final  -  \initial ] )$, if $n$ is large.

Averaging over the left-hand side of Ineq.~\eqref{eq:work-alpha-free-energyMAIN} 
yields the average work $\delta \langle W \rangle$ extracted per instance of the transformation.
The average over the right-hand side
approaches the change in $F_1$ [Eq.~\eqref{eq:freeenergy}]:
\begin{align}
\delta \<  W  \>
  \leq   \delta \< H_\sys  \>_{\initial}   -   T  \,  \delta S(\initial)   
       +   \sum_{j = 1}^c   \mu_j   \,   \delta  \<Q_{j_\sys}\>.
\end{align}
This bound is achievable with a reference system, as shown in 
\cite{aberg2014catalytic,korzekwa2015extraction}.

We have focused on the extraction of work defined by $\workf$.
One can extract, instead, an individual charge $Q_j$.
The second laws do not restrict single-charge extraction.
But extracting much of one charge $Q_j$
precludes the extraction of much of another charge, $Q_k$.
 In App.~\ref{section:SI_RT}, we discuss the tradeoffs amongst
 the extraction of different charges $Q_j$.

\section{Discussion}
\label{section:Discussion}

We have derived, via multiple routes, the form of the thermal state
of a system that has noncommuting conserved charges.
First, we regarded the system as part of a vast composite
that occupied an approximate microcanonical state.
Tracing out the environment yields a reduced state
that lies, on average, close to a thermal state of the expected form.
This microcanonical argument, 
with canonical typicality,
suggests that the \GGS{} is the fixed point of typical dynamics.
Defining a resource theory,
we showed that the \GGS{} is the only completely passive state
and is the only state preserved by free operations.
These physical derivations
buttress Jaynes's information-theoretic derivation
from the Principle of Maximum Entropy.

Our derivations also establish tools
applicable to quantum noncommutation in thermodynamics.
In the microcanonical argument,
we introduced an approximate microcanonical state $\Omega$.
This $\Omega$ resembles the microcanonical ensemble
associated with a fixed energy, a fixed particle number, etc.\@
but accommodates noncommuting charges.
Our complete-passivity argument relies on a little-explored resource theory for thermodynamics,
in which free unitaries conserve noncommuting charges.

We expect that the equilibrium behaviors predicted here may be observed in experiments.
Quantum gases have recently demonstrated 
equilibrium-like predictions about integrable quantum systems~\cite{Rigol07,Langen15}.

%
From a conceptual perspective, our work shows that notions 
previously considered relevant only to commuting charges---for example, microcanonicals
subspace---extend to noncommuting charges.
This work opens fully quantum thermodynamics
to analysis with familiar, but suitably adapted, technical tools.


%
%

\endgroup

\renewcommand{\bibsection}{\section*{\refname}}
\putbib[NatCommsRefs.bibolamazi] 
\end{bibunit}

%
%




%
%
%

\appendix

\renewcommand{\thesection}{\Alph{chapter}.\arabic{section}}
\renewcommand{\thesubsection}{\thesection \roman{subsection}}
\renewcommand{\thesubsubsection}{\thesubsection \alph{subsubsection}}

\makeatletter\@addtoreset{equation}{section}
\def\theequation{\thesection\arabic{equation}}

%
%
\chapter{Appendices for ``Jarzynski-like equality for the out-of-time-ordered correlator''}
\label{app:Jarz_like}
\begin{bibunit}



\begingroup
\newcommand{\W}{ \mathcal{W} }  
\newcommand{\C}{ F }  
\newcommand{\RegC}{ \C_\reg }  
\newcommand{\Dim}{ d }  
\newcommand{\TW}{ \tilde{W} } 
\newcommand{\Protocol}{ \mathcal{P} }  
\newcommand{\Protocoll}{ \tilde{ \mathcal{P} } }  
\newcommand{\RegProtocollOne}{ \tilde{ \mathcal{P} }_{\reg, 1} }
\newcommand{\RegProtocollTwo}{ \tilde{ \mathcal{P} }_{\reg, 2} }
\newcommand{\reg}{ {\text{reg}} }
\newcommand{\GW}{ G_\W }  
\newcommand{\GV}{ G_V }  
\newcommand{\gw}{ g_w }  
\newcommand{\gv}{ g_v }  
\newcommand{\gwP}{ g_{w'} }  
\newcommand{\gvP}{ g_{v'} }  
\newcommand{\DegenW}{ \alpha }  
\newcommand{\DegenV}{ \lambda }  
\newcommand{\Coupling}{ c }  
\newcommand{\Charac}{ \mathcal{G} }  
\newcommand{\NondegW}{ \tilde{\W} } 
\newcommand{\NondegV}{ \tilde{V} } 
\newcommand{\U}{ \mathcal{U} } 
\newcommand{\Prob}{ \mathscr{P} } 
\newcommand{\TProb}{ \tilde{ \Prob } } 
\newcommand{\weak}{ {\text{weak}} } 
\newcommand{\PWeak}{ \mathscr{P}_\weak } 
\newcommand{\WeakInt}{ \mathcal{I} }

%
%
\section{Weak measurement of 
the combined quantum amplitude $\tilde{A}_\rho$}
\label{section:Jarz_like_App_A}

$\tilde{A}_\rho$ [Eq.~\eqref{eq:TildeAExp}] 
resembles the Kirkwood-Dirac quasiprobability
for a quantum state~\cite{Kirkwood_33_Quantum,Dirac_45_On,Dressel_15_Weak}.
This quasiprobability has been inferred
from weak-measurement experiments~\cite{Lundeen_11_Direct,Lundeen_12_Procedure,Bamber_14_Observing,Mirhosseini_14_Compressive,Dressel_14_Understanding}.
Weak measurements have been performed on cold atoms~\cite{Smith_04_Continuous},
which have been proposed as platforms 
for realizing scrambling and quantum chaos~\cite{Swingle_16_Measuring,Yao_16_Interferometric,Danshita_16_Creating}.

$\tilde{A}_\rho$ can be inferred from many instances of a protocol $\Protocol_\weak$.
$\Protocol_\weak$ consists of a state preparation,
three evolutions interleaved with three weak measurements,
and a strong measurement.
The steps appear in Sec.~\ref{section:WeakMain}. 

I here flesh out the protocol,
assuming that the system, $S$, begins in
the infinite-temperature Gibbs state: $\rho = \id / \Dim$.
$\tilde{A}_\rho$ simplifies as in Eq.~\eqref{eq:TildeASimple}.
The final factor becomes $p_{w_3, \DegenW_{w_3} }  =  1 / \Dim$.
The number of weak measurements in $\Protocol_\weak$ reduces to two.
Generalizing to arbitrary $\rho$'s
is straightforward but requires lengthier calculations
and more ``background'' terms.

Each trial in the simplified $\Protocol_\weak$ consists of a state preparation,
three evolutions interleaved with two weak measurements,
and a strong measurement.
Loosely, one performs the following protocol:
Prepare $\ket{ w_3,  \DegenW_{w_3} }$.
Evolve $S$ backward under $U^\dag$.
Measure $\ketbra{ v_1,  \DegenV_{v_1} }{ v_1,  \DegenV_{v_1} }$ weakly.
Evolve $S$ forward under $U$.
Measure $\ketbra{ w_2,  \DegenW_{w_2} }{ w_2,  \DegenW_{w_2} }$ weakly.
Evolve $S$ backward under $U^\dag$.
Measure $\ketbra{ v_2,  \DegenV_{v_2} }{ v_2,  \DegenV_{v_2} }$ strongly.

Let us analyze the protocol in greater detail.
The $\ket{ w_3,  \DegenW_{w_3} }$ preparation and backward evolution yield
$\ket{ \psi }  =  U^\dag \ket{ w_3,  \DegenW_{w_3} }$.
The weak measurement of 
$\ketbra{ v_1,  \DegenV_{v_1} }{ v_1,  \DegenV_{v_1} }$ is implemented as follows:
$S$ is coupled weakly to an ancilla $\mathcal{A}_a$.
The observable $\NondegV$ of $S$ comes to be correlated
with an observable of $\mathcal{A}_a$.
Example $\mathcal{A}_a$ observables include
a pointer's position on a dial
and a component $\sigma_\ell$ of a qubit's spin (wherein $\ell = x, y, z$).
The $\mathcal{A}_a$ observable is measured projectively.
Let $x$ denote the measurement's outcome.
$x$ encodes partial information about the system's state.
We label by $( v_1,  \DegenV_{v_1} )$ the $\NondegV$ eigenvalue
most reasonably attributable to $S$
if the $\mathcal{A}_a$ measurement yields $x$.

The coupling and the $\mathcal{A}_a$ measurement
evolve $\ket{ \psi }$ under the Kraus operator~\cite{NielsenC10}
\begin{align}
   \label{eq:Mx}
   M_x  =  \sqrt{p_a(x)} \: \id
   + g_a(x)  \,  \ketbra{ v_1,  \DegenV_{v_1} }{ v_1,  \DegenV_{v_1} }   \, .
\end{align}
Equation~\eqref{eq:Mx} can be derived, e.g., from
the Gaussian-meter model~\cite{Dressel_15_Weak,Aharonov_88_How}
or the qubit-meter model~\cite{White_16_Preserving}.
The projector can be generalized to 
a projector $\Pi_{v_1}$ onto a degenerate eigensubspace.
The generalization may decrease exponentially
the number of trials required~\cite{BrianDisc}.
By the probabilistic interpretation of quantum channels,
the baseline probability $p_a(x)$ denotes
the likelihood that, in any given trial,
$S$ fails to couple to $\mathcal{A}_a$
but the $\mathcal{A}_a$ measurement yields $x$ nonetheless.
The detector is assumed, for convenience, to be calibrated such that 
\begin{align}
   \label{eq:Calibrate}
   \int dx \cdot x  \: p_a(x)  =  0 \, .
\end{align}
The small tunable parameter $g_a(x)$ 
quantifies the coupling strength.

The system's state becomes 
$\ket{ \psi' }  =  M_x U^\dag \ket{ w_3,  \DegenW_{w_3} }$,
to within a normalization factor.
$S$ evolves under $U$ as
\begin{align}
  \ket{ \psi' }  \mapsto  \ket{ \psi'' }
   =  U M_x U^\dag \ket{ w_3,  \DegenW_{w_3} } \, ,
\end{align}
to within normalization.
$\ketbra{ w_2,  \DegenW_{w_2} }{ w_2,  \DegenW_{w_2} }$ is measured weakly:
$S$ is coupled weakly to an ancilla $\mathcal{A}_b$.
$\NondegW$ comes to be correlated with
a pointer-like variable of $\mathcal{A}_b$.
The pointer-like variable is measured projectively.
Let $y$ denote the outcome.
The coupling and measurement evolve $\ket{ \psi'' }$ 
under the Kraus operator
\begin{align}
   \label{eq:My}
   M_y  =  \sqrt{ p_b(y) }  \:  \id
   +  g_b(y)  \,  \ketbra{ w_2,  \DegenW_{w_2} }{ w_2,  \DegenW_{w_2} }  \, .
\end{align}
The system's state becomes
$\ket{ \psi''' }  =  M_y U M_x U^\dag \ket{ w_3,  \DegenW_{w_3} } $,
to within normalization.
The state evolves backward under $U^\dag$.
Finally, $\NondegV$ is measured projectively.

Each trial involves two weak measurements
and one strong measurement.
The probability that the measurements yield the outcomes $x$, $y$, 
and $( v_2,  \DegenV_{v_2} )$ is
\begin{align}
   \label{eq:PWeak}
   \PWeak \LParen x, y,  ( v_2,  \DegenV_{v_2} ) \RParen =  
   | \langle v_2,  \DegenV_{v_2} | U^\dag M_y U M_x U^\dag 
   | w_3,  \DegenW_{w_3} \rangle |^2 \, .
\end{align}
Integrating over $x$ and $y$ yields
\begin{align}
   \label{eq:WeakInt}
   \WeakInt  :=
   \int dx \; dy \cdot x \, y \; \PWeak \LParen x,y, ( v_2,  \DegenV_{v_2} ) \RParen \, .
\end{align}
We substitute in for $M_x$ and $M_y$ 
from Eqs.~\eqref{eq:Mx} and~\eqref{eq:My},
then multiply out.
We approximate to second order in the weak-coupling parameters.
The calibration condition~\eqref{eq:Calibrate}
causes terms to vanish:
\begin{align}
   \label{eq:WeakInt2}
   \WeakInt  &  =  \int dx \; dy  \cdot  x \:  y  \;   \sqrt{ p_a(x) \: p_b(y)  }  
   \Big[  g_a (x) \, g_b(y)  \cdot \Dim
            \nonumber \\ &    \times  
            \tilde{A}_{\id / \Dim} ( w, v, \DegenW_w, \DegenV_v )
            +  \cc \Big] 
   +  \int dx \; dy  \cdot  x \:  y  \;
   \sqrt{ p_a(x) \: p_b(y)  }
   \nonumber \\ &  \times
   \Big[  g_a^*(x) \, g_b(y)  \,  
   \langle v_2,  \DegenV_{v_2} | U^\dag |  w_2,  \DegenW_{w_2}  \rangle 
   \langle  w_2,  \DegenW_{w_2}  | w_3,  \DegenW_{w_3} \rangle
   \nonumber \\ &  \times
   \LParen \langle v_2,  \DegenV_{v_2} | v_1,  \DegenV_{v_1} \rangle 
   \langle v_1,  \DegenV_{v_1} | U^\dag 
   | w_3,  \DegenW_{w_3} \rangle \RParen^*
   + \cc \Big] 
   \nonumber \\ & 
   + O \LParen g_a(x)^2 \, g_b(y) \RParen
   + O \LParen g_a(x) \, g_b(y)^2 \RParen \, .
\end{align}

The baseline probabilities $p_a(x)$ and $p_b(x)$
are measured during calibration.
Let us focus on the second integral.
By orthonormality, $\langle  w_2,  \DegenW_{w_2}  | w_3,  \DegenW_{w_3} \rangle 
= \delta_{ w_2 w_3}  \,  \delta_{ \DegenW_{w_2}  \DegenW_{w_3} }$,
and $\langle v_2,  \DegenV_{v_2} |  v_1,  \DegenV_{v_1}  \rangle  
=  \delta_{v_2  v_1}  \,  \delta_{ \DegenV_{v_2}  \DegenV_{v_1} }$.
The integral vanishes if 
$( w_3,  \DegenW_{w_3} ) \neq ( w_2,  \DegenW_{w_2} )$ 
or if $( v_2,  \DegenV_{v_2} ) \neq  ( v_1,  \DegenV_{v_1} )$.
Suppose that $( w_3,  \DegenW_{w_3} ) = ( w_2,  \DegenW_{w_2} )$ 
and $( v_2,  \DegenV_{v_2} ) = ( v_1,  \DegenV_{v_1} )$.
The second integral becomes
\begin{align}
   & \int dx \; dy  \cdot  x \:  y  \;
   \sqrt{ p_a(x) \, p_b(y) }  \:  
   \Big[ g_a^*(x) \, g_b(y) \,
   \nonumber \\ & \qquad \times
   | \langle v_2,  \DegenV_{v_2} | U^\dag |  w_3,  \DegenW_{w_3}  \rangle |^2
   + \cc \Big]  .
\end{align}
The square modulus, a probability,
can be measured via Born's rule.
The experimenter controls $g_a(x)$ and $g_b(y)$.
The second integral in Eq.~\eqref{eq:WeakInt2}
is therefore known.

From the first integral, we infer about   $\tilde{A}_{\id / \Dim}$.
Consider trials in which the couplings are chosen such that
\begin{align}
   \alpha  :=  
    \int dx \; dy  \cdot  x \:  y  \;
   \sqrt{ p_a(x) \, p_b(y) }  \:  
   g_a(x) \, g_b(y) \in \mathbb{R} \, .
\end{align}
The first integral becomes
$2 \alpha \, \Dim  \, \Re \LParen 
\tilde{A}_{\id / \Dim}  ( w, v, \DegenW_w, \DegenV_v ) \RParen$.
From these trials, one infers the real part of $\tilde{A}_{\id / \Dim}$.
Now, consider trials in which
$i \, \alpha  \in  \mathbb{R}$.
The first bracketed term becomes
$2 | \alpha | \, \Dim  \,  \Im \LParen 
\tilde{A}_{\id / \Dim}  ( w, v, \DegenW_w, \DegenV_v ) \RParen \, .$
From these trials, one infers the imaginary part of $\tilde{A}_{\id / \Dim}$.

$\alpha$ can be tuned between real and imaginary in practice~\cite{Lundeen_11_Direct}.
Consider a weak measurement in which the ancillas are qubits.
An ancilla's $\sigma_y$ can be coupled to a system observable.
Whether the ancilla's $\sigma_x$ or $\sigma_y$ is measured
dictates whether $\alpha$ is real or imaginary.

The combined quantum amplitude $\tilde{A}_\rho$ 
can therefore be inferred from weak measurements.
$\tilde{A}_\rho$ can be measured alternatively via interference.

%
%
\section{Interference-based measurement of the combined quantum amplitude
$\tilde{A}_\rho$}
\label{section:Interfere} 

I detail an interference-based scheme for measuring 
$\tilde{A}_\rho( w, v, \DegenW_w, \DegenV_v )$ 
[Eq.~\eqref{eq:TildeAExp}].
The scheme requires no reversal of the time evolution in any trial.
As implementing time reversal can be difficult,
the absence of time reversal can benefit 
OTOC-measurement schemes~\cite{Yao_16_Interferometric,Zhu_16_Measurement}.

I specify how to measure an inner product
$z  :=  \langle a | \U | b \rangle$,
wherein $a, b  \in \{  ( w_\ell,  \DegenW_{w_\ell } ),
 ( v_m ,  \DegenV_{v_m}  )  \}$ 
 and $\U  \in  \{ U, U^\dag \}$.
Then, I discuss measurements of the state-dependent factor
in Eq.~\eqref{eq:TildeAExp}.

The inner product $z$ is measured as follows.
The system $S$ is initialized 
to some fiducial state $\ket{f}$.
An ancilla qubit $\mathcal{A}$ is prepared in the state 
$\frac{1}{\sqrt{2} }( \ket{0} + \ket{1} )$.
The $+1$ and $-1$ eigenstates of $\sigma_z$
are denoted by $\ket{0}$ and $\ket{1}$.
The composite system $\mathcal{A}S$ begins in the state
$\ket{ \psi }  =  \frac{1}{\sqrt{2} } ( \ket{0} \ket{f}  +  \ket{1} \ket{f} )$.

A unitary is performed on $S$, conditioned on $\mathcal{A}$:
If $\mathcal{A}$ is in state $\ket{0}$, 
then $S$ is brought to state $\ket{b}$,
and $\U$ is applied to $S$.
If $\mathcal{A}$ is in state $\ket{1}$,
$S$ is brought to state $\ket{a}$.
The global state becomes
$\ket{ \psi' }  =  \frac{1}{ \sqrt{2} } [
   \ket{0} ( \U \ket{b} )  +  \ket{1} \ket{a} ) ] \, .$
A unitary $e^{ - i \theta \sigma_x }$
rotates the ancilla's state 
through an angle $\theta$ about the $x$-axis.
The global state becomes
\begin{align}
   \label{eq:PsiPP}
   \ket{ \psi'' }  & =  \frac{1}{ \sqrt{2} }  \Bigg[
   \left(  \cos  \frac{\theta}{2}  \ket{0}  
   -  i \sin \frac{ \theta }{ 2 }  \ket{1}   \right)  
   ( \U \ket{ b }  )
   \nonumber \\ & \qquad
   +  \left(  - i \sin \frac{ \theta }{ 2 }  \ket{0}  
   +  \cos  \frac{ \theta }{ 2 }  \ket{1} \right)
   \ket{a }  \Bigg]  \, .
\end{align}

The ancilla's $\sigma_z$ is measured,
and the system's $\{ \ket{ a } \}$ is measured.
The probability that the measurements
yield $+1$ and $a$ is
\begin{align} 
   \Prob(+1, a)  & =  \frac{1}{4} ( 1 - \sin \theta)
   \Bigg(  \cos^2  \frac{ \theta }{ 2 }  \:  | z |^2
   -  \sin \theta  \:  \Im  (z)
   +  \sin^2 \frac{ \theta }{ 2 }  \Bigg) \, .
\end{align}
The imaginary part of $z$ is denoted by $\Im(z)$.
$\Prob(+1, a)$ can be inferred from the outcomes of multiple trials.
The $| z |^2$, representing a probability, can be measured independently.
From the $| z |^2$ and $\Prob(+1, a)$ measurements,
$\Im (z)$ can be inferred.

$\Re(z)$ can be inferred from another set of interference experiments.
The rotation about $\hat{x}$ is replaced with
a rotation about $\hat{y}$.
The unitary $e^{-i \phi \sigma_y }$ implements this rotation,
through an angle $\phi$.
Equation~\eqref{eq:PsiPP} becomes
\begin{align}
   \ket{ \tilde{ \psi}'' } & =
   \frac{1}{ \sqrt{2} }  \Big[  
   \left(  \cos \frac{ \phi }{ 2 }  \,  \ket{0}  +  \sin \frac{ \phi }{2} \, \ket{1}  \right)
   ( \U \, \ket{b} )
   \nonumber \\ & \qquad \qquad +  
   \left(  - \sin \frac{\phi }{2}  \,  \ket{0}
   +  \cos \frac{ \phi }{2}  \,  \ket{1}  \right)  \ket{a}  
   \Big] \, .
\end{align}
The ancilla's $\sigma_z$ and the system's $\{ \ket{a} \}$
are measured.
The probability that the measurements yield $+1$ and $a$ is
\begin{align}
   \TProb(+1, a)  & =  \frac{1}{4}  ( 1 - \sin \phi )
   \Bigg(  \cos^2 \frac{\phi}{2}  \,  | z |^2
   \nonumber \\ & \qquad
   - \sin \phi \; \Re(z)  +  \sin^2  \frac{\phi}{2}  \Bigg) \, .
\end{align}
One measures $\TProb(+1, a)$ and $|z|^2$,
then infers  $\Re(z)$.
The real and imaginary parts of $z$ are thereby gleaned
from interferometry.

Equation~\eqref{eq:TildeAExp} contains the state-dependent factor
$M := \langle v_1,  \DegenV_{v_1}  |  
  \rho  U^\dag  |  w_3,  \DegenW_{w_3}  \rangle$.
This factor is measured easily if 
$\rho$ shares its eigenbasis with $\NondegW(t)$ or with $\NondegV$.
In these cases, $M$ assumes the form 
$\langle a | U^\dag | b \rangle \, p$.
The inner product is measured as above.
The probability $p$ is measured via Born's rule.
In an important subcase, $\rho$ is 
the infinite-temperature Gibbs state $\id / \Dim$.
The system's size sets $p = 1 / \Dim$.
Outside of these cases, $M$ can be inferred from quantum tomography~\cite{Paris_04_Q_State_Estimation}.
Tomography requires many trials but is possible in principle
and can be realized with small systems.

%

\endgroup

\putbib[Jarz_like_bib] 
\end{bibunit}

%
%
\chapter{Appendices for ``The quasiprobability behind the out-of-time-ordered correlator''}
\label{app:OTOC_Quasi}
\begin{bibunit}

\begingroup


\newcommand{\W}{ \mathcal{W} }  
\newcommand{\Dim}{ d }  
\newcommand{\DegenW}{ \alpha }  
\newcommand{\DegenV}{ \lambda }  
\newcommand{\Sites}{N}  
\newcommand{\Sys}{S}  
\newcommand{\td}{t_{\text{d}}}  
\newcommand{\Lyap}{\lambda_{\text{L}}}  
\newcommand{\KD}{ \tilde{p}_{\text{KD}} }  
\newcommand*{\OurKD}[1]{\tilde{A}_{#1}}  
\newcommand*{\SumKD}[1]{\tilde{ \mathscr{A} }_{#1}}  
\newcommand*{\ProjW}[1]{\Pi^{ \W }_{#1}}  
\newcommand*{\ProjWt}[1]{\Pi^{ \W(t) }_{#1}}  
\newcommand*{\ProjV}[1]{\Pi^{ V }_{#1}}  

\newcommand{\NondegW}{ \tilde{\W} }
\newcommand{\NondegV}{ \tilde{V} }
\newcommand{\reg}{ {\text{reg}} } 
\newcommand{\TOC}{F_\toc} 
\newcommand{\GW}{ G_\W }
\newcommand{\GV}{ G_V }
\newcommand{\gw}{ g_w }
\newcommand{\gv}{ g_v }
\newcommand{\gwP}{ g_{w'} }
\newcommand{\gvP}{ g_{v'} }
\newcommand{\Coupling}{c}
\newcommand{\Charac}{ \mathcal{G} }  
\newcommand{\U}{ \mathcal{U} }  
\newcommand{\weak}{ {\text{weak}} }
\newcommand{\target}{ {\text{target}} }
\newcommand{\WeakInt}{ \mathcal{I} }
\newcommand{\ParenW}{{(\W)}}
\newcommand{\ParenV}{{(V)}}
\newcommand{\Protocol}{\mathcal{P}}  
\newcommand{\Protocoll}{\mathcal{P}_{\W}}  
\newcommand{\ProtocolA}{\mathscr{P}_\Amp}  
\newcommand{\toc}{{\text{TOC}}}
\newcommand*{\TOCKD}[1]{\tilde{A}^\toc_{#1}}  
\newcommand{\A}{\mathcal{A}} 
\newcommand{\B}{ \mathcal{B} } 
\newcommand{\C}{ \mathcal{C} } 
\newcommand{\K}{ \mathcal{K} } 
\newcommand{\Oper}{\mathcal{O}} 
\newcommand{\Ops}{\mathscr{K}} 
\newcommand{\Opsb}{\bar{\mathscr{K}}} 
\newcommand{\Gest}{ \Gamma_{\text{est}} } 
\newcommand{\TU}{ \tilde{U} } 
\newcommand{\TRho}{ \tilde{\rho} } 
\newcommand{\Amp}{A} 

\newcommand*{\Unit}[1]{ \bm{ \hat{ #1 }} }  
\newcommand{\ParenA}{{(a)}}
\newcommand{\ParenB}{{(b)}}
\newcommand{\ParenK}{{(\Ops)}}
\newcommand{\ParenKB}{{( \Opsb )}}

\section{Mathematical properties of $P(W, W')$}
\label{section:P_Properties}

Summing $\OurKD{\rho}$, with constraints,
yields $P(W, W')$ [Eq.~\eqref{eq:PWWPrime}].
Hence properties of $\OurKD{\rho}$ (Sec.~\ref{section:TA_Props})
imply properties of $P(W, W')$.

\begin{property}
\label{property:P_Complex}
$P(W, W')$ is a map from
a composition of two sets of complex numbers
to the complex numbers:
$P \:  :  \:  \Set{ W }  \times  \Set{  W'  }  \to  \mathbb{C}$.
The range is not necessarily real:
$\mathbb{C}  \supset \mathbb{R}$.
\end{property}

Summing quasiprobability values can eliminate nonclassical behavior:
Interference can reduce quasiprobabilities' nonreality and negativity.
Property~\ref{prop:MargOurKD} consists of an example.
One might expect $P (W, W')$, a sum of $\OurKD{\rho} ( . )$ values,
to be real.
Yet $P(W, W')$ is nonreal in many numerical simulations
(Sec.~\ref{section:Numerics}).

\begin{property} \label{prop:MargP}

Marginalizing $P(W, W')$ over one argument
yields a probability if $\rho$ shares
the $\NondegV$ eigenbasis or the $\NondegW(t)$ eigenbasis.
\end{property}

Consider marginalizing Eq.~\eqref{eq:PWWPrime} over $W'$.
The $( w_2, \DegenW_{w_2} )$  and  $( v_1,  \DegenV_{v_1} )$
sums can be performed explicitly:
\begin{align}
   P(W)  & :=  \sum_{ W' }  P(W, W')
   \\ &  \label{eq:PW_Help1}
   = \sum_{ \substack{ ( v_2,  \DegenV_{v_2} ),  \\
                                     ( w_3,  \DegenW_{w_3} ) } }
   \langle w_3,  \DegenW_{w_3}  |  U  |  v_2,  \DegenV_{v_2}  \rangle
   \langle  v_2,  \DegenV_{v_2}  |  \rho  U^\dag  |
                w_3,  \DegenW_{w_3}  \rangle
   \nonumber \\ & \qquad \qquad \qquad   \times
   \delta_{W ( w_3^*  v_2^* ) }  \, .
\end{align}
The final expression is not obviously a probability.

But suppose that $\rho$ shares its eigenbasis with
$\NondegV$ or with $\NondegW(t)$.
Suppose, for example, that $\rho$ has
the form in Eq.~\eqref{eq:WRho}.
Equation~\eqref{eq:PW_Help1} simplifies:
\begin{align}
   \label{eq:PW_Help2}
   P(W)  & =
   \sum_{ \substack{ ( v_2,  \DegenV_{v_2} ),  \\
                                 ( w_3,  \DegenW_{w_3} ) } }
   p (  v_2,  \DegenV_{v_2} ;   w_3,  \DegenW_{w_3} )  \,
   \delta_{W ( w_3^*  v_2^* ) } \, .
\end{align}
The
\mbox{$p (  v_2,  \DegenV_{v_2} ;   w_3,  \DegenW_{w_3} )
:=  | \langle  w_3,  \DegenW_{w_3}  |  U  |
       v_2,  \DegenV_{v_2} \rangle |^2  \,
     p_{ v_2,  \DegenV_{v_2} }$}
denotes the joint probability that a $\NondegV$ measurement of $\rho$
yields $( v_2,  \DegenV_{v_2} )$
and, after a subsequent evolution under $U$,
a $\NondegW$ measurement yields $( w_3,  \DegenW_{w_3} )$.

Every factor in Eq.~\eqref{eq:PW_Help2} is nonnegative.
Summing over $W$ yields
a sum over the arguments of $\OurKD{\rho} ( . )$.
The latter sum equals one, by Property~\ref{prop:MargOurKD}:
$\sum_W  P(W)  =  1$.
Hence $P(W)  \in  [0, 1]$.
Hence $P(W)$ behaves as a probability.

We can generalize Property~\ref{prop:MargP} to arbitrary Gibbs states
$\rho = e^{ - H / T } / Z$,
using the regulated quasiprobability~\eqref{eq:RegKD2}.
The regulated OTOC~\eqref{eq:RegOTOC_def}
equals a moment of the complex distribution
\begin{align}
   \label{eq:RegP_def}
   & P_\reg ( W, W' )  :=
   \sum_{ \substack{ ( v_1,  \DegenV_{v_1} ),  ( w_2,  \DegenW_{w_2} ),
                                 ( v_2,  \DegenV_{v_2} )    ( w_3,  \DegenW_{w_3} ) } }
   \\  \nonumber  &
   \OurKD{\rho}^\reg  (  v_1,  \DegenV_{v_1} ;  w_2, \DegenW_{w_2} ;
   v_2,  \DegenV_{v_2}  ;  w_3,  \DegenW_{w_3} )  \,
   \delta_{ W ( w_3^* v_2^* ) }  \,  \delta_{ W' ( w_2  v_1 ) }  \, .
\end{align}
The proof is analogous to the proof of Theorem~1 in~\cite{YungerHalpern_17_Jarzynski}.

Summing over $W'$ yields
$P_\reg (W)  :=  \sum_{ W' }  P_\reg  (W, W')$.
We substitute in from Eq.~\eqref{eq:RegP_def},
then for $\OurKD{\rho}^\reg$ from Eq.~\eqref{eq:RegKD2}.
We perform the sum over $W'$ explicitly,
then the sums over $(w_2, \DegenW_{w_2} )$ and $( v_1 ,  \DegenV_{v_1} )$:
\begin{align}
   P_\reg (W)  =  \sum_{ \substack{ ( v_2,  \DegenV_{v_2} ) \\
                                                        ( w_3,  \DegenW_{w_3} ) } }
   | \langle  w_3,  \DegenW_{w_3}  |  \tilde{U} |
     v_2,  \DegenV_{v_2}  \rangle |^2  \,
   \delta_{ W ( w_3^*  v_2^* ) }  \, .
\end{align}
This expression is real and nonnegative.
$P_\reg(W)$ sums to one, as $P(W)$ does.
Hence $P_\reg(W)  \in  [ 0, \, 1 ]$ acts as a probability.

\begin{property}[Degeneracy of every $P(W, W')$
associated with $\rho = \id / \Dim$ and
with eigenvalue-$( \pm 1 )$ operators $\W$ and $V$]
\label{prop:P_Degen}

Let the eigenvalues of $\W$ and $V$ be $\pm 1$.
For example, let $\W$ and $V$ be Pauli operators.
Let $\rho = \id / \Dim$ be the infinite-temperature Gibbs state.
The complex distribution has the degeneracy
$P(1, -1 )  =  P(-1, 1)$.
\end{property}

Property~\ref{prop:P_Degen} follows from
(1) Eq.~\eqref{eq:SumKD_simple2} and
(2) Property~\ref{property:Syms} of $\OurKD{ ( \id / \Dim) }$.
Item (2) can be replaced with the trace's cyclicality.
We reason as follows:
$P(W, W')$ is defined in Eq.~\eqref{eq:PWWPrime}.
Performing the sums over the degeneracies
yields $\SumKD{ ( \id / \Dim) }$.
Substituting in from Eq.~\eqref{eq:SumKD_simple2} yields
\begin{align}
   \label{eq:P_Degen_Help1}
   P(W, W')  & =   \frac{1}{ \Dim }
   \sum_{ v_1 , w_2 , v_2 , w_3 }
   \Tr \left(  \ProjWt{w_3}  \ProjV{v_2}  \ProjWt{w_2}  \ProjV{v_1}  \right)
   \nonumber \\ & \qquad \qquad \qquad \; \times
   \delta_{W ( w_3^*  v_2^* ) }  \delta_{W' ( w_2  v_1 ) } \, .
\end{align}

Consider inferring $\OurKD{ ( \id / \Dim) }$ or $\SumKD{ ( \id / \Dim) }$
from weak measurements.
From one trial, we infer about four random variables:
$v_1,  w_2,  v_2$ and $w_3$.
Each variable equals $\pm 1$.
The quadruple $(v_1,  w_2,  v_2 ,  w_3)$ therefore
assumes one of sixteen possible values.
These four ``base'' variables are multiplied to form
the composite variables $W$ and $W'$.
The tuple $(W, W')$ assumes one of four possible values.
Every $(W, W')$ value can be formed from
each of four values of $(v_1,  w_2,  v_2 ,  w_3)$.
Table~\ref{table:P_Degen} lists the tuple-quadruple correspondences.

%
%
\begin{table*}[t]
\begin{center}
\begin{tabular}{|c|c|}
   \hline
        $(W, W')$
   &   $( v_1,  w_2,  v_2,  w_3 )$
   \\  \hline \hline
        $(1, 1)$
   &   $(1, 1, 1, 1),  (1, 1, -1, -1),  (-1, -1, 1, 1),  (-1, -1, -1, -1)$
   \\  \hline
        $(1, -1)$
   &   $(-1, 1, 1, 1),  (-1, 1, -1, -1),  (1, -1, 1, 1),  (1, -1, -1, -1)$
   \\  \hline
        $(-1, 1)$
   &   $(1, 1, -1, 1),  (1, 1, 1, -1),  (-1, -1, -1, 1),  (-1, -1, 1, -1)$
   \\  \hline
        $(-1, -1)$
   &   $(-1, 1, -1, 1),  (-1, 1, 1, -1),  (1, -1, -1, 1),  (1, -1, 1, -1)$
   \\  \hline
\end{tabular}
\caption{\caphead{Correspondence between
tuples of composite variables
and quadruples of ``base'' variables:}
From each weak-measurement trial, one learns about
a quadruple $( v_1,  w_2,  v_2,  w_3 )$.
Suppose that the out-of-time-ordered-correlator operators $\W$ and $V$
have the eigenvalues $w_\ell,  v_m  =  \pm 1$.
For example, suppose that $\W$ and $V$ are Pauli operators.
The quadruple's elements are combined into
$W := w_3^* v_2^*$ and $W'  :=  w_2 v_1$.
Each $(W, W')$ tuple can be formed from
each of four quadruples.}
\label{table:P_Degen}
\end{center}
\end{table*}

Consider any quadruple associated with
$(W, W')  =  (1, -1)$, e.g., $(-1, 1,  1,  1)$.
Consider swapping $w_2$ with $w_3$
and swapping $v_1$ with $v_2$.
The result, e.g., $(1, 1, -1, 1)$, leads to $(W, W')  =  (-1, 1)$.
This double swap amounts to a cyclic permutation
of the quadruple's elements.
This permutation is equivalent to
a cyclic permutation of the argument of
the~\eqref{eq:P_Degen_Help1} trace.
This permutation preserves the trace's value
while transforming the trace into $P(-1, 1)$.
The trace originally equaled $P(1, -1)$.
Hence $P(1, -1)  =  P(-1, 1)$.

\section{Retrodiction about the symmetrized composite observable $\tilde{\Gamma}  :=  i ( \K \ldots \A  -  \A \ldots \K )$}
\label{section:RetroK2}

Section~\ref{section:TA_retro} concerns retrodiction about
the symmetrized observable $\Gamma  :=  \K \ldots \A + \A \ldots \K$.
The product $\K \ldots \A$ is symmetrized also in
$\tilde{\Gamma}  :=  i ( \K \ldots \A  -  \A \ldots \K )$.
One can retrodict about $\tilde{\Gamma}$,
using $\Ops$-extended KD quasiprobabilities $\OurKD{\rho}^\ParenK$,
similarly to in Theorem~\ref{theorem:RetroK}.

The value most reasonably attributable retrodictively to
the time-$t'$ value of $\tilde{\Gamma}$ is
given by Eqs.~\eqref{eq:GammaW},~\eqref{eq:QuasiBayesLeft1},
and~\eqref{eq:QuasiBayesRt1}.
The conditional quasiprobabilities
on the right-hand sides of
Eqs.~\eqref{eq:QuasiBayesLeft2} and~\eqref{eq:QuasiBayesRt2} become
\begin{align}
   \label{eq:QuasiBayesLeft2Tilde}
   \tilde{p}_\rightarrow ( a, \ldots, k, f | \rho )
   =  \frac{ - \Im ( \langle f' | k \rangle  \langle k |  \ldots
   | a \rangle  \langle a |  \rho'  | f' \rangle ) }{
   \langle f' | \rho' | f' \rangle }
\end{align}
and
\begin{align}
   \label{eq:QuasiBayesRt2Tilde}
   \tilde{p}_\leftarrow ( k, \ldots, a, f | \rho )
   =  \frac{ \Im  ( \langle f' | a \rangle  \langle a |  \ldots
   | k \rangle  \langle k | \rho' | f' \rangle ) }{
   \langle f' | \rho' | f' \rangle }  \, .
\end{align}
The extended KD distributions become
\begin{align}
   \label{eq:Extend_KD_Left_Tilde}
   \OurKD{ \rho, \rightarrow }^\ParenK  ( \rho, a, \ldots, k , f )
   =  i  \langle f' | k \rangle  \langle k |  \ldots
   | a \rangle  \langle a |  \rho'  | f' \rangle
\end{align}
and
\begin{align}
   \label{eq:Extend_KD_Rt_Tilde}
   \OurKD{ \rho, \leftarrow }^\ParenK  ( \rho, k, \ldots, a , f )
   =  - i \langle f' | a \rangle  \langle a |  \ldots
   | k \rangle  \langle k | \rho | f' \rangle \, .
\end{align}

To prove this claim, we repeat the proof of Theorem~\ref{theorem:RetroK}
until reaching Eq.~\eqref{eq:Choose2}.
The definition of $\tilde{\Gamma}$ requires that
an $i$ enter the argument of the first $\Re$
and that a $-i$ enter the argument of the second $\Re$.
The identity $\Re ( i z )  =  - \Im (z)$, for $z \in \mathbb{C}$,
implies Eqs.~\eqref{eq:QuasiBayesLeft2Tilde}--\eqref{eq:Extend_KD_Rt_Tilde}.

\endgroup

\putbib[OTOC_Quasi_bib] 
\end{bibunit}

%
%
\chapter{Appendices for ``MBL-Mobile: Many-body-localized engine''}
\label{app:MBL_Mobile}
\begin{bibunit}



\begingroup

\newcommand{\JAvg}{ \langle J \rangle } 
\newcommand{\dTyp}{ \delta_{\text{typ}} } 
\newcommand{\VeryDeloc}{ {\text{v. \; deloc}} } 
\newcommand{\Deloc}{ {\text{deloc}} } 
\newcommand{\VeryLoc}{<} 
\newcommand{\Loc}{>} 
\newcommand{\HDim}{\mathcal{N}} 
\newcommand{\Sites}{N} 
\newcommand{\SitesTot}{ \Sites_\tot }
\newcommand{\NCross}{ n_{\text{cross}} }
\newcommand{\Cross}{{\text{Cross}}}
\newcommand{\Poisson}{ {\text{Poisson}} }
\newcommand{\etaLarge}{ \eta_{ {\text{large}} } }
\newcommand{\etaSmall}{ \eta_{ {\text{small}} } }
\newcommand{\high}{ {\text{high}} } 
\newcommand{\low}{ {\text{low}} } 
\newcommand{\Power}{ \mathscr{P} } 
\newcommand{\In}{ {\text{in}} } 
\newcommand{\zd}{ {\text{ZD}} } 
\newcommand{\diab}{{\text{diab}}}
\newcommand{\adiab}{{\text{adiab}}}
\newcommand{\loc}{ {\text{loc}} }
\newcommand{\MBL}{{\text{MBL}}}
\newcommand{\ETH}{{\text{GOE}}}
\newcommand{\Anderson}{ {\text{And}} }
\newcommand{\class}{ {\text{class}} }
\newcommand{\shallow}{ {\text{shallow}} }
\newcommand{\disorder}{ {\text{disorder}} }
\newcommand{\particle}{ {\text{particle}} }
\newcommand{\Otto}{{\text{Otto}}}
\newcommand{\QHO}{{\text{QHO}}}
\newcommand{\ideal}{ {\text{ideal}} }
\newcommand{\therm}{{\text{th}}}
\newcommand{\coupling}{g}
\newcommand{\tune}{{\text{tune}}}
\newcommand{\imperfect}{ {\text{imp}} }
\newcommand{\opt}{ {\text{opt}} }
\newcommand{\Carnot}{\text{Carnot}} 
\newcommand{\Wb}{W_{\text{b}}}  
\newcommand{\HTemp}{{\text{H}}} 
\newcommand{\CTemp}{ {\text{C}} }  
\newcommand{\THot}{T_\HTemp}  
\newcommand{\TCold}{T_\CTemp}  
\newcommand{\betaH}{\beta_\HTemp}  
\newcommand{\betaC}{\beta_\CTemp}  
\newcommand{\gap}{ \Delta }
\newcommand{\HScale}{\mathcal{E}}
\newcommand{\TGap}{ \tilde{\varepsilon} }  
\newcommand{\LZ}{{\text{LZ}}} 
\newcommand{\HalfLZ}{\text{frac-LZ}}  
\newcommand{\DB}{{\text{DB}}}  
\newcommand{\APT}{{\text{APT}}}  
\newcommand{\PDown}{ P_\downarrow }  
\newcommand{\PUp}{ P_\uparrow }  
\newcommand{\PDownn}{ \PDownnn^{\text{true}} }
\newcommand{\PDownnn}{ \mathcal{P}_\downarrow }
\newcommand{\PUppp}{ \mathcal{P}_\uparrow }  
\newcommand{\EjE}{ E_j }  
\newcommand{\EjM}{ E'_j }  
\newcommand{\disp}{ {\text{displ}} }  
\newcommand{\qubit}{{\text{qubit}}}  
\newcommand{\meso}{{\text{meso}}}  
\newcommand{\Sim}{{\text{sim}}}  
\newcommand{\ZH}{Z}    
\newcommand{\ZC}{Z'}
\newcommand{\dAvg}{\expval{\delta}}
\newcommand{\dAvgSub}{\dAvg}
\newcommand{\macro}{{\text{macro}}}
\newcommand{\deltaMBL}{\delta_-}
\newcommand{\Sys}{S}
\newcommand{\J}{ \mathcal{J} }
\newcommand{\JFar}{\J_{L \gg \xi}}
\newcommand{\JClose}{\J_{L \leq \xi}}
\newcommand{\Cp}{C_{\text{P}}} 
\newcommand{\Cv}{C_{\text{v}}}  
\newcommand{\A}{{\text{A}}}
\newcommand{\B}{{\text{B}}} 
\newcommand{\DOS}{\mu}  
\newcommand{\Err}{\epsilon}  
\newcommand{\Li}{{\text{Li}}}  
\newcommand{\bath}{{\text{bath}}}  
\newcommand{\LBath}{L_\bath}  
\newcommand{\cycle}{{\text{cycle}}}  
\newcommand{\HighOrd}{{\text{high-ord.}}}  

\newcommand{\Ell}{ { (\ell) } }
\newcommand{\ParenJ}{ { (j) } } 
\newcommand{\ParenE}{ { (E) } } 
\newcommand{\LL}{ { (L) } } 


%
%
%
\section{Quantitative assessment of the mesoscopic MBL Otto engine}
\label{section:PowerApp}

We asses the mesoscopic engine introduced in Sec.~\ref{section:Meso_main}.
Section~\ref{section:Notation_app} reviews and introduces notation.
Section~\ref{section:Small_params} introduces
small expansion parameters.
Section~\ref{section:PSWAP} reviews the partial swap~\cite{Ziman_01_Quantum,Scarani_02_Thermalizing},
used to model cold thermalization (stroke 2).
The average heat $\expval{ Q_2 }$ absorbed during stroke 2
is calculated in Sec.~\ref{section:Q2};
the average heat $\expval{ Q_4 }$ absorbed during stroke 4,
in Sec.~\ref{section:Q4};
the average per-trial power $\expval{ W_\tot }$,
in Sec.~\ref{section:WTot};
and the efficiency $\eta_\MBL$, in Sec.~\ref{section:AdiabaticEta}.
The foregoing calculations rely on
adiabatic tuning of the Hamiltonian.
Six diabatic corrections are estimated
in Sec.~\ref{section:App_Diab}.

\subsection{Notation}
\label{section:Notation_app}

We focus on one mesoscopic engine $\Sys$ of $\Sites$ sites.
The engine corresponds to a Hilbert space
of dimensionality $\HDim \sim 2^N$.
We drop the subscript from the Hamiltonian $H_\meso (t)$.
$H(t)$ is tuned between $H_\ETH$, which obeys the ETH, 
and $H_\MBL$, which governs an MBL system.
Unprimed quantities often denote properties of $H_\ETH$;
and primed quantities, properties of $H_\MBL$:
$E_j$ denotes the $j^\th$-greatest energy of $H_\ETH$;
and $E'_j$, the $j^\th$-greatest energy of $H_\MBL$.
$\delta_j$ denotes the gap just below $E_j$;
and $\delta'_j$, the gap just below $E'_j$.
When approximating the spectra as continuous,
we replace $E_j$ with $E$ and $E'_j$ with $E'$.

Though the energies form a discrete set,
they can approximated as continuous.
ETH and MBL Hamiltonians have Gaussian DOSs:
\begin{align}
   \label{eq:DOS_App}
   \DOS(E)  =  \frac{ \HDim }{ \sqrt{ 2 \pi  \Sites }   \;  \HScale }  \: 
   e^{ - E^2  /  (2 \Sites  \HScale^2 ) }  \, ,
\end{align}
normalized to 
$\int_{-\infty}^\infty  dE  \;  \DOS ( E )   =  \HDim$.
The unit of energy, or energy density per site, is $\HScale$.
We often extend energy integrals' limits to $\pm \infty$,
as the Gaussian peaks sharply about $E = 0$.
The local average gap $\dAvg_E  =  \frac{1}{ \DOS(E) }$ 
and the average gap $\dAvg  
:=  \frac{ \HDim }{ \int_{-\infty}^\infty  dE  \;  \DOS^2(E) }
=  \frac{2 \sqrt{ \pi \Sites }  \:  \HScale }{ \HDim }$
(footnote~\ref{footnote:dAvg}).

The average $H_\ETH$ gap, $\dAvg$, 
equals the average $H_\MBL$ gap, by construction.
$\dAvg$ sets the scale for work and heat quantities.
Hence we cast $Q$'s and $W$'s as
\begin{align}
   (\text{number})(\text{function of small parameters}) \dAvg  \, .
\end{align}

The system begins the cycle in the state 
$\rho ( 0 )  =  e^{ - \betaH H_\ETH } / Z$.
The partition function $Z :=  \Tr \left( e^{ - \betaH H_\ETH } \right)$
normalizes the state.
$\Wb$ denotes the cold bath's bandwidth.
We set $\hbar  =  \kB  =  1 \, .$

$H(t)$ is tuned at a speed 
$v  :=  \HScale \left\lvert  \frac{ d \alpha_t }{ dt }  \right\rvert$,
wherein $\alpha_t$ denotes the dimensionless tuning parameter.
$v$ has dimensions of 
$\text{energy}^2$, 
as in~\cite{Landau_Zener_Shevchenko_10}.
Though our $v$ is not defined identically to
the $v$ in~\cite{Landau_Zener_Shevchenko_10},
ours is expected to behave similarly.

\subsection{Small parameters}
\label{section:Small_params}

We estimate low-order contributions 
to $\expval{ W_\tot }$ and to $\eta_\MBL$
in terms of small parameters:
\begin{enumerate}[leftmargin=*]

   \item The cold bath has a small bandwidth:
   $\frac{ \Wb }{ \dAvg }  \ll  1$.

   \item  The cold bath is cold: $\betaC \Wb > 0$.

   \item  Also because the cold bath is cold, 
   $1  \gg  e^{ - \betaC  \Wb } \approx  0$, 
   and $\frac{1}{ \betaC, \dAvg }  \ll  1$.

   \item
   The hot bath is hot: $\sqrt{ \Sites }  \:  \betaH \HScale \ll 1$.
   This inequality prevents $\betaH$ from contaminating
   leading-order contributions to heat and work quantities.
   ($\betaH$ dependence manifests in factors of
   $e^{ - \Sites ( \betaH \HScale )^2 / 4 } \, .$)
   Since $\betaH \HScale  \ll  \frac{1}{ \sqrt{ \Sites } }$ and 
   $\frac{ \dAvg }{ \HScale }  \ll 1  \, ,$
   $\betaH \dAvg  =  ( \betaH \HScale) \left( \frac{ \dAvg }{ \HScale }  \right)  
      \ll  \frac{1}{ \sqrt{ \Sites } } \, .$
\end{enumerate}

We focus on the parameter regime in which
\begin{align}
   \label{eq:Regime}
   \TCold  \ll  \Wb  \ll \dAvg
   \qquad \text{and} \qquad
   \sqrt{ \Sites }  \:  \betaH \HScale  \ll  1  \, .
\end{align}
The numerical simulations (Sec.~\ref{section:Numerics_main}) 
took place in this regime.
We approximate to second order in 
$\frac{1}{ \betaC \dAvg } \, ,$
$\frac{ \Wb }{ \dAvg } \, ,$ and $\Sites ( \betaH \HScale )^2  \, .$
We approximate to zeroth order in the much smaller $e^{ - \betaC \Wb } \, .$

The diabatic corrections to $\expval{ W_\tot }$
involve three more small parameters.
$H(t)$ is tuned slowly:
$\frac{ \sqrt{ v } }{ \dAvg } \ll 1$.
The MBL level-repulsion scale $\deltaMBL$ 
(Appendix~\ref{section:ThermoLimitApp})
is very small:
$\frac{ \deltaMBL }{ \dAvg }  \ll  1 \, .$
The third parameter,
$\frac{ \dAvg }{ \HScale }  \ll  1$,
follows from $\dAvg  \sim  \frac{ \HScale }{ \HDim }$.

\subsection{Partial-swap model of thermalization}
\label{section:PSWAP}

Classical thermalization can be modeled with 
a \emph{probabilistic swap}, or \emph{partial swap,}
or \emph{$p$-SWAP}~\cite{Ziman_01_Quantum,Scarani_02_Thermalizing}.
Let a column vector $\vec{v}$ represent the state.
The thermalization is broken into time steps.
At each step, a doubly stochastic matrix $M_p$ operates on $\vec{v}$.
The matrix's fixed point is a Gibbs state $\vec{g}$.

$M_p$ models a probabilistic swapping out of $\vec{v}$ for $\vec{g}$:
At each time step, the system's state has a probability $1 - p$ of being preserved
and a probability $p \in [0, \: 1]$ of being replaced by $\vec{g}$.
This algorithm gives $M_p$ the form
$M_p  =  (1 - p) \id  +  p G$.
Every column in the matrix $G$ equals the Gibbs state $\vec{g}$.

We illustrate with thermalization across two levels.
Let $0$ and $\Delta$ label the levels, such that
$\vec{g} = \left( \frac{ e^{ - \beta \gap } }{ 1 + e^{ - \beta \gap } } \, ,
\frac{1}{ 1 + e^{ - \beta \gap } } \right)$:
\begin{align}
   M_p  =  \begin{bmatrix}
   1 - p  \;  \frac{ 1 }{ 1 + e^{ - \beta \gap } }  &
   p  \;  \frac{ e^{ - \beta \gap } }{ 1 + e^{ - \beta \gap } }  \\
   p  \;  \frac{ 1 }{ 1 + e^{ - \beta \gap } }   &
   1  -  p  \;  \frac{ e^{ - \beta \gap } }{ 1 + e^{ - \beta \gap } } 
   \end{bmatrix} \, .
\end{align}
The off-diagonal elements, or transition probabilities,
obey detailed balance~\cite{YungerHalpern_15_Introducing,Crooks_98}:
$\frac{ P( 0 \to \gap ) }{ P( \gap \to 0 ) }
=  e^{ - \beta \gap }$.

Repeated application of $M_p$ 
maps every state to $\vec{g}$~\cite{YungerHalpern_15_Introducing}:
$\lim_{n \to \infty} \left( M_p \right)^n  \vec{v}  =  \vec{g}$.
The parameter $p$ reflects the system-bath-coupling strength.
We choose $p = 1$: 
The system thermalizes completely at each time step.
(If $p \neq 1$, a more sophisticated model may be needed
for thermalization across $>2$ levels.)

%
%
%
\subsection{Average heat $\expval{ Q_2 }$ absorbed during stroke 2}
\label{section:Q2}

We calculate $\expval{ Q_2 }$ in four steps,
using the density operator's statistical interpretation
(see the caption of Fig.~\ref{fig:Compare_thermo_Otto_fig}).
Section~\ref{sec:Q2_OneTrial} focuses on one trial.
We average over two distributions in Sec.~\ref{sec:Q2_First_2_Avgs}:
(1) the probabilities that
cold thermalization changes or preserves the engine's energy
and (2) the Poisson gap distribution, 
$P_\MBL^\ParenE ( \delta )$.
We average with respect to the initial density operator,
$\rho ( 0 )  =  e^{ - \betaH H_\ETH } / \ZH$,
in Sec.~\ref{sec:Q2_Th_Avg}.

\subsubsection{Heat $Q_2$ absorbed during one trial}
\label{sec:Q2_OneTrial}

Let $j$ denote the $H_\ETH$ level on which
the engine begins.
Stroke 1 (adiabatic tuning) preserves the occupied level's index.
Let $Q_2^\ParenJ$ denote the heat absorbed during cold thermalization.
Suppose that the gap just above level $j$
is smaller than the cold bath's bandwidth: $\delta'_{j+1}  <  \Wb$.
The engine might jump upward, 
absorbing heat $Q_2^\ParenJ  =  \delta'_{j + 1}$.
Suppose that the gap just below level $j$
is small enough: $\delta'_j  <  \Wb$.
The engine might drop downward, 
absorbing $Q_2^\ParenJ  =  - \delta'_j$.
The engine absorbs no heat if it fails to hop:
\begin{align}
   \label{eq:Qj} 
   Q_2^\ParenJ  =  \begin{cases}
      \delta'_{j + 1} \, ,  &   \text{engine jumps} \\
      - \delta'_j  \, ,  &   \text{engine drops}  \\
      0 \, ,  &   \text{cold thermalization preserves engine's energy}  
   \end{cases}  \, .
\end{align}

\subsubsection{Averages with respect to 
cold-thermalization probabilities and gap distributions}
\label{sec:Q2_First_2_Avgs}

The discrete $E_j$ becomes a continuous $E$:
\begin{align}
   \label{eq:Qj_help1}
   \expval{  \expval{ Q_2(E) }_{ \substack{ \text{cold} \\ \text{therm.} } }  }_{\text{gaps}} 
   & =  \int_0^{ \Wb }  d \delta'_{j + 1}  \;  \delta'_{j + 1}  \:
   \mathcal{P} ( \text{$\Sys$ jumps}  \:  |  \:   \delta'_{j + 1} < \Wb )  \:
   \mathcal{P} ( \delta'_{j + 1} < \Wb  \, ;  \, \text{$\Sys$ does not drop} )
   \nonumber \\ & \quad
   +  
   \int_0^{ \Wb }  d \delta'_j  \; (- \delta'_j )  \:
       \mathcal{P} ( \text{$\Sys$ drops}  \:  |  \:  \delta'_j  <  \Wb )  \:
       \mathcal{P}  ( \delta'_j  <  \Wb  \, ;  \:  \text{$\Sys$ does not jump} )  \, .
\end{align}
Each $\mathcal{P}(a)$ denotes the probability that event $a$ occurs.
$\mathcal{P} ( a | b )$ denotes the conditional probability that, 
if an event $b$ has occurred, $a$ will occur.
$\mathcal{P} ( a ; b )$ denotes the joint probability
that $a$ and $b$ occur.

The p-SWAP model (Suppl. Mat.~\ref{section:PSWAP}) 
provides the conditional probabilities.
The Poisson distribution provides the probability that
a gap is small enough.
Each joint probability factorizes, e.g.,
$\mathcal{P} ( \delta'_{j + 1} < \Wb  \, ;  \, \text{$\Sys$ does not drop} )
=  \mathcal{P} ( \delta'_{j + 1} < \Wb )  \:
\mathcal{P} ( \text{$\Sys$ does not drop} )$.

The engine refrains from dropping if
(1) the gap below level $j$ is too large or if
(2) the gap below $j$ is small
but cold thermalization fails to drop the engine's state:
\begin{align}
   \label{eq:Qj_help2}
   & \mathcal{P} ( \text{$\Sys$ does not drop} )
   = \mathcal{P}( \delta'_j > \Wb )  
   +  \mathcal{P} \left( \text{$\Sys$ does not drop}  \:  |  
        \:  \delta'_{j + 1}  <  \Wb  \right)
       \mathcal{P} ( \delta'_{j + 1}  <  \Wb )  \, .
\end{align}
The gap has a probability 
\mbox{$\mathcal{P}( \delta'_j > \Wb )    =  1 + O \left( \frac{ \Wb }{ \dAvg } \right)$}
of being too large and a probability
$\mathcal{P} ( \delta'_{j + 1}  <  \Wb )  =  O \left( \frac{ \Wb }{ \dAvg } \right)$
of being small enough.\footnote{
\label{footnote:Small_gap_prob}
Any given gap's probability of being small enough to thermalize equals
\begin{align}
   \mathcal{P}( \delta \leq \Wb )
   & =  \frac{1}{ \HDim}  \int_{ E_\Min }^{ E_\Max }  dE  \;  \DOS (E)
   \int_{ 0 }^{ \Wb }  d \delta  \;  P_\MBL^\ParenE ( \delta )  
   \approx  \frac{1}{ \HDim}    \int_{ - \infty }^\infty  dE  \;  \DOS (E)  \,  
   \left[  1  -  e^{ - \DOS (E)  \Wb }  \right]  \, .
\end{align}
The first term evaluates to one. 
We Taylor-expand the exponential to first order, then integrate term by term:
\begin{align}
   \mathcal{P}( \delta \leq \Wb )
   & \approx  1  -  \Bigg[
   \frac{1}{ \HDim }  \int_{ - \infty}^\infty  dE  \;  \DOS (E)
   -  \frac{ \Wb }{ \HDim }  \int_{ -\infty }^\infty  dE  \;  \DOS^2 (E)
   +  O \left(  \frac{ \left( \Wb \right)^2 }{ \HDim }  
       \int_{ - \infty}^\infty   dE  \;  \DOS^3 (E)   \right)
   \Bigg] \\
   & =  \label{eq:Small_gap_prob}
   \frac{ \Wb }{ \dAvg }  
   +  O \left(  \left[  \frac{ \Wb }{ \dAvg }  \right]^2   \right)  \, .
\end{align}}
The detailed-balance probability 
$\mathcal{P} \left( \text{$\Sys$ does not drop}  \:  |  
        \:  \delta'_{j + 1}  <  \Wb  \right)$
is too small to offset the $O \left( \frac{ \Wb }{ \dAvg } \right)$ scaling
of $\mathcal{P} ( \delta'_{j + 1}  <  \Wb )$.
Hence the $O \left( \frac{ \Wb }{ \dAvg } \right)$ terms are negligible here:
Each multiplies, in Eq.~\eqref{eq:Qj_help1}, 
a $\delta'_{j + 1}$ that will average to $\sim \Wb$
and a $\mathcal{P} ( \text{$\Sys$ jumps}  \:  |  \:   \delta'_{j + 1} < \Wb )$
that will average to $\sim  \frac{ \Wb }{ \dAvg }$.
Each such compound term $\sim  \Wb \left( \frac{ \Wb }{ \dAvg } \right)^2
=  \dAvg  \left( \frac{ \Wb }{ \dAvg } \right)^3$.
We evaluate quantities only to second order in 
$\frac{ \Wb }{ \dAvg }  \ll  1 \, .$
Hence Eq.~\eqref{eq:Qj_help2} approximates to one.
A similar argument concerns the final factor in Eq.~\eqref{eq:Qj_help1}.
Equation~\eqref{eq:Qj_help1} becomes
\begin{align}
   \label{eq:Qj_help3}
   \expval{  \expval{ Q_2(E) }_{ \substack{ \text{cold} \\ \text{therm.} } }  }_{\text{gaps}} 
   & =  \int_0^{ \Wb }  d \delta'_{j + 1}  \;
   \delta'_{j + 1}  \;
   \frac{ e^{ - \betaC \delta'_{j+1} } }{ 1  +  e^{ - \betaC \delta'_{j+1} }  }  \;
   P_\MBL^\ParenE ( \delta'_{j + 1} )
   \nonumber \\ & \quad
   -  \int_0^{ \Wb }  d \delta'_j  \;
   \delta'_j  \;
    \frac{ 1 }{ 1  +  e^{ - \betaC \delta'_j }  }  \;
    P_\MBL^\ParenE ( \delta'_j )
   +  \DOS(E)  O  \left( \left[  \frac{ \Wb }{ \dAvg }  \right]^3  \right)  \, .
\end{align}

Computing the integrals is tedious
but is achievable by techniques akin to the Sommerfeld expansion~\cite{Ashcroft_76_Solid}.
The calculation appears in~\cite[App.~G 4]{NYH_17_MBL}
and yields
\begin{align}
   \label{eq:Qj_help4}
   \expval{  \expval{ Q_2(E) }_{ \substack{ \text{cold} \\ \text{therm.} } }  }_{\text{gaps}}
   & =  - \frac{1}{2}  \,   \DOS(E)  \,   ( \Wb )^2
   +  \frac{ \pi^2 }{ 6 }  \:  \frac{ \DOS(E) }{ ( \betaC )^2 }  
   +  \DOS (E)  \Bigg\{
   O  \left(  \left[  \DOS (E)  \,  \Wb  \right]^3  \right)
   \nonumber \\ &
   +  O  \left(  [ \DOS (E)  \,  \Wb ]  \,  \frac{ \DOS(E) }{ \betaC }  \:
                    e^{ - \betaC \Wb }  \right)    
   +  O \left(  \left[  \frac{ \DOS(E) }{ \betaC }  \right]^3  \right)  
   \Bigg\}   \, .
\end{align}
We have assumed that the engine cannot cold-thermalize down
two adjacent small gaps (from level $j+1$ to level $j-1$, 
wherein $\delta'_j ,  \delta'_{j - 1}  <  \Wb$). 
Such a gap configuration appears with probability 
$\propto  \DOS(E) [  \DOS(E)  \Wb  ]^2$.
Each gap contributes energy $\sim \Wb$ to the heat.
Hence double drops contribute to~\eqref{eq:Qj_help4}
at third order in $\DOS(E)  \Wb$.

\subsubsection{Thermal average with respect to $\rho(0)$}
\label{sec:Q2_Th_Avg}

We integrate Eq.~\eqref{eq:Qj_help4} over energies $E$,
weighted by the initial-state Gibbs distribution:
\begin{align}
   \label{eq:Q2_help1}
   \expval{ Q_2 }  
   & :=  \expval{  \expval{  \expval{ Q_2(E) }_{ \substack{ \text{cold} \\ \text{therm.} } }  }_{\text{gaps}}  }_{ \rho(0) } \\
   & =  \left(  - \frac{ ( \Wb )^2 }{ 2 }  
   +  \frac{ \pi^2 }{ 6 }  \: \frac{ 1 }{ ( \betaC )^2 }  \right)
   \int_{ -\infty }^\infty  dE  \;  \DOS^2(E)  \;
   \frac{ e^{ - \betaH E } }{ \ZH }
  +  \dAvg \Bigg\{  
  O  \left(  \left[  \frac{ \Wb }{ \dAvg }  \right]^3  \right)
  +  O  \left(  \frac{ \Wb }{ \dAvg }  \;   e^{ - \betaC \Wb }  \right) 
  \nonumber \\ & \qquad  
  +  O \left(  \left[  \frac{ \DOS(E) }{ \betaC }  \right]^3  \right)  
  \Bigg\} \, .
\end{align}
The DOS's sharp peaking about $E = 0$
justifies our approximation of
the energy integral as extending between $\pm \infty$.
We substitute in for the DOS from Eq.~\eqref{eq:DOS_App}:
\begin{align}
   \label{eq:Q2_help2}
   \expval{ Q_2 }  & =  
   \frac{ \HDim^2 }{ 2 \pi \Sites \HScale^2 }  \;
   \frac{1}{ \ZH }  \:
    \left(  - \frac{ ( \Wb )^2 }{ 2 }  
    +  \frac{ \pi^2 }{ 6 }  \:  \frac{ 1 }{ ( \betaC )^2 }  \right)
    \int_{ -\infty }^\infty  dE  \;  e^{ - E^2 / \Sites \HScale^2 }  \;
    e^{ - \betaH E }
    + O ( . )  \, .
\end{align}
We have abbreviated the correction terms.
The integral evaluates to $\sqrt{ \pi \Sites }  \:  \HScale  \,  
e^{ \Sites ( \betaH \HScale )^2 / 4 }$.
The partition function is
\begin{align}
   \label{eq:ZH}
   \ZH  =  \int_{ -\infty }^\infty  dE  \;  
   \DOS(E)  e^{ - \betaH E } 
   =  \HDim e^{ \Sites ( \betaH \HScale )^2 / 2 }  \, .
\end{align}
Substituting into Eq.~\eqref{eq:Q2_help2} yields
\begin{align}
   \label{eq:Q2_help3}
   \expval{ Q_2 }  & =  \frac{ \HDim }{ 2 \sqrt{ \pi \Sites }  \:  \HScale }  \:
   \left(  - \frac{ ( \Wb )^2 }{ 2 }  
            +  \frac{ \pi^2 }{ 6 }  \:  \frac{ 1 }{ ( \betaC )^2 }  \right)  \:
   e^{ - \Sites ( \betaH \HScale )^2 / 4 }  
   +  O ( . )  \\
   \label{eq:Q2_help3b}
   & =  \left(  - \frac{ ( \Wb )^2 }{ 2 \dAvg }  
                   +  \frac{ \pi^2 }{ 6 }  \:  \frac{ 1 }{ ( \betaC )^2 \dAvg }  \right)  \:
   e^{ - \Sites ( \betaH \HScale )^2 / 4 }
   +  \dAvg \Bigg\{ 
   O  \left(  \left[  \frac{ \Wb }{ \dAvg }  \right]^3  \right) 
   +  O  \left(  [ \DOS (E)  \,  \Wb ]  \,  \frac{ \DOS(E) }{ \betaC }  \:
                     e^{ - \betaC \Wb }  \right)  
   \nonumber \\ & \qquad   
   +  O \left(  \left[  \frac{ \DOS(E) }{ \betaC }  \right]^3  \right)
   +  O  \left(  \left[ \sqrt{ \Sites } \:  \betaH \HScale \right]^4  \right)  
   \Bigg\}   \, .
\end{align}
The prefactor was replaced with $\frac{1}{ \dAvg }$ via Eq.~\eqref{eq:dAvg_def}. 

Equation~\eqref{eq:Q2_help3} is compared with numerical simulations in Fig.~\ref{fig:num_Q2}.
In the appropriate regime (wherein $\Wb \ll \dAvg$ and $\TCold \ll \Wb$),
the analytics agree well with the numerics, to within finite-size effects.

In terms of small dimensionless parameters, 
\begin{align}
   \label{eq:Q2_help4}
   \expval{ Q_2 }   & =  \dAvg 
   \left[ - \frac{1}{2}  \left( \frac{ \Wb }{ \dAvg }  \right)^2
           +  \frac{ \pi^2 }{ 6 }  \:  \frac{ 1 }{  ( \betaC  \dAvg )^2 }  \right]
  \left[ 1  -  \frac{ \Sites }{ 4 }  \left( \betaH \HScale \right)^2  \right]
   +  O ( . )   \, .
\end{align}
The leading-order term is second-order.
So is the $\betaC$ correction; but 
$\frac{ 1 }{  ( \betaC  \dAvg )^2 }  
\ll  \left( \frac{ \Wb }{ \dAvg }  \right)^2$, 
by assumption [Eq.~\eqref{eq:Regime}].
The $\betaH$ correction is fourth-order---too small to include.
To lowest order, 
\begin{align}
   \label{eq:EDiff2b}   \boxed{
   \expval{ Q_2 }  \approx
   -  \frac{ \left( \Wb \right)^2 }{ 2 \dAvg }  } \, .
\end{align}


\begin{figure}
  \begin{subfigure}{0.3\textwidth}
    \centering
    \includegraphics[width=\textwidth]{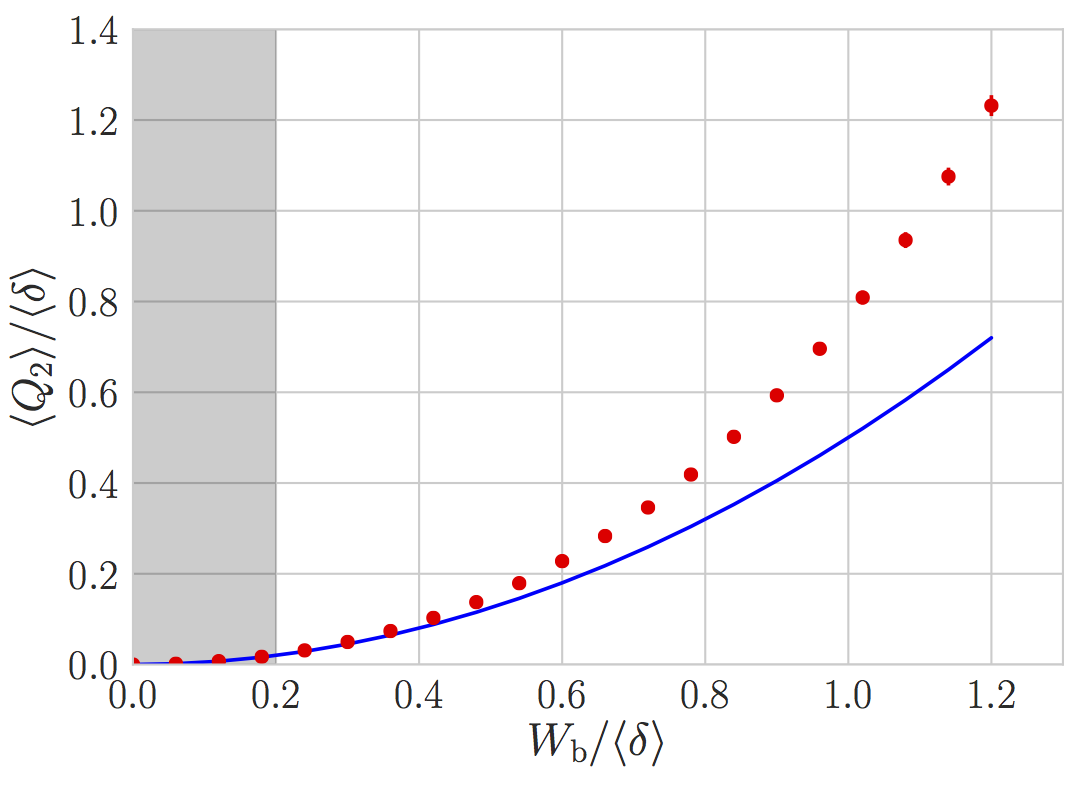}
    \caption{$\expval{Q_2}$ vs. $\Wb$    at $\TCold = 0$ and $\THot = \infty$}
    \label{fig:Q2_Wb}
  \end{subfigure}
  \begin{subfigure}{0.3\textwidth}
    \centering
    \includegraphics[width=\textwidth]{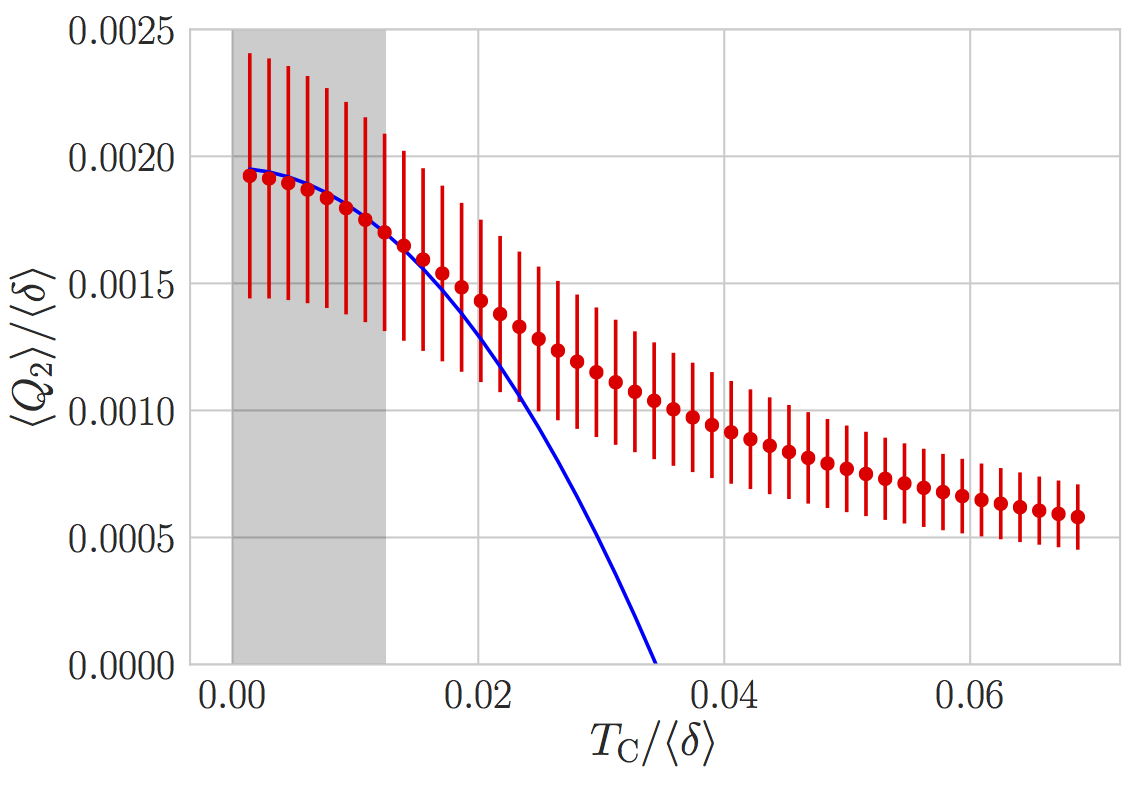}
    \caption{$\expval{Q_2}$ vs. $\TCold$ 
    at $\THot = \infty$ and    $\Wb = 2^{-4}\dAvg$}
    \label{fig:Q2_TC}
  \end{subfigure}
  \begin{subfigure}{0.3\textwidth}
    \centering
    \includegraphics[width=\textwidth]{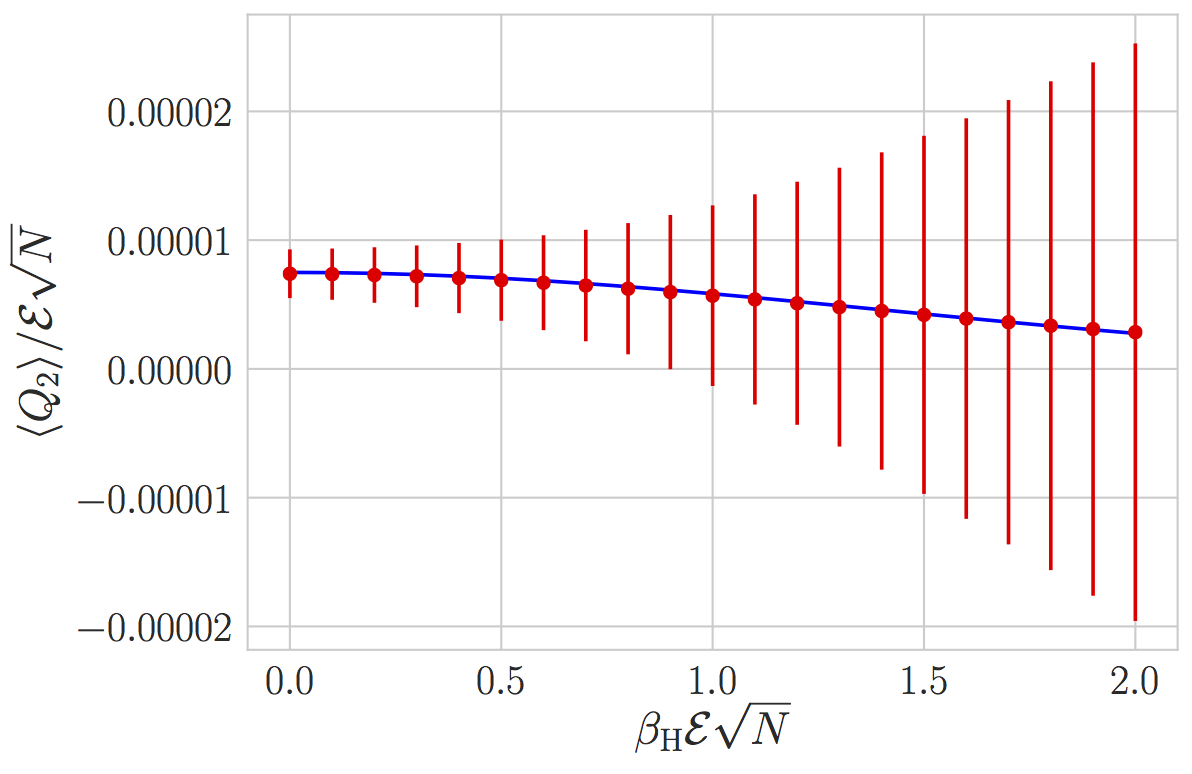}
    \caption{$\expval{Q_2}$ vs. $\betaH$ 
    at $\TCold = 0$ and    $\Wb = 2^{-4}\dAvg$}
    \label{fig:Q2_betaH}
  \end{subfigure}
  \caption{\caphead{Magnitude $| \expval{Q_2} |$ 
  of the average heat absorbed during cold thermalization (stroke 2) 
  as a function of 
    the cold-bath bandwidth $\Wb$ (\ref{fig:Q2_Wb}), 
    the cold-bath temperature $\TCold$ (\ref{fig:Q2_TC}), and 
    the hot-bath temperature $\THot = 1 / \betaH$ (\ref{fig:Q2_betaH}):}
    The blue lines represent the magnitude of 
    the analytical prediction~\eqref{eq:Q2_help3}.
    See Sec.~\ref{section:Numerics_main} for 
    other parameters and definitions.
    The analytics match the numerics' shapes,
    and the agreement is fairly close, in the appropriate limits
    (where $\frac{ \Wb }{ \dAvg } \ll 1$ and $\TCold/\dAvg \ll 1$, 
     in the gray shaded regions).
    The analytics systematically underestimate $\expval{ Q_2 }$ 
    at fixed $\Wb$, due to
    the small level repulsion at finite $\Sites$.
    The analytical prediction~\eqref{eq:Q2_help3} 
    substantially underestimates $\expval{ Q_2 }$ 
    when the cold-bath bandwidth is large, $\Wb \gtrsim \dAvg$.
    Such disagreement is expected:
    The analytics rely on $\frac{ \Wb }{ \dAvg }  \ll  1$,
    neglecting chains of small gaps $\delta'_j, \delta'_{j+1} \dots < \Wb$.
    Such chains proliferate as $\Wb$ grows.
    A similar reason accounts for the curve's 
    crossing the origin in Fig.~\ref{fig:Q2_TC}:
    We analytically compute $\expval{Q_2}$ only to second order in $\TCold/\dAvg$.    }
  \label{fig:num_Q2}
\end{figure}

\subsection{Average heat $\expval{ Q_4 }$ absorbed during stroke 4}
\label{section:Q4}

The $\expval{ Q_4 }$ calculation proceeds similarly to
the $\expval{ Q_2 }$ calculation.
When calculating $\expval{ Q_2 }$, however,
we neglected contributions from 
the engine's cold-thermalizing down
two small gaps.
Two successive gaps have a probability
$\sim \left( \frac{ \Wb }{ \dAvg }  \right)^2$
of being $< \Wb$ each.
Thermalizing across each gap produces heat $\leq \Wb$.
Each such pair therefore contributes
negligibly to $\expval{ Q_2 }$, as
$\dAvg  O  \left(  \left[  \frac{ \Wb }{ \dAvg }  \right]^3  \right)$.

We cannot neglect these pairs  
when calculating $\expval{ Q_4 }$.
Each typical small gap widens, during stroke 3, 
to size $\sim \dAvg \, .$
These larger gaps are thermalized across during stroke 4,
contributing at the nonnegligible second order, as
\mbox{ $\sim \dAvg O  \left(  \left[  \frac{ \Wb }{ \dAvg }  \right]^2  \right)$}
to $\expval{ Q_4 } \, .$
Chains of $\geq 3$ small MBL gaps contribute negligibly.

The calculation is tedious, appears in~\cite[App. G 5]{NYH_17_MBL},
and yields
\begin{align}
   \label{eq:Q4_Result}  \boxed{
   \expval{ Q_4 }  \approx  }  \;
   \expval{ Q_4^{ n{=}1 } }  +  \expval{ Q_4^{ n{=}2 } }
   \approx  \boxed{  
   \Wb  -  \frac{ 2 \ln 2 }{ \betaC }
   +  \frac{ ( \Wb )^2 }{ 2 \dAvg }
   +  4 \ln 2  \:  \frac{ \Wb }{ \betaC \dAvg }  }  \, .
\end{align}
The leading-order term, $\Wb$, is explained heuristically
below Eq.~\eqref{eq:WTotApprox2_Main}.

The leading-order $\betaC$ correction,
$- \frac{ 2 \ln 2 }{ \betaC }$, shows that
a warm cold bath
lowers the heat required to reset the engine.
Suppose that the cold bath is maximally cold: $T_\CTemp = 0$.
Consider any trial that $\Sys$ begins just above a working gap
(an ETH gap $\delta > \Wb$ that narrows to 
an MBL gap $\delta' < \Wb$).
Cold thermalization drops $\Sys$ deterministically to the lower level.
During stroke 4, $\Sys$ must absorb $Q_4 > 0$ 
to return to its start-of-trial state.
Now, suppose that the cold bath is only cool: $\TCold \gtrsim  0$.
Cold thermalization might leave $\Sys$ in the upper level.
$\Sys$ needs less heat, on average, to reset
than if $T_\CTemp = 0$.
A finite $T_\CTemp$ detracts from $\expval{ Q_4 }$.
The $+  4 \ln 2  \:  \frac{ \Wb }{ \betaC \dAvg }$
offsets the detracting. However, the positive correction
is smaller than the negative correction, 
as $\frac{ \Wb }{ \dAvg }  \ll  1 \, .$

A similar argument concerns $T_\HTemp  <  \infty$.
But the $\betaH$ correction is too small
to include in Eq.~\eqref{eq:Q4_Result}:
$\expval{ Q_4 }  \approx  \Wb
- \frac{ 2 \ln 2 }{ \betaC }  
+  \frac{ ( \Wb )^2 }{ 2 \dAvg }  \:  e^{ - ( \betaH \HScale )^2 / 4 }$.

Figure~\ref{fig:num_Q4} shows
Eq.~\eqref{eq:Q4_Result}, to lowest order in $\TCold$,
as well as the $\betaH$ dependence of $\expval{ Q_4 }$.
The analytical prediction is compared with numerical simulations.
The agreement is close, up to finite-size effects, 
in the appropriate regime
($\TCold \ll \Wb \ll \dAvg$).

\begin{figure}
  \begin{subfigure}{0.3\textwidth}
    \centering
    \includegraphics[width=\textwidth]{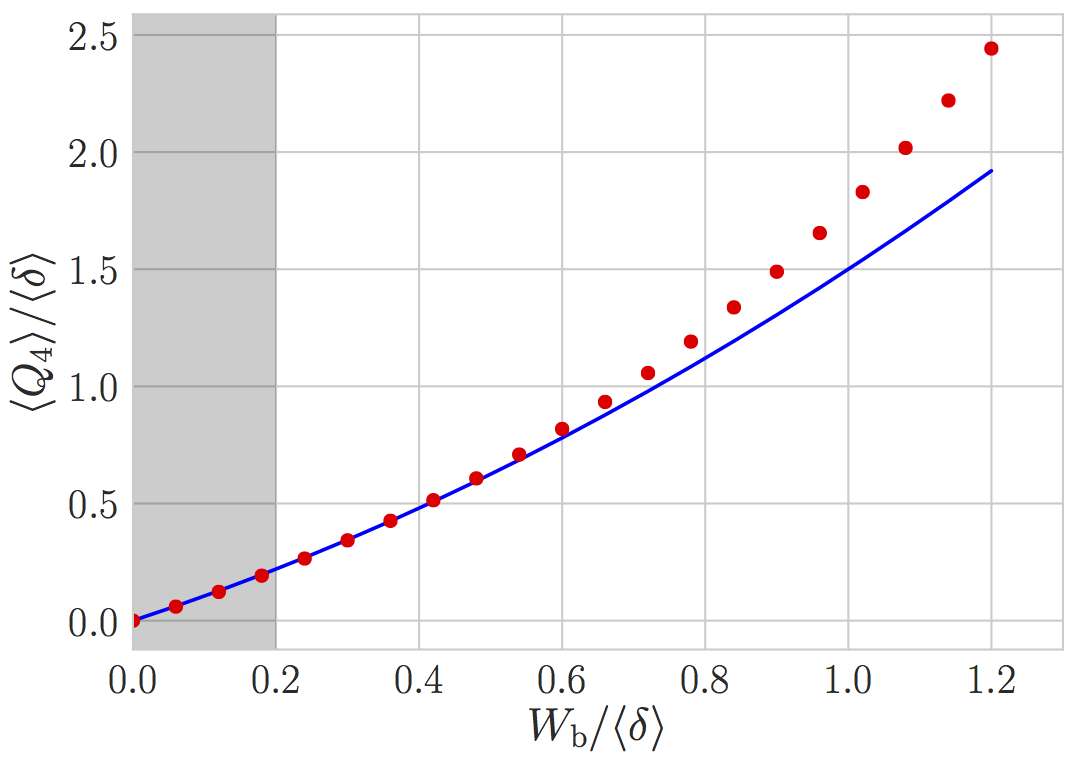}
    \caption{$\expval{Q_4}$ vs. $\Wb$ at 
    $\TCold = 0$ and $\THot = \infty$}
  \end{subfigure}
  \begin{subfigure}{0.3\textwidth}
    \centering
    \includegraphics[width=\textwidth]{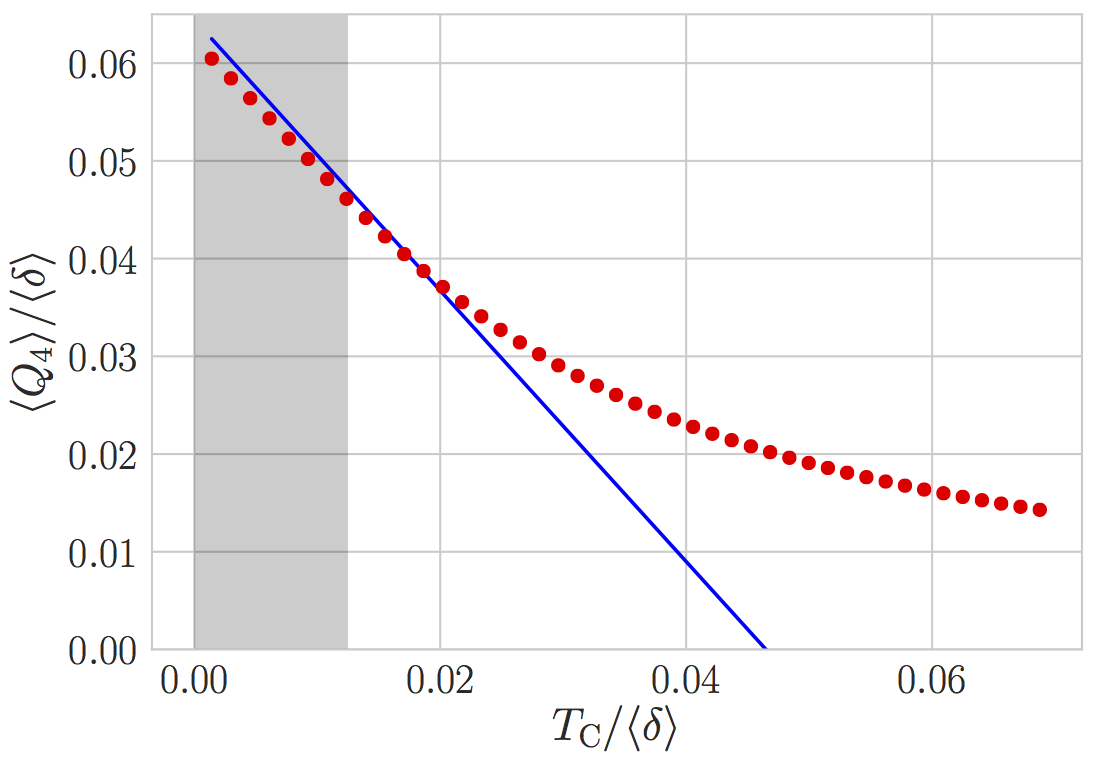}
    \caption{$\expval{Q_4}$ vs. $\TCold$ at 
    $\THot = \infty$ and 
    $\Wb = 2^{-4}\dAvg$}
  \end{subfigure}
  \begin{subfigure}{0.3\textwidth}
    \centering
    \includegraphics[width=\textwidth]{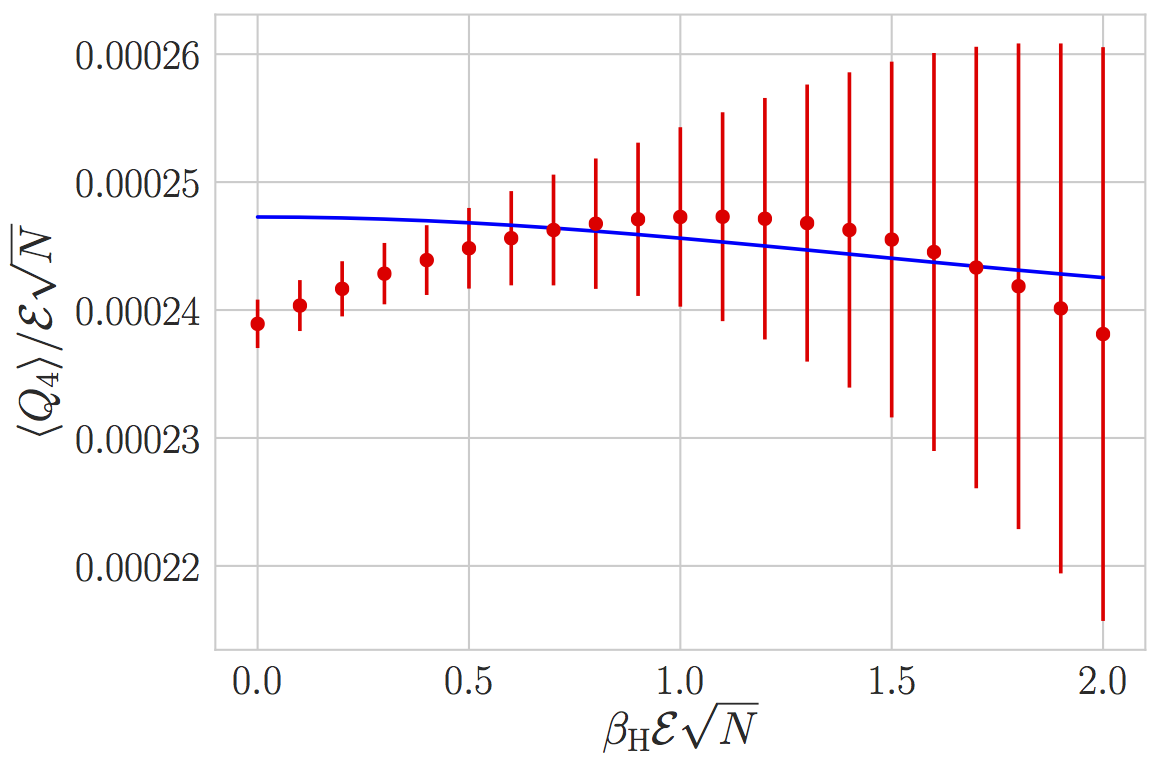}
    \caption{$\expval{Q_4}$ vs. $\betaH$ 
    at $\TCold = 0$ and
    $\Wb = 2^{-4}\dAvg$}
  \label{fig:Q4_betaH}
  \end{subfigure}
  \caption{\caphead{Average heat $\expval{Q_4}$ absorbed
    during hot thermalization (stroke 4) as a function of 
    the cold-bath bandwidth $\Wb$,
    the cold-bath temperature $\TCold$,
    and the hot-bath temperature $\THot = 1 / \betaH$:} 
    The blue lines represent the analytical prediction~\eqref{eq:Q4_Result}, 
    to lowest order in $\TCold$,
with the $\betaH$ dependence of $\expval{ Q_4 }$,
too small a correction to include in Eq.~\eqref{eq:Q4_Result}:
$\expval{ Q_4 }  \approx  \Wb
- \frac{ 2 \ln 2 }{ \betaC }  
+  \frac{ ( \Wb )^2 }{ 2 \dAvg }  \:  e^{ - ( \betaH \HScale )^2 / 4 }$.
    See Sec.~\ref{section:Numerics_main} for 
    other parameters and definitions.
    The analytics' shapes agree with the numerics',
    and the fit is fairly close, in the appropriate limits
    (where $e^{ - \betaC \Wb }  \ll  1$, $\frac{ 1 }{ \betaC \dAvg }  \ll  1$,
    and $\frac{ \Wb }{ \dAvg }  \ll  1$,    in the gray shaded regions).
    The predictions underestimate $\expval{ Q_4 }$;
    see the Fig.~\ref{fig:num_Q2} caption.
    Figure~\ref{fig:Q4_betaH} suggests that the numerics deviate significantly from the analytics: The numerics appear to depend on $\betaH$ via a linear term absent from the $\expval{ Q_4 }$ prediction. This seeming mismatch appears symptomatic of finite sample and system sizes.
  }
  \label{fig:num_Q4}
\end{figure}

\subsection{Per-cycle power $\expval{ W_\tot }$}
\label{section:WTot}

By the first law of thermodynamics,
the net work outputted by the engine equals
the net heat absorbed.
Summing Eqs.~\eqref{eq:Q4_Result} and~\eqref{eq:EDiff2b}
yields the per-trial power, or average work outputted per engine cycle:
\begin{align}
   \label{eq:WTotApprox2}
   \boxed{ \expval{ W_\tot } }
   =  \expval{ Q_2 }  +  \expval{ Q_4 }
   \boxed{ \approx    
   \Wb  -  \frac{ 2 \ln 2 }{ \betaC }
   +  4 \ln 2  \:  \frac{ \Wb }{ \betaC \dAvg } } \, .
\end{align}
The leading-order $\betaH$ correction is negative
and too small to include---of order 
$\left(  \frac{ \Wb }{ \dAvg }  \right)^2 
\Sites  \left( \betaH \HScale  \right)^2  \, .$
Equation~\eqref{eq:WTotApprox2} agrees well with the numerics in the appropriate limits ($\TCold \ll \Wb \ll \dAvg$) and beyond,
as shown in Fig.~\ref{fig:num_WTOT}.
The main text contains
the primary analysis of Eq.~\eqref{eq:WTotApprox2}.
Here, we discuss the $\expval{ Q_2 }$ correction, limiting behaviors,
and scaling.

The negative $\expval{ Q_2 }  =  - \frac{ \left( \Wb \right)^2 }{ \dAvg }$ 
detracts little from the leading term $\Wb$ of $\expval{ Q_4 }$:
$\frac{ ( \Wb )^2 }{ \dAvg } \ll \Wb$, 
since $\frac{ \Wb }{ \dAvg } \ll 1$.
The $\expval{ Q_2 }$ cuts down on the per-trial power little.

The limiting behavior of Eq.~\eqref{eq:WTotApprox2} makes sense:
Consider the limit as $\Wb \to 0$.
The cold bath has too small a bandwidth to thermalize the engine.
The engine should output no work.
Indeed, the first and third terms in Eq.~\eqref{eq:WTotApprox2} vanish,
being proportional to $\Wb$.
The second term vanishes
because $\betaC \to \infty$ more quickly than $\Wb \to 0 \, ,$
by Eq.~\eqref{eq:Regime}: The cold bath is very cold.

Equation~\eqref{eq:WTotApprox2} scales with the system size $\Sites$
no more quickly than $\sqrt{\Sites} / 2^\Sites$,
by the assumption 
$\Wb  \ll  \dAvg  \sim  \sqrt{\Sites} / 2^\Sites$.
This scaling makes sense:
The engine outputs work
because the energy eigenvalues meander
upward and downward in Fig.~\ref{fig:Compare_thermo_Otto_fig}
as $H(t)$ is tuned.
In the thermodynamic limit, levels squeeze together.
Energy eigenvalues have little room in which to wander,
and $\Sys$ outputs little work.
Hence our parallelization of fixed-length mesoscopic subengines
in the thermodynamic limit (Sec.~\ref{section:Thermo_limit_main}).

\begin{figure}
  \begin{subfigure}{0.3\textwidth}
    \centering
    \includegraphics[width=\textwidth]{MBLSZ30_wb-WTOT-L12}
    \caption{$\expval{W_\tot}$ vs. $\Wb$ 
    at $\TCold = 0$ and $\THot = \infty$}
  \end{subfigure}
  \begin{subfigure}{0.3\textwidth}
    \centering
    \includegraphics[width=\textwidth]{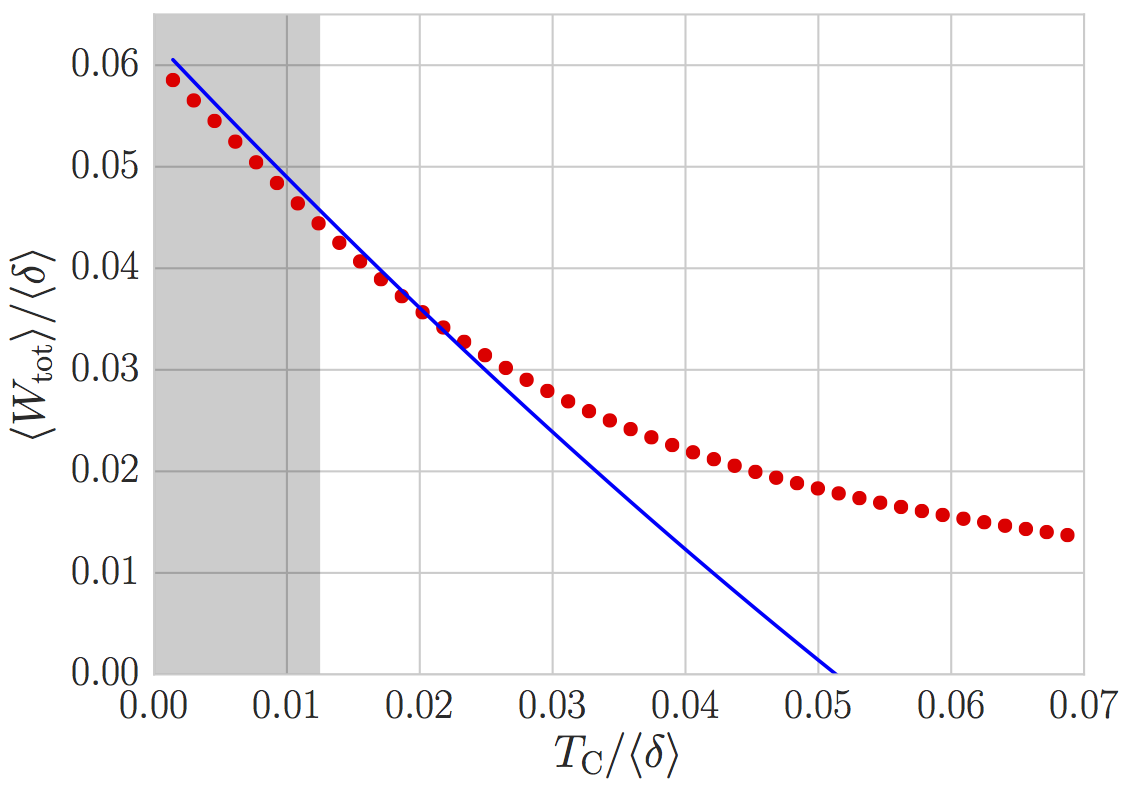}
    \caption{$\expval{W_\tot}$ vs. $\TCold$ 
    at $\THot = \infty$ and $\Wb = 2^{-4}\dAvg$}
  \end{subfigure}
  \begin{subfigure}{0.3\textwidth}
    \centering
    \includegraphics[width=\textwidth]{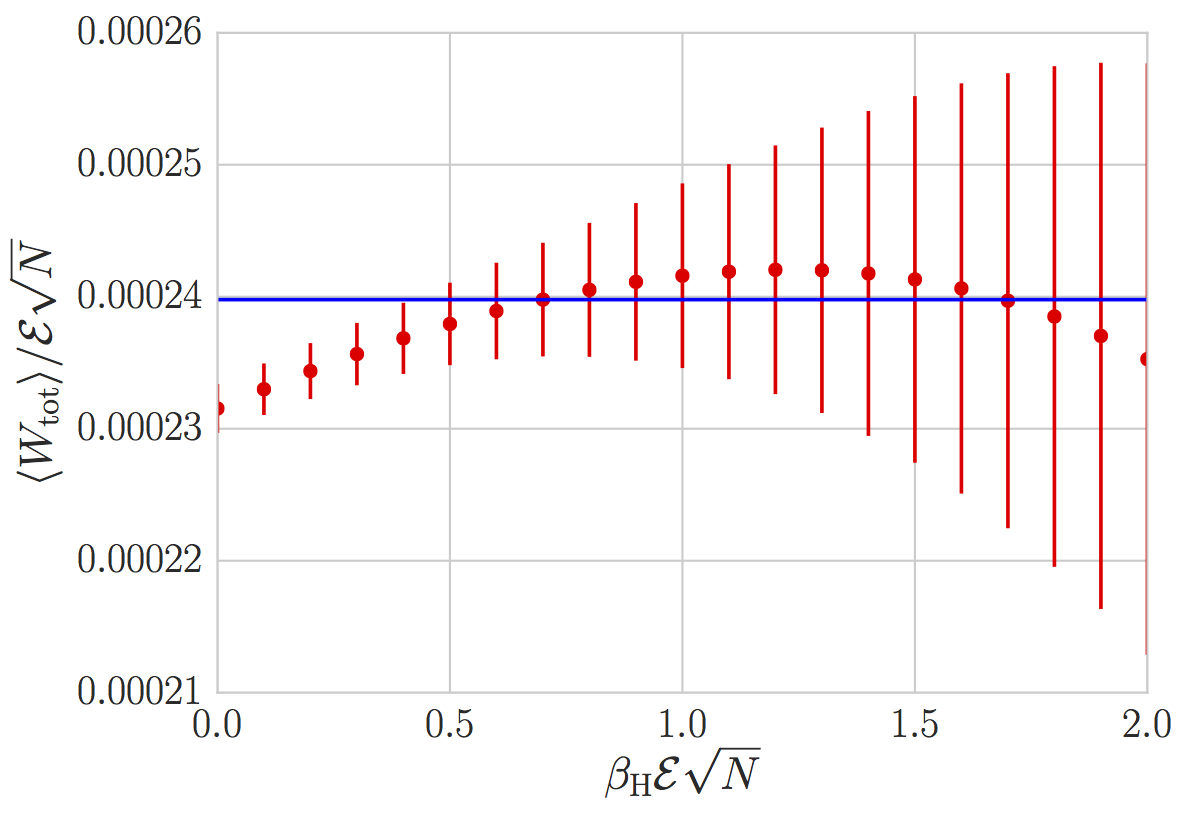}
    \caption{$\expval{W_\tot}$ vs. $\betaH$ 
    at $\TCold = 0$ and $\Wb = 2^{-4}\dAvg$}
  \label{fig:Wtot_betaH}
  \end{subfigure}
  \caption{\caphead{Per-cycle power $\expval{W_\tot}$ as a function of 
    the cold-bath bandwidth $\Wb$, 
    the cold-bath temperature $\TCold$,
    and the hot-bath temperature $\THot = 1 / \betaH$:}
    The blue lines represent the analytical prediction
    $\expval{ W_\tot }  \approx  \Wb  -  \frac{ 2 \ln 2 }{ \betaC }$:
    Eq.~\eqref{eq:WTotApprox2}, 
    to first order in $\frac{ \Wb }{ \dAvg }$ and in $\frac{1}{ \betaC \dAvg }$.
    See Sec.~\ref{section:Numerics_main} for 
    other parameters and definitions.
    The analytics largely agree with the numerics
    in the appropriate regime:
    $\frac{ \Wb }{\dAvg} \ll 1, \frac \TCold \dAvg \ll 1$ (in the gray shaded region).
    Outside that regime,
    the analytics underestimate $\expval{ W_\tot }$;
    see Fig.~\ref{fig:num_Q2} for analysis.
    Figure~\ref{fig:Wtot_betaH} suggests that the numerics depend on $\betaH$
    via a linear term absent from the analytical prediction;
    see the caption of Fig.~\ref{fig:Q4_betaH}.}
  \label{fig:num_WTOT}
\end{figure}

\subsection{Efficiency $\eta_\MBL$ in the adiabatic approximation}
\label{section:AdiabaticEta}


The efficiency is defined as
\begin{align}
   \label{eq:EtaDefAgain}
   \eta_\MBL  :=   \frac{ \expval{ W_\tot } }{ \expval{ Q_\In } }  \, .
\end{align}
The numerator is averaged separately from the denominator because
averaging $W_\tot$ over runs of one mesoscopic engine
is roughly equivalent to 
averaging over simultaneous runs of parallel subengines
in one macroscopic engine.
$\frac{ \expval{ W_\tot } }{ \expval{ Q_\In } }$ 
may therefore be regarded as
the $\frac{ W_\tot }{ Q_\In }$ of one
macroscopic-engine trial.

Having calculated $\expval{ W_\tot }$, 
we must identify $\expval{ Q_\In } \, .$
In most trials, the engine expels heat
$- Q_2 > 0$ during cold thermalization
and absorbs $Q_4 > 0$ during hot thermalization.
The positive-heat-absorbing-stroke is stroke 4, 
in the average trial:
\begin{align}
   \expval{ Q_\In }  
   =  \expval{ Q_4 }
   =   \expval{ W_\tot }  -  \expval{ Q_2 }
   = \label{eq:Q4Eval}
   \expval{ W_\tot }  
   \left( 1 - \frac{ \expval{ Q_2 } }{ \expval{ W_\tot } }  \right)
   =      \expval{ W_\tot }     \left( 1  +  \phi \right)    \, ,
\end{align}
wherein
\begin{align}
   \label{eq:Xi1}
   \phi   :=   -  \frac{  \expval{ Q_2 } }{ \expval{ W_\tot } }
   \approx  \frac{ ( \Wb )^2 }{ 2 \dAvg }   \;
   \frac{1}{ \Wb }
    \approx  \frac{\Wb }{ 2 \dAvg }  \, .
\end{align}

Substituting from Eq.~\eqref{eq:Q4Eval} 
into Eq.~\eqref{eq:EtaDefAgain} yields
\begin{align}
   \label{eq:ManyBodyEff}
   \boxed{ \eta_\MBL  \approx  }
   \frac{ \expval{ W_\tot } }{   \expval{ W_\tot } ( 1  +  \phi )  }
   \approx   1  -  \phi
   =  \boxed{    1  -  \frac{ \Wb  }{ 2 \dAvg }  } \,  .
\end{align}


Using suboptimal baths diminishes the efficiency. 
Addding $\betaC$-dependent terms
from Eq.~\eqref{eq:WTotApprox2} to $\expval{ W_\tot }$ yields
\begin{align}
  \label{eq:phi-corrections}
   \phi'  =   \frac{ \Wb}{ 2 \dAvg }  
   +  \frac{ \ln 2 }{ \betaC \dAvg }
   -  2 \ln 2  \:  \frac{ \Wb }{ \dAvg }  \:  \frac{1}{ \betaC \dAvg }  \, .
\end{align}
The $\betaH$ correction,
$1 - \frac{ \Wb }{ 2 \dAvg }  \:  e^{ - \Sites ( \betaH \HScale )^2 / 4 }$,
is too small to include.
The correction shares the sign of $\betaH$:
A lukewarm hot bath lowers the efficiency.

Expressions~\eqref{eq:ManyBodyEff} and~\eqref{eq:phi-corrections}
are compared with results from numerical simulations
in Fig.~\ref{fig:num_eta}.
The analytics agree with the numerics in the appropriate regime 
($\TCold \ll \Wb \ll \dAvg$).

\begin{figure}
  \begin{subfigure}{0.3\textwidth}
    \centering
    \includegraphics[width=\textwidth]{MBLSZ30_wb-eta-L12}
    \caption{$\eta_\MBL$ vs. $\Wb$ at $\TCold = 0$ and $\THot = \infty$}
  \end{subfigure}
  \begin{subfigure}{0.3\textwidth}
    \centering
    \includegraphics[width=\textwidth]{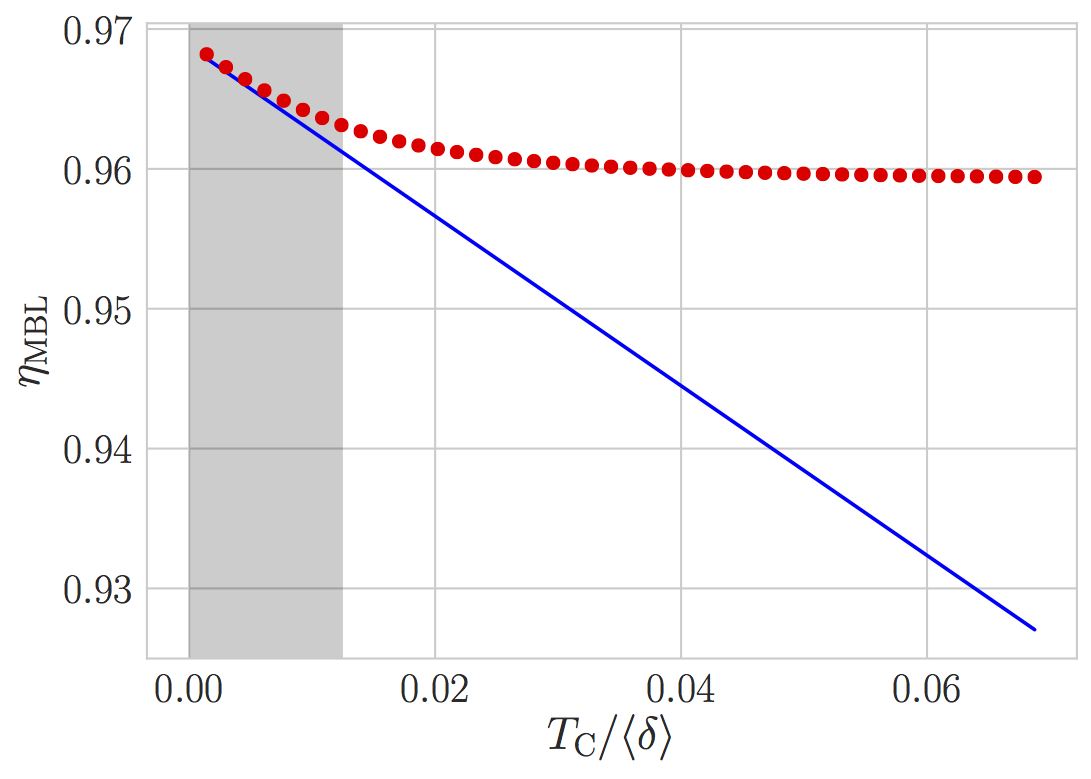}
    \caption{$\eta_\MBL$ vs. $\TCold$ at 
    $\THot = \infty$ and
    $\Wb \approx 10^{-4} \sqrt{\Sites}\HScale \approx 0.04\dAvg$}
  \end{subfigure}
  \begin{subfigure}{0.3\textwidth}
    \centering
    \includegraphics[width=\textwidth]{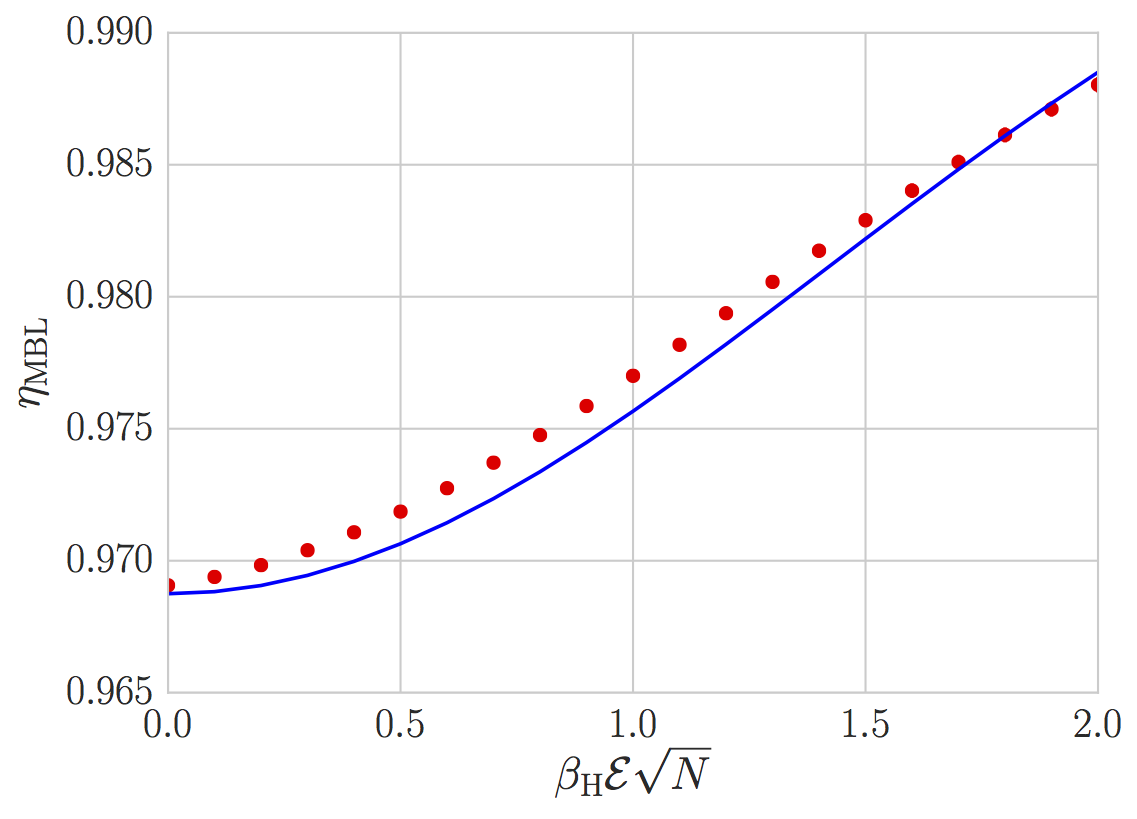}
    \caption{$\eta_\MBL$ vs. $\betaH$ 
    at $\TCold = 0$ and    $\Wb = 2^{-4}\dAvg$}
  \label{fig:Eta_betaH}
  \end{subfigure}
  \caption{\caphead{
    Efficiency $\eta_\MBL$ as a function of 
    the cold-bath bandwidth $\Wb$, 
    the cold-bath temperature $\TCold$,
    and the hot-bath temperature $\THot = 1 / \betaH$:}
    The blue lines represent the analytical predictions~\eqref{eq:ManyBodyEff}
    and~\eqref{eq:phi-corrections}.
    Figure~\eqref{fig:Eta_betaH} shows 
the leading-order $\betaH$ dependence of $\eta_\MBL$,
a correction too small to include in Eq.~\eqref{eq:phi-corrections}:
$1 - \frac{ \Wb }{ 2 \dAvg }  \:  e^{ - \Sites ( \betaH \HScale )^2 / 4 }$.
See Sec.~\ref{section:Numerics_main} for 
    other parameters and definitions.
    The analytics agree with the numerics fairly well
    in the appropriate regime 
    ($ \frac{ \Wb }{ \dAvg }  \ll 1$, $\frac{ \TCold }{ \dAvg }  \ll  1$,
    and $\sqrt{ \Sites }  \: \THot \HScale  \ll  1$).
    The analytics underestimate $\eta_\MBL$;
    see the Fig.~\ref{fig:num_Q2} caption.
  }
  \label{fig:num_eta}
\end{figure}

\subsection{Diabatic corrections}
\label{section:App_Diab}

We have approximated strokes 1 and 3 as quantum-adiabatic.
But the strokes proceed at a finite speed 
$v  :=  \HScale  \left\lvert   \frac{ d \alpha_t }{ dt }  \right\rvert$.
The engine may ``hop'' diabatically between energy eigenstates.
We estimate the work costs from three types of diabatic transitions,
introduced in Sec.~\ref{section:Diab_main}.
APT transitions are analyzed in Sec.~\ref{section:App_Diab_APT};
Landau-Zener (LZ) transitions, in Sec.~\ref{section:WLZ};
and fractional-LZ transitions, in Sec.~\ref{section:HalfLZ}.
The efficiency $\eta_\MBL$ is diabatically corrected 
in Sec.~\ref{eq:Diab_eff}.

We neglect the variation of the local average gap $\dAvg_E$ with energy.
The approximations facilitate this appendix's calculations,
which can require heavier machinery 
than the adiabatic approximation.
We aim to estimate just diabatic corrections' sizes
and scalings.

\subsubsection{Average work costs of APT transitions in the ETH phase:
$\expval{ W_{\APT, 1} }$ and $\expval{ W_{\APT, 3} }$}
\label{section:App_Diab_APT}

Consider tuning $H(t)$ near the start of stroke 1,
withinin the ETH phase.
(An analogous argument concerns the end of stroke 3.)
Let $\ket{ E_m (t) }$ denote the instantaneous 
$m^\th$ eigenstate of $H(t)$.
The perturbation couples together eigenstates of the initial Hamiltonian.
$\Sys$ can hop from its initial state, $\ket{ E_m ( t_i ) }$,
to some other energy eigenstate $n \neq m$.
The transition probability is denoted by $P_\APT ( n | m )$.
These transitions cost, on average,
work $\expval{ W_{ \APT, 1 } }$ during stroke 1
and work $\expval{ W_{\APT, 3} }$ during stroke 3.
We estimate $P_\APT ( n | m )$ from
an APT calculation in~\cite{DeGrandi_10_APT}.
We estimate $\expval{ W_{ \APT, 1 } }$,
then argue that $\expval{ W_{\APT, 3} }  \approx  \expval{ W_{ \APT, 1 } }$.

\textbf{Diabatic-hopping probability 
$\mathbf{P_{\text{\textbf{APT}}} (E_f - E_i)}$
from APT:}
In this section, we generalize from the engine $\Sys$
to a closed quantum system $\tilde{\Sys}$.
Let $H(t)$ denote a time-dependent Hamiltonian. 
The $m^\th$ instantaneous energy eigenstate is denoted by $\ket{ E_m (t) }$.
Let $\tilde{\Sys}$ begin in the state $\ket{ E_m ( t_i ) }$.
Let $V$ denote the term ``turned on'' in $H(t)$.
$V$ couples $H ( t_i )$ eigenstates together. 
The coupling transfers $\tilde{\Sys}$
to some $\ket{ E_n (t_f) }$
with probability $P_\APT ( n | m )$.

De Grandi and Polkovnikov calculate~\cite[Eq.~(20), p.~4]{DeGrandi_10_APT}
\begin{align}
   \label{eq:DeG_20}
   P_\APT( n | m ) &  \approx   \left( \frac{v}{ \HScale } \right)^2
   \Bigg[  \frac{ \Big\lvert  
                     \langle E_n(t) | \partial_{\alpha_t} | 
                                E_m(t) \rangle  |_{\alpha_{t_i}}
                     \Big\rvert^2 }{ [ E_n ( t_i )  -  E_m ( t_i ) ]^2 }
            +  \frac{ \Big\lvert  
                     \langle E_n (t) | \partial_{\alpha_t} | 
                                 E_m (t) \rangle  |_{\alpha_{t_f}}
                     \Big\rvert^2 }{ [ E_n (t_f)  -  E_m (t_f) ]^2 }
   \nonumber \\ & \qquad \qquad \quad
   -  2 \: \frac{ \langle E_n(t) | \partial_{\alpha_t} | 
                      E_m(t) \rangle |_{ \alpha_i } }{
                      E_n ( t_i )  -  E_m ( t_i ) }  \:
         \frac{ \langle E_n(t) | \partial_{\alpha_t} | 
                   E_m(t) \rangle |_{ \alpha_f } }{
                   E_n (t_f)  -  E_m (t_f) }  \:
         \cos ( \Delta \Theta_{nm} )  \Bigg]  \, .
\end{align}
De Grandi and Polkovnikov's $\lambda$ is 
our Hamiltonian-tuning parameter $\alpha_t$.  
Their speed $\delta$, which has dimensions of energy, equals our
$\frac{v}{ \HScale } \, .$ 
The $\Delta \Theta_{nm}$ denotes a
difference between two phase angles.

The final term in Eq.~\eqref{eq:DeG_20} results from interference.
This term often oscillates quickly and can be neglected~\cite{DeGrandi_10_APT}.
Furthermore, we will integrate $P_\APT ( n | m )$ over energies.
The integration is expected to magnify cancellations.

The second term in Eq.~\eqref{eq:DeG_20} shares 
the first term's form. 
The first term is evaluated at $t = t_i$;
the second term, at $t = t_f \, .$
The quantities evaluated at $t_i$ are close their $t_f$ counterparts,
as $H(t)$ obeys the ETH at all $t \in [ t_i , t_f ]$.
Equation~\eqref{eq:DeG_20} approximates to\footnote{
Equation~\eqref{eq:MatrixRatio} accounts for the greater frequency
with which APT transitions occur
in the ETH phase than in the MBL phase.
In the ETH phase, $| \langle E_n  |  V  |  E_m  \rangle |$ 
has a considerable size, $\sim \frac{1}{ \sqrt{ \HDim } }$, 
for most $(n, m)$ pairs~\cite{Srednicki_94_Chaos}.
In the MBL phase, few pairs correspond to a large numerator:
$| \langle E_n  |  V  |  E_m  \rangle |  \sim \frac{1}{ \HDim }$~\cite{Pal_Huse_10_MBL}.
The corresponding energies tend to lie far apart:
$| E_n  -  E_m |  \gg  | \langle E_n  |  V  |  E_m  \rangle |$. 
Most APT transition probabilities are therefore suppressed~\cite{Serbyn_15_Criterion}.}
\begin{align}
   \label{eq:Build_P_APT}
   P_\APT( n | m ) &  \sim  
   2  \left(  \frac{ v }{ \HScale }  \right)^2
   \frac{ \Big\lvert   \langle E_n(t) | \partial_{\alpha_t} | 
                             E_m(t) \rangle  |_{\alpha_{t_i}}
             \Big\rvert^2 }{ [ E_n ( t_i )  -  E_m ( t_i ) ]^2 }  \, .
\end{align}

The perturbation-matrix element comes from the Chain Rule
and from ~\cite[Eq.~(10)]{DeGrandi_10_APT}:
\begin{align}
   \langle E_n(t) | \partial_{\alpha_t} | E_m(t) \rangle
   =  \left\langle  E_n(t)   \left\lvert  
        \frac{ \partial t }{ \partial \alpha_t }  \;  \frac{ \partial }{ \partial t }  
        \right\rvert  E_m(t) \right\rangle
   =  \frac{ \HScale }{v}  \;  
       \langle E_n(t) | \partial_t | E_m(t) \rangle
   =  \label{eq:MatrixRatio}
       \frac{ \HScale }{v}  \left(   -  \frac{ v }{ \HScale } \:  
       \frac{ \langle E_n(t) | V | E_m(t) \rangle }{
       E_n(t)  -  E_m(t) }  \right)  \, .
\end{align}
The modulus $ | \langle E_n(t) | V | E_m(t) \rangle |$ 
scales as $1 / \sqrt{ \HDim }$ 
for ETH Hamiltonians~\cite{Srednicki_94_Chaos}.\footnote{
\label{footnote:APT_HDim}
One might worry that, when this mesoscale engine functions as 
a component of a macroscopic engine, 
the Hamiltonian will not obey the ETH. 
Rather, $H(t)$ will be MBL at all times $t$.
However, for the purposes of level-spacing statistics and operator expectation values on length scales of the order of the localization length, $L  \sim  \xi_\Loc$, 
$H(t)$ can be regarded as roughly ETH.
The shallowly localized Hamiltonian's key feature is some nontrivial amount of level repulsion.
The ETH gap distribution, encoding level repulsion, suffices as an approximation.
However, $ | \langle E_n(t) | V | E_m(t) \rangle |  \sim  \frac{1}{ \HDim }$
for a mesoscale subengine in the macroscopic engine~\cite{Pal_Huse_10_MBL}.}
We introduce an $\HScale$ for dimensionality:
$| \langle E_n(t) | \partial_{\alpha_t} | E_m(t) \rangle |
\sim  \frac{ \HScale }{ \sqrt{ \HDim }  \:  | E_n(t)  -  E_m(t) | } \, .$
Substituting into Eq.~\eqref{eq:Build_P_APT} yields
\begin{align}
   \label{eq:P_APT_0}
   P_\APT ( n | m )  \sim  
   \frac{ v^2 }{ \HDim  \,  [ E_n ( t_i )  -  E_m ( t_i ) ]^4 }  \, . 
\end{align}
We have dropped a two, due to our focus on scaling.
We will drop the time arguments.
This probability is an even function of the signed gap 
$E_n  -  E_m$:
Only the gap's size, not its direction,
affects the hopping probability.

$P_\APT ( n | m )$ is normalized to one, so
the right-hand side of Eq.~\eqref{eq:P_APT_0} 
makes sense only when $< 1$.
The right-hand side diverges if $E_n$ lies close to $E_m$.
But energies rarely lie close together in the ETH phase,
due to level repulsion.
Furthermore, slow tuning of $H(t)$ impedes diabatic transitions.
Hence we introduce a regularization factor $R$:
\begin{align}
   \label{eq:P_APT}
   P_\APT ( n | m )  \sim  \frac{ v^2 }{ \HDim 
   \left[  ( E_n - E_m )^2  +  R^2  \right]^2 }  \, .
\end{align}

In the worst case---when the right-hand side of Eq.~\eqref{eq:P_APT} 
is largest---$| E_n - E_m |$ is small.
The right-hand side then approximates to $\frac{ v^2 }{ \HDim R^4 }$,
which must $< 1$.
The regularization must obey
\begin{align}
   \label{eq:R_bound}
   R  >  \frac{ \sqrt{v} }{ \HDim^{ 1 / 4 } }  \, . 
\end{align}

How to choose a form for $R$ is unclear. 
We therefore leave $R$ unspecified temporarily.
We will compute $\expval{ W_\APT }$ in terms of $R$,
then survey the possible forms of $R$.
We will choose the worst-case form for $R$---the 
form that maximizes the average work cost $\expval{ W_\APT }$---consistent
with Ineq.~\eqref{eq:R_bound} and with the smallness of $v$.

%
%
%
\textbf{Average work cost  $\expval{ W_{ \APT , 1 } }$
of stroke-1 APT transitions:}
$\Sys$ begins stroke 1 in a temperature-$\THot$ Gibbs state.
We focus on $T_\HTemp < \infty$.
Most of the state's weight
lies below the energy band's center:
$\expval{ E_m }  \equiv
\Tr  \left(  \frac{ e^{ - \betaH H_\ETH } }{ \ZH }  \:  H_\ETH  \right) < 0$.
More levels lie above $\expval{ E_m }$ than below.
Hence $\Sys$ more likely hops upward than drops.
APT transitions draw the state toward maximal mixedness.

Let $\Sys$ begin on the energy-$E_m$ level.
A \emph{conditional density of states} contributes to
the probability that $\Sys$ hops to 
the energy-$E_n$ level.
$E_m$ has a negligible chance of lying within $< \dAvg$ of $E_n$,
due to level repulsion:
\begin{align}
   \label{eq:RelDOS}
   \DOS ( n | m )  \sim  \DOS ( E_n )  \:
   \frac{ | E_n  -  E_m | }{ \sqrt{ ( E_n  -  E_m )^2  +  \dAvg^2 } }  \, .
\end{align}
We approximate sums with integrals, 
replacing $E_m$ with $E$ and $E_n$ with $E'$:
\begin{align}
   \label{eq:W_HTemp_a}
   \expval{ W_{\APT, 1} }  & \sim
   \int_{ -\infty }^\infty  d E  \:  \frac{ e^{ - \betaH  E } }{ \ZH }  \:
   \DOS  \left( E  \right)  
   \int_{ -\infty }^\infty  d E'  \:
   \DOS ( E' | E )  \:  P_\APT \left( E' | E \right)  \cdot
   \left( E'  -  E  \right)  \, .
\end{align}
The partition function appears in Eq.~\eqref{eq:ZH};
the DOS, in Eq.~\eqref{eq:DOS_App};
and the APT hopping probability, in Eq.~\eqref{eq:P_APT}: 
\begin{align}
   \label{eq:W_APT1_a}
   \expval{ W_{\APT, 1} }  &  \sim
   \int_{ -\infty }^\infty  d E  \;  
   \frac{ e^{ - \betaH  E } }{ \ZH }  \;
   \left(  \frac{ \HDim }{ \sqrt{ 2 \pi \Sites  \HScale^2 } }  \;
            e^{ - \frac{ ( E )^2 }{ 2 \Sites  \HScale^2 } }  \right)  \\
   \nonumber & \quad \times
   \int_{ -\infty }^\infty   d E'  \;
   \left(  \frac{ \HDim }{ \sqrt{ 2 \pi \Sites  \HScale^2 } }  \;
            e^{ - \frac{ ( E' )^2 }{ 2 \Sites  \HScale^2 } }  \;
            \frac{ | E'  -  E | }{ \sqrt{ ( E'  -  E )^2  +  \dAvg^2 } }  \right)
   \left(  \frac{ v^2 }{ \HDim  \;  [ ( E'  -  E )^2  +  R^2 ]^2 }  \right)
   ( E'  -  E )  \, .
\end{align}
We change variables from $E$ and $E'$ to
$x  :=  E - E'$ and $y := E + E'$.
As $E = \frac{1}{2}  \:  ( x + y )$ and $E'  =  \frac{1}{2}  \:  ( y - x )$,
\begin{align} 
   \label{eq:W_APT1_b}
   \expval{ W_{\APT, 1} }  & \sim
   - \frac{ 1 }{8 \pi }  \:  \frac{ v^2  \,  \HDim }{ \Sites \HScale^2 }
   \int_{ - \infty }^\infty  dy  \;  \frac{ e^{ - \betaH y / 2 } }{ \ZH }  \;
   e^{ - y^2 / 4 \Sites \HScale^2 }
   \int_{ - \infty }^\infty  dx  \;  e^{ - \betaH x / 2 }  \;  
   e^{ - x^2 / 4 \Sites \HScale^2 }  \;
   \frac{ | x | x }{ \sqrt{ x^2  +  \dAvg^2 }  \:  ( x^2  +  R^2 )^2 }  \, .
\end{align}

We focus first on the $x$ integral, $\mathcal{I}$.
The regularization factor, $R$, is small.
(Later, we will see that all reasonable options for $R  \leq  \sqrt{ v}$,
which $\ll  \dAvg$ by assumption.)
Therefore, the integral peaks sharply around $x = 0$.
We Taylor-approximate the slowly varying numerator exponentials
to first order in $x$:
$e^{ - \betaH x / 2 }  \;  e^{ - x^2 / 4 \Sites \HScale^2 }
\sim  \left( 1 - \frac{ \betaH }{ 2 }  \:  x  \right)
\left(  1  -  \frac{ x^2 }{ 4 \Sites \HScale^2 }  \right)$.
The zeroth-order term vanishes by parity:
\begin{align}
   \label{eq:APT_I_a}
   \mathcal{I}  \sim  - \frac{ \betaH }{ \dAvg }
   \int_{ - \infty }^\infty  dx  \;
   \frac{ x^2 | x | }{ \sqrt{ x^2  +  \dAvg^2 }  \:   ( x^2  +  R^2 )^2 }
   =  -  \frac{ 2 \betaH }{ \dAvg }
   \int_0^\infty  dx  \;
   \frac{ x^3 }{ \sqrt{ x^2  +  \dAvg^2 }  \:  ( x^2  +  R^2 )^2 }  \, .
\end{align}
The final equality follows from the integrand's evenness.

The square-root's behavior varies between two regimes:
\begin{align}
   \frac{1}{ \sqrt{ x^2  +  \dAvg^2 } }  =
   \begin{cases}
      \frac{1}{ \dAvg }  +  O \left( \left[ \frac{ x }{ \dAvg } \right]^2  \right)  \, ,
         &  x  \ll  \dAvg   \\
      \frac{1}{x}  +  O  \left( \left[  \frac{ \dAvg }{ x } \right]^2  \right)  \, ,
         &  x  \gg \dAvg
   \end{cases}  \, .
\end{align}
We therefore split the integral:
\begin{align}
   \label{eq:APT_I_a2}
   \mathcal{I}  \sim  - \frac{ 2 \betaH }{ \dAvg }  \Bigg(
   \frac{1}{ \dAvg }  \int_0^{ \dAvg }  d x  \;  
   \frac{ x^3 }{ ( x^2  +  R^2 )^2 }
   +  \int_{ \dAvg }^\infty  dx  \;  \frac{1}{ x^2 }
   \Bigg)  \, .
\end{align}
We have dropped the $+ R^2$ from the second integral's denominator:
Throughout the integration range, $x \gg \dAvg$, which $\gg R$.
Integrating yields
\begin{align}
   \label{eq:APT_I_b}
   \mathcal{I}  \approx  -  \frac{ \betaH }{ \dAvg }  
   \log \left(  \frac{ \dAvg^2 }{ R^2 }  \right)  \, .
\end{align}

We have evaluated the $x$ integral in Eq.~\eqref{eq:W_APT1_b}.
The $y$ integral evaluates to
$\frac{ 2 \sqrt{ \pi \Sites } }{ \HDim }  \:  \HScale 
e^{ - \Sites ( \betaH \HScale )^2 / 4 }$.
Substituting into Eq.~\eqref{eq:W_APT1_b} yields
\begin{align}
   \expval{ W_{\APT, 1} }  & \sim  
   \left( - \frac{1}{ 8 \pi }  \:  \frac{ v^2  \,  \HDim }{ \Sites  \HScale^2 }  \right)
   \left[ - \frac{ 2 \betaH }{ \dAvg }  \:  \log \left( \frac{ \dAvg^2 }{ R^2 }  \right)  \right]
   \left( \frac{ 2 \sqrt{ \pi \Sites } }{ \HDim }  \:  \HScale 
           e^{ - \Sites ( \betaH \HScale )^2 / 4 } \right) \\
   \label{eq:W_APT1_c}
   & =  \frac{1}{ 2 \sqrt{ \pi } }  \:  \frac{ 1 }{ \sqrt{\Sites} }  \:
   \frac{ v^2  \betaH }{ \HScale  \dAvg }  \:
   \log \left( \frac{ \dAvg^2 }{ R^2 }  \right)  \,
   e^{ - \Sites ( \betaH \HScale )^2 / 4 }  \, .
\end{align}

The regularization $R$ appears only in the logarithm.
Hence the form of $R$ barely impacts $\expval{ W_{\APT, 1} }$.
Which forms can $R$ assume?
$R$ should be small in $v$
and should have dimensions of energy.
The only other relevant energy scales are $\dAvg$ and $\HScale$.\footnote{
$\deltaMBL$ is irrelevant, being a property of MBL systems.
This calculation concerns the ETH phase.}
We choose the ``worst-case'' $R$,
which leads to the greatest $\expval{ W_\APT, 1 }$
consistent with Ineq.~\eqref{eq:R_bound}
and with the smallness of $v$.
$\expval{ W_{\APT, 1} }$ is large when $R$ is small.
The possible regularizations small in $v$ are 
$\sqrt{v}$, $\frac{ v }{ \dAvg },$ and $\frac{ v}{ \HScale }$.
Consider substituting each value into Ineq.~\eqref{eq:R_bound}.
If $R \propto v$, Ineq.~\eqref{eq:R_bound} lower-bounds $v$.
Diabatic transitions should upper-bound, not lower-bound, the speed.
We therefore disregard $\frac{ v }{ \dAvg }$ and $\frac{ v}{ \HScale }$.
Substituting $R = \sqrt{v}$ into Ineq.~\eqref{eq:R_bound} yields
$1 > \frac{ 1 }{ \HDim^{ 1/4} }$, which is true. We therefore choose
\begin{align}
   \label{eq:R_choice}
   R  =  \sqrt{ v }  \, .
\end{align}

Consequently,
\begin{align}
   \label{eq:W_APT1_d}  \boxed{
   \expval{ W_{\APT, 1} }
   \sim   \frac{ 1 }{ \sqrt{\Sites} }  \:
   \frac{ v^2  \betaH }{ \HScale  \dAvg }  \:
   \log \left( \frac{ \dAvg^2 }{ v }  \right)  \,
   e^{ - \Sites ( \betaH \HScale )^2 / 4 }  }  \, .
\end{align}

\textbf{Average work cost $\expval{ W_{ \APT, 3 } }$
of stroke-3 APT transitions:}
In the lowest-order approximation,
(i) the stroke-1 tuning is adiabatic, and
(ii) cold thermalization transfers $\Sys$ across just one gap.
Let $\PDown$ ($\PUp$) denote the engine's probability of dropping (rising)
during cold thermalization.
The average work cost is
\begin{align}
   \label{eq:W_APT_3_a}
   \expval{ W_{\APT, 3} }  & \approx
   \sum_m   \frac{ e^{ - \betaH  E_m } }{ \ZH }
   \sum_n   \Bigg\{
   \int_{ - \Wb }^0  d \delta'  \;  P_\MBL ( | \delta' | )  \;  \PDown ( | \delta' | )  \;
   P_\APT ( n | m - 1 )  ( E_n  -  E_{m - 1} )
   \nonumber \\ & \quad
   +  \int_0^{ \Wb }  d \delta'  \;  P_\MBL ( \delta' )  \;  \PUp ( \delta' )  \;
   P_\APT ( n | m + 1 )  ( E_n  -  E_{m + 1 } )
   \nonumber \\ & \quad
   + \Bigg[  1  
   -  \int_{ - \Wb }^0  d \delta'  \;  P_\MBL ( | \delta' | )  \;  \PDown ( | \delta' | )
   -  \int_0^{ \Wb }  d \delta'  \;  P_\MBL ( \delta' )  \;  \PUp ( \delta' )  \Bigg]
   \nonumber \\ & \qquad \times 
   P_\APT ( n | m )   ( E_n  -  E_m )  \Bigg\}  \, .
\end{align}
We have artificially extended the gap variable $\delta'$ to negative values:
$\delta' < 0$ denotes a size-$|\delta'|$ gap just below level $m$.
The bracketed factor $[ 1 - \ldots ]$ represents the probability that
cold thermalization preserves the engine's energy.

Let us analyze $\expval{ W_{ \APT, 3} }$ physically.
Consider the $\TCold = 0$ limit, for simplicity.
On average over trials, the engine's state barely changes 
between strokes 1 and 3.
Tiny globules of weight drop across single gaps.
Hence most stroke-3 APT transitions look identical, on average over trials,
to the stroke-1 APT transitions:
$\expval{ W_{\APT, 3} }  \approx  \expval{ W_{\APT, 1} }
+ \text{(correction)}$.

The correction comes from the probability-weight globules.
APT transitions hop some globules 
off the bottoms of ``working gaps'' 
(Fig.~\ref{fig:Compare_thermo_Otto_fig}),
derailing trials that would have outputted $W_\tot \sim \dAvg$.
But other globules, which began stroke 3 elsewhere in the spectrum,
hop onto the bottoms of working gaps.
The globules hopping off roughly cancel with
the globules hopping on:
$\boxed{ \expval{ W_{ \APT, 3 } }  \approx   \expval{ W_{ \APT, 1 } } }$,
and $\boxed{  \expval{ W_{\APT} }  \approx  \expval{ W_{ \APT, 1 } } }$.

\subsubsection{Average work costs of Landau-Zener diabatic jumps:
$\expval{W_{\LZ, 1} }$ and $\expval{ W_{\LZ, 3} }$}
\label{section:WLZ}

Consider $H(t)$ within the MBL phase
(near, but not quite at, the end of stroke 1 or the start of stroke 3).
Two energy levels can wiggle toward each other and apart.
The wiggling has a probability
\begin{align}
   \label{eq:P_LZ}
   P_\LZ ( \gap )  \approx  e^{ - 2 \pi \left( \deltaMBL \right)^2 / v } 
\end{align}
of inducing a Landau-Zener transition~\cite{Landau_Zener_Shevchenko_10}.
$\deltaMBL$ roughly equals the size of 
the Hamiltonian-perturbation matrix element
that couples the wiggling-together states.

The average work cost vanishes by parity.
The engine's probability of hopping upward
equals its probability of dropping, by Eq.~\eqref{eq:P_LZ}.
Only hops to nearest neighbors have significant probabilities.
Hence the existence of more levels above $\expval{ H(t) }$
than below has no impact on $\expval{W_{\LZ, 1} }$.\footnote{
The imbalance impacted the $\expval{ W_{\APT} }$'s 
in Sec.~\ref{section:App_Diab_APT}.
There, we Taylor-approximated $e^{ - \betaH x }$ 
to first order in $x := E - E'$,
because $\Sys$ could hop across several levels.
The zeroth-order term vanished by parity.
The LZ calculation may be thought of as 
a truncation of the APT calculation at zeroth order,
because $\Sys$ can hop only one gap.
Put another way, in the APT calculation, the $E$ integral affected the $\gap$ integral,
preventing parity from sending the $\gap$ integral to zero.
Here, the integrals decouple.}
The upward hops' work cost cancels, on average,
with the drops' work cost:
$\boxed{  \expval{ W_{\LZ, 3} }  =  \expval{W_{\LZ, 1} }  =  0  } \, .$


%
%
%
\subsubsection{Average work costs of fractional-Landau-Zener diabatic jumps:
$\expval{ W_{{\text{frac-LZ}}, 1} }$ and $\expval{ W_{{\text{frac-LZ}}, 3} }$}
\label{section:HalfLZ}

%
%
\begin{figure}[tb]
\centering
\includegraphics[width=.45\textwidth, clip=true]{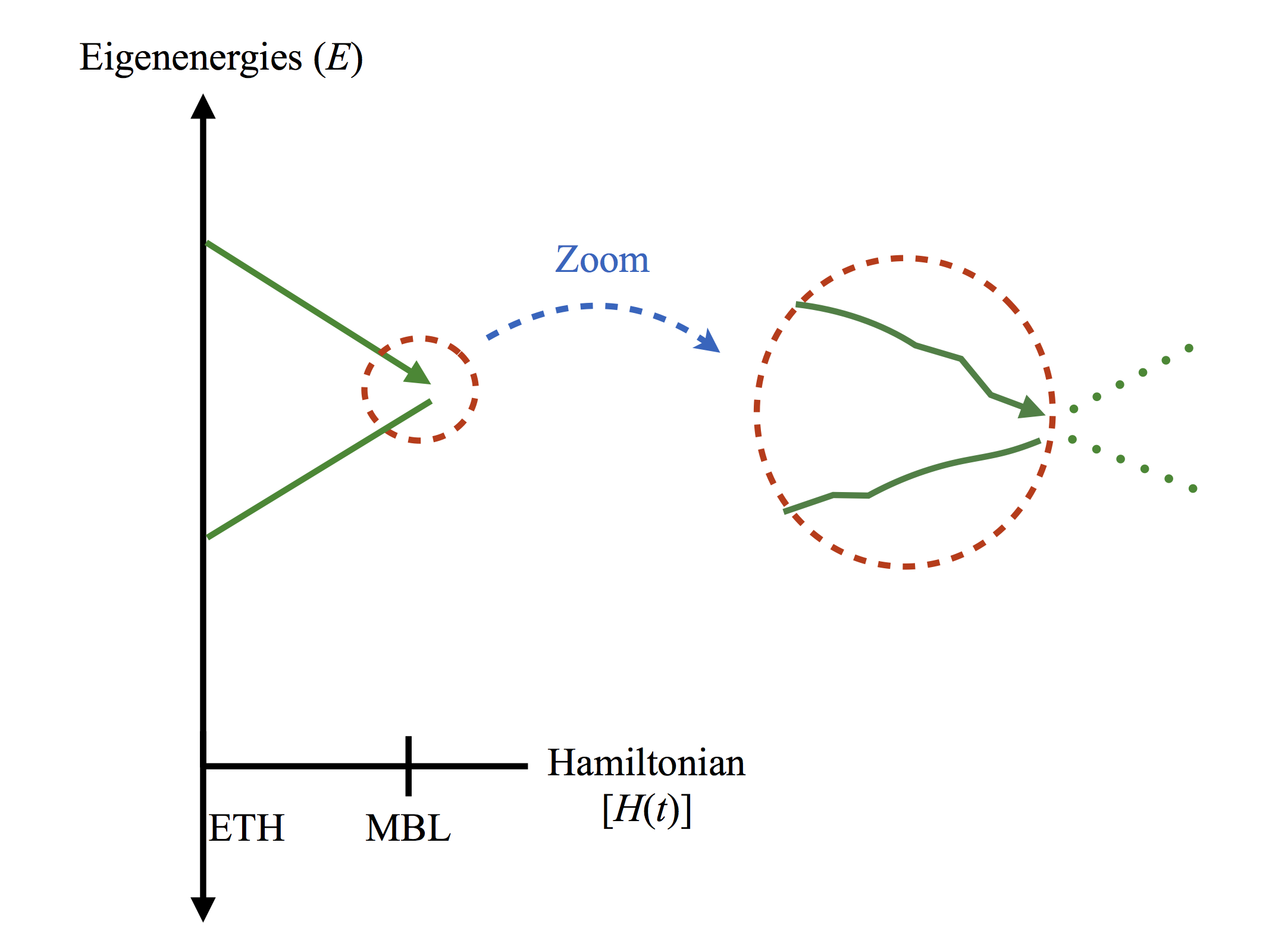}
\caption{\caphead{Fractional-Landau-Zener transition:} 
The straight solid green lines represent two eigenenergies.
The engine ideally occupies the upper level throughout stroke 1.
At the end of stroke 1, the energies approach each other. 
Zooming in on the approach shows that the lines are not straight,
but wiggle slightly.
A full Landau-Zener transition could occur
if the approaching lines came very close together
and then separated.
The green dotted lines illustrate the hypothetical separation.
Since the approaching energies do not separate,
the engine may undergo an approximate fractional-Landau-Zener transition.}
\label{fig:Half_LZ}
\end{figure}

A Landau-Zener transition can occur
when two energies begin far apart,
come together, suffer a mixing of eigenstates, and separate.
Eliminating the first or last step can induce
a \emph{fractional-Landau-Zener transition}.
Such transitions can occur at the end of stroke 1
(Fig.~\ref{fig:Half_LZ})
or the start of stroke 3.
We apply to these strokes the model in~\cite{DeGrandi_10_APT}.

\textbf{Modeling fractional-Landau-Zener transitions:}
De Grandi and Polkovnikov model 
an arbitrary portion of the LZ process
using APT~\cite[Sec. II A]{DeGrandi_10_APT}.
We conjugate their Hamiltonian [their Eq.~(21)]
by the Hadamard 
$\frac{1}{ \sqrt{2} } \left( \sigma^x  +  \sigma^z  \right)$:
\begin{align}
   \label{eq:H_LZ}
   H_{ \HalfLZ }  =  \deltaMBL  \,  \sigma^z
   +  v t  \,  \sigma^x  \, .
\end{align}
This Hamiltonian captures the basic physics
of growing energies 
and rotating eigenstates.
De Grandi and Polkovnikov's speed $\delta$ translates into our $v$.\footnote{
The significance of $\delta$ changes between the general APT discussion
and the fractional-LZ discussion in~\cite{DeGrandi_10_APT}.
In the latter discussion, $\delta$ has dimensions of time$^2$.}

De Grandi and Polkovnikov's time parameter $t  \in  [t_i , t_f]$.
In the ordinary Landau-Zener problem,
$t  \in  ( -\infty  ,  \infty )$.
We approximate $t  \in  (- \infty ,  0 ]$
at the end of stroke 1 and 
$t  \in  [ 0 ,  \infty )$ at the start of stroke 3.

The qubit's probability of hopping 
between eigenstates is~\cite[Eq.~(29)]{DeGrandi_10_APT}
\begin{align}
   \label{eq:Half_LZ_Prob_1}
   P_{\text{frac-LZ}}  &  \approx
   \frac{ v^2   \left( \deltaMBL \right)^2 }{ 16 } 
   \left( \frac{1}{ \left[ \left( \deltaMBL \right)^2  
                         +  \left(v t_i  \right)^2  \right]^3 }
          +  \frac{ 1}{  \left[ \left( \deltaMBL \right)^2  
                         +  \left( v  t_f  \right)^2  \right]^3 }
   \right) \\
   & = \label{eq:Half_LZ_Prob_2}
   \frac{ v^2   \left( \deltaMBL \right)^2 }{ 16 } 
   \left( \frac{1}{ \left( \text{Initial gap} \right)^6 }
           +  \frac{1}{ \left( \text{Final gap} \right)^6 }  \right)     \, .
\end{align}

We focus on stroke 3, which dominates $\expval{ W_{ \HalfLZ } }$.
The second fraction vanishes, since $t_f = \infty$.
Let $\gap'$ denote the gap with which stroke 3 starts.
We can no longer approximate the MBL ``working gaps''
as $\gap'  \in  [0,  \Wb ]$:
To avoid the $\gap'  =  0$ divergence, we refine our model.
In which trials do fractional-LZ transitions cost 
$W_{ \HalfLZ }  >  0$?
The trials that otherwise---in 
the absence of the transitions---would output $W_\tot > 0$.\footnote{
A fractional-LZ transition costs work of two types.
To describe them concretely, we suppose that the transition 
boosts the engine's energy at the start of stroke 3:
(1) $\Sys$ absorbs energy from the battery while hopping.
(2) After hopping, typically, $\Sys$ slides up an energy level,
like the top green line in Fig.~\ref{fig:Compare_thermo_Otto_fig}.
The sliding ``undoes'' the stroke-1 work extraction.
The average type-(1) work cost $\approx \Wb$.
The average type-(2) work cost $\approx \dAvg  \gg  \Wb$.
Hence $\expval{ W_\HalfLZ } \approx$ 
the type-(2) work.}
Most otherwise-successful trials involve 
gaps $\gap'  \sim  \Wb$.
Hence we integrate $\gap'$ from $\Err \Wb$ to $\Wb$, 
wherein $\Err  \in  \left(  \frac{ \deltaMBL }{ \Wb } ,  1  \right)$. 

\textbf{Simple approximation of $\exp{ W_\HalfLZ }$
and associated $v$ bound:}
The engine has a probability $\sim \frac{ \Err \Wb }{ \dAvg }$
of neighboring an MBL gap $\gap'  \lesssim  \Err \Wb$.
In the worst case, whenever the engine neighbors such a gap,
the engine suffers a stroke-3 fractional-LZ transition.
Suppose, for simplicity, that $\TCold = 0$.
Each such transition costs work $\sim \dAvg$
(the work that the trial would have outputted
in the transition's absence).
Hence gaps $\gap'  \lesssim  \Err \Wb$ cost, at most, work
\begin{align}
   \label{eq:Extra_fLZ_W}
   \frac{ \Err \Wb }{ \dAvg }  \cdot  \dAvg
   =  \Err \Wb   \, ,
\end{align}
on average.
This bound shows that $\expval{ W_{\HalfLZ} }$ is small.

Approximating dominant initial gaps with $\sim \Wb$
implies a condition on $v$
under which Eq.~\eqref{eq:Half_LZ_Prob_2} is justified.
The probability $P_{ \HalfLZ }$ must be normalized, so
$P_{ \HalfLZ }  \sim  \frac{ v ( \deltaMBL )^2 }{ 16  ( \Wb )^6 }
\leq 1$. Solving for the speed yields
\begin{align}
   \label{eq:Early_HalfLZ_v_bd}
   v  \leq  \frac{ 4 ( \Wb )^3 }{ \deltaMBL }  \, .
\end{align}
We can bound $v$, instead, by 
(1) estimating $\expval{ W_{ \HalfLZ } }$
and (2) demanding that 
fractional-LZ transitions cost less work than 
the engine outputs per ideal average cycle:
$\expval{ W_{ \HalfLZ } }  \ll  \expval{ W_\tot }$.
This inequality implies Ineq.~\eqref{eq:Early_HalfLZ_v_bd},
up to prefactors, we will find.
Hence~\eqref{eq:Half_LZ_Prob_2} leads to a self-consistent argument.

\textbf{Average work cost 
$\bm{\langle}  \mathbf{ W_{ \HalfLZ , 1} }  \bm{\rangle}$
of fractional-Landau-Zener diabatic transitions
at the end of stroke 1:}
These transitions cost zero work, on average, by symmetry:
$\boxed{  \expval{ W_{\HalfLZ, 1} } = 0  }$.
Suppose that $\Sys$ starts stroke 1 on the $j^\th$ energy level.
At the end of stroke 1, level $j$ as likely approaches level $j - 1$
as it approaches level $j + 1$.
A fractional-LZ transition as likely costs $W > 0$ 
as it costs $W < 0$.
Hence $\expval{W} = 0$.
This symmetry is absent from $\expval{ W_{\HalfLZ, 3} }$,
due to cold thermalization.

\textbf{Average work cost 
$\bm{\langle}  \mathbf{ W_{ \HalfLZ, 3} }  \bm{\rangle}$
of fractional-Landau-Zener diabatic transitions
at the start of stroke 3:}
%
%
%
%
%
%
$\Sys$ starts the trial of interest with the ETH eigenenergy $E$, 
which tuning maps to the MBL $E'$.
No diabatic transitions occur during stroke 1, 
in the lowest-order approximation.
$E'$ neighbors at most one small gap, to lowest order.
Cold thermalization hops $\Sys$ upward/downward
with probability $\frac{ 1 }{ 1 + e^{ \pm \betaC | \gap' | } }$.
As stroke 3 begins, $\Sys$ reverses across the gap
with probability $P_{ \HalfLZ } (\gap')$.
Cold thermalization has a probability 
$1  -  \PDownnn  -  \PUppp  \equiv  1  
-  \int_{ - \Wb }^0  d \gap'  \;
P_\MBL ( | \gap' | )  \;  \frac{ 1 }{ 1  +  e^{ - \betaC | \gap' | } }
-  \int_0^{ \Wb }  d \gap'  \;
P_\MBL ( \gap' )  \;  \frac{ 1 }{ 1  +  e^{ \betaC \gap' } }$
of preserving the engine's energy.
In this case, any stroke-3 fractional-LZ transition
costs $\expval{ W_{ \HalfLZ, 1} } = 0$. Hence
\begin{align}
   \expval{ W_{ \HalfLZ, 3} }
   & \approx  \int_{ -\infty }^\infty  d E  \;  
   \frac{ e^{ - \betaH   E } }{ \ZH }  \;     \DOS ( E ) 
   \Bigg[  \int_{ - \Wb }^{ - \Err \Wb }  d \gap'  \;
   P_\MBL ( | \gap' | )  \;  \frac{1}{ 1  +  e^{ - \betaC | \gap' | } }  \;
   P_{ \HalfLZ} ( \gap' )
   \nonumber \\ & \qquad \times
   \int_0^\infty  d \gap  \cdot \gap  \;  P_\ETH ( \gap )
   \nonumber \\ & 
   +  \int_{ \Err \Wb }^{ \Wb }  d \gap'  \;  P_\MBL ( \gap' )  \;
   \frac{ e^{ - \betaC \gap' } }{ 1  +  e^{ - \betaC \gap' } }  \;   
   P_{ \HalfLZ }  ( \gap' )
   \int_{ - \infty}^0  d \gap  \cdot  \gap  \;  P_\ETH ( | \gap | )  \Bigg]
   \nonumber \\ &
   + ( 1 - \PDownnn  -  \PUppp )  \expval{ W_{ \HalfLZ, 1} }   
   +  \Err \Wb   \, .
\end{align}
The final term is consistent with~\eqref{eq:Extra_fLZ_W}.

Computing the integral~\cite[App.~G 8 iii]{NYH_17_MBL} yields
\begin{align}
   \label{eq:Half_LZ_3_e}   & \boxed{
   \expval{ W_{ \HalfLZ , 3 } }
   \approx  \frac{1}{  80 \Err^5 }  \:  \frac{ v^2  ( \deltaMBL )^2 }{  ( \Wb )^5 }
   +  \Err \Wb }  \, .
\end{align}
By assumption, $\Err  <  1$. We will often assume that $\Err \approx \frac{1}{3}$.
Hence the final term in Eq.~\eqref{eq:Half_LZ_3_e}
is smaller than $\expval{ W_\tot }  \sim  \Wb$.

Equation~\eqref{eq:Half_LZ_3_e} implies an upper bound on $v$
of the form in Ineq.~\eqref{eq:Early_HalfLZ_v_bd}.
The Hamiltonian must be tuned slowly enough that
fractional-LZ transitions cost less work than
an ideal cycle outputs, on average:
$\expval{ W_{ \HalfLZ } }  \ll  \expval{ W_\tot }$.
The right-hand side roughly equals $\Wb$, by Eq.~\eqref{eq:WTotApprox2_Main}.
We substitute in for the left-hand side from Eq.~\eqref{eq:Half_LZ_3_e}.
Solving for the speed yields
$v  \ll  \sqrt{80  \Err^5 }  \:  
   \frac{ ( \Wb )^3 }{ \deltaMBL }   \, .$
For every tolerance $\Err \in (0, 1)$, there exist speeds $v$ 
such that the inequality is satisfied.
For simplicity, we suppose that $\Err  \approx  \frac{1}{3},$
such that the overall constant $\approx 1$.
The bound reduces to
\begin{align}
   \label{eq:Late_HalfLZ_v_bd}
   v  \ll  \frac{ ( \Wb )^3 }{ \deltaMBL }   \, .
\end{align}
This bound has the form of Ineq.~\eqref{eq:Early_HalfLZ_v_bd}.
Our approximation~\eqref{eq:Half_LZ_Prob_2} leads to
a self-consistent argument.

\subsubsection{Diabatic correction to the efficiency $\eta_\MBL$}
\label{eq:Diab_eff}

The efficiency has the form
\begin{align}
   \eta_\MBL  & :=  \frac{ \expval{ W_\tot } }{ \expval{ Q_\In } }
   =  \frac{ \expval{ W_\tot } }{ \expval{ W_\tot }  -  \expval{ Q_2 } }
   =  \frac{ \expval{ W_\tot } }{ \expval{ W_\tot }  
        \left( 1 - \frac{ \expval{ Q_2 } }{ \expval{ W_\tot } } \right) }  
\end{align}
Here, $\expval{ W_\tot }$ denotes the net work extracted per trial,
on average over trials.
[Earlier, $\expval{ W_\tot }$ denoted the average net work extracted
per trial in which $H(t)$ is tuned adiabatically.]
We Taylor-approximate to first order,
relabel as $\expval{ W_\tot^\adiab }$
the adiabatic approximation~\eqref{eq:WTotApprox2},
and denote by $\expval{ W_\diab }$
the average total per-cycle diabatic work cost:
$\eta_\MBL  
   =  1  +  \frac{ \expval{ Q_2 } }{ 
                        \expval{ W_\tot^\adiab }  -  \expval{ W_\diab } }  \, .$
Invoking $\expval{ W_\tot^\adiab }  \gg  \expval{ W_\diab } \, ,$
we Taylor-approximate again:
$\eta_\MBL  
   \approx  1 + \frac{ \expval{ Q_2 } }{ \expval{ W_\tot^\adiab } }
   \left( 1  +  \frac{ \expval{ W_\diab } }{ \expval{ W_\tot^\adiab } }  \right) \, .$

We relabel as $\eta_\MBL^\adiab$
the adiabatic estimate~\eqref{eq:ManyBodyEff}
of the efficiency:
$\eta_\MBL   \approx
   \eta_\MBL^\adiab  +  \expval{ W_\diab }
   \frac{ \expval{ Q_2 } }{ \expval{ W_\tot^\adiab }^2 }  \, .$
Substituting in from Eq.~\eqref{eq:EDiff2b}, 
and substituting in the leading-order term 
from Eq.~\eqref{eq:WTotApprox2}, yields
\begin{align}
   \label{eq:EtaDiab1}
   \eta_\MBL  \approx     \eta_\MBL^\adiab  
   -  \frac{ \expval{ W_\diab } }{ 2 \dAvg }
   \equiv  \eta_\MBL^\adiab  -  \phi_\diab \, .
\end{align}

For simplicity, we specialize to $\THot = \infty$ and $\TCold = 0 \, .$
The correction becomes
\begin{align}
   \label{eq:Diab_eff_app}  \boxed{
   \phi_\diab }  
   & =  \frac{ \expval{ W_\diab } }{ 2 \dAvg }
   \Bigg\lvert_{  \TCold = 0,    \THot = \infty  }  
   =  \frac{1}{ 2 \dAvg }     \expval{ W_{ \APT, 3 } }  
   \Big\rvert_{  \TCold = 0,    \THot = \infty  }  
   \boxed{ \approx  
   %
   \frac{1}{ 160 \Err^5 }  \:  \frac{ v^2  ( \deltaMBL )^2 }{ ( \Wb )^5  \dAvg }
   +  \frac{ \Err }{ 2 }  \,
   \frac{ \Wb }{ \dAvg }   }  \, ,
\end{align}
by Eqs.~\eqref{eq:W_APT1_d} and~\eqref{eq:Half_LZ_3_e}.

As expected, work-costing diabatic jumps 
detract from the efficiency slightly.
The first term is suppressed in 
in $\frac{ \sqrt{ v } }{ \dAvg }  \ll  1$ and in 
$\frac{ \deltaMBL }{ \dAvg }  \ll  1$.
The second term is suppressed in $\frac{ \Wb }{ \dAvg }  \ll  1$
and in a constant $\frac{ \Err }{2}  \approx  \frac{1}{6}$.

\section{Phenomenological model for 
the macroscopic MBL Otto engine}
\label{section:ThermoLimitApp}


The macroscopic MBL Otto engine
benefits from properties of MBL
(Sec.~\ref{section:Thermo_limit_main}):
localization and local level repulsion.
We understand these properties from 
(1) Anderson insulators~\cite{Anderson_58_Absence} and
(2) perturbation theory.
Anderson insulators are reviewed in Sec.~\ref{section:Anderson_Ham}.
Local level repulsion in Anderson insulators~\cite{Sivan_87_Energy} 
in the strong-disorder limit
is reviewed in Sec.~\ref{section:And_Repulsion}.
Section~\ref{section:Phenom_MBL_subapp}
extends local level repulsion to MBL.
Local level repulsion's application to the MBL engine
is discussed in Sec.~\ref{section:Repuls_Eng}.
Throughout this section, $\Sites$ denotes
the whole system's length.

\subsection{Anderson localization}
\label{section:Anderson_Ham}

Consider a 1D spin chain 
or, equivalently, lattice of spinless fermions.
An Anderson-localized Hamiltonian $H_\Anderson$ 
has almost the form of Eq.~\eqref{eq:SpinHam},
but three elements are removed:
(1) the $t$-dependence,
$Q \LParen h ( \alpha_t )  \RParen$, and
the interaction.
[The 
$\bm{\sigma}_j  \cdot  \bm{\sigma}_{j + 1}$
is replaced with 
$ \left( \sigma_j^+  \,  \sigma_{j + 1}^-   
+   \sigma_j^-  \,  \sigma_{j + 1}^+   \right)$.
The site-$j$ raising and lowering operators are denoted by
$\sigma_j^+ :=  \frac{1}{ 2 } \left( \sigma_j^x + i \sigma_j^y \right)$ and 
$\sigma_j^-  :=  \frac{1}{ 2 } \left( \sigma_j^x - i \sigma_j^y \right)$.]

Let $\ket{0}$ denote some reference state
in which all the spins point downward
(all the fermionic orbitals are empty).
In this section, we focus, for concreteness, on 
the properties of single-spin excitations 
relative to $\ket{0}$~\cite{Anderson_58_Absence,Sivan_87_Energy}.
The $\ell^\th$ excitation is represented, in fermionic notation, 
as $\sum\nolimits_x \psi_\ell(x)  \,   \sigma^+_{x_\ell} |0\rangle$. 
The single-excitation wave functions $\psi_\ell(x)$ are localized:
$x_\ell$ denotes the point at which the probability density
$| \psi_\ell (x) |^2$ peaks.
The wave function decays exponentially with the distance $|x - x_\ell |$
from the peak:
\begin{align}
   \label{eq:Eigenfxn}
   \psi_\ell(x)  \approx  \sqrt{ \frac{2}{ \xi_\Anderson } }  \;
   e^{ - | x - x_\ell | / \xi_\Anderson } \, .
\end{align}
The localization length varies with 
the Hamiltonian parameters as~\cite[App.~H 2]{NYH_17_MBL}
\begin{align}
   \label{eq:Xi}
   \xi_\Anderson  \sim  \frac{1}{ \ln h }  \, .
\end{align}

%
%
%
\subsection{Local level repulsion in Anderson insulators}
\label{section:And_Repulsion}

We begin with the infinitely localized limit $h \rightarrow \infty$.
We take $\HScale \rightarrow 0$ to keep 
the Hamiltonian's energy scale finite. 
The hopping terms can be neglected, 
and particles on different sites do not repel.
Single-particle excitations are localized on single sites.
The site-$i$ excitation corresponds to an energy $2 \HScale h h_i$. 
Since the on-site potentials $h \cdot h_i$ are uncorrelated,
neighboring-site excitations' energies are uncorrelated.

Let us turn to large but finite $h$. 
Recall that $h \cdot h_i$ is drawn uniformly at random from $[-h, \, h]$.
The uniform distribution has a standard deviation of
$\frac{h}{ \sqrt{3} }  \gg  1 \, .$
Therefore, $h | h_i - h_{i+1}| \gg 1$
for most pairs of neighboring sites.
The hopping affects these sites' wave functions and energies weakly.
But with a probability $\sim  \frac{1}{h}$, 
neighboring sites have local fields $h \cdot h_i$ and $h \cdot h_{i+1}$ 
such that $h |h_i - h_{i+1}| \ll 1$. 
The hopping hybridizes such sites. 
The hybridization splits the sites' eigenvalues by an amount
$\sim \sqrt{h^2 (h_i - h_{i+1})^2 + \HScale^2 } \geq \HScale$. 

Consider, more generally, two sites separated by a distance $L \, .$
Suppose that the sites' disorder-field strengths are separated by $< 1/h^L$.
(The upper bound approximates the probability amplitude associated with
a particle's hopping the $L$ intervening sites).
The sites' excitation energies and energy eigenfunctions 
are estimated perturbatively.
The expansion parameter is $1 / h \, .$
To zeroth order, the energies are uncorrelated
and (because $h | h_i - h_{i + L} |  <  1 / h^L$) are
split by $< \HScale /h^L \, .$
The eigenfunctions are hybridized at order $L \, .$
The perturbed energies are split by 
$\geq  \HScale/h^L  \sim  \HScale    e^{ - L / \xi_\Anderson }  \, .$ 
[Recall that $\xi_\Anderson  \sim  1/\ln h$, by Eq.~\eqref{eq:Xi}.]

Hence eigenstates localized on nearby sites have correlated energies:
\emph{The closer together sites lie in real space, 
the lower the probability that they correspond to similar energies.}
This conclusion agrees with global Poisson statistics: 
Consider a large system of $\Sites \gg 1$ sites.
Two randomly chosen single-particle excitations 
are typically localized a distance $\sim \Sites$ apart.
The argument above implies only that 
the energies lie $> \HScale  e^{ - \Sites /\xi_\Anderson }$ apart. 
This scale is exponentially smaller (in $\Sites$) than
the average level spacing $\sim \frac{ \HScale  h }{ \Sites }$ 
between single-particle excitations.\footnote{
\label{footnote:SinglePartE}
The average level spacing between single-particle excitations 
scales as $\sim 1 / \Sites$ for the following reason.
The reference state $\ket{0}$ consists of
$\Sites$ downward-pointing spins.
Flipping one spin upward yields a single-particle excitation.
$\Sites$ single-particle-excitation states exist,
as the chain contains $\Sites$ sites.
Each site has an energy $\sim \pm \HScale  h$, to zeroth order,
as explained three paragraphs ago.
The excitation energies therefore fill 
a band of width $\sim  \HScale h \, .$
An interval $\sim \frac{ \HScale  h }{ \Sites }$
therefore separates single-particle-excitation energies,
on average.}

We can quantify more formally the influence of hybridization
on two energies separated by $\omega$
and associated with eigenfunctions 
localized a distance $L$ apart.
The \emph{level correlation function} is defined as
\begin{equation}
   \label{eq:R}
   R(L, \omega)  :=  \frac{1}{ \Sites^2 } \sum_{i, n, n'} 
   |\langle 0 | \sigma^+_i | n \rangle|^2  \,
   |\langle 0 | \sigma^+_{i+L} | n' \rangle|^2  \,
   \delta(E_n - E_{n'} - \omega) - \tilde{\DOS} (\omega)^2  \, .
\end{equation}
The spatially averaged density of states at frequency $\omega$
is denoted by $\tilde{\DOS} (\omega) := \frac{1}{ \Sites } 
\sum_n |\langle 0 | \sigma^+_i | n \rangle|^2  \,\delta(E_n - \omega) \, .$ 
$|n\rangle$ and $|n'\rangle$ denote eigenstates,
corresponding to single-particle excitations relative to $|0\rangle$, associated with energies $E_n$ and $E_{n'}$. 
In the Anderson insulator, $R(L, \omega) \approx 0$ 
when $\omega \gg \HScale    e^{ -L/\xi_\Anderson }$: 
Levels are uncorrelated when far apart in space and/or energy. 
When energies are close ($\omega \ll \HScale    e^{ -L/\xi_\Anderson }$), 
$R(L, \omega)$ is negative. These levels repel (in energy space).

\subsection{Generalization to many-body localization}
\label{section:Phenom_MBL_subapp}

The estimates above can be extended 
from single-particle Anderson-localized systems 
to MBL systems initialized in arbitrary energy eigenstates
(or in position-basis product states).
$R(L, \omega)$ is formulated
in terms of matrix elements $\langle 0 | \sigma_i^+ | n \rangle$ 
of local operators $\sigma_i^+$. 
The local operators relevant to Anderson insulators 
have the forms of 
the local operators relevant to MBL systems.
Hence $R(L, \omega)$ is defined for MBL
as for Anderson insulators. 
However, $\ket{0}$ now denotes a generic many-body state.

Let us estimate the scale $\J_L$ of 
the level repulsion between MBL energies,
focusing on exponential behaviors.
The MBL energy eigenstates result from
perturbative expansions about Anderson energy eigenstates.
Consider representing the Hamiltonian as a matrix $\mathcal{M}$
with respect to the true MBL energy eigenbasis.
Off-diagonal matrix elements
couple together unperturbed states.
These couplings hybridize the unperturbed states,
forming corrections.
The couplings may be envisioned as
rearranging particles throughout a distance $L$.

MBL dynamics is unlikely to rearrange particles 
across considerable distances, due to localization.
Such a rearrangement is
encoded in an off-diagonal element 
$\mathcal{M}_{ij}$ of $\mathcal{M}$.
This $\mathcal{M}_{ij}$ must be small---suppressed exponentially in $L$.
$\mathcal{M}_{ij}$ also forces
the eigenstates' energies apart,
contributing to level repulsion~\cite[App.~F]{NYH_17_MBL}.
Hence the level-repulsion scale is suppressed exponentially in $L$:
\begin{align}
   \label{eq:J_L}
   \J_L  \sim  \HScale  e^{-L/\zeta }  \, ,
\end{align}
for some $\zeta \, .$
At infinite temperature, $\zeta$ must $<  \frac{1}{ \ln 2 }$ 
for the MBL phase to remain stable~\cite{mbmott}.
Substituting into Eq.~\eqref{eq:J_L} yields $\J_L  <  \frac{ \HScale }{ 2^L }$.
The level-repulsion scale is smaller than the average gap.

The size and significance of $\J_L$ depend on 
the size of $L$.
At the crossover distance $\xi$,
the repulsion $\J_L$ (between 
energy eigenfunctions localized a distance $\xi$ apart)
becomes comparable to
the average gap $\sim \frac{ \HScale }{ 2^{\xi} }$ between 
the eigenfunctions in the same length-${\xi}$ interval:
$\HScale  e^{ - {\xi} / \zeta }  
\sim  \frac{1}{e}  \,  \frac{\HScale}{ 2^{\xi} } \, .$
Solving for the crossover distance yields
\begin{align}
   \label{eq:Xi_MBL}
   \xi  \sim  \frac{1}{ \frac{1}{ \zeta }  -  \ln 2 } \, .
\end{align}
Relation~\eqref{eq:Xi_MBL} provides a definition
of the MBL localization length $\xi \, .$
[This $\xi$ differs from the Anderson localization length $\xi_\Anderson$,
Eq.~\eqref{eq:Xi}.]
Solving for $\zeta$ yields
\begin{align}
   \label{eq:Zeta_xi}
   \zeta  \sim  \frac{1}{ \frac{1}{ \xi }  +  \ln 2 }  \, .
\end{align}

The MBL Otto cycle involves two localization lengths
in the thermodynamic limit.
In the shallowly localized regime, 
$\xi  =  \xi_\Loc \, .$
Each eigenfunction has significant weight on $\xi_\Loc \sim 12$ sites,
in an illustrative example.
In the highly localized regime, $\xi  =  \xi_\VeryLoc \, .$
Eigenfunctions peak tightly, 
$\xi_\VeryLoc  \sim  1 \, .$

Suppose that the particles are rearranged across a large distance $L  \gg  \xi$.
The level-repulsion scale
\begin{align} 
   \label{eq:JFarExpn}  
   & \boxed{ \JFar 
   \sim  \HScale  e^{ - L / \xi }  \;  2^{ - L }  }  \, .
\end{align}
In the MBL engine's very localized regime,
wherein $\xi = \xi_\VeryLoc$,
if $L = \xi_\Loc$ equals one subengine's length,
$\JFar  =  \deltaMBL$.

Now, suppose that particles are rearranged across a short distance
$L  \lesssim  \xi$.
Random-matrix theory approximates this scenario reasonably
(while slightly overestimating the level repulsion).
We can approximate the repulsion between 
nearby-eigenfunction energies with
the average gap $\dAvg^\LL$ in the energy spectrum of a length-$L$ system:
\begin{align}
   \label{eq:JCloseExpn}  \boxed{
   \JClose   \sim    \dAvg^\LL
   \sim   \frac{ \HScale }{ 2^L }  } \, .
\end{align}

\subsection{Application of local level repulsion 
to the MBL Otto engine in the thermodynamic limit}
\label{section:Repuls_Eng}

Consider perturbing an MBL system locally.
In the Heisenberg picture, 
the perturbing operator spreads across a distance 
$L(t) \sim \zeta \ln (\HScale t)$~\cite{Nandkishore_15_MBL}.
(See also~\cite{Khemani_15_NPhys_Nonlocal}.)
The longer the time $t$ for which the perturbation lasts
the farther the influence spreads.

Consider tuning the Hamiltonian infinitely slowly,
to preclude diabatic transitions: $t \to \infty \, .$
Even if the Hamiltonian consists of spatially local terms,
the perturbation to each term
spreads across the lattice.
The global system cannot be subdivided into 
independent subengines.\footnote{
Granted, subengines are coupled together even if
the Hamiltonian is quenched infinitely quickly:
$H_\Sim(t)$ encodes a nearest-neighbor interaction, for example.
That interaction might be regarded as
coupling the edge of subengine $k$
with the edge of subengine $k + 1 \, .$
But subengines' edges may be regarded as ill-defined.
The sites definitively in subengine $k$, near subengine $k$'s center,
should not couple to the sites near subengine $\ell$'s center,
for any $\ell \neq k \, ,$
if the subengines are to function mostly independently.
Alternatively, one may separate subenegines with ``fallow'' buffer zones.}
The global system's average gap vanishes
in the thermodynamic limit: $\dAvg  \to  0 \, .$
The average gap sets the scale of 
one engine's per-cycle power, $\expval{ W_\tot }$.
Hence the per-cycle power seems to vanish in the thermodynamic limit:
$\expval{ W_\tot }  <  \dAvg  \sim  0 \, .$

But consider tuning the Hamiltonian at a finite speed $v$.
Dimensional analysis suggests that
the relevant time scale is $t  \sim  \frac{ \HScale }{ v } \, .$
Local perturbations affect a region of length
$\sim  L( \HScale / v ) \sim \zeta \ln (\HScale^2/v)$. 
On a length scale $L( \HScale / v )$,
global level correlations govern the engine's performance 
less than local level correlations do,
i.e., less than $R \LParen L( \HScale / v ), \omega \RParen$ does. 
This correlator registers level repulsion 
at a scale independent of $\Sites$. 
Finite-speed tuning enables local level repulsion
renders finite the average gap accessible to 
independent subengines,
the $\dAvg$ that would otherwise close in the thermodynamic limit.
Each mesoscale subengine therefore outputs $\expval{ W_\tot } > 0 \, .$

We can explain the gap's finiteness differently:
Suppose that the engine's state starts some trial 
with weight on the $j^\th$ energy level.
The eigenenergies wiggle up and down during stroke 1.
The $j^\th$ energy may approach the $(j - 1)^\th$.
Such close-together energies likely correspond to far-apart subengines.
If the levels narrowly avoided crossing,
particles would be rearranged across a large distance.
Particles must not be, as subengines must function independently.
Hence the engine must undergo a diabatic transition:
The engine's state must retain its configuration.
The engine must behave as though 
the approaching energy level did not exist.
Effectively removing the approaching level from 
available spectrum creates a gap in the spectrum.
One can create such a gap (promote such diabatic transitions)
by tuning the Hamiltonian at a finite $v$
(Suppl. Mat.~\ref{section:SpeedBounds}).

\section{Constraint 2 on cold thermalization:
Suppression of high-order-in-the-coupling energy exchanges}
\label{section:Tau_therm_virtual_app}


Section~\ref{section:Times_main} introduces 
the dominant mechanism by which 
the bath changes a subengine's energy.
The subengine energy change by an amount $\sim \Wb$,
at a rate $\sim \coupling$.
Higher-order processes can change the subengine energy
by amounts $> \Wb$ and operate at rates $O ( \coupling^\ell )$, 
wherein $\ell \geq 2$.
The subengine should thermalize across just small gaps.
Hence the rate-$\coupling^\ell$ processes must operate much more slowly
than the rate-$\coupling$ processes:
$\coupling$ must be small.
We describe the higher-order processes, 
upper-bound $\coupling$, and lower-bound $\tau_\therm$.

The higher-order processes can be understood as follows.
Let $H_\tot  =  H_\macro(\tau)  +  H_\bath  +  H_\inter$
denote the Hamiltonian that governs the engine-and-bath composite.
$H_\tot$ generates the time-evolution operator
$U(t)  :=  e^{ -i H_\tot t }$.
Consider Taylor-expanding $U(t)$.
The $\ell^\th$ term is suppressed in $\coupling^\ell$;
contains $2\ell$ fermion operators $c_j$ and $c_{j'}^\dag$; 
and contains $\ell$ boson operators $b_\omega$ and $b_{\omega'}^\dag$.
This term encodes 
the absorption, by the bath, of $\ell$ energy quanta of sizes $\leq \Wb$.
The subengine gives the bath a total amount $\sim \ell \Wb$ of heat.
The subengine should not lose so much heat.
Hence higher-order processes should occur much more slowly
than the rate-$\coupling$ processes:
\begin{align}
   \label{eq:High_ord_1}
   \tau_\HighOrd  \gg  \tau_\therm  \, .
\end{align}

Let us construct an expression for the left-hand side.
Which processes most urgently require suppressing?
Processes that change the subengine's energy by $\gtrsim \dAvg$.
Figure~\ref{fig:Compare_thermo_Otto_fig} illustrates why.
If the right-hand leg has length $\gtrsim \dAvg$,
the right-hand leg might be longer than the left-hand leg.
If the right-hand leg is longer,
the trial yields net negative work, $W_\tot < 0$.
The bath would absorb energy $\dAvg$ from a subengine
by absorbing $\sim \frac{ \dAvg }{ \Wb }$ packets
of energy $\sim \Wb$ each.
Hence the bath would appear to need to flip $\sim L  =  \frac{ \dAvg }{ \Wb }$ spins
to absorb energy $\sim \dAvg$.
(We switch from fermion language to spin language for convenience.)
However, the length-$L$ spin subchain
has a discrete effective energy spectrum.
The spectrum might lack a level associated with the amount
$\text{(initial energy)} - \dAvg$ of energy.
If so, the bath must flip more than $\frac{ \dAvg }{ \Wb }$ spins.
Local level correlations suggest that the bath must flip 
$\sim \xi_\Loc$ spins (Suppl. Mat.~\ref{section:ThermoLimitApp}).
Hence $L  =  \max \left\{  \frac{ \dAvg }{ \Wb } ,  \xi_\Loc  \right\}$.
Energy is rearranged across the distance $L$
at a rate $\propto  \coupling^L$.

Having described the undesirable system-bath interactions,
we will bound $\coupling$ via Fermi's Golden Rule, 
Eq.~\eqref{eq:FGR_Main}.
Let $\Gamma_{fi}  \sim  1 / \tau_\HighOrd$ now denote 
the rate at which 
order-$g^L$ interactions occur.
The bath DOS remains $\DOS_\bath ( E_{if} )  \sim  \frac{1}{ \Wb }$.
Let us estimate the matrix-element size $| \langle f | V | i \rangle |$.
The bath flips each spin at a rate $\coupling$
(modulo a contribution from the bath's DOS).
Flipping one spin costs an amount $\sim \HScale$ of energy, on average.
[$\HScale$ denotes the per-site energy density,
as illustrated in Eq.~\eqref{eq:SpinHam}.]
Hence $L$ spins are flipped at a rate 
$\sim  \HScale \left( \frac{ \coupling }{ \HScale } \right)^L$.
The initial $\HScale$ is included for dimensionality.
We substitute into Fermi's Golden Rule [Eq.~\eqref{eq:FGR_Main}],
then solve for the time:
\begin{align}
   \label{eq:High_ord_2}
   \tau_\HighOrd  
   \sim  \frac{ \Wb  \,  
   \HScale^{ 2 \left( L  -  1 \right) } }{
   \coupling^{ 2 L } }  \,
   \quad \text{wherein} \quad
   L  =  \max \left\{  \frac{ \dAvg }{ \Wb } ,  \:  \xi_\Loc  \right\}  \, .
\end{align}

We substitute from Eqs.~\eqref{eq:High_ord_2} and~\eqref{eq:Tau_therm_main}
into Ineq.~\eqref{eq:High_ord_1}.
Solving for the coupling yields
\begin{align}
   \label{eq:High_ord_3} 
   \coupling  \ll   \HScale^{ ( L - 2 ) / ( L - 1 ) }  \:  
   \deltaMBL^{ 1 / (L - 1) }  \, ,
   \quad \text{wherein} \quad
   L  =  \max \left\{  \frac{ \dAvg }{ \Wb } ,  \:  \xi_\Loc  \right\}  \, .
\end{align}
Substituting back into Eq.~\eqref{eq:Tau_therm_main} yields 
a second bound on $\tau_\therm$:
\begin{align}
   \label{eq:High_ord_4}  \boxed{
   \tau_\therm  \gg \Wb
   \left(  \frac{ \HScale }{ ( \deltaMBL )^L }  \right)^{2 / (L - 1) }  \, ,
   \quad \text{wherein} \quad
   L  =  \max \left\{  \frac{ \dAvg }{ \Wb } ,  \:   \xi_\Loc  \right\}   }  \, .
\end{align}

Let us express the bound in terms of localization lengths.
We set $\Wb  \sim  \frac{ \dAvg }{10}$, as usual.
We approximate $L \pm 1 \sim  L  \sim  \xi_\Loc$.
We substitute in for $\dAvg$ from Eq.~\eqref{eq:DAvg_Subeng}
and for $\deltaMBL$ from Eq.~\eqref{eq:deltaMBL}:
\begin{align}
   \label{eq:High_ord_5}  \boxed{
   \tau_\therm  \gg  \frac{1}{10  \HScale}  \:
   e^{2 \xi_\Loc / \xi_\VeryLoc }  \:
   2^{ 2 \xi_\Loc }  }  \, .
\end{align}
This inequality is looser than Ineq.~\eqref{eq:Lower_v_bd_main_HalfLZ}:
The no-higher-order-processes condition 
is less demanding than Markovianity.

\section{Optimization of the MBL Otto engine}
\label{section:Optimize}

Section~\ref{section:Thermo_limit_main} introduced
the macroscopic MBL engine.
This section provides background about identifies 
the engine's optimal parameter regime.
The Hamiltonian-tuning speed $v$ 
is bounded in Sec.~\ref{section:SpeedBounds};
the cycle time $\tau_\cycle$, in Sec.~\ref{section:Cycle_time_app};
and the cold-bath bandwidth $\Wb$, in Sec.~\ref{section:SpeedBounds}.

We focus on order-of-magnitude estimates 
and on exponential scaling behaviors.
$\Wb$ and $\betaH$ necessitate exceptions.
These quantities do not inherently scale in any particular ways,
unlike $\dAvg$ and $\deltaMBL$.
We choose $\Wb  \sim  \frac{1}{10}  \dAvg$,
in the spirit of Sec.~\ref{section:Order_main},
and $\betaH  \ll  \frac{1}{ \HScale \sqrt{ \Sites } }$,
in accordance with Suppl. Mat.~\ref{section:Small_params}.

The calculations in Suppl. Mat.~\ref{section:PowerApp}
concern one length-$\Sites$ mesoscale engine.
We translate the calculations into the thermodynamic limit approximately:
$\Sites$ is replaced with 
the subengine length $\xi_\Loc$. 
Energies such as $\expval{ W_\tot }$ are multiplied by
the number of subengines, $\propto \Sites_\macro$.
Granted, shallowly-localized-MBL energy spectra
do not obey $P_\ETH^\ParenE ( \delta )$.
This distribution can be replaced with, e.g.,
the Rosenzweig-Porter distribution~\cite{Altland_97_Perturbation}.
But $P_\ETH^\ParenE ( \delta )$ captures
the crucial physics, some level repulsion.


%
%
%
\subsection{Bounds on the Hamiltonian-tuning speed $v$}
\label{section:SpeedBounds}

As $H_\macro(t)$ is tuned,
the time-$t$ energy eigenstates become
linear combinations of the old eigenstates.
Levels narrowly avoid crossing.
The engine must have high probabilities of
(i) transitioning diabatically between energy eigenstates 
$\ket{ \psi_1 }$ and $\ket{ \psi_2 }$
coupled strongly by nonlocal operators,
so that subengines barely interact, and
(ii) transitionining adiabatically between $\ket{ \psi_1 }$ and $\ket{ \psi_2 }$
coupled strongly by local operators,
to approximate adiabatic ideal.
Requirement (i) lower-bounds $v$ (Sec.~\ref{section:Lower_v_bd}),
and (ii) upper-bounds $v$ (Sec.~\ref{section:Upper_v_bound}).

\subsubsection{Lower bound on $v$ from subengine independence}
\label{section:Lower_v_bd}


\begin{figure}[tb]
\centering
\includegraphics[width=.48\textwidth, clip=true]{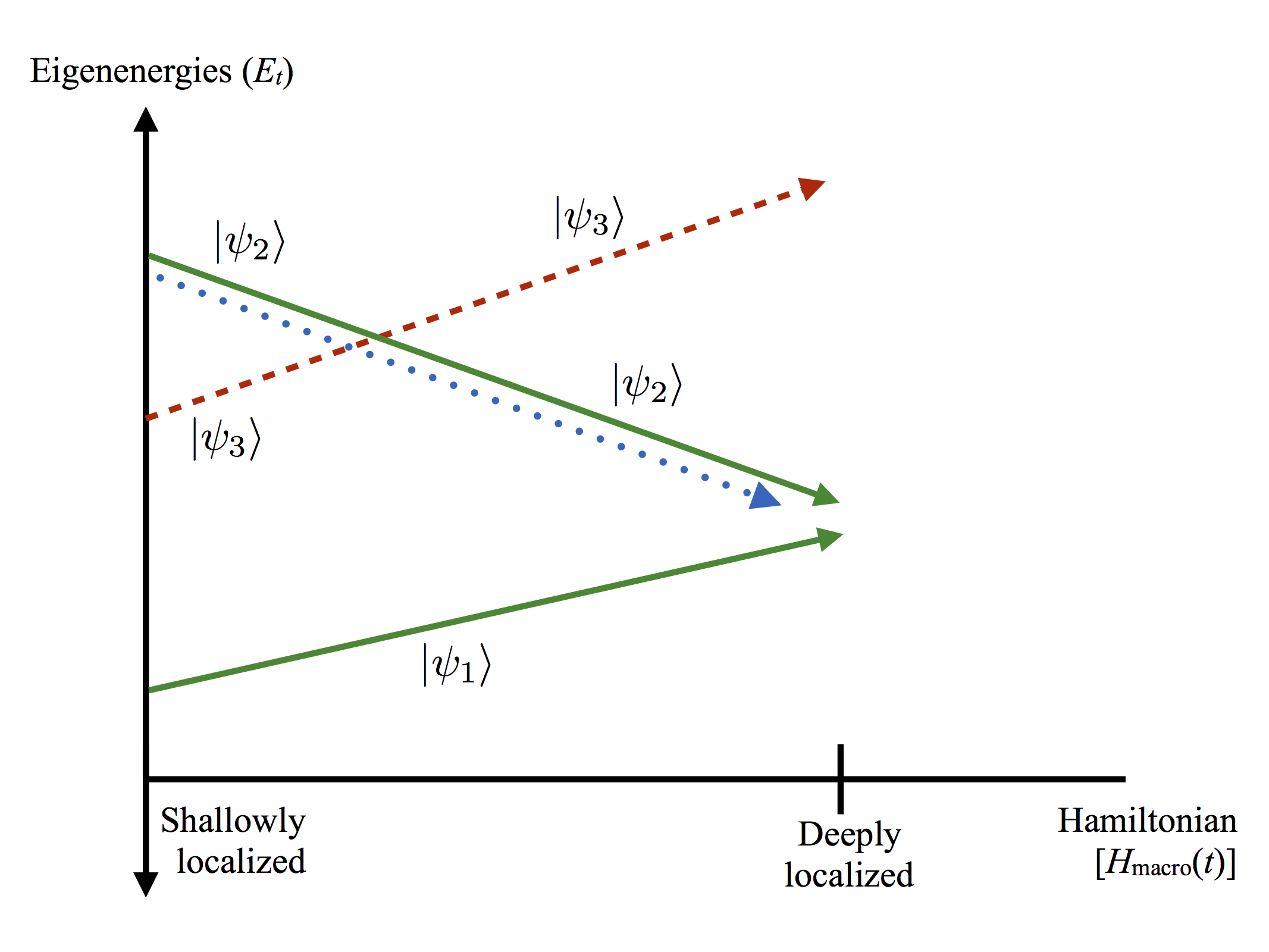}
\caption{\caphead{Desirable diabatic transition between energy eigenfunctions
localized in different subengines:}
The green, sloping solid lines represent elements 
$\ket{ \psi_1 }$ and $\ket{ \psi_2 }$
of the diabatic basis.
(The functional forms of the $\ket{ \psi_\ell }$'s remain constant:
Suppose that,  at some instant $t$, 
$\ket{ \psi_1 }$ equals some linear combination
$c_1 \ket{ \uparrow \ldots \uparrow } 
+ \ldots c_{2^\Sites} \ket{ \downarrow \ldots \downarrow }$
of tensor products of $\sigma_j^z$ eigenstates.
$\ket{ \psi_1 }$ equals that combination at all times.)
The dashed, red line represents an energy eigenstate $\ket{ \psi_3 }$
that turns into $\ket{ \psi_2 }$
via long-range rearrangements of much energy.
The eigenstates' energies change 
as the Hamiltonian is tuned.
The blue, dotted line represents 
a state desirable for the engine to occupy.}
\label{fig:Diabatic}
\end{figure}

Figure~\ref{fig:Diabatic} illustrates three energy eigenstates.
%
Let $L$ denote the scale of the distance over which
energy is rearranged
during a transition between $\ket{ \psi_2 }$ and $\ket{ \psi_3 }$.
If $L \geq 1.5 \xi_\Loc$, energy is transferred between subengines.\footnote{
One may separate neighboring subengines with ``fallow'' buffer zones.
Buffers would loosen the inequality to
$L \gg 1.5 \xi_\Loc + \text{(buffer length)}$.}
Subengines should evolve independently.
Hence the engine must have a low probability of transitioning 
from $\ket{ \psi_2 }$ to $\ket{ \psi_3 }$.
The crossing must have a high probability of being diabatic.

This demand can be rephrased in terms of work.
$\expval{ W_\tot }$ denotes the average work outputted
by one ideal subengine per cycle.
Let $\expval{ W_\adiab^\cost }$ denote the work cost
of undesirable adiabatic transitions 
incurred, on average, per subengine per cycle.
The cost must be much less than the extracted ideal:
\begin{align}
   \label{eq:Lower_v_demand}
   \expval{ W_\adiab^\cost }  \ll  \expval{ W_\tot } \, .
\end{align}
The right-hand side $\sim   \Wb$, to lowest order,
by Eq.~\eqref{eq:WTotApprox2_Main}.

Let us estimate the left-hand side.
We label as a ``close encounter''
an approach, of two levels,
that might result in an undesirable adiabatic transition.
The left-hand side of Ineq.~\eqref{eq:Lower_v_demand} has the form
\begin{align}
   \label{eq:W_adiab_cost_1}
   & \expval{ W_\adiab^\cost }  \approx
   \left(  \frac{ \text{Work cost} }{ \text{1 undesirable adiab. transition} }  \right)
   \left( \frac{ \text{Prob. of undesirable adiab. transition} }{ 
                    \text{1 close encounter} } \right)
   \\ &   \nonumber
   \times \left(  \frac{ \text{\# close encounters} }{ \text{1 tuning stroke} }  \right)
   \left(  \frac{ \text{Avg. \# strokes during which can lose work to adiab. transitions} }{ 
   \text{1 cycle} }  \right)  \, .
\end{align}
We estimate the factors individually.


We begin with the first factor, assisted by Fig.~\ref{fig:Num_undes_adiab}.
Suppose that the engine starts a tuning stroke
just above or below a working gap (on a green, solid line).
The engine might undesirably transition adiabatically
to a red, dashed line.
$\dAvgSub$ denotes the average gap 
in the part of the spectrum
accessible to an ideal mesoscale subengine
[Eq.~\eqref{eq:DAvg_Subeng}].
The red line likely originated, in the shallowly-MBL regime,
a distance $\sim (\const) \dAvgSub$ away.
Hence one undesirable adiabatic transition costs
$\sim  \dAvgSub$.

%
%
\begin{figure}[tb]
\centering
\includegraphics[width=.45\textwidth, clip=true]{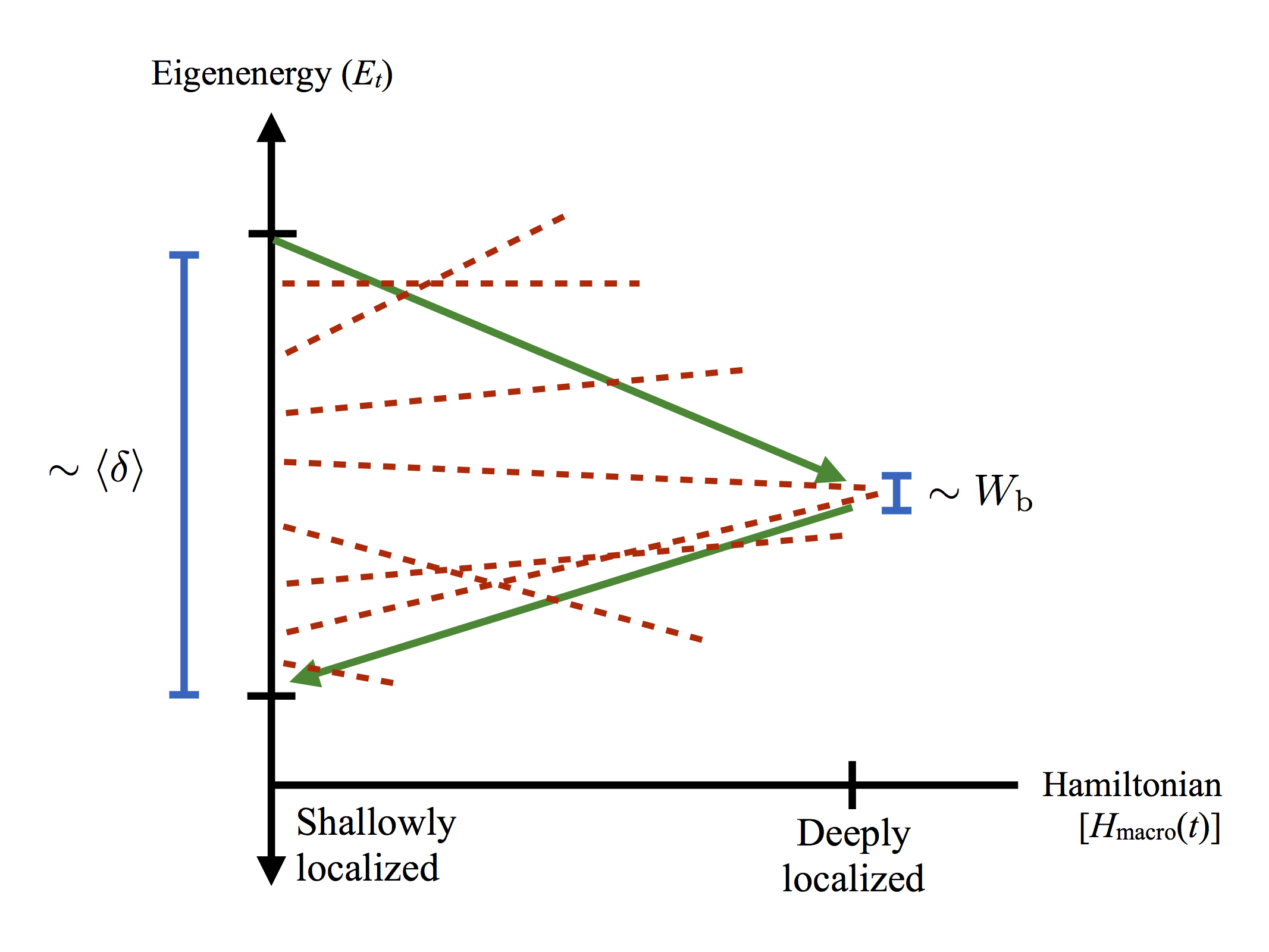}
\caption{\caphead{``Close encounters'' that might result in
undesirable adiabatic transitions:}
The sloping, green solid lines represent the top and bottom of a ``working gap.''
The red, dashed lines represent other energy levels.
Some cross (or anticross with) the working levels.
Each such ``close encounter'' should proceed diabatically.}
\label{fig:Num_undes_adiab}
\end{figure}

The Landau-Zener formula gives
the second factor in Eq.~\eqref{eq:W_adiab_cost_1}~\cite{Landau_Zener_Shevchenko_10}:
\begin{align}
   \label{eq:LZSxn}
   P_\adiab  
   =  1 - P_\diab
   =  1  -    e^{ - 2 \pi  \J^2  / v }
   \approx
2 \pi \frac{ \J^2 }{ v } \, .
\end{align}
$\J$ denotes the magnitude of the transition-matrix element
between the states.
$\J$ roughly equals the least size $\J_{L \sim 1.5 \xi_\Loc}$ 
reasonably attributable to any gap accessible to a subsystem
of length $L \sim 1.5 \xi_\Loc$ (Suppl. Mat.~\ref{section:Thermo_limit_main}).
The condition $L \sim 1.5 \xi_\Loc$ ensures that
the lefthand end of subengine $\ell$ fails to interact with
the middle of subengine $\ell \pm 1$ 
(Fig.~\ref{fig:Boxes_2}).\footnote{
If buffers separate the subengines, the condition becomes
\mbox{$L  >  1.5 \xi_\Loc$.} The lower bound on $v$ weakens.}
According to Eq.~\eqref{eq:JFarExpn},
\begin{align}
   \label{eq:LZ_J}  
   \boxed{ \J_{1.5 \xi_\Loc}
   \sim    \HScale  
   e^{ -1.5  \xi_\Loc  /  \xi(t) }  \:   2^{ - 1.5 \xi_\Loc }  }  \, .
\end{align}
$\xi( t )$ denotes the time-$t$ localization length.
Substituting into Eq.~\eqref{eq:LZSxn} yields
\begin{align}
   \label{eq:Prob_adiab}
   P_\adiab  
   \sim  \frac{ ( \J_{ 1.5  \xi_\Loc } )^2 }{ v }
   \sim  e^{ -3 \xi_\Loc  /  \xi(t) }  \;
  2^{ - 3 \xi_\Loc }  \;  \frac{ \HScale^2 }{ v } \, .  
\end{align}

%
%
\begin{figure}[tb]
\centering
\includegraphics[width=.45\textwidth, clip=true]{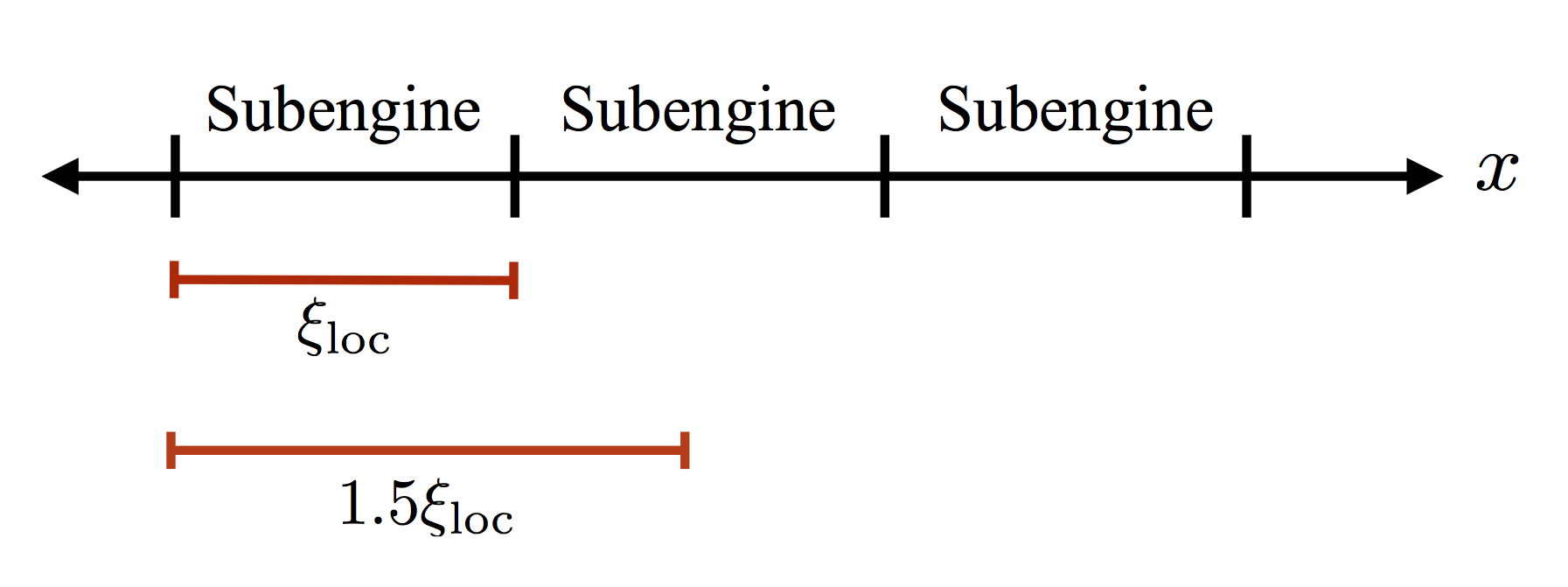}
\caption{\caphead{Condition forbidding subengines from interacting:}
The long black line represents the composite engine.
Each subengine has size $\xi_\Loc$,
the Hamiltonian's localization length in the shallow-localization regime.
Subengines must not interact: Consider particles on 
one subengine's left-hand side.
Those particles must not shift to the middle of any neighboring subengine,
across a distance $1.5 \xi_\Loc$.}
\label{fig:Boxes_2}
\end{figure}

To estimate the third factor in Eq.~\eqref{eq:W_adiab_cost_1},
we return to Fig.~\ref{fig:Num_undes_adiab}.
How many dashed, red lines cross the bottom green line? Roughly
\begin{align}
   \label{eq:Cross_prose}
   \frac{1}{2}  [ & \text{(\# red lines inside the working gap
   in the shallow-localization regime)}
   \nonumber \\ & 
   - \text{(\# red lines inside the working gap
   in the deep-localization regime)} ]  \, .
\end{align}
Let us estimate the first term.
When $H_\macro (t)$ is shallowly localized,
the working gap is of size
$\sim  \dAvgSub  \sim  \HScale  2^{ - \xi_\Loc }$
[Eq.~\eqref{eq:DAvg_Subeng}].
The DOS accessible to a size-$(1.5 \xi_\Loc)$ subsystem
is $\DOS_{(1.5 \xi_\Loc)} (E)  \sim  \frac{1}{ \dAvg^{ ( 1.5 \xi_\Loc ) } }
\sim  \frac{ 2^{1.5 \xi_\Loc } }{ \HScale }$.
Hence roughly
$\dAvgSub  \times  \DOS_{(1.5 \xi_\Loc)} (E)
   \sim  2^{ \xi_\Loc / 2 }$
red lines begin inside the working gap.

The second term in~\eqref{eq:Cross_prose} 
$\lesssim  \Wb  \times  \DOS_{(1.5 \xi_\Loc)} (E)$, 
as shown in Fig.~\ref{fig:Num_undes_adiab}.
By design, $\Wb  \ll  \dAvgSub$ (Suppl. Mat.~\ref{section:Small_params}).
Hence the second term in is much less than the first
and can be neglected.
Hence a subengine suffers about 
\begin{align}
   \label{eq:Num_close_encs}
   \frac{1}{2}  \,  \dAvgSub  \times  \DOS_{1.5 \xi_\Loc} (E)  
   \sim  \frac{ \dAvgSub }{ \dAvg^{ ( 1.5 \xi_\Loc ) } }
   \sim  2^{ \xi_\Loc / 2 }  
\end{align}
close encounters per stroke.

Finally, we estimate the last factor in Eq.~\eqref{eq:W_adiab_cost_1}.
Adiabatic transitions cost $\expval{ W_\adiab^\cost } > 0$
only during otherwise-successful trials---trials in which the subengine of interest 
would have outputted $W_\tot > 0$
in the absence of undesirable adiabatic transitions.
Why only otherwise-successful trials? 

Suppose, for simplicity, that $\TCold = 0$.
First, we argue that inter-subengine adiabatic transitions cost 
$\expval{ W } > 0$ during otherwise-successful trials.
Suppose that the engine starts a trial on 
the downward-sloping green line in Fig.~\ref{fig:Num_undes_adiab}.
During stroke 1, intersubengine adiabatic hops tend to lift the engine
to upward-sloping red, dashed lines.
Upward hops cost $W > 0$.
During stroke 3, the hops tend to lift the engine
to red lines that slope upward from right to left.
Such hops cost $W > 0$.
Hence cross-engine adiabatic hops during otherwise-successful trials
cost $\expval{ W } > 0$.

Now, we argue that intersubengine adiabatic hops
incurred during no-ops cost $\expval{ W } = 0$.
By ``no-op,'' we mean a trial during which, in the absence of undesirable hops,
the subengine of interest would output $W_\tot = 0$.
Suppose that the engine starts some trial on 
the bottom green line in Fig.~\ref{fig:Num_undes_adiab}.
The engine would slide up the bottom green line during stroke 1,
then slide downward during stroke 3: $W_\tot = 0$.
Interengine adiabatic hops during stroke 1 
tend to drop the engine to a red, dashed line, 
costing $W < 0$.
The hops during stroke 3 tend to raise the engine to a red, dashed line,
costing $W > 0$.
The two costs cancel each other, on average, by symmetry.
An analogous argument concern no-ops begun on a downward-sloping green line.
Hence interengine adiabatic hops during no-ops
cost $\expval{ W } = 0$.

We can now assemble the final factor in Eq.~\eqref{eq:W_adiab_cost_1}:
\begin{align}
   & \frac{ \text{Avg. \# strokes during which can lose work to adiab. transitions} }{
            \text{1 cycle} }  
   \nonumber \\ & \quad \approx  
   \left( \frac{ \text{2 strokes} }{ \text{1 otherwise successful trial} }  \right)
   \left( \frac{ \text{Prob. of success} }{ \text{1 hop-free trial} } \right) \\
   \label{eq:Num_strokes_ad}
   & \quad \approx  2  \:  \frac{ \Wb }{ \dAvgSub }
   \sim  \frac{ \Wb }{ \dAvgSub }  \, .
\end{align}
The final factor was estimated below Eq.~\eqref{eq:WTotApprox2_Main}.

We have estimated the factors in Eq.~\eqref{eq:W_adiab_cost_1}.
Substituting in from Eqs.~\eqref{eq:DAvg_Subeng},~\eqref{eq:Prob_adiab},~\eqref{eq:Num_close_encs}, and~\eqref{eq:Num_strokes_ad} yields
\begin{align} 
   \label{eq:W_adiab_cost_2}
   \expval{ W_\adiab^\cost}  
    \sim  \dAvgSub  \cdot  \frac{ ( \J_{ 1.5 \xi_\Loc } )^2 }{ v }
    \cdot  \frac{ \dAvgSub }{ \dAvg^{ ( 1.5 \xi_\Loc ) } }  
    \cdot  \frac{ \Wb }{ \dAvgSub } 
    =  \frac{ (  \J_{1.5 \xi_\Loc } )^2  \:  \Wb }{ v }  \:  
    \frac{ \dAvgSub }{ \dAvg^{ ( 1.5 \xi_\Loc ) } }  \, .
\end{align}
We substitute into Ineq.~\eqref{eq:Lower_v_demand} and solve for $v$:
\begin{align}
   \label{eq:LowerVBound}
   & \boxed{ v   \gg  (  \J_{1.5 \xi_\Loc } )^2   \:  
   \frac{ \dAvgSub }{ \dAvg^{ ( 1.5 \xi_\Loc ) } }
   \sim  e^{ - 3 \xi_\Loc  /  \xi(t) }  
   \;  2^{ - 2.5 \xi_\Loc }  \;  \HScale^2 }  \, .
\end{align}
The bound is twofold small in $\J_{1.5 \xi_\Loc}  \ll  \HScale$
and onefold large in $\frac{ \dAvgSub }{ \dAvg^{ ( 1.5 \xi_\Loc ) } }  > 1$.

Let us evaluate the bound in the very localized regime,
whose $\xi(t)  \sim  \xi_\VeryLoc$, 
and in the shallowly localized regime,
whose $\xi(t)  \sim  \xi_\Loc$.
If $\xi_\Loc  = 12$ and $\xi_\VeryLoc = 1$,
\begin{align}
   \label{eq:LowerVBound_num}  \boxed{
   v \gg \begin{cases}
      10^{ - 25 } \, \HScale ^2  \, ,  &  \text{very localized}  \\
      10^{ - 11 }  \:  \HScale^2  \, ,  &   \text{shallowly localized}
   \end{cases} }  \, .
\end{align}


%
%
%
\subsubsection{Upper bound on the Hamiltonian-tuning speed $v$}
\label{section:Upper_v_bound}

Undesirable diabatic transitions cost 
a total amount $\expval{ W_\diab^\cost }$ of work, on average
(Suppl. Mat.~\ref{section:App_Diab}).\footnote{
The average diabatic work cost was denoted 
by $\expval{ W_\diab^\cost }$ earlier.
The subscript is added here for emphasis and clarity.}
One ideal, adiabatic subengine outputs
$\expval{ W_\tot }  \sim  \Wb$ per trial, on average.
The requirement 
\begin{align}
   \label{eq:SmallCost}
   \expval{ W_\diab^\cost } \ll \expval{ W_\tot }
\end{align}
upper-bounds $v$. 
APT transitions dominate the left-hand side of Ineq.~\eqref{eq:SmallCost}
in the shallowly localized regime.
Fractional-Landau-Zener transitions dominate
in the very localized regime.

\textbf{Upper bound on $v$ in the shallowly localized regime:}
Substituting from Eq.~\eqref{eq:W_APT1_d}
into Ineq.~\eqref{eq:SmallCost} yields\footnote{
Equation~\eqref{eq:W_APT1_d} follows from
the $\sim \frac{1}{ \sqrt{ \HDim } }$ scaling of a matrix element
in the ETH phase.
The ETH phase features in the mesoscale-MBL-engine cycle
where shallowly localized MBL features in 
the thermodynamically-large-MBL-engine cycle.
In the MBL phase, the matrix element $\sim \frac{1}{ \HDim }$
(footnote~\ref{footnote:APT_HDim}).
Introducing the extra $\frac{1}{ \sqrt{ \HDim } }$ 
would loosen the bound~\eqref{eq:APT_v_bd_3} 
by a factor of $\sqrt{ \HDim }$.}
$\frac{ 1 }{ \sqrt{\Sites} }  \:
   \frac{ v^2  \betaH }{ \HScale  \dAvg }  \:
   \log \left( \frac{ \dAvg^2 }{ v }  \right)  \,
   e^{ - \Sites ( \betaH \HScale )^2 / 4 }
   \ll  \Wb$.
The $\frac{ 1 }{ \sqrt{\Sites} }$ and the log
contribute subdominant (nonexponential) factors.
The explicit exponential $\approx 1$,
since $\sqrt{ \Sites }  \:  \betaH \HScale  \ll 1$ by assumption:
\begin{align}
   \label{eq:APT_v_bd_2}
   \frac{ v^2  \betaH }{ \HScale  \dAvg }  \ll  \Wb
   \qquad \Rightarrow \qquad
   v  \ll  \sqrt{ \frac{ \dAvg  \Wb  \HScale }{ \betaH } }  \, .
\end{align}

We approximate $\Wb  \sim  \frac{ \dAvg }{ 10 }$.
Since $\sqrt{ \Sites }  \:  \betaH  \HScale  \ll 1$ by assumption,
$\frac{1}{ \betaH }  \gg  \sqrt{ \Sites }  \:  \HScale$.
We approximate $\frac{1}{ \betaH }  \sim  \Sites \HScale$.
We substitute into Ineq.~\eqref{eq:APT_v_bd_2} 
and ignore subdominant factors:
\begin{align}
   \label{eq:APT_v_bd_3}
   v  \ll  \dAvg   \HScale  \, .
\end{align}
This bound is looser than the small-parameter assumption
\begin{align}
   \label{eq:v_small_param}
   v  \ll  \dAvg^2  
   \sim  \frac{ \HScale^2 }{ \HDim^2 }
   \sim  2^{ - 2 \xi_\Loc }  \:  \HScale^2  \, .
\end{align}
in Suppl. Mat.~\ref{section:Small_params}.
APT transitions do not upper-bound the tuning speed painfully.

The upper bound~\eqref{eq:v_small_param} lies above
the lower bound~\eqref{eq:LowerVBound}.
The upper bound is suppressed only in $2^{ - 2 \xi_\Loc }$;
the lower bound, in 
$e^{ - 3 \xi_\Loc  /  \xi(t) }   \;  2^{ - 2.5 \xi_\Loc }$.
The upper bound $\sim 10^{ -7 }  \:  \HScale^2$, if $\xi_\Loc = 12$.
The lower bound $\sim 10^{ - 11}  \:  \HScale^2$ [Ineq.~\eqref{eq:LowerVBound_num}].
Therefore, the bounds are consistent with each other.

\textbf{Upper bound on $v$ in the deeply localized regime 
from fractional-LZ transitions:}
We have already derived the bound~\eqref{eq:Late_HalfLZ_v_bd}.
We assess the bound's size by expressing 
the right-hand side in terms of small parameters:
$v  \ll  \left(  \frac{ \Wb }{ \dAvg }  \right)^3  \frac{ \dAvg }{ \deltaMBL }  \:  \dAvg^2$.
The right-hand side is threefold suppressed in $\frac{ \Wb }{ \dAvg }  \ll 1$
and is large in $\frac{ \dAvg }{ \deltaMBL }  \gg  \frac{ \dAvg }{ \Wb }  \gg   1$.

We can express the bound in terms of localization lengths.
We substitute in for $\dAvgSub$ from Eq.~\eqref{eq:DAvg_Subeng}
and assume that
$\Wb  \sim  \frac{1}{10}  \dAvgSub$.
A $\deltaMBL$ expression follows from substituting
$\xi  =  \xi_\VeryLoc$ and $L  =  \xi_\Loc$ into Eq.~\eqref{eq:JFarExpn}:
\begin{align}
   \label{eq:deltaMBL}
   \deltaMBL  \sim  \HScale
   e^{ - \xi_\Loc / \xi_\VeryLoc }  \,
   2^{ - \xi_\Loc }  \, .
\end{align}
Inequality~\eqref{eq:Late_HalfLZ_v_bd} becomes
\begin{align}
   &  \label{eq:VUpperBd_2}  \quad  \boxed{
   v  \ll  \frac{ 1 }{ 10^3 }   \;
   e^{ \xi_\Loc / \xi_\VeryLoc }  \:  2^{ - 2 \xi_\Loc }
   \HScale^2  }  \, .
\end{align}

Let us check that this upper bound lies above the lower bound,
Ineq.~\eqref{eq:LowerVBound}.
The lower bound is suppressed in 
$e^{ - 3 \xi_\Loc / \xi_\VeryLoc }  \;  2^{ - 2.5 \xi_\Loc }$.
The upper bound is suppressed only in
$2^{ - 2 \xi_\Loc }$
and is large in 
$e^{ \xi_\Loc / \xi_\VeryLoc }$.
Hence $\text{(lower bound)}  \ll  \text{(upper bound)}$
by scaling. More concretely, substituting 
$\xi_\Loc = 12$ and $\xi_\VeryLoc = 1$ 
into Ineq.~\eqref{eq:Late_HalfLZ_v_bd} yields
\begin{align}
   \label{eq:VUpperBd_Num}  \quad  \boxed{
   v  \ll  10^{ - 5}  \:  \HScale^2 } \, .
\end{align}
This upper bound above below the lower bound,
$v \gg 10^{ - 25 }  \,  \HScale^2$
[Ineq.~\eqref{eq:LowerVBound_num}].
The bounds are consistent
and lie orders of magnitude apart.

%
%
%
\subsection{Time $\tau_\cycle$ required to implement a cycle}
\label{section:Cycle_time_app}

Different cycle segments must satisfy different bounds
on $v$ or on implementation time.
We (1) compare the bounds and 
(2) derive bounds on time from bounds on $v$:
\begin{enumerate}[leftmargin=*] 
   \item  \label{item:HalfLZ_bd}
   To suppress undesirable fractional-Landau-Zener transitions,
   the tuning speed must satisfy
   $v  \ll   \frac{ ( \Wb )^3 }{ \deltaMBL }
   \sim  \frac{ 1 }{ 10^3 }   \;
   e^{ \xi_\Loc / \xi_\VeryLoc }  \:  2^{ - 2 \xi_\Loc }   \HScale^2$
   in the deeply localized regime
   [Ineqs.~\eqref{eq:Late_HalfLZ_v_bd} and~\eqref{eq:VUpperBd_2}].
   This bound implies a bound on a time scale.
   Since $v  :=  \HScale  \left\lvert  \frac{ d \alpha_t }{ dt }  \right\rvert$,
   $\left\lvert \frac{ dt }{ d \alpha_t }  \right\rvert  =  \frac{ \HScale }{ v }$.
   Fear of fractional-LZ transitions limits $v$ during some part of stroke 3.
   Imagine that that part extends throughout stroke 3.
   $\alpha_t$ runs from 1 to 0, so stroke 3 lasts for a time
   \begin{align}
      \tau_{ \HalfLZ }  
      & =  \int_1^0   \frac{ dt }{ d \alpha_t }  \:  d \alpha_t
   =  \int_1^0  \left( -  \frac{ \HScale }{ v } \right)  d \alpha_t
   =  \frac{ \HScale }{ v }  \\
   & \label{eq:Tau_HalfLZ_1}
   \gg  \frac{ \deltaMBL  \HScale }{ ( \Wb )^3 }
   \sim  10^3  \:
   e^{ - \xi_\Loc / \xi_\VeryLoc }  \:  2^{ 2 \xi_\Loc } /  \HScale   \, .
   \end{align}
   If $\xi_\Loc  =  12$ and $\xi_\VeryLoc  =  1$,
      $\tau_{ \HalfLZ }  \sim  10^5  /  \HScale  \, .$

   %
   \item 
   Let $\tau_\APT$ denote the time for which
   fear of APT transitions governs $v$:
    $v \leq \dAvg^2   \sim  2^{ - 2 \xi_\Loc }  \:  \HScale^2$  
   [Ineq.~\eqref{eq:v_small_param}].
   $\tau_\APT$ includes stroke 1.
   Hence $\tau_\APT  \sim  \frac{\HScale}{ v }  
   \sim  2^{ 2 \xi_\Loc }  \,  \HScale
   \sim  10^7  /  \HScale$.
   The final expression follows from $\xi_\Loc = 12$.
   $\tau_\APT  \gg  \tau_{ \HalfLZ }$, so
   APT transitions bound the tuning time more stringently than
   fractional-LZ transitions do, if $\betaH > 0$.
   
   \item
   The engine thermalizes with the cold bath for a time
   $\tau_\therm  >  
   \frac{ \HScale^2 }{  \Wb  ( \deltaMBL )^2  }
   \sim  \frac{ 10 }{ \HScale }  \:
   e^{ 2 \xi_\Loc  / \xi_\VeryLoc }  \:  2^{ 3 \xi_\Loc }$
   [Ineqs.~\eqref{eq:Lower_v_bd_main_HalfLZ}].
   If $\xi_\Loc = 12$ and $\xi_\VeryLoc = 1$,
   $\tau_\therm  >  10^{22} / \HScale$.
   Cold thermalization lasts much longer than the Hamiltonian tunings,
   dominating the cycle time: $\tau_\cycle  \sim  \tau_\therm$.
   (Hot thermalization requires less time than cold,
   involving an ordinary bath bandwidth.)
   
\end{enumerate}

\subsection{Bounds on the cold-bath bandwidth $\Wb$}
\label{section:WbBounds}

$\Wb$ must be large enough
to couple nearby energies, deep in the MBL phase,
accessible to a subengine.
Hence $\Wb > \deltaMBL$,
estimated in Eq.~\eqref{eq:deltaMBL}.
$\Wb$ must be small enough to couple only levels
whose energies likely separate during stroke 3,
such that subengines output $\expval{ W_\tot } > 0$.
$\Wb$ must be less than the average level spacing $\dAvg$
accessible to a subengine [Eq.~\eqref{eq:DAvg_Subeng}].
Hence
\begin{align}
   \label{eq:WbBounds_App}
   \boxed{  \deltaMBL  <  \Wb  \ll  \dAvg  }
   \, ,  \quad  \text{or}  \quad
   \boxed{   \HScale  e^{ -  \xi_\Loc / \xi_\VeryLoc }  \;  2^{ -  \xi_\Loc }
   <  \Wb  \ll
    \frac{ \HScale }{ 2^{ \xi_\Loc} }  } \, .
\end{align}

\section{Numerical simulations of the MBL Otto engine}
\label{section:MBLNumApp}

We simulated one 12-site mesoscale engine at half-filling.
(We also studied other system sizes, to gauge finite-size effects.)
The random-field Heisenberg Hamiltonian~\eqref{eq:SpinHam}
governed the system.
We will drop the subscript from $H_\Sim(t)$.


Call the times at which the strokes end $t = \tau, \tau', \tau'',$ and $\tau'''$.
For each of $N_{\text{reals}} \sim 1,000$ disorder realizations, we computed the whole density matrix $\rho(t)$ at $t = 0, \tau, \tau', \tau'', \tau'''$. (See Suppl. Mat.~\ref{section:MBLNumApp:adiabatic} and~\ref{section:MBLNumApp:coldtherm} for an explanation of how.) 
The engine's time-$t$ internal energy is
$E(t) = \Tr\LParen  H(t)\rho(t)  \RParen \, .$
The quantities of interest are straightforwardly
\begin{align}
  & \langle W_1\rangle = E(0) - E(\tau)  \, ,  \quad
  \langle W_3\rangle = E(\tau''') - E(\tau'')  \, ,  \\
  & \langle Q_2\rangle = E(\tau'')  - E(\tau')  \, ,  \quad \text{and} \quad
  \langle Q_4\rangle = E(0)  - E(\tau''')  \, .
\end{align}
We disorder-average these quantities before dividing to compute the efficiency,
$\eta_\MBL  =  1 -  \frac{ \expval{ W_1 }  +  \expval{ W_3 } }{
\expval{ Q_4 } }  \, .$

\subsection{Scaling factor}
\label{section:MBLNumApp:scale}

We wish to keep the DOS constant through the cycle. 
To fix $\DOS(E)$, we rescale the Hamiltonian by a factor $Q \LParen h ( \alpha_t )  \RParen$.
We define $Q^2  \LParen h( \alpha_t )  \RParen$ as
the disorder average of the variance of the unrescaled DOS:
\begin{align}
   \label{eq:Q2_def}
  Q^2  \LParen h( \alpha_t )  \RParen 
  &:=   \Bigg\langle
  \Bigg(\frac 1 {\HDim} \sum_{j = 1}^{\HDim} E_j^2\Bigg)
  - \Bigg( \frac 1 {\HDim}  \sum_{j = 1}^\HDim E_j\Bigg)^2
  \Bigg\rangle_{\text{disorder}}
  =
  \Bigg\langle
  \frac{1}{ \HDim }  \Tr \LParen \tilde{H}^2 ( t ) \RParen 
  - \Bigg(\frac{1}{ \HDim }\Tr \LParen \tilde{H} ( t )  \RParen  \Bigg)^2
  \Bigg\rangle_{\text{disorder}}   \, .  \\
\end{align}
The $\tilde{H}(t)$ denotes an unrescaled variation on the random-field Heisenberg Hamiltonian $H(t)$ of Eq.~\eqref{eq:SpinHam}:
\begin{equation}
  \tilde{H}(t) :=  \HScale  \left[
      \sum_{j = 1}^{\Sites - 1}   
      \bm{\sigma}_j  \cdot  \bm{\sigma}_{j+1}
      +  h ( \alpha_t )  \sum_{j = 1}^\Sites
      h_j     \sigma_j^z  \right]  \, .
\end{equation}

To compute $Q^2  \LParen h ( \alpha_t )  \RParen$, 
we rewrite the unrescaled Hamiltonian as
\begin{equation}
  \tilde{H}(t) =   \HScale  \left[
  2\sum_{j = 1}^{\Sites - 1} \left(  \sigma^+_j\sigma^-_{j+1} +  \hc  \right)
  + \sum_{j = 1}^{\Sites - 1} \sigma^z_j\sigma^z_{j+1} 
  + h( \alpha_t ) \sum_{j = 1}^{\Sites} h_j \sigma^z_j  \right] \, .
\end{equation}
We assume that $\Sites$ is even, and we work at half-filling.
The $\frac{ \Sites }{ 2 }$-particle subspace has dimensionality
$\HDim = {\Sites \choose \Sites/2}  \, .$

Let us calculate some operator traces that we will invoke later.
Let $X := \prod_{j = 1}^\Sites\sigma^x$ denote the global spin-flip operator.
For any operator $A$ such that $X^\dag AX = -A$,
\begin{equation}
  \Tr (A) = \Tr \left( X^\dag A X \right) = - \Tr (A)  \, .
\end{equation}
We have used the evenness of $\Sites$,
which implies the invariance of the half-filling subspace under $X$. Also,
$\Tr (A) = 0$.
In particular,
$0 = \Tr ( \sigma^z_j ) = \Tr ( \sigma^z_{j}\sigma^z_{j'}\sigma^z_{j''} )$,
if $j \ne j' \ne j''$.

Traces of products of even numbers of $\sigma^z$ factors
require more thought:
\begin{align}
   \label{eq:TraceHelp1}
  \Tr ( \sigma^z_j\sigma^z_{j+1} ) &=
      (\text{\# states $j, j+1 = \uparrow\uparrow$}) 
      + (\text{\# states $j, j+1 = \downarrow\downarrow$}) 
      - 2(\text{\# states $j, j+1 = \uparrow\downarrow$}) 
      \notag\\
      &= {{\Sites - 2} \choose {\Sites/2 - 2}} + {{\Sites - 2} \choose {\Sites/2}} - 2 {{\Sites - 2} \choose {\Sites/2 - 1}}\notag\\
      &= - \HDim \frac 1 {\Sites - 1}\, .
\end{align}
Similarly,
\begin{align}
   \label{eq:TraceHelp2}
  \Tr \left( [\sigma^+_j\sigma^-_j] [\sigma^-_{j+1}\sigma^+_{j+1}]  \right) 
  &= \Tr \left( [\sigma^-_j\sigma^+_j] [\sigma^+_{j+1}\sigma^-_{j+1}]  \right) 
  \nonumber \\ &
  = (\text{\# states $j, j+1 = \uparrow\downarrow$}) 
  = {{\Sites - 2} \choose {\Sites/2 - 1}}  \\
  &= \HDim \frac{\Sites}{4(L-1)} \, ,
\end{align}
and 
\begin{align}
   \label{eq:TraceHelp3}
  \Tr \left( \sigma^z_j\sigma^z_{j+1}\sigma^z_{j'}\sigma^z_{j'+1} \right)
  &= (\text{\# states $j, j+1, j', j'+1 = \uparrow\uparrow\uparrow\uparrow$})
  + { 4 \choose 2}(\text{\# states $j, j+1, j', j'+1 = \uparrow\uparrow\downarrow\downarrow$})\notag\\
  &\quad+ (\text{\# states $j, j+1, j', j'+1 = \downarrow\downarrow\downarrow\downarrow$})\notag\\
  &\quad- {4 \choose 1}(\text{\# states $j, j+1, j', j'+1 = \uparrow\uparrow\uparrow\downarrow$})
  - {4 \choose 1}(\text{\# states $j, j+1, j', j'+1 = \uparrow\downarrow\downarrow\downarrow$})\notag\\
  &=  {{\Sites - 4} \choose {\Sites/2 - 4}}
  + 6 {{\Sites - 4} \choose {\Sites/2 - 2}}
  +   {{\Sites - 4} \choose {\Sites/2 }} 
  - 6 {{\Sites - 4} \choose {\Sites/2 - 3}}
  - 6 {{\Sites - 4} \choose {\Sites/2 - 1}}\notag\\
  &= \HDim \frac 3 {(\Sites - 1)(\Sites - 3)}  \, ,
\end{align}
wherein the first equality's combinatorial factors come from 
permutations on sites $j$, $j+1$, $j'$, and $j'+1$.

Assembling these pieces, we find
$\Tr \LParen \tilde{H} (t) \RParen 
  = \HScale  \sum_{j = 1}^{ \Sites -1} \Tr \left( \sigma^z_j\sigma^z_j \right) 
  = -  \HScale \HDim.$
Next, we compute $\Tr \LParen \tilde{H}^2(t) \RParen$:
\begin{align}
  \tilde{H} ^2(t)  &=  \HScale^2  \Bigg[
  4\sum_{j}^{\Sites-1} (\sigma^+_j\sigma^-_j)(\sigma^-_{j+1}\sigma^+_{j+1})
  + 4\sum_{j}^{\Sites-1} (\sigma^-_j\sigma^+_j)(\sigma^+_{j+1}\sigma^-_{j+1}) 
  + \sum_{j, j' = 1}^{\Sites-1}\sigma^z_j\sigma^z_{j+1}\sigma^z_{j'}\sigma^z_{j'+1}
  \nonumber \\ & \qquad \quad
  + h^2 ( \alpha_t )  \sum_{j = 1}^\Sites h_j^2
  + (\text{traceless terms})  \Bigg]  \\
  &=  \HScale^2  \Bigg[
  4\sum_{j}^{\Sites-1} (\sigma^+_j\sigma^-_j)(\sigma^-_{j+1}\sigma^+_{j+1})
  + 4\sum_{j}^{\Sites-1} (\sigma^-_j\sigma^+_j)(\sigma^+_{j+1}\sigma^-_{j+1})
  + \sum_{j = 1}^{\Sites-1}  \id
  + \sum_{j = 1}^{\Sites-2}\sigma^z_j\sigma^z_{j+2}
    \nonumber \\ & \qquad \quad
  + \sum_{j=1}^{\Sites-3}\sum_{j' = j+2}^{\Sites - 1}\sigma^z_j\sigma^z_{j+1}\sigma^z_{j'}\sigma^z_{j'+1} + h(\alpha_t)^2 ( \alpha_t )  \sum_{j = 1}^\Sites h_j^2
  + (\text{traceless terms})  \Bigg]  \, .
\end{align}
We take the trace, using Eqs.~\eqref{eq:TraceHelp1},~\eqref{eq:TraceHelp2}, and~\eqref{eq:TraceHelp3}:
\begin{equation}
  \Tr \LParen  \tilde{H}^2(t)  \RParen 
  = \HDim\Bigg[3\Sites - 1 + \frac{\Sites - 2}{\Sites - 1} + h^2 \sum_{j = 1}^\Sites h_j^2\Bigg] \, .
\end{equation}
We disorder-average by taking $h_j^2 \mapsto \int_0^1 dh_j h_j^2 = \frac 1 3$:
\begin{equation}
  \Big\langle\Tr (H^2(t))\Big\rangle_{\text{disorder}} = \HDim\Bigg[3\Sites - 1 + \frac{\Sites - 2}{\Sites - 1} + \Sites \frac{h^2}{3} \Bigg] \, .
\end{equation}
Substituting into Eq.~\eqref{eq:Q2_def},
we infer the rescaling factor's square:
\begin{equation}
  \label{num:rescale:disorder-averaged}
  Q^2  \LParen h( \alpha_t )  \RParen
  = 3  \Sites - 2 + \frac{\Sites - 2}{\Sites - 1} + \Sites \frac{h^2}{3}  \, .
\end{equation}

Our results are insensitive to the details of $Q$.
The width of the DOS in one disorder realization will differ from the disorder average~\eqref{num:rescale:disorder-averaged}. 
Moreover, that difference will vary as we tune $h(\alpha_t)$, 
because the disorder affects only one term.
The agreement between the analytics,
in which $\DOS (E)$ is assumed to remain constant in $t$,
and the numerics is therefore comforting:
The engine is robust against small variations in the rescaling.

\subsection{Representing states and Hamiltonians}
We structured our software to facilitate two possible extensions.
First, the Hamiltonian tuning may be generalized to arbitrary speeds.
Second, the cold bath might be modeled more realistically,
as coupling to the engine only locally.

We represent the state of one mesoscopic MBL Otto engine with a density matrix
$\rho \in \mathbb{C}^{\HDim\times\HDim} \, ,$
and the Hamiltonian with a matrix
$H \in \mathbb{C}^{\HDim\times\HDim} \, ,$
relative to the basis 
$\Set{ \ket{ s_1 } , \ldots,  \ket{ s_{\HDim }  } } 
=  \Set{  \ket{ \uparrow \ldots \uparrow } ,  \ldots, 
\ket{ \downarrow \ldots \downarrow } }$ 
of products of $\sigma^z$ eigenstates.
We track the whole density matrix, rather than just the energy-diagonal elements, with an eye toward the coherent superpositions that diabatic corrections create.
For an $\Sites$-site chain at half-filling,
$\HDim = {\Sites \choose \Sites/2} 
  \simeq \sqrt{\frac 2 {\pi \Sites}}  \:  2^{\Sites}  \, .$

\subsection{Strokes 1 and 3: Tuning}
\subsubsection{Adiabatic evolution}
\label{section:MBLNumApp:adiabatic}

The $(l, m )$ entry of the initial-state density matrix is
\begin{equation}
  \rho(0)_{lm} =   \bra{ s_l }  \frac 1 Z e^{-\betaH H(0)}   \ket{ s_m }
  = \frac 1 Z \sum_j e^{-\betaH E_j (0) }  
  \braket{s_l}{E_j(0)}\braket{E_j(0)}{s_m}  \, .
\end{equation}
The $j^\th$ eigenstate of $H(0)$, associated with energy $E_j(0)$,
is denoted by $\ket{E_j(0)}$.
We approximate the time evolution from $0$ to $\tau$ (during stroke 1) as adiabatic.
The evolution therefore does not move weight between levels:
\begin{equation}
  \rho(\tau)_{lm} = \frac{1}{ Z } 
  \sum_j e^{-\betaH E_j (0)} \braket{s_l}{E_j(\tau)}
  \braket{E_j(\tau)}{s_m}  \, .
\end{equation}
If we represented our density matrix relative to an instantaneous energy eigenbasis, simulating the time evolution would be trivial: We would reinterpret the diagonal matrix $\rho$ as being diagonal with the same elements in a new basis.
However, we wish to represent $\rho(t)$ relative to the $\sigma_j^z$ product basis.
This representation enhances the code's flexibility, facilitating future inclusion of diabatic evolutions and a more detailed model of cold thermalization.
To represent $\rho(t)$ relative to the $\sigma_j^z$ product basis, we note that 
\begin{equation}
  \rho(\tau)_{lm} = \sum_{j} \braket{s_l}{E_j(\tau)} \bra{E_j(0)}\rho(0)\ket{E_j(0)} \braket{E_j(\tau)}{s_m}
  = [U(\tau,0)\rho(0) U(\tau,0)^\dag]_{lm}  \, .
\end{equation}
We have defined a time-evolution matrix
$U(\tau, 0) \in \mathbf{C}^{\HDim\times\HDim}$ by
$U(\tau, 0)_{lm} = \sum_j\braket{s_l}{E_j(\tau)} \braket{E_j(0)}{s_m}  \, .$
This matrix is easily computed via
exact diagonalization of $H(0)$ and $H( \tau )$. 

We can compute the density matrix $\rho( \tau'' )$ at the end of stroke 3 
(the tuning from MBL to GOE) 
from the density matrix $\rho ( \tau' )$ at the end of stroke 2 
(the cold-bath thermalization) similarly:
$\rho(\tau'') = U(\tau'', \tau') \rho(\tau') U(\tau'', \tau')^\dag  \, .$
The time-evolution matrix $U(\tau'', \tau') \in \mathbf{C}^{\HDim\times\HDim}$ 
is given by
$U(\tau'', \tau')_{lm} = \sum_j\braket{s_l}{E_j(0)} \braket{E_j(\tau)}{s_m}  \, .$
[Recall that $H ( \tau'' ) = H(0)$ and $H ( \tau' ) = H ( \tau )$.]

\subsubsection{Diabatic (finite-time) evolution}
We simulate a stepwise tuning---that is, we take
\begin{equation}
  \alpha(t) = (\delta t) \lfloor v t/(\delta t) \rfloor  \;.
\end{equation}
To do this, we compute a time-evolution unitary for the whole stroke
by chaining together the unitaries for each timestep: so for stroke 1
\begin{equation}
  U(\tau,0; v,\delta t) =  e^{-iH(\tau - \delta t)\delta t} e^{-iH(\tau - 2\delta t)\delta t} \dots e^{-iH(0)\delta t}
\end{equation}
with the number of timesteps set by the speed.
We use timestep $\delta t = 0.405 \dAvg$, but our results are not sensitive to timestep.

In judging the effectiveness of the engine at finite tuning speed, we must estimate the level-repulsion scale $\delta_-$.
We do this by diagonalizing $10^6$ disorder realizations at the relevant $h_1 = 20,\ L = 8$, plotting a histogram of the gaps (Fig.~\ref{fig:estimate-delta-}, and visually estimating the point at which the distribution turns over. Our results are not sensitive to this value.

\begin{figure}[h]
  \centering
  \includegraphics[width=0.6\textwidth]{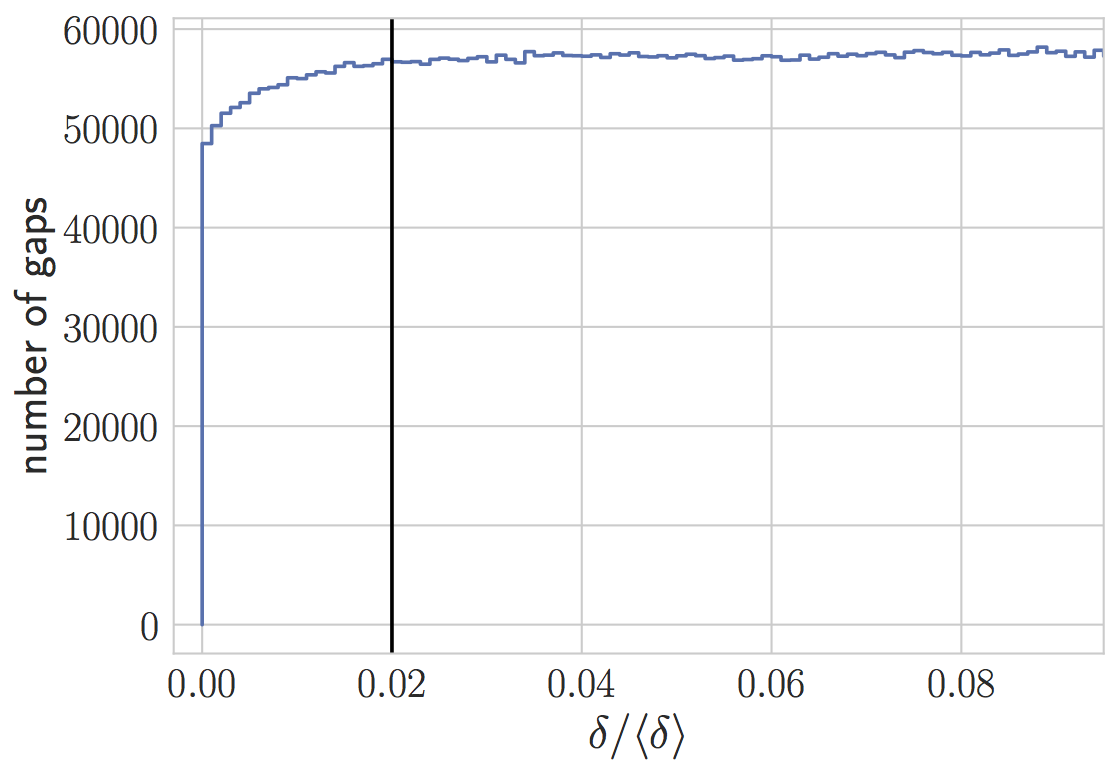}
  \caption{Level-spacing distribution for $10^6$ disorder realizations of the random-field Heisenberg model at field $h_1 = 20$ and system-size $L = 8$ (blue line), with estimate (vertical black line) for the level-repulsion parameter $\delta_-$.}
  \label{fig:estimate-delta-}
\end{figure} 
\subsection{Stroke 2: Thermalization with the cold bath}
\label{section:MBLNumApp:coldtherm}

During stroke 2, the system thermalizes with 
a bandwidth-$\Wb$ cold bath.
We make three assumptions.
First, the bandwidth cutoff is hard:
The bath can transfer only amounts $< \Wb$ of energy at a time. 
Therefore, the cold bath cannot move probability mass 
between adjacent levels separated by just one gap $\delta' > \Wb$. 
Second, the bath is Markovian.
Third, the system thermalizes for a long time.
The bath has time to move weight across 
sequences of small gaps $\delta'_j, \delta'_{j + 1}, \ldots < \Wb$.

We can implement thermalization as follows.
First, we identify sequences of levels connected by small gaps.
Second, we reapportion weight amongst the levels 
according to a Gibbs distribution.

%
%
\begin{figure}[tb]
\centering
\includegraphics[width=.4\textwidth, clip=true]{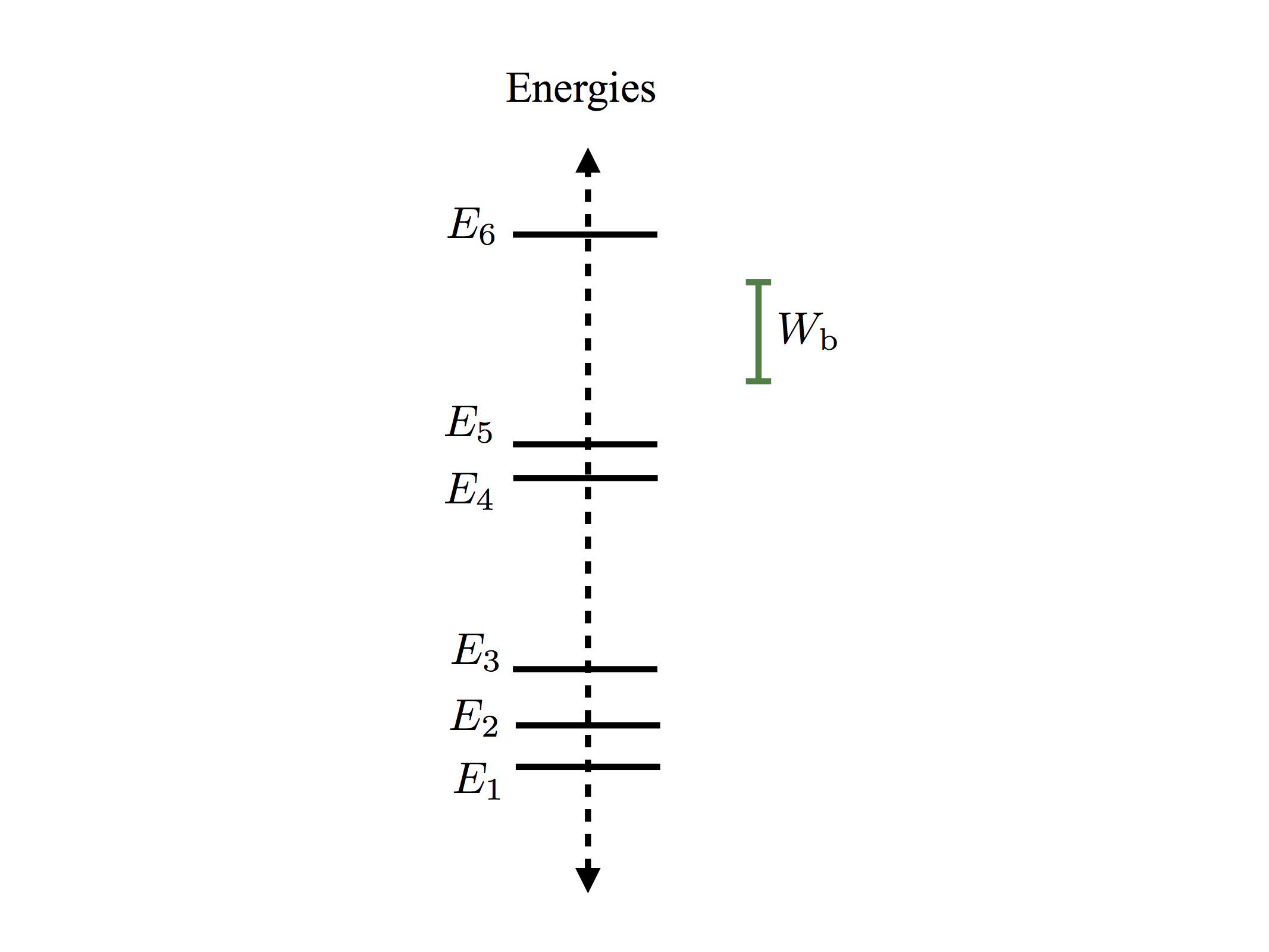}
\caption{\caphead{Energies of a cold-thermalized system:}
We illustrate our implementation of cold thermalization
with this example chain of six energies.
The cold bath has a bandwidth of size $\Wb$, depicted in green.}
\label{fig:Therm_ex}
\end{figure}

Suppose, for example, that the MBL Hamiltonian $H ( \tau )$
contains the following chain of six energies, $E_1, \dots, E_6$,
separated from its surrounding levels by large gaps (Fig.~\ref{fig:Therm_ex}):
\begin{align}
  (E_2 - E_1), (E_3 - E_2) < \Wb  \, ,  \quad
  (E_5 - E_4) < \Wb  \, ,  \quad  \text{and}   \quad
  (E_4 - E_3), (E_6 - E_5) > \Wb  \, . 
\end{align}
We suppress the time arguments to simplify notation.
Before thermalization, 
the density operator is diagonal with respect to the energy basis:
$\rho ( \tau ) = \sum_j \rho_j \ketbra{E_j}{E_j} \, .$
The weight on level $j$ is denoted by $\rho_j$.
Thermalization maps
\begin{align}
  \rho ( \tau ) \mapsto \rho ( \tau' ) & =
      \frac{\rho_1 + \rho_2 + \rho_3}
           {e^{-\betaC E_1} + e^{-\betaC E_2} + e^{-\betaC E_3}}
           \Big(     e^{-\betaC E_1} \ketbra{E_1}{E_1}
                 + e^{-\betaC E_2} \ketbra{E_2}{E_2}
                 + e^{-\betaC E_3} \ketbra{E_3}{E_3}  \Big)  \notag\\
      & \quad +
      \frac{\rho_4 + \rho_5}
           {e^{-\betaC E_4} + e^{-\betaC E_5}}
           \Big(     e^{-\betaC E_4} \ketbra{E_4}{E_4}
               + e^{-\betaC E_5} \ketbra{E_5}{E_5}  \Big) 
      + \rho_6 \ketbra{E_6}{E_6}  \, .
\end{align}

\section{Comparison with competitor Otto engines:
Details and extensions}
\label{section:CompetitorApp}


This section contains elaborates on the bandwidth engine 
(Sec.~\ref{section:Bandwidth_engine_app})
and on an MBL engine tuned between
equal-strength disorder realizations (Sec.~\ref{section:DisorderEngine}).
Section~\ref{section:DisorderEngine} compares with
an MBL engine thermalized with an ordinary-bandwidth cold bath.

\subsection{Details: Comparison with bandwidth engine}
\label{section:Bandwidth_engine_app}

Section~\ref{section:Bandwidth_engine_main} introduced
the accordion-like ``bandwidth engine.''
To work reasonably, we claimed, that engine 
must not undergo diabatic hops.
Can the bandwidth engine not withstand several hops---say, 
through $0.02 \HDim_\macro$ levels?

No, because the ground state pulls away from the rest of the spectrum 
as $\Sites_\macro$ grows.
Suppose, for simplicity, that $\TCold = 0$ and $\THot = \infty$.
The bandwidth engine starts stroke 1 
in $\rho(0)  =  \id / \HDim_\macro$.
Diabatic hops preserve $\rho(t)$ during stroke 1, on average:
The engine as likely hops upward as drops.
Cold thermalization drops the engine to the ground state
(plus an exponentially small dusting of higher-level states).
The ground-state energy is generically extensive.
Hence the engine absorbs 
$\expval{ Q_2 }_\macro  \sim  - \Sites_\macro$,
on average.
Suppose that, during stroke 3, the engine jumps up 
through 2\% of the levels.
The engine ends about two standard deviations 
below the spectrum's center,
with average energy $\sim \sqrt{\Sites_\macro }$.
While returning to $\THot = 0$ during the average stroke 4, 
the bandwidth engine absorbs
$\expval{ Q_4 }_\macro  \sim  \sqrt{ \Sites_\macro }$.
The average outputted work 
$\expval{ W_\tot }_\macro  =  \expval{Q_4}_\macro  +  \expval{ Q_2 }_\macro  
\sim  \sqrt{ \Sites_\macro }  -  \Sites_\macro$.
As $\Sites_\macro$ grows, 
$\expval{ W_\tot }_\macro$ shrinks, then goes negative.
A few diabatic jumps threaten the bandwidth engine's 
ability to output $\expval{ W_\tot } > 0$.

The bandwidth engine's $v$ must decline 
as $\Sites_\macro$ grows also because 
the typical whole-system gap 
$\dAvg_\macro \sim  \frac{ \HScale }{ \HDim_\macro }$ shrinks.
The smaller the gaps, the greater the likelihood
that a given $v$ induces hops.
As $\dAvg_\macro \to 0$, $v$ must $\to 0$.
The MBL Otto cycle proceeds more quickly,
due to subengines' parallelization 
(Suppl. Mat.~\ref{section:SpeedBounds}).

\subsection{Details: Comparison with MBL engine tuned
between same-strength disorder realizations}
\label{section:DisorderEngine}

This engine was introduced in Sec.~\ref{section:Equal_h_main}.
Here, we estimate the probabilities that $\Sys$ and $\tilde{\Sys}$
undergo worst-case trials.
$W_\tot < 0$ if an engine traverses, clockwise, a trapezoid
whose shorter vertical leg lies leftward of
its longer vertical leg (Fig.~\ref{fig:Compare_thermo_Otto_fig}).
$\Sys$ and $\tilde{S}$ have equal probabilities
of traversing trapezoids
whose right-hand legs are short.
$\tilde{S}$ has a greater probability than $\Sys$
of traversing a trapezoid
whose left-hand leg is short:
The left-hand $\tilde{S}$ leg represents a gap
in a Hamiltonian 
as localized as
the right-hand Hamiltonian.
The two $\tilde{S}$ Hamiltonians 
have the same gap statistics.
Hence $\tilde{S}$ has a nontrivial probability
of starting any given trial
atop a small gap $\Delta < \Wb$
that widens to $\Delta' \in (\Delta, \, \Wb)$.
$\tilde{S}$ has a higher probability of traversing
a worst-case trapezoid.

Suppose, for simplicity, that 
$T_\HTemp = \infty$ and $T_\CTemp = 0$.
The probability that any given $\Sys$ trial outputs $W_\tot < 0$ is
\begin{align}
   \label{eq:PWorstS0}  
   p_\worst  & \approx  
   \text{(Prob. that the left-hand gap $<$ the right-hand gap)} 
   \\ \nonumber & \quad \times
   \text{(Prob. that the right-hand gap is small enough 
   to be cold-thermalized)}   \\
   \label{eq:PWorstS}
   & \approx  
   \text{(Prob. that the left-hand gap $< \Wb$)}  \times
   \frac{ \Wb }{ \dAvg } \, .
\end{align}
The initial factor is modeled by the area of a region
under the $P_\ETH^\ParenE (\delta)$ curve.
The region stretches from $\delta = 0$ to $\delta = \Wb$.
We approximate the region as a triangle
of length $\Wb$ and height $\frac{\pi}{2}  \,  \frac{ \Wb }{ \dAvg^2 }  \,
e^{ - \frac{ \pi }{ 4 }  \,  \left( \Wb \right)^2 / \dAvg^2 }
\sim  \frac{ \Wb }{ \dAvg^2 }$,
[$\delta \approx \Wb$, 
Eq.~\eqref{eq:P_ETH_Main}, and
$\frac{ \Wb }{ \dAvg }  \ll  1$].
The triangle has an area of 
$\frac{1}{2}  \cdot   \Wb  \cdot
\frac{\pi}{2}  \,  \frac{ \Wb }{ \dAvg^2 }
\sim  \left(  \frac{ \Wb }{ \dAvg }  \right)^2$.
Substituting into Eq.~\eqref{eq:PWorstS} yields
\begin{align}
   \label{eq:PWorstS2}
   p_\worst  
   \sim   \left(  \frac{ \Wb }{ \dAvg }  \right)^3  \, .
\end{align}

Let $\tilde{p}_\worst$ denote
the probability that any given $\tilde{S}$ trial
outputs $W_\tot < 0$.
$\tilde{p}_\worst$ shares the form of Eq.~\eqref{eq:PWorstS}.
The initial factor approximates to
the area of a region under the $P_\MBL^\ParenE (\delta)$ curve.
The region extends from $\delta = 0$ to $\delta = \Wb$.
The region resembles a rectangle
of height $P_\MBL^\ParenE (0)  \approx \frac{1}{ \dAvg }$.
Combining the rectangle's area, $\frac{ \Wb }{ \dAvg }$,
with Eq.~\eqref{eq:PWorstS} yields
\begin{align}
   \label{eq:PWorstSPrime}
   \tilde{p}_\worst  &    \sim  
   \left(  \frac{ \Wb }{ \dAvg }  \right)^2 \, .
\end{align}
Since $\frac{ \Wb }{ \dAvg }  \ll  1$,
$p_\worst  \ll  \tilde{p}_\worst \, .$\footnote{
The discrepancy is exaggerated if
the exponent in Eq.~\eqref{eq:PWorstS2} rises,
if the left-hand $\Sys$ Hamiltonian
is modeled with a Gaussian ensemble other than the GOE.
The Gaussian unitary ensemble (GUE) 
contains an exponent of 4;
the Gaussian symplectic ensemble (GSE),
an exponent of 6.
Different ensembles model different symmetries.}

%
%
%
\subsection{Comparison with an MBL Otto engine
whose cold bath has an ordinary bandwidth}

Small-bandwidth baths have appeared elsewhere~\cite{Khaetskii_02_Electron,Gopalakrishnan_14_Mean,Parameswaran_17_Spin,Woods_15_Maximum,Perry_15_Sufficient}.
But suppose that realizing them poses challenges.
Let $\tilde{S}$ denote an ordinary-bandwidth MBL subengine;
and $\Sys$, the small-$\Wb$ subengine.
$\tilde{S}$ has a greater probability of traversing 
a quadrilateral, as in Fig.~\ref{fig:Compare_thermo_Otto_fig},
in any given trial.
But during the average quadrilateral traversal, 
$\Sys$ outputs more work than $\tilde{S}$.

An engine can output work
only during a quadrilateral traversal,
in the adiabatic approximation.
An engine fails to traverse a quadrilateral
by failing to cold-thermalize.
Cold thermalization fails if $\Wb < \delta'$,
the gap just below the level occupied by the engine
at the end of stroke 1.
The larger the $\Wb$, the more likely the engine thermalizes.

Let us estimate the engines'
quadrilateral-traversal probabilities.
Suppose, for simplicity, that $T_\HTemp = \infty$ and $T_\CTemp = 0$.
$\Sys$ has a probability $\approx \frac{ \Wb }{ \dAvg }$ 
of starting a trial
on one side of a gap $\Delta$
that shrinks to a $\Delta' < \Wb$.
The $T_\CTemp = 0$ bath drops the subengine's energy.
Hence $\Sys$'s probability of traversing a quadrilateral 
$\approx  \frac{ \Wb }{ \dAvg }$.
$\tilde{S}$ has a probability $\approx 1$ of starting a trial
just above a gap $\Delta$ that shrinks to a $\Delta' < \Wb$.
Hence $\tilde{S}$ much more likely traverses a quadrilateral
than $\Sys$ does: $1  \gg  \frac{ \Wb }{ \dAvg }$.

But $\tilde{S}$ outputs less work per average quadrilateral traversal.
Cold thermalization likely shifts $\Sys$ one level downward.
During strokes 1 and 3, $\Sys$'s energy likely declines
more than it rises (Fig.~\ref{fig:Compare_thermo_Otto_fig}).
$\Sys$ likely outputs $W_\tot > 0$.
In contrast, cold thermalization can shift $\tilde{S}$ to any energy level.
The energies less likely splay out during stroke 3.
If $\tilde{S}$ traverses a quadrilateral,
it outputs average work $\approx$
(average right-hand gap) - (average left-hand gap)
$\approx \dAvg  -  \dAvg  =  0$.

\endgroup

\renewcommand{\bibsection}{\section*{\refname}}
\putbib[MBL_bib,MBL_bib2] 
\end{bibunit}

\chapter{Appendices for ``Microcanonical and resource-theoretic derivations of the thermal state of a quantum system with noncommuting charges''}
\label{app:Noncommq}
\begin{bibunit}

\begingroup


\def\batt{ {\text{W}} }
\def\bath{ {\text{R}} }
\def\sys{ {\text{S}} }
\def\anc{ {\text{A}} }
\def\cat{ {\text{X}} }

\def\id{\mathbbm{1}}

\newcommand{\Wtran}[2]{W_{#1\rightarrow#2}}
\def\qtot{Q_{i_{\text{tot}}}}
\def\final{\rho_{\text{S}}'}  

\def\GTOlong{Non-Abelian Thermal Operations}
\def\GTOlongSing{Non-Abelian Thermal Operation}
\def\GTO{NATO}
\def\GGS{NATS}
\def\GGSlong{Non-Abelian Thermal State}



\def\toto{\mathop{\to}\limits^{TO}}
\def\pij{p_{i\to j}}
\def\dmin{D_{\min}}
\def\dmax{D_{\max}}
\def\<{\langle}
\def\>{\rangle}
\def\ot{\otimes}
\def\ermax{E_R^{\max}}
\def\esmax{E_S^{\max}}
\def\ecal{\mathcal{E}}
\def\hcal{\mathcal{H}}
\def\ecalr{{\ecal_R}}
\def\etaE{\eta_{E-E_S}^R}
\def\etaEprim{\eta'_{E-E_S}}
\def\ideta{\eta}
\def\rhor{\rho_R}
\def\rhos{\rho_S}
\def\rhors{\rho_{RS}}
\def\energy{E_0}

\def\rhoinr{\rho^{\text{in}}_R}
\def\rhoins{\rho^{\text{in}}_S}
\def\rhoinw{\rho^{\text{in}}_W}
\def\rhoinc{\rho^{\text{in}}_C}
\def\rhooutr{\rho^{\text{out}}_R}
\def\rhoouts{\rho^{\text{out}}_S}
\def\rhooutw{\rho^{\text{out}}_W}
\def\rhooutsc{\rho^{\text{out}}_{SC}}
\def\rhooutsw{\rho^{\text{out}}_{SW}}
\def\rhooutc{\rho^{\text{out}}_C}
\def\rhozeror{\rho^{0}_R}
\def\rhozeros{\rho^{0}_S}
\def\rhozeroc{\rho^{0}_C}
\def\sgibbs{\rho_S^\beta}
\def\rgibbs{\rho_R^\beta}
\def\cgibbs{\rho_C^\beta}
\def\ancgibbs{\rho_{\text{anc}}^\beta}
\def\rscgibbs{\rho_{\text{RSC}}^\beta}
\def\scgibbs{\rho_{\text{SC}}^\beta}

\def\twirl{\mathcal{T}}
\def\funny{{f-smooth}}
\def\funnyh{{f-smooth}}
\def\Funnyh{{F-smooth}}
\def\fh{\tilde H}
\def\Funny{{F-smooth}}
\def\frk{{\text{frk}}}
\def\ftail{{\text{ftail}}}
\def\tail{{\text{tail}}}
\def\rk{{\text{rk}}}
\def\rkd{{\text{rk}}^\delta}

\def\stipp{\bar{p}^{(\delta)}}
\def\stipq{\bar{q}^{(\delta)}}
\def\stipf{\bar{f}^{(\delta)}}
\def\flatp{\mathbf{p}^{(\delta)}}
\def\flatq{\mathbf{q}^{(\delta)}}
\def\flatf{\mathbf{f}^{(\delta)}}
\def\flattest{flattest}
\def\stippest{steeppest}

\newcommand{\be}{\begin{eqnarray} \begin{aligned}}
\newcommand{\ee}{\end{aligned} \end{eqnarray} }
\newcommand{\benn}{\begin{eqnarray*} \begin{aligned}}
\newcommand{\eenn}{\end{aligned} \end{eqnarray*} }

\newcommand{\ben}{\begin{eqnarray} \begin{aligned}}
\newcommand{\een}{\end{aligned} \end{eqnarray} }

\newcommand{\hmax}{\mathrm{H}_{\max}}
\newcommand{\hminp}{\mathrm{H}_{\min}}
\newcommand{\h}{\mathrm{H}}
\newcommand{\bc}{\begin{center}}
\newcommand{\ec}{\end{center}}
\newcommand{\half}{\frac{1}{2}}
\newcommand{\ran}{\rangle}
\newcommand{\lan}{\langle}
\newcommand{\im}{\mathbbmss{1}}
\newcommand{\idop}{\mathcal{I}}
\newcommand{\nullmat}{\mathbf{0}}
\newcommand{\fid}{\mathcal{F}}
\newcommand{\Cn}{\mathbb{C}}
\newcommand{\re}{\mathop{\mathbb{R}}\nolimits}
\newcommand{\natnum}{\mathop{\mathbb{N}}\nolimits}
\newcommand{\esp}{\mathbb{E}}
\newcommand{\cov}{\mathop{\mathrm{Cov}} \nolimits}
\newcommand{\var}{\mathop{\mathrm{Var}} \nolimits}
\newcommand{\mse}{\mathop{\mathrm{MSE}} \nolimits}
\newcommand{\rlprt}{\mathop{\mathrm{Re}} \nolimits}
\newcommand{\nth}{\mathrm{th}}
\newcommand{\imprt}{\mathop{\mathrm{Im}}\nolimits}
\newcommand{\tr}{\mathop{\mathsf{tr}}\nolimits}
\newcommand{\eval}[2]{\left. #1 \right \arrowvert _{#2}}
\newcommand{\norm}[1]{\left\| #1\right \|}
\newcommand{\smalloh}{\mathrm{o}}
\newcommand{\bigoh}{\mathrm{O}}
\newcommand{\pr}{\prime}
\newcommand{\di}{\text{d}}
\newcommand{\trans}[1]{#1^{\top}}
\newcommand{\iu}{i}
\newcommand{\vc}[1]{#1}				
\newcommand{\abs}[1]{\left|#1 \right|}				
\newcommand{\der}[1]{\frac{\partial}{\partial #1}}
\newcommand{\e}{\mathrm{e}}
\newcommand{\upa}{\uparrow}
%
\newcommand{\beq}{\begin{eqnarray} \begin{aligned}}
\newcommand{\eeq}{\end{aligned} \end{eqnarray} }
\newcommand{\bea}{\begin{array}}
\newcommand{\eea}{\end{array}}

\newcommand{\bee}{\begin{enumerate}}
\newcommand{\eee}{\end{enumerate}}
\newcommand{\bei}{\begin{itemize}}
\newcommand{\eei}{\end{itemize}}

\newcommand{\dna}{\downarrow}
\newcommand{\ra}{\rightarrow}
\newcommand{\hil}{\mathcal{H}}
\newcommand{\kil}{\mathcal{K}}
\newcommand{\allrhos}{\mathcal{M}}
\newcommand{\allrhosin}[1]{\mathcal{S}(#1)}
\newcommand{\pop}[1]{\mathcal{M}_{+}(#1)}
\newcommand{\cpmap}{\mathcal{E}}
\newcommand{\uop}{\mathcal{U}}
\newcommand{\su}{\mathfrak{su}}
\newcommand{\finproof}{$\Box$}
\newcommand{\textmath}{}
\newcommand{\wt}{\widetilde}
\newcommand{\wavyline}[3]{
\multiput(#1,#2)(4,0){#3}{
\qbezier(0,0)(1,1)(2,0)
\qbezier(2,0)(3,-1)(4,0)}}
\newcommand{\spann}{\mathop{\mathrm{span}}\nolimits}  

\newcommand{\wigglyline}[3]{
\multiput(#1,#2)(1,0){#3}{
\qbezier(0,0)(0.25,0.25)(0.5,0)
\qbezier(0.5,0)(0.75,-0.25)(1,0)}}

\newcommand{\allsigmasin}[1]{\tilde{\mathcal{S}}(#1)}
\newcommand{\instr}{\mathcal{N}}
\newcommand{\dfdas}{\stackrel{\textrm{\tiny def}}{=}}
\newcommand{\argmax}{\mathop{\mathrm{argmax}}\nolimits}
\newcommand{\bin}{\textrm{Bin}}
\newcommand{\topr}{\stackrel{P}{\to}}
\newcommand{\todst}{\stackrel{D}{\to}}
\newcommand{\lil}{\mathcal{L}}
\newcommand{\myabstract}[1]{\begin{quote}{\small  #1 }\end{quote}}
\newcommand{\myacknowledgments}{\begin{center}{\bf Acknowledgments}\end{center}\par}
\newcommand{\otherfid}{\mathsf{F}}
\newcommand{\mypacs}[1]{\begin{flushleft} {\small PACS numbers: #1}\end{flushleft}}
\newcommand{\basedon}[1]{\begin{flushleft} {\small This chapter is based on #1.}\end{flushleft}}
\newcommand{\onbased}[1]{\begin{flushleft} {\small  #1 is based on this chapter.}\end{flushleft}}
\newcommand{\ds}{\displaystyle}
\newcommand{\co}{\textrm{co}}
\newcommand{\eg}{\epsilon}
\newcommand{\zg}{\theta}
\newcommand{\ag}{\alpha}
\newcommand{\bg}{\beta}
\newcommand{\dg}{\delta}
\newcommand{\mc}{\mathcal}

\def\Real{\mathbb{R}}
\def\Complex{\mathbb{C}}
\def\Natural{\mathbb{N}}
\def\id{\mathbb{I}}

\def\01{\{0,1\}}
\newcommand{\ceil}[1]{\lceil{#1}\rceil}
\newcommand{\floor}[1]{\lfloor{#1}\rfloor}
\newcommand{\eps}{\varepsilon}
\newcommand{\outp}[2]{|#1\rangle\langle#2|}
\newcommand{\proj}[1]{|#1\rangle\langle#1|}

\newcommand{\inp}[2]{\langle{#1}|{#2}\rangle} 

\newcommand{\rank}{\operatorname{rank}}

\newcommand{\mX}{\mathcal{X}}
\newcommand{\mY}{\mathcal{Y}}
\newcommand{\mB}{\mathcal{B}}

\newenvironment{sdp}[2]{
\smallskip
\begin{center}
\begin{tabular}{ll}
#1 & #2\\
subject to
}
{
\end{tabular}
\end{center}
\smallskip
}

\newcommand{\ens}{\mathcal{E}}
\newcommand{\Y}{|\mY|}
\newcommand{\psucRAC}{P^{\text{uncert}}}
\newcommand{\psuc}{P^{\text{succ}}}
\newcommand{\pgame}{{P^{\text{game}}}}
\newcommand{\pgameMAX}{{P^{\text{game}}_{\rm max}}}
\newcommand{\cE}{\mathcal{E}}
\newcommand{\cU}{\mathcal{U}}
\newcommand{\sx}{\mathcal{S}_x}
\newcommand{\hmin}{\ensuremath{H}_{\infty}}

\newcommand{\stateSet}{\mathscr{S}}

\newcommand{\secchsh}{\ref{sec:chsh}}
\newcommand{\nn}[1]{\textcolor{magenta}{nelly: #1}}
\newcommand{\mcomment}[1]{{\sf [#1]}\marginpar[\hfill !!!]{!!!}}
\newcommand{\steph}[1]{{\textcolor{blue}{steph: #1}}}
\newcommand{\jono}[1]{\textcolor{blue}{#1}}
\newcommand{\mh}[1]{\textcolor{red}{mh: #1}}
\newcommand{\pc}[1]{\textcolor{green}{\tt Piotr: #1}}
\newcommand{\fer}[1]{\textcolor{brown}{\tt Fernando: #1}}
\newcommand{\xstr}{{\mathbf{x}_{s,a}}}
\newcommand{\str}{{\mathbf{x}}}
\newcommand{\vstr}{x}
\newcommand{\vxstr}{{x_{s,a}}}
\newcommand{\xstrsub}{s,a}
\newcommand{\probun}[1]{P^{\text{un}}(#1)}
\newcommand{\probsteer}[1]{P^{\text{st}}(#1)}

\newcommand{\prob}[1]{p(#1)}

\newcommand{\good}{\mathcal{G}}
\newcommand{\bstate}[1]{\sigma_{\xstrsub}^{#1}}
\newcommand{\xor}{XOR}
\newcommand{\avgb}{\rho}
\newcommand{\icaus}{I_{caus}}
\newcommand{\irac}{I_{rac}}
\newcommand{\unop}{Q}
\newcommand{\supl}{Appendix}
\newcommand{\stateset}{\mathcal{F}_{s}}
\newcommand{\badproj}{\Pi_{bad}}

\newcommand{\setA}{\mathcal{A}}
\newcommand{\setB}{\mathcal{B}}
\newcommand{\setS}{\mathcal{S}}
\newcommand{\setT}{\mathcal{T}}
\newcommand{\secxor}{I}

\newcommand{\dist}[2]{\mathcal{D}(#1\rangle#2)}

\def\<{\langle}
\def\>{\rangle}
\def\ot{\otimes}
\def\ermax{E_R^{\max}}
\def\esmax{E_S^{\max}}
\def\ecal{\mathcal{E}}
\def\hcal{\mathcal{H}}
\def\ecalr{{\ecal_R}}
\def\etaE{\eta_{E-E_{\text{S}}}}
\def\rhor{\rho_{\text{R}}}  
\def\rhos{\rho_{\text{S}}}  
\def\nats{\gamma_{\mathbf{v}}}
\def\gibbs{\rho_{\text{R}}^\beta}  
\def\gibbsS{\gamma_{\text{S}}}  
\def\gibbsBatt{\gamma_{ \batt }}  
\def\gibbsSBatt{\gamma_{\sys{\batt}}}  
\newcommand\gibbsParam[1]{\gamma_{#1}}
\def\final{\rho_{\text{S}}'}  
\def\initialBatt{\rho_{ \batt }}  
\def\finalBatt{{\rho'_{ \batt }}}  
\def\ergotropymin{F_\epsilon^{min}}
\def\ergotropymax{F_\epsilon^{max}}
\def\gmin{G_\epsilon^{min}}
\def\gmax{G_\epsilon^{max}}
\def\supthermomaj{C}
\def\supdistillation{D}
\def\supformation{D}
\def\supparadigm{A}
\def\supprocesses{E}

\def\rhodec{\omega}
\def\s{\,\,\,\,}
\def\dmin{D_{min}}
\def\dmax{D_{max}}
\def\dmine{D^\epsilon_{min}}
\def\dmaxe{D^{\epsilon}_{max}}
\def\wit{\psi_W}
\def\witi{0}
\def\initial{\rho_\sys}
\def\Hw{\hat{W}}
\def\Hin{H}
\def\Hout{H'}
\def\Zout{Z'}
\def\mainsection{Main Section}
\newcommand{\alfree}[1]{F_\alpha(#1,\gibbs)}
\newcommand{\qalfree}{{\hat F}}
\newcommand{\qalfreesimple}{{\tilde F}}
\def\fmin{F_{\text{min}}}
\def\fmax{F_{\text{max}}}
\def\tauout{\tau'}
\def\ep{\epsilon}
\newcommand{\sgn}{\operatorname{sgn}}
\def\qrenyi{S}
\def\qrenyisimple{\tilde S}
\def\trumpd{\text{D_{work}}}
\def\catalyst{\sigma}
\def\expbound{\exp(-\Omega(\sqrt{\log(N)}))} 

\def\gibbsin{\rho_\beta^{(0)}}
\def\gibbsout{\rho_\beta^{(1)}}

\def\genFs{generalized free energies}
\def\workf{\mathcal{W}}

\newcommand{\newreptheorem}[2]{%
\newenvironment{rep#1}[1]{%
 \def\rep@title{#2 \ref{##1} (restatement)}%
 \begin{rep@theorem}}%
 {\end{rep@theorem}}}
\makeatother

\newreptheorem{thm}{Theorem}
\newreptheorem{lem}{Lemma}

%
%
\def\topfraction{0.5}
\def\bottomfraction{0}

%
%
\let\oldparagraph\paragraph
\def\paragraph#1{%
  \smallskip%
  \par\noindent{\textbf{#1}}\quad
}

%
%
\let\tr\Tr


\def\topfraction{1}

The microcanonical, dynamical, and resource-theory arguments are detailed below.

%
%
%
%
\section{Microcanonical derivation of the \GGS{}'s form}
  \label{section:SI_Micro}

Upon describing the set-up, we will define an approximate microcanonical subspace $\mathcal{M}$.
Normalizing the projector onto $\mathcal{M}$ yields
an approximate microcanonical state $\Omega$.
Tracing out most of the system from $\Omega$ 
leads, on average, to a state close to the \GGSlong{} $\gamma_{ \mathbf{v} }$.
Finally, we derive conditions under which $\mathcal{M}$ exists.

%
%
%
%
\paragraph{Set-up:} 
Consider a system $\mathcal{S}$ associated with 
a Hilbert space $\mathcal{H}$ of dimension $d := \dim (\mathcal{H})$.
Let $H \equiv Q_0$ denote the Hamiltonian.
We call observables denoted by 
$Q_1, \ldots, Q_c$ ``charges.''
Without loss of generality, we assume that the $Q_j$'s form a linearly independent set.
The $Q_j$'s do not necessarily commute with each other.
They commute with the Hamiltonian
if they satisfy a conservation law,
\begin{align}
   [H, Q_j]  =  0
   \; \:  \forall j = 1, \ldots, c.
\end{align}
This conservation is relevant to dynamical evolution,
during which the \GGS\ may arise as the equilibrium state.
However, our microcanonical derivation does not rely on conservation.

%
%
%
%
\paragraph{Bath, blocks, and approximations to charges:}
Consider many copies $n$ of the system $\mathcal{S}$. 
Following Ogata~\cite{Ogata11}, we consider 
an average $\tilde{Q}_j$, over the $n$ copies, of each charge $Q_j$ (Fig.~\ref{fig:OgataProofSetup} of the main text):
\begin{align}
   \tilde{Q}_j  :=
   \frac{1}{n}  \sum_{ \ell = 0 }^{ n - 1 }
   \id^{ \otimes \ell }  \otimes  Q_j  \otimes  \id^{ \otimes (n - 1 - \ell) }.
\end{align}


In the large-$n$ limit, the averages $\tilde{Q}_j$
are approximated by observables $\tilde{Y}_j$ that commute~\cite[Theorem 1.1]{Ogata11}:
\begin{align}
   & \| \tilde{Q}_j  -  \tilde{Y}_j  \|_\infty  
   \leq  \epsilon_{\mathrm{O}}(n)  \to  0,
   \text{ and } \\
   & [ \tilde{Y}_j,  \tilde{Y}_k ]  =  0
   \; \; \forall j, k  = 0, \ldots, c.
\end{align}
The $\tilde{Y}_j$'s are defined on $\mathcal{H}^{ \otimes n}$, 
$\| \cdot \|_\infty$ denotes the operator norm, and $\epsilon_{\mathrm{O}}(n)$ denotes a
  function that approaches zero as $n\to\infty$.

Consider $m$ blocks of $n$ copies of $\mathcal{S}$,
i.e., $N = nm$ copies of $\mathcal{S}$.
We can view one copy as the system of interest and 
$N - 1$ copies as a bath.
Consider the average, over $N$ copies, of a charge $Q_j$:
\begin{align}   
\label{eq:Ogata1}
   \bar{Q}_j :=
   \frac{1}{N}  \sum_{ \ell = 0}^{ N - 1}  
      \id^{\otimes \ell }  \otimes  Q_j  \otimes  \id^{ \otimes (N - 1 - \ell) }.
\end{align}
This $\bar{Q}_j$ equals also
the average, over $m$ blocks, of the block average $\tilde{Q}_j$:
\begin{align}
\label{eq:Ogata2}
   \bar{Q}_j   = 
   \frac{1}{m}  \sum_{ \lambda = 0}^{m - 1}
   \id^{ \otimes \lambda n }  \otimes  \tilde{Q}_j  \otimes  \id^{\otimes [ N - n (\lambda + 1) ] }.
\end{align}

Let us construct observables $\bar{Y}_j$ that approximate the $\bar{Q}_j$'s and that commute:
$[ \bar{Y}_j,  \bar{Y}_k ]  =  0$, and
$\| \bar{Q}_j  -  \bar{Y}_j  \|_\infty  \leq  \epsilon$
for all $m$.
Since $\tilde{Y}_j$ approximates the $\tilde{Q}_j$ in Eq.~\eqref{eq:Ogata2},
we may take
\begin{align}
   \bar{Y}_j  =
   \frac{1}{m}  \sum_{ \lambda = 0}^{m - 1}
   \id^{ \otimes \lambda n }  \otimes  \tilde{Y}_j  \otimes  \id^{\otimes [ N - n (\lambda + 1) ] }.
\end{align}

%
%
%
%
%
%
%
%

\paragraph{Approximate microcanonical subspace:}
Recall the textbook derivation
of the form of the thermal state
of a system that exchanges commuting charges with a bath.
The composite system's state occupies a microcanonical subspace.
In every state in the subspace, every whole-system charge, including the energy, 
has a well-defined value.
Charges that fail to commute 
might not have well-defined values simultaneously.
But, if $N$ is large, the $\bar{Q}_j$'s nearly commute;
they can nearly have well-defined values simultaneously.
This approximation motivates our definition
of an approximate microcanonical subspace $\mathcal{M}$.
If the composite system occupies any state in $\mathcal{M}$,
one has a high probability of being able to predict
the outcome of a measurement of any $\bar{Q}_j$.

%
%
%
%
\begin{definition}
\label{def:approx-microcanonical-subspace}
For $\eta,\eta',\epsilon, \delta, \delta' > 0$, an
\emph{$(\epsilon, \eta, \eta', \delta, \delta')$-approximate microcanonical (a.m.c.) subspace}
$\mathcal{M}$ of $\mathcal{H}^{\otimes N}$
associated with observables $Q_j$ and
with approximate expectation values $v_j$
consists of the states $\omega$ for which the
probability distribution over the possible outcomes of 
a measurement of any $\bar{Q}_j$ peaks sharply about $v_j$. 
More precisely, we denote by $\Pi_j^\eta$ the projector onto 
the direct sum of the eigensubspaces of $\bar Q_j$ 
associated with the eigenvalues in the interval $[v_j-\eta\Sigma(Q_j),v_j+\eta\Sigma(Q_j)]$.
Here, $\Sigma(Q) = \lambda_{\max}(Q)-\lambda_{\min}(Q)$ is the
spectral diameter of an observable $Q$.
$\mathcal{M}$ must satisfy the following conditions:
\begin{enumerate}
   
   \item  
   \label{item:Def1A}
   Let $\omega$ denote any state, defined on $\mathcal{H}^{\otimes N}$,
   whose support lies in $\mathcal{M}$. 
   A measurement of any $\bar{Q}_j$ is likely to
   yield a value near $v_j$:
   \begin{align}
      \label{eq:Ogata3}
      {\text{supp}} (\omega)  \subset  \mathcal{M}
      \quad  \Rightarrow  \quad
      \Tr ( \omega  \Pi_j^\eta )  \geq  1 - \delta  
      \; \: \forall j.
   \end{align}
   
   \item
   \label{item:Def1B}
   Conversely, consider any state $\omega$, defined on $\mathcal{H}^{\otimes N}$,
   whose measurement statistics peak sharply. 
   Most of the state's probability weight lies in $\mathcal{M}$:
   \begin{align}
      \label{eq:Ogata3B}
      \Tr ( \omega  \Pi_j^{\eta'} )  \geq  1 - \delta'  \; \:  \forall j
      \quad  \Rightarrow \quad
      \Tr ( \omega  P)  \geq  1 - \epsilon,
   \end{align}
   wherein $P$ denotes the projector onto $\mathcal{M}$.
   
\end{enumerate}

\end{definition}

This definition merits two comments.
First, $\mathcal{M}$ is the trivial (zero) subspace 
if the $v_j$'s are inconsistent, i.e., if no state 
$\rho$ satisfies $\Tr ( \rho \, Q_j )  =  v_j  \; \;  \forall j$.
Second, specifying $(\eta,\eta',\epsilon, \delta, \delta')$
does not specify a unique subspace.
The inequalities enable multiple approximate microcanonical subspaces to satisfy 
Definition~\ref{def:approx-microcanonical-subspace}.
The definition ensures, however, that any two such subspaces
overlap substantially.

%
%
%
%
\medskip\noindent
\textbf{The approximate microcanonical subspace leads to the \GGS{}:}
Let us show that Definition~\ref{def:approx-microcanonical-subspace} exhibits the property
desired of a microcanonical state:
The reduced state of each subsystem is close to the \GGS{}.

We denoted by $P$ the projector onto the approximate microcanonical subspace $\mathcal{M}$.
Normalizing the projector yields the approximate microcanonical state
$\Omega  :=  \frac{1}{\Tr (P)  }P$.
Tracing out all subsystems but the $\ell^{\text{th}}$ yields
$\Omega_\ell  :=  \Tr_{0, \ldots, \ell - 1, \ell + 1, \ldots, N-1} ( \Omega )$.

We quantify the discrepancy between $\Omega_\ell$ and the \GGS{}
with the relative entropy:
\begin{align}
   \label{eq:RelEnt}
   D ( \Omega_\ell \| \gamma_{ \mathbf{v} } )
   :=  - S( \Omega_\ell )
        -  \Tr \Big(  \Omega_\ell    \log  ( \gamma_{ \mathbf{v} } )  \Big).
\end{align}
wherein $S ( \Omega_\ell )  :=  - \Tr \Big( \Omega_\ell  \log  ( \Omega_\ell )  \Big)$
is the von Neumann entropy.
The relative entropy is lower-bounded by the trace norm,
which quantifies quantum states' distinguishability~\cite{HiaiOT81}:
\begin{align}
  \label{eq:PinskerIneq-2}
  D(\Omega_\ell\Vert\gamma_{\mathbf v})
  \geq \frac12 \norm{\Omega_\ell-\gamma_{\mathbf v}}_1^2.
\end{align}

%

\begin{theorem}
  \label{thm:microcan-implies-GGS}
  Let $\mathcal{M}$ denote an $(\epsilon,\eta,\eta',\delta,\delta')$-approximate
  microcanonical subspace of $\mathcal{H}^{\otimes N}$
  associated with the $Q_j$'s and the $v_j$'s,
  for $N \geq [2   \norm{Q_j}_\infty^2/(\eta^2) ]\log(2/\delta')$.
  The average, over the $N$ subsystems, 
  of the relative entropy between 
  each subsystem's reduced state $\Omega_\ell$ and the \GGS{} is small:
  \begin{align}
     \label{eq:AvgRelEnt}
      \frac{1}{N} \sum_{ \ell = 0}^{ N - 1} D ( \Omega_\ell \| \gamma_{ \mathbf{v} } )
       \leq \theta + \theta'.
  \end{align}
  This $\theta = \mathrm{(const.)} / \sqrt{N}$
  is proportional to a constant dependent on $\epsilon$, on the $v_j$'s, and on $d$.
  This  $\theta'   =   (c+1)   \mathrm{(const.)}   
  (\eta   +   2 \delta   \cdot   \max_j \{ \norm{Q_j}_\infty   \})$
  is proportional to a constant dependent on the $v_j$'s.
\end{theorem}

\begin{proof}
  We will bound each term in the definition~\eqref{eq:RelEnt} of the relative entropy $D$.
  The von Neumann-entropy term $S( \Omega_\ell )$,
  we bound with Schumacher's theorem for typical subspaces.
  The cross term is bounded, by the definition of 
  the approximate microcanonical subspace $\mathcal{M}$,
  in terms of the small parameters that quantify the approximation.

  First, we lower-bound the dimensionality of $\mathcal{M}$ in terms of
  $\epsilon,\eta,\eta',\delta$,  and  $\delta'$.  
  Imagine measuring some $\bar{Q}_j$ 
  of the composite-system state $\gamma_{\mathbf v}^{\otimes N}$.
  This is equivalent to 
  measuring each subsystem's $Q_j$,
  then averaging the outcomes.
  Each $Q_j$ measurement would yield a random outcome
  $X^j_\ell   \in   [\lambda_{\min}({Q_j}),\lambda_{\max}({Q_j})]$, 
  for $\ell=0,\ldots,N-1$.  
  The average of these $Q_j$-measurement outcomes
  is tightly concentrated around $v_j$, by Hoeffding's Inequality~\cite{Hoeffding63}:
  \begin{align}
      1   -   \Tr \,\bigl(   \gamma_{\mathbf v}^{\otimes N}    \Pi_j^\eta   \bigr) 
      &= \operatorname{Pr}\left\{ \Bigl\lvert 
            \frac1N   \sum_{\ell = 0}^{N - 1}    X^j_\ell - v_j
            \rvert > \eta\Sigma(Q_j) \right\}  \\
      &\leq 2 \exp\left( - 2\eta^2 N \right) \\
      &\leq \delta',
  \end{align}
  for large enough $N$.
  From the second property in Definition~\ref{def:approx-microcanonical-subspace}, 
  it follows that
  $\Tr\,\bigl(\gamma_{\mathbf v}^{\otimes N} P\bigr) \geq 1-\epsilon $.
  Hence $\mathcal{M}$ is a high-probability subspace of
  $\gamma_{\mathbf v}^{\otimes N}$.  

  By Schumacher's Theorem, or by the stronger~\cite[Theorem I.19]{PhdWinter1999}, 
  \begin{align}
    S(\Omega) 
    = \log  \Big( \dim (P) \Big) 
    &\geq N S\bigl(\gamma_{\mathbf v}\bigr)
    - (\mathrm{const.})   \sqrt{N}   \\
    &= N S\bigl(\gamma_{\mathbf v}\bigr) - N\theta,
      \label{eq:lower-bound-approx-microcan-subspace-dim}
  \end{align}
  wherein $\theta  :=  (\mathrm{const.})  /  \sqrt N$.
  The constant depends on $\epsilon$,
  $d$, and the charge values $v_j$.
  The entropy's subadditivity implies that
  $S(\Omega) \leq    \sum_{\ell = 0}^{N - 1}   S(\Omega_\ell)$. 
  Combining this inequality with Ineq.~\eqref{eq:lower-bound-approx-microcan-subspace-dim} yields
  \begin{align}
    S\bigl(\gamma_{\mathbf v}\bigr) - \theta
    \leq \frac1N    \sum_{\ell = 0}^{N - 1}    S(\Omega_\ell).
    \label{eq:lower-bound-entropy-of-average-Omegal}
  \end{align}
  
  The support of $\Omega$ lies within $\mathcal{M}$:
  $\operatorname{supp} ( \Omega )  \subset  \mathcal{M}$.   Hence 
  $\Tr(\Omega   \,   \Pi^\eta_j)   =   1   \geq   1-\delta$ for all $j$. 
  Let  $\bar\Omega  :=   \frac1N    \sum_{\ell = 0}^{N - 1}   \Omega_\ell$.
  We will bound the many-copy average
  \begin{align} 
     w_j  := \Tr(  Q_j      \,  \bar \Omega)  
     & =  \label{eq:wj1}
            \frac1N    \sum_{\ell = 0}^{N - 1}    \Tr( \Omega_\ell   \,   Q_j)  \\
     & =   \label{eq:wj2}
             \Tr( \Omega  \, \bar Q_j).
  \end{align}
  Let us bound this trace from both sides.
  Representing $\bar Q_j = \sum_{q} q\,\Pi^q_j$
  in its eigendecomposition,
  we upper-bound the following average:
  \begin{align}
     \Tr(\Omega  \,   \bar Q_j)
     & =   \sum_q q    \Tr  \left(\Omega  \,  \Pi^q_j  \right)  \\
     &  \leq [   v_j+\eta\Sigma(Q_j)   ]   
          \Tr  \left(\Omega  \,  \Pi^\eta_j  \right)
           + \norm{Q_j}_\infty   \Tr  \Big(   \Omega       \left[ \id-\Pi^\eta_j \right]  \Big)  \\
     &  \label{eq:Bound1}
         \leq v_j + \norm{Q_j}_\infty (\eta+\delta).
  \end{align} 
  We complement this upper bound with a lower bound:
  \begin{align}
     \Tr(\Omega    \,   \bar Q_j)
     &  \geq   [  v_j-\eta\Sigma(Q_j)  ]     
                    \Tr   \left(\Omega  \,  \Pi^\eta_j   \right) -
                  \norm{Q_j}_\infty   \Tr   \Big( \Omega    \left[ \id-\Pi^\eta_j \right]   \Big)   \\
     &  \label{eq:Bound2}
        \geq   [v_j-\eta\Sigma(Q_j)  ]  (1-\delta) - \norm{Q_j}_\infty\delta.
  \end{align}
  Inequalities~\eqref{eq:Bound1} and~\eqref{eq:Bound2} show that
  the whole-system average $w_j$ is close to the single-copy average $v_j$:
  \begin{align}
     \xi_j   \label{eq:xijDefn}
     := \abs{w_j - v_j}    
     &  =   \abs{\Tr   (\Omega     \,    \bar Q_j   ) - v_j}   \\
     & \label{eq:xijBound}
         \leq  (\eta + 2\delta) \norm{Q_j}_\infty .
  \end{align}
  %

  Let us bound the average relative entropy. By definition,
  \begin{align}
     \label{eq:DExpn}
    \frac1N   \sum_{\ell = 0}^{N - 1} 
         D\,(\Omega_\ell\Vert\gamma_{\mathbf v})
    = -\frac1N   \sum_{\ell = 0}^{N - 1}   \left[ S(\Omega_\ell) 
        +     \Tr  \Big(  \Omega_\ell   \log  ( \gamma_{\mathbf v} )   \Big)   \right].
  \end{align}
  Let us focus on the second term.
  First, we substitute in the form of $\gamma_{\mathbf{v}}$ 
  from Eq.~\eqref{maintext-eq:GGS} of the main text.
  Next, we substitute in for $w_j$, using Eq.~\eqref{eq:wj1}.
  Third, we substitute in $\xi_j$, using Eq.~\eqref{eq:xijDefn}.
  Fourth, we invoke the definition of $S( \gamma_{ \mathbf{v} } )$,
  which we bound with Ineq.~\eqref{eq:lower-bound-entropy-of-average-Omegal}:
  \begin{align}
    -\frac1N   \sum_{\ell = 0}^{N - 1}
        & \Tr  \,   \Big(   \Omega_\ell   \log  ( \gamma_{\mathbf v} )  \Big) \\
    & =  \frac1N   \sum_{\ell = 0}^{N - 1}
        \Big[ \log (Z)   
        +  \sum_{j = 0}^c   \mu_j   \Tr   (\Omega_\ell  \,   Q_j)  \Big]  \\
    & = \log Z +   \sum_{j = 0}^c   \mu_j w_j  \\
    & \leq \log Z +   \sum_{j = 0}^c   \mu_j v_j 
        +   \sum_{j = 0}^c   \abs{\mu_j}\xi_j   \\
    &= S(\gamma_{\mathbf v}) 
          +  \sum_{j = 0}^c   \abs{\mu_j}\xi_j  \\
    &  \leq  \frac{1}{N}   \sum_{ \ell  =  0}^{N - 1}   S ( \Omega_\ell )  
          +  \theta   +   \sum_{j = 0}^c   \abs{\mu_j}\xi_j.
  \end{align}
  Combining this inequality with Eq.~\eqref{eq:DExpn} yields
  \begin{align}
    \hspace{1em}&\hspace{-1em}%
    \frac1N    \sum_{\ell = 0}^{N - 1} 
                   D\,(   \Omega_\ell\Vert\gamma_{\mathbf v}   )
    \leq   \theta   +   \sum_{j = 0}^c   \abs{\mu_j}\xi_j \\
    &  \leq  \theta  +    (c+1)  \,  \left( \max_j  | \mu_j | \right)  
                                         \left( \max_j \xi_j \right)    \\
    &  \label{eq:LateDBound}
        \leq  \theta  +    (c+1)  \,  \left( \max_j  | \mu_j | \right) 
        \left[ (\eta  +  2\delta) \cdot  
                \max_j  \left\{   \| Q_j \|_\infty \right\}  \right].
  \end{align}
  The final inequality follows from Ineq.~\eqref{eq:xijBound}.
  Since the $v_j$'s determine the $\mu_j$-values,
  $(c+1) \left( \max_j  | \mu_j | \right)$ is a constant determined by the $v_j$'s.
  The final term in Ineq.~\eqref{eq:LateDBound}, therefore, is upper-bounded by 
  $\theta'  =   (c+1) \mathrm{(const.)}   
  (\eta   +   2 \delta)   \cdot    \max_j   \left\{ \norm{Q_j}_\infty  \right\}$.
\end{proof}

\paragraph{Existence of an approximate microcanonical subspace:}
Definition~\ref{def:approx-microcanonical-subspace} does not reveal
under what conditions  an approximate microcanonical subspace $\mathcal{M}$ exists.
We will show that an $\mathcal{M}$ exists
for $\epsilon, \eta, \eta', \delta, \delta'$ that can approach zero simultaneously,
for  sufficiently large $N$.
First, we prove the existence of a microcanonical subspace for commuting observables.
Applying this lemma to the $\tilde{Y}_j$'s
shows that $\mathcal{M}$ exists for noncommuting observables.

%
%
%
%
\begin{lemma}
\label{lemma:MicroSubspaceCom}
  Consider a Hilbert space $\mathcal{K}$ with commuting observables
  $X_j$, $j=0,\ldots,c$. For all $\epsilon, \eta, \delta > 0$
  and for sufficiently large $m$, there exists an
  $\left(\epsilon, \eta, \eta'{=}\eta, \delta, \delta'{=}\frac{\epsilon}{c+1}\right)$-approximate
  microcanonical subspace $\mathcal{M}$ of $\mathcal{K}^{\otimes m}$ 
  associated with the observables ${X}_j$ 
  and with the approximate expectation values $v_j$.
\end{lemma}

\begin{proof}
Recall that 
\begin{equation} 
  \bar{X}_j = \frac{1}{m} \sum_{\lambda=0}^{m-1}
   \id^{\otimes \lambda} \otimes    X_{ j }    \otimes \id^{\otimes (m-1-\lambda)}
\end{equation}
is the average of $X_j$ over the $m$ subsystems.
Denote by
\begin{equation}
  \Xi_j^\eta   := \bigl\{ v_j-\eta \leq \bar{X}_j \leq v_j+\eta \bigr\}
\end{equation}
the projector onto the direct sum of the $\bar X_j$ eigenspaces 
associated with the eigenvalues in $[v_j-\eta,v_j+\eta]$.
Consider the subspace $\mathcal{M}_{\text{com}}^\eta$
projected onto by all the 
$X_j$'s.
The projector onto $\mathcal{M}_{\text{com}}^\eta$ is
\begin{align}
   \label{align:UpsilonProd}
   P_{\text{com}}  :=  \Xi_0^\eta   \,  \Xi_1^\eta  \cdots  \Xi_c^\eta.
\end{align}


Denote by $\omega$ any state whose support lies in   $\mathcal{M}_{\text{com}}^\eta$.
Let us show that $\omega$ satisfies the inequality in~\eqref{eq:Ogata3}.
By the definition of $P_\mathrm{com}$, 
${\text{supp}}( \omega )   \subset   {\text{supp}} ( \Xi_j^\eta )$.
Hence  $\Tr \left( \omega   \Xi_j^\eta   \right)   =   1   \geq 1-\delta$.

Let us verify the second condition in Definition~\ref{def:approx-microcanonical-subspace}.
Consider any eigenvalue $\bar{y}_j$ of $\bar{Y}_j$, for each $j$.
Consider the joint eigensubspace, shared by the $\bar{Y}_j$'s,
associated with any eigenvalue $\bar{y}_1$ of $\bar{Y}_1$,
with any eigenvalue $\bar{y}_2$ of $\bar{Y}_2$, etc.
Denote the projector onto this eigensubspace of $\mathcal{H}^{\otimes N}$ by
$\mathcal{P}_{\bar y_1, \cdots, \bar y_c}$.

Let $\delta'   =   \frac{\epsilon}{c+1}$.
Let   $\omega$ denote any state, defined on  $\mathcal{H}^{\otimes N}$, for which
$\Tr  \left(   \omega   \,   \Xi_j^\eta   \right) \geq 1-\delta'$, 
for all $j=0,\ldots,c$.
The left-hand side of the second inequality in~\eqref{eq:Ogata3B} reads, 
$\Tr \left(   \omega   P_\mathrm{com}   \right)$.
We insert the resolution of identity
$ \sum_{\bar y_0,  \ldots, \bar y_c}    \mathcal{P}_{\bar y_0  \ldots  \bar y_c}$
into the trace.
The property $\mathcal{P}^2  =  \mathcal{P}$
of any projector $\mathcal{P}$
enables us to square each projector.
Because $[\mathcal{P}_{\bar y_0  \ldots   \bar y_c},   P_\mathrm{com}]=0$,
\begin{align}
  \Tr   \left(  \omega   P_\mathrm{com}\right)
  & = \Tr   \left(   \sum_{\bar y_0,  \ldots, \bar y_c}    \mathcal{P}_{\bar y_0  \ldots  \bar y_c} 
         \omega    \mathcal{P}_{\bar y_0   \ldots   \bar y_c} P_\mathrm{com} \right)   \\
  & =: \Tr \left(   \omega'    P_\mathrm{com}   \right),
\end{align}
wherein $\omega'  :=  \sum_{\bar y_0,  \ldots, \bar y_c}    \mathcal{P}_{\bar y_0  \ldots  \bar y_c} 
         \omega    \mathcal{P}_{\bar y_0   \ldots   \bar y_c}$ 
is $\omega$ pinched with the complete set 
$\{   \mathcal{P}_{\bar y_0\bar y_1 \ldots \bar y_c}\}$
of projectors~\cite{Hayashi02}.  
By this definition of $\omega'$,
$\Tr \left(   \omega'    \,    \Xi_j^\eta   \right)
= \Tr \left(   \omega   \,    \Xi_j^\eta   \right)
\geq   1-\delta'$, and
$[\omega',   \Xi_j^\eta]=0$.
For all $j$,  therefore,
\begin{align}
   \omega'    \,  \Xi_j^\eta 
   = \omega' - \omega'\left(\id-\Xi_j^\eta   \right) 
   =: \omega' -   \Delta_j,
\end{align}
wherein
\begin{align}
   \Tr ( \Delta_j  )
   = \Tr \left(   \omega'   \left[   \id-\Xi_j^\eta   \right]   \right) 
   \leq \delta'.
\end{align}
Hence
\begin{align}
   \Tr  \left(   \omega'   P_{\text{com}}   \right)
   &   =   \Tr \left(   \omega'   \,   \Xi_0^\eta      \,    \Xi_1^\eta
            \cdots   \Xi_c^\eta   \right)  \\
   &  \geq \Tr \left(   \left[  \omega'   -   \Delta_0\right]  
                               \Xi_1^\eta   \cdots\Xi_c^\eta\right)   \\
   %
   %
   & \geq \Tr \left(   \omega'   \,   \Xi_1^\eta\cdots\Xi_c^\eta\right)  - \delta'   \\
   & \geq \Tr \left(   \omega'   \right) - (c+1)\delta'   \\
   & = 1 - (c+1)\delta'
     =   1  -  \epsilon.
\end{align}
As $\omega$ satisfies~\eqref{eq:Ogata3B},
$\mathcal{M}_\text{com}^\eta$ is an
$(\epsilon, \eta, \eta'{=}\eta, \delta, \delta'{=}\frac{\epsilon}{c+1})$-approximate 
microcanonical subspace.
\end{proof}

Lemma~\ref{lemma:MicroSubspaceCom} proves the existence of an approximate microcanonical subspace 
$\mathcal{M}_{\text{com}}^\eta$ for the $\tilde{Y}_j$'s defined on $\mathcal{K} = \mathcal{H}^{\otimes n}$
and for sufficiently large $n$.
In the subsequent discussion, we denote
by $\Upsilon_j^\eta$ the projector onto the direct sum of the $\bar Y_j$ eigenspaces 
associated with the eigenvalues in $[v_j-\eta\Sigma(\tilde{Y}_j),v_j+\eta\Sigma(\tilde{Y}_j)]$.
Passing from $\tilde{Y}_j$ to $\tilde{Q}_j$ to $Q_j$,
we now prove that the same $\mathcal{M}_{\text{com}}^\eta$ is an
approximate microcanonical subspace for the $Q_j$'s.

%
%
%
%
\begin{theorem}
  \label{thm:mocrocononocol}
  Under the above assumptions, 
  for every $\epsilon > (c+1)\delta' > 0$, $\eta > \eta' > 0$,
  $\delta > 0$, and all sufficiently large $N$, there exists an
  $(\epsilon, \eta, \eta', \delta, \delta')$-approximate
  microcanonical subspace $\mathcal{M}$ of $\mathcal{H}^{\otimes N}$ 
  associated with the observables $Q_j$ 
  and with the approximate expectation values $v_j$.  
\end{theorem}

\begin{proof}
Let $\hat{\eta} = (\eta+\eta')/2$.
For a constant $C_{\text{AP}} > 0$ to be determined later,
let $n$ be such that $\epsilon_{\text{O}} = \epsilon_{\mathrm{O}}(n)$
from Ogata's result~\cite[Theorem~1.1]{Ogata11} is small enough so that
$\eta > \hat{\eta}+C_{\text{AP}}\epsilon_{\text{O}}^{1/3}$
and $\eta' < \hat{\eta}-C_{\text{AP}}\epsilon_{\text{O}}^{1/3}$,
as well as such that 
$\hat{\delta} = \delta - C_{\text{AP}}\epsilon_{\text{O}}^{1/3} > 0$
and such that 
$\hat{\delta}' = \delta' + C_{\text{AP}}\epsilon_{\text{O}}^{1/3} \leq \frac{\epsilon}{c+1}$.

Choose $m$ in Lemma~\ref{lemma:MicroSubspaceCom} large enough
such that an $(\epsilon,\hat{\eta},\hat{\eta}'{=}\hat{\eta},\hat{\delta},\hat{\delta}')$-approximate
microcanonical subspace $\mathcal{M} := \mathcal{M}_{\text{com}}$ associated with the 
commuting $\tilde{Y}_j$ exists, with approximate expectation values $v_j$.

Let $\omega$ denote a state defined on $\mathcal{H}^{\otimes N}$.
We will show that,
if measuring the $\bar{Y}_j$'s of $\omega$ yields sharply peaked statistics,
measuring the $\bar{Q}_j$'s yields sharply peaked statistics.
Later, we will prove the reverse
(that sharply peaked $\bar{Q}_j$ statistics imply 
sharply peaked $\bar{Y}_j$ statistics).

Recall from Definition~\ref{def:approx-microcanonical-subspace} that
$\Pi^\eta_j$ denotes the projector onto 
the direct sum of the $\bar Q_j$ eigenstates 
associated with the eigenvalues in $[v_j-\eta\Sigma(Q_j),  \:  v_j+\eta\Sigma(Q_j)]$.
These eigenprojectors are discontinuous functions of the observables.
Hence we look for better-behaved functions.
We will approximate the action of $\Pi^\eta_j$ by using 
\begin{align}
   \label{eq:CtsFxn}
   f_{\eta_0,\eta_1}(x)  :=  \begin{cases}
               1,  &  x  \in  [ - \eta_0,  \eta_0 ]   \\
               0,  &  | x |  >  \eta_1
               \end{cases},
\end{align}
for $\eta_1>\eta_0>0$.
The Lipschitz constant of $f$ is bounded by
$\lambda := \frac{1}{\eta_1 - \eta_0} \in \mathbb{R}$.  

The operator
$f_{\eta_0\Sigma(Q_j),\eta_1\Sigma(Q_j)}(\bar Q_j - v_j\id)$
approximates the projector $\Pi_j^{\eta_0}$.
Indeed, as a matrix, $f_{\eta_0\Sigma(Q_j),\eta_1\Sigma(Q_j)}( \bar{Q}_j - v_j \id )$ 
is sandwiched between the projector $\Pi_j^{\eta_0}$,
associated with a width-$\eta_0$ interval around $v_j$,
and a projector $\Pi_j^{\eta_1}$ 
associated with a width-$\eta_1$ interval of eigenvalues.
$f_{\eta,\eta}$ is the indicator function on the interval $[-\eta,\eta]$.
Hence $\Pi_j^\eta = f_{\eta\Sigma(Q_j),\eta\Sigma(Q_j)}(\bar Q_j - v_j\id )$.
Similarly, we can regard 
$f_{\eta_0\Sigma(Q_j),\eta_1\Sigma(Q_j)}( \bar{Y}_j - v_j \id )$ as 
sandwiched between $\Upsilon_j^{\eta_0}$ 
and $\Upsilon_j^{\eta_1}$.

Because $\bar{Q}_j$ is close to $\bar{Y}_j$, $f( \bar{Q}_j )$ is close to $f ( \bar{Y}_j )$:
Let $n$ be large enough so that, by~\cite[Theorem 1.1]{Ogata11},
$\| \bar{Q}_j - \bar{Y}_j \|_\infty \leq \epsilon_\mathrm{O}$. 
By~\cite[Theorem 4.1]{AleksandrovP10},
\begin{multline}
   \label{eq:fsClose}
   \|  f_{ \eta_0\Sigma(Q_j),  \eta_1\Sigma(Q_j) }( \bar{Y}_j - v_j\id )   
      -   f_{ \eta_0\Sigma(Q_j),  \eta_1\Sigma(Q_j) } ( \bar{Q}_j - v_j\id )  \|_\infty
\\
  \leq  \kappa_\lambda,
\end{multline}
wherein $\kappa_\lambda  =  C_{\text{AP} }  \lambda  \epsilon_\mathrm{O}^{2/3}$
and $C_{\text{AP}}$ denotes a universal constant.
Inequality~\eqref{eq:fsClose} holds because 
$f$ is $\lambda$-Lipschitz and bounded, 
so the H\"{o}lder norm in~\cite[Theorem 4.1]{AleksandrovP10} 
is proportional to $\lambda$.

Let us show that, 
if measuring the $\bar{Y}_j$'s of $\omega$ yields sharply peaked statistics, 
then measuring the $\bar{Q}_j$'s yields sharply peaked statistics, 
and vice versa.
First, we choose $\eta_0=\eta$, $\eta_1=\eta + \epsilon_\mathrm{O}^{1/3}$, 
and $\lambda=\epsilon_\mathrm{O}^{-1/3}$ such that 
$\kappa   :=   \kappa_\lambda
=   C_\mathrm{AP}\epsilon_\mathrm{O}^{1/3}$.
By the ``sandwiching,''
\begin{align}
   \label{eq:Direc1A}
   \Tr   \left( \omega  \Pi_j^{\eta+\epsilon_\mathrm{O}^{1/3}} \right)   
   & \geq   \Tr \left(  \omega  f_{\eta_0\Sigma(Q_j),\eta_1\Sigma(Q_j)} 
                      \left[ \bar{Q}_j   -  v_j  \id \right]  \right).
\end{align}
To bound the right-hand side, we invoke Ineq.~\eqref{eq:fsClose}:
\begin{align}
   \kappa  
   &  \geq   \|  f_{ \eta_0\Sigma(Q_j),  \eta_1\Sigma(Q_j) }( \bar{Y}_j - v_j\id )   
                     \nonumber   \\ & \qquad 
                     -   f_{ \eta_0\Sigma(Q_j),  \eta_1\Sigma(Q_j) } ( \bar{Q}_j - v_j\id )  \|_\infty  \\
   &  \geq  \Tr \Big(  f_{ \eta_0\Sigma(Q_j),  \eta_1\Sigma(Q_j) }( \bar{Y}_j - v_j\id )   
                               \nonumber   \\ & \qquad 
                     -   f_{ \eta_0\Sigma(Q_j),  \eta_1\Sigma(Q_j) } ( \bar{Q}_j - v_j\id ) \Big)   \\
   &  \geq  \Tr \Big(  \omega  \Big[  
                               f_{ \eta_0\Sigma(Q_j),  \eta_1\Sigma(Q_j) }( \bar{Y}_j - v_j\id )   
                               \nonumber   \\ & \qquad 
                     -   f_{ \eta_0\Sigma(Q_j),  \eta_1v } ( \bar{Q}_j - v_j\id ) 
                               \Big]  \Big).
\end{align}
Upon invoking the trace's linearity, we rearrange terms:
\begin{align}
   \Tr  \Big(  & \omega   f_{ \eta_0\Sigma(Q_j),  \eta_1\Sigma(Q_j) } ( \bar{Q}_j - v_j\id )   \Big)
   \\ &   \geq 
   \Tr  \Big(  \omega   f_{ \eta_0\Sigma(Q_j),  \eta_1\Sigma(Q_j) }( \bar{Y}_j - v_j\id )   \Big)
   -   \kappa  \\
   &   \geq  \label{eq:Direc1B}
         \Tr  \left(  \omega  \Upsilon^\eta_j   \right)   -   \kappa.
\end{align}
The final inequality follows from the ``sandwiching'' property of 
$f_{ \eta_0,  \eta_1 }$.
Combining Ineqs.~\eqref{eq:Direc1A} and~\eqref{eq:Direc1B} yields
a bound on fluctuations in $\bar{Q}_j$ measurement statistics
in terms of fluctuations in $\bar{Y}_j$ statistics:
\begin{align}
   \label{eq:tromegaPij-geq-tromegaUpsilonj}
   \Tr  \left( \omega  \,  \Pi_j^{\eta+\epsilon_\mathrm{O}^{1/3}}   \right)
   \geq   \Tr  \left(  \omega    \Upsilon^\eta_j   \right)   -   \kappa.
\end{align}

Now, we bound fluctuations in $\bar{Y}_j$ statistics
with fluctuations in $\bar{Q}_j$ statistics.
If    $\eta_0   =   \eta-\epsilon_\mathrm{O}^{1/3}$;    $\eta_1=\eta$;
$\lambda=\epsilon_\mathrm{O}^{-1/3}$, as before, and
$\kappa=\kappa_\lambda=C_\mathrm{AP}\epsilon_\mathrm{O}^{1/3}$, then
\begin{align}
  \Tr   \left(\omega \Upsilon_j^{\eta}\right)
   \geq   \Tr  \left( \omega  \Pi_j^{\eta-\epsilon_\mathrm{O}^{1/3}}   \right)  -  \kappa.  
 \label{eq:tromegaUpsilonj-geq-tromegaPij}
\end{align}
Using Ineqs.~\eqref{eq:tromegaPij-geq-tromegaUpsilonj} and~\eqref{eq:tromegaUpsilonj-geq-tromegaPij}, 
we can now show that $\mathcal{M} := \mathcal{M}_\mathrm{com}^{\hat{\eta}}$ is an
approximate microcanonical subspace for the observables
$Q_j$ and the approximate charge values $v_j$.
In other words, $\mathcal{M}$ is an
approximate microcanonical subspace for the observables $\tilde{Q}_j$.

First, we show that $\mathcal{M}$ satisfies the first condition in Definition~\ref{def:approx-microcanonical-subspace}.
Recall that $\mathcal{M}_\mathrm{com}^\eta$ is an
$\left( \epsilon, \eta, \eta'{=}\eta, \delta, \delta'{=}\frac{\epsilon}{c} \right)$-approximate
microcanonical subspace for the observables $\tilde{Y}_j$ with the approximate charge values $v_j$,
for all $\epsilon,\eta,\delta>0$ and for large enough $m$ (Lemma~\ref{lemma:MicroSubspaceCom}).
Recall that $N=nm$.
Choose $\delta   =   \hat\delta-\kappa>0$.
Let $\omega$ denote any state, defined on $\mathcal{H}^{\otimes N}$, 
whose support lies in
$\mathcal{M}  =  \mathcal{M}_\mathrm{com}^{\eta}$. 
Let   $\hat\eta = \eta+\epsilon_\mathrm{O}^{1/3}$. 
By the definitions of $\omega$ and $\mathcal{M}$, 
$\Tr    \left(  \omega \Upsilon_j^{\eta}   \right)   =   1   \geq 1-\delta$.
By Ineq.~\eqref{eq:tromegaPij-geq-tromegaUpsilonj}, therefore,
\begin{align}
   \Tr   \left(   \omega   \Pi_j^{\hat\eta}   \right) 
   \geq \Tr  \left(   \omega\Upsilon_j^{\eta}   \right)  -   \kappa 
   \geq    1   -   \delta-\kappa 
   =   1 - \hat\delta.
\end{align}
Hence $\mathcal{M}$ satisfies Condition~\ref{item:Def1A} in Definition~\ref{def:approx-microcanonical-subspace}.

To show that $\mathcal{M}$ satisfies Condition~\ref{item:Def1B}, let
$\hat\eta'=\eta - \epsilon_\mathrm{O}^{1/3}$, and let
$\hat\delta'
=   \delta'-   \kappa   
=   \frac{\epsilon}{c}   -   C_{\mathrm{AP}}\epsilon_\mathrm{O}^{1/3} 
>0$. 
Let   $\omega$ in $\mathcal{H}^{\otimes N}$ satisfy
$\Tr   \left(   \omega \Pi_j^{\hat\eta'}   \right)   \geq 1-\hat\delta'$ for all $j$. 
By Ineq.~\eqref{eq:tromegaUpsilonj-geq-tromegaPij},
\begin{align}
   \Tr \left(   \omega\Upsilon_j^{\eta}   \right)   
   \geq 1 - \hat\delta'-\kappa
   =1-\delta'.
\end{align}
By Condition~\ref{item:Def1B} in the definition of $\mathcal{M}_\mathrm{com}^\eta$,
therefore, at least fraction $1 - \epsilon$ of the probability weight of $\omega$ 
lies in $\mathcal{M}_\mathrm{com}^\eta   =   \mathcal{M}$:
$\Tr   \left(   \omega P_\mathrm{com}   \right)   \geq 1   -   \epsilon$. 
As $\mathcal{M}$ satisfies Condition~\ref{item:Def1B}, 
$\mathcal{M}$ is an $(\epsilon,\hat\eta,\hat\eta',\hat\delta,   \hat\delta')$-approximate microcanonical subspace.
\end{proof}

This derivation confirms physically the information-theoretic 
maximum-entropy derivation. By ``physically,'' we mean, 
``involving the microcanonical form of a composite system's state
and from the tracing out of an environment.''
The noncommutation of the charges $Q_j$ required us to define 
an approximate microcanonical subspace $\mathcal{M}$.
The proof of the subspace's existence, under appropriate conditions,
crowns the derivation.

The physical principle underlying this derivation
is, roughly, the Correspondence Principle.
The $Q_j$'s of one copy of the system $\mathcal{S}$ fail to commute with each other.
This noncommutation constitutes quantum mechanical behavior.
In the many-copy limit, however,
averages $\bar{Q}_j$ of the $Q_j$'s 
are approximated by commuting $\bar{Y}_j$'s, whose
existence was proved by Ogata~\cite{Ogata11}.
In the many-copy limit,
the noncommuting (quantum) problem reduces approximately
to the commuting (classical) problem.

We stress that the approximate microcanonical subspace $\mathcal{M}$
corresponds to a set of observables $Q_j$ and a set of values $v_j$.
Consider the subspace $\mathcal{M}'$ associated with
a subset of the $Q_j$'s and their $v_j$'s.
This $\mathcal{M}'$ differs from $\mathcal{M}$.
Indeed, $\mathcal{M}'$ typically has a greater dimensionality than $\mathcal{M}$,
because fewer equations constrain it.
Furthermore, consider a linear combination 
$Q' = \sum_{j=0}^c \mu_j Q_j$. 
The average $\bar{Q'}$ of $N$ copies of $Q'$
equals $\sum_{j=0}^c \mu_j \bar{Q}_j$.
The approximate microcanonical subspace $\mathcal{M}$ of the whole
set of $Q_j$'s has the property that 
all states that lie mostly on it 
have sharply defined values near 
$v' = \sum_{j=0}^c \mu_j v_j$.
Generally, however, our $\mathcal{M}$ is not 
an approximate microcanonical subspace for $Q'$, or a selection
of $Q'$, $Q''$, etc., unless these primed operators span the same set of observables as the $Q_j$'s.

%
%
%
%
\section{Dynamical considerations}
\label{section:SI_Dynamics}

Inequality~\eqref{maintext-eq:typical-2} of the main text is derived as follows:
Let us focus on $\|  \rho_\ell  -  \gamma_{ \mathbf{v} }  \|_1$.
Adding and subtracting $\Omega_\ell$ to the argument,
then invoking the Triangle Inequality, yields
\begin{align}
   \| \rho_\ell  -  \gamma_{ \mathbf{v} } \|_1
   \leq   \|  \rho_\ell  -  \Omega_\ell \|_1  +  \|  \Omega_\ell  -  \gamma_{ \mathbf{v} }  \|_1.
\end{align}
We average over copies $\ell$ 
and average (via $\langle . \rangle$) over pure whole-system states $\ket{ \psi }$.
The first term on the right-hand side is bounded 
in Ineq.~\eqref{maintext-eq:typical} of the main text:
\begin{align}
   \label{eq:Typical1}
   \biggl\langle    \frac1N    \sum_{\ell=0}^{N-1} 
         \| \rho_\ell   -    \gamma_{\mathbf{v}}  \|_1    \biggr\rangle
   \leq   
          \frac{d}{ \sqrt{D_M} }
          +   \left\langle   \frac1N    \sum_{\ell=0}^{N-1}
           \|  \Omega_\ell  -  \gamma_{ \mathbf{v} }  \|_1   \right\rangle.
\end{align}
To bound the final term, we invoke Pinsker's Inequality [Ineq.~\eqref{eq:PinskerIneq-2}],
$\|  \Omega_\ell  -  \gamma_{ \mathbf{v} }  \|_1
\leq   \sqrt{ 2 D( \Omega_\ell  ||  \gamma_{ \mathbf{v} } ) }$.
Averaging over $\ell$ and over states $\ket{\psi}$ yields
\begin{align}
   \left\langle   \frac1N    \sum_{\ell=0}^{N-1}
      \|  \Omega_\ell  -  \gamma_{ \mathbf{v} }  \|_1
      \right\rangle
   & \leq   \left\langle   \frac1N    \sum_{\ell=0}^{N-1}
            \sqrt{ 2 D( \Omega_\ell  ||  \gamma_{ \mathbf{v} } ) }
            \right\rangle   \\
   &  \leq  \left\langle   
               \sqrt{ \frac2N    \sum_{\ell=0}^{N-1}
               D( \Omega_\ell  ||  \gamma_{ \mathbf{v} } ) }
               \right\rangle,
\end{align}
wherein $D$ denotes the relative entropy.
The second inequality follows from the square-root's concavity.
Let us double each side of Ineq.~\eqref{eq:AvgRelEnt},
then take the square-root:
\begin{align}
   \sqrt{ \frac{2}{N} \sum_{ \ell = 0}^{ N - 1} 
    D ( \Omega_\ell \| \gamma_{ \mathbf{v} } ) }
    \leq  \sqrt{ 2 (\theta + \theta') }.
\end{align}
Combining the foregoing two inequalities,
and substituting into Ineq.~\eqref{eq:Typical1}, yields Ineq.~\eqref{maintext-eq:typical-2} of the main text.

%
%
%
%
\section{Derivation from complete passivity and resource theory}
  \label{section:SI_RT}

An alternative derivation of 
the thermal state's form relies on {complete passivity}.
One cannot extract work from any number of copies of 
the thermal state via any energy-preserving unitary~\cite{PuszW78,Lenard78}.  
We adapt this argument to noncommuting conserved charges.
The {\GGSlong} is shown to be the completely passive ``free'' state 
in a  thermodynamic resource theory.

%

{Resource theories} are models, developed in quantum information theory, for scarcity.
Using a resource theory, one can calculate the value attributable to a quantum state
by an agent limited to performing only certain operations,
called ``free operations.''
The first resource theory described pure bipartite entanglement~\cite{HorodeckiHHH09}.
Entanglement theory concerns how one can manipulate entanglement,
if able to perform only local operations and classical communications.
The entanglement theory's success led to 
resource theories for asymmetry~\cite{BartlettRS07}, for stabilizer codes in quantum computation~\cite{VeitchMGE14}, for coherence~\cite{WinterY15}, for quantum Shannon theory~\cite{DevetakHW05}, and for thermodynamics, amongst other settings.

Resource-theoretic models for heat exchanges were constructed recently~\cite{janzing_thermodynamic_2000,FundLimits2}.
The free operations, called ``thermal operations,''
conserve energy.
How to extend the theory to other conserved quantities was noted in~\cite{FundLimits2}.
The commuting-observables version of the theory was defined and analyzed in~\cite{YungerHalpernR14,YungerHalpern14},
which posed questions about modeling noncommuting observables.
We extend the resource theory 
to model thermodynamic exchanges of noncommuting observables.
The free operations that define this theory,
we term ``\GTOlong'' (\GTO).
This resource theory is related to that in~\cite{Imperial15}.
We supplement earlier approaches with 
a work payoff function,
as well as with a reference frame associated with a non-Abelian group.

This section is organized as follows.
First, we introduce three subsystems and define work.
Next, we define \GTO.
The \GTO{} resource theory leads to the \GGS{} via two routes:
\begin{enumerate}
   \item 
   The \GGS{} is completely passive:
   The agent cannot extract work from
   any number of copies of $\gamma_{ \mathbf{v}}$.

   \item 
   The \GGS{} is the state preserved by \GTO{}, the operations that require no work.
\end{enumerate}
The latter condition leads to ``second laws'' for thermodynamics 
that involves noncommuting conserved charges.
The second laws imply the maximum amount of work extractable
from a transformation between states.

%
%
%
%
\paragraph{Subsystems:}
To specify a physical system in this resource theory, one specifies
a Hilbert space, a density operator, a Hamiltonian,
and operators that represent the system's charges.
To specify the subsystem $S$ of interest, for example,
one specifies a Hilbert space $\mathcal{H}$;
a density operator $\rho_\sys$; a Hamiltonian $H_\sys$;
and charges $Q_{1_\sys}, \ldots, Q_{c_\sys}$.

Consider the group $G$ formed from elements of the form 
$e^{i  \boldsymbol{\mu}   \cdot   \mathbf{Q}}$.
  Each $Q_j$ can be viewed as a generator.  $G$ is non-Abelian if the $Q_j$'s fail to
commute with each other.  Following~\cite{kitaev2014super}, we assume that $G$ is a
compact Lie group.
The compactness assumption is satisfied if the system's Hilbert space is 
finite-dimensional.
(We model the reference frame's Hilbert space as infinite-dimensional for convenience.
Finite-size references can implement the desired protocols with arbitrary
fidelity~\cite{kitaev2014super}.)

We consider three systems, apart from $S$:
First, $R$ denotes a reservoir of free states.
The resource theory is nontrivial, we prove, if and only if
the free states have the \GGS{}'s form.
Second, a battery $W$ stores work.
$W$ doubles as a non-Abelian reference frame.
Third, any other ancilla is denoted by $A$.

The Hamiltonian
$H_{\text{tot}} := H_\sys   +   H_\bath   +   H_{ \batt }+H_\anc$
governs the whole system.
The $j^{\text{th}}$ whole-system charge has the form
$Q_{j_{\text{tot}}} := Q_{j_\sys}  +  Q_{j_\bath}  +  Q_{j_{ \batt }}  +  Q_{j_\anc}$.  
Let us introduce each subsystem individually.

\paragraph{Battery:}
We define work by modeling the system that stores the
work. In general, the mathematical expression for thermodynamic work
depends on which physical degrees of freedom a system has.
A textbook example concerns a gas,
subject to a pressure $p$,
whose volume increases by an amount $dV$.
The gas performs an amount $dW  =  p \, dV$ of work.
If a force $F$ stretches a polymer through a displacement $dx$,
$dW  =  - F \, dx$.
If a material's magnetization decreases by an amount $dM$
in the presence of a strength-$B$ magnetic field,
$dW = B\,dM$.
%


We model the ability to convert,
into a standard form of work, 
a variation in some physical quantity. 
The model consists of an observable called a ``payoff function.''  
The payoff function is defined as
\begin{align}
  \workf := \sum_{j=0}^c\mu_j Q_{j}\ .
  \label{eq:SuppInf-workfunc}
\end{align}
We generally regard the payoff function as an observable of the battery's.
We can also consider the $\workf$ of the system of interest. 
If the system whose $\workf$ we refer to is not obvious from context, 
we will use a subscript.
For example, $\workf_\batt$ denotes the battery's work function. 

One might assume that the battery exchanges only finite amounts of charges.
Under this assumption,
a realistically sized battery can implement the desired protocols with perfect fidelity~\cite{kitaev2014super}.

%
%
%
%
\paragraph{Work:}
We define as {average extracted work} $W$ 
the difference in expectation value of the payoff function $\workf$:
\begin{align}
  \label{eq:Work}
  W := \Tr   \left(\rho_{ \batt }' \workf\right) - \Tr   \left(\rho_{ \batt } \workf\right)\, .
\end{align}
The battery's initial and final states
are denoted by $\rho_{ \batt }$ and $\rho_{ \batt }'$.
If the expectation value increases, then $W>0$, and work
has been extracted from the system of interest.  
Otherwise, work has been expended.

We focus on the average work extracted in the asymptotic limit:
We consider processing many copies of the system,
then averaging over copies.
Alternatively, one could focus on 
one instance of the transformation.
The deterministic or maximal guaranteed work would quantify the protocol's efficiency
better than the average work would~\cite{dahlsten2011inadequacy,del2011thermodynamic,FundLimits2,aaberg2013truly}.

%
%
%
%
\paragraph{Reference frame:}
Reference frames have appeared in 
the thermodynamic resource theory for heat exchanges~\cite{BrandaoHORS13,aberg2014catalytic,korzekwa2015extraction}.
We introduce a non-Abelian reference frame
into the thermodynamic resource theory for noncommuting conserved charges.
Our agent's reference frame
carries a representation of the $G$ associated with the charges~\cite{kitaev2014super,BRS-refframe-review}.

The reference frame expands the set of allowed operations
from a possibly trivial set. 
A superselection rule restricts the free operations, as detailed below.
Every free unitary $U$ conserves (commutes with) each charge. 
The system charges $Q_{j_\sys}$ might not commute with each other.
In the worst case, 
the $Q_{j_\sys}$'s share no multidimensional eigensubspace.
The only unitary that conserves all such $Q_{j_\sys}$'s 
is trivial: $U  \propto  \id$.

A reference frame ``frees up'' dynamics,
enabling the system to evolve nontrivially.
A free unitary can fail to commute with a $Q_{j_\sys}$
while preserving $Q_{j_{\text{tot}}}$.
This dynamics transfers charges between the system and the reference frame.

Our agent's reference frame doubles as the battery.
The reference frame and battery are combined for simplicity, 
to reduce the number of subsystems under consideration.

%
%
%
%
\paragraph{Ancillas:}
The agent could manipulate extra subsystems, called ``ancillas.''
A list $(\rho_\anc,   H_\anc,   Q_{1_\anc},   \ldots,   Q_{c_\anc})$
specifies each ancilla $A$.
Any ancillas evolve cyclically under free operations.
That is, \GTO{} preserve the ancillas' states, $\rho_\anc$.
If \GTO{} evolved ancillas acyclically, the agent could ``cheat,''
extracting work by degrading an ancilla~\cite{brandao2013second}.

Example ancillas include catalysts.
A {catalyst} facilitates a transformation 
that could not occur for free in the catalyst's absence~\cite{brandao2013second}.
Suppose that a state 
$S  =  (\rho_\sys,  H_\sys,  Q_{1_\sys},  \ldots,  Q_{c_\sys})$ cannot transform into a state 
$\tilde{S}  =  (\tilde{\rho}_\sys,  \tilde{H}_\sys,  \tilde{Q}_{1_\sys},  \ldots,  \tilde{Q}_{c_\sys})$ 
by free operations:
$S  \not\mapsto \tilde{S}$.
Some state $X  =  (\rho_\cat,  H_\cat,  Q_{1_\cat},  \ldots,  Q_{c_\cat})$ might enable 
$S  \otimes  X  \mapsto  \tilde{S}  \otimes  X$
to occur for free.
Such a facilitated transformation is called a ``catalytic operation.''

%
%
%
%
\paragraph{\GTOlong{}:}
\GTO{} are the resource theory's free operations.
\GTO{} model exchanges of heat and of charges that might not commute with each other.
\begin{definition}
\label{definition:GTO}
Every \GTOlongSing{} (\GTO{}) consists of the following three steps.
Every sequence of three such steps forms a \GTO{}:
\begin{enumerate}
\item 
\label{free}
Any number of free states $(\rho_\bath, H_\bath, Q_{1_\bath},  \ldots,  Q_{c_\bath})$ can be added.

\item 
         \label{carry-out-U}
         Any unitary $U$ that satisfies the following conditions
         can be implemented on the whole system:
	\begin{enumerate} 
	\item $U$ preserves energy: $[U,H_{\text{tot}}]=0$. \label{econ}
	
	\item $U$ preserves every total charge: 
	         $[U,  Q_{j_{\text{tot}}}]=0  \;  \: \forall j = 1,  \ldots, c$. \label{qcon}
	         		
	\item Any ancillas return to their original states:
	         $\Tr_{\backslash A}  (U\rho_\mathrm{tot} U^\dagger)   =   \rho_\anc$.
	\end{enumerate}
	
\item Any subsystem can be discarded (traced out). \label{trace}

\end{enumerate}
\end{definition}

Conditions~\ref{econ} and~\ref{qcon} ensure that 
the energy and the charges are conserved.
The allowed operations are $G$-invariant,
or symmetric with respect to the non-Abelian group $G$.
Conditions~\ref{econ} and~\ref{qcon} do not significantly restrict the allowed operations,
if the agent uses a reference frame. 
Suppose that the agent wishes to implement, on $S$,
some unitary $U$ that fails to commute with some $Q_{j_\sys}$. 
$U$ can be mapped to a whole-system unitary $\tilde{U}$
that conserves $Q_{j_{\text{tot}}}$.
The noncommutation represents the transfer of charges to the battery,
associated with work.


The construction of $\tilde{U}$ from $U$
is described in~\cite{kitaev2014super}.
(We focus on the subset of free operations analyzed in~\cite{kitaev2014super}.)
Let $g, \phi  \in  G$ denote any elements of the symmetry group.
Let $T$ denote any subsystem (e.g., $T = S, W$).
Let $V_{\text{T}}(g)$ denote a representation,
defined on the Hilbert space of system $T$, of $g$.
Let $\ket{ \phi }_{\text{T}}$ denote a state of $S$
that transforms as the left regular representation of $G$:
$V_{\text{T}}(g)  \ket{\phi }_{\text{T}}  =  \ket{ g \phi }_{\text{T}}$.
$U$ can be implemented on the system $S$ of interest
by the global unitary
\begin{align}
   \tilde{U}  :=  \int  d \phi  \:
   \ketbra{\phi}{\phi}_\batt    \otimes
   [ V_\sys ( \phi )  \,  U  \,  V_\sys^{-1} (\phi) ].
  \label{eq:kitaev-unitary-construction}
\end{align}

The construction~\eqref{eq:kitaev-unitary-construction} does not increase 
the reference frame's entropy 
if the reference is initialized to $\ket{\phi=1}_\batt$.  
This nonincrease keeps the extracted work ``clean''~\cite{Skrzypczyk13,aaberg-singleshot,BrandaoHNOW14}. 
No entropy is ``hidden'' in the reference frame $W$.
$W$ allows us to implement the unitary $U$, 
providing or storing the charges 
consumed or outputted by the system of interest.

%
%
%
%
\subsection{A zeroth law of thermodynamics: Complete passivity of the \GGSlong{}}

Which states $\rho_\bath$ should the resource-theory agent access for free?
The free states are the only states from which
work cannot be extracted via free operations.
We will ignore $S$ in this section,
treating the reservoir $R$ as the system of interest.

%
%
%
%
\paragraph{Free states in the resource theory for heat exchanges:}
Our argument about noncommuting charges will mirror 
the argument about extracting work 
when only the energy is conserved.
Consider the thermodynamic resource theory for energy conservation.
Let $H_\bath$ denote the Hamiltonian of $R$. 
The free state $\rho_\bath$ has the form
$\rho_\bath = e^{-\beta H_\bath}/Z$~\cite{BrandaoHNOW14,YungerHalpernR14}.  
This form follows from the canonical ensemble's completely passivity 
and from the nonexistence of any other completely passive state.
Complete passivity was introduced in~\cite{PuszW78,Lenard78}.

\begin{definition}[Passivity and complete passivity]
Let $\rho$ denote a state governed by a Hamiltonian $H$.
$\rho$ is passive with respect to $H$
if no free unitary $U$ 
can lower the energy expectation value of $\rho$:
\begin{align}
   \label{eq:Passive}
   \not\exists  \, U  \:  :  \:
   \Tr \left(  U \rho U^\dag \, H  \right)
   <  \Tr  \left( \rho H \right).
\end{align}
That is, work cannot be extracted from $\rho$
by any free unitary.
If work cannot be extracted from any number $n$
of copies of $\rho$, $\rho$ is completely passive with respect to $H$:
\begin{align}
   \label{eq:ComPassive}
    \forall n = 1, 2, \ldots,  \quad
    \not\exists  \, U  \:  :  \:
   \Tr \left(  U \rho^{ \otimes n} U^\dag \, H  \right)
   <  \Tr  \left( \rho^{ \otimes n} H \right).
\end{align}
\end{definition}
\noindent A free $U$ could lower the energy expectation value
only if the energy expectation value
of a work-storage system increased.
This transfer of energy
would amount to work extraction.

Conditions under which $\rho$ is passive
have been derived~\cite{PuszW78,Lenard78}:
Let $\lbrace p_i\rbrace$ and $\lbrace E_i\rbrace$ denote the eigenvalues of $\rho$ and $H$. 
$\rho$ is passive if
\begin{enumerate}
   \item
   $[\rho,H]=0$ and 
   \item
   $E_i > E_j$ implies that $p_i \leq p_j$   for all $i,j$.
\end{enumerate}
One can check that $e^{ - \beta H_\bath } / Z$ is completely passive
with respect to $H_\bath$.

No other states are completely passive (apart from the ground state).
Suppose that the agent could access 
copies of some $\rho_0  \neq  e^{ - \beta H_\bath } / Z$.
The agent could extract work via thermal operations~\cite{brandao2013second}.
Free (worthless) states could be transformed into a (valuable) resource for free.
Such a transformation would be unphysical, 
rendering the resource theory trivial, in a sense.
(As noted in ~\cite{Lostaglio2015PRX_coherence}, if a reference frame is not allowed,
the theory might be nontrivial  in that 
creating superpositions of energy eigenstates would not be possible).


%
%
%
%
\paragraph{Free states in the resource theory of \GTOlong{}:}
We have reviewed the free states in the resource theory for heat exchanges.
Similar considerations characterize the resource theory for noncommuting charges $Q_j$.
The free states, we show, have the \GGS{}'s form.
If any other state were free,
the agent could extract work for free.

%
%
\begin{theorem} \label{zeroth}
%
There exists an $m > 0$ such that 
a \GTO{} can extract a nonzero amount of chemical work from
$(\rho_\bath)^{\otimes m}$
if and only if  
$\rho_\bath \neq e^{-\beta\,( H_\bath +\sum_j \mu_j  Q_{j_\bath} )}/Z$  
for some $\beta  \in  \mathbb{R}$.
\end{theorem}

\begin{proof}
We borrow from~\cite{pusz_passive_1978,Lenard78}
the proof that canonical-type states,
and only canonical-type states,
are completely passive.
We generalize complete passivity with respect to a Hamiltonian $H$
to complete passivity with respect to the work function $\workf$.

Every free unitary preserves every global charge. 
Hence the lowering of the expectation value
of the work function $\workf$ of a system
amounts to transferring work from the system to the battery:
\begin{align}
   \label{eq:Preserve}
   \Delta  \Tr (\workf_{ \batt }   \rho_{ \batt })   
   =   -  \Delta \Tr (\workf_\bath   \rho_\bath ).
\end{align}
Just as $e^{ - \beta H } / Z$ is completely passive with respect to $H$~\cite{pusz_passive_1978,Lenard78},
the \GGS{} is completely passive with respect to $\workf_{ \bath }$ for some $\beta$.

Conversely, if $\rho_\bath$ is not of the \GGS{} form, 
it is not completely passive with respect to $\workf_\bath$. 
Some unitary $U_{\bath^{\otimes m}}$ 
lowers the energy expectation value of $\rho_\bath^{\otimes m}$,
$\Tr    ( U_{\bath^{\otimes m}}   [\rho_\bath^{\otimes m} ]
        U_{\bath^{\otimes m}}^\dagger \workf_{\bath^{\otimes m}} )
< \Tr    ( \rho_{\bath}^{\otimes m}\workf_{\bath^{\otimes m}} )$,
for some great-enough $m$.
A joint unitary defined on $R^{\otimes m}$ and $W$ 
approximates $U_{R^{\otimes m}}$ well 
and uses the system $W$ as a reference frame
[Eq.~\eqref{eq:kitaev-unitary-construction}].  
This joint unitary conserves every global charge.  
Because the expectation value of $\workf_{\bath^{\otimes m}}$ decreases, 
chemical work is transferred to the battery.
\end{proof}

%

%
%
%
%
The \GGS{} is completely passive with respect to $\workf_\bath$
but not necessarily with respect to each charge $Q_j$.
The latter lack of passivity was viewed as problematic in~\cite{Imperial15}.
The lowering of the \GGS{}'s $\<Q_j\>$'s
creates no problems in our framework, 
because free operations cannot lower the \GGS{}'s $\<\workf\>$.
The possibility of extracting charge of a desired type $Q_j$,
rather than energy, is investigated also in~\cite{teambristol}.

For example,
let the $Q_j$'s be the components $J_j$ of the spin operator $\mathbf{J}$.
Let the $z$-axis point in the direction of $\boldsymbol{\mu}$, 
and let $\mu_z>0$:
\begin{align}
   \sum_{j = 1}^3   \mu_j  J_j
   \equiv   \mu_z  J_z.
\end{align}
The \GGS{} has the form
$\rho_\bath  =  e^{ - \beta (  H_{ \bath }  -  \mu_z J_{z_{ \bath }} ) } / Z$.
This $\rho_\bath$ shares an eigenbasis with $J_{z_{ \bath }}$.
Hence the expectation value of the battery's $J_x$ charge vanishes:
 $\Tr ( \rho_\bath   J_{x_{ \bath }} )   =   0$.
 A free unitary, defined on $R$ and $W$,
can rotate the spin operator that appears in the exponential of $\rho_\bath$.
Under this unitary, the eigenstates of $\rho_\bath$ 
become eigenstates of $J_{x_{ \bath }}$.
$\Tr ( J_x   \rho_\bath )$ becomes negative;
work appears appears to be extracted ``along the $J_x$-direction''
from $\rho_\bath$.
Hence the \GGS{} appears to lack completely passivity.
The unitary, however, extracts no chemical work: 
The decrease in
$\Tr(\rho_\bath J_{x_{ \bath }} )$ is compensated for by an increase in
$\Tr ( \rho_{ \bath } J_{z_{ \bath }} )$.

Another example concerns the charges $J_i$ and
$\rho_\bath  =  e^{ - \beta (  H_{\bath}  -  \mu_z J_{z_{ \bath }} ) } / Z$.
No amount of the charge $J_z$ can be extracted from $\rho_\bath$.
But the eigenstates of $-J_z$ are inversely populated:
The eigenstate $\ket{ z }$ 
associated with the low eigenvalue $-\frac{\hbar}{2}$ of $-J_z$ 
has the small population $e^{- \beta \hbar / 2}$.
The eigenstate $\ket{ -z }$
associated with the large eigenvalue $\frac{\hbar}{2}$ of $-J_z$
has the large population $e^{ \beta \hbar / 2}$.
Hence the charge $-J_z$ can be extracted from $\rho_\bath$.
This extractability does not prevent $\rho_\bath$ from being completely passive,
according our definition.
Only the extraction of $\mathcal{W}$ corresponds to chemical work.
The extraction of just one charge does not.

The interconvertibility of types of free energy
associated with commuting charges
was noted in~\cite{YungerHalpern14}.
Let $Q_1$ and $Q_2$ denote commuting charges, and let
$\rho_\bath   =   e^{-\beta(H_{ \bath }  -  \mu_1 Q_{1_{ \bath }} -  \mu_2 Q_{2_{ \bath }})}$.
One can extract $Q_1$ work at the expense of $Q_2$ work, 
by swapping $Q_1$ and $Q_2$ 
(if an allowed unitary implements the swap).

%
%
%
%
\subsection{\GTOlong{} preserve the \GGSlong{}.}
The \GGS{}, we have shown, is the only completely passive state.
It is also the only state preserved by \GTO{}. 
\begin{theorem} \label{gibbspreserving}
Consider the resource theory, defined by \GTO{}, associated with a fixed $\beta$.
Let each free state be specified by
$(\rho_{ \bath },   H_\bath,   Q_{1_\bath},   \ldots,   Q_{c_\bath})$,
wherein $\rho_{ \bath }  :=  e^{-\beta\,( H_\bath   -  \sum_{j = 1}^c   \mu_j   Q_{j_\bath}  )}/Z$.
Suppose that the agent has access to the battery, 
associated with the payoff function~\eqref{eq:SuppInf-workfunc}.
The agent cannot, at a cost of $\langle   \mathcal{W}   \rangle   \leq 0$, 
transform any number of copies of free states into any other state.
In particular, the agent cannot change the state's $\beta$ or $\mu_j$'s.
\end{theorem}

\begin{proof}
Drawing on Theorem~\ref{zeroth},
we prove Theorem~\ref{gibbspreserving} by contradiction.
Imagine that some free operation
could transform some number $m$ of copies
of $\nats := e^{-\beta\,(H_\bath   -   \sum_j\mu_j Q_{j_\bath})}/Z$
into some other state $\nats'$:
$\nats^{ \otimes m }  \mapsto  \nats'$.
($\nats'$ could have a different form from
  the \GGS{}'s.
  Alternatively, $\nats'$ could have the same form 
  but have different $\mu_j$'s or a different $\beta$.)
$\nats'$ is not completely passive.
Work could be extracted from 
some number $n$ of copies of $\nats'$,
by Theorem~\ref{zeroth}.
By converting copies of $\nats$ into copies of $\nats'$,
and extracting work from copies of $\nats'$,
the agent could extract work from $\nats$ for free.
But work cannot be extracted from $\nats$,
by Theorem~\ref{zeroth}.
Hence $\nats^{ \otimes m}$ must not be convertible into 
any $\nats' \neq \nats$,
for all $m = 1, 2, \ldots$.
\end{proof}

%
%
%
\paragraph{Second laws:}
Consider any resource theory defined by operations that preserve some state,
e.g., states of the form
$e^{-\beta\,( H_\bath   -  \sum_{j = 1}^c   \mu_j   Q_{j_\bath}  )}/Z$.
Consider any distance measure on states 
that is contractive under the free operations.
Every state's distance from the preserved state  $\rho_{ \bath }$
decreases monotonically under the operations. 
\GTO{} can be characterized with
any distance measure from $\rho_{ \bath }$
that is contractive under completely positive trace-preserving maps.
We focus on the R\'{e}nyi divergences,
extending the second laws developed in~\cite{brandao2013second}
for the resource theory for heat exchanges.

To avoid excessive subscripting, we alter our notation for the \GGS{}.
For any subsystem $T$,
we denote by $\gibbsParam{T}$ the \GGS{} relative 
to  the fixed $\beta$, to the fixed $\mu_j$'s,
and to the Hamiltonian $H_T$ and the charges $Q_{1_T}, \ldots, Q_{c_T}$
associated with $T$.
For example, 
$\gibbsSBatt  :=  e^{ - \beta [ (H_\sys  +  H_{ \batt }) 
+  \sum_{j = 1}^c  \mu_j  (Q_{j_\sys}  +  Q_{j_{ \batt }} ) ] }  / Z$
denotes the \GGS{} associated with the system-and-battery composite.

We define the generalized free energies 
\begin{equation}
   F_\alpha(\initial,   \gibbsS)   
   :=   k_{\text{B}} T  D_\alpha(\initial   \|\gibbsS)   -   \kB T  \log (Z).
   \label{eq:genfree}
\end{equation}
The classical R{\'e}nyi divergences $D_\alpha(\initial\|\gibbsS)$ are defined as 
\begin{equation}
   D_\alpha(\initial   \|   \gibbsS)
   := \frac{\sgn(\alpha)}{\alpha-1} \log 
   \left(   \sum_k    p_k^\alpha    q_k^{1-\alpha}   \right),
   \label{eq:renyidivergence-2}
\end{equation}
wherein $p_k$ and $q_k$ denote the probabilities 
of the possible outcomes
of measurements of the work function $\workf$
associated with $\initial$ and with $\gibbsS$.
The state $\initial$ of $S$ is
compared with the \GGS{} associated with $H_\sys$ and with the $Q_{j_\sys}$'s.

The $F_\alpha$'s generalize the thermodynamic free energy.
To see how, we consider transforming $n$ copies $(\initial)^{\otimes n}$ of a state $\initial$.
Consider the asymptotic limit, similar to the thermodynamic limit,
in which $n \to \infty$.
Suppose that the agent has some arbitrarily small, nonzero probability $\varepsilon$
of failing to achieve the transformation.
$\varepsilon$ can be incorporated into any $F_\alpha$ via ``smoothing''~\cite{brandao2013second}.
The smoothed $F^\varepsilon_\alpha$ per copy of $\initial$
approaches $F_1$ in the asymptotic limit~\cite{brandao2013second}:
\begin{align}
  \lim_{n\rightarrow\infty}   \frac{1}{n}   
   F^\varepsilon_\alpha &  \Big(  
        (\initial)^{\otimes n},    ( \gibbsS )^{\otimes n}    \Big)
   =   F_1  ( \initial )   \\ 
  &  =  \< H_\sys  \>_{\initial}   -   TS(\initial)   
       +   \sum_{j = 1}^c   \mu_j   \<Q_{j_\sys}\>.
\end{align}
This expression resembles the definition
$F  :=  E - TS  +  \sum_{j = 1}^c  \mu_j  Q_j$
of a thermodynamic free energy $F$.
In terms of these generalized free energies, we formulate second laws.

%
%
\begin{proposition}
  In the presence of a heat bath 
  of inverse temperature $\beta$ and chemical potentials $\mu_j$, 
  the free energies   $F_\alpha(\initial,   \gibbsS)$ 
  decrease monotonically:
  \begin{align}
     F_{\alpha}(\initial,\gibbsS) \geq F_{\alpha}(\final,\gibbsS)
     \; \: \forall \alpha\geq 0,
  \end{align}
  wherein $\initial$ and $\final$ denote the system's initial and final states.
If    
\begin{align}
   &   [\workf_\sys,   \final]   = 0   \quad {\text{and}}
   \nonumber \\ &  
   F_{\alpha}(\initial,  \gibbsS) \geq F_{\alpha}(\final,\gibbsS)
\;   \:   \forall   \alpha \geq 0,
\end{align}
some catalytic \GTO\ maps $\initial$ to $\final$.
\end{proposition}

The $F_\alpha(\initial,   \gibbsS)$'s are called ``monotones.''
Under \GTO{}, the functions cannot increase.
The transformed state approaches the \GGS{} or retains its distance.

Two remarks about extraneous systems are in order.
First, the second laws clearly govern operations
during which no work is performed on the system $S$.
But the second laws also govern work performance:
Let $SW$ denote the system-and-battery composite.
The second laws govern the transformations of $SW$.
During such transformations,
work can be transferred from $W$ to $S$.

Second, the second laws govern transformations
that change the system's Hamiltonian. 
An ancilla facilitates such transformations~\cite{HO-limitations}.
Let us model the change, via external control,
of an initial Hamiltonian $H_\sys$ into $H_\sys'$.
Let $\gibbsS$ and ${\gibbsS}'$ denote the \GGS{}s
relative to $H_\sys$ and to $H_\sys'$.
The second laws become
\begin{align}
   F_{\alpha}( \initial,   \gibbsS ) 
   \geq F_{\alpha}(  \final,   {\gibbsS}'  )
   \; \: \forall \alpha   \geq   0.
\end{align}

%
%
%
%
\paragraph{Extractable work:}
In terms of the free energies, we can bound the work 
extractable from a resource state via \GTO{}.
Unlike in the previous section,
we consider the battery $W$ separately from the system $S$ of interest.
We assume that $W$ and $S$ initially occupy a product state. 
(This assumption is reasonable for the idealised, infinite-dimensional battery
we have been considering.
As we will show, the assumption can be dropped when we focus on average work.)
Let $\initialBatt$ and $\finalBatt$
denote the battery's initial and final states.
For all $\alpha$,
\begin{align}
  F_\alpha(\initial   \otimes   \initialBatt,   \gibbsSBatt)
  \geq F_\alpha(\final   \otimes   \finalBatt,   \gibbsSBatt).
\end{align}
Since
$F_\alpha(   \initial   \otimes   \initialBatt,   \gibbsSBatt) 
= F_\alpha(\initial,   \gibbsS) +
F_\alpha   \left(\initialBatt,   \gibbsBatt   \right)$,
\begin{align}
     F_\alpha   \left( \finalBatt,   \gibbsBatt   \right)
     -   F_\alpha   \left( \initialBatt,   \gibbsBatt   \right)
      \leq 
     F_\alpha(\initial,   \gibbsS) -  F_\alpha(\final,   \gibbsS).
     \label{eq:work-alpha-free-energy}
\end{align}

If the battery states 
$\initialBatt$ and $\finalBatt$ 
are energy eigenstates,
the left-hand side of Ineq.~\eqref{eq:work-alpha-free-energy} 
represents the work extractable during one implementation of the protocol.
Hence the right-hand side  
bounds the  work extractable during the transition $\initial   \mapsto   \final$. 
This bound is a necessary condition
under which work can be extracted.

%
%

When $\alpha=1$, 
we need not assume that $W$ and $S$ occupy a product state.
The reason is that subadditivity implies
$F_1(\rho_{\text{SW}},\gamma_{\text{SW}})
\leq F_1(\rho_\sys,   \gamma_\sys) + F_1(\rho_{W},\gamma_\batt)$. 
$F_1$ is the relevant free energy 
if only the average work is important.


\paragraph{Quantum second laws:}
As in \cite{brandao2013second}, additional laws can be derived 
in terms of quantum R{\'e}nyi divergences~\cite{HiaiMPB2010-f-divergences,Muller-LennertDSFT2013-Renyi,WildeWY2013-strong-converse,JaksicOPP2012-entropy}. 
These laws provide extra constraints
if $\initial$ (and/or $\final$) has coherences 
relative to the $\workf_\sys$ eigenbasis.
Such coherences would prevent $\initial$ from commuting
with the work function.
Such noncommutation is a signature of truly quantum behavior.
Two quantum analogues of $F_\alpha(\initial,   \gibbsS)$ are defined as
\begin{equation}
   \qalfreesimple_\alpha(\initial,   \gibbsS)
   :=   k_{\text{B}} T \frac{{\text{sgn}}(\alpha)}{\alpha-1} 
        \log   \Big(  \Tr \left(\initial^\alpha (\gibbsS)^{1-\alpha} \right)   \Big)
        -   k_{\text{B}} T   \log (Z)
\end{equation}
and 
\begin{align}
   \qalfree_\alpha(\initial,   \gibbsS)   & :=
   \kB   T   \frac{1}{\alpha-1}\log 
   \left(   \Tr   \left( (\gibbsS)^{\frac{1-\alpha}{2\alpha}} 
             \initial (\gibbsS)^{\frac{1-\alpha}{2\alpha}}   \right)^\alpha \right) 
   \nonumber \\  & \qquad
   -   k_{\text{B}}T   \log (Z).
\end{align}
The additional second laws have the following form.

\begin{proposition} 
\GTO{} can transform $\initial$ into $\final$ only if
\begin{align}
   & \qalfree_\alpha(\initial,\gibbsS) 
   \geq \qalfree_\alpha(\final,\gibbsS)
   \quad  \forall    \alpha   \geq \frac12,
   \\ 
   & \qalfree_\alpha(\gibbsS , \initial) \geq \qalfree_\alpha(\gibbsS,   \initial)
   \quad   \forall   \alpha   \in   \left[   \frac12, 1 \right], 
   \quad {\text{and}}  
   \\ 
   & \qalfreesimple_\alpha(\initial, \gibbsS) 
   \geq \qalfreesimple_\alpha(\final,\gibbsS)
   \quad  \forall  \alpha  \in  [0,  2] .
\end{align}
\end{proposition}
\noindent These laws govern transitions during which the Hamiltonian changes via an ancilla, 
as in~\cite{HO-limitations}.

\endgroup

\putbib[NatCommsRefs.bibolamazi] 
\end{bibunit}

%
%
%



\end{document}